# SEMICONDUCTOR SPINTRONICS

**Jaroslav Fabian,**[1,a] **Alex Matos-Abiague**[a], **Christian Ertler**[a], **Peter Stano,**[2,a] **Igor Žutić**[b]
[a] *Institute for Theoretical Physics, University of Regensburg, 93040 Regensburg, Germany*
[b] *Department of Physics, State University of New York at Buffalo, Buffalo NY, 14260, USA*

Spintronics refers commonly to phenomena in which the spin of electrons in a solid state environment plays the determining role. In a more narrow sense spintronics is an emerging research field of electronics: spintronics devices are based on a spin control of electronics, or on an electrical and optical control of spin or magnetism. While metal spintronics has already found its niche in the computer industry—giant magnetoresistance systems are used as hard disk read heads—semiconductor spintronics is yet to demonstrate its full potential. This review presents selected themes of semiconductor spintronics, introducing important concepts in spin transport, spin injection, Silsbee-Johnson spin-charge coupling, and spin-dependent tunneling, as well as spin relaxation and spin dynamics. The most fundamental spin-dependent interaction in nonmagnetic semiconductors is spin-orbit coupling. Depending on the crystal symmetries of the material, as well as on the structural properties of semiconductor based heterostructures, the spin-orbit coupling takes on different functional forms, giving a nice playground of effective spin-orbit Hamiltonians. The effective Hamiltonians for the most relevant classes of materials and heterostructures are derived here from realistic electronic band structure descriptions. Most semiconductor device systems are still theoretical concepts, waiting for experimental demonstrations. A review of selected proposed, and a few demonstrated devices is presented, with detailed description of two important classes: magnetic resonant tunnel structures and bipolar magnetic diodes and transistors. In view of the importance of ferromagnetic semiconductor materials, a brief discussion of diluted magnetic semiconductors is included. In most cases the presentation is of tutorial style, introducing the essential theoretical formalism at an accessible level, with case-study-like illustrations of actual experimental results, as well as with brief reviews of relevant recent achievements in the field.

[1]E-mail address: jaroslav.fabian@physik.uni-regensburg.de

[2]Currently at the Research Center for Quantum Information, Institute of Physics, Slovak Academy of Sciences, Bratislava, Slovakia.

## Contents

# I. Introduction

## A. Semiconductor spintronics

In a narrow sense spintronics refers to spin electronics, the phenomena of spin-polarized transport in metals and semiconductors. The goal of this applied spintronics is to find effective ways of controlling electronic properties, such as the current or accumulated charge, by spin or magnetic field, as well as of controlling spin or magnetic properties by electric currents or gate voltages. The ultimate goal is to make practical device schemes that would enhance functionalities of the current charge-based electronics. An example is a spin field-effect transistor, which would change its logic state from ON to OFF by flipping the orientation of a magnetic field.

In a broad sense spintronics is a study of spin phenomena in solids, in particular metals and semiconductors and semiconductor heterostructures. Such studies characterize electrical, optical, and magnetic properties of solids due to the presence of equilibrium and nonequilibrium spin populations, as well as spin dynamics. These fundamental aspects of spintronics give us important insights about the nature of spin interactions—spin-orbit, hyperfine, or spin exchange couplings—in solids. We also learn about the microscopic processes leading to spin relaxation and spin dephasing, microscopic mechanisms of magnetic long-range order in semiconductor systems, topological aspects of mesoscopic spin-polarized current flow in low-dimensional semiconductor systems, or about the important role of the electronic band structure in spin-polarized tunneling, to name a few.

Processes relevant for spintronics are summarized in Fig. I.1. All three processes are equally important, though the hierarchy starts naturally with spin injection, as a way to introduce nonequilibrium spin into a conductor. If you take a piece of iron and aluminum, connect the two in series and make electrical current flow through them, you have likely achieved electrical spin injection, see Fig. I.1 top. If electrons flow from the iron, where most electrons are spin polarized (there are more spin up, say, than spin down electrons), to the aluminum, the spin is accumulated in aluminum, the result of spin injection. If the current is reversed and electrons flow from the aluminum into the iron, the spin is taken from the aluminum and we speak of spin extraction.[3] We understand these processes reasonably well, at least for the most studied cases of highly degenerate charge-neutral electronic systems. In non-degenerate semiconductors, for example, spin injection may be absent due to space charges and electron population statistics. What we call the standard theory of spin injection, as well as of spin transport and spin-dependent tunneling is presented in detail in this text.

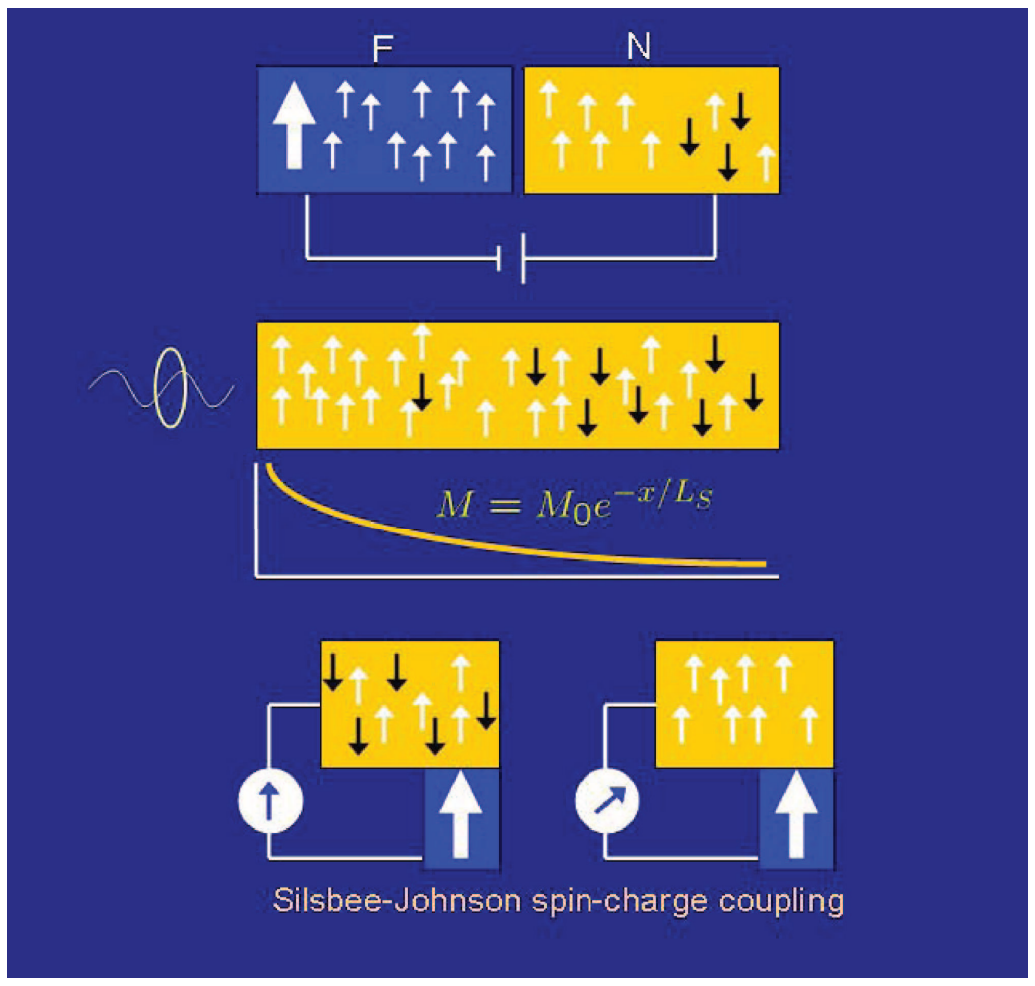

Fig. I.1. Successful spintronics applications need to satisfy three basic requirements: efficient spin injection or spin generation (top), whereby spin is injected from (here) a ferromagnetic into a nonmagnetic conductor, reasonably long spin (magnetization, $M$) diffusion, at least tens of nanometers, and possibility of efficient spin manipulation (middle), and, finally, spin detection, here illustrated by the Silsbee-Johnson spin-charge coupling. Spin detection, if performed by spin-to-resistance conversion, is at the heart of spintronics devices.

Once the spin is injected, we need to manipulate it or control it. This is usually achieved by applying an external magnetic field to rotate the spin, although the presence of spin-orbit coupling allows one to control spin electronically. Indeed, the spin-orbit coupling in semiconductor heterostructures can be tailored by voltage gates on the top of the heterostructures, allowing to control the spin by voltage. We still need to find practical ways to do that; understanding the spin-orbit interactions is crucial. This article present detailed derivations of the effective Hamiltonians describing the spin-orbit interactions in the most studied classes of semiconductors and their heterostructures—the so-called Dresselhaus and Bychkov-Rashba Hamiltonians.

The injected spin has to survive sufficiently long, and travel sufficiently far, to transfer information between the injected point and the point of detection. The transfer is inhibited by irreversible processes of spin relaxation and spin dephasing. These processes arise due to the combined actions of the spin-orbit interaction and momentum relaxation. The former provides spin flips or spin rotations, the latter gives irreversible time evolution. The interaction of spin with a solid-state environment is a complex process whose description relies on effective perturbative approximations. Such a formalism is introduced here, along with the most relevant spin relaxation mechanisms in semiconductors and in important classes of tailored semiconductor superstructures–lateral quantum dots which are potentially important for spin-based quantum information processing.

Finally, the spin has to be detected. Even if you pass current from the aluminum to the iron, you have to prove that spin-polarized electrons indeed accumulate in the aluminum. This is a highly nontrivial task. In Fig. I.1 the detection scheme is based on the Silsbee-Johnson spin-charge coupling. This coupling is the inverse of the spin injection. In a spin injection electrical current drives spin-polarized electrons from a ferromagnetic metal to a nonmagnetic conductor. In a spin-charge coupling an electrical contact between a ferromagnet and a nonmagnetic conductor containing a nonequilibrium spin population results in electrical current (or electromotive force in an open circuit). The presence of the electron spin can then be detected electrically. Other frequently encountered ways of detecting spin include a spin-valve effect, in which the injected spin-polarized electrons enter a detecting ferromagnetic electrode with an efficiency given by the relative orientation of the injecting and detecting electrodes, or optical detection in which spin-polarized electrons recombine with unpolarized holes and emit circularly polarized light

[3] Similar statements should be always taken with caution; in real materials much depends on the specific electronic band structure as well as on the properties of the interface. Extraction may be masked by spin accumulation due to reflection from the ferromagnet, for example.

which can be analyzed.

This tutorial style review presents the mainstream knowledge of semiconductor spintronics. However, we had to omit many important developments as the field is growing at enormous rates for one review to cover it all. Below we give two experimental discoveries, both defining the current state-of-the-art, whose underlying physics is not discussed in the article, but which are inspirational in demonstrating how much new fundamental spin physics has been learned from investigating spin-polarized transport in semiconductors.

Fascinating fundamental discoveries have been made relating to what is now called the spin Hall effect (or, rather, effects). This effect was proposed decades ago by D'yakonov and Perel' (1971a), who suggested that passing en electrical current through a conductor will result in a spin accumulation at the edges of the conductor transverse to the current flow, due to spin-dependent scattering off impurities (Mott scattering). Because of the spin-orbit coupling induced either by the impurities or by the host lattice, electrons with a drift velocity along the sample scatter preferably left if, say, their spin is up, and right, if their spin is down. The difference in the scattering probabilities for the two spin orientations is typically small, say, 10 ppm, but even this small difference leads to spin currents transverse to the electron drift motion. In a finite sample, the spin currents at the edges need to be balanced by opposing diffusive currents, which can be set up if there is spin accumulation at the edges, forming a gradient of the spin density.

The experimental discovery of this effect was reported by Kato *et al.* (2004) and Wunderlich *et al.* (2005). We present the experiment of Kato *et al.* (2004) in Fig. I.2. The sample is a GaAs slab lightly doped with silicon donors. Electrical current flows along the sample, subject to the electric field $E = 10$ mV $\mu\text{m}^{-1}$, directed from bottom up. The spatially resolved magnetization of the sample is detected by the magneto-optical scanning Kerr spectroscopy, with a micron resolution. The measurements were performed at 30 K. As is seen from Fig. I.2, electron motion in one direction leads to transverse spin accumulation, as predicted by D'yakonov and Perel'. While the observed spin polarization is rather small, below 0.01%, this beautiful experiment presents a fundamental discovery about the nature of the coupling of spin and charge motion in electronic systems. A popular account of the spin Hall effect can be found in (Sih *et al.*, 2005).

The above experiment demonstrates what is now called the extrinsic spin Hall effect, which is due to the spin-orbit scattering by impurities as well as due to the nonequilibrium electronic population set up to give the electric current. Study of another class of spin Hall effects, called intrinsic[4], was initiated by Murakami *et al.* (2003). The intrinsic spin Hall effect relies on the spin-orbit description of the underlying band structure and results from a spin-dependent deformation of the electron wave functions due to the electric field which gives the electric current. While the extrinsic spin Hall effect disappears in the absence of impurities (the clean system limit), the intrinsic spin Hall effect is still present. Despite being essentially a single-electron phenomenon, the spin Hall effect has attracted wide theoretical attention. We refer the reader to recent review articles for more details (Schliemann, 2006; Engel *et al.*, 2007).

The other example we present is again about generating spin flows, albeit by different mechanisms, depicted in Fig. I.3. In certain classes of semiconductors the crystal symmetry allows coupling of axial and polar vectors (such systems, the prominent example is GaAs, are also called gyrotropic[5]). Such two vectors are spin and current, or spin and momentum, for axial and polar, respectively. If the two can be coupled in a linear way, several fascinating phenomena result, known as the spingalvanic effects (Ganichev *et al.*, 2001; Ganichev and Prettl, 2003; Ganichev *et al.*, 2002). When a THz photon is absorbed by a gyrotropic system, the absorption probability depends on both the spin and the momentum (Ganichev *et al.*, 2006). In Fig. I.3 a, the spin up electron has a higher probability of being excited by the photon to end up with a positive momentum, than with a negative one; the necessary momentum conservation is facilitated by phonons. As a result the excited spin up electrons prefer to move to the right. On the contrary, excited spin down electrons prefer moving to the left, as required by time reversal symmetry (reversing both spin and momentum leads to the same result). The net result is a pure spin current, with no net charge current flowing. This effect can be observed indirectly[6] by applying a small magnetic field to give a tiny imbalance between spin up and spin down electrons, transforming the spin into a charge current. The experimental observation of this charge current is shown in Fig. I.4, demonstrating what is called zero-bias spin separation (Ganichev *et al.*, 2006) stressing that no applied voltage is necessary to drive pure spin currents, separating spin up and spin down electrons. Similar effects arise from the spin-dependent energy relaxation phenomena, illustrated in Fig. I.3 b. Simply heating up the electron gas (keeping the lattice temperature lower so that energy relaxation occurs) in a uniform gyrotropic system with no magnetic fields applied and with no electric currents flowing results in a pure spin current.

[4]The terms extrinsic and intrinsic SHE were introduced by Sinova *et al.* (2004).

[5]A real life example is a bicycle ride or the action of a corkscrew, in which torque results in a linear momentum.

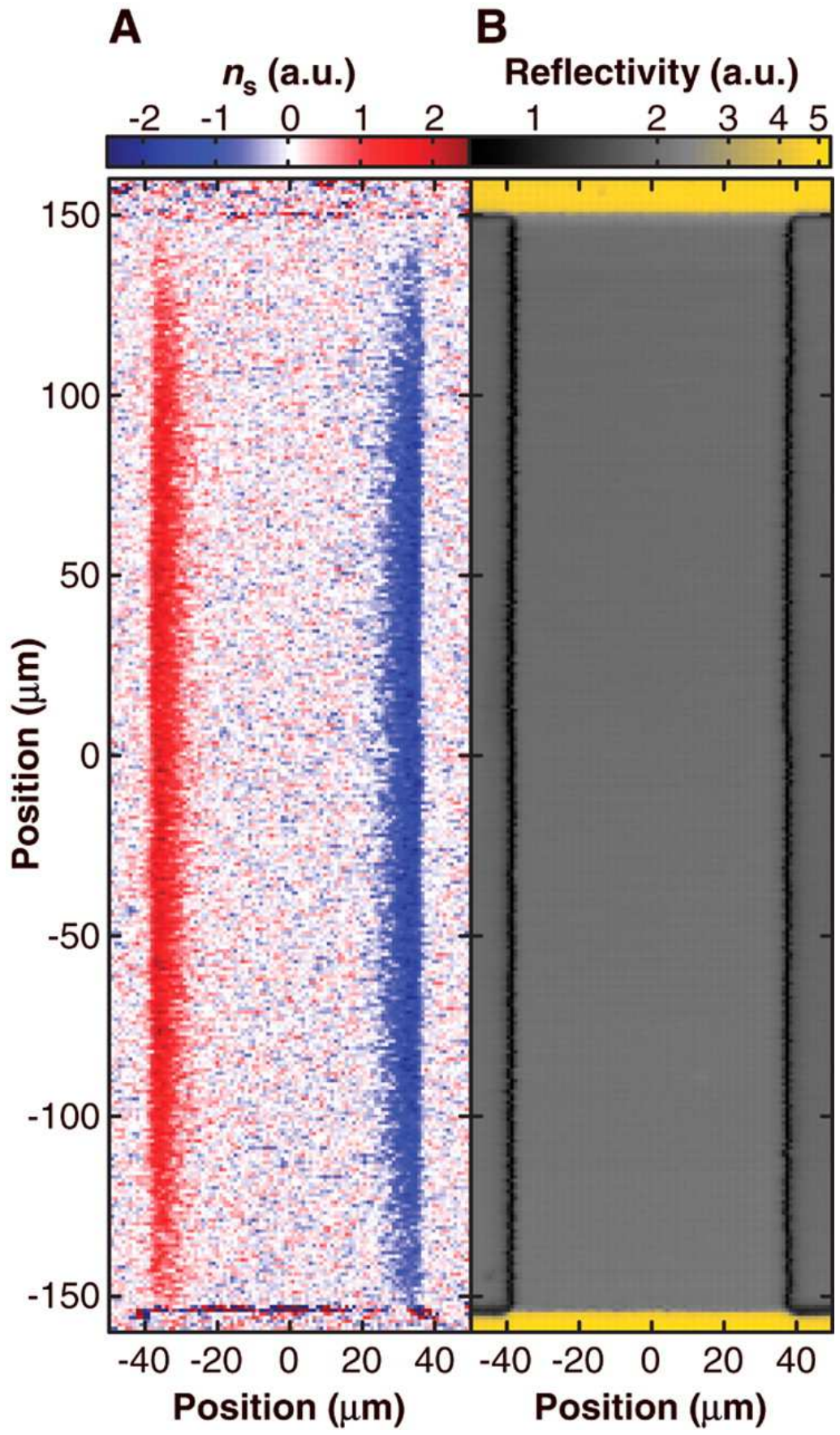

Fig. I.2. (A) Two-dimensional image, obtained by magneto-optic Kerr spectroscopy, of the spin polarization at the edges of a GaAs sample. Red is for positive spin (out or the page), blue for negative (Sih *et al.*, 2005). (B) Spatially resolved reflectance, showing the edges of the sample. The yellow metal contacts are also visible. From Y. K. Kato *et al., Science* **306**, *1910 (2004). Reprinted with permission from AAAS.*

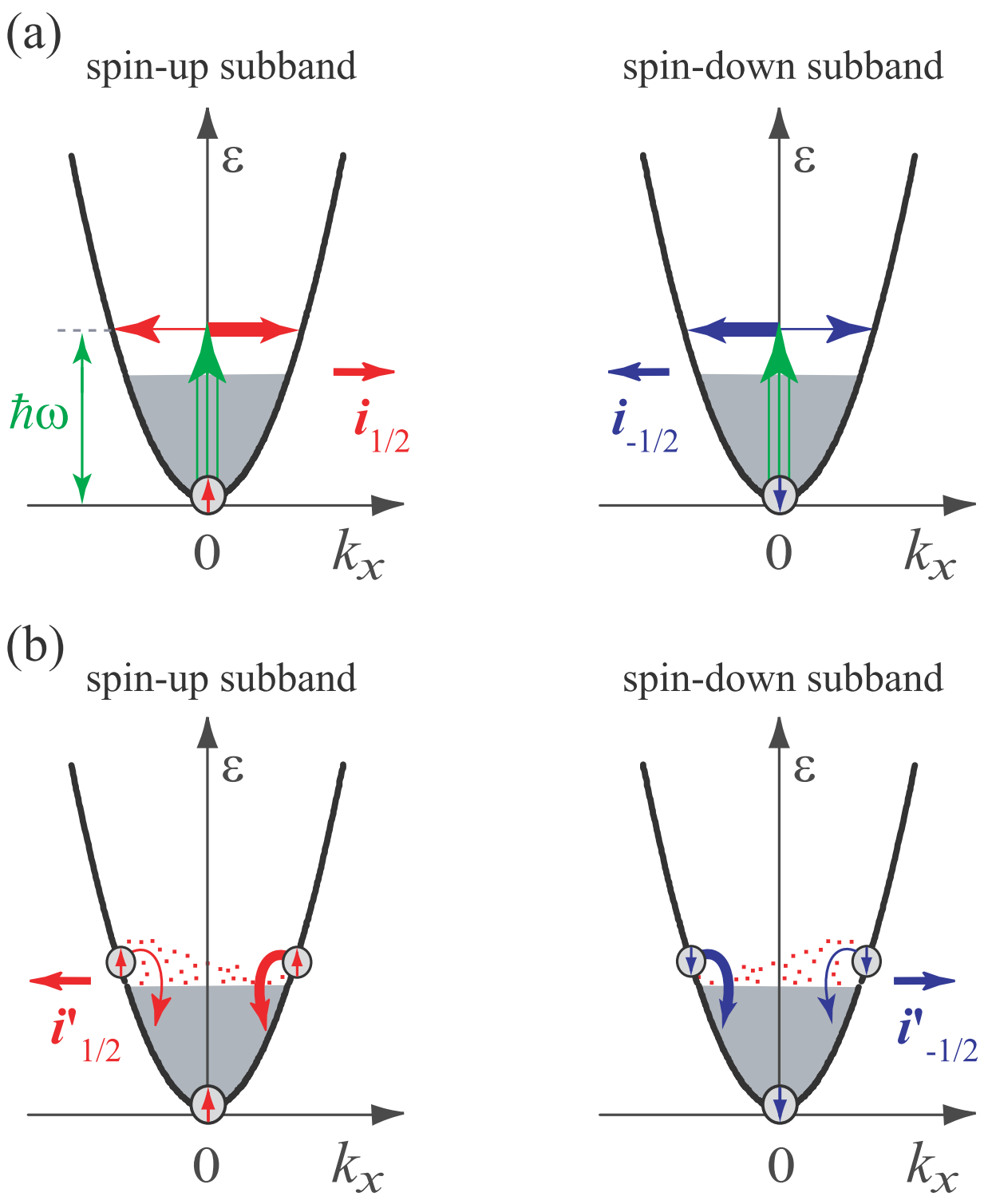

Fig. I.3. Where spin current comes from in zero-bias spin separation. (a) Spin-dependent absorption of THz radiation. Electrons with spin up are more likely to be excited (here) into the positive momentum states, spin down electrons are more likely to go into negative momentum states. (b) Spin-dependent electron thermalization. Excited electrons emit phonons, to equilibrate with the colder lattice. Spin up electrons of a positive momentum lose energy (and momentum) faster than those of a negative momentum; opposite is true for spin down electrons. Both (a) and (b) result in pure spin currents. Reprinted by permission from Macmillan Publishers Ltd: *Nature. S. D. Ganichev et al., Nature Physics* **2**, *609 (2006), copyright 2007.*

[6]Unlike electric current which is directly observable, we do not have means to detect spin current, only spin polarization which is manifested by magnetization, for example.

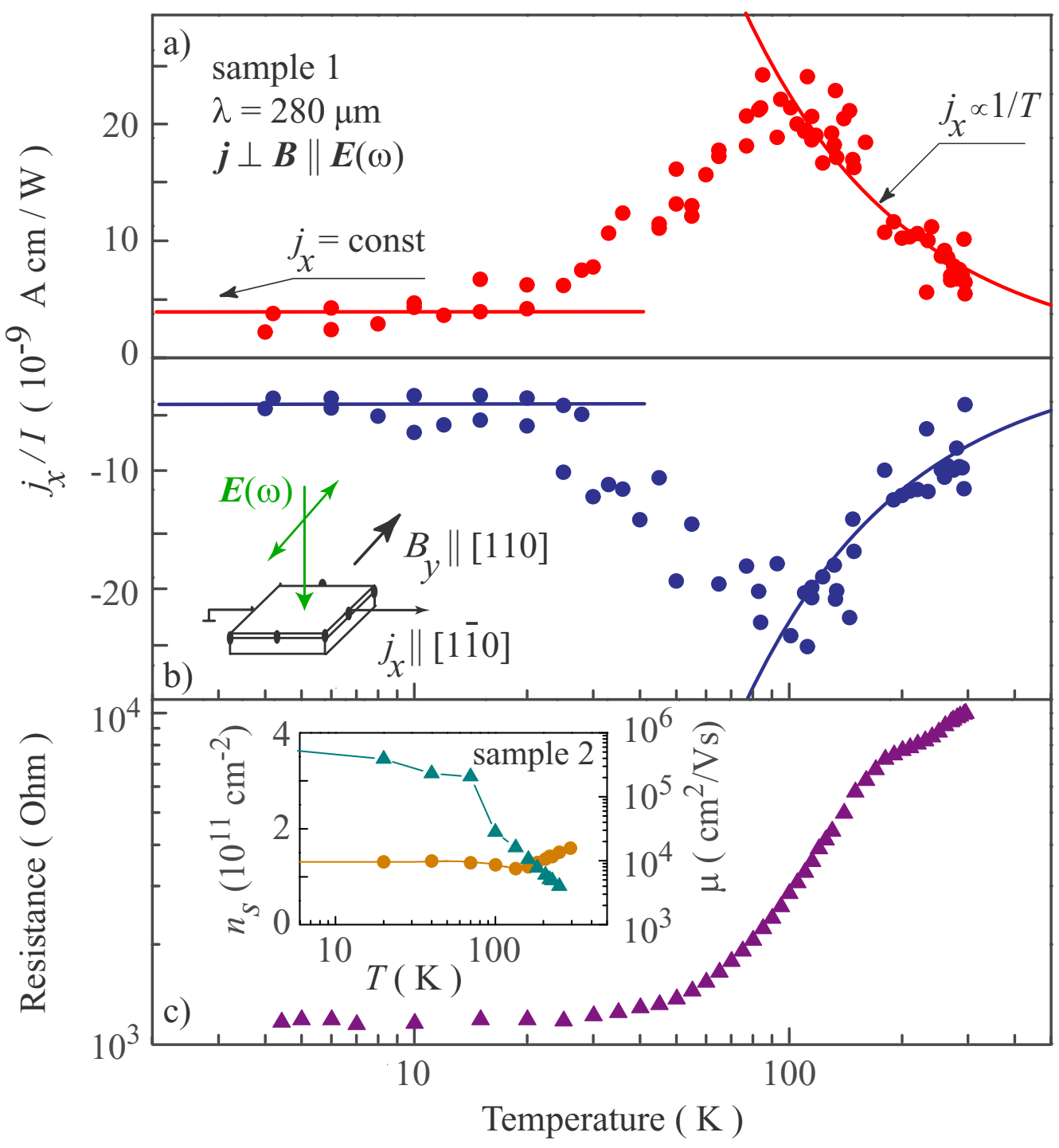

Fig. I.4. Measured charge current due to THz radiation excitation of a GaAs sample, in the presence of a spin-polarizing magnetic field along $y$; the scheme and the definition of directions are in (b). The charge current versus temperature is shown in (a) and (b), while the resistance versus temperatures is plotted in (c); the inset there shows the carrier density, $n_s$, and mobility, $\mu$. The magnetic field, $B_y$, in (a) has the opposite sign from that in (b), resulting in current reversal. Reprinted by permission from Macmillan Publishers Ltd: *Nature*. *S. D. Ganichev et al., Nature Physics* **2**, *609 (2006), copyright 2007.*

There have been several other novel fundamental physical phenomena discovered in the course of investigations of spin transport, not covered in detail in this text. Apart from the above mentioned spin Hall effects and spingalvanic phenomena, the list would include spin transfer torque (Brataas *et al.*, 2006), spin coherent transport and dynamics in low-dimensional mesoscopic semiconductor systems (Zaitsev *et al.*, 2005; Bardarson *et al.*, 2007; Nikolic *et al.*, 2005, 2006; Smirnov *et al.*, 2007), spin-dependent quantum interference effects in Aharonov-Bohm rings (Frustaglia *et al.*, 2001; Frustaglia and Richter, 2004; Hentschel *et al.*, 2004; Frustaglia *et al.*, 2004; Nikolic *et al.*, 2005; Souma and Nikolic, 2005, 2004; Mal'shukov *et al.*, 2002), observed experimentally in (König *et al.*, 2006), Zitterbewegung of conduction electrons (Schliemann *et al.*, 2005, 2006), spin ratchets (Scheid *et al.*, 2006; Pfund *et al.*, 2006), or the spin Coulomb drag effect (D'Amico and Vignale, 2000, 2003; Weber *et al.*, 2005; Tse and Das Sarma, 2007; Badalyan *et al.*, 2007).

The present text, which is part tutorial and part review, relies heavily on the comprehensive review (Žutić *et al.*, 2004), which should serve as the complementary reference; many important works omitted here are described in that reference. For nice popular accounts of spintronics we refer the reader to (Das Sarma, 2001; Awschalom and Kikkawa, 1999; Awschalom *et al.*, 2002); spintronics perspectives can be found in various theme articles (Fabian and Das Sarma, 1999b; Das Sarma *et al.*, 2000c,b,a, 2003a, 2001; Žutić *et al.*, 2007; Awschalom and Flatté, 2007; Wolf *et al.*, 2001; Rashba, 2006; Ohno, 2002; Johnson, 2005; Žutić *et al.*, 2006a; Flatté, 2007).

As part of the Introduction, we recall below the elementary physics of the electron spin, as well as introduce two not widely known gedanken (thought) experiments, by the founders of quantum mechanics, about the impossibility of measuring directly the spin of a free electron, and on the impossibility of performing a Stern-Gerlach experiment with electron beams. Although such arguments may not be appreciated by many as being perhaps too vague to be valid in general, they are intellectually appealing for pointing out fundamental underlying physics and should be a standard knowledge in spin physics.

## B. The spin magnetic moment of a free electron

The spin of a free electron gives rise to a magnetic moment, opposite to the spin direction, of magnitude:

$$\mu = \frac{g_0}{2}\mu_B, \tag{I.1}$$

where $g_0$ is the so-called electron g-factor and $\mu_B = e\hbar/2m$ is the Bohr magneton; $m$ is the electron mass. The g-factor is

$$g_0 = 2(1 + \frac{\alpha}{2\pi} + \ldots) \approx 2.0023. \tag{I.2}$$

The value of 2 comes from the Dirac equation, while the rest is the so-called anomalous contribution, proportional to the fine structure constant, $\alpha = e^2/\hbar c 4\pi\epsilon_0$, arising from quantum electrodynamic corrections. In solids, conduction electrons can have the g-factor very different from the free electron case. In simple metals the deviations are not very dramatic (usually less than a percent), but in semiconductors the values can be an order of magnitude larger ($g \approx -50$ in InSb[7]) or smaller ($g \approx -0.44$ in GaAs); g-factor can even approach zero in specially engineered semiconductor heterostructures. The g-factors of conduction electrons are strongly affected by the spin-orbit interaction due to the lattice ions.

Since we often learn about the electron spin via the spin magnetic moment, it is instructive to bring forward two thought experiments, due to Bohr, Pauli, and Mott, as presented in (Mott, 1929; Mott and Massey, 1965), on the impossibility of detecting spin magnetic moments of free electrons (as opposed to electrons confined, say, to atomic shells). While the arguments are inspirational, they should be taken with a grain of salt. Indeed, the magnetic moment of a free electron has been measured, to great precision [see, for example, (Van Dyck *et al.*, 1986)], while the generality of the thought experiments has been questioned (Batelaan *et al.*, 1997; Garraway and Stenholm, 1999). Nevertheless, it appears impractical to perform a Stern-Gerlach experiment with electron beams. It was even remarked that, “Such attempts have the same challenges as “thought” experiments for constructing perpetual-motion machines.” (Kessler, 1985).

[7] The negative value of the g-factor means that the magnetic moment of the electron is parallel (as opposed to antiparallel for free electrons) to the spin direction.

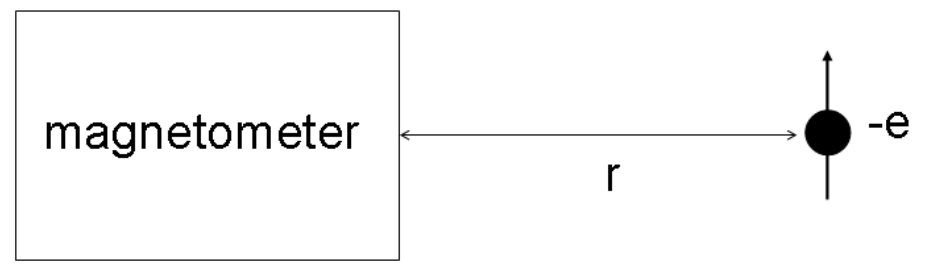

Fig. I.5. An electron moving with velocity $v$ at a distance $r$ from the magnetometer.

It remains to be seen if Stern-Gerlach experiments are also prohibited with conduction electrons in solids. It has been proposed that a transient Stern-Gerlach-like spin separation could be observed in a drift-diffusive electronic motion in a realistic metal or semiconductor (Fabian and Das Sarma, 2002), or that spin separation can be engineered in a ballistic, two-dimensional electron gas (Wrobel *et al.*, 2001, 2004).

We now present the two arguments, the first on the impossibility of measuring the spin magnetic moment of a free electron, the second on the impossibility to perform a Stern-Gerlach experiment with electrons. The common theme in both is that, if the motion of electrons can be described by trajectories, the effects of the electron spin are masked by the cyclotron motion due to the Lorentz force.

### B.1 Can the spin magnetic moment of a free electron be detected?

Let an electron move with velocity $v$ relative to a magnetometer placed at a distance $r$ from the electron, as in Fig. I.5. The magnetometer detects the magnetic field due to both the electron spin magnetic moment and due to the electron's orbital motion. The spin magnetic dipolar moment gives rise, at the place of the magnetometer, to the field of magnitude,

$$B_{\text{spin}} \approx \frac{\mu_0}{4\pi}\frac{\mu_B}{r^3}, \tag{I.3}$$

while the orbital contribution gives, [8]

$$B_{\text{orb}} \approx \frac{\mu_0}{4\pi}\frac{ev}{r^2}. \tag{I.5}$$

We have denoted by $\mu_0$ the permeability of free space. Since $\mu_B = e\hbar/2m$, the orbital contribution can be rewritten as

$$B_{\text{orb}} \approx \frac{\mu_0}{4\pi}\mu_B\frac{2p/\hbar}{r^2}, \tag{I.6}$$

[8] This estimate can be obtained from the Biot-Savart law for the magnetometer at $\mathbf{r} = 0$,

$$\mathbf{B}(0) = \frac{\mu_0}{4\pi}\int d^3\mathbf{r}'\mathbf{J}(\mathbf{r}') \times \left(-\frac{\mathbf{r}'}{r'^3}\right), \tag{I.4}$$

by substituting $\mathbf{J}(\mathbf{r}') = -e\mathbf{v}\delta(\mathbf{r}' - \mathbf{r})$ for the current density at point $\mathbf{r}'$ due to the electron at $\mathbf{r}$, and assuming that the velocity is perpendicular to the radius vector connecting the magnetometer and the electron. Here $\delta(\mathbf{r})$ is the Dirac delta-function.

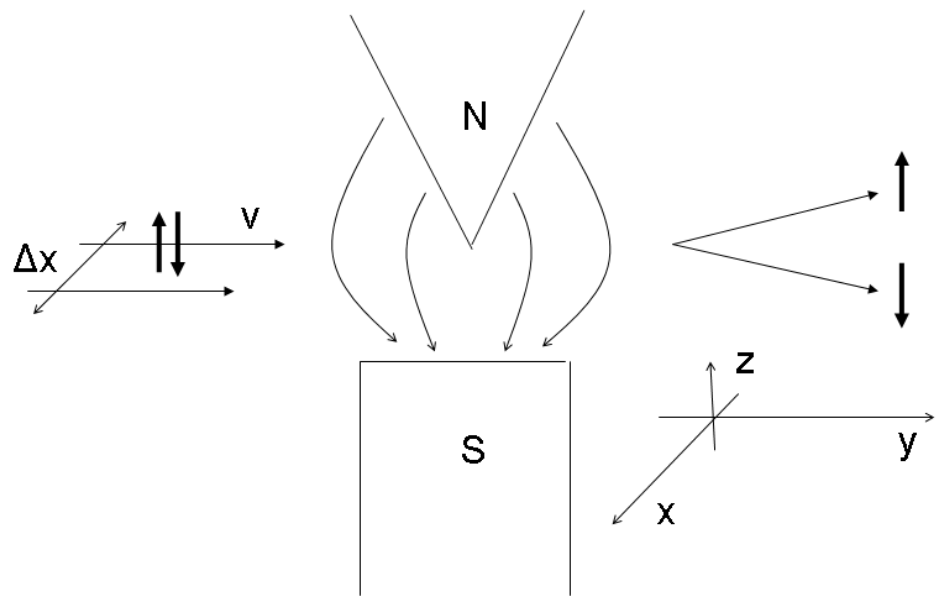

Fig. I.6. Scheme of a Stern-Gerlach apparatus for spatially separating spin up and spin down electrons. The magnetic field and its gradient are in the z-direction. The electron beam has necessarily a final width $\Delta x$ transverse to the beam's velocity $v$.

where $p = mv$ is the electron's momentum.

If we can measure precisely both the momentum $p$ and the distance $r$, we can separate the two contributions to the magnetic field and determine the value of the Bohr magneton, which is the measure of the spin magnetic moment. If the uncertainty in the measurement of $r$ is $\Delta r$, the Heisenberg principle restricts the uncertainty of the momentum to at least $\Delta p \approx \hbar/\Delta r$. In order to obtain $\mu_B$ from the resolved $B_{\rm spin}$, we need to know $r$ with the precision $\Delta r \ll r$. This restriction leads to the main uncertainty in $B_{\rm orb}$ as,

$$\Delta B_{\rm orb} \approx \frac{\mu_0}{4\pi}\mu_B \frac{2\Delta p/\hbar}{r^2} \gtrsim \frac{\mu_0}{4\pi}\mu_B \frac{2}{r^2 \Delta r}. \tag{I.7}$$

Since $r$ needs to be known precisely,

$$\Delta B_{\rm orb} \gtrsim B_{\rm spin}, \tag{I.8}$$

that is, the uncertainty in measuring the orbital contribution to the magnetic field of a moving electron is larger than the spin contribution itself. We cannot resolve $B_{\rm spin}$ and detect the spin magnetic moment $\mu_B$, for a free electron. We invite the reader to present arguments how the above reasoning changes if electrons are confined.

### B.2 Can Stern-Gerlach experiments be used to polarize electron beams?

The argument now is only a bit more subtle than the previous reasoning about the impossibility to detect the spin magnetic moment of a free electron. In essence the uncertainty relation leads to an unexpected strong contribution of the Lorentz force which masks the spin splitting of the electron beam.

Take a beam of spin unpolarized electrons arriving at an opening in a magnet in which there is a gradient of the magnetic field. Let the direction of motion be $y$, the direction of the quantizing magnetic field $z$, while $x$ be the transverse direction to both. The scheme of the apparatus is in Fig. I.6.

We have a magnetic field $B_z$ in the z-direction, with the spatial derivative $\partial B_z/\partial z$. Since the free electron has its magnetic moment opposite to its spin, the spin up, $s = 1/2$, electrons would prefer to go in the direction of increasing $B_z$, while the spin down, $s = -1/2$, electrons prefer to go in the opposite direction, leading to spin separation. The spin-dependent force $F_z$ acting on the electrons in the z-direction due to the magnetic field gradient is,

$$F_{\rm spin,z} = -\mu_B g_0 s \left(\frac{\partial B_z}{\partial z}\right). \tag{I.9}$$

If this were the whole story, the free electrons arriving at the magnet would split into two beams, depending on their spin, as is observed with beams of neutral atoms with unpaired electron spins.

It is the electron charge that prohibits observation of spin splitting of an electron beam. The electrons moving with velocity $\mathbf{v}$ in magnetic field $\mathbf{B}$ feel the orbital Lorentz force,

$$\mathbf{F}_{\rm orb} = -e\mathbf{v} \times \mathbf{B}. \tag{I.10}$$

If the magnetic field were in the z-direction only, the orbital force would affect the motion transverse to the spin-splitting direction, not masking the spin separation. However, the presence of the magnetic field gradient necessitates the presence of another component of the magnetic field, perpendicular to the z-direction. Indeed, any magnetic field is sourceless: $\nabla \cdot \mathbf{B} = 0$. If we suppose that $B_y = 0$ (taking a symmetric magnet, for example), we then need that,

$$\frac{\partial B_x}{\partial x} = -\frac{\partial B_z}{\partial z}. \tag{I.11}$$

At a small distance $x$ away from the center of the electron beam, the x-component of the magnetic field is then roughly

$$B_x(x) \approx -\frac{\partial B_z}{\partial z} x \tag{I.12}$$

The electron beam has necessarily a finite transverse width, $\Delta x$; otherwise the uncertainty in the transverse velocity would be infinite. The electrons at the edges of the beam feel the Lorenz force due to $B_x(\Delta x)$, which pushes those electrons along the z-direction:

$$F_{{\rm orb},z} = ev_y B_x(\Delta x) \approx -ev_y \frac{\delta B_z}{\partial z} \Delta x. \tag{I.13}$$

The electrons at one side of the beam, say at $x = |\Delta x|$, will be pushed down, while those at the other side, at $x = -|\Delta x|$, will be pushed up. For the Stern-Gerlach effect to be observed, these orbital motions must be weaker than the spin-splitting motion; the following inequality must hold:

$$F_{\rm spin,z} \gg F_{\rm orb,z}. \tag{I.14}$$

This condition leads us immediately to the requirement that,

$$mv_y \Delta x \ll \hbar. \tag{I.15}$$

The uncertainty principle only says that the width of our beam and the uncertainty in the transverse directions are connected by

$$m\Delta v_x \Delta x \approx \hbar, \tag{I.16}$$

which leads to the following requirement for the Stern-Gerlach experiment:

$$\Delta v_x \gg v_y. \tag{I.17}$$

In other words, the beam would not be a beam at all, since it would spread to the transverse direction faster than it is moving forward ; the confining slit for such a beam in the transverse direction would be much smaller than is the de Broglie wavelength: $\Delta x \ll \hbar/p_y$. This demonstrates the impossibility of observing spin-splitting with conventional Stern-Gerlach apparatus.

## C. Overview

Essential spintronic concepts are described in four main chapters. Chapter II deals with electrical spin injection. This chapter is most elementary. Concepts such as spin injection efficiency, spin-charge coupling, the Hanle effect, or spin-resolved Andreev reflection, are explained. The treatment is based on the drift-diffusion model which is accessible to readers with undergraduate knowledge of physics. The chapter contains also examples of spin injection into GaAs, the material of choice for study electrical spin injection as well as other spin-related properties. However, very recently spin injection has been achieved also in silicon, the traditional information-age material. These experiments are also covered as a nice example, apart from their great importance to the emerging spintronics technology, of the spin-valve and Hanle effects. Chapter II also introduces a large class of phenomena occuring in ferromagnet/insulator/ferromagnet structures, known under the name of tunneling magnetoresistance (TMR). We also decribe spin-orbit induced tunneling anisotropies in ferromagnet/semiconductor systems, known as tunneling anisotropic magnetoresistance (TAMR).

Chapter III works out details of spin-orbit coupling in semiconductors. We describe the effects of the coupling on the energy states in semiconductors with and without a center of inversion symmetry, and derive essentially from scratch, explaining and using what are called the $k \cdot p$ and envelope function theories the most important effective Hamiltonians describing spin dynamics in bulk semiconductors (Dresselhaus Hamiltonian) as well as in semiconductor heterostructures (Bychkov-Rashba Hamiltonian).

Chapter IV is devoted to spin relaxation and spin dynamics. We give a simple model for spin dynamics in the presence of environment, and present most relevant spin relaxation mechanisms in semiconductors. As case studies we have chosen to describe recent experiments on the influence of the electron-electron interactions on the spin relaxation in semiconductor quantum wells, as well as earlier spin resonant and more recent spin injection experiments measuring electron spin lifetime in silicon. This chapter also contain a discussion of the important experiments as well as theoretical concepts of spin dynamics and spin coherence in lateral semiconductor quantum dots, from the perspective of using the electron spin in these systems for quantum information processing.

Finally, Chapter V introduces basic concepts of proposed and demonstrated spintronic devices, focusing on a large class of magnetic resonant tunneling structures (magnetic resonant diodes and digital magnetoresistance circuits) as well as on the class of devices known as bipolar spintronic devices (again diodes and transistors). Since materials issues are critical in building new spintronics systems, we have also included, in a tutorial way, a brief discussion of relevant diluted magnetic semiconductors (DMS) outlining the most widely used mean-field theoretical method of calculating the Curie temperatures in bulk and heterostructure systems based on DMS.

## II. Spin injection and spin-dependent tunneling

This section deals with electrical spin injection from a ferromagnetic electrode into a nonmagnetic conductor, which can be a metal or a semiconductor. Since the essentials of spin injection are by now well understood, we offer below what we believe to be a description suitable for a senior undergraduate level. For this tutorial part to be self-contained, we introduce the necessary background material related to drift, diffusion, chemical potentials, and charge transport. Spin transport adds to this picture spin-resolved parameters, spin current and spin accumulation, spin relaxation, as well as spin dynamics. Readers familiar with the basic concepts can directly proceed to Sec. D. for the description of what we call the standard model of spin injection.

### A. Particle drift and diffusion

Consider electrons undergoing random walk in one dimension. The electrons move with the velocity $v$ a distance $l$, before they switch to a new direction. The time of flight is $\tau = l/v$. We apply electric field $E$ which, if not strong enough to significantly change the velocity $v$, leads to a biased random walk. The requirement on the field is,

$$|\Delta v| = |eE\tau/m| \ll v, \tag{II.1}$$

where $\Delta v$ is the velocity gain during a single step.

The above model is a good (one dimensional) first approximation to what happens in real metals and semiconductors in which free electrons perform random walk due to scattering off impurities, phonons, or boundaries. The step size $l$ is the mean free path and $\tau$ is the momentum relaxation time, as indicated in Fig. II.1.

The average velocity, $v_{\rm av}$, of electrons is markedly different from $v$. In a simple model the time evolution for the average velocity is,

$$\dot{v}_{\rm av} = -\frac{eE}{m} - \frac{v_{\rm av}}{\tau}, \tag{II.2}$$

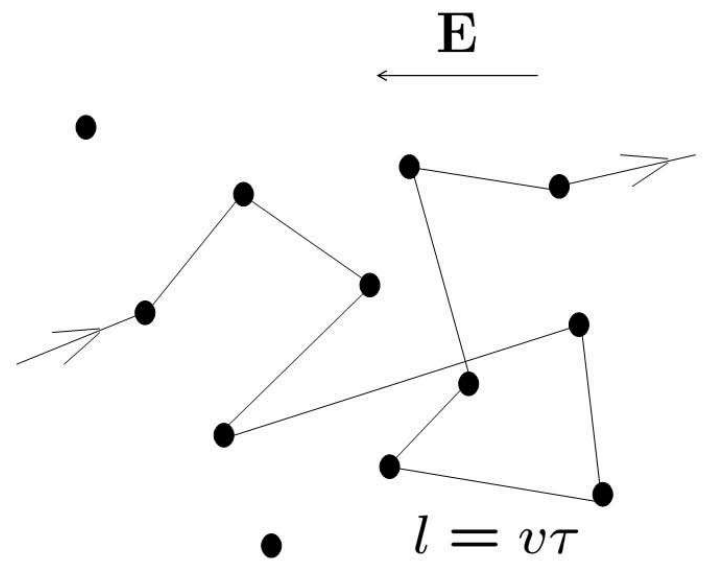

Fig. II.1. Illustration of the drift-diffusive transport of electrons in a disordered solid, in the presence of an electric field. Scattering by impurities or phonons causes electrons to change their direction of motion, while the electric field forces them in one direction. In this simplified picture the mean free path, $l$, is roughly the average distance between the impurities.

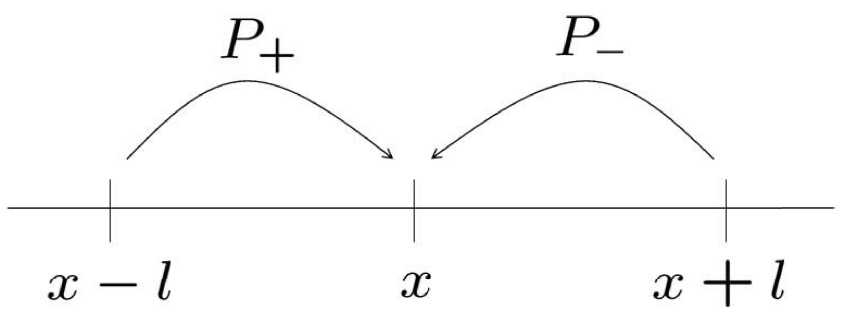

Fig. II.2. Biased random walk is represented by probabilities $p_+$ of moving right and $p_-$ of moving left, a step size of $l$ over time step of $\tau$.

where $m$ is the electron mass and the last term describes frictional effects of the scattering. In a steady state regime the average velocity is called the drift velocity, $v_d$, which can be determined from the above equation by putting $\dot{v}_{av} = 0$:

$$v_d = -\frac{e\tau}{m}E. \tag{II.3}$$

A reasonable value for $v_d$ is 1 cm/s. In a typical metal, $v \approx 10^6$ m/s, so the condition $v_d \ll v$ is well satisfied. Noting that $\Delta v$ has the same magnitude as $v_d$, the requirement for the biased random walk, Eq. (II.1), is fulfilled.

Let us continue with our random walk model. We are interested in the time evolution of the spatial profile of the density of random walkers. Let the density at time $t$ and position $x$ be $n(x,t)$. We consider $N_0$ electrons and assume that they cannot be created or destroyed.[9] This gives the normalization condition

$$N_0 = \int_{-\infty}^{\infty} n(x,t)dx, \tag{II.4}$$

to be valid at all times. At time $t$ the density of electrons at position $x$ is given by the densities at $x-l$ and $x+l$ at the previous time step $t-\tau$; see Fig. II.2. If the probability for electrons to move to the right is $p_+$ and to the left $p_-$, satisfying the condition, $p_+ + p_- = 1$, the following balance equation follows:

$$n(x,t) = n(x-l, t-\tau)p_+ + n(x+l, t-\tau)p_-. \tag{II.5}$$

If the random walk is unbiased ($E = 0$), then $p_+ = p_- = 1/2$. In general the two probabilities differ. We denote $\Delta p = p_+ - p_-$. We will see that $\Delta p \ll 1$ for conduction electrons.

Expand the densities on the right hand side of Eq. (II.5) in a Taylor series around $(x,t)$ for infinitesimal $l$ and $\tau$, keeping terms up to $O(l^2)$ and $O(\tau)$:

$$n(x,t) = n(x,t) - l\Delta p\frac{\partial n(x,t)}{\partial x} + \frac{1}{2}l^2\frac{\partial^2 n(x,t)}{\partial x^2} - \tau\frac{\partial n(x,t)}{\partial t}. \tag{II.6}$$

It follows that,

$$\tau\frac{\partial n}{\partial t} = \frac{1}{2}l^2\frac{\partial^2 n}{\partial x^2} - l\Delta p\frac{\partial n}{\partial x}. \tag{II.7}$$

[9] In semiconductors electrons can recombine with holes and this condition need not be valid.

We now introduce

$$D = \frac{1}{2}\frac{l^2}{\tau} = \frac{1}{2}v^2\tau, \tag{II.8}$$

and call it diffusivity (or the diffusion coefficient). This parameter describes the rate of electron diffusion. The units of $D$ are m$^2$·s$^{-1}$; in semiconductor physics we usually use cm$^2 \cdot s^{-1}$. For a metal at room temperature, $\tau \sim 10^{-14}$ s, giving $D \sim 10^{-2}$ m$^2$·s$^{-1}$. We will see that the second term on the right hand side of Eq. (II.7) describes drift, and that,

$$v_d = \frac{l}{\tau}\Delta p = v\Delta p \ll v, \tag{II.9}$$

from which the condition $\Delta p \ll 1$ follows. Using the new notation we finally obtain

$$\frac{\partial n}{\partial t} = D\frac{\partial^2 n}{\partial x^2} - v_d\frac{\partial n}{\partial x}, \tag{II.10}$$

which is the drift-diffusion partial differential equation describing the time evolution of the electron density profile $n$.

We will now justify the name drift-diffusion. Physically drift should mean that the whole electron density moves with a constant velocity $v_d$. Let us calculate the velocity of the average position of electrons:

$$\overline{x}(t) = \frac{1}{N_0}\int_{-\infty}^{\infty} xn(x,t)dx. \tag{II.11}$$

The time derivative is

$$\begin{aligned}
\dot{\overline{x}} &= \frac{1}{N_0}\frac{d}{dt}\int_{-\infty}^{\infty} xn(x,t)dx = \frac{1}{N_0}\int_{-\infty}^{\infty} x\frac{\partial n(x,t)}{\partial t}dx && \text{(II.12)}\\
&= \frac{1}{N_0}\int_{-\infty}^{\infty} xD\frac{\partial^2 n}{\partial x^2}dx - \frac{1}{N_0}v_d\int_{-\infty}^{\infty} x\frac{\partial n}{\partial x}dx && \text{(II.13)}\\
&= v_d. && \text{(II.14)}
\end{aligned}$$

We have used integration per parts and the physical requirement that the density vanishes at infinity. In particular,

$$\int_{-\infty}^{\infty} x\frac{\partial^2 n}{\partial x^2}dx = x\frac{\partial n}{\partial x}|_{-\infty}^{\infty} - \int_{-\infty}^{\infty}\frac{\partial n}{\partial x}dx = 0, \tag{II.15}$$

$$\int_{-\infty}^{\infty} x\frac{\partial n}{\partial x}dx = xn|_{-\infty}^{\infty} - \int_{-\infty}^{\infty} ndx = -N_0. \tag{II.16}$$

We have shown that the velocity with which the average position of the electron density moves is $v_d$. It remains to demonstrate diffusion, that is, that the variance of the density evolves linearly with time. We leave as an exercise to show that

$$\sigma^2 = \overline{x^2} - \overline{x}^2 = 2Dt, \tag{II.17}$$

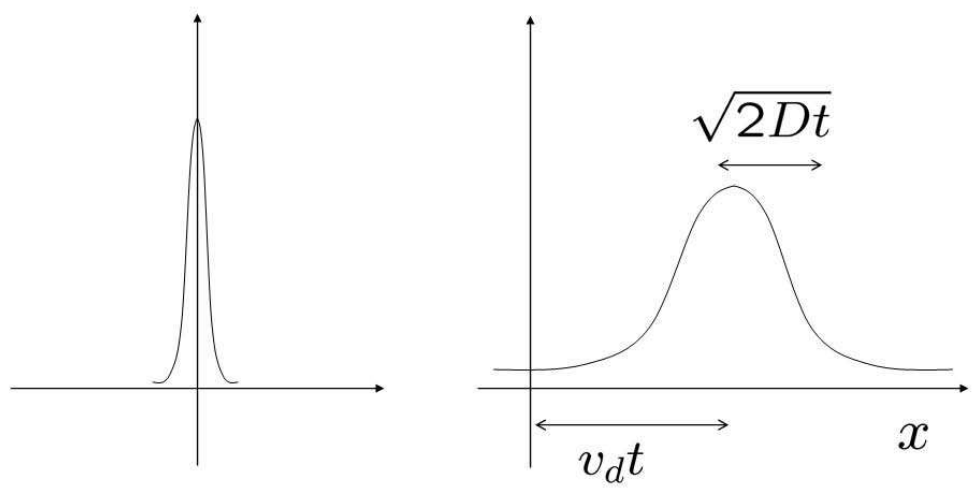

Fig. II.3. The initial delta-like distribution widens at a rate proportional to $\sqrt{t}$, due to diffusion, while its center moves linearly, as $t$, due to drift.

assuming a density profile with zero variance at $t = 0$. The standard deviation of the average electron position is then

$$\sigma(t) = \sqrt{2Dt}. \tag{II.18}$$

This result conforms to what one usually learns about random walks, namely, that the variance after $N$ steps is

$$\sigma^2 = l^2 N. \tag{II.19}$$

Considering that $N = t/\tau$ and $D = l^2/2\tau$ we obtain Eq. (II.18). In $d = 2$ and $d = 3$ dimensions the variance is $\sqrt{4Dt}$ and $\sqrt{6Dt}$, respectively.

The time evolution of the electron density thus constitute a drift due to the presence of the electric field and scattering induced friction, as well as diffusion which is due to scattering. A typical evolution of the electron density looks like the one depicted in Fig. II.3.

It is instructive to see how the diffusion arises from the scaling considerations of the diffusion equation. Take $v_d = 0$. We first notice that the typical length scale of diffusion goes as $\sim \sqrt{t}$. This leads to the guess that,

$$n(x,t) = \frac{N_0}{\sqrt{t}} f(\frac{x}{\sqrt{t}}), \tag{II.20}$$

where $f(\xi)$ is a function of one variable, $\xi = x/\sqrt{t}$, subject to the normalization condition

$$\int_{-\infty}^{\infty} f(\xi) d\xi = 1. \tag{II.21}$$

Substituting this guess to the diffusion equation, which is a partial differential equation, we obtain an ordinary differential equation for $f$:

$$f'' + \frac{1}{2}\xi f + f = 0. \tag{II.22}$$

The solution, with the boundary condition of $f(0) = 1$, is

$$f(\xi) = \frac{1}{\sqrt{4\pi D}} e^{-\xi^2/4D}. \tag{II.23}$$

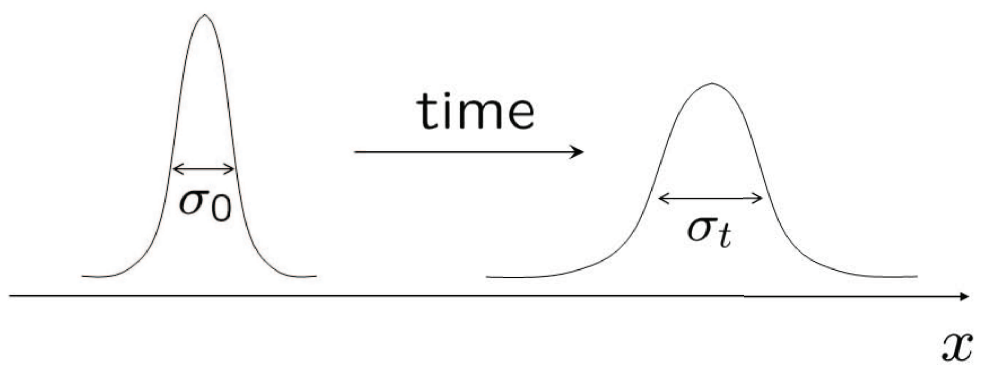

Fig. II.4. Spread of the probability distribution in time due to particle diffusion.

This solves our original problem:

$$n(x,t) = \frac{N_0}{\sqrt{4\pi Dt}} e^{-x^2/4Dt}, \tag{II.24}$$

satisfying the initial condition $n(x,0) = N_0\delta(x)$. The form of $n(x,t)$ is that of the normal (gaussian) probability distribution with the variance $\sigma^2 = 2Dt$, once again demonstrating the diffusion character of the solution; see Eq. (II.18) and Fig. II.4 for illustration.

The electric field enters through the drift velocity $v_d$ which we saw to originate the bias $\Delta p$. We have found that the drift velocity is proportional to the electric field:

$$v_d = -\frac{e\tau}{m}E. \tag{II.25}$$

We call the magnitude of the proportionality coefficient the electron mobility $\mu$:

$$\mu = \frac{e\tau}{m}. \tag{II.26}$$

The units of mobility are $m^2$/Vs (again, in semiconductor physics we customarily use $cm^2$/Vs). For a metal $\mu \sim 10^{-3}$ $m^2$/Vs. Typical electric fields inside a metallic conductor are $E = v_d/\mu \sim$ 10 V/m.

The drift-diffusion equation now reads[10]

$$\frac{\partial n}{\partial t} + \frac{\partial}{\partial x}\left[-\mu n E - D\frac{\partial n}{\partial x}\right] = 0. \tag{II.27}$$

We introduce the electron (particle) current, given by the expression in the brackets:

$$J = -\mu n E - D\frac{\partial n}{\partial x.} \tag{II.28}$$

The first term describes drift, the second diffusion current. The electrical (charge) current is

$$j = -eJ = \sigma E + eD\frac{\partial n}{\partial x}, \tag{II.29}$$

[10]The spin injection model to be discussed later assumes charge neutrality. This means that $dE/dx = 0$, from Gauss'law.

where,

$$\sigma = e\mu n, \tag{II.30}$$

is the conductivity. Typical values for metals at room temperature are $[1\mu\Omega\mathrm{cm}]^{-1}$. For semiconductors the magnitudes are orders of magnitude less, due to the lower electronic density. The inverse conductivity is resistivity,

$$\rho = 1/\sigma. \tag{II.31}$$

There is a nice relation between the diffusivity and mobility. The dimensions of the two quantities suggest that the ratio $eD/\mu$ has the dimension of energy. Indeed,

$$\frac{eD}{\mu} = \frac{ev^2\tau/2}{e\tau/m} = \frac{1}{2}mv^2, \tag{II.32}$$

getting the electrons' kinetic energy. For degenerate electrons the energy is roughly the Fermi energy, $\varepsilon_F$, while for nondegenerate electrons we get the thermal energy, $k_BT$. The above relation, also called Einstein's, will be given a more precise form later.

It remains to justify our choice for the current $J$. We find,

$$\frac{\partial n}{\partial t} + \frac{\partial J}{\partial x} = 0, \tag{II.33}$$

which is the continuity equation expressing the conservation of particles. The particle density can increase in a certain volume only if there is more current flowing in than out; in other words, when $\nabla \cdot \mathbf{J} < 0$. Above equation says the same in one dimension.

## B. Spin drift and diffusion

Consider now electrons which can be labeled as spin up and spin down. The total number of electrons is assumed to be preserved. If the electron densities are $n_\uparrow$ and $n_\downarrow$ for the spin up and spin down states, the total particle density is,

$$n = n_\uparrow + n_\downarrow, \tag{II.34}$$

while the spin density is,

$$s = n_\uparrow - n_\downarrow. \tag{II.35}$$

We have already found the drift-diffusion equation for $n$. Can we find one for $s$ as well?

We have spin up and spin down electrons performing random walk, as before. However, we will now allow for spins to be flipped and assign the probability of $w$ that a spin is flipped in the time of $\tau$, so that the spin flip rate is $w/\tau$. We have the diagram of Fig. II.5.

We will assume that $w \ll 1$. This is well justified for conduction electrons, as we will see in the chapter of spin relaxation, Sec. IV. The actual spin flip probability during the relaxation time $\tau$ is typically $10^{-3}$ to $10^{-6}$, so that electrons need to experience thousands scatterings before spin flips.

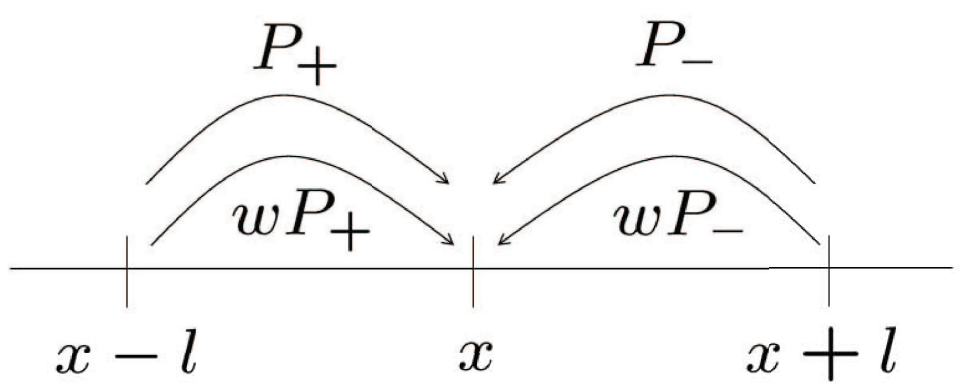

Fig. II.5. Random walk scheme with indicated spin-flip probabilities $w$.

Let us write the balance equation for the spin up density, and make the Taylor expansion around $(x, t)$:

$$\begin{aligned} n_\uparrow(x,t) &= n_\uparrow(x-l, t-\tau)(1-w)p_+ + n_\uparrow(x+l, t-\tau)(1-w)p_- + && \text{(II.36)} \\ &+ n_\downarrow(x-l, t-\tau)wp_+ + n_\downarrow(x+l, t-\tau)wp_- && \text{(II.37)} \\ &\approx n_\uparrow(1-w) - l\frac{\partial n_\uparrow}{\partial x}\Delta p + \frac{1}{2}l^2\frac{\partial^2 n_\uparrow}{\partial x^2} - \tau\frac{\partial n_\uparrow}{\partial t} + n_\downarrow w. && \text{(II.38)} \end{aligned}$$

This leads to the following drift-diffusion equations for $n_\uparrow$ and, similarly, for $n_\downarrow$:

$$\begin{aligned} \frac{\partial n_\uparrow}{\partial t} &= D\frac{\partial^2 n_\uparrow}{\partial x^2} - v_d\frac{\partial n_\uparrow}{\partial x} - w(n_\uparrow - n_\downarrow), && \text{(II.39)} \\ \frac{\partial n_\downarrow}{\partial t} &= D\frac{\partial^2 n_\downarrow}{\partial x^2} - v_d\frac{\partial n_\downarrow}{\partial x} - w(n_\downarrow - n_\uparrow). && \text{(II.40)} \end{aligned}$$

Adding the two equations we obtain the drift diffusion equation for the density $n$, Eq. (II.27); subtracting them we get the drift-diffusion equation for $s$:

$$\frac{\partial s}{\partial t} = D\frac{\partial^2 s}{\partial x^2} - v_d\frac{\partial s}{\partial x} - \frac{s}{\tau_s}. \qquad \text{(II.41)}$$

Here,

$$\frac{1}{\tau_s} = \frac{2w}{\tau}, \qquad \text{(II.42)}$$

describes the spin relaxation; $\tau_s$ is the spin relaxation time. Why is spin relaxation twice as large as spin flip? Because each spin flip contributes to relaxation of *both* spin up and spin down, so that spin relaxation is twice as fast.

Let us write the spin drift-diffusion equation in terms of mobility:

$$\frac{\partial s}{\partial t} = D\frac{\partial^2 s}{\partial x^2} + \mu E\frac{\partial s}{\partial x} - \frac{s}{\tau_s}. \qquad \text{(II.43)}$$

We can, as before, write it in the form of a continuity equation,

$$\frac{\partial s}{\partial t} + \frac{\partial}{\partial x}\left(-\mu E s - D\frac{\partial s}{\partial x}\right) = -\frac{s}{\tau_s}. \qquad \text{(II.44)}$$

The expression in the brackets is identified as the spin (particle) current,

$$J_s = -\mu s E - D\frac{\partial s}{\partial x}. \tag{II.45}$$

This allows us to write the spin continuity equation as,

$$\frac{\partial s}{\partial t} + \frac{\partial J_s}{\partial x} = -\frac{s}{\tau_s}. \tag{II.46}$$

The right hand side represents the spin relaxation with the rate $s/\tau_s$. The spin in a given volume can decrease either by spin current flowing away from the volume, or by spin relaxation. If there would be an equilibrium spin, $s_0$ in the sample, we would need to replace the right-hand side by $-(s - s_0)/\tau_s$. At this point we introduce the spin (charge) current,

$$j_s = -eJ_s = \sigma_s E + eD\frac{\partial s}{\partial x}, \tag{II.47}$$

where the spin conductivity is introduced as

$$\sigma_s = e\mu s. \tag{II.48}$$

We also define the density spin polarization

$$P_n = \frac{n_\uparrow - n_\downarrow}{n} = \frac{s}{n}, \tag{II.49}$$

as well as the current spin polarization,

$$P_j = \frac{j_\uparrow - j_\downarrow}{j} = \frac{j_s}{j}, \tag{II.50}$$

which will be useful in our model of spin injection.

Let us now solve the spin drift-diffusion equation in three cases of interest, neglecting spin drift. For a worked out problem in which spin drift plays important role, see Sec. E.2.

**Spin diffusion for E=0.** Consider a spin density which, at $t = 0$, is all concentrated at $x = 0$:

$$s(x, 0) = S_0\delta(x), \tag{II.51}$$

where $S_0$ is the total spin at $t = 0$. The solution of the spin diffusion equation,

$$\frac{\partial s}{\partial t} = D\frac{\partial^2 s}{\partial x^2} - \frac{s}{\tau_s}, \tag{II.52}$$

with the given initial condition is,

$$s(x, t) = \frac{S_0}{\sqrt{4\pi Dt}} e^{-x^2/4Dt} e^{-t/\tau_s}. \tag{II.53}$$

The first exponential term describes spin diffusion, the second spin relaxation—the total spin decays as

$$S = \int_{-\infty}^{\infty} s(x,t)dx = S_0 e^{-t/\tau_s}. \tag{II.54}$$

How far does the spin diffuse till it relaxes? The answer is given by the standard deviation of the spin distribution at time $t = \tau_s$:

$$\sigma = \sqrt{2D\tau_s} = \sqrt{2}L_s, \tag{II.55}$$

where we introduced the *spin diffusion length*[11]

$$L_s = \sqrt{D\tau_s}. \tag{II.56}$$

In semiconductors, $L_s \sim \mu$m, while in metals, in which the diffusivity is larger (since the electron velocity at the Fermi level is larger), $L_s \sim 0.1$ mm. In $d$ dimensions the standard deviation is $\sigma = \sqrt{2d}L_s$.

**Steady state diffusion at E=0.** We now look at the steady state into which spin diffusion settles given certain sources of spin. We need to solve

$$D\frac{\partial^2 s}{\partial x^2} = \frac{s}{\tau_s}, \tag{II.57}$$

which can be written in the more familiar term

$$\frac{\partial^2 s}{\partial x^2} = \frac{s}{L_s^2}. \tag{II.58}$$

Suppose there is a source spin at $x = 0$, given by $s(0) = s_0$ and $s(\infty) = 0$. The spin density in the interval $(0, \infty)$ is,

$$s(x) = s_0 e^{-x/L_s}. \tag{II.59}$$

The spin spreads to a distance $L_s$ from its source at $x = 0$, as illustrated in Fig. II.6.

**Steady state spin pumping.** Suppose that instead of a spin source we have a given spin current at $x = 0$: $J_s(0) = -D\partial s/\partial x|_{x=0} = J_{s0}$. The solution to the spin diffusion equation with this boundary condition is,

$$s(x) = J_{s0}\frac{L_s}{D}e^{-x/L_s}. \tag{II.60}$$

The spin at $x = 0$ is,

$$s(0) = J_{s0}\frac{L_s}{D}. \tag{II.61}$$

[11] Often one speaks of the spin relaxation length. We find it useful to reserve the term of spin diffusion length for the case of purely diffusive motion, while the term spin relaxation length for more general cases, such as drift or magnetic field influenced decay of the spin density. See, for example, Sec. E.2.

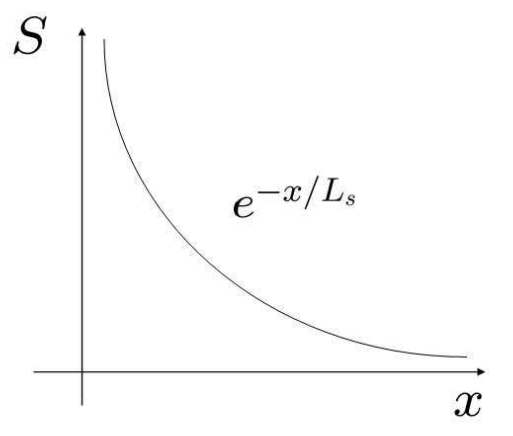

Fig. II.6. Spin decays exponentially in space with the characteristic length $L_s$, the spin diffusion length.

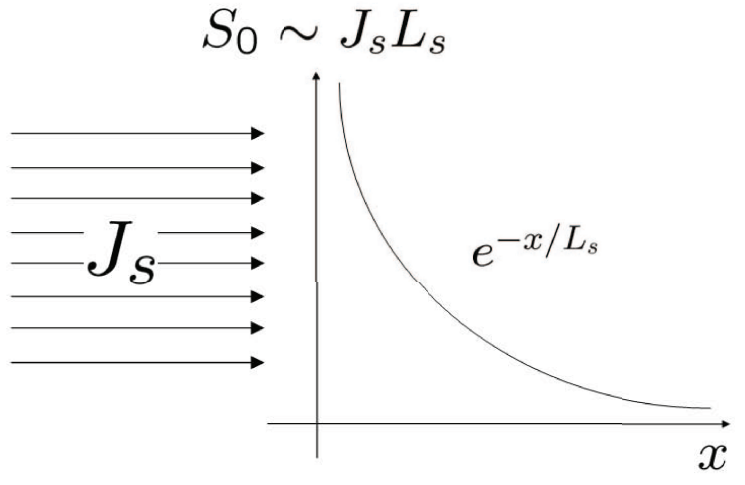

Fig. II.7. If spin flows from the ferromagnetic, $x < 0$ region into a nonmagnetic region at $x > 0$, the accumulated spin at the interface is proportional to the product of the spin current and the spin diffusion length. The spin decays exponentially in the nonmagnetic region.

This spin density is called spin accumulation,[12] as it results from spin injection (say, from a ferromagnetic metal at $x < 0$). In this case spin injection is spin pumping: the spin accumulation is proportional to the spin injection intensity (pumping), while it is also proportional to the spin diffusion length. The more one pumps and the less the spin relaxes, the more spin accumulation can be achieved. The above model, illustrated in Fig. II.7 is a simplest description of spin injection.

Several generalizations of this approach can be made by considering additional effects of spin-orbit coupling (Tse *et al.*, 2005; Pershin, 2004), transient effects arising from the Boltzmann equation (Villegas-Lelovsky, 2006a,b) and Monte Carlo simulations (Saikin *et al.*, 2003).

### C. Quasichemical potentials $\mu$ and $\mu_s$.

Consider a Fermi gas in equilibrium. Let the density of the gas is $n_0$, given by the chemical potential $\eta$. If the minimum of the band energy is taken to be $\varepsilon = 0$, the electron density is

[12]More conventially the term spin accumulation is often reserved for the nonequilibrium spin quasichemical potential, see Sec. C.

related to the chemical potential by the following expressions:

$$
\begin{aligned}
n_0 &= 2\frac{1}{V}\sum_{\mathbf{k}} f_0(\varepsilon_{\mathbf{k}} - \eta) &\text{(II.62)}\\
&= 2\frac{1}{V}\frac{V}{(2\pi)^3}\int d^3k f_0(\varepsilon_{\mathbf{k}} - \eta) &\text{(II.63)}\\
&= \int_0^\infty d\varepsilon f_0(\varepsilon - \eta)\frac{2}{(2\pi)^3}\int d^3k\delta(\eta - \varepsilon_{\mathbf{k}}) &\text{(II.64)}\\
&= \int_0^\infty d\varepsilon g(\varepsilon) f_0(\varepsilon - \eta). &\text{(II.65)}
\end{aligned}
$$

Here $f_0$ is the Fermi-Dirac distribution

$$
f_0(\varepsilon - \eta) = \left[e^{(\varepsilon-\eta)/k_BT} + 1\right]^{-1}, \tag{II.66}
$$

and $g(\varepsilon)$ is the density of states per unit volume,

$$
g(\varepsilon) = \frac{2}{(2\pi)^3}\int d^3k\delta(\epsilon - \epsilon_{\mathbf{k}}). \tag{II.67}
$$

The factor of two in the above formulas comes from the spin degeneracy.

For a degenerate Fermi gas, $\eta \approx \varepsilon_F$, and the Fermi-Dirac distribution resembles the step function,

$$
\begin{aligned}
f_0(\varepsilon - \eta) &\approx \theta(\eta - \varepsilon), &\text{(II.68)}\\
-\frac{\partial f}{\partial \varepsilon} = \frac{\partial f}{\partial \eta} &\approx \delta(\varepsilon - \eta). &\text{(II.69)}
\end{aligned}
$$

For a nondegenerate Fermi gas, $\varepsilon \gg \eta + k_BT$, the Fermi-Dirac distribution has the functional form of the Maxwell-Boltzmann statistics,

$$
f_0(\varepsilon - \eta) \approx e^{-(\varepsilon-\eta)/k_BT}. \tag{II.70}
$$

In general we can write in the functional form that

$$
n_0 = n_0(\eta). \tag{II.71}
$$

Figure II.8 illustrates the difference between degenerate and nondegenerate statistics. Metals and heavily doped semiconductors obey degenerate statistics, while low doped semiconductors are well described by nondegenerate statistics.

We are going to generalize the above description to weakly nonequilibrium situations. Suppose there is a static electric field $E = -\nabla\phi$ in our conductor and still no current flows. We are still at equilibrium. This is not possible to achieve in metals (as is known from elementary electrodynamics), but the field can exist in inhomogeneously doped semiconductors. In fact, the diodes work as current rectifiers, or semiconductor solar cells as current generators because an equilibrium electric field is established between a p-doped (filled with acceptors) and n-doped (filled with donors) semiconductor regions. How does the electron density change in the presence of such a field? The chemical potential $\eta$ must be uniform, since we are still in equilibrium. The only thing that changes is electron's energy, which is reflected in the Fermi-Dirac distribution. The state counting is otherwise unaffected: the electrons occupy the band states $\varepsilon$ with the density of states $g(\varepsilon)$, but at each state the total electron energy is $\varepsilon - e\phi$. This gives

$$n(\mathbf{r}) = \int_0^\infty d\varepsilon g(\varepsilon) f_0(\varepsilon - e\phi - \eta). \qquad \text{(II.72)}$$

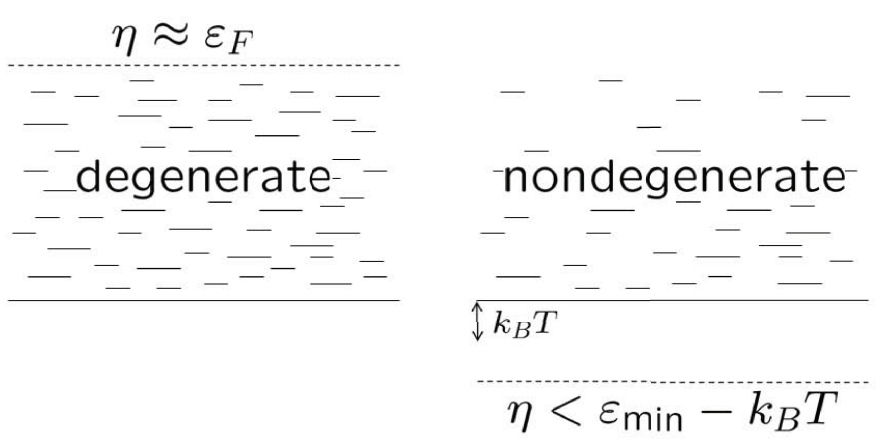

Fig. II.8. Degenerate semiconductors are like metals: the chemical potential $\eta$ lies close to the Fermi energy $\varepsilon_F$. Nondegenerate semiconductors obey the Boltzmann statistics, for which $(\varepsilon_{\rm min} - \mu)/k_BT \gg 1$, implying that the occupation numbers of the single-particle levels in the conduction band are much less unity. Here $\varepsilon_{\rm min}$ is the minimum of the conduction band.

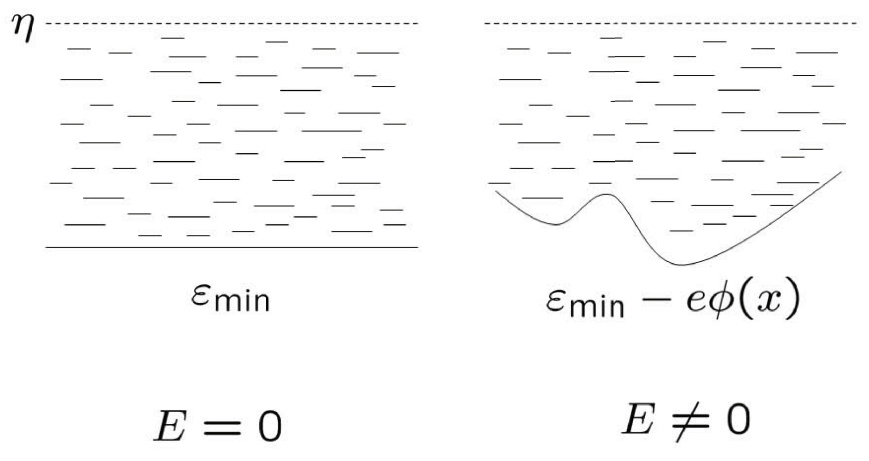

Fig. II.9. Application of an electric field, under the conditions of thermal equilibrium, changes locally the energy levels, but the chemical potential $\eta$ is still uniform. No current flows.

In other words,

$$n(\mathbf{r}) = n_0(\eta + e\phi). \qquad \text{(II.73)}$$

The higher is the potential, the higher is the electron density, as illustrated in Fig. II.9.

Let us see an important consequence of this functional form. The electric current,

$$\begin{aligned} j &= \sigma E + eD\nabla n & \text{(II.74)} \\ &= -\sigma\nabla\phi + eD\frac{\partial n_0}{\partial \eta} e\nabla\phi & \text{(II.75)} \\ &= \nabla\phi\left(-\sigma + e^2 D\frac{\partial n_0}{\partial \eta}\right), & \text{(II.76)} \end{aligned}$$

must vanish in equilibrium. This gives

$$\sigma = e^2 D \frac{\partial n_0}{\partial \eta}, \tag{II.77}$$

which is the general form of Einstein's relation. This relation is an example of a much broader class of so-called fluctuation-dissipation relations, which describe the linear relationships between fluctuation and dissipation strengths. In our case the fluctuation strength is given by diffusivity $D$, which measures the fluctuations of the velocity (recall that $D$ is a measure of $v^2$), while the dissipation is represented by $\sigma$, which measures energy dissipation due to Joule heating. For degenerate electrons in the absence of electric field

$$\frac{\partial n_0}{\partial \eta} = \int_0^\infty d\varepsilon \, g(\varepsilon) \frac{\partial f_0}{\partial \eta} \tag{II.78}$$

$$= g(\eta) \approx \frac{k_F^3}{\varepsilon_F} \approx \frac{n_0}{\varepsilon_F}. \tag{II.79}$$

Here we introduced the Fermi wave vector, $k_F$, by $\varepsilon_F = \hbar^2 k_F^2/2m$. We thus recover the simple estimate for the ratio of $eD/\mu$ found earlier. The reader should repeat the above calculation for a nondegenerate electron gas.

Relax now the equilibrium condition and allow the current to flow. The chemical potential is no longer uniform. In fact, there is no guarantee that the quantity such as the chemical potential makes sense. In most cases, however, we can assume that, in general, nonequilibrium electron distribution $f$ will depend on the electron state only through its energy, as momentum relaxation proceeds on a faster scale. This temporal coarse graining allows us to write for the electron distribution function in the momentum space,

$$f_{\mathbf{k}} \approx f(\varepsilon_{\mathbf{k}}) = f_0(\varepsilon_{\mathbf{k}} - e\phi - \eta - e\mu), \tag{II.80}$$

where $\mu = \mu(x)$ is a spatially dependent addition to the chemical potential, often called quasi-chemical potential; do not confuse it with mobility. We thus attempt to describe the current flow in the system by writing,

$$\eta \to \eta + e\mu(x). \tag{II.81}$$

We then have,

$$n(x) = n_0(\eta + e\mu + e\phi). \tag{II.82}$$

The current becomes,

$$j = -\sigma \nabla \phi + eD \nabla n \tag{II.83}$$

$$= \nabla \phi \left( -\sigma + e^2 D \frac{\partial n_0}{\partial \eta} \right) + e^2 D \frac{\partial n_0}{\partial \eta} \nabla \mu. \tag{II.84}$$

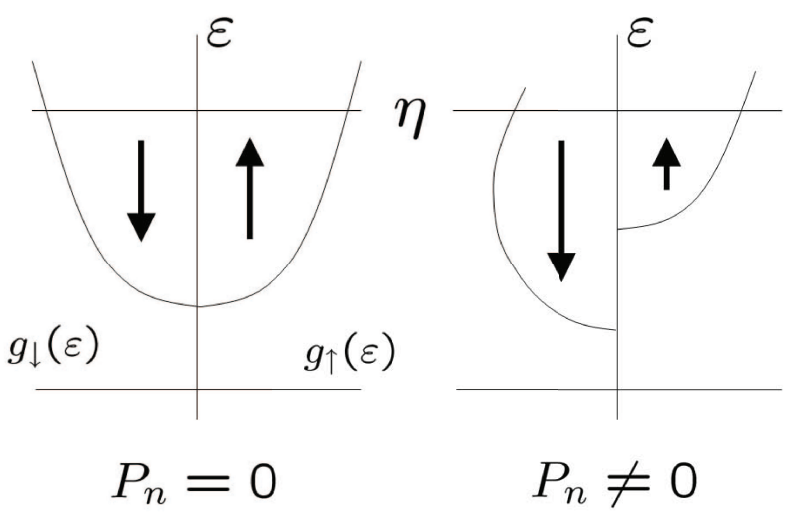

Fig. II.10. In a nonmagnetic conductor (left), there is equal number of spin up and spin down electrons in equilibrium. In a ferromagnetic conductor, the densities are different due to the exchange splitting causing the minima of the two spin bands to be displaced. The spin of the larger electronic density is called majority spin; the spin of the smaller density is called minority spin. However, usually the density of states at the Fermi level is higher for the minority than for the majority electrons.

We can now invoke Einstein's relation, Eq. (II.77), and finally write[13]

$$j = \sigma \nabla \mu. \tag{II.85}$$

The current is driven by the gradient of the quasichemical potential which describes both drift and diffusion terms. We will see that this reformulation of the problem greatly simplifies the problem of electric spin injection.

**Generalization of the drift-diffusion formalism, to ferromagnetic conductors.** To proceed further we need to briefly discuss the essential electronic characteristics of ferromagnetic metals or semiconductors. Figure II.10 compares nonmagnetic and ferromagnetic conductors (assumed to be degenerate). The most dramatic situation occurs when the majority band is filled. The metallic behavior is due to the minority band only. Such metals are called half-metallic ferromagnets. See Fig. II.11.

The difference between the densities of states, $g_\uparrow$ and $g_\downarrow$, at the Fermi level, as well as the Fermi velocities for the majority and minority spins, is essential. This difference transcends to the differences in the relaxation times, mean free paths, mobilities, diffusivities, or conductivities. If we also allow for different quasichemical potentials $\mu_\uparrow$ and $\mu_\downarrow$, describing the possibility that there is a nonequilibrium spin in the system, we can write

$$j_\uparrow = \sigma_\uparrow \nabla \mu_\uparrow, \tag{II.86}$$

$$j_\downarrow = \sigma_\downarrow \nabla \mu_\downarrow. \tag{II.87}$$

[13]We have established Einstein's relation only for the equilibrium case. There is no guarantee the relation holds in general. In our framework of the linear regime (current proportional to $E$ and $\nabla n$), all deviations from the relation would be at least linear (explicitly) in $\nabla \mu$ and go beyond our linear regime description. Note that the linear character of our drift-diffusion equation does not mean that the current is linearly proportional to voltage. Significant deviations from linear I-V characteristics can occur if the electron density depends on the field, even within the linear response framework.

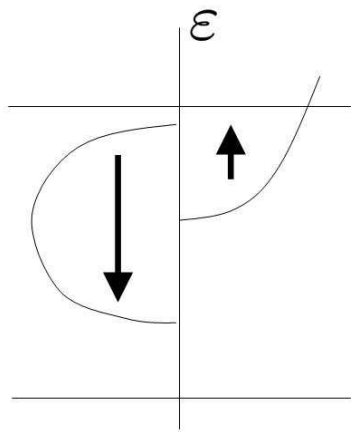

Fig. II.11. A half-metallic ferromagnet has zero density of states at the Fermi level, for one of the spin states. The spin polarization of the density of states at the Fermi level, $P_g$, is 100%.

Let us make the following definitions:

$$g = g_\uparrow + g_\downarrow \quad , \quad g_s = g_\uparrow - g_\downarrow \tag{II.88}$$

$$\sigma = \sigma_\uparrow + \sigma_\downarrow \quad , \quad \sigma_s = \sigma_\uparrow - \sigma_\downarrow \tag{II.89}$$

$$\mu = \frac{1}{2}(\mu_\uparrow + \mu_\downarrow) \quad , \quad \mu_s = \frac{1}{2}(\mu_\uparrow - \mu_\downarrow) \tag{II.90}$$

$$D = \frac{1}{2}(D_\uparrow + D_\downarrow) \quad , \quad D_s = \frac{1}{2}(D_\uparrow - D_\downarrow). \tag{II.91}$$

The charge and spin currents can now be written as,

$$j = j_\uparrow + j_\downarrow = \sigma \nabla \mu + \sigma_s \nabla \mu_s \tag{II.92}$$

$$j_s = j_\uparrow - j_\downarrow = \sigma_s \nabla \mu + \sigma \nabla \mu_s. \tag{II.93}$$

In a ferromagnet, in which $\sigma_s \neq 0$, a nonequilibrium spin gradient causes charge current, while spin current flows due to an applied bias. The reason is simple: since there are more, say, spin up than spin down electrons, a bias causes different spin up and spin down currents, and thus both charge and spin current. In normal conductors, at least at small spin polarizations at which $\sigma_s \approx 0$, this is not possible.

*From now on we will deal with degenerate conductors.* Not that we do not know how to calculate spin injection with nondegenerate electrons in semiconductors, but we do not have a "universal" analytical model for such cases due to complication from charging effects and the need to solve, in addition, Poisson's equation. The model for degenerate electrons, on the other hand, has useful analytical solutions in an important case of negligible charging.

For degenerate conductors the deviations from the chemical potential can be considered small, since it is only the electrons at the Fermi level which contribute to the currents. Then we can expand,

$$n_\uparrow = n_{\uparrow 0}(\eta + e\mu_\uparrow + e\phi) \approx n_{\uparrow 0} + \frac{\partial n_{\uparrow 0}}{\partial \eta}(e\mu_\uparrow + e\phi), \tag{II.94}$$

$$n_\downarrow = n_{\downarrow 0}(\eta + e\mu_\downarrow + e\phi) \approx n_{\downarrow 0} + \frac{\partial n_{\downarrow 0}}{\partial \eta}(e\mu_\downarrow + e\phi). \tag{II.95}$$

Recognizing that $\partial n_{\uparrow 0}/\partial \eta = g_\uparrow$, and similarly for spin down, for degenerate electrons, we have,

$$n_\uparrow = n_{\uparrow 0} + g_\uparrow e\mu_\uparrow + g_\uparrow e\phi, \tag{II.96}$$
$$n_\downarrow = n_{\downarrow 0} + g_\downarrow e\mu_\uparrow + g_\uparrow e\phi. \tag{II.97}$$

The electron density is,

$$n = n_\uparrow + n_\downarrow = n_0 + eg(\mu + \phi) + eg_s\mu_s. \tag{II.98}$$

We now impose *local charge neutrality*, namely, the condition of $n = n_0$. This condition is well satisfied in metals and very heavily doped semiconductors, in which the charge is screened on the atomic scales. This condition eliminates the electric potential from the problem, by relating it with the quasichemical potentials:

$$g(\mu + \phi) + g_s\mu_s = 0. \tag{II.99}$$

In a normal conductor, in which $g_s = 0$, we obtain $\mu = -\phi$, which is the obvious condition for local charge neutrality, that is, $n(\eta + e\mu + e\phi) = n_0(\eta)$.

Let is look at the spin density. We have,

$$s = s_0 + eg_s(\mu + \phi) + eg\mu_s \tag{II.100}$$
$$= s_0 + 4e\mu_s \frac{g_\uparrow g_\downarrow}{g}, \tag{II.101}$$

where we have used the local charge neutrality condition to eliminate $\mu+\phi$. The above expression gives for the nonequilibrium spin density,

$$\delta s = s - s_0 = 4e\mu_s \frac{g_\uparrow g_\downarrow}{g}. \tag{II.102}$$

The nonequilibrium spin density is proportional to the quasichemical potential $\mu_s$. We call either the nonequilibrium spin, $\delta s$, or the spin quasichemical potential, $\mu_s$, *spin accumulation.* In a normal conductor, $\delta s = s = e\mu_s g$, which is the number of electron states in the interval of $e\mu_s$ at the Fermi level.

In the following, the cental quantity of interest will be the current spin polarization, $P_j = j_s/j$. Recall that the density spin polarization $P_n = s/n$, while we introduce, in addition, the conductivity spin polarization,

$$P_\sigma = \sigma_s/\sigma. \tag{II.103}$$

From the expression for charge current, Eq. (II.92), we extract $\nabla\mu$:

$$\nabla\mu = \frac{1}{\sigma}\left(j - \sigma_s\nabla\mu_s\right). \tag{II.104}$$

We substitute this gradient to obtain the spin current:

$$j_s = \sigma_s\nabla\mu + \sigma\nabla\mu_s \tag{II.105}$$
$$= P_\sigma j + 4\nabla\mu_s \frac{\sigma_\uparrow\sigma_\downarrow}{\sigma}. \tag{II.106}$$

For the current spin polarization this gives,

$$P_j = P_\sigma + \frac{1}{j} 4\nabla\mu_s \frac{\sigma_\uparrow\sigma_\downarrow}{\sigma}. \tag{II.107}$$

This equation shows that in order to have a significant current spin polarization, we need to establish a large gradient of the spin quasichemical potential.

We still need to figure out the equation for $\mu_s$. Recall that in a steady state in a normal conductor we have derived the continuity equation for spin, see Eq. (II.46),

$$\nabla j_s = e\frac{s}{\tau_s}. \tag{II.108}$$

This equation describes a nonequilibrium spin $s$. In a ferromagnetic conductor we need to modify this equation to

$$\nabla j_s = e\frac{\delta s}{\tau_s}, \tag{II.109}$$

which describes the decay of the nonequilibrium spin $\delta s \to 0$, or the spin $s \to s_0$. We then have,

$$\begin{aligned} \nabla j_s &= e\frac{\delta s}{\tau_s} = 4e^2\mu_s \frac{g_\uparrow g_\downarrow}{g}\frac{1}{\tau_s} && \text{(II.110)} \\ &= \nabla\left(P_\sigma j + \nabla\mu_s \frac{4\sigma_\uparrow\sigma_\downarrow}{\sigma}\right) && \text{(II.111)} \\ &= 4\frac{\sigma_\uparrow\sigma_\downarrow}{\sigma}\nabla^2\mu_s. && \text{(II.112)} \end{aligned}$$

We have used Eq. (II.102) and the fact that in a steady state the electric current is continuous, $\nabla j = 0$. The above gives the desired diffusion equation for $\mu_s$:

$$\nabla^2\mu_s = \frac{\mu_s}{L_s^2}, \tag{II.113}$$

where the generalized spin diffusion length, $L_s$, is

$$L_s = \sqrt{\overline{D}\tau_s}, \tag{II.114}$$

and the generalized diffusivity

$$\overline{D} = \frac{g}{g_\uparrow/D_\downarrow + g_\downarrow/D_\uparrow}. \tag{II.115}$$

In a normal conductor, $\overline{D} = D$. In our formalism, $L_s$ is a phenomenological parameter.

### D. Standard model of spin injection: F/N junctions

What we call the standard model of spin injection has its roots in the original proposal of Aronov (1976). The thermodynamics of spin injection has been developed by Johnson and Silsbee, who also formulated a Boltzmann-like transport model for spin transport across ferromagnet/nonmagnet (F/N) interfaces (Johnson and Silsbee, 1987, 1988). The theory of spin injection has been further developed in (van Son *et al.*, 1987; Valet and Fert, 1993; Fert and Jaffres, 2001; Hershfield and Zhao, 1997; Schmidt *et al.*, 2000; Fabian *et al.*, 2002b; Žutić *et al.*, 2002; Rashba, 2000, 2002; Vignale and D'Amico, 2003; Fert *et al.*, 2007; Žutić *et al.*, 2006b). In the following we adopt the treatment of Rashba (2000, 2002), using the notation from Žutić *et al.* (2004) where the mapping between the formulations of the spin injection problem by Johnson-Silsbee and Rashba is given.

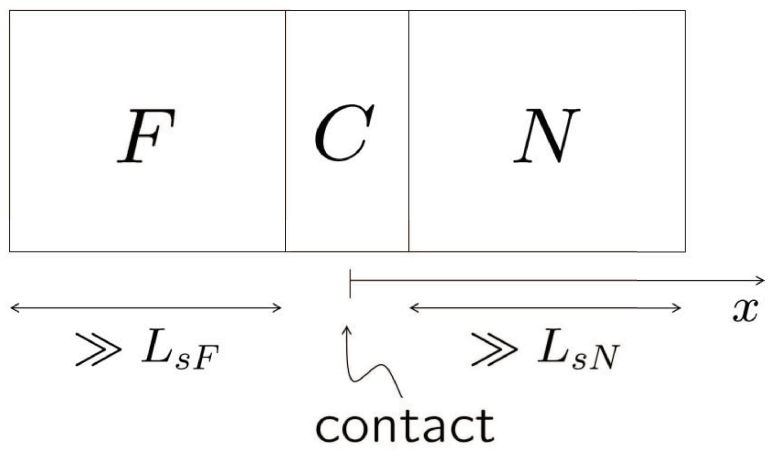

Fig. II.12. Scheme of our spin-injection geometry. The ferromagnetic conductor (F) forms a junction with the nonmagnetic conductor (N). The contact region (C) is assumed to be infinitely narrow, forming the discontinuity at $x = 0$. It is assumed that the physical widths of the conductors are greater than the corresponding spin diffusion lengths.

Our goal is to find the current spin polarization, $P_j(0)$, which determines the spin accumulation, $\mu_{sN}(0)$, in the normal conductor. We will assume that the lengths of the ferromagnet and the nonmagnetic regions are greater than the corresponding spin diffusion lengths. The spin injection scheme is illustrated in Fig. II.12. We assume that at the far ends of the junction, the nonequilibrium spin vanishes. We now look at the three regions separately.

### D.1 Ferromagnet

Adapting Eq. (II.107), the current spin polarization at $x = 0$ in the ferromagnet is,

$$P_{jF}(0) = P_{\sigma F} + \frac{1}{j} 4 \frac{\sigma_{F\uparrow}\sigma_{F\downarrow}}{\sigma_F} \nabla \mu_{sF}(0). \tag{II.116}$$

To obtain $\mu_{sF}(0)$, we need to look at the diffusion equation, Eq. (II.113), in the ferromagnet,

$$\nabla^2 \mu_{sF} = \frac{1}{L_{sF}^2} \mu_{sF}. \tag{II.117}$$

The solution, with the boundary condition, $\mu_{sF}(-\infty) = 0$, is,

$$\mu_{sF}(x) = \mu_{sF}(0) e^{x/L_{sF}}, \tag{II.118}$$

so that,

$$\nabla \mu_{sF} = \frac{\mu_{sF}}{L_{sF}}. \tag{II.119}$$

We then obtain for the spin polarization of the current,

$$P_{jF}(0) = P_{\sigma F} + \frac{1}{j}\frac{\mu_{sF}(0)}{R_F}, \tag{II.120}$$

where we have introduced an effective resistance of the ferromagnet,

$$R_F = \frac{\sigma_F}{4\sigma_{F\uparrow}\sigma_{F\downarrow}} L_{sF}. \tag{II.121}$$

This is not the electrical resistance of the region, only an effective resistance that appear in the spin-polarized transport and is roughly equal to the actual resistance of the region of size $L_{sF}$.[14] We finally obtain,

$$P_{jF}(0) = P_{\sigma F} + \frac{1}{j}\frac{\mu_{sF}(0)}{R_F}. \tag{II.122}$$

### D.2 Nonmagnetic conductor

Since in the nonmagnetic conductor $P_\sigma = 0$, and $\sigma_{N\uparrow} = \sigma_{N\downarrow} = \sigma_N/2$, we have,

$$P_{jN}(0) = \frac{1}{j}\sigma_N \nabla\mu_{sN}(0). \tag{II.123}$$

Similarly as in the ferromagnetic case, we need to solve the diffusion equation,

$$\nabla^2\mu_{sN} = \frac{1}{L_{sN}^2}\mu_{sN}, \tag{II.124}$$

now with the boundary condition of $\mu_{sN}(\infty) = 0$. The solution is

$$\mu_{sN} = \mu_{sN}(0)e^{-x/L_{sN}}; \tag{II.125}$$

the gradient,

$$\nabla\mu_{sN} = -\frac{1}{L_{sN}}\mu_{sN}. \tag{II.126}$$

The current spin polarization in the nonmagnetic conductor then becomes,

$$P_{jN}(0) = -\frac{1}{j}\frac{\mu_{sN}(0)}{R_N}, \tag{II.127}$$

where

$$R_N = \frac{L_{sN}}{\sigma_N}, \tag{II.128}$$

is the effective resistance of the nonmagnetic region. Once we would know $P_{jN}(0)$, we would also know the spin accumulation,

$$\mu_{sN}(0) = -jP_{jN}(0)R_N. \tag{II.129}$$

Spin accumulation is proportional to the spin current, $j_{sN}(0) = jP_{jN}(0)$, which pumps the spin into the system, as well as to the effective resistance, $R_N = L_{sN}/\sigma_N$—the greater is the spin diffusion length, the greater is the spin accumulation.

[14] If $\sigma_{F\uparrow} = \sigma_{F\downarrow} = \sigma/2$, as in nonmagnetic conductors, the effective resistance becomes $R_F = L_{sF}/\sigma$, which is the resistance of a unit cross sectional area; to get the ohmic resistance we would need to divide $R_F$ by the actual area.

### D.3 Contact

The advantage of the quasichemical potential model over continuous drift-diffusion equations for charge and spin current, is in describing the spin-polarized transport across the contact region at $x = 0$. Since we have a single point, we cannot define gradients to introduce currents. Instead, we resort to the discontinuity of the chemical potential across the contact and write,

$$j_\uparrow = \Sigma_\uparrow \left[\mu_{\uparrow N}(0) - \mu_{\uparrow F}(0)\right] = \Sigma_\uparrow \Delta\mu_\uparrow(0), \tag{II.130}$$

$$j_\downarrow = \Sigma_\downarrow \left[\mu_{\downarrow N}(0) - \mu_{\downarrow F}(0)\right] = \Sigma_\downarrow \Delta\mu_\downarrow(0). \tag{II.131}$$

Here we introduced spin-dependent contact conductances, not conductivities as in the bulk F and N regions, $\Sigma_\uparrow$ and $\Sigma_\downarrow$. In terms of charge and spin currents, this gives,

$$j = \Sigma\Delta\mu(0) + \Sigma_s\Delta\mu_s(0), \tag{II.132}$$

$$j_s = \Sigma_s\Delta\mu(0) + \Sigma\Delta\mu_s(0). \tag{II.133}$$

The conductance $\Sigma = \Sigma_\uparrow + \Sigma_\downarrow$, while the spin conductance $\Sigma_s = \Sigma_\uparrow - \Sigma_\downarrow$. Eliminating $\Delta\mu(0)$ from the equation for $j$ and substituting to $j_s$, we obtain,

$$j_s = P_\Sigma j + \frac{\Delta\mu_s(0)}{R_c}, \tag{II.134}$$

where the conductance spin polarization is,

$$P_\Sigma = \frac{\Sigma_\uparrow - \Sigma_\downarrow}{\Sigma}, \tag{II.135}$$

and the effective resistance of the contact is,

$$R_c = \frac{\Sigma}{4\Sigma_\uparrow\Sigma_\downarrow}. \tag{II.136}$$

Finally, for the current spin polarization, $j_s/j$, at the contact, we obtain,

$$P_{jc} = P_\Sigma + \frac{1}{j}\frac{\Delta\mu_s(0)}{R_c}. \tag{II.137}$$

### D.4 Spin injection and spin extraction

We have three equations for $P_j(0)$, Eqs. (II.122), (II.127), and (II.137), and five unknown quantities: $P_{jF}(0)$, $P_{jN}(0)$, $P_{jc}(0)$, $\mu_{sF}(0)$, and $\mu_{sN}(0)$. We need further physical assumptions to eliminate two unknown parameters. This assumption, which is an approximation, is the spin current continuity at the contact:

$$P_j = P_{jF}(0) = P_{jN}(0) = P_{jc}. \tag{II.138}$$

The above equalities are justified if spin-flip scattering (Galinon *et al.*, 2005; Bass and Prat Jr., 2007) can be neglected in the contact. For contacts with paramagnetic impurities, we would need to take into account contact spin relaxation which would lead to spin current discontinuity.

This assumption of the low rate of spin flip scattering at the interface should also be carefully reconsidered when analyzing room temperature spin injection experiments (Garzon *et al.*, 2005; Godfrey and Johnson, 2006).

Using the spin current continuity equations, we can solve our algebraic system and readily obtain for the spin injection efficiency,

$$P_j = \frac{R_F P_{\sigma F} + R_c P_\Sigma}{R_F + R_c + R_N} = \langle P_\sigma \rangle_R. \tag{II.139}$$

The spin injection efficiency is the conductivity spin polarization $P_\sigma$ averaged over the three regions, weighted by the effective resistances. In our linear regime the spin injection efficiency does not depend on the current. Equation (II.139) is the central result of the standard model of electrical spin injection.

What is the spin accumulation? We have earlier found that,

$$\mu_{sN}(0) = -jP_j R_N. \tag{II.140}$$

If $j < 0$, so that electrons flow from F to N, the spin accumulation is positive, $\mu_{sN}(0) > 0$; we speak of *spin injection*. If $j > 0$, the electrons flow from N to F, and $\mu_{sN}(0) < 0$; we speak of *spin extraction*. If we look at the density spin polarization, $P_n = s/n$, we get for the density spin polarization in the nonmagnetic region,

$$P_n(0) = e\mu_{sN}(0)\frac{g_N}{n} = -jeR_N \frac{g_N}{n} P_j. \tag{II.141}$$

The density spin polarization is roughly equal to the fraction of electrons in the energy interval of $jeR_N$ (the voltage drop at the distance of $L_{sN}$) times $P_j$. Since the injected spin polarization is proportional to the charge current, the electrical spin injection is an example of *spin pumping*.[15]

### D.5 The equivalent circuit of F/N spin injection

The standard model of spin injection can be summarized by the equivalent electrical circuit shown in Fig. II.13. Spin up and spin down electrons form parallel channels for electric current. Each region of the junction is characterized by its own effective resistance, determined by the spin diffusion lengths in the bulk regions, or by the spin-dependent conductances in the contact (Jonker *et al.*, 2003).

It is a simple exercise, left to the reader, to show that the equivalent circuit in Fig. II.13 leads to Eq. (II.139) for the current spin polarization, $P_j = (I_\uparrow - I_\downarrow)/I$.

### D.6 Quasichemical potentials, nonequilibrium resistance, and spin bottleneck

The spin quasichemical potential exhibits a drop at the contact region. This drop follows from Eq. (II.137):

$$\Delta\mu_s(0) = \mu_{sN}(0) - \mu_{sF}(0) = jR_c(P_j - P_\Sigma). \tag{II.142}$$

[15] Spin pumping in optical orientation refers to spin orientation by absorption of circularly polarized light of an n-doped semiconductor. Since the excited spin polarization is shared by the existing Fermi sea of electrons, the more intense the light the more spin polarization. In contrast, in p-doped samples absorption of circularly polarized light results in electron spin polarization that is independent of the light intensity. See (Žutić *et al.*, 2004).

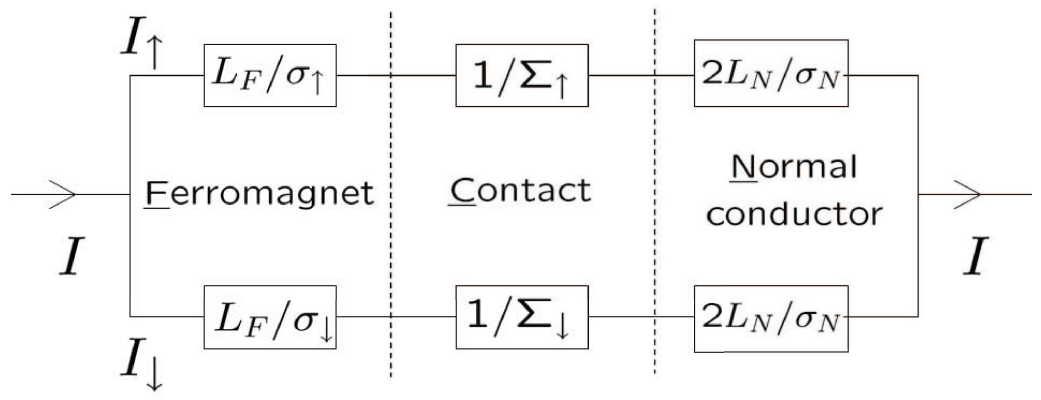

Fig. II.13. The equivalent circuit of the standard model of spin injection in F/N junctions. The ferromagnet, contact, and normal conductor regions are identified. The electric current splits into the spin up and spin down components, each passing through the corresponding spin-resolved resistors.

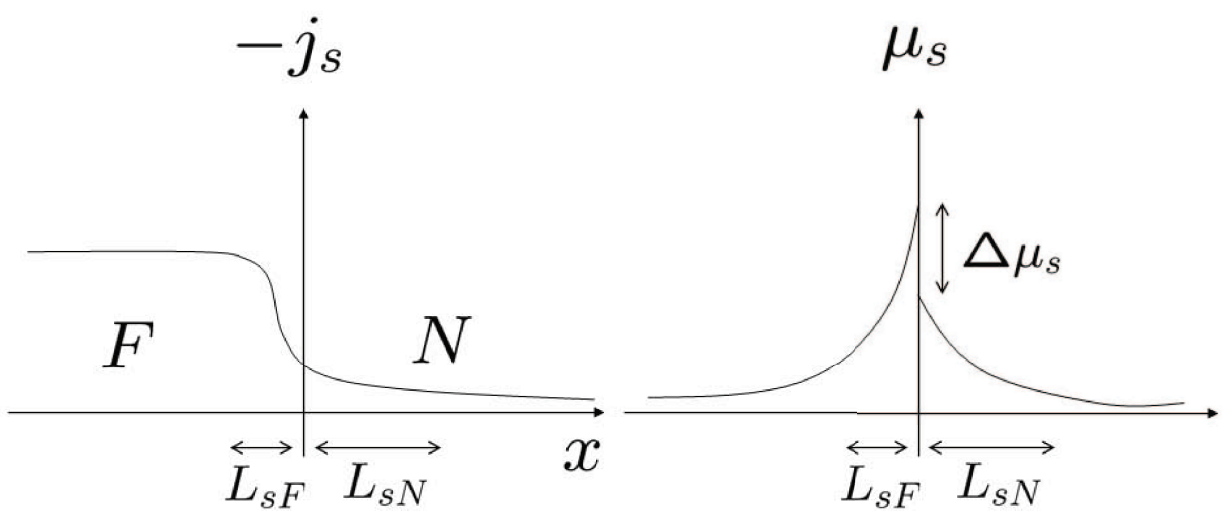

Fig. II.14. The left figure illustrates the profile of the spin current in the F/N junction, for $j < 0$, that is, the spin injection regime. The spin current is assumed continuous at the contact, $x = 0$. The right figure illustrates the nonequilibrium spin quasichemical potentials, under the same conditions. The potentials exhibit discontinuity proportional to the electric current, contact resistance, and spin current polarization. The nonequilibrium spin properties decay on the length scales of the corresponding spin diffusion lengths.

Since $P_j$ is a materials parameter, the drop of the spin quasichemical potential across the contact changes sign with flipping the direction of the charge current. The spatial profile of the spin current density, $j_s$, as well as that of $\Delta\mu_s$, is illustrated in Fig. II.14.

Thus far we looked at spin properties. Is there anything useful to be learned from the charge quasichemical potential $\mu$ which we swapped earlier for the local charge neutrality? We have already found, see Eq. (II.104), that

$$\nabla\mu = \frac{1}{\sigma}\left(j - \sigma_s \nabla\mu_s\right) = \frac{j}{\sigma} - P_\sigma \nabla\mu_s. \tag{II.143}$$

For the ferromagnet,

$$\nabla\mu_F = \frac{j}{\sigma_F} - P_{\sigma F}\nabla\mu_{sF}, \tag{II.144}$$

while for the normal conductor,

$$\nabla\mu_N = \frac{j}{\sigma_N}. \tag{II.145}$$

Integrating the above equations for $\mu_F$ and $\mu_N$, we obtain

$$\mu_F(-\infty) - \mu_F(0) = -j\tilde{R}_F - P_{\sigma F}\left[\mu_{sF}(-\infty) - \mu_{sF}(0)\right], \quad \text{(II.146)}$$

$$\mu_N(\infty) - \mu_N(0) = j\tilde{R}_N, \quad \text{(II.147)}$$

Here we introduced,

$$\tilde{R}_F = \int_{-\infty}^{0} dx \frac{1}{\sigma_F} \equiv \frac{L_F}{\sigma_F}, \quad \text{(II.148)}$$

which is the actual resistance (of a unit cross-sectional area) of the ferromagnetic region of size $L_F \gg L_{sF}$, taken to be infinity in the arguments of the quasichemical potentials; recall that $R_F$ is an effective resistance of a region of length $L_{sF}$. Similarly, $\tilde{R}_N$ is the actual resistance of the nonmagnetic conductor.

Now, $\mu_{sF}(-\infty) = 0$, as we assume no spin accumulation at the left end of the ferromagnet. By subtracting Eq. (II.146) from Eq. (II.147), we get,

$$[\mu_N(\infty) - \mu_F(-\infty)] - [\mu_N(0) - \mu_F(0)] = j\tilde{R}_N + j\tilde{R}_F - P_{\sigma F}\mu_{sF}(0). \quad \text{(II.149)}$$

Normally we would expect Ohm's law for our junction in the form,

$$\mu_N(\infty) - \mu_F(-\infty) = \left(\tilde{R}_N + \tilde{R}_F + \frac{1}{\Sigma}\right) j, \quad \text{(II.150)}$$

that is, the total drop of the quasichemical potential (or voltage for a charge neutral conductor) is given by the serial resistance of the three regions times the charge current. Instead, we see that in the presence of spin accumulation the junction resistance acquires a correction $\delta R$, modifying Ohm's law as,

$$\mu_N(\infty) - \mu_F(-\infty) = \left(\tilde{R}_N + \tilde{R}_F + \frac{1}{\Sigma} + \delta R\right) j. \quad \text{(II.151)}$$

Substituting this nonequilibrium Ohm's law into Eq. (II.149) we obtain,

$$\left(\frac{1}{\Sigma} + \delta R\right) j = -P_{\sigma F}\mu_{sF}(0) + \mu_N(0) - \mu_F(0). \quad \text{(II.152)}$$

At the contact,

$$\Delta\mu(0) = \mu_N(0) - \mu_F(0) = \frac{j}{\Sigma} - P_\Sigma \Delta\mu_s(0), \quad \text{(II.153)}$$

so that there is an additional resistance due to the spin accumulation $\Delta\mu_s(0)$. Since $\Delta\mu_s(0)$ is proportional to $j$, see Eq. (II.142), the quasichemical potential drop at the contact is

$$\Delta\mu(0) = \frac{j}{\Sigma} - jP_\Sigma(P_j - P_\Sigma)R_c. \quad \text{(II.154)}$$

Substituting the above into Eq. (II.152) gives the nonequilibrium resistance due to spin accumulation,

$$\delta R = -P_\Sigma(P_j - P_\Sigma)R_c - P_{\sigma F}R_F(P_j - P_{\sigma F}). \quad \text{(II.155)}$$

Substituting for the spin injection efficiency $P_j$ from Eq. (II.139) we finally obtain,

$$\delta R = \frac{R_N(P_\Sigma^2 R_c + P_{\sigma F}^2 R_F) + R_F R_c (P_{\sigma F} - P_\Sigma)^2}{R_F + R_c + R_N}. \tag{II.156}$$

What is important, this resistance correction is always positive: $\delta R > 0$. One can also obtain this nonequilibrium spin resistance from the equivalent circuit model, of Fig. II.13, by calculating $\delta R = R - L_{sF}/\sigma_F - L_{sN}/\sigma_N$, with $R$ being the circuit resistance. We leave this exercise to the reader.

Why does the additional resistance appear? Spin accumulation leads to nonequilibrium spins in the ferromagnet as well as in the contact. The nonequilibrium spin causes spin diffusion, driving the spin away from the contact. Since in the ferromagnet, as well as in the spin-polarized contact, any spin current gives charge current (due to the nonvanishing $P_\sigma$), this spin flow causes electron flow which is oriented opposite to the electron flow due to the external battery. This opposition to the charge current, which does not depend on the direction of the current flow, and manifests itself as the additional resistance $\delta R$, is called the *spin bottleneck* effect (Johnson, 1991).

We will now consider two important limits of the spin injection model: transparent and tunnel contacts.

### D.7 Transparent contact

By a transparent contact we mean that the following condition is satisfied:

$$R_c \ll R_N, R_F. \tag{II.157}$$

In this limit the following equations characterize the F/N junction:

$$P_j = \frac{R_F}{R_N + R_F} P_{\sigma F}, \tag{II.158}$$

$$\delta R = \frac{R_N R_F}{R_N + R_F} P_{\sigma F}^2, \tag{II.159}$$

$$P_{nN}(0) = -ejR_N \frac{g_N}{n_N} P_j. \tag{II.160}$$

Recall that $P_n = s/n$ is the spin polarization of the electron density. The spin injection depends on the spin properties of the ferromagnet. If the F and N regions are equally conducting, $R_N \approx R_F$, then the spin injection efficiency is high:

$$P_j \approx P_{\sigma F}. \tag{II.161}$$

This is the usual case of a spin injection from a ferromagnetic metal to a normal metal, or from a ferromagnetic (magnetic) semiconductor to a normal semiconductor.[16]

---

[16]When we say that $R_N$ and $R_F$ should be similar, we need to decipher this statement from the definition of the effective resistance:

$$\frac{L_{sF}}{\sigma_F} \approx \frac{L_{sN}}{\sigma_N}. \tag{II.162}$$

Typically $\sigma_N$ is about an order of magnitude greater than $\sigma_F$, but $L_{sF}$ is at least an order of magnitude smaller than $L_{sN}$; $R_F$ is then somewhat smaller than $R_N$, although $R_F/R_N$ is still a significant fraction of one.

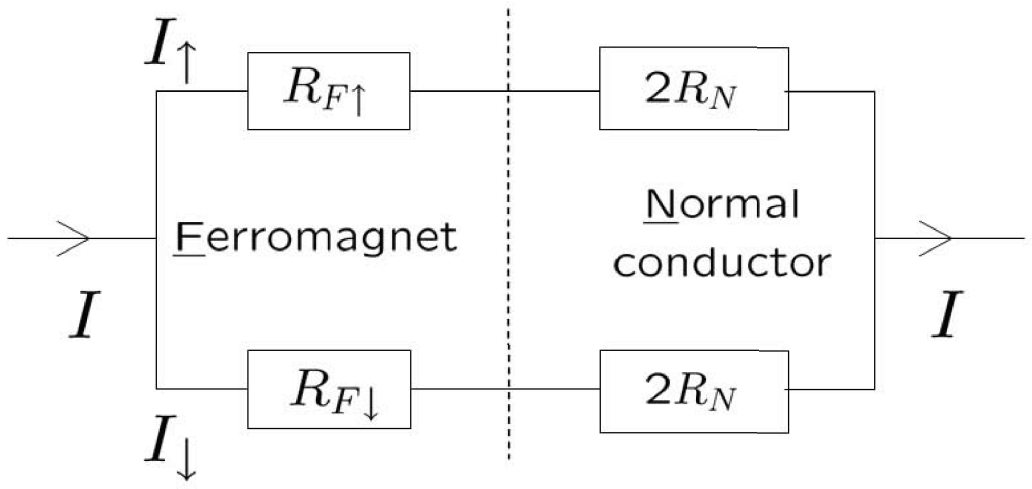

Fig. II.15. The equivalent circuit for a transparent contact, $R_c \ll R_N, R_F$. If, in addition, $R_N \gg R_F$, the voltage drop occurs mainly in the nonmagnetic region, and $I_\uparrow = I_\downarrow$, so that no spin-polarized current flows.

What happens if we are to inject spin from a ferromagnetic metal to a nonmagnetic semiconductor? Since the semiconductor has a much smaller carrier density than the ferromagnetic metal, we then have $R_N \gg R_F$, in which case

$$P_j \approx \frac{R_F}{R_N} P_{\sigma F} \ll P_{\sigma F}, \quad \text{(II.163)}$$

$$\delta R \approx R_F P_{\sigma F}^2 \approx R_F, \quad \text{(II.164)}$$

$$P_{nN}(0) \approx -ejR_F \frac{g_N}{n_N} P_{\sigma F} \approx -\frac{eV_F}{\varepsilon_{FN}} P_{\sigma F}, \quad \text{(II.165)}$$

where $V_F$ is the voltage drop in $F$ along $L_{sF}$ and $\varepsilon_{FN}$ is the Fermi energy of the nonmagnetic conductor. The spin injection efficiency is greatly reduced! The inhibited spin injection efficiency in such a (highly idealized) ferromagnet/semiconductor junction is termed the *conductivity mismatch* problem (Schmidt *et al.*, 2000), since it comes from the largely different conductivities of the ferromagnetic metal and the semiconductor.

The basic physics is nicely illustrated by the equivalent circuit, shown in Fig. II.15. If $R_N \gg R_{F\uparrow}, R_{F\downarrow}$, the voltage drop along this parallel-resistor circuit is $V \approx 2R_N I_\uparrow \approx 2R_N I_\downarrow$, so that $I_\uparrow \approx I_\downarrow$. The currents in the spin up and spin down channels are determined solely by the effective resistance of the semiconductor since this is where all the voltage drops. As a result, $P_j \ll P_{\sigma F}$, meaning that spin injection is inefficient. Similarly inhibited is the spin bottleneck effect: $\delta R \approx R_F \ll R_N$. What about the injected spin polarization, $P_{nN}(0)$, given now by Eq. (II.165)? Since the Fermi energy of the semiconductor is much smaller than that of a metal, the spin polarization is not strongly reduced, compared to an all-metallic F/N transparent junction. Still, the magnitude is tiny: $P_{nN}(0) \ll 1$.[17]

[17]Unlike in metals, injected spin polarization in semiconductors is in general significant, with $P_n$ reaching possibly tens of percent (not in transparent junctions, though). In principle, our theory does not directly apply to such cases since we have assumed only small perturbations from the spin-unpolarized equilibrium, as well as strictly non-spin polarized materials parameters in the $N$ region. In particular, $P_{\sigma N}$ is in general a nonnequilibrium quantity, which is finite since there is a different conductivity for spin up and down electrons in $N$ if the corresponding Fermi levels differ significantly.

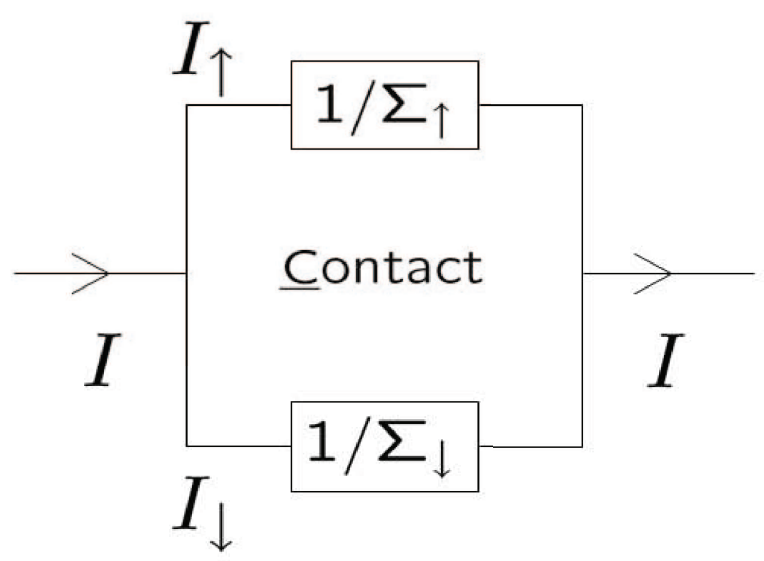

Fig. II.16. The equivalent circuit for a tunnel contact, $R_c \gg R_N, R_F$. The voltage drops mainly at the contact, so $I_\uparrow \neq I_\downarrow$, and spin-polarized current flows.

### D.8 Tunnel contact

If the contact resistance is the largest resistance in the junction, $R_c \gg R_F, R_N$, we call the contact a tunnel contact. The following equations define the spin injection in tunnel contacts:

$$P_j = P_\Sigma, \tag{II.166}$$

$$\delta R = \frac{R_N R_c P_\Sigma^2 + R_F R_c (P_{\sigma F} - P_\Sigma)^2}{R_c}, \tag{II.167}$$

$$P_{nN}(0) = -ej\frac{g_N}{n_N} R_N P_\Sigma. \tag{II.168}$$

All the interesting quantities related to spin depend on $P_\Sigma$. The contact acts as a spin filter.

The addition of the high resistive contact enables to inject spin from a ferromagnetic metal into a semiconductor, resolving the conductivity mismatch problem discussed in the previous section. Indeed, if $R_N \gg R_F$, we have

$$P_j = P_\Sigma, \tag{II.169}$$

$$\delta R \approx R_N P_\Sigma^2, \tag{II.170}$$

$$P_{nN}(0) \approx -je\frac{g_N}{n_N} R_N P_\Sigma = -\frac{eV_N}{\varepsilon_{FN}} P_\Sigma. \tag{II.171}$$

Here $V_N$ is the voltage drop in the $N$ region along the length of $L_{sN}$. The spin injection is still determined by the contact conductance polarization,[18] while the nonequilibrium resistance, $\delta R$, is roughly equal to the effective resistance of the semiconductor, $R_N \gg R_F$. In the conductivity mismatch problem, the nonequilibrium resistance was much smaller, $\delta R \approx R_F$. Finally, the injected spin polarization is now also much greater than for a transparent contact, since $L_{sN} \gg L_{sF}$, so that $V_N \gg V_F$ (see Eq. (II.165)).

The spin filtering is nicely seen in the equivalent circuit of Fig. II.16. The voltage drop across the circuit is $V = I_\uparrow/\Sigma_\uparrow = I_\downarrow/\Sigma_\downarrow$, so that $P_j = P_\Sigma$.

[18] In a tunnel junction between a ferromagnetic and nonmagnetic conductor, the electron transmission through the tunnel barrier depends, in general, on the electron spin. This gives rise to $P_\Sigma \neq 0$ and spin filtering.

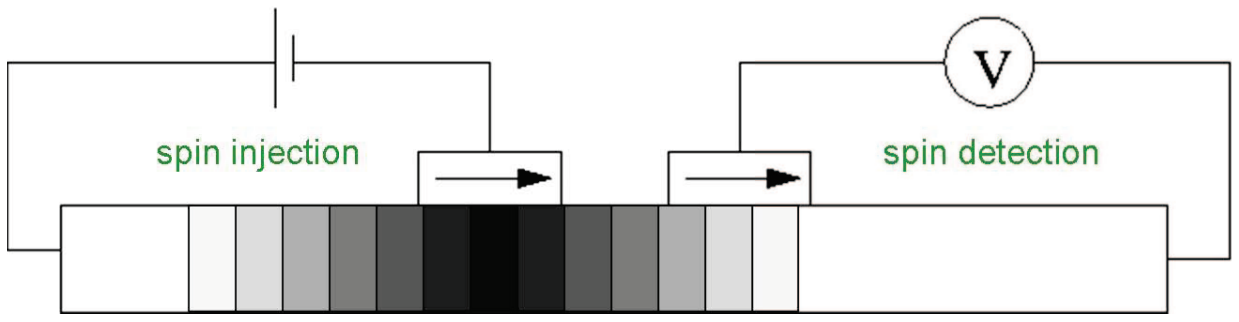

Fig. II.17. The Johnson-Silsbee non-local spin injection and detection scheme. Spin is injected through one F/N junction. The spin detection is done by a different F/N junction, by the Silsbee-Johnson spin-charge coupling. Spin diffusion from the injector is indicated by the different shades of grey.

Finally, let us look at the ferromagnetic region, in which the spin accumulation is

$$\mu_{sF}(0) = R_F j(P_j - P_{\sigma F}). \tag{II.172}$$

If $P_{\sigma F} \ll P_j \approx P_\Sigma$, then the accumulation is negative, $\mu_{sF}(0) < 0$, for electrons flowing from the ferromagnet to the nonmagnetic conductor, $j < 0$. On the other hand, the accumulation is positive in the nonmagnetic conductor: $\mu_{sN}(0) = -R_N j P_\Sigma > 0$. If the contact spin-filtering dominates, in the electrical spin injection setup there is a spin extraction from the ferromagnet and a spin injection into the nonmagnetic conductor.

It is a nice exercise for the reader to consider, instead of an infinite nonmagnetic conductor, a finite $N$ region of width $w$. What is the effective resistance, $R_N$, in this case?

### D.9 Silsbee-Johnson spin-charge coupling

In electrical spin injection we drive spin-polarized electrons from a ferromagnet into a nonmagnetic conductor. As a result, nonequilibrium spin accumulates in the nonmagnetic conductor. The opposite is also true: If a spin accumulation is generated in a nonmagnetic conductor that is in a proximity of a ferromagnet, a current flows in a closed circuit, or an electromotive force (emf) appears in an open circuit. This inverse effect is called the *Silsbee-Johnson spin-charge coupling*. This coupling was first proposed by Silsbee (1980) and experimentally demonstrated by Johnson and Silsbee (1985) in the first electrical spin injection experiment.

In this section we will calculate the emf resulting from the spin-charge coupling, using the standard model of spin injection. Our physical system is shown in Fig. II.17. Spin is injected by the left ferromagnetic electrode, and detected by the right one, making it a non-local measurement. The injected spin diffuses in all directions (here left and right), unlike for the charge current. The nonequilibrium spin at the right ferromagnetic electrode is picked-up by the Silsbee-Johnson spin-charge coupling, producing a measurable emf in the right circuit. [19]

Consider an F/N junction with a special boundary condition: a nonequilibrium spin is maintained, by whatever means,[20] at the far right boundary of the nonmagnetic conductor:

$$\mu_{sN}(\infty) \neq 0. \tag{II.173}$$

At the far left boundary of the ferromagnetic region, the spin is assumed to be in equilibrium:

$$\mu_{sF}(-\infty) = 0. \tag{II.174}$$

[19] A similar approach was used recently to detect a spin Hall effect in metals (Valenzuela and Tinkham, 2006).

[20] The source could be electrical spin injection or optical spin orientation, for example

One of our goals is to find the spatial profile of the spin accumulation inside the junction. Our main goal is then to calculate the induced electromotive force, defined by

$$\mathrm{emf} = \mu_N(\infty) - \mu_F(-\infty), \tag{II.175}$$

under the condition of no electrical current flow (open circuit):

$$j = 0. \tag{II.176}$$

The emf represents the drop of the quasichemical potential, $\mu$, across the junction. If such a drop is present, the system acts as a battery: by closing the circuit, charge current flows. In electrical and spin equilibrium, the quasichemical potential drop must vanish.

We have found earlier, see Eq. (II.99), that the local charge neutrality requires that in degenerate conductors,

$$\phi(x) = -\mu(x) - P_g \mu_s(x). \tag{II.177}$$

In the ferromagnet,

$$\phi_F(x) = -\mu_F(x) - P_{gF}\mu_{sF}(x), \tag{II.178}$$

while in the nonmagnetic conductor,

$$\phi_N(x) = -\mu_N(x), \tag{II.179}$$

since $P_{gN} = 0$. Since $\mu_{sF}(-\infty) = 0$, the drop of the chemical potential is equal the (minus) drop of the electrostatic potential:

$$\mathrm{emf} = \mu_N(\infty) - \mu_F(-\infty) = \phi_F(-\infty) - \phi_N(\infty). \tag{II.180}$$

The emf can be detected as a voltage drop.[21]

From the drift-diffusion model, Eq. (II.144), we have,

$$\nabla\mu_F = \frac{j}{\sigma_F} - P_{\sigma F}\nabla\mu_{sF} = -P_{\sigma F}\nabla\mu_{sF}, \tag{II.181}$$

since $j = 0$. Integrating this equation in the F region, from $-\infty$ to 0, and recalling that $\mu_{sF}(-\infty) = 0$, the following holds,

$$\mu_F(-\infty) - \mu_F(0) = -P_{\sigma F}\left[\mu_{sF}(-\infty) - \mu_{sF}(0)\right] = P_{\sigma F}\mu_{sF}(0). \tag{II.182}$$

Similarly for the $N$ region,

$$\mu_N(\infty) - \mu_N(0) = 0. \tag{II.183}$$

[21] If there is a charge (also called space charge) in a conductor at equilibrium, there is an additional electrostatic voltage drop–an equilibrium one–due to this built-in charge. In such cases the emf vanishes, it is not given by the (minus) voltage drop, since there is a voltage drop even in equilibrium; no charge current would flow if we closed the circuit. A most prominent example of this is the p-n junction. Only if the voltage drop is accompanied by a drop in the quasichemical potential, can we speak of a nonequilibrium situation giving rise to emf.

There is a drop of the quasichemical potential in the F region, due to the spin-polarization of the conductivity, while the quasichemical potential is constant over the N region. Then

$$\text{emf} = \mu_N(\infty) - \mu_F(-\infty) = \Delta\mu(0) - P_{\sigma F}\mu_{sF}(0). \tag{II.184}$$

Recall that the drop of the quasichemical potential across the contact region is,

$$\Delta\mu(0) = \mu_N(0) - \mu_F(0). \tag{II.185}$$

We need to determine both the quasichemical potential drop, $\Delta\mu(0)$, at the contact, as well as the spin accumulation in the ferromagnet, $\mu_{sF}(0)$, in order to calculate the emf.

Let us first calculate the drop of the electrostatic potential across the contact. The charge neutrality gives,

$$\Delta\phi(0) = \phi_N(0) - \phi_F(0) = -\Delta\mu(0) + P_{gF}\mu_{sF}(0). \tag{II.186}$$

Recall the drift diffusion model at the contact:

$$j = \Sigma\Delta\mu(0) + \Sigma_s\Delta\mu_s(0), \tag{II.187}$$
$$j_s = \Sigma_s\Delta\mu(0) + \Sigma\Delta\mu_s(0). \tag{II.188}$$

Eliminate $\Delta\mu_s(0)$ from the second equation, substitute for the first, and use that $j = 0$ to obtain the quasichemical potential drop at the contact,

$$\Delta\mu(0) = -R_c P_\Sigma j_s(0), \tag{II.189}$$

where $R_c$ is the effective contact resistance given by Eq. (II.136).

We then have,

$$\Delta\phi(0) = R_c P_\Sigma j_s(0) + P_{gF}\mu_{sF}(0). \tag{II.190}$$

We have now connected the voltage drop across the contact to the spin current, $j_s$, and the spin accumulation in the ferromagnet, $\mu_{sF}$, at the contact.

With the boundary condition, Eq. (II.173), the following equation describes the spatial profile of $\mu_s$:

$$\mu_{sN}(x) = \mu_{sN}(\infty) + [\mu_{sN}(0) - \mu_{sN}(\infty)]\, e^{-x/L_{sN}}. \tag{II.191}$$

The value of $\mu_{sN}(0)$ is yet to be determined. The above equation gives,

$$\nabla\mu_{sN}(0) = -\frac{1}{L_{sN}}\left[\mu_{sN}(0) - \mu_{sN}(\infty)\right]. \tag{II.192}$$

Using the condition of $j = 0$, we then obtain the following set of equations for the spin currents at $x = 0$:

$$j_{sN}(0) = -\frac{1}{R_N}\left[\mu_{sN}(0) - \mu_{sN}(\infty)\right], \tag{II.193}$$
$$j_{sF}(0) = \frac{1}{R_F}\mu_{sF}(0), \tag{II.194}$$
$$j_{sc} = \frac{1}{R_c}\Delta\mu_s(0). \tag{II.195}$$

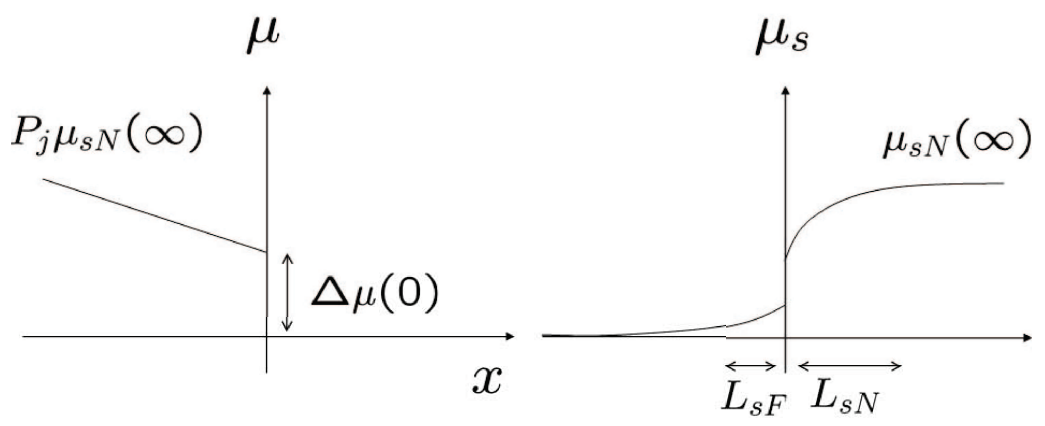

Fig. II.18. Sketch of the profile for the quasichemical potential $\mu$ (left) and the spin accumulation $\mu_s$ (right).

Assuming again that the spin is conserved across the interface at $x = 0$, that is,

$$j_s = j_{sF}(0) = j_{sc} = j_{sN}(0), \tag{II.196}$$

we obtain for $\Delta\phi(0)$ by substituting for $j_s = j_{sF}(0)$ from Eq. (II.194) into Eq. (II.190) :

$$\Delta\phi(0) = \left(\frac{R_c P_\Sigma + R_F P_{gF}}{R_F}\right) \mu_{sF}(0). \tag{II.197}$$

The quasichemical potential, $\mu_{sN}(0)$, can be obtained by eliminating $j_s$ and $\mu_{sN}(0)$ from the Eqs. (II.193), (II.194), and (II.195):

$$\mu_{sF}(0) = \frac{R_F}{R_F + R_c + R_N}\, \mu_{sN}(\infty) < \mu_{sN}(\infty). \tag{II.198}$$

This allows to write the spin current at the contact as,

$$j_s(0) = \frac{1}{R_F + R_c + R_N} \mu_{sN}(\infty). \tag{II.199}$$

Substituting for the spin current in Eq. (II.189), we obtain for the quasichemical potential drop,

$$\Delta\mu(0) = -\frac{R_c P_\Sigma}{R_F + R_c + R_N}\, \mu_{sN}(\infty). \tag{II.200}$$

The drop of the quasichemical potential across the contact is due to the spin filtering effect of the contact. If the contact conductance were spin-independent, the chemical potential would be continuous at $x = 0$. Furthermore, substituting for the spin quasichemical potential from Eq. (II.198) into the equation for the electrical voltage drop, Eq. (II.197), yields,

$$\Delta\phi(0) = \frac{R_c P_\Sigma + R_F P_{gF}}{R_F + R_c + R_N}\, \mu_{sN}(\infty). \tag{II.201}$$

The electrostatic potential drop across the contact is due to the spin polarization of the ferromagnet as well as due to the spin filtering effects of the contact. The profiles for the quasichemical and spin quasichemical potentials across the junction are shown in Fig. II.18. We leave as an exercise for the reader to calculate the profile of the spin current across the junction in the regime of the spin-charge coupling.

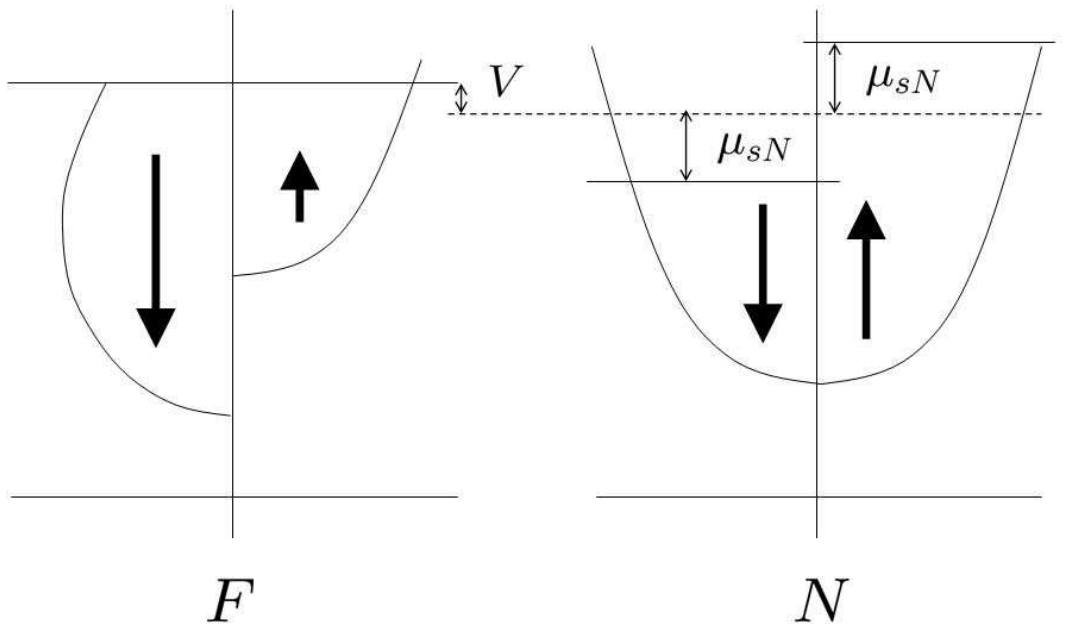

Fig. II.19. Toy model of the Silsbee-Johnson spin-charge coupling. The ferromagnetic region (F) has exchange split bands. The nonmagnetic conductor (N) has nonequilibrium spin of accumulation $\mu_{sN}$. The applied voltage $V$ gives an offset of the quasichemical potentials in F and N.

We are now ready to write the formula for the emf. Substituting Eq. (II.198) for $\mu_{sF}(0)$ and Eq. (II.200) for $\Delta\mu(0)$ into Eq. (II.184), the spin-induced emf becomes,

$$\text{emf} = -\frac{R_F P_{\sigma F} + R_c P_{\Sigma}}{R_F + R_c + R_N}\,\mu_{sN}(\infty) = -P_j\,\mu_{sN}(\infty). \tag{II.202}$$

Recall that $P_j$ is the spin injection efficiency, given by Eq. (II.139). If there is a positive spin accumulation at $x = \infty$ so that $\mu_{sN}(\infty) > 0$, and if the junction spin polarization is also positive, $P_j > 0$, the emf is negative: the quasichemical potential in the F region is greater than that in the N region. To preserve charge neutrality, opposite holds for the electrostatic potential: there is a higher electrostatic potential in N than in F. If an external circuit is added into our system, maintaining the spin accumulation at $x = \infty$, the current would flow from N to F (electrons from F to N) through the external circuit, and from F to N (electrons from N to F) through our system,[22] completing the circuit. Reversing either $P_j$ or $\mu_{sN}(\infty)$ (not both), changes the sign of the emf, in a spin-valve fashion. In an open circuit electrons create opposite surface charges at the far ends of the F and N regions: the equilibrium is reached by electrons flowing from N to F until an equilibrium electric field (and the associated voltage drop $\Delta\phi$) is established to prevent further flow of electrons.

The spin-charge coupling, given by Eq. (II.202), says that there is an emf developed if an equilibrium spin, here $P_j$, is in electrical contact with a nonequilibrium spin, here $\mu_{sN}(\infty)$. This effect allows detection of nonequilibrium spin, by putting a ferromagnetic electrode over the region of spin accumulation. By measuring the emf across this junction, we obtain information about the spin in the nonmagnetic conductor.

The spin-charge coupling can also be understood from the following toy model of the electronic structure of F and N regions, as illustrated in Fig. II.19. There is a nominal voltage $V$ applied across the junction, while a nonequilibrium spin in the nonmagnetic region is described by the spin quasichemical potential $\mu_{sN}$. The tunneling current across the junction is proportional to the transmission probability, the product of the densities of states, and the voltage drop. The spin up and spin down currents thus are

$$I_\uparrow \sim t_\uparrow g_N g_{F\uparrow}(V + \mu_{sN}), \tag{II.203}$$
$$I_\downarrow \sim t_\downarrow g_N g_{F\downarrow}(V - \mu_{sN}), \tag{II.204}$$

where $t_\sigma$ is the tunneling probability for electrons with spin $\sigma$. Let us denote as $t = (t_\uparrow + t_\downarrow)/2$ and $t_s = (t_\uparrow - t_\downarrow)/2$. The total current then is

$$I = I_\uparrow + I_\downarrow \sim g_N(tg_F + t_s g_{sF})V + g_N(tg_{sF} + t_s g_F)\mu_{sN}. \tag{II.205}$$

[22]This is the signature effect of a battery: electrons in the battery are driven, by what one could term *spin affinity*, against the electrical force.

The current flows even in the absence of external bias, $V = 0$:

$$I \sim (tg_{sF} + t_s g_F)\mu_{sN}. \tag{II.206}$$

This spin-charge coupling current is proportional to the equilibrium materials parameters of the ferromagnet and the contact, and due to the spin accumulation in the nonmagnetic conductor. In an open circuit we put $I = 0$, and the resulting voltage $V$ is the emf:

$$\text{emf} = -\frac{P_{gF} + P_t}{1 + P_t P_{gF}}\mu_{sN}. \tag{II.207}$$

Here $P_t$ is the spin polarization of the transmission rates, and $P_g$ is the spin polarization of the density of states. If a conducting region with an equilibrium magnetization (characterized by $P_g$) is in proximity with a conducting region of a nonequilibrium polarization ($\mu_{sN}$), an emf is generated. The same holds if the contact between the conductors is a spin-filter.

## E. Spin dynamics

### E.1 Drift-diffusion model for spin dynamics

Let us come back to our random walk model of spin-polarized electrons, introduced in Sec. B. We can generalize the model by considering the possibility of the rotation of the electron spins, due to the presence of an external magnetic field $\mathbf{B}$. A spin $\mathbf{s}$ undergoes a precession according to the equation,

$$\frac{d\mathbf{s}}{dt} = \mathbf{s} \times \boldsymbol{\omega}_\mathbf{0}. \tag{II.208}$$

Here $\boldsymbol{\omega}_\mathbf{0} = \gamma\mathbf{B}$, with $\gamma$ denoting the gyromagnetic ratio, is the directed Larmor frequency. If at time $t$ the spin is $\mathbf{s}$, at time $t + \tau$, where $\tau \ll 1/\omega_0$, the spin will be,

$$\mathbf{s}(t + \tau) = \mathbf{s}(t) + \mathbf{s}(t) \times \boldsymbol{\omega}_\mathbf{0}\tau. \tag{II.209}$$

The spin changes by an amount proportional to the product of the Larmor frequency and the time $\tau$ of precession. The absolute value of the product gives the change of the phase of a spin precessing transverse to the magnetic field.

Suppose that the electrons undergo random walk, with the step size $l$ over a time step $\tau$, as depicted in Fig. II.20. The spin at point $x$ at time $t + \tau$ will be given by the sum of the spins at $x+l$ and $x-l$, at time $t$, each rotated about $\omega_0$ over the time step $\tau$ (the electron spin precesses while the electrons wait for their turn to jump) and decreased by the fraction of $\tau/\tau_s$ due to spin relaxation:[23]

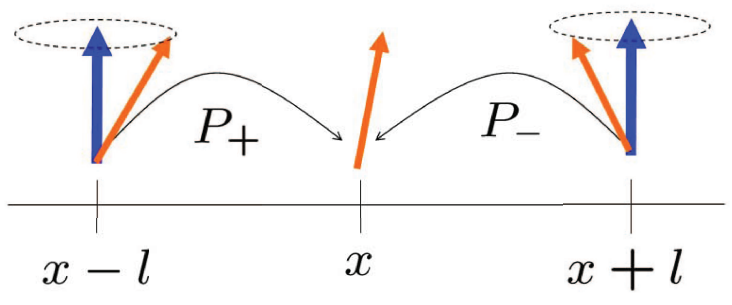

Fig. II.20. Spin precession and random walk. Electrons have probabilities $p_+$ and $p_-$ to jump either right or left. If an applied magnetic field (blue vertical arrow) is present, the electrons' spins precess. At $x$ the spin is a weighted average of the spin at $x-l$ and $x+l$ points, at the time of jump.

$$\begin{aligned}\mathbf{s}(x,t+\tau) &= p_+\left[\mathbf{s}(x-l,t)+\mathbf{s}(x-l,t)\times\boldsymbol{\omega_0}\tau-\mathbf{s}(x-l,t)\frac{\tau}{\tau_s}\right] \\ &+ p_-\left[\mathbf{s}(x+l,t)+\mathbf{s}(x+l,t)\times\boldsymbol{\omega_0}\tau-\mathbf{s}(x+l,t)\frac{\tau}{\tau_s}\right]. \qquad \text{(II.210)}\end{aligned}$$

As before, we denote by $p_+$ and $p_-$ the probability for the electrons to jump right and left, respectively. We can expand the left-hand side in Taylor series around $t$, and the right-hand side in Taylor series around $x$, in full analogy with the discussion in Sec. B., and obtain:

$$\frac{\partial \mathbf{s}}{\partial t} = \mathbf{s}\times\boldsymbol{\omega_0} + D\nabla^2\mathbf{s} + \mu E\nabla\mathbf{s} - \frac{\mathbf{s}}{\tau_s}. \qquad \text{(II.211)}$$

The above is the drift-diffusion equation for spin dynamics. The first term on the right-hand side describes spin precession, while the rest describe spin diffusion, spin drift, and spin relaxation, respectively. In principle, the last term describing spin relaxation should read $(\mathbf{s}-\mathbf{s}_0)/\tau_s$, since spin relaxation yields equilibrium spin, $\mathbf{s}_0$. For small magnetic fields in degenerate conductors the tiny equilibrium spin can be neglected from the treatment of dynamical effects.

Can we write a continuity equation for spin including spin dynamics? If we suppose charge neutrality, so that the electric field is constant, we can rewrite Eq. (II.211) in the form,

$$\frac{\partial \mathbf{s}}{\partial t} - \nabla\left[\mu E\mathbf{s} - D\nabla\mathbf{s}\right] = \mathbf{s}\times\boldsymbol{\omega_0} - \frac{\mathbf{s}}{\tau_s}. \qquad \text{(II.212)}$$

The expression inside the brackets is the generalized spin current,[24]

$$\mathbf{J}_s = -\mu E\mathbf{s} - D\nabla\mathbf{s}. \qquad \text{(II.213)}$$

[23] We will see in the chapter on spin relaxation that there is a need to introduce one spin relaxation time for the spin parallel to $\mathbf{B}$, called longitudinal time $T_1$, and a different spin relaxation time for the spin perpendicular to $\mathbf{B}$, called transverse time $T_2$. In most conducting solids at small magnetic fields (say, smaller than a tesla), the two times are roughly equal; we call them here commonly $\tau_s$.

[24] In principle the spin current forms a tensor, as there are three directions for the current for three orientations of the spin. Here we have a one-dimensional model, so the only directions for the current come from the spin. There is a different current for each spin direction. This one-dimensional model is general enough to describe most existing spin-injection experiments with spin dynamics.

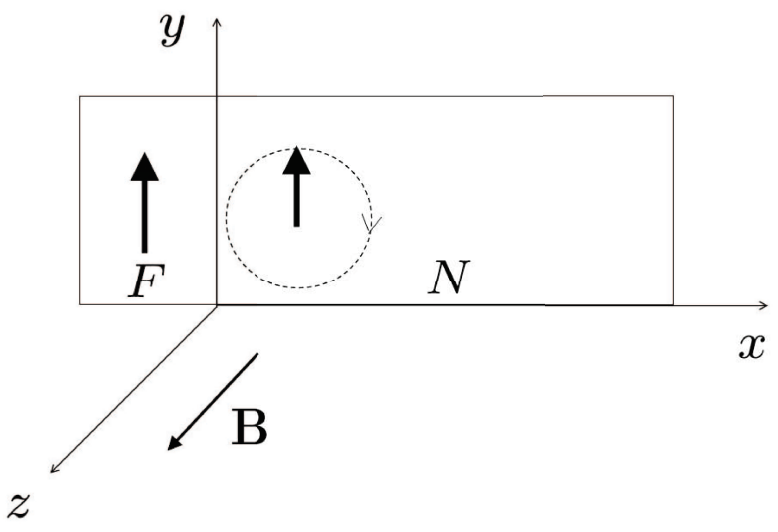

Fig. II.21. Spin injection geometry for the Hanle effect. Electron spins in the ferromagnet (F) are oriented along $y$. An applied magnetic field in the $z$ direction causes spin precession of the injected electrons in the nonmagnetic (N) region.

The corresponding spin charge current then is,

$$\mathbf{j}_s = -e\mathbf{J}_s = e\mu E\mathbf{s} + eD\nabla\mathbf{s}. \tag{II.214}$$

Finally, the continuity equation reads,

$$\frac{\partial \mathbf{s}}{\partial t} + \nabla\mathbf{J}_s = \mathbf{s}\times\boldsymbol{\omega_0} - \frac{\mathbf{s}}{\tau_s}. \tag{II.215}$$

Spin dynamics is taken into account by the first term on the right-hand side.

## E.2 Hanle effect

The Hanle effect refers to the dependence of the spin accumulation on magnetic field applied perpendicular to the injected spin. At weak magnetic fields the spin accumulation decreases with the field, while at large fields one observes coherent decaying oscillations due to spin precession. The decay of the spin accumulation at large magnetic fields is due to the drift-diffusive character of the electron transport. The spin probe, at some distance from the point of spin injection, detects the average spin. This spin comes from electrons that diffused from the point of spin injection. As different electrons have different transit times, their spin precession angles will differ. If this difference becomes comparable to the Larmor period, the average spin at the probe will be zero. This explain why at large transverse magnetic fields the detected spin vanishes.

The model most widely employed in the experimental literature to describe the Hanle effect uses spin diffusion, spin precession, and spin relaxation processes to describe the spin density at a distance $x$ from the point of spin injection. Suppose we inject spin $s_y$ along the $x$ direction, and apply magnetic field along the $z$ direction, as depicted in Fig. II.21. The magnetic field applies a torque on the spin, whose $x$ component, $s_x$, we wish to detect. Let the injected spin density be $s_{y0}$, the electron diffusion $D$ and drift velocity $v_d$, and the Larmor frequency $\omega_0$. The time it takes for the average electron to move to $x$ is $t = x/v_d$. Because of the processes of spin relaxation and spin precession, the spin component along the $x$ direction of the diffusing electrons is given, up to the overall scale,

$$s_x(x,t) \sim \frac{1}{\sqrt{4\pi Dt}} e^{-(x-v_d t)^2/4Dt} e^{-t/T_2}\sin(\omega_0 t). \tag{II.216}$$

Above we use $T_2$ (we should use $T_2^*$ if inhomogeneous broadening is present) to denote spin dephasing time, since we consider spin dynamics of transvere spin. The sine function describes the spin dynamics corresponding to the boundary and initial condition of vanishing $s_x$ at $x = 0$ and $t = 0$. Since all the electrons present at the detection point, $x$, need to be counted, we have to integrate over all the transport times $t$, for the fixed $x$, to obtain the measured spin:

$$s_x(x) \sim \int_0^\infty dt\, s_x(x,t) = \int_0^\infty dt\, \frac{1}{\sqrt{4\pi Dt}} e^{-(x-v_d t)^2/4Dt} e^{-t/T_2} \sin(\omega_0 t). \tag{II.217}$$

Depending on the size of our spin probe, we can integrate $s_x(x)$ along the probe dimension to get the result suitable for the experimental setup. Equation (II.217) is most commonly used to extract $T_2$ and $\omega_0$ (to get the g-factor) from the experimental results. The input parameters, $D$ and $v_d$ can be obtained from charge transport measurements.

Let us consider the Hanle effect more carefully, in the context of electrical spin injection (Johnson and Silsbee, 1988). We will study the spin injection geometry shown in Fig. II.21. Spin, initially along $y$, is injected from the ferromagnetic conductor into the nonmagnetic one. The contact is at $x = 0$. In the $N$ region, the injected spin precesses about the direction of the external magnetic field $\mathbf{B}$, which points in the $z$ direction. Our goal is to find the spin accumulation (and spin polarization) at a distance $x > 0$ away from the point of injection.

We will solve the spin dynamics drift-diffusion equation, (II.211), with the following boundary conditions:

$$\mathbf{s}(\infty) = 0, \tag{II.218}$$

valid for our $N$ region which is greater than the spin diffusion length;

$$j_{sx}(0) = j_{sz}(0) = 0;\ j_{sy}(0) = j_{s0}, \tag{II.219}$$

expressing spin current conservation at the contact at $x = 0$. Only the spin current for electron spins oriented along $y$ (see Fig. II.21) is present. For our treatment below, we take $j_{s0}$ to be a parameter. In order to determine it from materials constants, we would need to apply the standard spin injection model, properly generalized to take into account spin dynamics. Since the value of this parameter does not influence the functional form of the spin profile, which is what is interesting in experiments on the Hanle effect, we do not go into the detailed considerations. Forced to give an order of magnitude, the spin current at $x = 0$ would be something like,

$$j_{s0} \approx P_{nF} j, \tag{II.220}$$

where $j$ is the charge current and $P_{nF}$ is the density spin polarization in the ferromagnet. That the above equation is only a rough approximation, we learned from the standard spin injection model in Sec. D.4. Another parameter, which also should be determined self-consistently from the generalized standard model, is the electric field, $\mathbf{E}$, taken to be constant here due to the assumed charge neutrality. This is also useful since the electric drift field can be controlled in experiments to qualitatively change the picture of the Hanle effect.

We will be interested in the steady-state effects, meaning that the time derivatives are to be taken zero. Our starting equations are,

$$\begin{aligned} \dot{s}_x &= s_y\omega_0 + Ds_x'' - v_d s_x' - s_x/\tau_s = 0, &\text{(II.221)}\\ \dot{s}_y &= -s_x\omega_0 + Ds_y'' - v_d s_y' - s_y/\tau_s = 0, &\text{(II.222)}\\ \dot{s}_z &= Ds_z'' - v_d s_z' - s_z/\tau_s = 0, &\text{(II.223)} \end{aligned}$$

subject to the boundary conditions,

$$s_x(\infty) = s_y(\infty) = s_z(\infty) = 0, \tag{II.224}$$

describing the vanishing of spin at the far end of the nonmagnetic conductor, and,

$$\begin{aligned} -v_d s_x'(0) + Ds_x'(0) &= 0, &\text{(II.225)}\\ -v_d s_y'(0) + Ds_y'(0) &= j_{s0}/e, &\text{(II.226)}\\ -v_d s_z'(0) + Ds_z'(0) &= 0, &\text{(II.227)} \end{aligned}$$

reflecting the spin current conservation at the contact region. Above we have introduced the drift speed, $v_d = -\mu E$.

Since the spin component $s_z$ is not coupled to the other two, the solution for its profile is simple:

$$s_z(x) = 0. \tag{II.228}$$

This reflects the fact that there is no external torque acting on the injected $s_y(0)$ to precess away from the $xy$ plane. As for the $x$ and $y$ components, let us write the drift-diffusion equations for them in a more transparent form:

$$\begin{aligned} s_x'' &-2\kappa\frac{s_x'}{L_s} - \frac{s_x}{L_s^2} + \frac{s_y}{L_s^2}(\omega_0\tau_s) = 0, &\text{(II.229)}\\ s_y'' &-2\kappa\frac{s_y'}{L_s} - \frac{s_y}{L_s^2} - \frac{s_x}{L_s^2}(\omega_0\tau_s) = 0, &\text{(II.230)} \end{aligned}$$

where we used that $D = L_s^2\tau_s$ and we have denoted by

$$\kappa = \frac{L_d}{2L_s}, \tag{II.231}$$

a dimensionless parameter measuring the strength of the drift over diffusion; the spin drift length above is defined by,

$$L_d = v_d\tau_s, \tag{II.232}$$

which is the length over which electrons drift before they lose their spin memory. Similarly, we write for the boundary conditions,

$$\begin{aligned} &-2\kappa\tfrac{1}{L_s}s_x(0) + s_x'(0) = 0, &\text{(II.233)}\\ &-2\kappa\tfrac{1}{L_s}s_y(0) + s_y'(0) = \tfrac{j_{s0}}{eD}. &\text{(II.234)} \end{aligned}$$

Searching for the solutions of Eqs. (II.229) and (II.230) in the form of,

$$s_x(x) = A_x e^{-\alpha x/l_s}, \quad \text{(II.235)}$$
$$s_y(x) = A_y e^{-\alpha x/l_s}, \quad \text{(II.236)}$$

gives the following condition on $\alpha$,

$$\begin{vmatrix} \alpha^2 + 2\alpha\kappa - 1 & \omega_0\tau_s \\ -\omega_0\tau_s & \alpha^2 + 2\alpha\kappa - 1 \end{vmatrix} = 0. \quad \text{(II.237)}$$

Solving for the determinant equation we obtain four distinct complex solutions for $\alpha$. Selecting the two of them with a real positive part, and solving quite a few algebraic equations to find out the coefficients $A_x$ and $A_y$ to satisfy the boundary conditions at $x = 0$, we finally obtain the spin profiles (note that in the spin injection regime the spin charge current, $j_s$, is negative, so that electrons go from F to N):

$$s_x(x) = \frac{-j_{s0}L_s}{eD} e^{-\alpha_1 x/L_s} \left[ \frac{2\kappa + \alpha_1}{(2\kappa + \alpha_1)^2 + \alpha_2^2} \sin\left(\frac{\alpha_2}{L_s}x\right) + \frac{\alpha_2}{(2\kappa + \alpha_1)^2 + \alpha_2^2} \cos\left(\frac{\alpha_2}{L_s}x\right)\right], \quad \text{(II.238)}$$

$$s_y(x) = \frac{-j_{s0}L_s}{eD} e^{-\alpha_1 x/L_s} \left[ \frac{2\kappa + \alpha_1}{(2\kappa + \alpha_1)^2 + \alpha_2^2} \cos\left(\frac{\alpha_2}{L_s}x\right) - \frac{\alpha_2}{(2\kappa + \alpha_1)^2 + \alpha_2^2} \sin\left(\frac{\alpha_2}{L_s}x\right)\right]. \quad \text{(II.239)}$$

Here,

$$\alpha_1 = \frac{1}{\sqrt{2}}\sqrt{1 + \kappa^2 + \sqrt{(1+\kappa^2)^2 + (\omega_0\tau_s)^2}} - \kappa, \quad \text{(II.240)}$$
$$\alpha_2 = \frac{1}{\sqrt{2}}\sqrt{-1 - \kappa^2 + \sqrt{(1+\kappa^2)^2 + (\omega_0\tau_s)^2}}, \quad \text{(II.241)}$$

Parameter $\alpha_1$ describes the effective spin relaxation, while $\alpha_2$ describes the effective spin precession. Indeed, the effective spin relaxation length is,

$$L_{s,\text{eff}} = \frac{L_s}{\alpha_1}, \quad \text{(II.242)}$$

while the effective period length of the Larmor spin precession is,

$$L_0 = \frac{2\pi L_s}{\alpha_2}. \quad \text{(II.243)}$$

In order for Hanle oscillations to be visible, one should ideally have $L_{s,\text{eff}} \gtrsim L_0$. To summarize, three parameters determine the functional form (up to the overall scale) of the spin profiles in the Hanle effect: $\kappa$, $\omega_0\tau_s$, and $L_s$. Alternatively, one can consider fitting experiments with the effective parameters $\alpha_1/L_s$, $\alpha_2/L_s$, and $\kappa$.

We also give the expression for the profile of the magnitude of the spin, $s = (s_x^2 + s_y^2)^{1/2}$:

$$s(x) = \frac{j_{s0} L_s}{eD} \frac{\exp(-\alpha_1 x / L_s)}{\sqrt{(2\kappa + \alpha_1)^2 + \alpha_2^2}}. \tag{II.244}$$

In practice, the spin detector covers a finite length over $x$. In order to convert the above calculated spin profiles into the measured quantities, one needs to integrate the profile over the detector's length. By changing the magnetic or the drift fields one then observes the changes in the integrated profile, extracting from those useful parameters such as the spin relaxation length $L_s$ or the spin relaxation time $\tau_s$.

Another word of caution about applying the above obtained spin profiles to spin-valve experiments. Equations (II.238) and (II.239) were obtained for the case of vanishing of spin at $x \to \infty$. In a spin-valve experiment one needs to apply a different boundary condition, that of continuous spin current at the point $x = L$ in which the second ferromagnetic electrode, into which electrons are injected, is placed. This boundary condition requires using all four different solutions of the determinant equation, (II.237). If $L \lesssim L_s$, the role of $L_s$ will be played by $L$, to some extent. It is straightforward to use the simple method above in specific device settings, to obtain reliable quantitative fits and extract useful parameters.

In order to give physical meaning to the obtained formulas describing the Hanle effect, we discuss two important cases: diffusion dominated and drift dominated Hanle effect.

**Diffusion dominated Hanle effect.** The diffusion regime is characterized by

$$\kappa \ll 1, \tag{II.245}$$

meaning that the spin drift length, $L_d$, is much smaller than the spin diffusion length, $L_s$. This is the typical regime in which the Hanle effect is observed in metals or in heavily doped semiconductors, in which the electric field is rather small due to the large number of free carriers. We have,

$$\alpha_1 = \frac{1}{\sqrt{2}} \sqrt{1 + \sqrt{1 + (\omega_0 \tau_s)^2}}, \tag{II.246}$$

$$\alpha_2 = \frac{1}{\sqrt{2}} \sqrt{-1 + \sqrt{1 + (\omega_0 \tau_s)^2}}. \tag{II.247}$$

At small magnetic fields, when $\omega_0 \tau_s \ll 1$, parameter $\alpha_1 \to 1$, while $\alpha_2 \to 0$. In the high magnetic-field limit they both go as $\alpha_{1,2} \sim \sqrt{\omega_0 \tau_s / 2}$.

The effective spin relaxation length becomes,

$$L_{s,\text{eff}} = \sqrt{2} \frac{L_s}{\sqrt{1 + \sqrt{1 + (\omega_0 \tau_s)^2}}}. \tag{II.248}$$

At large magnetic fields, the effective spin relaxation length decreases due to rapid spin precession. The effective period length of the spin precession is,

$$L_0 = \sqrt{2} \frac{2\pi L_s}{\sqrt{-1 + \sqrt{1 + (\omega_0 \tau_s)^2}}}. \tag{II.249}$$

At large magnetic fields the period decreases as

$$L_0 \to \sqrt{2}\frac{2\pi L_s}{\sqrt{\omega_0 \tau_s}} = 2\sqrt{2}\pi\sqrt{\frac{D}{\omega_0}}, \tag{II.250}$$

which can be interpreted as the Larmor length—roughly the distance over which spin diffuses in the Larmor period $2\pi/\omega_0$. In the diffusive limit, the spin magnitude decreases with distance $x$, as well as with the magnetic field, according to

$$s(x) = \frac{j_{s0} L_s}{eD} \frac{\exp(-\alpha_1 x/L_s)}{[1 + (\omega_0 \tau_s)^2]^{1/4}}. \tag{II.251}$$

Finally, we give the effective period of the Larmor frequency, for a given point in space, $x = x_0$. In experiments one usually varies magnetic field for a fixed position of the spin probe. As the field varies, due to the spin precession, peaks and dips are obtained in the probed spin as a function of magnetic field. These periodic structures diminish at large fields (or large Larmor frequencies), due to the increased effective spin relaxation and a decrease in the overall spin magnitude. In the diffusive regime, the periodicity of $2\pi n$ occurs for frequencies (up to an offset),

$$\omega_{0n} \approx \frac{8\pi^2 n^2 D}{x_0^2}. \tag{II.252}$$

The fact that the period frequencies increase quadratically with increasing $n$ comes from the diffusive character of motion: the characteristic time increases quadratically with the covered distance. The above relation says that $n$ is roughly the number of Larmor diffusion lengths, $(D2\pi/\omega_0)^{1/2}$, in $x_0$.

The diffusion-dominated Hanle effect is illustrated in Fig. II.22. For the parameters used, the effective spin relaxation length is about $L_s$, while the effective period length is $L_0 \approx 4L_s$. Since $L_s \lesssim L_0$, the oscillations decay very fast and only one cycle, already quite reduced, is seen. As a function of magnetic field, the spin decays at large magnetic fields due to the diffusion, as described in the introduction to this section. Diffusion causes different transit times for electrons, so if the spread in the transit times due to diffusion becomes comparable to the Larmor period, average spin at a given point $x$ is greatly reduced. The $s_y$ spin density decays initially, as a function of magnetic field, decreasing its magnitude to about one half of the initial one, at the larmor frequency such that $\omega_0 \approx 1/\tau_s$. This decay shape can be used to obtain an estimate on the spin relaxation time $\tau_s$.

**Drift dominated Hanle effect.** The drift dominated regime is defined by

$$\kappa \gg 1, \tag{II.253}$$

meaning that the spin drift length is much greater than the spin diffusion length. This regime occurs in Hanle effect experiments on very clean samples, such as the silicon samples discussed in Sec. F.2. In this regime we can expand about $1/\kappa$, and obtain,

$$\alpha_1 = \frac{1}{2\kappa}, \tag{II.254}$$

$$\alpha_2 = \frac{\omega_0 \tau_s}{2\kappa}. \tag{II.255}$$

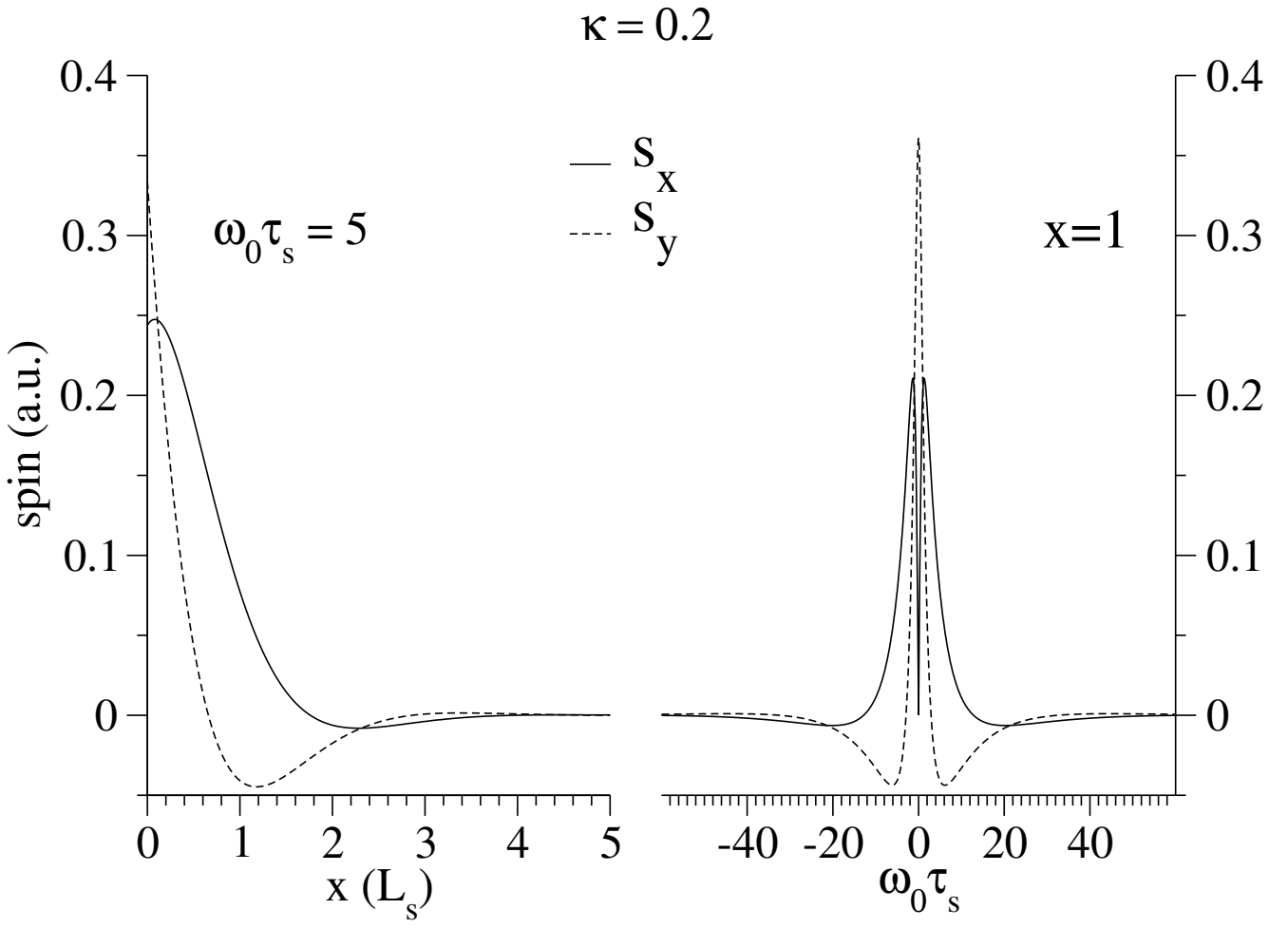

Fig. II.22. Diffusion-dominated Hanle effect. The left graph shows the spin profiles, while the right graph shows the magnetic field dependence of the spins, at a fixed point, a distance $x = 1L_s$ away from the spin injection point. The distance is in the units of the spin diffusion length. The parameters are $\kappa = 0.2$, meaning that $L_d = 0.4L_s$ and, for the left graph, $\omega_0\tau_s = 5$, meaning that there is about one Larmor precession (note that the period is $2\pi/\omega_0$ in the time interval of spin relaxation time) in the time of $\tau_s$. Hanle oscillations decay rapidly on both graphs, at large distances and at large magnetic fields, due to the effective drift and magnetic field induced spin relaxation.

The effective spin relaxation length is then

$$L_{s,\text{eff}} = L_d = v_d\tau_s, \tag{II.256}$$

equal to the spin drift length. This has simple interpretation. Spin is dragged by the electric field in the sample, increasing effectively the spin relaxation length beyond the spin diffusion length, as noticed already by Aronov (1976).

The effective spin precession period in the drift regime is,

$$L_0 = \frac{2\pi L_d}{\omega_0\tau_s} = \frac{2\pi v_d}{\omega_0}. \tag{II.257}$$

Again, the interpretation is straightforward. The period length of the spin precession is the distance over which electrons drift over time of the Larmor period, $2\pi/\omega_0$.

As we did in the diffusive regime, we also present the Larmor frequencies at which we can expect the signal to exhibit peaks or dips, $2\pi n = \alpha_{2n}x_0/L_s$, given by (up to an offset),

$$\omega_{0n} \approx 2\pi n\frac{v_d}{x_0}. \tag{II.258}$$

The peaks (or dips) occur in magnetic fields at these frequencies. Since the drift produces characteristic times linearly proportional to the distance, the increase of $\omega_{0n}$ with increasing $n$ is also linear, simply given by the number of full Larmor periods over the transit time, $x_0/v_d$.

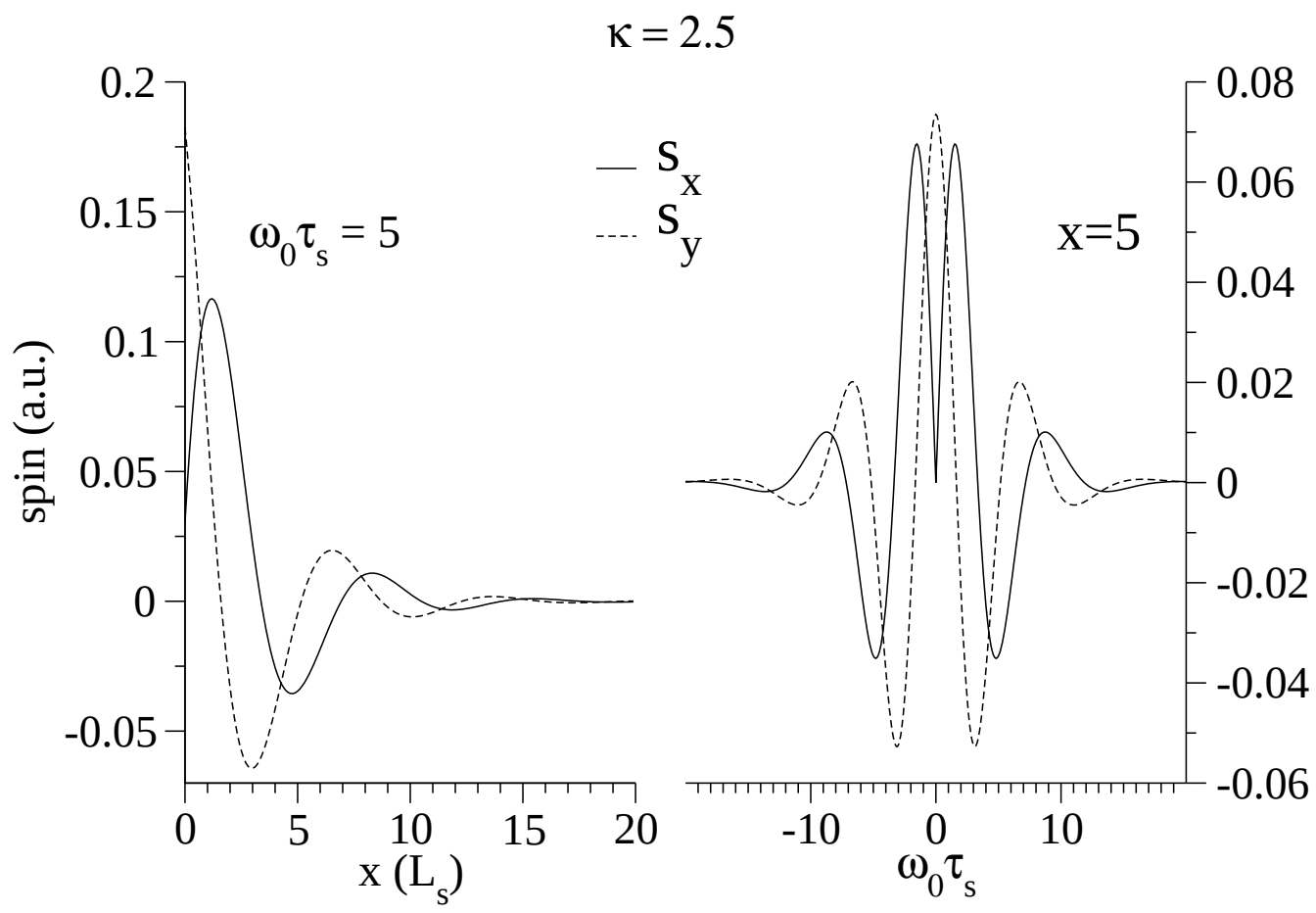

Fig. II.23. Drift-dominated Hanle effect. The left graph shows the spin profiles, while the right graph shows the magnetic field dependence of the spins, at a fixed point, a distance $x = 5L_s$ away from the spin injection point. The distance is in the units of the spin diffusion length. The parameters are $\kappa = 2.5$, meaning that $L_d = 5L_s$ and, for the left graph, $\omega_0\tau_s = 5$, meaning that there is about one Larmor precession, in the time of $\tau_s$. Note that the period is $2\pi/\omega_0$ in the time interval of spin relaxation time. Hanle oscillations are seen in both graphs, decaying at large distances and at large magnetic fields, due to the effective, drift and magnetic field induced spin relaxation.

The drift-dominated Hanle effect is illustrated in Fig. II.23. Hanle oscillations are nicely observed, decaying in space with the effective drift-induced spin relaxation length of $L_{s,\text{eff}} = 5L_s$. The periodic decaying oscillations have the period of $2\pi$, as seen immediately from the $s_y$ component, for example. At $x = 0$ it is mostly $s_y$, due to the spin injection, which is excited, while $s_x$ is phase shifted. The dependence on the magnetic field also shows oscillations, with the period roughly $2\pi$. The spin dependence on $B$ for negative $B$ (pointing along $-z$) is a mirror image of the positive field curve, reflecting the opposite direction of the spin precession. We will see that similar Hanle plots have been observed in the experiments on the spin injection into silicon, described in Sec. F.2.

## F. Spin injection into semiconductors

Electrical spin injection into semiconductors, such as GaAs, has been demonstrated to be highly efficient Fiederling *et al.* (1999); Ohno *et al.* (1999); Jonker *et al.* (2000); Mattana *et al.* (2003). Those early investigations are covered in detail by Žutić *et al.* (2004). Here we show two recent examples. The first is the spin injection across an Fe/GaAs interface, imaged by scanning probe Kerr rotation microscopy, the second is the very recent milestone—injection and detection of spin into silicon.

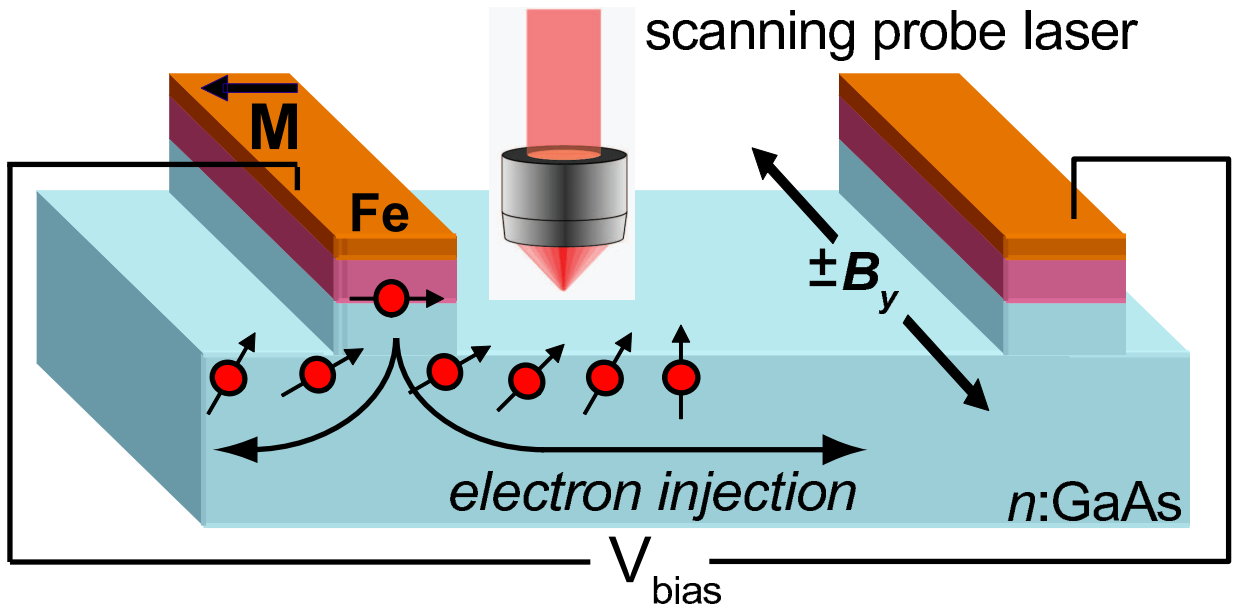

Fig. II.24. Scheme of electrical spin injection device allowing spin imaging. Electrons are injected from the left electrode into n-doped GaAs, with the electron and spin flow indicated. A transverse magnetic field $B_y$ is applied to rotate the spins. The scanning probe laser detects the vertical component of the spin by the magneto-optical Kerr effect. Reused with permission from S. A. Crooker et al., Journal of Applied Physics, 101, 081716 (2007). Copyright 2007, American Journal of Physics. The magnetization direction of the injecting electrode is opposite to that in the published version, to reflect the experiment discussed in the text. Courtesy of S. A. Crooker.

### F.1 Visualizing spin injection

We will illustrate electrical spin injection across ferromagnet/semiconductor interfaces on the example of a lateral Fe/GaAs system, analogous to the lateral spin-valve structures intially studied by Johnson and Silsbee (1985). The particular choice of materials is motivated by the high quality of the lattice matched Fe/GaAs and extensive previous studies showing an efficient spin injection (Hanbicki *et al.*, 2002, 2003; Wunnicke *et al.*, 2002) in these junctions. A close connection between the first-principles calculations and atomic resolution interface imaging revealed that the increase in the spin injection efficiency is due to the abruptness as well as due to the chemical and structural coherence of the annealed interface (Erwin *et al.*, 2002; Zega *et al.*, 2006). Beautiful images of electrically injected spin have been obtained by (Crooker *et al.*, 2005, 2007) and by (Kotissek *et al.*, 2007), following earlier imaging of optically injected spin (Crooker and Smith, 2005); see also (Stephens *et al.*, 2004).

The scheme of the device used by Crooker *et al.* (2005) is shown in Fig. II.24. The standard spin-valve setting comprises two ferromagnetic electrodes, made of Fe, epitaxially grown on top of heavily n-doped GaAs layers, which are then grown on the GaAs substrate. The contacts between the electrodes and heavily-doped layers form Schottky barriers, overcoming the conductivity mismatch problem, see Secs. D.7 and D.8, allowing efficient spin injection into the substrate semiconductor. The length of the GaAs channel for electric current is 300 $\mu$m. The GaAs substrate is doped with $2 \times 10^{16}$ cm$^{-3}$ silicon donors. This doping lies at the metal-to-insulator transition region for GaAs, in which longest spin relaxation times, up to 200 ns, have been found (Žutić *et al.*, 2004; Dzhioev *et al.*, 2002; Kikkawa and Awschalom, 1998; Oestreich *et al.*, 2005; Schreiber *et al.*, 2007). As the injected electrons diffuse through GaAs, their spins precess due to the external magnetic field, applied perpendicular to the injected spin direction. The spin component (magnetization) out of the plane of the substrate is measured by the Kerr rotation spectroscopy (see Sec. IV.E.1 for the description of how Kerr and Faraday magneto-optic spectroscopies work), with the spatial resolution of a few microns. The experiment has been performed at 4 K, and the estimated injected spin polarization is 5-10 %. The extracted spin lifetime $\tau_s$, from the Hanle effect measurements (see Sec. E.2) is about 150 ns, in line what has been seen by other methods (Žutić *et al.*, 2004) in this doping regime.

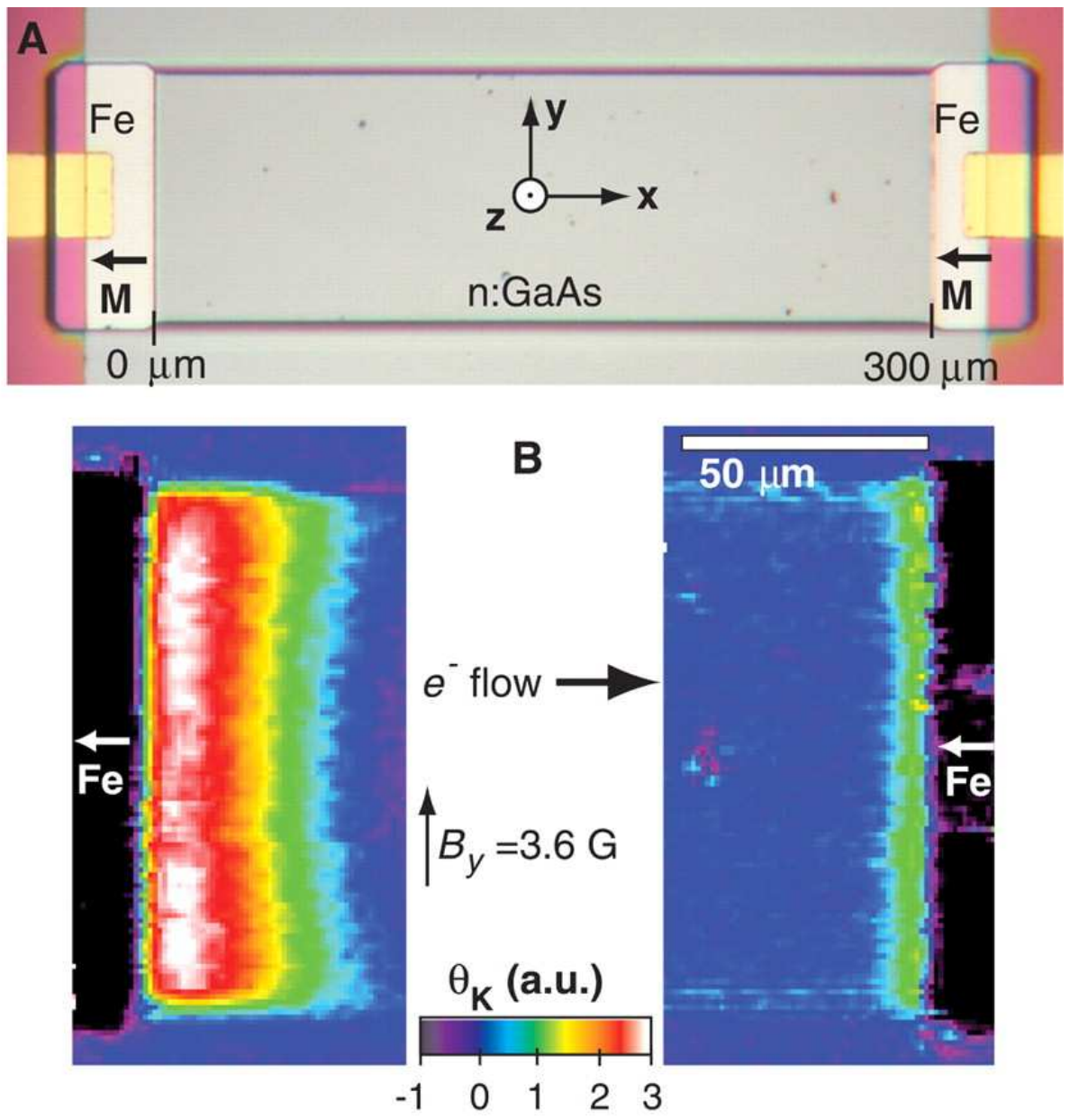

Fig. II.25. (A) Photomicrograph of the electrical spin injection device used to image spin in lightly n-doped GaAs. The reference coordinate system is indicated. (B) Images of the Kerr rotation angle, $\theta_K$. Positive angle means positive spin along the $z$ direction. The source-drain voltage is $V_b = 0.4$ V, the transverse magnetic field $B_y = 3.6$ gauss. The spin injected at the source contact (left) is antiparallel to the magnetization $M$; similarly for the spin accumulated at the drain contact (right). From S. A. Crooker *et al., Science* **309**, *2191 (2005). Reprinted with permission from AAAS.*

The micrograph of the structure is shown in Fig. II.25 A. Spin-polarized electrons are injected from the left ferromagnetic electrode whose magnetization is directed to the left (along $-x$). This means that the magnetization of the majority electrons points along $-x$ (which is the easy axis for the magnetization of the structure, [011] crystallographic direction), while the minority electron magnetization points along $x$. As the $g$-factor in Fe is positive,[25] $g \approx 2.09$ (Frait, 1977), the majority electron spin points along $x$, and the minority electron spin points along $-x$. In the GaAs, the $z$-component of the electron spin (or magnetization, see footnote 25) is probed by the Kerr angle. The positive Kerr angle $\theta_K$ translates into a positive electron spin.

Figure II.25 B is the image taken by the Kerr probe. Electrons injected into the GaAs substrate from Fe source electrode have their spins oriented antiparallel to the magnetization of the electrode. This is seen from the fact that the Kerr angle is positive close to the source electrode.

[25]For a positive $g$-factor the spin and magnetization have opposite directions. For a negative g-factor semiconductor, such as GaAs, the two vectors point in the same direction.

Recall that a magnetic dipole $\mathbf{m}$ feels torque $\mathbf{N} = \mathbf{m} \times \mathbf{B}$ in a magnetic field $\mathbf{B}$. Therefore a spin (magnetization in GaAs) pointing along $x$ will feel a torque along $z$, due to the magnetic field $B_y$. This torque rotates the electron spin towards positive $z$. If the spin were injected along $-x$, the rotation would be towards $-z$. We can then conclude that the injected spin points along the *spin* of the majority electrons in Fe, that is, it is the majority electron spin which is injected from Fe into GaAs. Simple considerations would predict that minority spin would be injected: indeed, one expects that while the majority electrons have larger density, their density of states at the Fermi level is smaller than that for the minority electrons. Since in tunneling it is the density of states that matter, one would expect that minority electron spin would be injected in excess of that of the majority one. The fact that the experiment sees the majority spin injection shows that band-structure effects and specific tunnel structures play important qualitative role in spin injection.

This conclusion is even more emphasized by looking at the spin accumulation close to the drain electrode in Fig. II.25 B. While during most of the bulk transport spin polarization along $z$ is close to zero, due to the spin diffusion length of about 50 $\mu$m, there is appreciable spin polarization at the drain Fe electrode. Since the Kerr angle is again positive, one concludes that the spin orientation is along $x$, similar to the injected spin by the source. In contrast, the standard model of spin injection results in spin extraction if electrons are driven from a nonmagnetic conductor into a ferromagnet, see Sec. D.4. It is then evident that the spin polarization at the drain interface is due to the reflection of the electrons from the ferromagnetic electrode. The subtle effects of the ferromagnetic Schottky barriers and the related spin extraction in the forward biased contacts, such as the one at the drain contact in Fig. II.25 or the MnAs/GaAs Schottky barrier studied in (Stephens *et al.*, 2004), have been further investigated experimentally (Lou *et al.*, 2006), and theoretically (Dery and Sham, 2007; Adagideli *et al.*, 2006; Osipov and Bratkovsky, 2005; Bauer *et al.*, 2005; Bratkovsky and Osipov, 2004; Osipov *et al.*, 2005; Bauer *et al.*, 2004). The prediction of spin extraction (Žutić *et al.*, 2002) shows already on the example of simple magnetic p-n junction important implications of the nonlinear regime (Žutić *et al.*, 2002; Fabian *et al.*, 2002b); at low biases there is no spin injection across a magnetic p-n junctions. For example, as observed experimentally, the spin polarization generated in the semiconductor can strongly depend on the applied bias, in contrast to the description based on the equivalent resistor scheme which leads to bias-independent spin polarization.

### F.2 Spin injection into silicon

Spin injection into silicon is a great challenge. There are several reasons for that. First, unlike GaAs, silicon is an indirect band gap semiconductor, so optical orientation or luminescence are ineffective for spin polarization or spin detection (Žutić *et al.*, 2004). One needs to consider alternative approaches, such as proposed by Žutić *et al.* (2006b). Second, it appears that silicon, due to materials reasons, does not easily form suitable spin-preserving contacts with common ferromagnetic metals, so that relatively complicated material processing is required (Min *et al.*, 2006; Žutić, 2006). Third, it is difficult to find an appropriate ferromagnetic semiconductor which would ensure high-quality interfaces with silicon and could be used as an efficient spin injecting electrode.[26] Again, such a material, GaMnAs, exists for GaAs.

[26]There is an initial support that Mn-doped chalcopyrite semiconductors could be possible candidates as they are lattice matched with silicon (Cho *et al.*, 2002; Ishida *et al.*, 2003; Erwin and Žutić, 2004). Similarly, while GaMnAs/Si

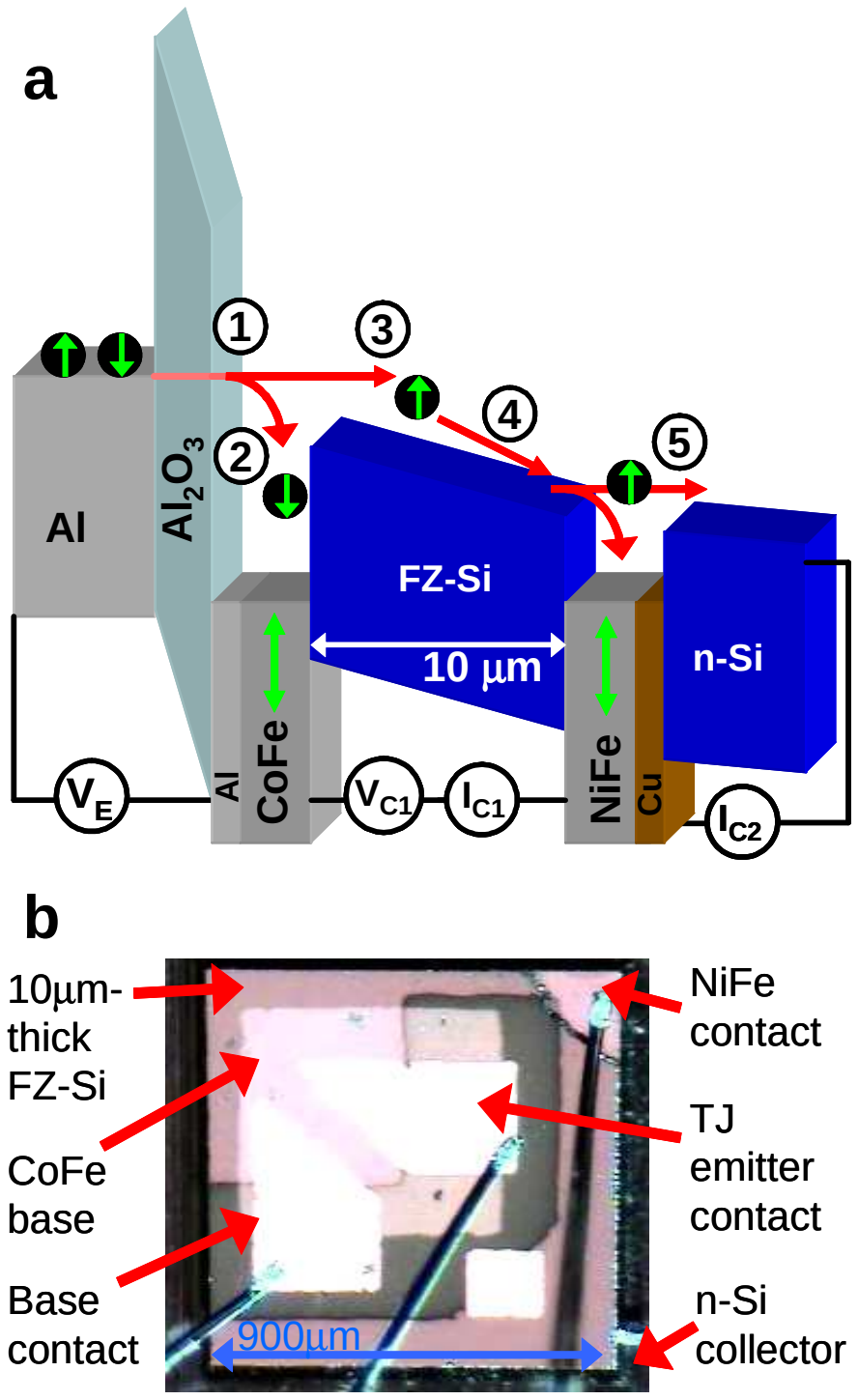

Fig. II.26. Scheme of the device demonstrating electrical spin injection into silicon. (a) Electronic band diagram of the hot-electron spin valve. The emitter voltage, $V_e$, is fixed, while the first, $I_{c1}$, and second, $I_{c2}$, collector currents are measured, as a function of the orientation of the magnetizations of the two ferromagnetic layers, as well as of the first collector voltage $V_{c1}$. (b) A micrograph of the actual device. Symbol FZ stands for float-zone, while TJ is brief for a tunnel junction. Reprinted by permission from Macmillan Publishers Ltd: *Nature. I. Appelbaum et al., Nature* **447**, *295 (2007), copyright 2007.*

Nevertheless, injecting spin into silicon and making a useful device application, such as a spin MOSFET, has remained a holy grail of semiconductor spintronics. These goals have been elusive for almost a decade, until the recent work of Appelbaum *et al.* (2007) reported a breakthrough in achieving robust electrical spin injection into pure silicon [see a popular account of this discovery in (Žutić and Fabian, 2007)]. Robust spin injection into silicon has also been achieved recently by Jonker *et al.* (2007), from an iron emitter over an $Al_2O_3$ barrier.

The electrical spin injection scheme of (Appelbaum *et al.*, 2007) is reproduced in Fig. II.26. Spin unpolarized electrons are injected from the aluminum (Al) emitter, through the aluminum oxide ($Al_2O_3$) barriers into, what could be termed base, the ferromagnetic layer CoFe. Passing electric current, electrons of the minority spin (those producing magnetization opposite to the magnetization of the material[27]) lose their energy much more rapidly than the electrons of the junctions with 4% lattice mismatch have already been fabricated (Zhao *et al.*, 2002), it remains to be seen if they could enable spin injection into silicon.

[27]For a positive g-factor, the electron magnetic moment is opposite to the electron spin. The minority electrons then

majority spin (those whose magnetization is in the direction of the material's magnetization). The ferromagnetic layer acts as an efficient spin filter, letting only electrons with the majority electron spin passing through. These spin-polarized electrons enter the silicon. The silicon in the experiment is of high purity (also called float-zone silion; float-zone is a bulk crystal-growth technique), undoped, 10 micron thick. The electrons traversing the silicon, which can be considered as the first collector, enter the second ferromagnetic filter, here NiFe, which works the same way on the spins as the first one. The electrons which pass ballistically through the second (NiFe/Cu) metal base, enter the second collector, here an n-doped silicon.

The variables in the experiment are the magnetization orientations of the two ferromagnetic filters, as well as the voltage drop $V_{c1}$ across the silicon. The observed quantity is the second collector current, $I_{c2}$. Since the device is based on hot (not thermalized) electrons,[28] the current $I_{c2}$ is not sensitive to the variations of $V_{c1}$: the current is limited by the supply of the electrons through the oxide barrier; the electrons in the silicon are simply swept to the second collector. The magnetization orientations allow to detect the spin-valve effect,[29] while changes of $V_{c1}$, in combination with an external magnetic field, allow to probe the Hanle effect. Both the spin-valve and the Hanle effect were demonstrated in the experiment, which was done at 85 K.

The spin-valve effect is observed by variations of the collector current, $I_{c2}$, between parallel and antiparallel magnetizations of the spin filter ferromagnets. The relative change of $I_{c2}$ in the experiment, for the parallel and antiparallel orientations, was about $2\%$. Subsequent variations of the experiment reported much higher, $35\%$ (Huang *et al.*, 2007d) and even $115\%$ (Huang *et al.*, 2007b) changes in the collector current. While in the original experiment, shown in Fig. II.26, the ballistic spin filtering in the ferromagnetic layers introduced spin-polarized electrons into the silicon, the newer variants of the experiment employ interface polarization in the emitter electrode. In addition to higher spin-valve signals, this modification also yields higher output currents, as the ballistic transport is through nonmagnetic Al and Cu, rather than through the shorter mean free path ferromagnetic CoFe layers. The spin-valve signal is a signature of the presence of spin-polarized electrons, as injected from the emitter, at the second collector. The electrons travel 10 microns through silicon without significantly losing their spin orientation. In Huang *et al.* (2007b), for example, the electron spin polarization is inferred to be remarkably high, at least $38\%$, after traversing 10 microns. Most recently, Huang *et al.* (2007a) demonstrated coherent spin-polarized transport across an entire silicon wafer, a remarkable 350 micron-long journey, translating to spin relaxation times as large as 200 ns (at 85 K), see Sec. IV.E.2.

More subtle, but also more conclusive, proof of electrical spin injection constitutes the observation of the Hanle effect, see Sec. E.2. In the Hanle effect an injected spin precesses in an applied transverse magnetic field. Due to diffusion, different electrons arrive at the collector in different times, so that the average spin at the collector in general diminishes with increasing magnetic field. If the Larmor period becomes comparable to the spread of the transit times, the average spin at the collector vanishes. The contribution of spin relaxation to the decay of the spin-precession signal is much weaker than the diffusive contribution, in these experiments. At smaller magnetic fields one can see oscillating spinvalve signals, as a function of the field, due have their spin in the direction of the magnetization of the material.

[28] In the device the hot electrons after tunneling lose energy due to phonons and equilibrate to the conduction band minimum in less than 100 nm.

[29] The spin-valve effect refers to the reduction of the electrical current by making the orientations of the ferromagnetic electrodes antiparallel. The "valve" opens for the parallel orientation.

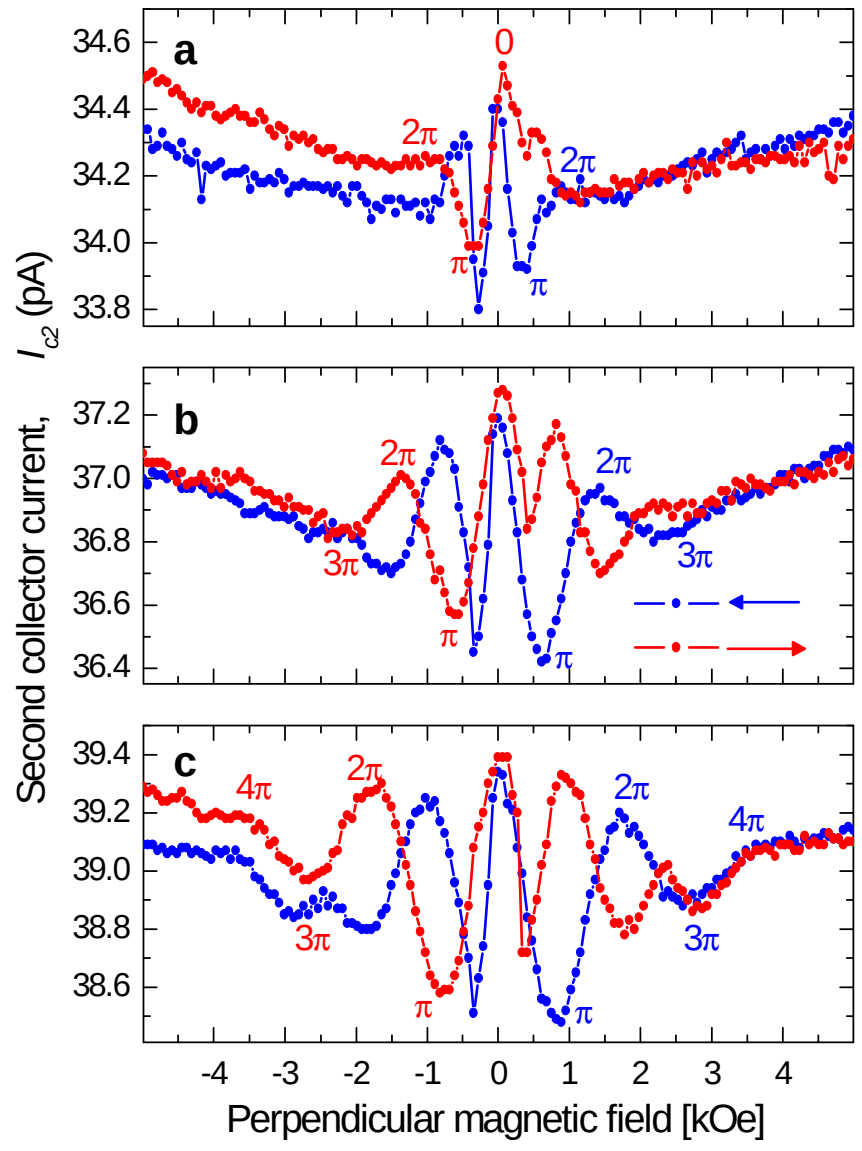

Fig. II.27. Demonstration of the Hanle effect in silicon. The emitter voltage is $V_e = -1.8$ V. The second collector current, $I_{c2}$, is plotted as the function of a perpendicular (to the orientation of the injected spin) magnetic field. While (a) is at zero $V_{c1}$, figures (b) and (c) correspond to $V_{c1} = 0.5$ V and 1.0 V, respectively. The extracted effective drift fields for the three cases are 400 V/cm, 900 V/cm, and 1400 V/cm (Huang *et al.*, 2007c). The red and blue lines are for forward (left to right) and subsequent reverse (right to left) sweeps of the magnetic field, respectively. Reprinted by permission from Macmillan Publishers Ltd: *Nature. I. Appelbaum et al., Nature* **447**, *295 (2007), copyright 2007.*

to spin precession. This is shown in Fig. II.27. Different graphs correspond to different drift velocities, which are controlled by the bias $V_{c1}$. The larger is $V_{c1}$, the larger is the drift velocity; the higher is the drift field, the smaller is the transit time. A larger magnetic field is then needed for the spins to make a full precession, see Eq. (II.258). This is the reason why the peaks corresponding to the full cycles (multiples of $2\pi$) shift to higher magnitudes as the drift field increases. Up to two full cycles are observed in II.27c. Half cycles (odd multiples of $\pi$) correspond to a spin flip, giving a reduced collector current, resulting in the observed dips. In a demonstration of spin transport across a 350 $\mu$m thick silicon wafer (Huang *et al.*, 2007a), oscillations up to four cycles were observed.

In Fig. II.27 shows that the Hanle oscillations depend on the direction of the magnetic field sweep. The reason is that the transverse field has a small in-plane (defined by the planes of the ferromagnetic electrodes in which their magnetizations lie) component of this external field capable of switching the magnetizations of the two ferromagnetic layers parallel or antiparallel. Starting with the red curve on the negative side, the two magnetizations are parallel. As one crosses through zero, they becom antiparallel, phase-shifting the oscillations (they are not mirror-reflected). Similarly for the blue curve which starts at the positive end. In the ideal case the red and the blue curves would be mirror images, which is roughly what is observed.

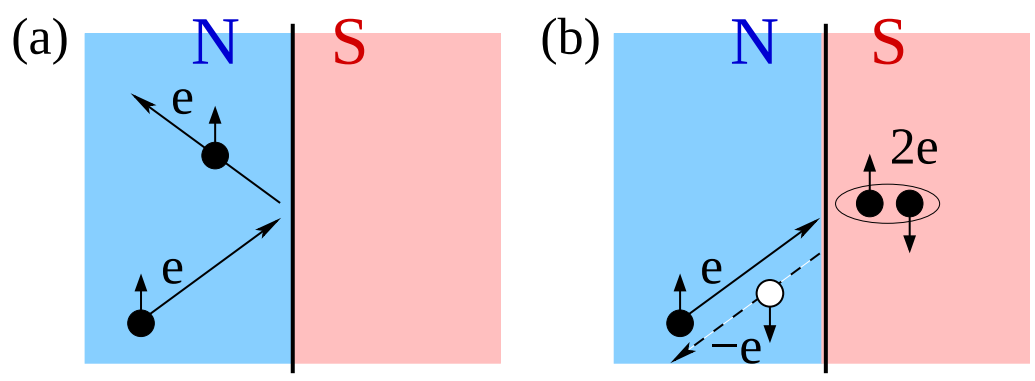

Fig. II.28. Specular and Andreev reflection at normal metal (N)/superconductor (S) interface. In contrast to the classical or specular reflection in (a), there is a change in the charge of the reflected particle undergoing Andreev reflection (b). The incident electron is reflected as a hole which retraces the initial trajectory (retro-reflection) and the two electrons are transferred to the superconducting region. One can also note that from the energy conservation in Andreev reflection an incident electron, slightly above the Fermi level will be accompanied by another electron of opposite spin slightly below the Fermi level. The transfer of the second electron below the Fermi level into a superconductor is equivalent to a reflected hole moving away from the N/S interface, as shown in (b).

This first observation of electrical spin injection into silicon is an important milestone for integrating silicon with spintronics. Two important challenges remain: electrically injecting spin into doped (n- or p-type) silicon, and making useful device spintronic structures based on silicon. The two challenges are related, as electronic devices employ doped silicon, in which current is carried by thermal electrons or holes. Electronic as well as spin properties of undoped and doped silicon are also very different. Spin relaxation, for example, is expected to increase rapidly with doping, but also with temperature, see Sec. IV.E.2. Furthermore, doped samples screen electric fields, generating different conditions at contacts with other materials. As for useful devices, it remains to be seen whether the nice spin-valve properties of the hot-electron device of Appelbaum *et al.* (2007) survive in more conventional device settings. The important question, "Can we make a useful silicon spin MOSFET?", remains open.

### G. Andreev reflection at superconductor/semiconductor interfaces

As we have seen from the previous discussion, a carrier spin polarization is an important quantity that determines spin injection efficiency and magnetoresistive effects such as the giant magnetoresistance (GMR) and the tunneling magnetoresistance (TMR), see Sec. H. Here we focus on a particular method for measuring spin polarization which relies on the scattering process at an interface with a superconductor, known as the Andreev reflection (Andreev, 1964; Deutscher, 2005). Before we discuss the relevant situation of the spin-polarized transport in junctions which combine ferromagnets and superconductors, it is helpful to briefly review a simpler case of the charge transport across a normal conductor (N) and superconductor (S) interface.

#### G.1 Conventional Andreev reflection

There are two ways in which electrons reflect off an N/S interface. Figure II.28 (a) shows classical (specular) reflection at an interface which is similar to a ball bouncing of a wall. In this case an electron (depicted as a solid circle) approaching the interface is reflected with the same charge ("e") and the same spin ("up arrow"). No electrical current is transferred to the superconductor. Figure II.28 (b) shows the Andreev reflection which is inherent to superconducting interfaces. In this case an electron approaching the interface is reflected backwards and converted into a hole (the absence of an electron is depicted as an empty circle) with opposite charge ("-e") and opposite spin ("down arrow"). In addition, when the Andreev reflection occurs two electron charges ("2e") are transferred across the interface into the superconductor. This pair of electrons with oppositely oriented spins, known also as the Cooper pair, carries dissipationless electrical current in the superconductor. Similar considerations apply also for incident holes so an incident electron (hole) of spin $\lambda$ is reflected as a hole (electron) belonging to the opposite spin subband $\overline{\lambda}$, back to the nonsuperconducting region, while a Cooper pair is transferred to the superconductor. This is a phase-coherent scattering process in which the reflected particle carries the information about both the phase of the incident particle and the macroscopic phase of the superconductor (Lambert and Raimondi, 1998; Pannetier and Courtois, 2000). Andreev reflection is thus responsible for a proximity effect where the phase correlations are introduced to a nonsuperconducting material (Demler *et al.*, 1997; Halterman and Valls, 2001; Buzdin, 2005; Tokuyasu *et al.*, 1988; Bergeret *et al.*, 2005; Izyumov *et al.*, 2002; Braude and Nazarov, 2007; Zareyan *et al.*, 2002).

In a superconducting region there is typically a finite energy gap for quasiparticle excitations, implying vanishing quasiparticle density of states at small enough energies. In a uniform superconductor such a gap in the excitation spectrum can be related to the absolute value of the so-called pair potential $\Delta$ (de Gennes, 1989), which couples electron and hole contributions to the superconducting wave function, as we have seen on the example of Andreev reflection. For simplicity, we will mostly focus on Andreev reflection with $\Delta$ being a spin singlet with orbital $s$-wave symmetry, referred also as conventional pairing. However, superconducting pairing can also include other orbital symmetries (such as $d$-wave with nodes in the superconducting gap), spin triplet pairing, and even coexistence with ferromagnetism (Linder *et al.*, 2007). Instead of electron and hole quasiparticles in the normal metals, there are electron-like and hole-like quasiparticles in the superconductor showing predominantly electron or hole character. It is then instructive to note a similarity between the two-component charge transport in a N/S junctions (for electron-like and hole-like quasiparticles) and spin-polarized transport in F/N junctions (for spin $\uparrow$, $\downarrow$), which both lead to current conversion, accompanied with the additional boundary resistance (Blonder *et al.*, 1982; van Son *et al.*, 1987). In a N/S junction the Andreev reflection is responsible for the conversion between the normal current and the supercurrent, characterized by the superconducting coherence length, while in the F/N case a conversion between the spin-polarized and unpolarized current is characterized by the spin diffusion length.

A convenient description for transport in superconducting junctions is provided by the Bogoliubov-de Gennes equations (de Gennes, 1989; de Jong and Beenakker, 1995; Žutić and Valls, 2000),

$$\begin{bmatrix} H_\lambda & \Delta \\ \Delta^* & -H^*_{\overline{\lambda}} \end{bmatrix} \begin{bmatrix} u_\lambda \\ v_{\overline{\lambda}} \end{bmatrix} = E \begin{bmatrix} u_\lambda \\ v_{\overline{\lambda}} \end{bmatrix}. \tag{II.259}$$

Here $H_\lambda$ is the single particle Hamiltonian for spin $\lambda = \uparrow, \downarrow$; $\overline{\lambda}$ denotes a spin opposite to $\lambda$, $E$ is the excitation energy, and $u_\lambda$ and $v_{\overline{\lambda}}$ are the electron-like quasiparticle and hole-like quasiparticle amplitudes, respectively. With the appropriate matching of the wave functions at the boundaries (interfaces) between different regions, Eq. (II.259) can describe a spatial variation of both superconducting properties (pair potential) as well as magnetic properties (exchange coupling or spin splitting). Furthermore, it is also straightforward to include the spin flip and spin-dependent interfacial scattering in Eq. (II.259) (Žutić and Das Sarma, 1999).

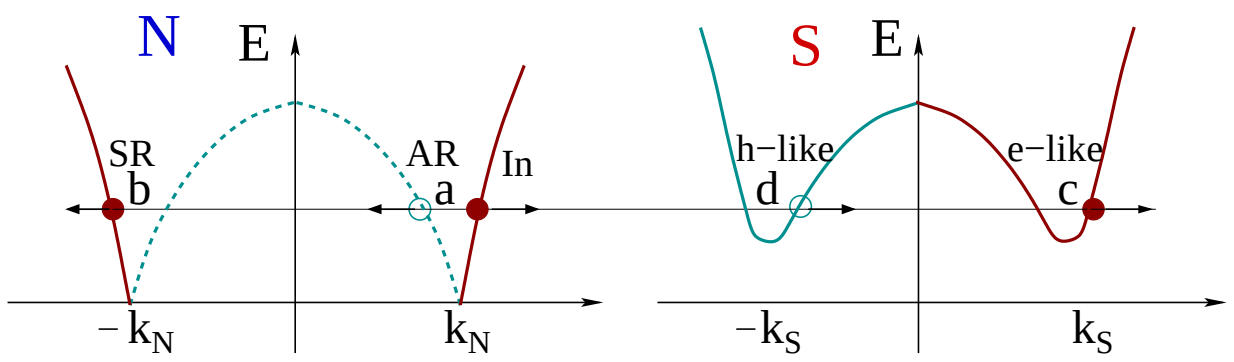

Fig. II.29. Scattering processes at an N/S interface. A sketch of the excitation energy as a function of the wave vector is given in the N and S regions (in the latter the spectrum is modified by the opening of a superconducting gap). Incident electrons (In) undergo Andreev reflection (AR) with amplitude $a$, and specular (classical) reflection (SR) with amplitude $b$, electron-like transmission with amplitude $c$ and hole-like transmission with amplitude $d$. In each scattering process, arrows depict the corresponding direction of the group velocity.

We can now consider a simple case of a one-dimensional charge transport across N/S junctions in which we take N and S semi-infinite regions ($x < 0$ and $x > 0$, respectively) separated by a planar interface at $x = 0$. For small applied bias $V$ we can approximate that the magnitude of wave vectors are equal to the Fermi wave vectors $k_N$, $k_S$, in the N and S regions, respectively. The appropriate scattering processes, for an incident plane wave quasiparticle in the N region, are depicted in Fig. II.29. For a step-like pair potential, which vanishes identically in the N-region, the two-component wave function can be expressed in the N- and S-region as

$$\begin{aligned}\psi_N &= \begin{bmatrix} 1 \\ 0 \end{bmatrix} e^{ik_N x} + a \begin{bmatrix} 0 \\ 1 \end{bmatrix} e^{ik_N x} + b \begin{bmatrix} 1 \\ 0 \end{bmatrix} e^{-ik_N x}, \\ \psi_S &= c \begin{bmatrix} u_0 \\ v_0 \end{bmatrix} e^{ik_S x} + d \begin{bmatrix} v_0 \\ u_0 \end{bmatrix} e^{-ik_S x}, \end{aligned} \tag{II.260}$$

where $u_0$ and $v_0$ are the coherence factors given by $u_0^2 = 1 - v_0^2 = (1/2)[1 + \sqrt{E^2 - \Delta_0^2}/E]$; $\Delta_0$ is the superconducting gap (in this case it coincides with the absolute value of the pair potential). Similar results were obtained in an early work by Griffin and Demers (1971) who solved the Bogoliubov-de Gennes equations with barrier (a square or a $\delta$-function) of a varying strength at an N/S interface. They obtained a result which interpolates between the clean and the tunneling limit. Blonder *et al.* (1982) have used a analogous approach, now known as the Blonder-Tinkham-Klapwijk (BTK) method where the two limits correspond to $Z \to 0$ and $Z \to \infty$, respectively, and $Z$ is the strength of the $\delta$-function barrier. Coefficients $a$, $b$, $c$, and $d$ are obtained by matching the wave functions $\psi_N$ and $\psi_S$ at the interface, while the discontinuity of their spatial derivatives is proportional to the strength of the interfacial $delta$-function potential. A more detailed procedure of an analogous wave function matching is discussed in Sec. H.3. The transparency of the Griffin-Demers-BTK approach makes it also suitable to study ballistic spin-polarized transport and spin injection even in the absence of a superconducting region (Hu and Matsuyama, 2001; Matsuyama *et al.*, 2002; Hu *et al.*, 2001; Heersche *et al.*, 2001; Žutić, 2002; Božović and Radović, 2002).

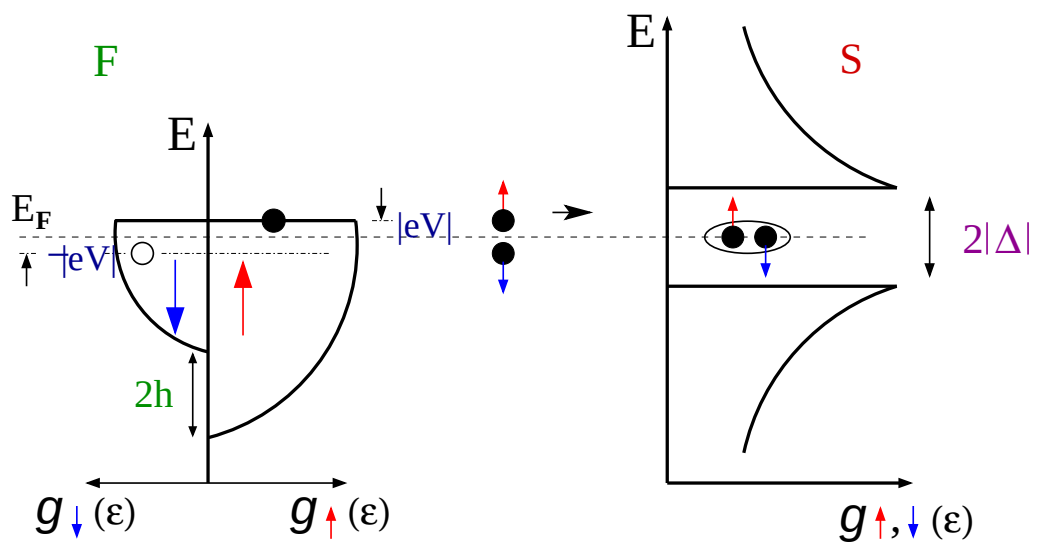

Fig. II.30. Schematic illustration of Andreev reflection at the ferromagnet/superconductor (F/S) interface showing the density of states $g_{\uparrow,\downarrow}$ in the normal region and in the superconductor with the energy gap $|\Delta|$. In the F-region, where the exchange energy leads to a spin splitting $2h$, only a fraction of incident electrons with spin up will be able to find a partner of opposite spin and contribute to the charge transfer by entering the superconductor and forming a Cooper pair.

From the preceding discussion it is possible to infer several properties of Andreev reflection, in particular its influence on the charge transport across N/S junctions. The probability for Andreev reflection at low biases ($|eV| \lesssim \Delta_0$) is related to the square of the normal state transmission and could be ignored for low transparency junctions with conventional superconductors (having $s$-wave symmetry pair potential), since the specular reflection will be dominant with the corresponding probability $|b|^2 \rightarrow 1$, recall Eq. (II.260)]. Conductance measurements in such N/S junctions would give a vanishingly small value at low biases and low temperatures since there is no mechanism for charge transfer into a superconducting region. In contrast, for high transparency junctions ($Z \rightarrow 0$) single-particle tunneling vanishes ($|c|^2$, $|d|^2 \rightarrow 0$) at low biases and temperatures—the Andreev reflection dominates.

### G.2 Spin-polarized Andreev reflection

If we consider a superconducting junction with a spin-polarized region, we need to generalize our description of scattering processes which now have spin-dependent scattering amplitudes and the Fermi wave vectors [recall Eq. (II.260)]. For spin-polarized carriers, with different populations of the spin subbands, as shown in Fig. II.30, only a fraction of the incident electrons from the majority subband will have a minority subband partner in order to be Andreev reflected. This can be quantified at zero bias and for transparent contacts ($Z = 0$), in terms of the total number of scattering channels (for each $k_{\|}$), $N_\lambda = k_{F\lambda}^2 A/4\pi$, at the Fermi level. Here A is the point contact area and $k_{F\lambda}$ is the spin-resolved Fermi wave vector. A spherical Fermi surface in F and S region, with no (spin-averaged) Fermi velocity mismatch, is assumed. When S is in the normal state the zero temperature conductance corresponds to the so-called Sharvin conductance (Sharvin, 1965), arising in the ballistic transport between two bulk regions connected by a contact (an orifice or a narrow and short constriction) with the radius much smaller than the mean free path, $a \ll l$:

$$G_{FN} = \frac{e^2}{h}(N_\uparrow + N_\downarrow). \tag{II.261}$$

In a 3D geometry this expression is equivalent to the inverse of the Sharvin resistance, $R^{-1}_{\rm Sharvin}$, given by

$$R_{\rm Sharvin} = \frac{4\rho l}{3\pi a^2} = \left[\frac{e^2}{h}\frac{k_F^2 A}{2\pi}\right]^{-1}, \tag{II.262}$$

where $h/e^2 \approx 25.81$ k$\Omega$ is the quantum of resistance per spin, $\rho$ is the resistivity, $A$ the contact area, and $k_F$ is the Fermi wave vector.

In the superconducting state all of the $N_\downarrow$ and only $(N_\downarrow/N_\uparrow)N_\uparrow$ scattering channels contribute to Andreev reflection across the F/S interface and transfer charge $2e$, yielding (de Jong and Beenakker, 1995),

$$G_{FS} = \frac{e^2}{h}\left(2N_\downarrow + \frac{2N_\downarrow}{N_\uparrow}N_\uparrow\right) = 4\frac{e^2}{h}N_\downarrow. \tag{II.263}$$

The suppression of the normalized zero bias conductance at $V = 0$ and $Z = 0$, (de Jong and Beenakker, 1995)

$$G_{FS}/G_{FN} = 2(1 - P_C) \tag{II.264}$$

with the increase in the spin polarization,

$$P_C \to (N_\uparrow - N_\downarrow)/(N_\uparrow + N_\downarrow), \tag{II.265}$$

was used as a sensitive transport probe to detect the spin polarization in a point contact (Soulen Jr. *et al.*, 1998). A potential advantage of this technique is the detection of spin polarization in a much wider range of materials than those which can be grown for detection in ferromagnetic tunnel junctions employing the spin-valve effect (Žutić *et al.*, 2004). From Eq. (II.264) one could expect that the reduction of the normalized conductance can be used to infer the spin polarization for a range of ferromagnetic materials. In particular, the conductance should be vanishingly small for half-metallic ferromagnets with $P_C \to 1$, see Sec. C. While $CrO_2$ indeed showed nearly vanishing low-bias conductance (consistent with $P_C > 0.9$), NiMnSb, one of the originally predicted half-metals (de Groot *et al.*, 1983), revealed only a partial spin polarization ($P_C \approx 0.5$) (Soulen Jr. *et al.*, 1998), suggesting that the theoretical concept of a half-metal needs to be carefully verified. A similar study, using a thin film nanocontact geometry (Upadhyay *et al.*, 1998), emphasized the importance of fitting the conductance data over a wide range of applied bias, not only at $V = 0$, in order to extract the spin polarization of the F region more precisely.

A large number of experimental results using the spin-polarized Andreev reflection has since been reported (Nadgorny *et al.*, 2001; Parker *et al.*, 2002; Bourgeois *et al.*, 2001; Miyoshi *et al.*, 2005, 2006; Clifford and Coey, 2006; D'yachenko *et al.*, 2006; Raychaudhuri *et al.*, 2003; Maekawa (Ed.), 2006) focusing mostly on conventional superconductors to determine spin polarization in metallic ferromagnets including also MnAs (Panguluri *et al.*, 2003), important for semiconductor spintronics as it can be grown in high quality junctions with GaAs. These studies provide valuable information about materials that could be used as efficient spin injectors or promising magnetic regions in magnetic tunnel junctions. At the time when there were only limited number of studies of novel (III,Mn)V ferromagnetic semiconductors, it was suggested that their spin polarization and interfacial properties can be studied in superconducting junctions (Žutić and Das Sarma, 1999). Similar measurements were performed on (Ga,Mn)As using the point contact technique (Panguluri *et al.*, 2005) and by growing (Ga,Mn)As heterostructures with Ga chosen as the superconducting region (Braden *et al.*, 2003). The results of Braden *et al.* (2003) are shown in Fig. II.31 suggesting very highly polarized material ($P > 0.8$), consistent with the small conductance observed at low bias. Andreev reflection was also used to measure the spin polarization in narrow-band gap (In,Mn)Sb (Panguluri *et al.*, 2004) and EuS (Ren *et al.*, 2007), from a class of ferromagnetic semiconductors known to be effective spin filters (Esaki *et al.*, 1967; Moodera *et al.*, 2007).

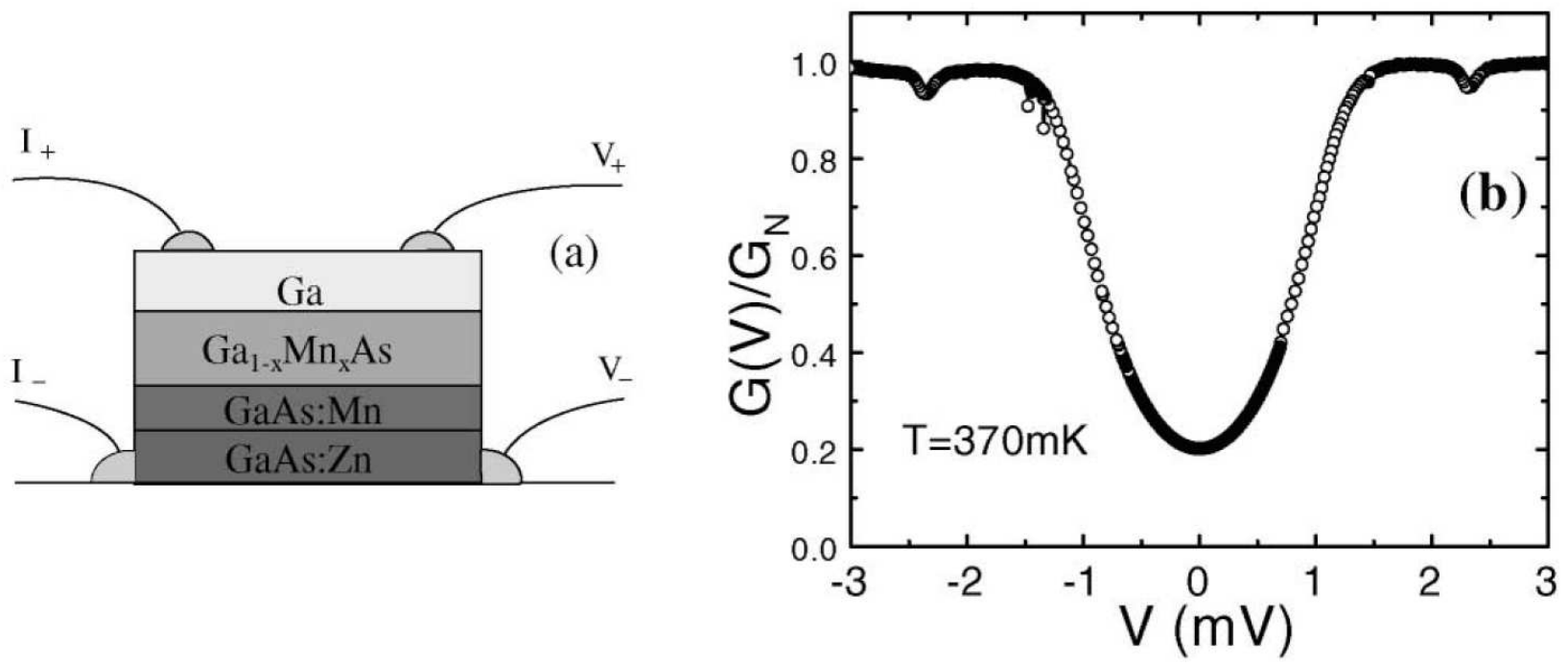

Fig. II.31. (a) A scheme of the voltage and current probes for transport measurements in Ga/(Ga,Mn)As-based junction grown by molecular beam epitaxy. (b) Measured normalized conductance spectrum of a Ga/(Ga,Mn)As junction shows a strongly suppressed Andreev reflection arising from the high spin polarization and junction transparency from Braden *et al.*, *Phys. Rev. Lett.* **91**, 0566602 (2003). Reprinted by permission of the American Physical Society, copyright (2003).

While the Griffin-Demers-BTK-approach, generalized to include the effects of spin-polarized transport, is valuable in estimating the measured spin polarization of ferromagnetic conductors, for quantitative analysis there are many additional factors not contained in the simple expression of Eq. (II.264). In the case of ferromagnetic semiconductors such as (Ga,Mn)As and (In,Mn)As, the carriers are spin-polarized holes and the appropriate spin-orbit coupling would need to be included. When the Fermi surface is not spherical, one has to specify what type of spin polarization is experimentally measured (Xia *et al.*, 2002; Mazin, 1999) and the care is needed even to calculate the charge transport in superconducting junctions (Žutić and Mazin, 2005). The roughness or the size of F/S interface may lead to a diffusive contribution to the transport (Jedema *et al.*, 1999; Fal'ko *et al.*, 1999; Mazin *et al.*, 2001). As a caution for possible difficulties in analyzing experimental data, we mention some subtleties which arise even for a simple model of spherical Fermi surfaces used to describe both F and S regions. Unlike for charge transport in N/S junctions (Blonder and Tinkham, 1983) in the Griffin-Demers-BTK approach, the Fermi velocity mismatch between the F and the S region does not simply increase the value of the effective $Z$. Specifically, at $Z = V = 0$ and normal incidence it is possible to have perfect transparency even when all the Fermi velocities differ, satisfying $(v_{F\uparrow}v_{F\downarrow})^{1/2} = v_S$, where $v_S$ is the Fermi velocity in a superconductor (Žutić and Das Sarma, 1999; Žutić and Valls, 1999, 2000). In other words, unlike in Eq. (II.264), the spin polarization (nonvanishing exchange energy) can *increase* the subband conductance, for fixed Fermi velocity mismatch. Conversely, at a fixed exchange energy an increase in Fermi velocity mismatch could increase the subgap conductance. Similar results were also obtained when F and S regions were separated by a quantum dot (Zhu *et al.*, 2001; Feng and Xiong, 2003; Zeng *et al.*, 2003) and even in a 1D tight-binding model with no spin polarization (Affleck *et al.*, 2000). In a typical interpretation of a measured conductance, complications can then arise in trying to disentangle the influence of the parameters $Z$, $P_C$, $\Delta_0$, and the Fermi velocity mismatch, from the nature of the point contacts (Kikuchi *et al.*, 2001) and from the role of inelastic scattering (Auth *et al.*, 2003). It was shown that several different combinations of $Z$, $P_C$, and $\Delta_0$, could provide conductance fits of similar accuracy (Panguluri *et al.*, 2005; Bugoslavsky *et al.*, 2005).

Conventional superconductors typically offer only a narrow temperature window for studying spin polarization, while Andreev reflection is limited to junctions with small $Z$-values. Both of these limitations are relaxed in junctions with high temperature superconductors (HTSC's), but the lack of our understanding of these materials makes such structures more a test ground for fundamental physics rather then a quantitative tool to determine spin polarization (Chen *et al.*, 2001; Vas'ko *et al.*, 1998; Chen *et al.*, 2005; Luo *et al.*, 2005; Visani *et al.*, 2007). There are also several important differences with the studies using conventional low temperature superconductors in N/S junctions. The superconducting pairing symmetry no longer yields an isotropic energy gap. The sign change of the pair potential can result in $G(V = 0) > 0$ for $T \to 0$ even for a strong tunneling barrier and give rise to zero bias conductance peak (Hu, 1994; Tanaka and Kashiwaya, 1995; Wei *et al.*, 1998; Hu, 1998; Kashiwaya and Tanaka, 2000; Sengupta *et al.*, 2001). Similar situation also pertains to F/S junctions with HTSC's. Even at large interfacial barrier (large $Z$) interference effects between the quasi-electron and quasi-hole scattering trajectories which feel, respectively, pair potentials of different sign lead to the formation of Andreev bound states which gives rise to a large conductance near zero bias (Žutić and Valls, 1999; Zhu *et al.*, 1999; Žutić and Valls, 2000; Kashiwaya *et al.*, 1999; Hu and Yan, 1999; Vodopyanov, 2005; Dong *et al.*, 2001). The suppression of a zero bias conductance peak, measured by a scanning tunneling microscope was recently used to detect injected spin into a high-temperature superconductors (Ngai *et al.*, 2004). In multiple F/S junctions Andreev bound states can also be formed with conventional superconductors and the transport properties are strongly modified with the change of the polarization of F-regions or carriers injected in S-regions (Fominov, 2003; Cayssol and Montambaux, 2004, 2005; Yamashita *et al.*, 2005; Petković *et al.*, 2006; Žikić and Dobrosavljević-Grujić, 2007; Zhao and Sauls, 2007).

## H. Spin-dependent tunneling in heterojunctions

### H.1 Tunneling magnetoresistance (TMR)

In 1975 Jullière reported the first results concerning an experiment performed on an Fe/Ge/Co junction, i.e., a junction made of a semiconducting slab, sandwiched between two ferromagnetic leads (Jullierè, 1975). The experiment showed a dependence of the resistance on whether the mean magnetization of the two ferromagnetic films were oriented in a parallel or antiparallel configuration. Although in both cases the electrons tunnel through the same Ge semiconducting barrier, leading to a high resistance, the measured resistance was higher in the case of the antiparallel alignment. This phenomenon, in which the resistance of a magnetic tunnel junction (MTJ) depends on the relative orientation of the magnetization in the ferromagnetic leads, was termed the tunneling magnetoresistance (TMR) effect. The size of the tunneling magnetoresistance is characterized by the quantity[30]

$$TMR = \frac{R_{AP} - R_P}{R_P} = \frac{G_P - G_{AP}}{G_{AP}}, \tag{II.266}$$

where $R_P$ ($G_P$) and $R_{AP}$ ($G_{AP}$) correspond to the resistance (conductance) measured in parallel [see Figs. II.32(a)] and antiparallel [see Figs. II.32(b)] configurations, respectively. The TMR observed by Jullière in Fe/Ge/Co junctions was about $14\%$ (Jullierѐ, 1975) but it was only observable at liquid He temperatures (and never reproduced). The first reproducible TMR was demonstrated by Maekawa and Gäfvert (1982) who observed a strong correlation between the tunnel conductance and the magnetization process in Ni/NiO/ferromagnet junctions with Ni, Fe, or Co as the counter electrode. The measured values of the TMR were, however, still very small at room temperature. It was not until 1995 that, triggered by the success of the giant magnetoresistance (GMR) and with the advent of superior fabrication techniques, ferromagnet/insulator/ferromagnet structures were revisited and a large room-temperature TMR ($\sim 18\%$) was observed (Miyazaki and Tezuka, 1995; Moodera *et al.*, 1995).[31] The discovery of the room-temperature TMR opened the possibility of using MTJs for fundamental studies of surface magnetism and room-temperature spin polarization in diverse ferromagnetic electrodes and produced a resurgence in interest in the study of MTJs.

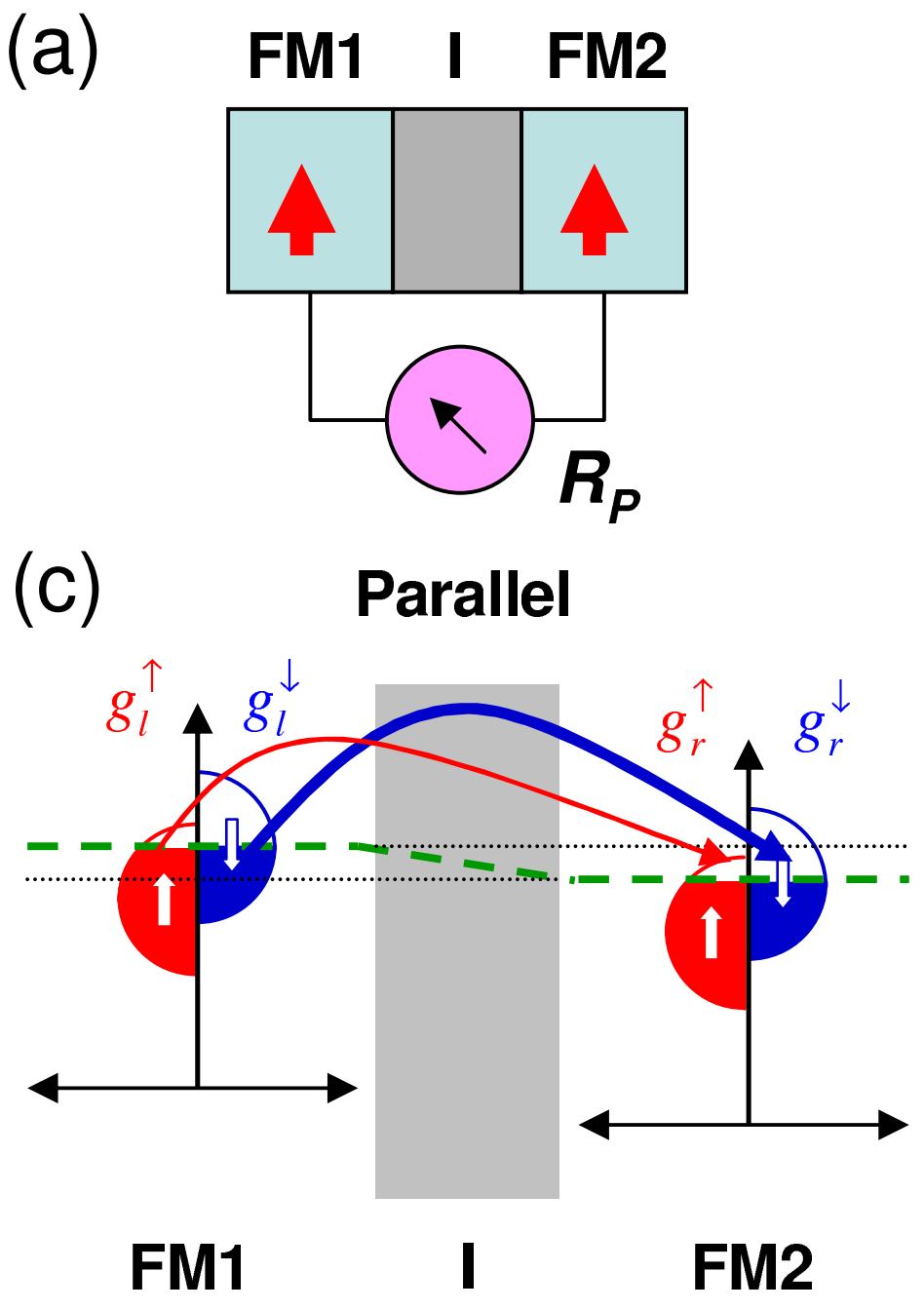

Fig. II.32. Schematics of parallel (a,c) and antiparallel (b,d) configurations for a tunnel junction composed of two ferromagnetic electrodes FM1 and FM2 separated by an isolating (I) barrier. The density of states corresponding to a spin-$\sigma$ particle in the left and right ferromagnetic electrodes are denoted by $g_l^\sigma$ and $g_r^\sigma$, respectively. Assuming the energy and spin are conserved during tunneling, the total current can be decomposed into spin-up (red arrows) and spin-down (blue arrows) contributions whose magnitudes are indicated by the thickness of their corresponding arrows in (c) and (d). By comparing (c) and (d) one can see that the total current in the parallel configuration is larger, leading to the smaller tunneling resistance.

[30] Alternatively, another definition of the TMR, i.e., $TMR = (R_{AP} - R_P)/R_{AP}$ is also widely used in the literature.

Furthermore, the TMR effect can be employed as widely as the GMR effect [e.g., highly sensitive magnetic-field sensors, magnetic read heads, spin-valve transistors, etc (Žutić *et al.*, 2004; Hirota *et al.*, 2002; Parkin, 2002; Dietzel, 2003)] but with the advantage of providing higher magnetoresistive signal amplitudes. The most important, presently discussed application of the TMR effect is, however, in the realization of magnetic random-access memories (MRAM) (Hirota *et al.*, 2002; Parkin, 2002; Slaughter *et al.*, 2003). The basic idea is to combine the non-volatility[32] of magnetic data storage with the short access times of present day random-access memories (DRAM). Thus, the recharging of the capacitors required for the periodic refreshing of the information in a DRAM is not needed in a MRAM device. Magnetic random-access memories are already commercially available (for more information visit www.freescale.com/mram). Main key features of these new devices are their high performance (with symmetrical read and write timing), small size and scalability for future technologies, nonvolatility (with virtually unlimited read-write endurance), low leakage, and low voltage capability. Specific details on the performance of the available MRAMs compared to other kind of memories can be found at www.freescale.com/mram.

The investigation of transport properties in all-semiconductor structures with magnetic semiconductor electrodes [e.g. (Ga,Mn)As/AlAs/(Ga,Mn)As and GaMnAs/GaAs/GaMnAs MTJs] has recently attracted much attention (see Sec. V.B.). In particular, TMR ratios larger than $250\%$ has been observed in (Ga,Mn)As/AlAs/(Ga,Mn)As MTJs (Mattana *et al.*, 2003; Chiba *et al.*, 2004; Elsen *et al.*, 2006). The use of such structures would simplify integration with the nowadays semiconductor-based electronics. Other systems of renewed interest for the study of the TMR are the ferromagnet/semiconductor/ferromagnet tunneling junctions (MacLaren *et al.*, 1999; Gustavsson *et al.*, 2001; Guth *et al.*, 2001; Kreuzer *et al.*, 2002; Zenger *et al.*, 2004; Moser *et al.*, 2006; Popescu *et al.*, 2004, 2005). In general, the presence of semiconductors in the MTJs introduces spin-orbit related effects, which can result in novel phenomena such as the tunneling anisotropic magnetoresistance (TAMR) effect. This effect refers to the dependence of the magnetoresistance on the absolute orientation of the magnetization with respect to crystallographic directions (Gould *et al.*, 2004a; Rüster *et al.*, 2005; Saito *et al.*, 2005; Brey *et al.*, 2004). Unlike the TMR, the TAMR is observed even in MTJs in which only one of the electrodes is magnetic (Gould *et al.*, 2004a; Moser *et al.*, 2007). The TAMR effect is discussed in Sec. H.4.

[31] Although both the TMR and GMR effects are based on the difference of magnetoresistance for parallel and antiparallel configurations of the ferromagnetic electrodes, they are different phenomena with different properties. The TMR effect occurs in MTJs composed of an insulating or semiconductor barrier sandwiched between two magnetic electrodes, while the GMR effect refers to trilayer structures consisting of two magnetic layers separated by a non-magnetic metallic spacer. Due to the distinct nature of the spacer (an insulator for TMR and a metal for GMR), the factors that influence the TMR and GMR effects are quite different.

[32] Non-volatile devices are those in which the information stored as a magnetic bit (i.e., a magnetic domain) is conserved over an extended period of time (typically $\geq 10$ years).

### H.2 Jullière's model

The main physical picture of the TMR effect can be understood on the basis of spin-polarized tunneling (Jullierè, 1975). Assume that the electrons tunnel without spin flip. In such a situation, the spin is conserved during tunneling and an initial spin-up (spin-down) electron in one electrode can only tunnel to an unoccupied spin-up (spin-down) final state in the other electrode. This simple observation gives a direct qualitative explanation of the origin of the TMR in a system consisting of two half metals (with only one spin direction at the Fermi level) separated by a thin insulating barrier (see Fig. II.33). In such a system, in the parallel (P) configuration, transmission from one electrode to the other occurs due to tunneling and the resistance is finite. On the contrary, in the antiparallel (AP) configuration, due to the lack of available states for the incident spins in the counter electrode, transmission is not allowed and the resistance of the junction becomes infinite (see Fig. II.33). This gives, for the half-metallic electrodes, an infinite TMR.

The TMR arises from the imbalance between the number of spin-up and spin-down electrons contributing to the tunneling current. In the more general case of a ferromagnet/insulator/ferromagnet junction the resistance is finite in both the parallel and antiparallel configurations. The application of a small bias $V_{\rm bias}$ defines the energy window for the electrons contributing to the tunneling current [see Figs. II.32 (c) and (d)]. Since the spin is conserved during the tunneling, the total current can be decomposed into spin-up (red arrows) and spin-down (blue arrows) contributions whose magnitudes are indicated by the thickness of their corresponding arrows in Figs. II.32(c) and (d). From the comparison between these two figures one can see that the total current in the parallel configuration is larger and, therefore leads to the smaller tunneling resistance.

Jullière's model can be deduced from the tunnel Hamiltonian approach (Maekawa *et al.*, 2002) (see also Sec. V.A.4), which consists in expressing the system Hamiltonian as

$$\mathrm{H} = \mathrm{H}_l + \mathrm{H}_r + \mathrm{H}_t, \tag{II.267}$$

where $\mathrm{H}_l$ and $\mathrm{H}_r$ are, respectively, the Hamiltonians of the left and right electrodes and $\mathrm{H}_t$ describes the tunneling process. Within this approach, the tunneling rate $\Gamma^{\sigma}_{l\to r}(V)$ at which electrons with spin $\sigma$ are transferred from the left to the right electrodes can be estimated from Fermi's golden rule as

$$\Gamma^{\sigma}_{l\to r}(V) = \frac{4\pi^2}{h}\sum_{\mathbf{k},\boldsymbol{\kappa}} |t_\sigma(\mathbf{k},\boldsymbol{\kappa})|^2 f(\epsilon_{\mathbf{k}\sigma})[1-f(\epsilon_{\boldsymbol{\kappa}\sigma})]\delta(\epsilon_{\mathbf{k}\sigma}-\epsilon_{\boldsymbol{\kappa}\sigma}+eV), \tag{II.268}$$

where $\epsilon_{\mathbf{k}\sigma}$ and $\epsilon_{\boldsymbol{\kappa}\sigma}$ are the one-electron energies measured from the Fermi level in the left and right electrodes, respectively. The corresponding Fermi-Dirac distribution functions are $f(\epsilon) = 1/[\exp(\epsilon/k_BT)+1]$, with $\epsilon=\epsilon_{\mathbf{k}\sigma}$ or $\epsilon=\epsilon_{\boldsymbol{\kappa}\sigma}$. The tunneling probability for a particle with spin $\sigma$ is determined by the module square of the tunneling matrix elements,

$$|t_\sigma(\mathbf{k},\boldsymbol{\kappa})|^2 = |\langle\Psi^{\sigma}_r(\boldsymbol{\kappa})|H_t|\Psi^{\sigma}_l(\mathbf{k})\rangle|^2, \tag{II.269}$$

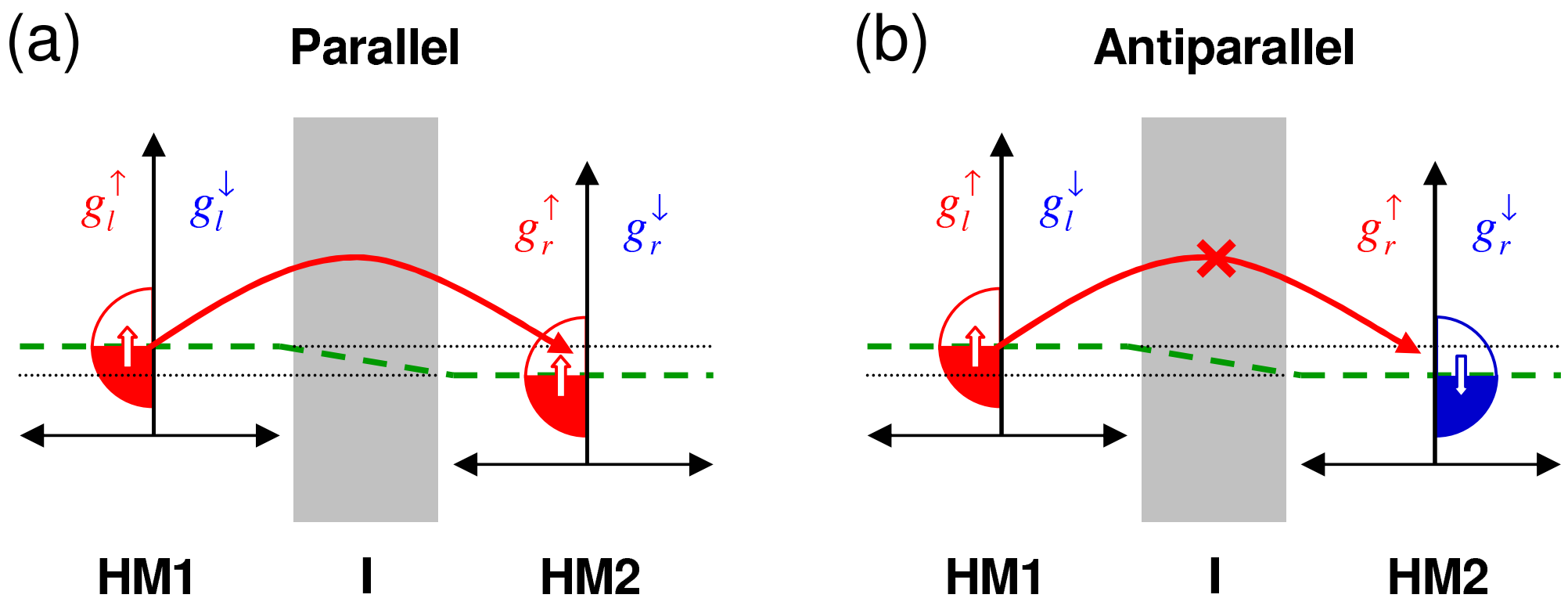

Fig. II.33. Schematics of the TMR effect for a MTJ composed of two half-metallic (HM) electrodes separated by an insulating (I) barrier. (a) Parallel configuration with $G_P > 0$. (b) Antiparallel configuration with $G_{AP} = 0$. The density of states corresponding to a spin-$\sigma$ particle in the left and right HM electrodes are denoted by $g_l^\sigma$ and $g_r^\sigma$, respectively. The transmission from one electrode to the other is only possible in the parallel configuration, resulting in an infinite TMR ratio [see Eq. (II.266)].

where $\Psi_l^\sigma(\mathbf{k})$ and $\Psi_r^\sigma(\boldsymbol{\kappa})$ are the wave functions describing electron states in the left and right electrodes, respectively. The Dirac delta function in Eq. (II.268) reflects the energy conservation, while $f(\epsilon_{\mathbf{k}\sigma})$ and $[1 - f(\epsilon_{\boldsymbol{\kappa}\sigma})]$ determine the occupancy and unoccupancy of states in the left and right electrodes, respectively. By introducing the densities of states $g_i^\sigma(\epsilon)$ $(i = l, r)$ for the spin-$\sigma$ particles in the left $(l)$ and right $(r)$ electrodes, one can transform the sums over $\mathbf{k}$ and $\boldsymbol{\kappa}$ in Eq. (II.268) into energy integrals. One then obtains

$$\Gamma_{l\to r}^\sigma(V) = \frac{4\pi^2}{h} T \int_{-\infty}^{\infty} g_l^\sigma(\epsilon - eV) g_r^\sigma(\epsilon) f(\epsilon - eV)[1 - f(\epsilon)] d\epsilon, \tag{II.270}$$

where we have assumed the tunneling matrix elements are spin and wave vector independent and $T = |t_\sigma(\mathbf{k}, \boldsymbol{\kappa})|^2$. Usually, the tunneling probability is assumed to be a constant such that $T \propto \exp(-2\xi d)$, where $\xi = \sqrt{2m_0\Phi}/\hbar$ characterizes the decay of the wave function in the barrier and $\Phi$ is the barrier height (measured from the Fermi level). Similarly, the tunneling rate $\Gamma_{r\to l}^\sigma(V)$ at which electrons with spin $\sigma$ are transferred from the right to the left electrode is given by

$$\Gamma_{r\to l}^\sigma(V) = \frac{4\pi^2}{h} T \int_{-\infty}^{\infty} g_l^\sigma(\epsilon - eV) g_r^\sigma(\epsilon) f(\epsilon)[1 - f(\epsilon - eV)] d\epsilon. \tag{II.271}$$

The difference between the forward and backward tunneling rates gives the steady-state current through the junction,

$$I_\sigma(V) = e[\Gamma_{l\to r}^\sigma(V) - \Gamma_{r\to l}^\sigma(V)], \tag{II.272}$$

which, considering Eqs. (II.270) and (II.271), reduces to (Maekawa *et al.*, 2002)

$$I_\sigma(V) = \frac{4e\pi^2}{h} T \int_{-\infty}^{\infty} g_l^\sigma(\epsilon - eV) g_r^\sigma(\epsilon)[f(\epsilon - eV) - f(\epsilon)] d\epsilon. \tag{II.273}$$

The total current $I = I_\uparrow + I_\downarrow$ is the sum of the spin-up and spin-down currents.

In the low bias regime, where the applied voltage $V$ is much smaller than the band width and the density of states is approximately constant one may obtain an approximate expression for $I_\sigma(V)$ by expanding Eq. (II.273) in powers of $V$. The result, to first order in the bias voltage, reads

$$I_\sigma(V) = G_\sigma V;\; G_\sigma \approx -\frac{4\pi^2 e^2}{h} T \int_{-\infty}^{\infty} g_l^\sigma(\epsilon) g_r^\sigma(\epsilon) \frac{\partial f(\epsilon)}{\partial \epsilon} d\epsilon. \tag{II.274}$$

For low temperatures $f(\epsilon) \approx \Theta(\epsilon_F - \epsilon)$ ([here $\Theta(x)$ represents the Heaviside step function], and Eq. (II.274) reduces to

$$G_\sigma \approx \frac{4\pi^2 e^2}{h} T g_l^\sigma g_r^\sigma, \tag{II.275}$$

with the densities of states $g_i^\sigma = g_i^\sigma(\epsilon_F)$ evaluated at the Fermi energy $\epsilon_F$. Note that Eq. (II.275) reflects the fact that at low temperatures and small voltages, the transport properties of the system are determined by those states at the Fermi level. For the case of parallel alignment we have,

$$G_P = G_P^\uparrow + G_P^\downarrow \propto \left( g_l^\uparrow g_r^\uparrow + g_l^\downarrow g_r^\downarrow \right), \tag{II.276}$$

while for the antiparallel configuration,

$$G_{AP} = G_{AP}^\uparrow + G_{AP}^\downarrow \propto \left( g_l^\uparrow g_r^\downarrow + g_l^\downarrow g_r^\uparrow \right), \tag{II.277}$$

where $g_i^\uparrow$ and $g_i^\downarrow$ are the density of states at the Fermi level for the majority and minority spin bands in the $i$th electrode, respectively. From Eqs. (II.266), (II.276), and (II.277) one finds for the TMR ratio,

$$TMR = \frac{R_{AP} - R_P}{R_P} = \frac{G_P - G_{AP}}{G_{AP}} \approx \frac{2P_{gl}P_{gr}}{1 - P_{gl}P_{gr}}. \tag{II.278}$$

Here we have introduced the spin polarization of the density of states at the Fermi level,

$$P_{gi} = \frac{g_i^\uparrow - g_i^\downarrow}{g_i^\uparrow + g_i^\downarrow};\; i = l, r \tag{II.279}$$

of the $i$th electrode.[33] By using Eq. (II.278) one can obtain information about the spin polarization of the electrodes by measuring the tunneling magnetoresistance in the parallel and antiparallel configurations. For example, if one (or both) of the electrodes is not spin polarized (i.e., when $P_{gl}P_{gr} = 0$), then the TMR vanishes. On the other hand the TMR becomes infinite when both electrodes are fully spin-polarized, i.e., when $P_{gl} = P_{gr} = 1$ (as in the case of half metallic electrodes, considered at the beginning of this section). Such a half-metallic behavior is rare, but some materials like the oxides $CrO_2$ and $Fe_3O_4$ (Shang *et al.*, 1998) and $La_{0.7}Sr_{0.3}MnO_3$ (De Teresa *et al.*, 1999b,a) appear to nearly exhibit this amazing property.

[33]The spin polarization of the density of states at the Fermi level, Eq. II.279, refers to the parallel configuration.

Usually the TMR ratio is positive (in such a case the effect is called normal). However, the inverse effect with a negative TMR ratio has also been observed (De Teresa *et al.*, 1999b,a). The inverse TMR effect occurs when the magnetic electrodes on both sides of the barrier have spin polarizations with opposite sign, i.e., when $P_{gl}P_{gr} < 0$ [see Eq. (II.278)].

The above model and the expression for the TMR ratio given in Eq. (II.278) are called Jullière's model and Jullière's formula, respectively. In spite of its relative simplicity, Jullière's model can explain in many cases the experimental trends and has continued to be used for interpreting the spin polarization in various MTJs.

In an approach complementary to Jullière's, Slonczewski (1989) considered a ferromagnet/insulator/ferromagnet junction as a single quantum mechanical system in a free-electron picture. The details of Slonczewski's model is discussed in the following section.

### H.3 Slonczewski's model

In Jullière's model the tunneling matrix elements are assumed to be constant, being the same for spin-up and spin-down electrons, i.e., the wave function in the barrier region is assumed to be spin- and $\mathbf{k}$-independent. A different approach, free of these restricting assumptions, was proposed by Slonczewski (1989), who considered the exact wave function in the barrier within the free-electron model. Slonczewski's model considers coherent tunneling and assumes the conservation of the in-plane wave vector during tunneling. This model is, therefore, particularly relevant for tunneling through epitaxially grown MTJs.

In the original work of (Slonczewski, 1989) a symmetric ferromagnet/insulator/ ferromagnet junction with a square barrier and constant effective masses along the heterostructure was considered. We discuss here a generalization of Slonczewski's model to the case of asymmetric junctions with a position dependent effective mass (Bratkovsky, 1997).

We note that the expression for the current given by Eq. (II.273) was derived from Fermi's golden rule in Eq. (II.268), which neglects the details of the wave function within the barrier region. A more convenient expression for the tunnel current including these details can be obtained by noting that the current density of spin-$\sigma$ particles from the left to the right electrode can be written as

$$I^{\sigma}_{l\to r}(V) = \sum_{\mathbf{k}} j^{\sigma}_{\mathbf{k}} f(\epsilon_{\mathbf{k}\sigma} + eV), \tag{II.280}$$

where the $z$-component of the current density carried by an incident from the left particle with spin $\sigma$ and wave vector $\mathbf{k}$ is given by

$$j^{\sigma}_{\mathbf{k}} = -\frac{e}{\Omega} T_{\sigma}(\epsilon_{\mathbf{k}\sigma}, \mathbf{k}_{\|}) v^{\sigma}_{z}. \tag{II.281}$$

Here $\Omega$ and $v^{\sigma}_{z}$ are the volume and $z$-component of the velocity, respectively. The tunneling probability, $T_{\sigma}(\epsilon_{\mathbf{k}\sigma}, \mathbf{k}_{\|})$, has to be calculated including the details of the wave function in the barrier region. Taking into account that $v^{\sigma}_{z} = \hbar^{-1}\partial\epsilon_{\mathbf{k}\sigma}/\partial k_z$ and approximating the sum over $\mathbf{k}$ by an integral one obtains from Eqs. (II.280) and (II.281) the relation

$$I^{\sigma}_{l\to r}(V) = -\frac{e}{(2\pi)^2 h} \int d\epsilon d^2\mathbf{k}_{\|} T_{\sigma}(\epsilon, \mathbf{k}_{\|}) f(\epsilon + eV). \tag{II.282}$$

Similarly, one obtains that the current flowing from the right to the left electrode is given by $J^{\sigma}_{r\to r} = J^{\sigma}_{l\to r}(0)$. The net tunnel current $J_\sigma(V) = J^{\sigma}_{l\to r}(V) - J^{\sigma}_{r\to l}$ for the $\sigma$ channel is, therefore,

$$I_\sigma(V) = \frac{e}{(2\pi)^2 h}\int d\epsilon d^2\mathbf{k}_\parallel T_\sigma(\epsilon, \mathbf{k}_\parallel)[f(\epsilon) - f(\epsilon + eV)]. \tag{II.283}$$

The total current flowing along the heterojunction is the sum of the spin-up and spin-down contributions, i.e., $I(V) = I_\uparrow(V) + I_\downarrow(V)$. In the limit of low temperatures and small biases, one can simplify Eq. (II.283) by following the same approximate procedure used in obtaining Eq. (II.275). The result is the linear response relation, $I_\sigma = G_\sigma V$, in which the conductance per unit area is given by

$$G_\sigma \approx \frac{e^2}{(2\pi)^2 h}\int d^2\mathbf{k}_\parallel T_\sigma(\epsilon_F, \mathbf{k}_\parallel). \tag{II.284}$$

It is often convenient to use

$$\int \frac{d^2\mathbf{k}_\parallel}{(2\pi)^2}... = \int \rho_\parallel(E_\parallel)dE_\parallel..., \tag{II.285}$$

where $E_\parallel = \hbar^2 k_\parallel^2/(2m)$ and $\rho_\parallel(E_\parallel) = \rho_\parallel = m/(2\pi\hbar^2)$ and $m$ are the 2D density of states and electron effective mass of the source electrode, respectively. With the help of Eq. (II.285), and assuming $T_\sigma(\epsilon_F, \mathbf{k}_\parallel) = T_\sigma(\epsilon_F, k_\parallel)$, we can rewrite Eq. (II.284) as

$$G_\sigma \approx \frac{e^2}{h}\rho_\parallel \int_0^{E_\parallel^{max}} T_\sigma(\epsilon_F, E_\parallel)dE_\parallel, \tag{II.286}$$

where the integration limit $E_\parallel^{max}$ is determined as the maximum value of $E_\parallel$ allowed for propagating modes.

Slonczewski's model assumes a free-electron approximation for the spin polarized conduction electrons together with a Stoner approach for the ferromagnetic electrodes. The single-particle Hamiltonian in the $i$th region of the heterojunction may be written as

$$\mathrm{H} = -\frac{\hbar^2\nabla^2}{2m_i^*} + V_i - \frac{\Delta_i}{2}\mathbf{n}_i \cdot \boldsymbol{\sigma};\ i = l, c, r \tag{II.287}$$

where $m_i^*$ and $V_i$ are the effective mass and potential energy, respectively, in the left ($i = l$), central ($i = c$), and right ($i = r$) regions. The exchange energy $\Delta_i$ vanishes in the central, non-magnetic barrier, i.e., $\Delta_c = 0$. The unit vectors $\mathbf{n}_l$ and $\mathbf{n}_r$ define the magnetization direction in the left and right ferromagnetic electrodes, respectively, and $\boldsymbol{\sigma}$ is a vector whose components are the Pauli matrices. We assume that the magnetization in the left electrode is oriented along the $x$ axis [$\mathbf{n}_l = (1, 0, 0)$], while the magnetization in the right electrode points in the direction $\mathbf{n}_r = (\cos\phi, \sin\phi, 0)$, where $\phi = \angle(\mathbf{n}_l, \mathbf{n}_r)$.

Assuming that the in-plane wave vector $\mathbf{k}_\parallel$ is conserved during tunneling, the motion along the growth direction ($z$) can be decoupled from the other spatial degrees of freedom. The $z$ component of the scattering state describing a spin-$\sigma$ particle in the left ($z \leq 0$) ferromagnetic electrode with eigenenergy $E$ is given by

$$\Psi_\sigma^{(l)}(z) = \frac{e^{ik_\sigma z}}{\sqrt{k_\sigma}}\chi_\sigma^{(l)} + r_{\sigma,\sigma}e^{-ik_\sigma z}\chi_\sigma^{(l)} + r_{\sigma,-\sigma}e^{-ik_{-\sigma} z}\chi_{-\sigma}^{(l)}, \tag{II.288}$$

where

$$\chi_\sigma^{(l)} = \frac{1}{\sqrt{2}}\begin{pmatrix} 1 \\ \sigma \end{pmatrix}, \tag{II.289}$$

represents the spinor corresponding to the spin parallel ($\sigma = 1$) or antiparallel ($\sigma = -1$) to the magnetization direction $\mathbf{n}_l = (1, 0, 0)$ in the left, ferromagnetic electrode, and

$$k_\sigma = \sqrt{k_{\sigma 0}^2 - \frac{2m_l^* E_\parallel}{\hbar^2}};\ k_{\sigma 0} = \sqrt{\frac{2m_l^*}{\hbar^2}\left(E - V_l + \sigma\frac{\Delta_l}{2}\right)}, \tag{II.290}$$

is the corresponding $z$ component of the wave vector for a spin-$\sigma$ particle in the left electrode with $E_\parallel = \hbar^2 k_\parallel^2/(2m_l^*)$. The first term of the sum in the right-hand side of Eq. (II.288) corresponds to a spin-$\sigma$ incident plane wave having unit incident particle flux in the left electrode, while the second and third terms describe reflected plane waves with spins $\sigma$ and $-\sigma$, respectively.

In the central, non-magnetic region ($0 < z < d$) we have evanescent plane waves given by

$$\Psi_\sigma^{(c)}(z) = \sum_{j=\pm 1}\left(A_{\sigma,j}e^{-qz} + B_{\sigma,j}e^{qz}\right)\chi_j^{(l)}. \tag{II.291}$$

Here $d$ denotes the barrier width and

$$q = \sqrt{q_0^2 + \frac{2m_l^* E_\parallel}{\hbar^2}};\ q_0 = \sqrt{\frac{2m_c^*}{\hbar^2}\left(V_c - E\right)}. \tag{II.292}$$

Note that since the barrier is non-magnetic, we have taken, without loss of generality, the spin quantization axis in the central region to be the same as the spin quantization axis of the incoming (from the left electrode) wave, i.e., $\chi_\sigma^{(c)} = \chi_\sigma^{(l)}$. In the right ($z \geq d$), ferromagnetic electrode the scattering states are composed of two transmitted plane waves with spins parallel and antiparallel to $\mathbf{n}_r$, i.e.,

$$\Psi_\sigma^{(r)}(z) = t_{\sigma,\sigma}e^{i\kappa_\sigma(z-d)}\chi_\sigma^{(r)} + t_{\sigma,-\sigma}e^{i\kappa_{-\sigma}(z-d)}\chi_{-\sigma}^{(r)}, \tag{II.293}$$

where

$$\chi_\sigma^{(r)} = \frac{1}{\sqrt{2}}\begin{pmatrix} 1 \\ \sigma e^{i\phi} \end{pmatrix}, \tag{II.294}$$

represents the spinor corresponding to the spin parallel ($\sigma = 1$) or antiparallel ($\sigma = -1$) to the magnetization direction $\mathbf{n}_r = (\cos\phi, \sin\phi, 0)$ in the right, ferromagnetic electrode, and

$$\kappa_\sigma = \sqrt{\kappa_{\sigma 0}^2 - \frac{2m_l^* E_\parallel}{\hbar^2}};\ \kappa_{\sigma 0} = \sqrt{\frac{2m_r^*}{\hbar^2}\left(E - V_r + \sigma\frac{\Delta_r}{2}\right)}, \tag{II.295}$$

is the corresponding $z$ component of the wave vector for a spin-$\sigma$ particle in the right electrode.

The eight coefficients $r_{\sigma,\pm\sigma}$, $A_{\sigma,\pm1}$, $B_{\sigma,\pm1}$, and $t_{\sigma,\pm\sigma}$ present in Eqs. (II.288) - (II.293) have to be determined as the solutions of the system of eight linear equations resulting from the continuity of the probability current across the interfaces, i.e.,

$$\Psi_\sigma^{(i)}(z_{ij}) = \Psi_\sigma^{(j)}(z_{ij}); \quad \frac{1}{m_i^*}\frac{\partial\Psi_\sigma^{(i)}}{\partial z}\bigg|_{z_{ij}} = \frac{1}{m_j^*}\frac{\partial\Psi_\sigma^{(j)}}{\partial z}\bigg|_{z_{ij}}; \; i,j = l,c,r; \; \sigma = \pm 1. \quad \text{(II.296)}$$

Here $z_{ij}$ represents the position of the interface between the $i$th and $j$th regions. The transmissivity of an incoming spin-$\sigma$ particle can be computed from the relations

$$T_\sigma(E,E_\parallel) = \left(\frac{m_l^*}{m_r^*}\right)\mathrm{Re}[\kappa_\sigma|t_{\sigma,\sigma}|^2+\kappa_{-\sigma}|t_{\sigma,-\sigma}|^2] = 1-\mathrm{Re}[k_\sigma|r_{\sigma,\sigma}|^2+k_{-\sigma}|r_{\sigma,-\sigma}|^2]. \quad \text{(II.297)}$$

Here the real-part functions Re[...] express the fact that only the propagating modes contribute to the tunneling. The coefficients $t_{\sigma,\sigma}$ and $t_{\sigma,-\sigma}$ can be found by directly solving the system of equations resulting from Eq. (II.296) or by using the transfer matrix method described in Sec. V.A.3. Simplified analytical expressions for these coefficients are found in the limit $qd \gg 1$. In such a case one finds the following approximate relation for the tunneling coefficients,

$$t_{\sigma,\sigma'} \approx \frac{-2im_c^* m_r^* q\sqrt{k_\sigma}}{(m_r^* q - im_c^*\kappa_\sigma)(m_l^* q - im_c^* k_\sigma)}\left(1+\sigma\sigma' e^{-i\phi}\right)e^{-qd}, \quad \text{(II.298)}$$

which is valid to first order in $\exp(-qd)$. Substituting Eq. (II.298) into (II.297) one obtains the transmission probability,

$$\begin{aligned} T_\sigma(E,E_\parallel) \;&\approx\; \frac{8m_l^* m_r^* m_c^{*2} k_\sigma(\kappa_\sigma+\kappa_{-\sigma})(m_r^{*2}q^2+m_c^{*2}\kappa_\sigma\kappa_{-\sigma})}{(m_l^{*2}q^2+m_c^{*2}k_\sigma^2)(m_r^{*2}q^2+m_c^{*2}\kappa_\sigma^2)(m_r^{*2}q^2+m_c^{*2}\kappa_{-\sigma}^2)} \\ &\times\; \left[1+\frac{(\kappa_\sigma-\kappa_{-\sigma})(m_r^{*2}q^2-m_c^{*2}\kappa_\sigma\kappa_{-\sigma})}{(\kappa_\sigma+\kappa_{-\sigma})(m_r^{*2}q^2+m_c^{*2}\kappa_\sigma\kappa_{-\sigma})}\cos\phi\right]e^{-2qd}. \end{aligned} \quad \text{(II.299)}$$

The conductance for the spin-$\sigma$ channel can now be calculated by using Eqs. (II.286) and (II.299). Since $T_\sigma(\epsilon_F, E_\parallel)$ is a complicated function of $E_\parallel$, the exact integration in Eq. (II.286) has to be performed numerically. One can, however, obtain an approximate analytical expression for the conductance in the case of a high potential barrier. In such a case $k_\parallel \ll q_0$ and one can approximate Eq. (II.292) as follows

$$q \approx q_0\left[1+\frac{1}{2}\left(\frac{k_\parallel}{q_0}\right)^2\right] = q_0 + \frac{m_l^* E_\parallel}{q_0\hbar^2}. \quad \text{(II.300)}$$

Introducing the new integration variable $\zeta = 2m_l^* E_\parallel d/(q_0\hbar^2)$, which is dimensionless, and taking into account Eq. (II.300) one can see that because of the presence of the exponential factor $\exp(-\zeta)$ in the transmissivity [see Eq. (II.299)], the main contribution to the integral in Eq. (II.286) comes from the vicinity of $\zeta \approx 0$. Within this approximation Eq. (II.286) transforms into

$$G_\sigma \approx \frac{e^2}{h}\left(\frac{q_0\hbar^2}{2m_l^* d}\right)\rho_\parallel T_\sigma(\epsilon_F,0) = \frac{e^2 q_0}{8\pi^2\hbar d}T_\sigma(\epsilon_F,0). \quad \text{(II.301)}$$

The total conductance, $G = G_\uparrow + G_\downarrow$, is obtained from Eqs. (II.299) and (II.301). The result is

$$G(\phi) \approx G_0 \left(1 + P_{gl}^{eff} P_{gr}^{eff} \cos\phi\right), \tag{II.302}$$

where

$$G_0 = \frac{2e^2 q_{_F} e^{-2q_{_F} d}}{\pi h d} \prod_{i=l,r} \left[\frac{m_i^* m_c^* (k_{i\uparrow} + k_{i\downarrow})(m_i^{*2} q_{_F}^2 + m_c^{*2} k_{i\uparrow} k_{i\downarrow})}{(m_i^{*2} q_{_F}^2 + m_c^{*2} k_{i\uparrow}^2)(m_i^{*2} q_{_F}^2 + m_c^{*2} k_{i\downarrow}^2)}\right], \tag{II.303}$$

and

$$P_{gi}^{eff} = \frac{(k_{i\uparrow} - k_{i\downarrow})}{(k_{i\uparrow} + k_{i\downarrow})} \frac{(m_i^{*2} q_{_F}^2 - m_c^{*2} k_{i\uparrow} k_{i\downarrow})}{(m_i^{*2} q_{_F}^2 + m_c^{*2} k_{i\uparrow} k_{i\downarrow})};\ (i = l, r), \tag{II.304}$$

is the effective spin polarization of the $i$th electrode. In Eqs. (II.303) and (II.304) we have introduced the notations $q_{_F} = q_0(\epsilon_F)$, $k_{l\sigma} = k_{\sigma 0}(\epsilon_F)$, and $k_{r\sigma} = \kappa_{\sigma 0}(\epsilon_F)$ $(\sigma = \uparrow, \downarrow)$.

In order to include the relative orientation of the magnetization directions in the ferromagnetic electrodes, we can generalize the TMR ratio defined in Eq. (II.266) as

$$TMR(\phi) = \frac{R(\phi) - R(0)}{R(0)} = \frac{G(0) - G(\phi)}{G(\phi)}. \tag{II.305}$$

Note that $G(0) = G_P$ and $G(\pi) = G_{AP}$ and, consequently, the standard definition of the TMR [see Eq. (II.266)] is recovered when $\phi = \pi$.

From Eqs. (II.302) - (II.305) one obtains for the angular dependent TMR,

$$TMR(\phi) \approx \frac{P_{gl}^{eff} P_{gr}^{eff} (1 - \cos\phi)}{1 + P_{gl}^{eff} P_{gr}^{eff} \cos\phi}, \tag{II.306}$$

which in the particular case $\phi = \pi$ reduces to

$$TMR(\pi) \approx \frac{2 P_{gl}^{eff} P_{gr}^{eff}}{1 - P_{gl}^{eff} P_{gr}^{eff}}. \tag{II.307}$$

This relation is similar to Jullière's formula [see Eq. (II.278)] but now the spin polarizations $P_{gl}$ and $P_{gr}$ of Jullière's model are replaced by the effective spin polarizations $P_{gl}^{eff}$ and $P_{gr}^{eff}$, respectively. Furthermore, considering that the density of state $g_i^\sigma \propto k_{i\sigma}$ and using Eqs. (II.279) and (II.304) one finds the effective spin polarizations $P_{gi}^{eff}$ and the spin polarizations $P_{gi}$ are related as follows:

$$P_{gi}^{eff} = P_{gi} \frac{(m_i^{*2} q_{_F}^2 - m_c^{*2} k_{i\uparrow} k_{i\downarrow})}{(m_i^{*2} q_{_F}^2 + m_c^{*2} k_{i\uparrow} k_{i\downarrow})};\ (i = l, r). \tag{II.308}$$

Thus, in Slonczewski's model the spin polarizations $P_{gi}$ of Jullière's model appear to be modified by a factor which depends on the electron penetration to the barrier, the effective masses in the different regions, and on the product $k_{i\uparrow} k_{i\downarrow}$. One can see from Eq. (II.308) that $|P_{gi}^{eff}| \leq |P_{gi}|$.

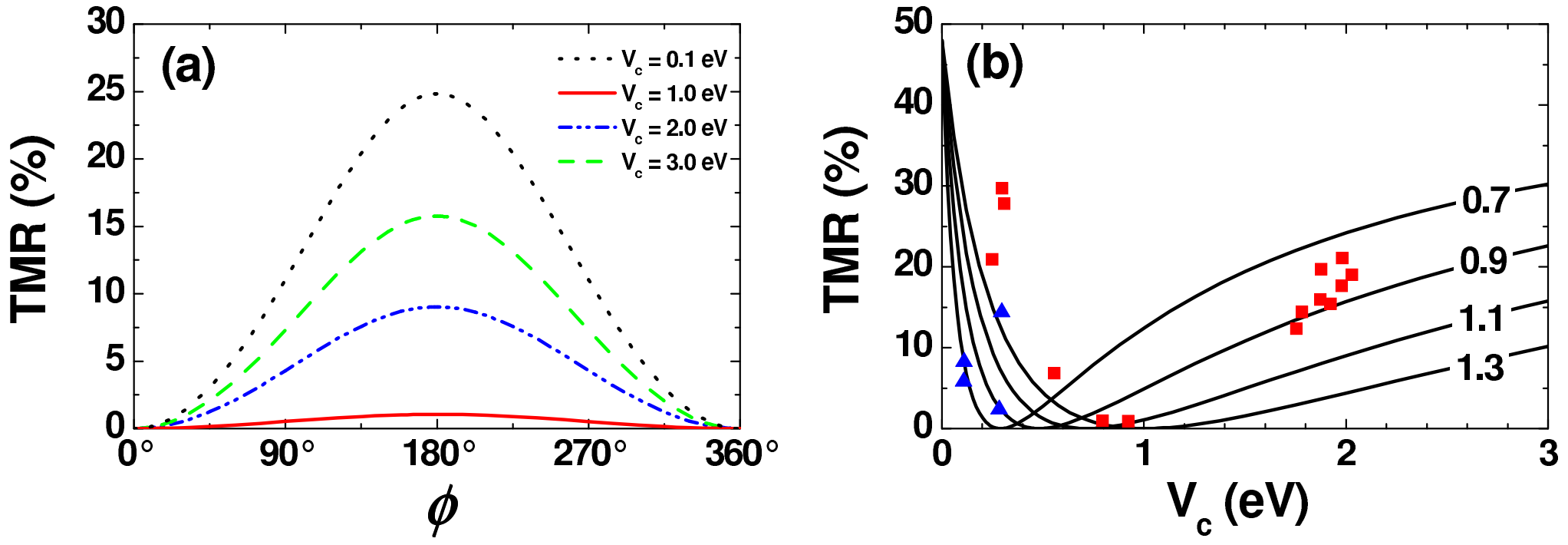

Fig. II.34. (a) Dependence of the TMR on the relative direction of the magnetizations of the Fe electrodes of an $\mathrm{Fe/Al_2O_3/Fe}$ MTJ for typical values of the barrier height $V_c$. (b) TMR dependence on the barrier height $V_c$ for $\phi = \pi$. Solid lines correspond to the theoretical results obtained from Eq. (II.307). Symbols represent experimental data taken from Tezuka and Miyazaki (1998a). The values of $k_{l\uparrow} = k_{r\uparrow}$ have been fixed to $0.7$ Å$^{-1}$, $0.9$ Å$^{-1}$, $1.1$ Å$^{-1}$, and $1.3$ Å$^{-1}$ (see labels on the curves), while $k_{l\downarrow} = k_{r\downarrow}$ have been evaluated by using the experimental value of $0.44$ for the spin polarization in Fe (Tedrow and Meservey, 1973).

On the other hand, for a tunnel junction such that $m_i^{*2} q_F^2 < m_c^{*2} k_{i\uparrow} k_{i\downarrow}$ we have $P_{gi} P_{gi}^{eff} < 0$, i.e., in such a system the electron spins reverse at the $i$th ferromagnet/insulator interface.

As an example, we consider an $\mathrm{Fe/Al_2O_3/Fe}$ heterojunction. Typical parameters for such a magnetojunction are $m_l^* = m_r^* \approx m_0$, $m_c^* \approx 0.4 m_0$ (Bratkovsky, 1997), $k_{l\uparrow} = k_{r\uparrow} = 1.09$ Å$^{-1}$, and $k_{l\downarrow} = k_{r\downarrow} = 0.42$ Å$^{-1}$. The dependence of the TMR on the relative orientation of the magnetizations of the Fe layers is shown in Fig. II.34(a) for typical values of the barrier height (measured from the Fermi level) and the thickness $d = 20$ Å. The TMR dependence on the barrier height $V_c$ for $\phi = \pi$ is shown in Fig. II.34(b). The values of $k_{l\uparrow} = k_{r\uparrow}$ have been fixed, while $k_{l\downarrow} = k_{r\downarrow}$ have been evaluated by using experimental results of spin polarization (Tedrow and Meservey, 1973). The overall agreement between the theoretical calculations and experimental results (Tezuka and Miyazaki, 1998a) is satisfactory.

Due to its relative simplicity, Slonczewski's approach has been extensively applied to different kind of magnetic junctions (Bratkovsky, 1997; Li *et al.*, 1998; Zhang and Levy, 1998; Qi *et al.*, 1998; Li *et al.*, 2004; Jin *et al.*, 2006). However Slonczewski's model is not easily generalizable beyond the simple parabolic bands model. In order to overcome such a limitation, Mathon (1997) has proposed a tight-binding extension of Slonczewski's model to a realistic band structure. The details of such a model, which combine Landauer's linear-response and Kubo formalisms together with the Green's functions technique are given in (Mathon, 1997; Moodera *et al.*, 1999; Moodera and Mathon, 1999).

An important effect from the viewpoint of applications is the bias dependence of the TMR. A significant decrease of the TMR with increasing bias voltage has been experimentally observed (Jullierè, 1975; Moodera *et al.*, 1995; Tezuka and Miyazaki, 1998b,a; Lu *et al.*, 1998; Sun *et al.*, 1998; Zhang and White, 1998; Miyazaki, 2002). This bias voltage dependence of the TMR originates, in part, from the changes in the barrier shape induced by the electric field. Increasing the voltage increases the overall conductance, leading to a decreasing of the TMR. Such a behavior can be qualitative explained by extending the Slonczewski model beyond the linear response approach (Liu and Guo, 2000; Xiang *et al.*, 2002). However, a more realistic approximation requires the inclusion of the details of the band structure (Heiliger *et al.*, 2005) and other factors such as excitation of magnons (Zhang *et al.*, 1997), energy dependence of spin polarization, and impurity and phonon scattering (Bratkovsky, 1997), which may also play a role in the decrease of the TMR with the bias. Many-body effects (Hong *et al.*, 2002) as well as the influence of the roughness and disorder (Xu *et al.*, 2006a) on the TMR have also been theoretically studied. Another interesting effect observed in MTJs is the inversion of the TMR (De Teresa *et al.*, 1999a; Sharma *et al.*, 1999; Moser *et al.*, 2006; Heiliger *et al.*, 2005, 2006; Tiusan *et al.*, 2004), i.e., the change in sign of the TMR ratio when varying the bias voltage. The origin of the bias induced inversion of the TMR has been associated with resonant tunneling via localized states (Tsymbal *et al.*, 2003) or with bias induced changes in the sign of the spin polarization of one of the two electrodes (De Teresa *et al.*, 1999a; Sharma *et al.*, 1999) (see also the discussion at the end of Sec. (H.2)). However, more investigations are still needed for a better understanding of the mechanisms causing the bias induced inversion of the TMR.

It has been observed that the TMR decreases when increasing temperature (Moodera *et al.*, 1999; Moodera and Mathon, 1999; Miyazaki, 2002). The decay of TMR with temperature has been attributed to the spin-flip scattering of tunneling electrons from magnetic impurities in the barrier and to a reduction of the magnetic moment in the ferromagnet due to the excitation of magnons (collective spin excitations). The presence of magnetic impurities in the tunneling barrier results in a temperature-dependent conductance[34] which was used, including both spin-dependent and spin-flip scattering, for fitting the decay of the TMR with temperature (Inoue and Makeawa, 1999; Jansen and Moodera, 2000; Miyazaki, 2002). On the other hand, hot electrons localized at ferromagnet/insulator interfaces were predicted to create magnons near the interfaces.[35] The creation (annihilation) of a magnon in the collision with an electron flips the electron spin. Since the magnons are created near the ferromagnet/barrier interface, the electron whose spin has been reversed undergoes tunneling to the other electrode. The effect of such spin-flip scattering on the TMR is analogous to the effects of spin-flip scattering from impurities in the barrier and leads, eventually, to decreasing of the TMR with increasing temperature (Moodera *et al.*, 1995). Regarding the specific temperature dependence of the TMR, it has been shown (Zhang *et al.*, 1997) that in the limit of a small bias, the conductance $G(T)$ at a temperature $T$ deviates from its zero-temperature value $G(0)$ according to $G(T) - G(0) \propto T \ln T$. This result was obtained by using the Hamiltonian of the exchange between itinerant $s$ and nearly localized $d$ electrons. A different temperature dependence related to the decrease of the surface magnetization (Pierce *et al.*, 1982; Pierce and Celotta, 1984), $M(T)/M(0) \propto T^{-3/2}$, was suggested by Moodera *et al.* (1998). Such a temperature dependence attributed to magnons was also obtained for the TMR (MacDonald *et al.*, 1998). Furthermore, an extra contribution to the decrease of the TMR with temperature is expected to originate from the spin-independent part of $G(T)$ (Shang *et al.*, 1998).

The upper limit of the TMR ratio measured at room temperature in MTJs using aluminium oxide as the barrier layer is about $70\%$ (Wang *et al.*, 2004). To overcome this limit novel MTJs utilizing magnesium oxide as the tunnel barrier has been considered. Aluminium oxide is an amorphous material with a disordered arrangement of atoms. Consequently, the electrons are scattered during tunneling. In contrast, since magnesium oxide forms a crystalline barrier with a well ordered atomic structure, the electrons can tunnel straight through the barrier without being scattered. Theoretical studies have predicted that crystalline tunnel barriers may give rise to large TMR values (Mavropoulos *et al.*, 2000). Recently, *ab initio* calculations considering coherent tunneling through perfectly ordered, Fe(001)/MgO(001)/Fe(001) single-crystalline MTJs predicted extremely large TMR ratios of about 1000% (Butler *et al.*, 2001; Mathon and Umerski, 2001). Motivated by the great impact that such a giant TMR could have on various information storage technologies,[36] intense research activities on MTJs with single-crystalline barriers have been carried out in the past several years. The history of the improvement of the TMR in MTJs with a MgO barrier at room temperature is displayed in Fig. II.35. The largest, up-to- date, value (410%) of room temperature TMR has been measured in fully epitaxial Co(001)/MgO(001)/Co(001) MTJs with metastable bcc Co(001) electrodes (Yuasa *et al.*, 2006).

[34]This phenomenon is usually referred to as the zero-bias anomalies (Anderson, 1966; Appelbaum, 1966; Duke, 1969).
[35]Magnons were observed (Tsui *et al.*, 1971) in Ni/NiO/Pb tunnel junctions.

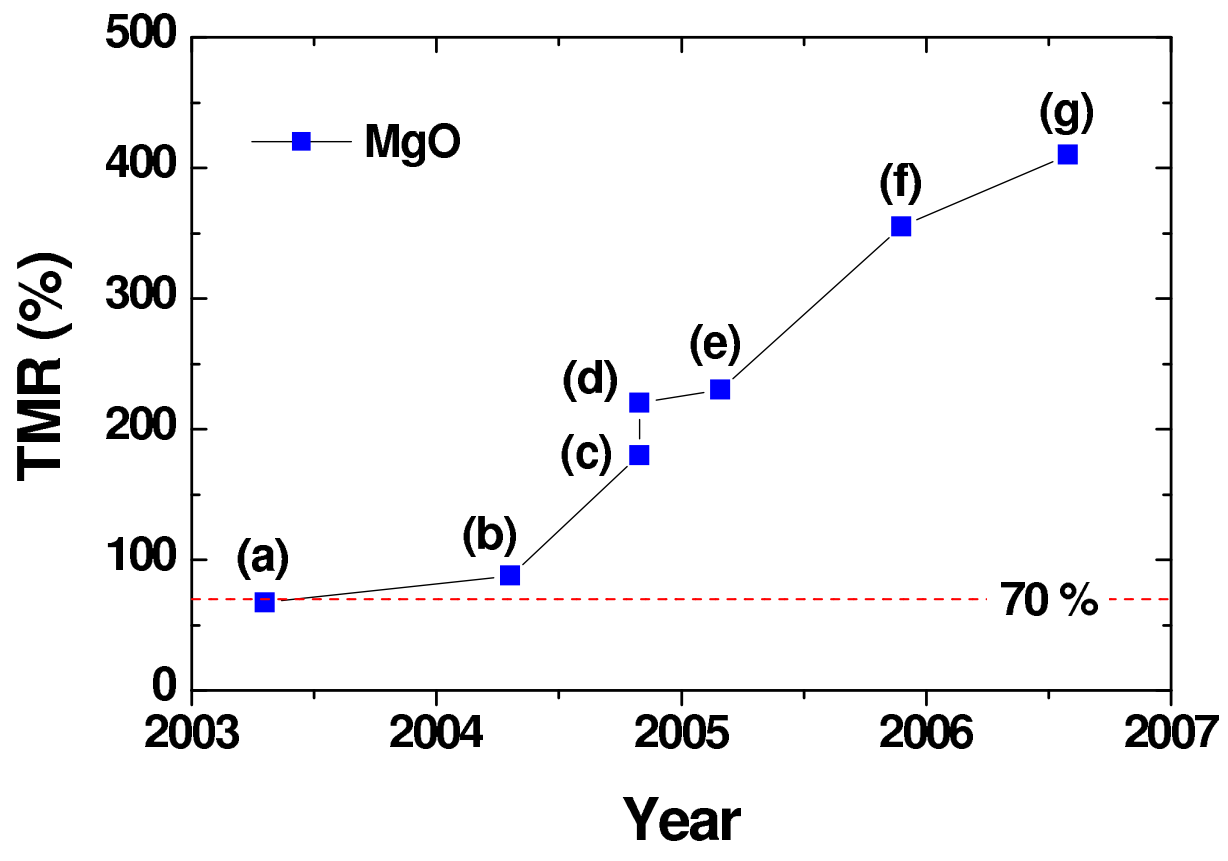

Fig. II.35. History of improvement of room temperature TMR ratio in MTJs with a MgO barrier. (a) Faure-Vincent *et al.* (2003). (b) Yuasa *et al.* (2004a). (c) Yuasa *et al.* (2004b). (d) Parkin *et al.* (2004). (e) Djayaprawira *et al.* (2005). (f) Ikeda *et al.* (2005). (g) Yuasa *et al.* (2006). For comparison, a dashed line indicating the largest TMR ratio (70%) measured in MTJs with an $Al_2O_3$ barrier (Wang *et al.*, 2004) has been included.

As an alternative to the difficult fabrication of ultrathin ($<$ 1 nm) oxide barriers (Rippard *et al.*, 2002), ferromagnet/semiconductor/ferromagnet tunnel junctions have also been considered. Some of these structures have been grown epitaxially, and the amplitude of the TMR ratio has be investigated for different crystallographic orientations of a ferrromagnet/semiconductor interface. Electronic structure calculations have been reported (MacLaren *et al.*, 1999) for Fe/ZnSe/Fe tunnel junctions. These calculations predicted a large TMR (up to $\sim$ 1000%), increasing with increasing of the barrier thickness. However, the TMR measured in Fe/ZnSe/$Fe_{0.85}Co_{0.15}$ was limited to $T < 50$ K, reaching $15\%$ at 10 K for junctions of higher resistance and lower defect density (Gustavsson *et al.*, 2001). A TMR ratio of about $5\%$ at room temperature was measured in MTJs with a ZnS barrier (Guth *et al.*, 2001). Experimental investigations of the TMR in Fe/GaAs/Fe tunnel junctions have recently been reported (Kreuzer *et al.*, 2002; Zenger *et al.*, 2004; Moser *et al.*, 2006). A TMR up to $1.7\%$ was observed in such MTJs (Zenger *et al.*, 2004; Moser *et al.*, 2006). A fully relativistic generalization of the Landauer-Büttiker formalism implemented and applied to the study of the TMR and spin-dependent transport in Fe/GaAs/Fe tunnel junctions (Popescu *et al.*, 2004, 2005) demonstrate a strong influence of the spin-orbit coupling inside the barrier on the TMR.

[36]For example, magnetic random access memories (MRAMs) with read performance an order of magnitude greater than current prototypes.

Much expectation has also been brought out by the possibility of using all-semiconductor MTJs with magnetic semiconductor electrodes (for a more detailed discussion see Secs. V.C. and V.D.). Compared to the conventional all-metal MTJs, the use of all-semiconductor MTJs would simplify integration with existing, conventional semiconductor-based electronics and allow for more flexibility in the design and fabrication of quantum structures. Large TMR ratios have been measured at low temperatures in epitaxially grown (Ga,Mn)As/AlAs/(Ga,Mn)As tunnel junctions, varying from about $70\%$ (Tanaka and Higo, 2001) up to $290\%$ (Mattana *et al.*, 2003; Chiba *et al.*, 2004; Elsen *et al.*, 2006). Recent theoretical investigations of the TMR in GaMnAs/GaAs/GaMnAs MTJs (Saffarzadeh and Shokri, 2006; Sankowski *et al.*, 2007) appear to explain the main features of the experimental results. The TMR in double and multi-barrier tunnel junctions with magnetic semiconductor electrodes is discussed in details in Sec. V.C.

### H.4 Tunneling anisotropic magnetoresistance (TAMR)

As discussed in the previous section, the TMR relies on the different spin polarizations at the Fermi energy in the ferromagnetic electrodes and strongly depends on the relative but not on the absolute magnetization directions in the ferromagnets. Therefore it cames as a surprise that the tunneling magnetoresistance may also strongly depend on the absolute orientation of the magnetization in the ferromagnets with respect to crystallographic directions (Gould *et al.*, 2004a; Rüster *et al.*, 2005; Saito *et al.*, 2005; Brey *et al.*, 2004). The phenomenon was termed tunneling anisotropic magnetoresistance (TAMR) (Gould *et al.*, 2004a; Brey *et al.*, 2004). The tunneling magnetoresistance in GaMnAs/GaAlAs/GaMnAs tunnel junctions was theoretically investigated by (Brey *et al.*, 2004). These authors predicted that, as a result of the strong spin-orbit interaction, the tunneling magnetoresistance depends on the angle between the current flow direction and the orientation of the electrode magnetization. Thus, a difference between the tunneling magnetoresistances in the in-plane (i.e., magnetization in the plane of the magnetic layers) and out of plane configurations of up to $6\%$ was predicted for large values of the electrode spin polarization (Brey *et al.*, 2004). Even more intriguing is the observation of the TAMR effect in tunnel junctions such as (Ga,Mn)As/$Al_2O_3$/Au (Gould *et al.*, 2004a) and Fe/GaAs/Au (Moser *et al.*, 2007) sandwiches, where only one of the layers is magnetic, and for which the TMR effect is absent. In these experiments, differences in the tunneling magnetoresistance are observed when the in-plane magnetization is rotated in the ferromagnetic layer. In what follows we concentrate on the TAMR in junctions with in-plane magnetization. In this case, in analogy to the TMR, one can define the TAMR ratio as

$$\mathrm{TAMR}_{[\mathrm{ref}]} = \frac{R(\phi) - R_{[\mathrm{ref}]}}{R_{[\mathrm{ref}]}}, \tag{II.309}$$

Tab. II.1. Experimental values of the TAMR ratio for different systems.

| System | TAMR (%) | Temperature (K) | Reference |
|---|---|---|---|
| (Ga,Mn)As/AlOx/Au | 2.7 | 4.2 | (Gould *et al.*, 2004a) |
| | 0 | 30 | |
| (Ga,Mn)As/AlOx/(Ga,Mn)As | 150 000 | 1.7 | (Rüster *et al.*, 2005) |
| | 300 | 4.2 | |
| (Ga,Mn)As/ZnSe/(Ga,Mn)As | 10 | 2 | (Saito *et al.*, 2005) |
| | 8.5 | 20 | |
| (Ga,Mn)As | 60 | 2 | (Giddings *et al.*, 2005) |
| lateral nanoconstriction | 0 | 15 | |
| Fe/GaAs/Au | 0.4 | 4.2 | (Moser *et al.*, 2007) |
| | 0.3 | 100 | (Lobenhofer *et al.*, 2007) |

where [ref] indicates the crystallographic direction taken as a reference (say, an easy, or hard axis) and $R(\phi)$ denotes the tunneling magnetoresistance measured when the magnetization direction in the ferromagnet forms and angle $\phi$ with respect to the reference direction [ref]. In particular, when the magnetization directions are parallel to [ref], we have $R(\phi = 0) = R_{[\mathrm{ref}]}$.

The first experimental measurement of the TAMR effect was reported by Gould *et al.* (2004a) who observed a spin-valve effect in (Ga,Mn)As/AlOx/Au heterojunctions with a TAMR ratio of about $2.7\%$. Experimental investigations of the TAMR in (Ga,Mn)As/GaAs/(Ga,Mn)As (Rüster *et al.*, 2005) and (Ga,Mn)As/ZnSe/(Ga,Mn)As (Saito *et al.*, 2005) tunnel junctions in which both electrodes are ferromagnetic have recently been reported. In the case of (Ga,Mn)As/ZnSe/(Ga,Mn)As the TAMR ratio was found to decrease with increasing temperature, from about $10\%$ at 2 K to $8.5\%$ at 20 K (Saito *et al.*, 2005). This temperature dependence of the TAMR is more dramatic in the case of (Ga,Mn)As/GaAs/(Ga,Mn)As, for which a TAMR ratio of order of a few hundred percent at 4 K was amplified to $150\ 000\%$ at 1.7 K (Rüster *et al.*, 2005). This huge amplification of the TAMR was suggested to originate from the opening of an Efros-Shklovskii gap (Efros and Shklovskii, 1975) at the Fermi energy when crossing the metal-insulator transition (Rüster *et al.*, 2005). A further investigation supporting such a suggestion has recently been reported (Pappert *et al.*, 2006). In addition to the above mentioned investigations involving vertical tunneling devices based on (Ga,Mn)As the TAMR has also been studied in (Ga,Mn)As lateral nanoconstrictions (Giddings *et al.*, 2005).

Beyond the area of currently low Curie temperature ferromagnetic semiconductors, the TAMR has been investigated in systems such as Fe/vaccum/Au (Chantis *et al.*, 2007) and Fe/GaAs/Au (Moser *et al.*, 2007; Matos-Abiague and Fabian, 2007) tunnel junctions, ferromagnetic metal break junctions (Bolotin *et al.*, 2006), and CoPt systems (Shick *et al.*, 2006). A summary of the size of the TAMR ratios experimentally measured in different systems is given in Tab. II.1.

An interesting issue emerging from the TAMR effect measured in (Ga,Mn)As based, vertical tunneling devices is that from symmetry considerations one would expect the existence of fourfold in-plane magnetic easy axes in the (Ga,Mn)As electrode. Experimentally, however, an uniaxial anisotropy is usually observed (Gould *et al.*, 2004a; Rüster *et al.*, 2005; Saito *et al.*, 2005). This phenomenon was related to the anisotropy of the density of states with respect to the magnetization direction, which results from the strong SOI in the ferromagnetic semiconductor valence band (Gould *et al.*, 2004a; Rüster *et al.*, 2005; Pappert *et al.*, 2006). However, the nature and details of the underlying mechanism producing the uniaxial anisotropy observed in the TAMR remains a puzzle. In fact, it has become clear that the responsible mechanisms for the uniaxial anisotropy of the TAMR could be different in different systems. In (Ga,Mn)As/$Al_2O_3$/Au (Gould *et al.*, 2004a) and (Ga,Mn)As/GaAs/(Ga,Mn)As (Rüster *et al.*, 2005; Pappert *et al.*, 2006) heterojunctions as well as in (Ga,Mn)As nanoconstrictions (Giddings *et al.*, 2005) the uniaxial anisotropy of the TAMR has been theoretically modelled by phenomenologically introducing an in-plane uniaxial strain in the range $0.1\% - 0.2\%$. On the other hand, Bolotin *et al.* (2006) have proposed that the TAMR effect in ferromagnetic metal break junctions originates from mesoscopic quantum interferences which depend on the orientation of the magnetization and result in mesoscopic fluctuations of the conductance and the spin-dependent local density of states. Alternatively, it has been proposed (Matos-Abiague and Fabian, 2007) that the uniaxial anisotropy of the TAMR in epitaxial ferromagnet/semiconductor/normal metal tunnel junctions originates from the interference of Dresselhaus and Bychkov-Rashba spin-orbit interactions.

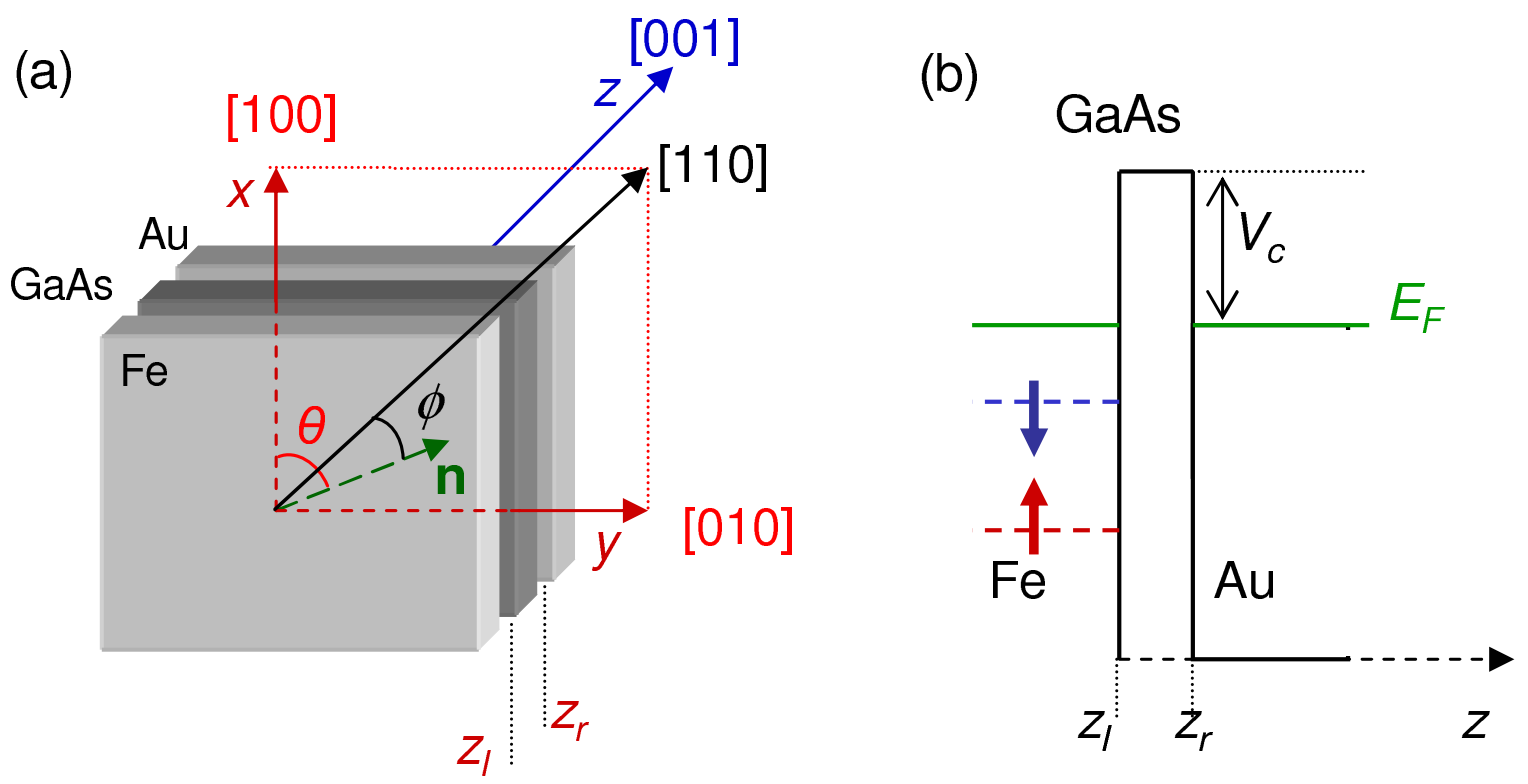

Fig. II.36. (a) Profile of a Fe/GaAs/Au MTJ grown in the $[001]$ crystallographic direction. The magnetization direction in the Fe layer is determined by the vector $\mathbf{n}$. (b) Schematics of the potential profile along the growth direction of the heterojunction. A Stoner model with two channels (majority and minority) is assumed for the Fe layer.

The TAMR in epitaxial Fe/GaAs/Au MTJs grown in the $[001]$ crystallographic direction (see Fig. II.36) has recently been investigated (Moser *et al.*, 2007; Matos-Abiague and Fabian, 2007). The experimentally observed tunneling magnetoresistance exhibits a two-fold symmetry with respect to the magnetization direction in the ferromagnet. It was also shown that the symmetry axis of the TAMR can be changed by varying the bias voltage (Moser *et al.*, 2007).[37] In order to explain the experimental findings, a model was proposed (Moser *et al.*, 2007; Matos-Abiague and Fabian, 2007), in which the two-fold symmetry of the TAMR observed in epitaxial ferromegnet/semiconductor/normal metal junctions originates from the interface-induced $C_{2v}$ symmetry of the SOI arising from the interference of Dresselhaus and Bychkov-Rashba spin-orbit couplings (see Sec. F.2). This symmetry, which is imprinted in the tunneling probability becomes apparent in the contact with a magnetic moment. Some details about such a model are discussed below.

[37] Note that this bias dependence of the TAMR suggests that strain is not the relevant mechanisms causing the TAMR effect in Fe/GaAs/Au.

Consider a ferromagnet/semiconductor/normal-metal tunnel heterojunction grown in the $[001]$ direction. The semiconductor is assumed to lack bulk inversion symmetry (zinc-blende semiconductors are typical examples). The bulk inversion asymmetry of the semiconductor together with the structure inversion asymmetry of the heterojunction give rise to the Dresselhaus (see Sec. F.) and Bychkov-Rashba (see Sec. III.E.2) SOIs, respectively. The interference of these two spin-orbit couplings leads to a net, anisotropic SOI with a $C_{2v}$ symmetry which is transferred to the tunneling magnetoresistance when the electrons pass through the semiconductor barrier. The model Hamiltonian describing the tunneling across the heterojunction reads

$$\mathrm{H} = \mathrm{H}_0 + \mathrm{H}_{_{BR}} + \mathrm{H}_{_D}. \tag{II.310}$$

Here

$$\mathrm{H}_0 = -\frac{\hbar^2\nabla^2}{2m_i^*} + V_i - \frac{\Delta_i}{2}\mathbf{n}_i \cdot \boldsymbol{\sigma};\ i = l, c, r \tag{II.311}$$

where $m_i^*$ and $V_i$ are the effective mass and potential energy, respectively, in the ferromagnet ($i = l$), semiconductor ($i = c$), and normal-metal ($i = r$) regions. The exchange energy $\Delta_i$ vanishes in the central (semiconductor) and right (normal-metal) regions, i.e, $\Delta_c = \Delta_r = 0$.[38] The unit vector $\mathbf{n}_l = (\cos\theta, \sin\theta, 0)$ defines the magnetization direction with respect to the [100] direction (see Fig. II.36), and $\boldsymbol{\sigma}$ is a vector whose components are the Pauli matrices. Note that Eq. (II.311) is the analog of Eq. (II.287) but for the case of a single magnetic layer. In the experiment the reference axis is taken as the [110] direction (referred to the GaAs crystallographic directions), which is the hard axis of magnetization in the Fe layer. Therefore, it is convenient to express the magnetization direction relative to the [110] crystallographic axis through the introduction of the new angle $\phi = \theta - \pi/4$ (see Fig. II.36). Thus, one can write $\mathbf{n}_l = (\cos(\phi + \pi/4), \sin(\phi + \pi/4), 0)$, where $\phi$ gives the magnetization direction with respect to the [110] direction.

It is worth noting that the form of Bychkov-Rashba-like and Dresselhaus-like SOIs in metal/semicondutor interfaces are not known.[39] However, since the metal/semiconductor interfaces of the investigated MTJ have the same $C_{2v}$ symmetry as a zinc-blende/zinc-blende interface (see Sec. G.), one can assume that both interfaces can be phenomenologically described by the same kind of SOI Hamiltonian. In what follows we adopt such an assumption and consider that the in-plane wave vector $\mathbf{k}_{\parallel}$ is conserved throughout the heterostructure.

[38] We consider here the case of a weak external magnetic field. Therefore, orbital effects as well as the Zeeman splitting in the semiconductor and normal metal are neglected.

[39] We use here the terminology Bychkov-Rashba-like and Dresselhaus-like SOIs to emphasize that, in addition to the conventional Bychkov-Rashba and Dresselhaus spin-orbit terms, there is a contribution to the SOI resulting from the interface inversion asymmetry (IIA). Such a contribution has the $C_{2v}$ symmetry of the metal/semiconductor interface (see Sec. III.G.), which is the same symmetry resulting from the interference of the conventional Bychkov-Rashba and Dresselhaus SOIs. Therefore, one can write the resulting SOI as an interaction having the same form of interfering Bychkov-Rashba and Dresselhaus terms but with new, renormalized spin-orbit parameters, which now account for the IIA too. In what follows we keep the usual terminology referring to Bychkov-Rashba and Dresselhaus SOIs but these interactions should be interpreted as explained above.

The Dresselhaus SOI including the bulk and interface contributions is given by Eqs. (III.148) - (III.150). The Dresselhaus parameter $\gamma(z)$ has a finite value $\gamma$ in the semiconductor region, where the bulk inversion asymmetry is present, and vanishes elsewhere, i.e., $\gamma^{(l)} = \gamma^{(r)} = 0$ and $\gamma^{(c)} = \gamma$.

The Bychkov-Rashba SOI due to the interface inversion asymmetry is incorporated in the model through a term of the form of Eq. (III.97). For the system here studied, the Bychkov-Rashba SOI contribution inside the semiconductor can be neglected and one is left only with the interface contributions

$$\mathrm{H}_{BR} = \frac{1}{\hbar} \sum_{i=l,r} \alpha_i (\sigma_x p_y - \sigma_y p_x) \delta(z - z_i), \tag{II.312}$$

where, $\alpha_l$ ($\alpha_r$) denotes the SOI strength at the left (right) interface $z_l = 0$ ($z_r = d$). These parameters are not known for metal/semiconductor interfaces and have to be extracted from experimental results or from ab initio calculations. For the case of an Fe/GaAs/Au, it was shown that the size of the TAMR is dominated by the parameter $\alpha_l$ (Matos-Abiague and Fabian, 2007). Then, since the values of the TAMR are not very sensitive to the changes of $\alpha_r$ one can set this parameter, without loss of generality, to zero. This leaves $\alpha_l$ as a single fitting parameter when comparing to experiment.

The $z$ component of the scattering states in the left (ferromagnetic) region [eigenstates of the Hamiltonian (II.310)] with eigenenergy $E$ can be written as

$$\Psi_\sigma^{(l)} = \frac{e^{ik_\sigma z} \chi_\sigma}{\sqrt{k_\sigma}} + r_{\sigma,\sigma} e^{-ik_\sigma z} \chi_\sigma + r_{\sigma,-\sigma} e^{-ik_{-\sigma} z} \chi_{-\sigma}, \tag{II.313}$$

where

$$\chi_\sigma^{(l)} = \frac{1}{\sqrt{2}} \begin{pmatrix} 1 \\ \sigma e^{i\left(\phi + \frac{\pi}{4}\right)} \end{pmatrix}, \tag{II.314}$$

represents a spinor corresponding to a spin parallel ($\sigma = 1$) or antiparallel ($\sigma = -1$) to the magnetization direction $\mathbf{n}_l = (\cos(\phi + \pi/4), \sin(\phi + \pi/4), 0)$ in the ferromagnet and $k_\sigma$ is the corresponding $z$ component of the wave vector in the left region. In the central (semiconductor) region one has (Perel' *et al.*, 2003; Wang *et al.*, 2005),

$$\Psi_\sigma^{(c)} = \sum_{i=\pm} (A_{\sigma,i} e^{q_i z} + B_{\sigma,i} e^{-q_i z}) \chi_i^{(c)}, \tag{II.315}$$

where $q_\pm = (1 \mp 2m_c \gamma k_\| / \hbar^2)^{-1/2} q_0$ (with $q_0$ being the $z$ component of the wave vector in the barrier in the absence of SOI) and

$$\chi_\pm^{(c)} = \frac{1}{\sqrt{2}} \begin{pmatrix} 1 \\ \sigma e^{i\varphi} \end{pmatrix}, \tag{II.316}$$

are spinors corresponding to spins parallel (+) and antiparallel ($-$) to the direction $\mathbf{k}_\| \times \mathbf{z}$ [$\mathbf{k}_\| = k_\| (\cos\varphi, \sin\varphi, 0)$], which is the spin quantization direction in the barrier. In the right (normal metal) region ($z \geq 0$) the scattering states read,

$$\Psi_\sigma^{(r)} = t_{\sigma,\sigma} e^{i\kappa_\sigma (z-d)} \chi_\sigma + t_{\sigma,-\sigma} e^{i\kappa_{-\sigma}(z-d)} \chi_{-\sigma}, \tag{II.317}$$

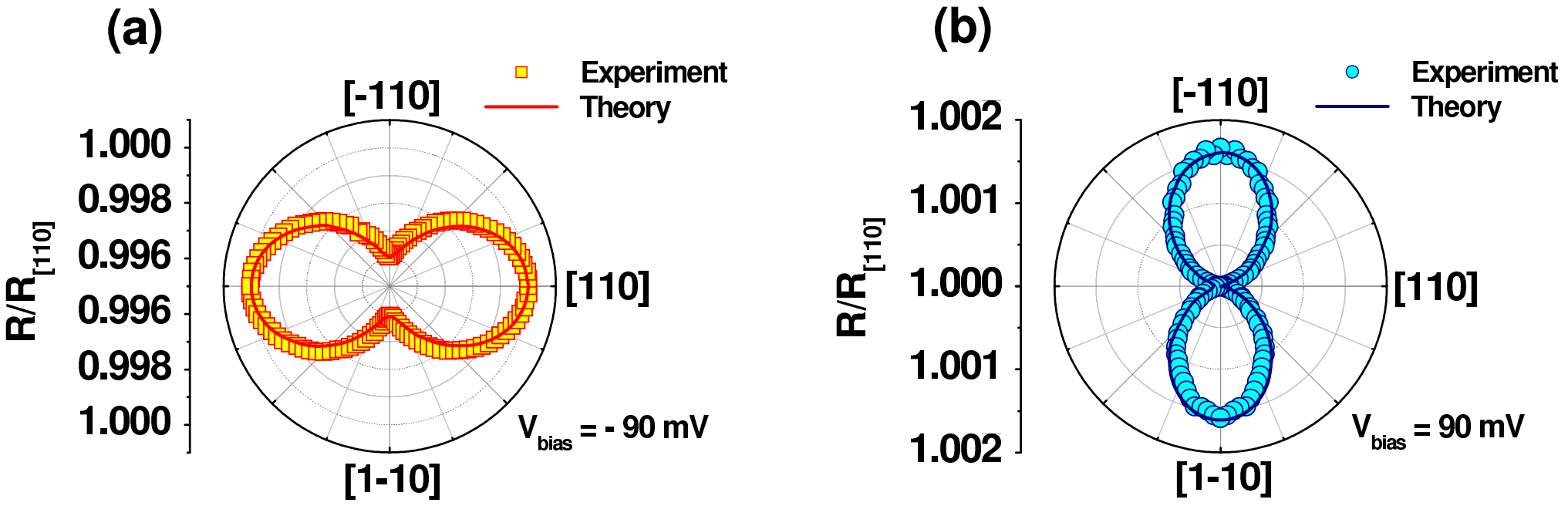

Fig. II.37. Angular dependence of the magnetoresistance $R$ normalized to the value $R_{[110]}$ for bias voltages $V_{\rm bias} = -90$ mV (a) and $V_{\rm bias} = 90$ mV (b). Solid lines corresponds to the theoretical results, while symbols represent the experimental data taken from (Moser *et al.*, 2007). The experiment was performed at 4.2 K, in a saturation magnetic field $|\mathbf{B}| = 0.5$ T. The presence of a twofold symmetry with uniaxial anisotropy is apparent. Furthermore, the symmetry axis of the magnetoresistance is flipped when changing the bias from $-90$ mV to 90 mV.

where $\kappa_\sigma$ is the corresponding $z$ component of the wave vector in the right region. The expansion coefficients in Eqs. (II.313) - (II.317) can be found by applying the matching conditions given in Eq. (III.152) and (III.153) at each interface. Once the wave function is determined, the particle transmissivity can be calculated according to Eq. (II.297). Making use of Eqs. (II.283) and (II.284) one can compute the total current $I(V) = I_\uparrow(V) + I_\downarrow(V)$ and the corresponding conductance (within linear response), respectively.

The TAMR refers to the changes of the tunneling magnetoresistance ($R$) when varying the magnetization direction $\mathbf{n}$ of the magnetic layer with respect to a fixed axis. Here we assume the magnetoresistance $R_{[110]}$, measured when $\mathbf{n}$ points in the [110] crystallographic direction (i.e., when $\phi = 0$) as a reference. The TAMR is then given by [see Eq. (II.309)]

$$\text{TAMR}_{[110]}(\phi) = \frac{R(\phi) - R(0)}{R(0)} = \frac{I(0) - I(\phi)}{I(\phi)} = \frac{G(0) - G(\phi)}{G(\phi)}, \tag{II.318}$$

where $\phi$ is the angle between the magnetization direction $\mathbf{n}_l = (\cos(\phi + \pi/4), \sin(\phi + \pi/4), 0)$ and the [110] crystallographic axis. We also find it useful to define the tunneling anisotropic spin polarization (TASP) as

$$\text{TASP}_{[110]}(\phi) = \frac{P(0) - P(\phi)}{P(\phi)}. \tag{II.319}$$

The TASP measures the changes in the tunneling spin polarization (Maekawa *et al.*, 2002; Žutić *et al.*, 2004; Perel' *et al.*, 2003) $P = (I_\uparrow - I_\downarrow)/I$ [which is a measurable quantity (Žutić *et al.*, 2004) accounting for the polarization efficiency of the transmission] when rotating the in-plane magnetization in the ferromagnet.

A polar plot of the tunneling magnetoresistance as a function of the angle $\phi$ between the magnetization direction in the Fe electrode and the reference direction [110] is displayed in Fig. II.37. The resistance has been normalized to the value $R_{[110]}$, which is the resistance measured when the magnetization in the Fe layer points in the $[110]$ direction. Symbols and solid lines represent the experimental data and theoretical calculations, respectively. The tunneling magnetoresistance exhibits a $C_{2v}$ symmetry[40] evidencing the existence of uniaxial anisotropy. The anisotropy depends on the applied bias voltage. When the bias voltage is reversed from $-90$ mV to $90$ mV, the symmetry axes of the angular dependence of the magnetoresistance are rotated by $90°$.

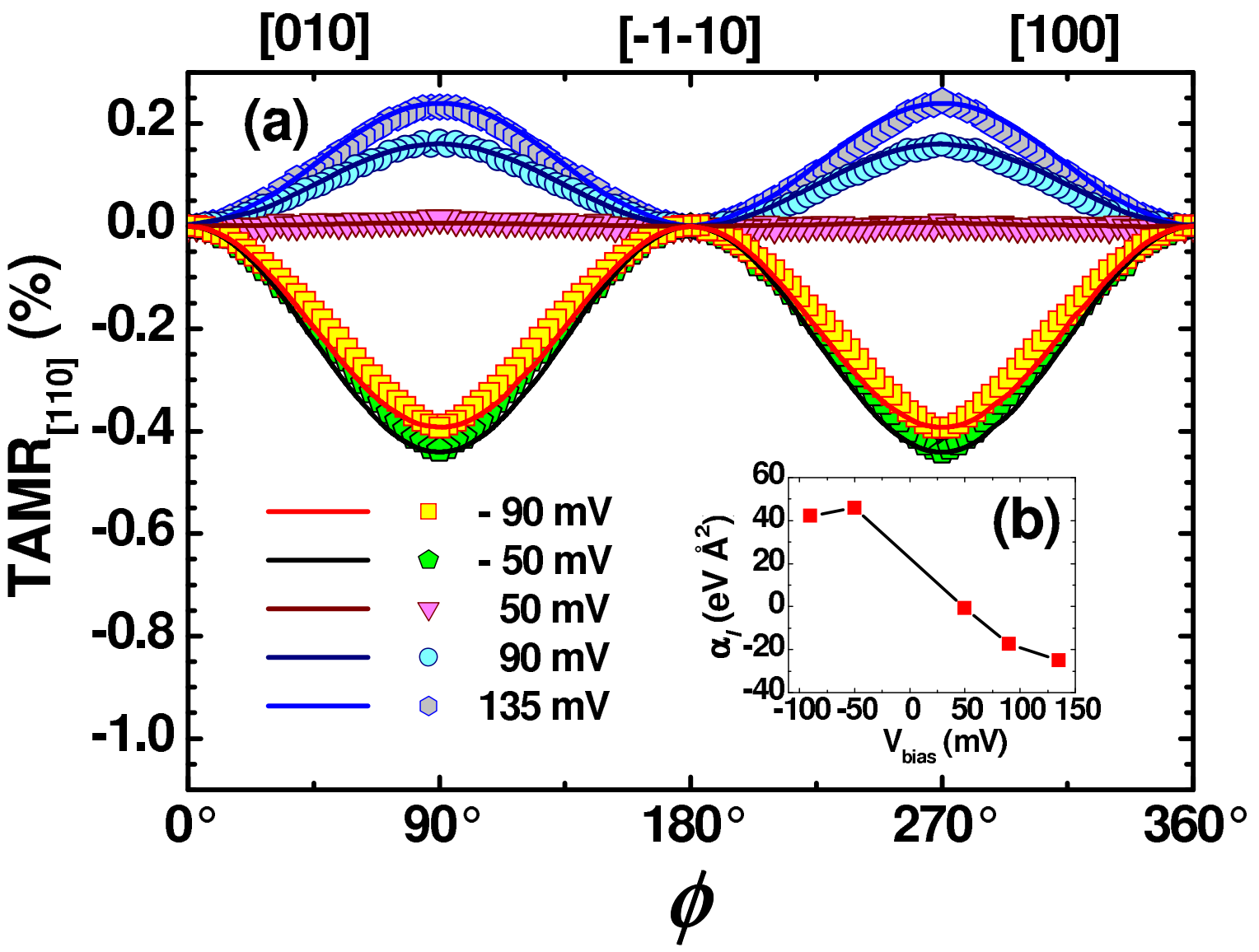

Fig. II.38. (a) Angular dependence of the TAMR in a Fe/GaAs/Au tunnel heterojunction for different values of the bias voltage $V_{\mathrm{bias}}$. Solid lines corresponds to the theoretical results while symbols represent the experimental data (conveniently mirrored) as deduced from (Moser *et al.*, 2007). The values of the phenomenological parameter $\alpha_l$ have been determined by fitting the theory to the experimental values of the ratio $R_{[\bar{1}10]}/R_{[110]}$ for each value of $V_{\mathrm{bias}}$. The extracted bias dependence of $\alpha_l$ is shown in (b).

A comparison between the theoretical results obtained within the model above discussed and the experimental data for the TAMR ratio in an Fe/GaAs/Au MTJ (Moser *et al.*, 2007) is displayed in Fig. II.38(a) for different values of the bias voltage. For the calculations we took the value $m_c = 0.067\ m_0$ (see Tab. III.6) for the electron effective mass in the central (GaAs) region. The barrier width and height (measured from the Fermi energy) are, respectively, $d = 80$ Å and $V_c = 0.75$ eV, corresponding to the experimental samples (Moser *et al.*, 2007). The Dresselhaus parameter in the GaAs region was taken as $\gamma = 24\,\text{eV Å}^3$ (see Tab. III.6). For the Fe layer a Stoner model with the majority and minority spin channels having Fermi momenta $k_{F\uparrow} = 1.05 \times 10^8\ \text{cm}^{-1}$ and $k_{F\downarrow} = 0.44 \times 10^8\ \text{cm}^{-1}$ (Wang *et al.*, 2003), respectively, was assumed. The Fermi momentum in Au was taken as $\kappa_F = 1.2 \times 10^8\ \text{cm}^{-1}$ (Ashcroft and Mermin, 1976). We consider the case of relatively weak magnetic fields (specifically, $B = 0.5\,\text{T}$). At high magnetic fields, say, several tesla, the model here discussed is no longer valid as it does not include cyclotron effects relevant when the cyclotron radius becomes comparable to the barrier width.

[40] Note that the twofold symmetry of the magnetoresistance resembles that of the pattern of the effective magnetic field arising from the interference of Bychkov-Rashba and Dresselhaus SOIs (see Fig. III.10 in Sec. F.2).

The agreement between theory and experiment is very satisfactory. The values of the phenomenological parameter $\alpha_l$ are determined by fitting the theory to the experimental value of the ratio $R_{[\bar{1}10]}/R_{[110]}$. This is enough for the theoretical model to reproduce the *complete* angular dependence of the TAMR, demonstrating the robustness of the model. The bias dependence of $\alpha_l$ can be extracted by performing the same fitting procedure for the available experimental data corresponding to different bias voltages. The results are shown in Fig. II.38(b), from which one estimates the value $\alpha_l \approx 23$ meV Å$^2$ at zero bias. This value is comparable to the corresponding value of the interface Bychkov-Rashba parameter $\alpha_{int} = \beta_{InAs} - \beta_{GaAs} \approx 27$ meV Å$^2$ obtained from Tab. III.6 for an InAs/GaAs interface. Selecting InAs/GaAs for comparison with Fe/GaAs we only wish to show that the fitted values have reasonable magnitude, not differing too much from known values in semiconductor interfaces. Interestingly $\alpha_l$ changes sign at a bias slightly below 50 mV. This bias induced change of the interface Bychkov-Rashba parameter results in an inversion of the TAMR [see Fig. II.38(a)]. Similar behavior is reported by ab initio calculations on Fe surfaces, where only Bychkov-Rashba SOI is present (Chantis *et al.*, 2007).

The robustness of the theoretical model can be understood from the following simplified picture of the TAMR effect. The SOI term $\mathrm{H}_{SO} = \mathrm{H}_D + \mathrm{H}_{BR}$ can be written [see Eqs. (III.149) and (II.312)] as a Zeeman-like term $\mathrm{H}_{SO} \sim -\hat{\mathbf{B}}_{eff} \cdot \boldsymbol{\sigma}$ with the effective magnetic field

$$\hat{\mathbf{B}}_{eff}(\mathbf{k}_{\|}) = (-\alpha_l \delta(z) k_y + \gamma k_x \partial_z^2, \alpha_l \delta(z) k_x - \gamma k_y \partial_z^2, 0), \tag{II.320}$$

where, for the sake of qualitative argument, we neglect the interface Dresselhaus contributions. Performing the average of $\hat{\mathbf{B}}_{eff}$ over the unperturbed (in the absence of SOI) states of the system one obtains the following general form of the spin-orbit magnetic field

$$\mathbf{w}(\mathbf{k}_{\|}) = (\tilde{\alpha}_l k_y - \tilde{\gamma} k_x, -\tilde{\alpha}_l k_x + \tilde{\gamma} k_y, 0), \tag{II.321}$$

where $\tilde{\alpha}_l = \alpha_l f_\alpha(k_{\|})$ and $\tilde{\gamma} = \gamma f_\gamma(k_{\|})$, with $f_\alpha(k_{\|})$ and $f_\gamma(k_{\|})$ being real functions of $k_{\|} = |\mathbf{k}_{\|}|$. The spin-orbit field $\mathbf{w}(\mathbf{k}_{\|})$ becomes anisotropic in the $\mathbf{k}_{\|}$-space with a $C_{2v}$ symmetry [see Fig. III.9(c)] when both $\alpha_l$ and $\gamma$ have finite values.[41] It characterizes the amount of $\mathbf{k}_{\|}$-dependent precession of the electron spin during the tunneling. A vector plot of the field $\mathbf{w}(\mathbf{k}_{\|})$ is sketched in Fig. II.39. A polar plot (solid line) of $|\mathbf{w}(\mathbf{k}_{\|})|/k_{\|}$ is also included for comparison. The uniaxial anisotropy of the TAMR can be explained as the result of the different amount of precession the incident spins experience under the influence of $\mathbf{w}(\mathbf{k}_{\|})$, depending on their initial orientation, which is determined by the magnetization direction (see Fig. II.39).

For a given $\mathbf{k}_{\|}$ there are only two preferential directions in the system, defined by $\mathbf{n}$ and $\mathbf{w}(\mathbf{k}_{\|})$. Therefore, the anisotropy of a scalar quantity such as the total transmissivity $T(E, \mathbf{k}_{\|}) = T_{\uparrow}(E, \mathbf{k}_{\|}) + T_{\downarrow}(E, \mathbf{k}_{\|})$ can be obtained as a perturbative expansion in powers of $\mathbf{n} \cdot \mathbf{w}(\mathbf{k}_{\|})$, since the SOI is much smaller than the other relevant energy scales in the system. The total transmissivity is then given, up to second order in the anisotropy, by

$$T(E, \mathbf{k}_{\|}) \approx T^{(0)}(E, k_{\|}) + T^{(1)}(E, k_{\|})[\mathbf{n} \cdot \mathbf{w}(\mathbf{k}_{\|})] + T^{(2)}(E, k_{\|})[\mathbf{n} \cdot \mathbf{w}(\mathbf{k}_{\|})]^2. \tag{II.322}$$

[41] Note that $\mathbf{w}(\mathbf{k}_{\|})$ in Eq. (II.321) has the same form as the effective magnetic field $\mathbf{B}_{eff}(\mathbf{k})$ in Eq. (III.156). Therefore all the discussion in Sec. F.2 concerning $\mathbf{B}_{eff}(\mathbf{k})$ also applies here.

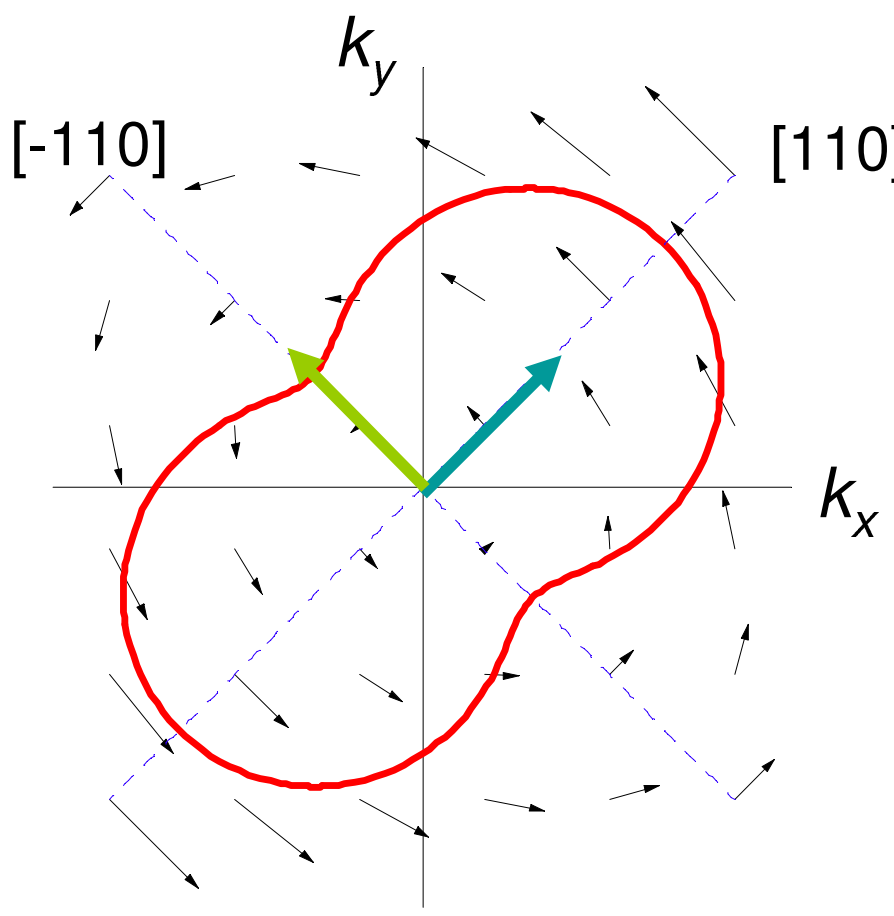

Fig. II.39. Schematics of the anisotropy of the spin-orbit magnetic field $\mathbf{w}(\mathbf{k}_{\parallel})$ [see Eq. (II.321)]. Thin arrows represent a vector plot of the field $\mathbf{w}(\mathbf{k}_{\parallel})$. The solid line is a polar plot of $|\mathbf{w}(\mathbf{k}_{\parallel})|/k_{\parallel}$. The thick arrows indicate two different orientations of the magnetization. The spin-orbit field is oriented in the [-110] ([110]) direction at the points of high (low) spin-orbit field, where the field amplitude reaches a maximum (minimum). When the magnetization points along the [110] direction the incident spins precess during tunneling around the direction [-110] of the high field but no precession around the low field direction ([110]) occurs. In contrast, when the magnetization is oriented along [-110], the incident spins precess around the low but not around the high field direction. This difference in the amount of spin precession results in different tunneling transmissivities depending on the magnetization direction and, consequently, in the uniaxial anisotropy of the TAMR.

Substituting Eq. (II.322) into the linear-response expression for the conductance [see Eq. (II.284)] one obtains,

$$G = \frac{e^2}{(2\pi)^2 h}\left(\langle T^{(0)}(E_F, k_{\parallel})\rangle + \langle T^{(2)}(E_F, k_{\parallel})[\mathbf{n}\cdot\mathbf{w}(\mathbf{k}_{\parallel})]^2\rangle\right), \tag{II.323}$$

where $\langle ...\rangle$ represents average over $\mathbf{k}_{\parallel}$. Note that the first order term vanishes after averaging over $\mathbf{k}_{\parallel}$, since $\mathbf{w}(\mathbf{k}_{\parallel}) = -\mathbf{w}(-\mathbf{k}_{\parallel})$.

Considering that $\mathbf{n}_l = (\cos(\phi + \pi/4), \sin(\phi + \pi/4), 0)$ and taking into account Eq. (II.321) one obtains from Eq. (II.323),

$$G = G^{(0)} + \Delta G_{so}, \tag{II.324}$$

where,

$$G^{(0)} = \frac{e^2}{(2\pi)^2 h}\langle T^{(0)}(E_F, k_{\parallel})\rangle \tag{II.325}$$

is the (unperturbed) conductance in absence of SOI and,

$$\Delta G_{so} = \frac{e^2}{(2\pi)^2 h}\left(\frac{1}{2}\langle[\tilde{\alpha}_l^2 + \tilde{\gamma}^2]T^{(2)}(E_F, k_{\parallel})k_{\parallel}^2\rangle + \langle\tilde{\alpha}_l\tilde{\gamma}T^{(2)}(E_F, k_{\parallel})k_{\parallel}^2\rangle\cos(2\phi)\right), \tag{II.326}$$

gives the SOI correction to the conductance. In obtaining Eq. (II.326) we have made use of the following symmetry relations

$$\langle T^{(2)}(E_F, k_\|)\tilde{\alpha}_l^2 k_x^2\rangle = \langle T^{(2)}(E_F, k_\|)\tilde{\alpha}_l^2 k_y^2\rangle = \frac{1}{2}\langle T^{(2)}(E_F, k_\|)\tilde{\alpha}_l^2 k_\|^2\rangle, \tag{II.327}$$

$$\langle T^{(2)}(E_F, k_\|)\tilde{\gamma}^2 k_x^2\rangle = \langle T^{(2)}(E_F, k_\|)\tilde{\gamma}^2 k_y^2\rangle = \frac{1}{2}\langle T^{(2)}(E_F, k_\|)\tilde{\gamma}^2 k_\|^2\rangle, \tag{II.328}$$

and

$$\langle T^{(2)}(E_F, k_\|)\tilde{\alpha}_l\tilde{\gamma} k_x k_y\rangle = \langle T^{(2)}(E_F, k_\|)\tilde{\alpha}_l^2 k_x k_y\rangle = \langle T^{(2)}(E_F, k_\|)\tilde{\gamma}^2 k_x k_y\rangle = 0, \tag{II.329}$$

which result from the fact that $T^{(2)}(E_F, k_\|)$, $\tilde{\alpha}_l$, and $\tilde{\gamma}$ depend only on $k_\| = \sqrt{k_x^2 + k_y^2}$ and are, therefore, even functions of $k_x$ and $k_y$ and symmetric under the interchange of $k_x$ and $k_y$.

Using Eq. (II.324) one can rewrite the TAMR ratio [see Eq. (II.318) as

$$\mathrm{TAMR}_{[110]} = \frac{G^{(0)} + \Delta G_{so}(0)}{G^{(0)} + \Delta G_{so}(\phi)} - 1. \tag{II.330}$$

For the system here investigated $\Delta G_{so}(\phi)/G^{(0)} \ll 1$ and one can expand Eq. (II.330) in powers of $\Delta G_{so}(\phi)/G^{(0)}$ and obtain

$$\mathrm{TAMR}_{[110]} \approx \frac{\Delta G_{so}(0) - \Delta G_{so}(\phi)}{G^{(0)}}. \tag{II.331}$$

Considering Eqs. (II.326) and (II.331) and taking into account that $\tilde{\alpha}_l \propto \alpha_l$ and $\tilde{\gamma} \propto \gamma$, one arrives to the relation (Moser *et al.*, 2007)

$$\mathrm{TAMR}_{[110]} \propto \alpha_l\gamma[\cos(2\phi) - 1]. \tag{II.332}$$

The angular dependence in Eq. (II.332) is consistent with that found experimentally, as well as that obtained from the full theoretical calculations [see Fig. II.38(a)]. One can clearly see from Eq. (II.332) that bias-induced changes of the sign of the Bychkov-Rashba parameter $\alpha_l$ lead to an inversion of the TAMR. When $\alpha_l\gamma = 0$, the two-fold TAMR is suppressed. To put in words, the Bychkov-Rashba (or Dresselhaus) term alone cannot explain the observed $C_{2v}$ symmetry. The suppression of the TAMR is approximately realized in Fig. II.38(a) for the case of a bias voltage of 50 mV.

In Fig. II.40, we show the angular dependence of the TASP [see Eq. (II.319)] for different values of the bias voltage. The anisotropy of the tunneling spin polarization indicates that the amount of transmitted and reflected spin at the interfaces depends on the magnetization direction in the Fe layer, resulting in an anisotropic spin local density of states at the Fermi surface and showing spin-valve-like characteristics.

Finally, we want to stress that although at low temperatures (say 4 K) the TAMR ratio measured in Fe/GaAs/Au appears to be quite small (about $0.4\%$) in comparison with the corresponding values reported for (Ga,Mn)As based tunnel junctions (from $10\%$ to a few hundred percent), at temperatures as high as 100 K the TAMR effect in Fe/GaAs/Au is only reduced to nearly $0.3\%$ (Lobenhofer *et al.*, 2007) while it completely vanishes in the case of (Ga,Mn)As based tunnel junctions. This robustness of the TAMR to the increasing of temperature suggests that the observation of the TAMR effect in Fe/GaAs/Au heterojunctions may still be possible at room temperature. From the theoretical point of view, a sizable room temperature TAMR effect has also been predicted for CoPt structures (Shick *et al.*, 2006). However, to the best of our knowledge, no experiment considering such structures has been reported.

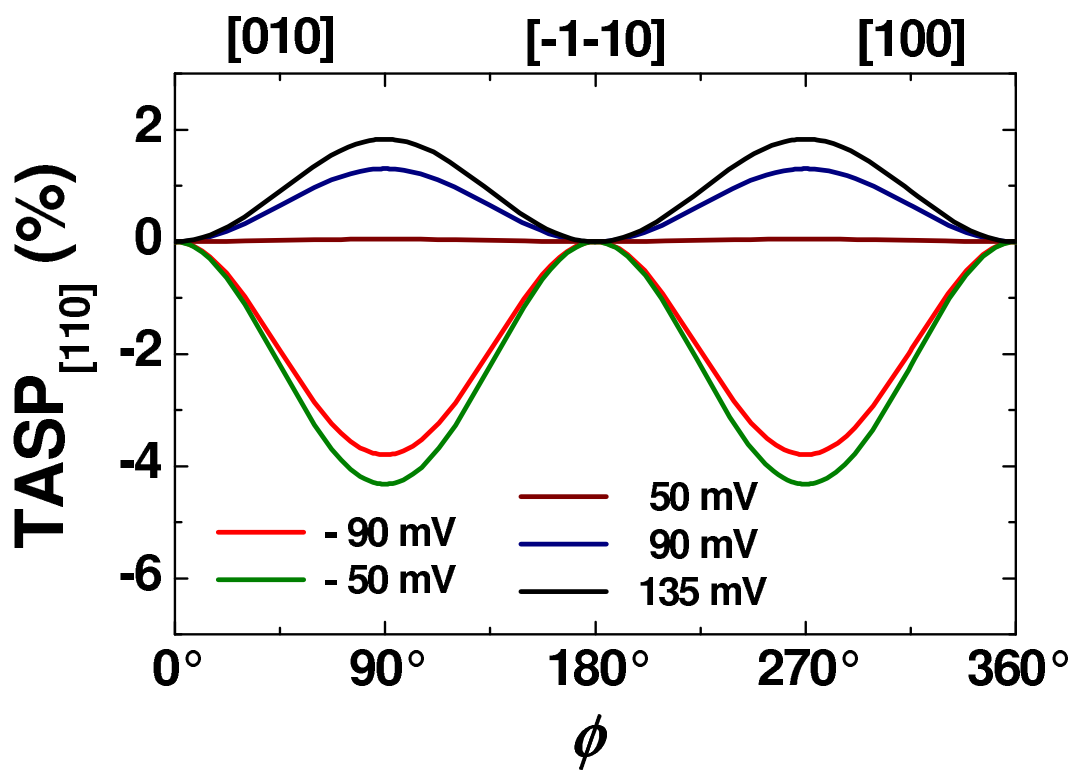

Fig. II.40. Angular dependence of the TASP for different values of the bias voltage.

## III. Spin-orbit coupling in semiconductors

It is well known from special relativity that the motion of an electron in an electric field results in a kinematic effect in which part of the electric field is seen as a magnetic field in the electron's rest frame (Jackson, 1998). The interaction of the electron spin with the electric field (via the associated magnetic field in the electron's rest frame) is called the spin-orbit interaction (SOI), which has the general form,[42]

$$\mathrm{H}_{so} = \frac{\hbar}{4m_0^2c^2}\mathbf{p}\cdot(\boldsymbol{\sigma}\times\boldsymbol{\nabla}V), \tag{III.1}$$

where $m_0$ is the free electron mass, $c$ is the velocity of light, $\boldsymbol{\sigma} = (\sigma_x, \sigma_y, \sigma_z)$ is a vector which components are the Pauli matrices, and $V$ is the electric potential. In Eq. (III.1), $\mathbf{p}$ represents the canonical momentum. In presence of an external magnetic field $\mathbf{B} = \nabla \times \mathbf{A}$, $\mathbf{p}$ should be replaced by the kinetic momentum $\mathbf{P} = \mathbf{p} + e\mathbf{A}$. In the case of atoms, for example, the spin-orbit interaction refers to the interaction of the electron spin with the average Coulomb field of the nuclei and other electrons. Similarly, the SOI in solids is determined by the interaction of the electron spin with the average electric field corresponding to the periodic crystal potential. Other internal or external electric fields can produce additional SOI terms.

### A. Semiconductors with space inversion symmetry

Elemental semiconductors such as silicon have spatial inversion symmetry. For such semiconductors states of a given momentum $\mathbf{k}$ are doubly degenerate:

$$\varepsilon_{\mathbf{k}\uparrow} = \varepsilon_{\mathbf{k}\downarrow}. \tag{III.2}$$

Indeed, time reversal symmetry requires that

$$\varepsilon_{\mathbf{k}\uparrow} = \varepsilon_{-\mathbf{k}\downarrow}, \tag{III.3}$$

since $\mathbf{k} \to -\mathbf{k}$ and $\sigma =\uparrow$ changes to $\sigma =\downarrow$ upon time reversal. Making space inversion means taking $\mathbf{k} \to -\mathbf{k}$, leaving the spins unchanged. Equation (III.2) follows.

How do the Bloch states look like in the presence of spin-orbit coupling? We will show that the degenerate Bloch states, corresponding to the lattice wave vector $\mathbf{k}$, can be written as (Elliott, 1954),

$$\Psi_{\mathbf{k},n\uparrow}(\mathbf{r}) = \left[a_{\mathbf{k}n}(\mathbf{r})|\uparrow\rangle + b_{\mathbf{k}n}(\mathbf{r})|\downarrow\rangle\right]e^{i\mathbf{k}\cdot\mathbf{r}}, \tag{III.4}$$

$$\Psi_{\mathbf{k},n\downarrow}(\mathbf{r}) = \left[a^*_{-\mathbf{k}n}(\mathbf{r})|\downarrow\rangle - b^*_{-\mathbf{k}n}(\mathbf{r})|\downarrow\rangle\right]e^{i\mathbf{k}\cdot\mathbf{r}}. \tag{III.5}$$

Usually we can select the two states such that $|a_{\mathbf{k}n}| \approx 1$ while $|b_{\mathbf{k}n}| \ll 1$, due to the weak spin orbit coupling; this justifies calling the two above states "spin up" and "spin down". We need only to prove that these states have the same energy. Denote by $\hat{K}$ the operator of time reversal [see, for example, (Schiff, 1968), p. 233],

$$\hat{K} = -i\sigma_y \hat{C}, \tag{III.6}$$

where $\hat{C}$ is the operator of complex conjugation. Since the Hamiltonian in the presence of spin-orbit coupling is time invariant, the state,

$$\hat{K}\Psi_{\mathbf{k},n\uparrow} = [a^*_{\mathbf{k}n}(\mathbf{r})|\downarrow\rangle - b^*_{\mathbf{k}n}(\mathbf{r})|\uparrow\rangle]\, e^{-i\mathbf{k}\cdot\mathbf{r}}, \tag{III.7}$$

has the same energy as the state $\Psi_{\mathbf{k},n\uparrow}$. The space inversion, here $\mathbf{k} \to -\mathbf{k}$, then leads us to Eq. (III.5). This completes the proof.

[42] A derivation of the SOI starting with the relativistic Dirac equation can be found in standard text books of quantum mechanics (Sakurai, 1963; Davydov, 1976). Alternatively, a semiclassical derivation of the SOI and the Thomas precession is given by (Jackson, 1998).

### B. Semiconductors without space inversion symmetry

We have seen in the previous section that the combined action of spatial and time inversion symmetries in centrosymmetric semiconductors leads to the double spin degeneracy of the electronic states [see Eq. (III.2)]. There are two ways of breaking this degeneracy. The obvious one is to break the time reversal symmetry by the application of an external magnetic field. The other, somehow more subtle, is to break the spatial inversion symmetry.[43] In the following sections we present within some details, how the lack of inversion symmetry in semiconductors can result, indeed, in a spin-orbit interaction that breaks the spin degeneracy. A mechanism of SOI originates from the structure inversion asymmetry (SIA) of the confining potential (Ohkawa and Uemura, 1974; Bychkov and Rashba, 1984a,b) and is, usually, referred to as the Bychkov-Rashba SOI (Bychkov and Rashba, 1984a,b). The other mechanism of interest here is the so-called Dresselhaus SOI (Dresselhaus, 1955), which is due to the bulk inversion asymmetry of the constituent material itself.

### C. Spin-orbit interaction in semiconductor heterostructures: A qualitative picture

In order to have a simple, intuitive picture of the origin of the spin-orbit interaction in semiconductor heterostructures we focus here, without loss of generality, on the case of a biased, symmetric semiconductor quantum well grown in the $z$ direction (see Fig. III.1). In such semiconductor heterostructures in addition to the SOI resulting from the microscopic, lattice-periodic crystal potential there is a SOI corresponding to other electric fields that may be present in the system (e.g., external and/or built in electric fields). The SOI resulting from external and/or built in electric fields is usually referred to as the Bychkov-Rashba SOI (Bychkov and Rashba, 1984a,b) and will be discussed in more details in Sec. E. In what follows, we assume that the SOI due to the crystal potential has already been incorporated into the band structure of the system and concentrate on the analysis of the effects of the external and/or built in electric fields on the spins of the conduction electrons. In particular, we address the question of which electric field is responsible for the Bychkov-Rashba SOI in the conduction band.

Consider the lowest conduction subband of the quantum well. One could naively think that the average electric field, $\langle E_c(z)\rangle_c = \langle\Psi_c|E_c(z)|\Psi_c\rangle$ (here $\Psi_c$ is the $z$ component of the wave function of the first conduction subband of the quantum well), of the biased conduction band profile of the quantum well (see Fig. III.1) results in a SOI for the conduction electrons. This is, however, not the case. In fact, one can demonstrate that when a position independent effective mass is considered, the average electric field vanishes, $\langle E_c(z)\rangle_c = 0$ (Zawadzki and Pfeffer, 2001). Indeed, the average value of the time derivative of the momentum $\dot{p}_z$ in the bound state $|\Psi_c\rangle$ of the Hamiltonian $H_c$ ($H_c$ describes the quantum well grown in the $z$ direction) is given by Ehrenfest's relation

$$\langle \dot{p}_z\rangle_c = \langle\Psi_c|\dot{p}_z|\Psi_c\rangle = \frac{1}{i\hbar}\langle\Psi_c|[p_z, H_c]|\Psi_c\rangle. \tag{III.8}$$

[43] The spatial inversion symmetry can be broken by applying an external electric field. Non-centrosymmetric systems such as bulk zinc-blende semiconductors lack inversion symmetry even in the absence of any external field.

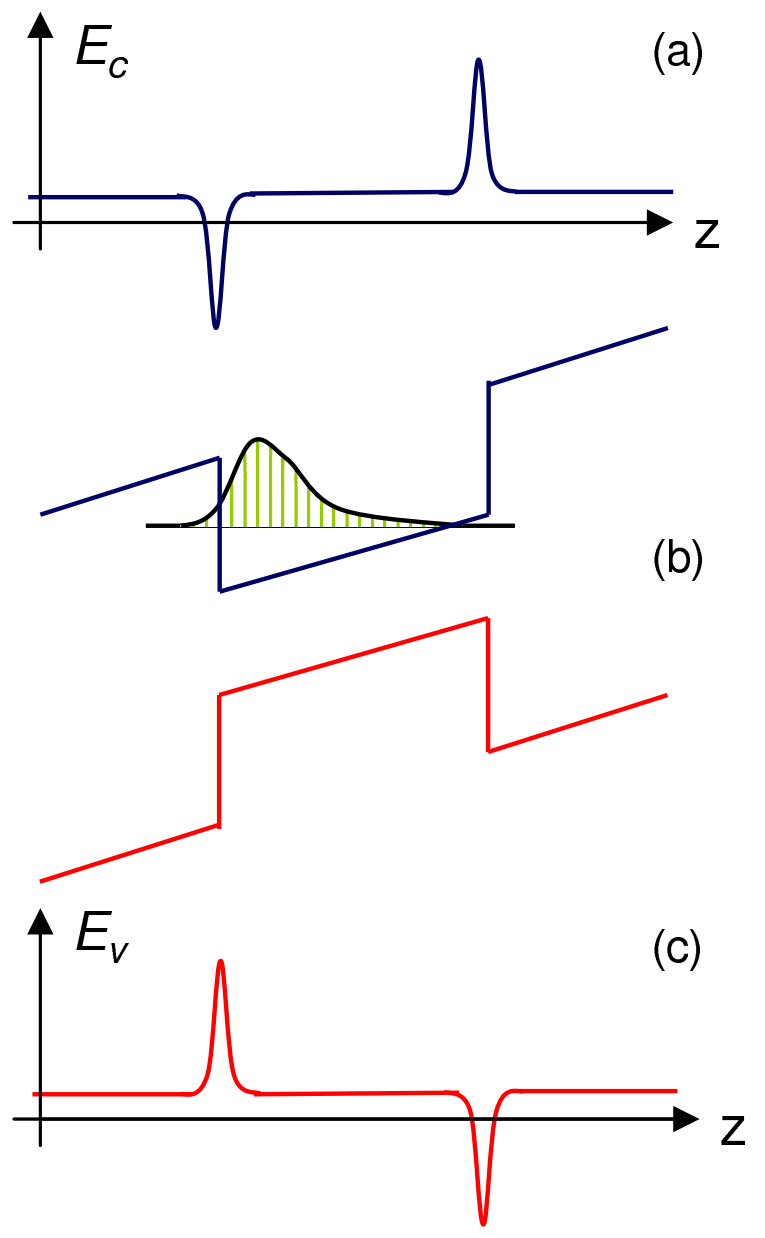

Fig. III.1. Schematics of the potential profiles of the conduction and valence bands (b) and their corresponding electric fields $E_c$ and $E_v$ [(a) and (c), respectively] for a biased quantum well. The electron wave function bound in the QW is sketched in (b). The negative contribution of $E_c(z)$ at the left interface [see (a)] is weighted with a large probability amplitude of the wave function at this interface [see (b)] and, exactly (ignoring the position dependence of the effective mass) cancels the remaining positive contribution after averaging. The average of $E_v(z)$ with the conduction electron wave function, however, does not vanish [see (b) and (c)].

Since $H_c$ is Hermitian and $|\Psi_c\rangle$ is a bound state (i.e., with a square integrable wave function) one obtains, after expanding the commutator $[p_z, H_c] = p_z H_c - H_c p_z$ in Eq. (III.8), that the average $\langle\dot{p}_z\rangle_c = 0$. On the other hand, the Hamiltonian describing the quantum well can be written as

$$H_c = \frac{p_z^2}{2m^*} + eV_c(z), \tag{III.9}$$

where, for simplicity, we have assumed a constant electron effective mass $m^*$. It follows from Eqs. (III.8) and (III.9) that

$$\langle \dot{p}_z \rangle_c = \frac{1}{i\hbar} \langle [p_z, H_c] \rangle_c = \frac{e}{i\hbar} \langle [p_z, V_c(z)] \rangle_c = -\left\langle e \frac{\partial V_c(z)}{\partial z} \right\rangle_c = \langle e E_c(z) \rangle_c. \tag{III.10}$$

Since, as obtained above, $\langle \dot{p}_z \rangle_c = 0$, one obtains from Eq. (III.10) a vanishing average electric field $\langle E_c(z) \rangle_c = 0$. The vanishing of $\langle E_c(z) \rangle_c$ is qualitatively explained in Fig. III.1. The negative contribution of $E_c(z)$ at the left interface is weighted with a large probability amplitude of the wave function at this interface and exactly cancels the remaining positive contributions after averaging.

The fact that $\langle E_c(z) \rangle_c = 0$, generated intense discussions about the nature of the electric field causing the Bychkov-Rashba SOI (Ohkawa and Uemura, 1974; Ando *et al.*, 1982; Bychkov and Rashba, 1984a,b; Lassnig, 1985; Malcher *et al.*, 1986; Lommer *et al.*, 1988; Sobkowicz, 1990; de Andrada e Silva *et al.*, 1994, 1997; Pfeffer and Zawadzki, 1995; Zawadzki and Pfeffer, 2001; Winkler, 2003, 2004a). In particular, it was argued that since $\langle E_c(z) \rangle_c = 0$ the Bychkov-Rashba spin splitting in the conduction band should be considerably small (Ando *et al.*, 1982). It was observed later that when an electron penetrates regions of different effective mass, the average electric field may be non-zero (Malcher *et al.*, 1986; Lommer *et al.*, 1988). In such a case one has to rewrite Eq. (III.9) as

$$H_c = -\frac{\hbar^2}{2} \frac{\partial}{\partial z} \left( \frac{1}{m^*(z)} \frac{\partial}{\partial z} \right) + e V_c(z), \tag{III.11}$$

which, in turn, leads to the following expression,

$$\langle \dot{p}_z \rangle_c = \frac{1}{i\hbar} \langle [p_z, H_c] \rangle_c = \langle e E_c(z) \rangle_c + \langle F_m \rangle_c = 0. \tag{III.12}$$

Here, in addition to the average electric force $\langle e E_c(z) \rangle_c$ there is an average force contribution due to the mass gradient (Malcher *et al.*, 1986; Zawadzki and Pfeffer, 2001),

$$\langle F_m \rangle_c = \left\langle \frac{\hbar^2}{2} \frac{\partial}{\partial z} \left[ \left( \frac{\partial}{\partial z} \frac{1}{m^*(z)} \right) \frac{\partial}{\partial z} \right] \right\rangle = -\frac{\hbar^2}{2} \int_{-\infty}^{\infty} \left| \frac{\partial \Psi_c}{\partial z} \right|^2 \left( \frac{\partial}{\partial z} \frac{1}{m^*(z)} \right) dz. \tag{III.13}$$

In the case of a quantum well with interfaces located at $z_{lc} = -d/2$ and $z_{cr} = d/2$ and piecewise continuous effective mass

$$m^*(z) = m^{*(l)} \Theta(-z - d/2) + m^{*(c)} \Theta(d/2 - |z|) + m^{*(r)} \Theta(z - d/2), \tag{III.14}$$

where $\Theta(x)$ represent the Heaviside step function, the wave function $\Psi_c$ obeys the following boundary conditions

$$\Psi_c|_{z_{ij}^-} = \Psi_c|_{z_{ij}^+}\,; \quad \frac{1}{m^{*(i)}} \frac{\partial \Psi_c}{\partial z}\bigg|_{z_{ij}^-} = \frac{1}{m^{*(j)}} \frac{\partial \Psi_c}{\partial z}\bigg|_{z_{ij}^+}\,; \; i, j = l, r, c; \tag{III.15}$$

which result from the continuity of the probability current across the interfaces. In the equations above $m^{*(l)}$, $m^{*(c)}$, and $m^{*(r)}$ represent the effective mass in the *l*eft barrier, well (*c*entral), and right barrier regions, respectively. From Eqs. (III.13) - (III.15) one obtains

$$\langle F_m \rangle_c = \frac{\hbar^2}{2} \left[ \left(m^{*(c)} - m^{*(r)}\right) \left( \left| \frac{1}{m^*} \frac{\partial \Psi_c}{\partial z} \right|^2 \right) \Bigg|_{z=-\frac{d}{2}} - \left(m^{*(c)} - m^{*(l)}\right) \left( \left| \frac{1}{m^*} \frac{\partial \Psi_c}{\partial z} \right|^2 \right) \Bigg|_{z=\frac{d}{2}} \right], \tag{III.16}$$

where, in virtue of Eq. (III.15), the values at $z = z_{ij}$ $(i, j = l, c, r)$ can be evaluated either using $m^{*(i)}$ and the limit $z = z_{ij}^-$, or $m^{*(j)}$ and the limit $z = z_{ij}^+$. Since $\langle F_m \rangle_c$ is, in general, different from zero, the average electric field $\langle E_c(z) \rangle_c = -\langle F_m \rangle_c / e$ does not vanish when the effective mass discontinuities are considered. The values of $\langle F_m \rangle_c$ lead, however, to small effects and underestimate the Bychkov-Rashba SOI in systems such as GaAs/AlGaAs (Malcher *et al.*, 1986). Thus, although in this case the average electric field $\langle E_c(z) \rangle_c = -\langle F_m \rangle_c / e$ does not vanish exactly, it contributes only in a small fraction to the Bychkov-Rashba SOI in the conduction band.

The controversial issue concerning the origin of the Bychkov-Rashba SOI was resolved by Lassnig (1985), who showed that the Bychkov-Rashba SOI in the conduction band results from the electric field in the valence band. This important observation can be qualitatively understood by extending our previous analysis and including the effects of the biased valence band potential profile on the conduction electrons. The situation is qualitatively sketched in Figs. III.1(b) and (c), from where it results clear that the average valence band electric field $\langle E_v(z) \rangle_c = \langle \Psi_c | E_v(z) | \Psi_c \rangle$ seen by the conduction electrons does not vanish [we note that $V_v(z)$ is not a part of the Hamiltonian in Eq. (III.9) and therefore the Ehrenfest's relation Eq. (III.10) does not apply to the case of $\langle E_v(z) \rangle_c$]. Thus, it is the average field $\langle E_v(z) \rangle_c \neq 0$ that causes the spin-orbit splitting of the conduction electron states. The coupling between conduction and valence band states is then a key ingredient that has to be considered when investigating the SOI in semiconductor heterostructures.

The above analysis can be extended to the case of a symmetric quantum well (see Figs. III.2). Unlike in the asymmetric structure in Fig. III.1, for the symmetric case both $\langle E_c(z) \rangle_c$ and $\langle E_v(z) \rangle_c$ vanish and the Bychkov-Rashba SOI is suppressed [see Figs. III.2]. For this reason, it is often concluded that the structure inversion asymmetry imposed by external and/or build in potentials constitutes a precondition for the presence of the Bychkov-Rashba SOI. We note however that even in perfectly symmetric quantum wells one may still observe spin-orbit related effects due to intersubband coupling (see Sec. E.2) (Bernardes *et al.*, 2006).

## D. Band structure of semiconductors

### D.1 The $\mathbf{k}.\mathbf{p}$ approximation

The $\mathbf{k}.\mathbf{p}$ approximation is a powerful technique for evaluating band structures of bulk semiconductors as well as of semiconductor heterostructures (Long, 1968; Singh, 1993; Bastard, 1998; Rössler, 2004; Yu and Cardona, 2001). In the $\mathbf{k}.\mathbf{p}$ method one can compute the semiconductor band structure in the vicinity of a given point $\mathbf{k}_0$ in the reciprocal space, for which the band structure is already known. Thus, knowing the band structure at some point $\mathbf{k}_0$ and using perturbation theory one can describe the bands away from that point. In order to take advantage of the symmetry properties when evaluating matrix elements, the $\mathbf{k}_0$ is usually taken at a high symmetry point.

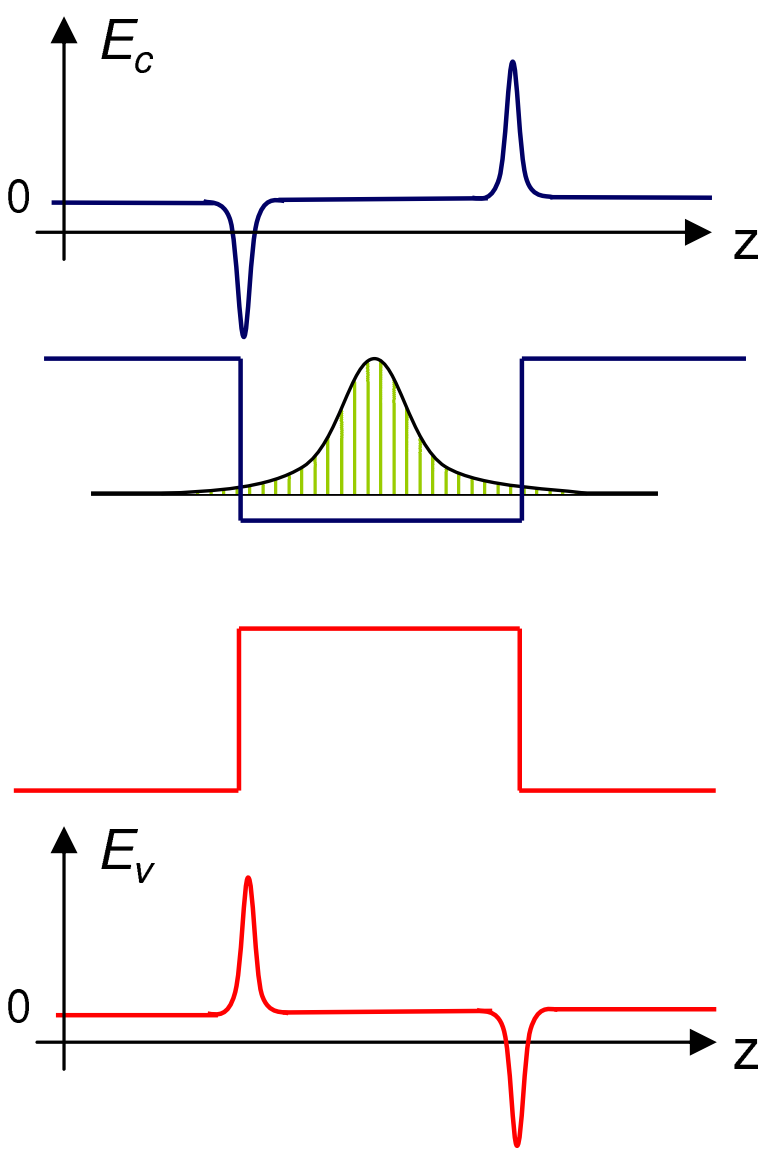

Fig. III.2. Schematics of the potential profiles of the conduction and valence bands and their corresponding electric fields $E_c$ and $E_v$ for a symmetric QW. Both average fields $\langle E_c(z)\rangle_c$ and $\langle E_v(z)\rangle_c$ vanish.

The Schrödinger equation for the Bloch functions $\Psi_{\nu\mathbf{k}}(\mathbf{r})$ in the microscopic lattice-periodic crystal potential $V_0(\mathbf{r})$ is given by

$$H_0\Psi_{\nu\mathbf{k}}(\mathbf{r}) = E_\nu(\mathbf{k})\Psi_{\nu\mathbf{k}}(\mathbf{r}); \quad H_0 = \frac{\mathbf{p}^2}{2m_0} + V_0(\mathbf{r}), \tag{III.17}$$

where $m_0$ represents the free-electron mass and $\nu$ denotes the band index. The Bloch functions can be written as $\Psi_{\nu\mathbf{k}}(\mathbf{r}) = e^{i\mathbf{k}.\mathbf{r}}u_{\nu\mathbf{k}}(\mathbf{r})$. One then obtains from Eq. (III.17) the following equation,

$$\left[\frac{\mathbf{p}^2}{2m_0} + V_0(\mathbf{r}) + \frac{\hbar^2k^2}{2m_0} + \frac{\hbar}{m_0}\mathbf{k}\cdot\mathbf{p}\right]|\nu\mathbf{k}\rangle = E_\nu(\mathbf{k})|\nu\mathbf{k}\rangle, \tag{III.18}$$

which contains only the lattice-periodic parts $u_{\nu\mathbf{k}}(\mathbf{r}) = \langle\mathbf{r}|\nu\mathbf{k}\rangle$ of the Bloch functions. The set $\{|\nu\mathbf{k}_0\rangle\}$ of lattice-periodic parts of the Bloch functions for a fixed value of $\mathbf{k}_0$ constitutes a complete and orthonormal basis. The idea of the $\mathbf{k}.\mathbf{p}$ method consists in expanding the general solutions, away from the known band edge solutions at $\mathbf{k} = \mathbf{k}_0$, in this basis set $\{|\nu\mathbf{k}_0\rangle\}$. In many semiconductors important for spintronics, such as GaAs, the valence band maximum and conduction band minimum occur at the $\Gamma$ point ($\mathbf{k} = \mathbf{0}$) and, consequently, it is the band structure in the vicinity of this high symmetry point what determines many of the properties of semiconductor materials (Singh, 1993; Bastard, 1998; Rössler, 2004; Yu and Cardona, 2001). Therefore, to take advantage of the symmetry properties of the band edge Bloch functions at the $\Gamma$ point, the band structure of semiconductors is usually expanded around $\mathbf{k} = 0$ (note, however, that the application of the method to any other expansion point $\mathbf{k}_0 \neq \mathbf{0}$ is straightforward). The corresponding basis set, including the spin degree of freedom, is then composed by the state vectors

$$|\nu\mathbf{0}\sigma\rangle = |\nu\mathbf{0}\rangle \otimes |\sigma\rangle, \tag{III.19}$$

which are the solutions of Eq. (III.18) for $\mathbf{k} = 0$. Here $|\sigma\rangle$ ($\sigma =\uparrow, \downarrow$) represents the spin eigenstates. In what follows we use the simplified notation,

$$|\nu\mathbf{0}\sigma\rangle = |\nu\sigma\rangle. \tag{III.20}$$

The general solutions of Eq. (III.18), away from the $\Gamma$ point, can be written as

$$|\nu\mathbf{k}\sigma\rangle = \sum_{\nu'} \sum_{\sigma'=\uparrow,\downarrow} c_{\nu\nu'\sigma'}(\mathbf{k})|\nu'\sigma'\rangle. \tag{III.21}$$

Substituting Eq. (III.21) into Eq. (III.18) and multiplying the resulting expression from the left by $\langle\nu''\sigma|$ one obtains, the following eigenvalue problem

$$\sum_{\nu'} \left\{ \left[ E_{\nu'}(\mathbf{0}) + \frac{\hbar^2 k^2}{2m_0} \right] \delta_{\nu''\nu'} + \frac{\hbar}{m_0}\mathbf{k}\cdot\mathbf{p}_{\nu''\nu'} \right\} c_{\nu\nu'\sigma}(\mathbf{k}) = E_{\nu\sigma}(\mathbf{k}) c_{\nu\nu''\sigma}(\mathbf{k}), \tag{III.22}$$

for determining the expansion coefficients $c_{\nu\nu'\sigma'}(\mathbf{k})$ and the eigenvalues $E_{\nu\sigma}(\mathbf{k})$. The momentum matrix elements in Eq. (III.22) are defined as

$$\mathbf{p}_{\nu''\nu'} = \langle\nu''|\mathbf{p}|\nu'\rangle. \tag{III.23}$$

Approximate solutions of Eq. (III.22) can be found in the vicinity of $\mathbf{k} = \mathbf{0}$ by treating the $\mathbf{k}.\mathbf{p}$ terms within perturbation theory (Singh, 1993; Bastard, 1998). This procedure is satisfactory only for small values of $k = |\mathbf{k}|$ and certainly fails when the change in $E_{\nu\sigma}(\mathbf{k})$ becomes of the order of the gap between the bands at the $\Gamma$ point.[44] A better approach is to solve Eq. (III.22) by exact diagonalization in a reduced Hilbert space (i.e., considering a finite number of basis functions). This is the foundation of the Kane model, (Kane, 1957, 1980) to be discussed in Sec. D.2.

Regarding the band edge Bloch states $|\nu\sigma\rangle$, it is worth making some remarks. These states are the starting point for both the $\mathbf{k}.\mathbf{p}$ and envelope function approximations (see Sec. D.2) and are assumed to be known. The precise form of the band edge Bloch states is, in principle, needed for evaluating the momentum matrix elements in Eq. (III.23). In practice, however, the detailed form of these states is not known.[45] Fortunately, it turns out, from symmetry considerations, that only a few matrix elements do not vanish. These non-vanishing matrix elements are then considered as phenomenological parameters to be found from experimental measurements.

[44]In such a case high order perturbation theory is required, which is quite cumbersome.

[45]The form of the band edge Bloch states can be determined from *ab initio* and/or tight binding calculations in which the microscopic details of the system are considered. This is, however, out of the scope of the present discussion.

Consider that the crystal states form as the lattice is hypothetically constructed by bringing the isolated atoms together. One can separate the electrons of an isolated atom into two groups: valence electrons and core electrons. The core electrons are those in the completely filled inner atomic shells and are mostly localized around the nuclei. The valence electrons are those in the outermost partially filled atomic shells.[46] The main properties of the semiconductors are determined by the valence electrons rather than by the mostly localized core electrons. For this reason we neglect the inner shell electrons in our further analysis. In diamond and zinc-blende lattices (these are the typical structures of semiconductors) each atom in the lattice is surrounded by its four valence electrons arranged in hybridized tetrahedral orbitals with each orbital lobe containing an electron pair shared with the neighboring atom. For most solids, it is a reasonable approximation to assume that the orbitals of each atom in the crystal overlap with those of its nearest neighbors only. Since in diamond and zinc-blende lattices there are two atoms per unit cell, the atomic orbitals of the two atoms overlap to form new orbitals. The orbitals on adjacent atoms can combine in two different ways to produce either a *bonding* or an *antibonding* composite orbital. In the case of diamond lattices, where the two adjacent atoms are identical, the bonding orbitals correspond to symmetric (with respect to the interchange of the two atoms) orbitals. The composite symmetric wave function provides a high probability of the electrons occupying the region between the atoms, promoting the formation of a covalent bond. On the other hand antibonding orbitals correspond to antisymmetric orbitals with a node of the wave function at the point midway between the atoms. This tends to exclude the electrons from the region between the atoms and prevents the formation of a covalent bond of low enough energy. Therefore the energy of the antibonding orbitals is higher than the energy of the corresponding bonding orbitals. In the case of zinc-blende lattices, because of the lack of inversion symmetry about the midpoint between the two different atoms in the unit cell, the bonding and antibonding orbitals refer to the combinations of orbital wave functions having the same and opposite signs, respectively. As a result of the orbital overlap in a solid, the bonding and antibonding orbitals are broadened into bands. Those occupied by the electrons form valence bands while the empty ones form conduction bands. (see Fig. III.3). Usually, the bonding orbitals of semiconductors are filled with electrons and become the valence bands while the antibonding orbitals give rise to the conduction bands.

In the absence of spin-orbit interaction, the bands originating from the overlap of the $s$ ($p$) atomic orbitals describing the outermost atomic shells are characterized by $s$-like ($p$-like) band states. In what follows we consider an approximation in which only the antibonding $s$-like state $|S\rangle$ in the lowest conduction band[47] and the topmost bonding $p$-like valence band states $|X\rangle$, $|Y\rangle$, and $|Z\rangle$ are taken into account, while the interaction with other bands is neglected. The $|S\rangle$, $|X\rangle$, $|Y\rangle$, and $|Z\rangle$ states transform as s, $p_x$, $p_y$, and $p_z$, respectively. They can be written in terms of the eigenstates $\phi_{l,m_l} = |lm_l\rangle$ of the orbital angular momentum (Singh, 1993; Bastard, 1998;

[46]In the atoms composing semiconductors the outermost partially filled atomic orbitals are s and p type. For example, the electron configuration of a Si atom is $1s^2 2s^2 2p^6 3s^2 3p^2$. In crystalline Si the $1s$, $2s$, and $2p$ orbitals are completely occupied while the outer $3s$ and $3p$ are only partially filled.

[47]Such an approximation holds for zinc-blende semiconductors such as GaAs but may not be valid for some diamond semiconductors such as Si, in which the lowest conduction band corresponds to an antibonding $p$-like state.

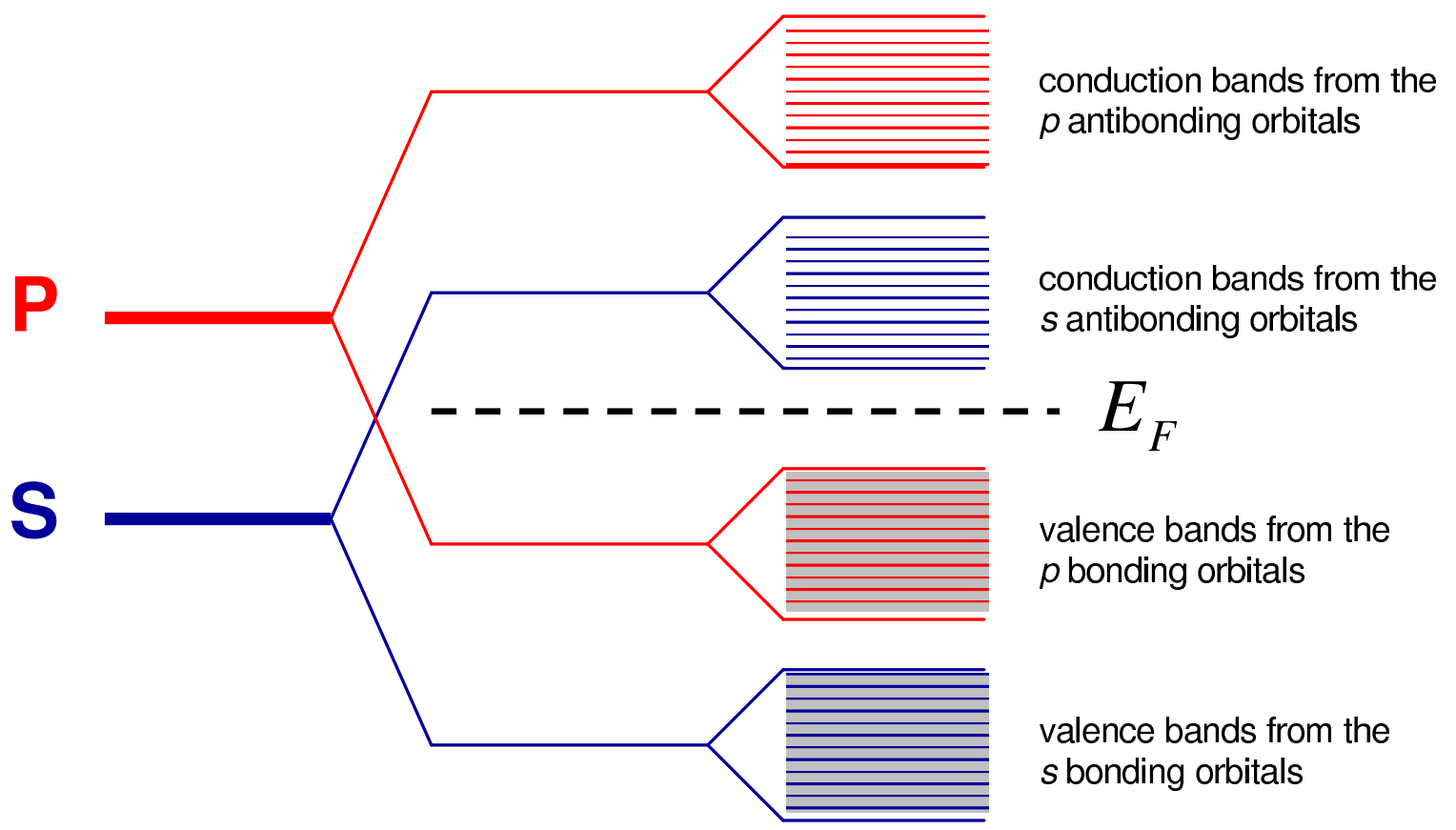

Fig. III.3. Schematics of the evolution of the atomic $s$ and $p$ orbitals into valence and conduction bands in a semiconductor. The Fermi energy is represented by $E_F$. Note that this schematics is not general. In Si, for example, the lowest conduction band corresponds to an antibonding $p$-like state.

Rössler, 2004; Yu and Cardona, 2001):

$$|S\rangle = \phi_{0,0};\ |X\rangle = \frac{1}{\sqrt{2}}(\phi_{1,-1} - \phi_{1,1});\ |Y\rangle = \frac{i}{\sqrt{2}}(\phi_{1,-1} + \phi_{1,1});\ |Z\rangle = \phi_{1,0}. \quad \text{(III.24)}$$

Conversely, one may also write

$$\phi_{0,0} = |S\rangle;\ \phi_{1,1} = -\frac{1}{\sqrt{2}}(|X\rangle + i|Y\rangle);\ \phi_{1,-1} = \frac{1}{\sqrt{2}}(|X\rangle - i|Y\rangle);\ \phi_{1,0} = |Z\rangle. \quad \text{(III.25)}$$

Relations of the type in Eqs. (III.24) and (III.25) are widespread in the literature [see, for instance, (Singh, 1993)]. Strictly speaking, these equalities are not valid, in general.[48] One must interpret the equality symbols in Eqs. (III.24) and (III.25) as a notation expressing the fact that the states in the left- and right-hand sides transform in the same way under symmetry operations. Having this in mind, the notation in Eqs. (III.24) and (III.25) is appropriate, as far as we are not interested in the explicit evaluation of the band edge Bloch functions but rather in their symmetry properties.

The momentum matrix elements involving the band edge Bloch states can be estimated by using Eq. (III.24) and the well-known properties of the states $\phi_{l,m_l} = |lm_l\rangle$ (Davydov, 1976; Sakurai, 1994). From Eq. (III.17) one has

$$[H_0, \mathbf{r}] = H_0\mathbf{r} - \mathbf{r}H_0 = \frac{i\hbar}{m_0}\mathbf{p}. \quad \text{(III.26)}$$

Taking into account that $H_0|\nu\sigma\rangle = E_\nu(\mathbf{0})|\nu\sigma\rangle$ ($\nu = S, X, Y, Z$; $\sigma = \uparrow, \downarrow$), one obtains from Eq. (III.26),

$$\frac{\hbar}{m_0}\langle\nu\sigma|\mathbf{p}|\nu'\sigma'\rangle = i\,[E_{\nu'}(\mathbf{0}) - E_\nu(\mathbf{0})]\,\langle\nu|\mathbf{r}|\nu'\rangle\delta_{\sigma\sigma'}. \quad \text{(III.27)}$$

[48]The exact wave functions corresponding to the $|S\rangle$, $|X\rangle$, $|Y\rangle$, and $|Z\rangle$ states are complicated functions not only on the angular but also on the radial degrees of freedom.

Considering the parity of the band edge estates and the degeneracy of the $p$-like band edge states $[E_{\rm p}(\mathbf{0}) = E_X(\mathbf{0}) = E_Y(\mathbf{0}) = E_Z(\mathbf{0})]$ at the $\Gamma$ point, it follows from Eq. (III.26) that the only non-vanishing momentum matrix elements involving the band edge Bloch states are

$$P_0 = \frac{\hbar}{m_0}\langle S\sigma|p_x|X\sigma\rangle = \frac{\hbar}{m_0}\langle S\sigma|p_y|Y\sigma\rangle = \frac{\hbar}{m_0}\langle S\sigma|p_z|Z\sigma\rangle. \tag{III.28}$$

As mentioned above, within the present model, the precise form of the band edge Bloch states is not known and, consequently, the values of $P_0$ can not be obtained from Eq. (III.28). Instead, $P_0$ is assumed as a phenomenological parameter to be extracted from experiment. Another phenomenological parameter that appears in the theoretical approximation is the energy,

$$E_0 = E_{\rm s}(\mathbf{0}) - E_{\rm p}(\mathbf{0}), \tag{III.29}$$

corresponding to the fundamental gap between the conduction and valence bands at the $\Gamma$ point.

In the analysis above we have omitted the effects of the spin-orbit interaction defined in Eq. (III.1). The spin-orbit interaction mixes the spin states $|\sigma\rangle$. Therefore, the spin is not a good quantum number when the SOI is included in Eq. (III.18). Although one may still use the basis set $\{|\nu\sigma\rangle\}$ ($\nu = S, X, Y, Z$; $\sigma = \uparrow, \downarrow$), it is more convenient to use what we call here an *intelligent* basis in which the spin-orbit matrix elements are already diagonal. This will be discussed in details in the following section.

### D.2 The envelope function approximation (EFA)

The envelope function approximation (EFA) can be seen as a sort of generalization of the $\mathbf{k}.\mathbf{p}$ to the case in which slowly varying, on the length scale of the lattice constant, electric and magnetic fields are present (Burt, 1992; Foreman, 1993; Bastard, 1981, 1998; Winkler, 2003; Yu and Cardona, 2001). The requirement of slowly varying fields [others than the lattice-periodic potential $V_0(\mathbf{r})$] may appear as a strong limitation for the validity of the EFA when modelling semiconductor heterostructures. In fact, the theoretical models of semiconductor heterostructures usually consider effective step-like potentials (for both electron and holes) resulting from the position-dependent band edges (see Fig. III.1). In spite of this controversial issue, it has been shown that the EFA can describe electron and hole states in quantum wells in excellent agreement with the experiment (Bastard, 1981, 1998). This suggests that the assumption of slowly varying potentials (even though they are usually modelled as step-like potentials) may still be a good ansatz. Furthermore, advanced derivations of the EFA show that even in cases where the potential is not slowly varying the application of the EFA can still be justified (Burt, 1992; Foreman, 1993).

Consider now the general Schrödinger equation

$$\left[\frac{\mathbf{P}^2}{2m_0} + V_0(\mathbf{r}) + V(\mathbf{r}) + \frac{g_0\mu_B}{2}\boldsymbol{\sigma}\cdot\mathbf{B} + \frac{\hbar}{4m_0^2c^2}\mathbf{P}\cdot(\boldsymbol{\sigma}\times\boldsymbol{\nabla}V_0)\right]\Psi_{n\mathbf{k}}(\mathbf{r}) = E_{n\mathbf{k}}\Psi_{n\mathbf{k}}(\mathbf{r}), \tag{III.30}$$

describing the electrons in the crystal, in the presence of the potential $V(\mathbf{r})$ (this potential refers to any, other than the crystal, potential $V_0(\mathbf{r})$, e.g., the potential resulting from the position-dependent band edges in a heterostructure or the potential of external electric fields) and an external magnetic field $\mathbf{B}$. The first term in the left-hand side of Eq. (III.30) corresponds to the electron kinetic energy, with $\mathbf{P} = \mathbf{p} + e\mathbf{A}(\mathbf{r})$ being the operator of the kinetic momentum (here $e$, $\mathbf{p}$, and $\mathbf{A}(\mathbf{r})$ are the electron charge, canonical momentum operator, and vector potential, respectively). The fourth term describes the Zeeman interaction with the external magnetic field $\mathbf{B}$. The gyromagnetic factor of the free electron is $g_0 = 2$. The fifth term corresponds to the spin-orbit interaction [see Eq. (III.1)].

We recall that in the presence of SOI the spin quantum number $\sigma$ is, by itself, not a good quantum number. Therefore, we have used in Eq. (III.30) only a common index $n$ for the orbital motion and the spin degree of freedom.

Following the $\mathbf{k}.\mathbf{p}$ method, one can, in principle, use the basis set $\{|\nu\sigma\rangle\}$ [see Eq. (III.19)]. However, as mentioned in Sec. D.1, it is more convenient to use an *intelligent* basis in which the matrix elements of the SOI become diagonal. Thus, in place of using directly the basis set $\{|\nu\sigma\rangle\}$ we will use the *intelligent* basis set $\{|m\rangle\}$ composed by the state vectors

$$u_m = |m\rangle = \sum_{\nu=S,X,Y,Z} \sum_{\sigma=\uparrow,\downarrow} h_{m\nu\sigma}|\nu\sigma\rangle. \tag{III.31}$$

Since the *intelligent* states $|m\rangle$ must reflect the symmetry of the bands, one can divide the *intelligent* basis set $\{|m\rangle\}$ into the subsets $\{|m\rangle_{\rm s}\}$ and $\{|m\rangle_{\rm p}\}$ composed by s-like and p-like intelligent states, respectively. Thus, for the s-like intelligent states the expansion in Eq. (III.31) reduces to,

$$|m\rangle_{\rm s} = \sum_{\sigma=\uparrow,\downarrow} h_{ms\sigma}|S\sigma\rangle, \tag{III.32}$$

while for the p-like intelligent states one obtains

$$|m\rangle_{\rm p} = \sum_{\nu=X,Y,Z} \sum_{\sigma=\uparrow,\downarrow} h_{m\nu\sigma}|\nu\sigma\rangle. \tag{III.33}$$

The expansion coefficients obey the following relations

$$\sum_{\sigma=\uparrow,\downarrow} h^*_{m's\sigma} h_{ms\sigma} = \delta_{mm'}; \quad \sum_{\nu=X,Y,Z} \sum_{\sigma=\uparrow,\downarrow} h^*_{m'\nu\sigma} h_{m\nu\sigma} = \delta_{mm'}, \tag{III.34}$$

which can be obtained from Eqs. (III.32) and (III.33), respectively, by considering the orthogonality of the basis sets $\{|\nu\sigma\rangle\}$ and $\{|m\rangle\}$. In Eq. (III.34) $\delta_{mm'}$ represents the Kronecker symbol.

The appropriate expansion coefficients $h_{m\nu\sigma}$ can be found by requiring that the SOI be diagonal in the *intelligent* basis. The details of such a procedure will be explained latter on in this section.

Expanded in the intelligent basis, the eigenstates $|n\mathbf{k}\rangle$ [$\Psi_{n\mathbf{k}}(\mathbf{r}) = \langle\mathbf{r}|n\mathbf{k}\rangle$] of Eq. (III.30) can be written as

$$|n\mathbf{k}\rangle = \sum_m f_{nm}(\mathbf{k},\mathbf{r})|m\rangle. \tag{III.35}$$

The position dependence of the expansion coefficients $f_{nm}$ accounts for the spatial dependence of the external and/or build in electric fields. These coefficients modulate the fast oscillations of the lattice-periodic functions $\langle \mathbf{r}|m\rangle$. Since this modulation varies slowly on the length scale of the lattice constant, the expansion coefficients $f_{nm}$ are called envelope functions (Bastard, 1981, 1998).

After some algebra one obtains that the operators in the Hamiltonian in Eq. (III.30) acts on the summands in Eq. (III.35) as follows,

$$\mathbf{P}^2 f_{nm}|m\rangle = (\mathbf{P}^2 f_{nm})|m\rangle + 2(\mathbf{p}f_{nm})\cdot(\mathbf{p}|m\rangle) + f_{nm}(\mathbf{p}^2|m\rangle) + 2ef_{nm}\mathbf{A}\cdot(\mathbf{p}|m\rangle), \quad \text{(III.36)}$$

$$\mathbf{P}\cdot(\boldsymbol{\sigma}\times\boldsymbol{\nabla}V_0)f_{nm}|m\rangle = (\mathbf{P}f_{nm})\cdot(\boldsymbol{\sigma}\times\boldsymbol{\nabla}V_0)|m\rangle + f_{nm}\mathbf{p}\cdot(\boldsymbol{\sigma}\times\boldsymbol{\nabla}V_0)|m\rangle, \quad \text{(III.37)}$$

and

$$(\boldsymbol{\sigma}\cdot\mathbf{B})f_{nm}|m\rangle = f_{nm}\mathbf{B}\cdot(\boldsymbol{\sigma}|m\rangle). \quad \text{(III.38)}$$

By substituting Eq. (III.35) in Eq. (III.30) and multiplying from the left by $\langle m'\mathbf{0}|$ one obtains the following eigenvalue problem for the envelope functions

$$\sum_m \left\{ \left[\frac{\mathbf{P}^2}{2m_0} + V(\mathbf{r})\right]\delta_{m'm} + \frac{\mathbf{P}}{m_0}\cdot\mathbf{p}_{m'm} + \triangle_{m'm} + E^{(0)}_{m'm} + \frac{g_0\mu_B}{2}\left(\mathbf{S}_{m'm}\cdot\mathbf{B}\right)\right\} f_{nm} = E_{n\mathbf{k}}f_{nm'}, \quad \text{(III.39)}$$

where

$$\mathbf{p}_{m'm} = \left\langle m' \left| \mathbf{p} + \frac{\hbar}{4m_0c^2}(\boldsymbol{\sigma}\times\boldsymbol{\nabla}V_0) \right| m \right\rangle, \quad \text{(III.40)}$$

$$\triangle_{m'm} = \frac{\hbar}{4m_0^2c^2}\langle m'|\mathbf{p}\cdot(\boldsymbol{\sigma}\times\boldsymbol{\nabla}V_0)|\rangle, \quad \text{(III.41)}$$

$$\mathbf{S}_{m'm} = \langle m'|\boldsymbol{\sigma}|m\rangle, \quad \text{(III.42)}$$

and

$$E^{(0)}_{m'm} = \langle m'\,|H_0|\,m\rangle\,, \quad \text{(III.43)}$$

with $H_0$ given by Eq. (III.17). In obtaining Eq. (III.39) we have used the relations in Eqs. (III.36) - (III.38). We have also taken advantage of the slow variations of the envelope functions, $V(\mathbf{r})$, $\mathbf{A}(\mathbf{r})$, and $\mathbf{B}$ on the lattice constant scale and have taken them out of the corresponding integrations over the unit cell.

The contribution of the term $\langle m'|\hbar(4m_0c^2)^{-1}(\boldsymbol{\sigma}\times\boldsymbol{\nabla}V_0)|m\rangle$ in Eq. (III.40) is much smaller than the momentum matrix elements. Therefore, we can neglect this term and use the approximation (Kane, 1980; Lassnig, 1985; Singh, 1993; Bastard, 1998; Winkler, 2003),

$$\mathbf{p}_{m'm} \approx \langle m'|\mathbf{p}|m\rangle. \quad \text{(III.44)}$$

The matrix element $E^{(0)}_{m'm}$ can be calculated by substituting the Eqs. (III.32) and (III.33) into Eq. (III.43) and taking into account that the states $|\nu\sigma\rangle$ are eigenstates of $H_0$ with eigenenergies $E_\nu(\mathbf{0})$. The results are,

$$E^{(0)}_{m'm} = \sum_{\sigma=\uparrow,\downarrow} h^*_{m's\sigma} h_{ms\sigma} E_{\rm s}(\mathbf{0}) = E_{\rm s}(\mathbf{0})\delta_{m'm}, \tag{III.45}$$

and

$$E^{(0)}_{m'm} = \sum_{\nu=X,Y,Z}\sum_{\sigma=\uparrow,\downarrow} h^*_{m'\nu\sigma} h_{m\nu\sigma} E_\nu(\mathbf{0}) = E_{\rm P}(\mathbf{0})\delta_{m'm}, \tag{III.46}$$

for matrix elements involving s-like and p-like intelligent states, respectively. The matrix elements mixing s-like and p-like intelligent states vanish. In obtaining Eqs. (III.45) and (III.46) we took into consideration Eq. (III.34) and the degeneracy of the $|S\sigma\rangle$ states [with eigenenergies $E_{\rm s}(\mathbf{0})$] and the $|X\sigma\rangle$, $|Y\sigma\rangle$, and $|Z\sigma\rangle$ states [with eigenenergies $E_{\rm p}(\mathbf{0}) = E_{_X}(\mathbf{0}) = E_{_Y}(\mathbf{0}) = E_{_Z}(\mathbf{0})$] at the $\Gamma$ point and in the absence of the SOI. Considering Eqs. (III.45) and (III.46), the eigenvalue problem in Eq. (III.39) can be rewritten as,

$$\sum_m \left\{ \left[ E^{(0)}_m + \frac{\mathbf{P}^2}{2m_0} + V(\mathbf{r}) \right] \delta_{m'm} + \frac{\mathbf{P}}{m_0}\cdot \mathbf{p}_{m'm} + \triangle_{m'm} + \frac{g_0\mu_B}{2}\left(\mathbf{S}_{m'm}\cdot\mathbf{B}\right) \right\} f_{nm} = E_{n\mathbf{k}} f_{nm'}, \tag{III.47}$$

where $E^{(0)}_m = E_{\rm s}(\mathbf{0})$ for s-like intelligent states and $E^{(0)}_m = E_{\rm P}(\mathbf{0})$ for p-like intelligent states.

A better approximation than the direct application of perturbation theory to solve Eq. (III.47) was formulated by E. O . Kane (1957,1980). Kane observed that in many situations the inclusion of only a few adjacent bands is enough for capturing the main physical features. He then proposed to work in a reduced Hilbert space in which Eq. (III.47) can be solved exactly. Thus, in the Kane model one takes full account of the $\mathbf{P}.\mathbf{p}_{m'm}$[49] and spin-orbit interactions only for the most relevant bands, whereas the contributions of the remote bands are treated perturbatively. In the original version[50] of the Kane model, which applies to any diamond or zinc blende type materials, one works in a reduced Hilbert space with an 8-dimensional *intelligent* basis set. The eight state vectors $|m\rangle$ of the *intelligent* basis are in turn linear combinations of the eight band edge Bloch states $|\nu\sigma\rangle$ [see Eq. (III.31)].

The spin-orbit interaction is large only near a lattice atom, where $\boldsymbol{\nabla} V_0$ is appreciable and the crystal potential $V_0$ is at least roughly spherically symmetric. One can then rewrite Eq. (III.1) in a similar way as for the isolated atoms,(Davydov, 1976; Yu and Cardona, 2001; Long, 1968)

$$\mathrm{H}_{so} = \frac{\hbar}{4m_0^2c^2}\frac{1}{r}\frac{dV_0}{dr}\mathbf{L}\cdot\mathbf{S}, \tag{III.48}$$

where $\mathbf{L} = \mathbf{r}\times\mathbf{p}$ and $\mathbf{S}$ represent the orbital angular momentum and the spin angular momentum operators, respectively. Averaging out the radial degree of freedom, one can rewrite Eq. (III.48)

[49] Note that the $\mathbf{P}.\mathbf{p}_{m'm}$ interaction in Eq. (III.47) is a generalization of the $\mathbf{k}.\mathbf{p}$ interaction to the case of a finite magnetic field. The canonical wave vector $\mathbf{k}$ is then substituted by the kinetic wave vector $\mathbf{P}/\hbar = \mathbf{k} + e\mathbf{A}/\hbar$.

[50] The original Kane model includes only 8 bands. A further generalization including 14 bands is usually referred to as the extended Kane model (see Sec. F.).

as

$$H_{so} = \lambda \mathbf{L} \cdot \mathbf{S}, \tag{III.49}$$

where $\lambda$ is a phenomenological parameter. The total angular momentum is $\mathbf{J} = \mathbf{L} + \mathbf{S}$, and consequently $\mathbf{J}^2 = \mathbf{L}^2 + \mathbf{S}^2 + 2\mathbf{L} \cdot \mathbf{S}$. Thus, in terms of the total angular momentum, Eq. (III.49) can be rewritten as

$$H_{so} = \frac{\lambda}{2}(\mathbf{J}^2 - \mathbf{L}^2 - \mathbf{S}^2). \tag{III.50}$$

It follows from Eq. (III.50) that the *intelligent* basis is nothing but a basis whose components are the eigenfunctions of $\mathbf{J}^2$ (note that if a function is an eigenfunction of $\mathbf{J}^2$, then it is also an eigenfunction of $\mathbf{L}^2$, $\mathbf{S}^2$, $J_z$, $L_z$, and $S_z$). Thus, in the *intelligent* basis $H_{so}$ becomes diagonal and its expectation value can easily be evaluated as

$$\langle H_{so} \rangle = \frac{\lambda \hbar^2}{2}[j(j+1) - l(l+1) - s(s+1)], \tag{III.51}$$

where $j$, $l$, and $s$ are the quantum numbers corresponding to the operators $\mathbf{J}^2$, $\mathbf{L}^2$, and $\mathbf{S}^2$, respectively.

The eigenfunctions of $(\mathbf{J}^2, J_z)$ can be found by applying the rule for the addition of angular momenta (Davydov, 1976; Sakurai, 1994). Since in the present case $\mathbf{J} = \mathbf{L} + \mathbf{S}$, one can write the eigenstates of $(\mathbf{J}^2, J_z)$ as

$$|l\ s\ j\ m_j\rangle = \sum_{m_s=-s}^{s} (l\ s\ m_j - m_s\ m_s|j\ m_j)|l\ m_j - m_s\rangle \otimes |m_s\rangle, \tag{III.52}$$

where $|l\ m_j - m_s\rangle$ are the eigenstates of $(\mathbf{L}^2, L_z)$ with quantum numbers $(l, m_l = m_j - m_s)$, $(l\ s\ m_j - m_s|j\ m_j)$ are the Clebsch-Gordon coefficients, and $m_j$ and $m_s$ are the quantum numbers corresponding to $J_z$ and $S_z$, respectively. For spin-$\frac{1}{2}$ particles $s = 1/2$ and Eq. (III.52) reduces to

$$|l\ \frac{1}{2}\ j\ m_j\rangle = \sum_{m_s=\pm\frac{1}{2}} (l\ \frac{1}{2}\ m_j - m_s\ m_s|j\ m_j)|l\ m_j - m_s\rangle \otimes |m_s\rangle. \tag{III.53}$$

The Clebsch-Gordon coefficients for $j = l \pm 1/2$ ($j > 0$) and $m_s = \pm 1/2$ can be calculated from the following relations (Davydov, 1976; Sakurai, 1994)

$$(l\ \ \frac{1}{2}\ \ m_j \mp \frac{1}{2}\ \ \pm\frac{1}{2}|l \pm \frac{1}{2}\ \ m_j) = \sqrt{\frac{l + m_j + 1/2}{2l+1}}, \tag{III.54}$$

$$\pm(l\ \ \frac{1}{2}\ \ m_j \pm \frac{1}{2}\ \ \mp\frac{1}{2}|l \pm \frac{1}{2}\ \ m_j) = \sqrt{\frac{l - m_j + 1/2}{2l+1}}. \tag{III.55}$$

Within the eight-bands Kane model $l = 0, 1$. The $s$-like states ($l = 0$) and the $p$-like states ($l = 1$) correspond to different bands. Therefore, it is enough to specify the band and the quantum numbers $j$ and $m_j$ for completely determining the basis vectors of the Kane model.

For this reason the notation $|l\ s\ j\ m_j\rangle$ is commonly shortened to $|j\ m_j\rangle$. From now on we will assume such an abbreviated notation.

The eight *intelligent* basis vectors of the Kane model are found from Eqs. (III.53)-(III.55). The $s$-like states ($l = 0$) consist of the $\Gamma_{6c}$ doublet[51]

$$|\frac{1}{2}\ \frac{1}{2}\rangle = \phi_{0,0} \otimes |\uparrow\rangle, \tag{III.56}$$

$$|\frac{1}{2}\ -\frac{1}{2}\rangle = \phi_{0,0} \otimes |\downarrow\rangle. \tag{III.57}$$

This doublet, corresponding to the lowest conduction bands, has $\langle H_{so}\rangle = 0$ [see Eq. (III.51)]. The $p$-like states ($l = 1$) consist of the $\Gamma_{8v}$ quadruplet

$$|\frac{3}{2}\ \frac{3}{2}\rangle = \phi_{1,1} \otimes |\uparrow\rangle, \tag{III.58}$$

$$|\frac{3}{2}\ \frac{1}{2}\rangle = \sqrt{\frac{2}{3}}\phi_{1,0} \otimes |\uparrow\rangle + \sqrt{\frac{1}{3}}\phi_{1,1} \otimes |\downarrow\rangle, \tag{III.59}$$

$$|\frac{3}{2}\ -\frac{1}{2}\rangle = \sqrt{\frac{1}{3}}\phi_{1,-1} \otimes |\uparrow\rangle + \sqrt{\frac{2}{3}}\phi_{1,0} \otimes |\downarrow\rangle, \tag{III.60}$$

$$|\frac{3}{2}\ -\frac{3}{2}\rangle = \phi_{1,-1} \otimes |\downarrow\rangle, \tag{III.61}$$

and the $\Gamma_{7v}$ doublet

$$|\frac{1}{2}\ \frac{1}{2}\rangle = -\sqrt{\frac{1}{3}}\phi_{1,0} \otimes |\uparrow\rangle + \sqrt{\frac{2}{3}}\phi_{1,1} \otimes |\downarrow\rangle, \tag{III.62}$$

$$|\frac{1}{2}\ -\frac{1}{2}\rangle = -\sqrt{\frac{2}{3}}\phi_{1,-1} \otimes |\uparrow\rangle + \sqrt{\frac{1}{3}}\phi_{1,0} \otimes |\downarrow\rangle. \tag{III.63}$$

In the absence of SOI the $\Gamma_{8v}$ quadruplet describing the light and heavy holes and the $\Gamma_{7v}$ doublet are energy degenerate at the $\Gamma$ point. However, when the effects of SOI are included, one finds from Eq. (III.51) the values $\langle H_{so}\rangle_{\Gamma_{8v}} = \hbar^2\lambda/2$ and $\langle H_{so}\rangle_{\Gamma_{7v}} = -\hbar^2\lambda$ for the $\Gamma_{8v}$ quadruplet and $\Gamma_{7v}$ doublet, respectively. This leads to the spin-orbit splitting,

$$\Delta_0 = \langle H_{so}\rangle_{\Gamma_{8v}} - \langle H_{so}\rangle_{\Gamma_{7v}} = \frac{3\hbar^2\lambda}{2}, \tag{III.64}$$

which is a phenomenological parameter of the Kane model.

In Eqs. (III.56)-(III.63) we have used the shorthand notations $|m_s = \frac{1}{2}\rangle = |\uparrow\rangle$, $|m_s = -\frac{1}{2}\rangle = |\downarrow\rangle$, and $\phi_{l,m_l} = |l\ m_l\rangle$. Using the relations in Eq. (III.25) one can express the *intelligent* basis defined in Eqs. (III.56)-(III.63) in terms of the band edge Bloch basis. The corresponding energies at the $\Gamma$ point and zero magnetic field can easily be computed from Eqs. (III.47) and (III.51). Taking the zero of the energy scale at the bottom of the $\Gamma_{6c}$ conduction band [i.e., making $E_s(\mathbf{0}) = 0$], the energy of the degenerate $\Gamma_{8v}$ valence bands is given by $E_s(\mathbf{0}) = -E_0$

[51] The notation $\Gamma_6$, $\Gamma_8$, etc, originates from group theory classification according to the irreducible representations of the symmetry group of the crystal, which determines the way the wave functions with wave vector $\mathbf{k}$ at the center of the Brillouin zone ($\Gamma$ point) transform (Yu and Cardona, 2001). In addition, the subindexes $c$ and $v$ indicate conduction and valence bands, respectively.

Tab. III.1. Basis functions of the Kane model and their corresponding energies at the $\Gamma$ point and zero magnetic field. The zero of the energy scale is taken at the bottom of the $\Gamma_{6c}$ conduction band. The phenomenological parameters $E_0$ and $\triangle_0$ correspond to the fundamental gap and spin-orbit splitting, respectively (see Fig. III.4).

| Band | $\|m\rangle$ | Basis functions | Energy |
|---|---|---|---|
| $\Gamma_{6c}$ | $\|\frac{1}{2}\ \frac{1}{2}\rangle$ | $\|S\uparrow\rangle$ | $0$ |
| | $\|\frac{1}{2}\ -\frac{1}{2}\rangle$ | $\|S\downarrow\rangle$ | $0$ |
| $\Gamma_{8v}$ | $\|\frac{3}{2}\ \frac{3}{2}\rangle$ | $-\frac{1}{\sqrt{2}}\|(X+iY)\uparrow\rangle$ | $-E_0$ |
| | $\|\frac{3}{2}\ \frac{1}{2}\rangle$ | $\sqrt{\frac{2}{3}}\|Z\uparrow\rangle-\frac{1}{\sqrt{6}}\|(X+iY)\downarrow\rangle$ | $-E_0$ |
| | $\|\frac{3}{2}\ -\frac{1}{2}\rangle$ | $\sqrt{\frac{2}{3}}\|Z\downarrow\rangle+\frac{1}{\sqrt{6}}\|(X-iY)\uparrow\rangle$ | $-E_0$ |
| | $\|\frac{3}{2}\ -\frac{3}{2}\rangle$ | $\frac{1}{\sqrt{2}}\|(X-iY)\downarrow\rangle$ | $-E_0$ |
| $\Gamma_{7v}$ | $\|\frac{1}{2}\ \frac{1}{2}\rangle$ | $-\frac{1}{\sqrt{3}}\|Z\uparrow\rangle-\frac{1}{\sqrt{3}}\|(X+iY)\downarrow\rangle$ | $-(E_0+\triangle_0)$ |
| | $\|\frac{1}{2}\ -\frac{1}{2}\rangle$ | $\frac{1}{\sqrt{3}}\|Z\downarrow\rangle-\frac{1}{\sqrt{3}}\|(X-iY)\uparrow\rangle$ | $-(E_0+\triangle_0)$ |

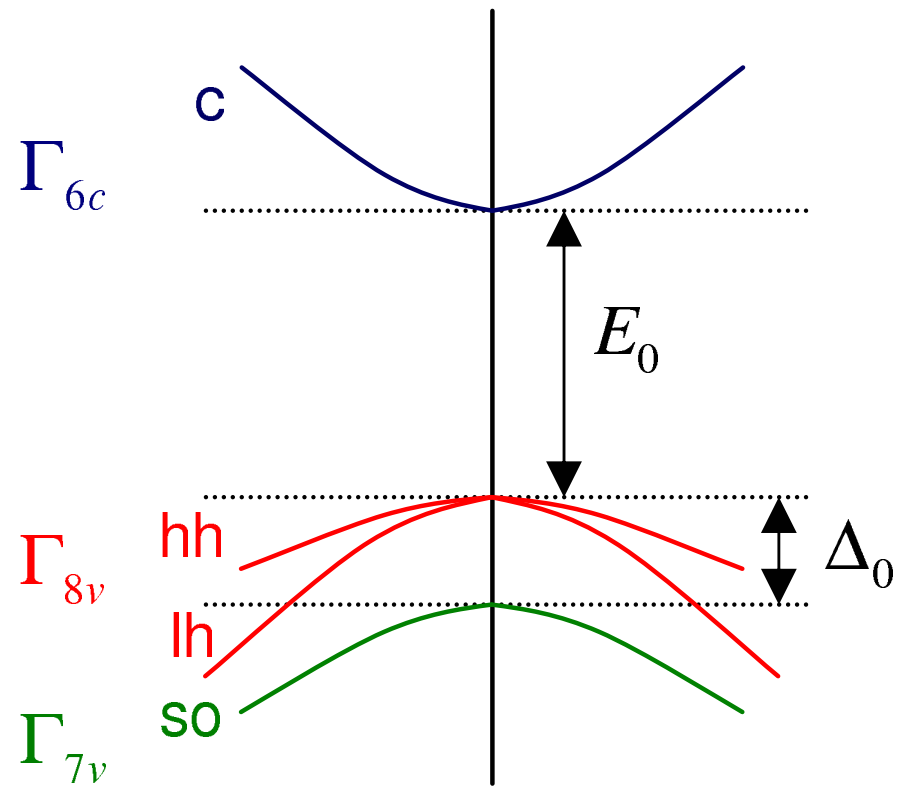

Fig. III.4. Schematics of a III-V semiconductor band structure obtained within the Kane model. Symbols indicate conduction (c), heavy-hole (hh), light-hole (lh), and spin split-off (so) bands.

[see Eq. (III.29)]. The energy of the spin split-off $\Gamma_{7v}$ bands is $E_{\mathrm{p}}(\mathbf{0})-\triangle_0$. The results are summarized in Tab. III.1. Furthermore, a schematics of a III-V semiconductor band structure obtained within the Kane model is displayed in Fig. III.4.

Using the basis set given in Tab. III.1, one can rewrite Eq. (III.47) as follows

$$\mathrm{Hf}=\mathrm{Ef}, \tag{III.65}$$

where $\mathbf{f}=(f_{n1},f_{n2},...,f_{n8})^T$ and

$$\mathrm{H}=\begin{pmatrix}\mathrm{H}_c & \mathrm{H}_{cv}\\ \mathrm{H}_{vc} & \mathrm{H}_v\end{pmatrix}. \tag{III.66}$$

Taking the zero of the energy scale at the bottom of the conduction band, the matrix blocks describing the conduction ($\Gamma_{6c}$) and valence ($\Gamma_{8c}$ and $\Gamma_{7v}$) bands can be written as

$$\mathrm{H}_c = \begin{pmatrix} \frac{\mathbf{P}^2}{2m_0} + V(\mathbf{r}) + \frac{g_0\mu_B}{2}B_z & \frac{g_0\mu_B}{2}B_- \\ \frac{g_0\mu_B}{2}B_+ & \frac{\mathbf{P}^2}{2m_0} + V(\mathbf{r}) - \frac{g_0\mu_B}{2}B_z \end{pmatrix} \tag{III.67}$$

and

$$\begin{aligned} \mathrm{H}_v = & \, [V(\mathbf{r}) - E_0]\mathbf{1}_{6\times 6} \\ + & \, g_0\mu_B \begin{pmatrix} \frac{1}{2}B_z & \frac{1}{2\sqrt{3}}B_- & 0 & 0 & \frac{1}{\sqrt{6}}B_- & 0 \\ \frac{1}{2\sqrt{3}}B_+ & \frac{1}{6}B_z & \frac{1}{3}B_- & 0 & -\frac{\sqrt{2}}{3}B_z & \frac{1}{6}B_- \\ 0 & \frac{1}{3}B_+ & -\frac{1}{6}B_z & \frac{1}{2\sqrt{3}}B_- & -\frac{\sqrt{2}}{6}B_+ & -\frac{\sqrt{2}}{3}B_z \\ 0 & 0 & \frac{1}{2\sqrt{3}}B_+ & -\frac{1}{2}B_z & 0 & -\frac{1}{\sqrt{6}}B_+ \\ \frac{1}{\sqrt{6}}B_+ & -\frac{\sqrt{2}}{3}B_z & -\frac{\sqrt{2}}{6}B_- & 0 & -\frac{\triangle_0}{g_0\mu_B} - \frac{1}{6}B_z & -\frac{1}{6}B_- \\ 0 & \frac{1}{\sqrt{6}}B_+ & -\frac{\sqrt{2}}{3}B_z & -\frac{1}{\sqrt{6}}B_- & -\frac{1}{6}B_+ & -\frac{\triangle_0}{g_0\mu_B} + \frac{1}{6}B_z \end{pmatrix} \end{aligned} \tag{III.68}$$

respectively. Here, and in what follows, $\mathbf{1}_{m\times m}$ represents the $(m \times m)$ unit matrix. On the other hand, the matrices

$$\mathrm{H}_{cv} = \begin{pmatrix} -\frac{1}{\sqrt{2}}\frac{P_0}{\hbar}P_+ & \sqrt{\frac{2}{3}}\frac{P_0}{\hbar}P_z & \frac{1}{\sqrt{6}}\frac{P_0}{\hbar}P_- & 0 & -\frac{1}{\sqrt{3}}\frac{P_0}{\hbar}P_z & -\frac{1}{\sqrt{3}}\frac{P_0}{\hbar}P_- \\ 0 & -\frac{1}{\sqrt{6}}\frac{P_0}{\hbar}P_+ & \sqrt{\frac{2}{3}}\frac{P_0}{\hbar}P_z & \frac{1}{\sqrt{2}}\frac{P_0}{\hbar}P_- & -\frac{1}{\sqrt{3}}\frac{P_0}{\hbar}P_+ & \frac{1}{\sqrt{3}}\frac{P_0}{\hbar}P_z \end{pmatrix} \tag{III.69}$$

and

$$\mathrm{H}_{vc} = \begin{pmatrix} -\frac{1}{\sqrt{2}}\frac{P_0}{\hbar}P_- & 0 \\ \sqrt{\frac{2}{3}}\frac{P_0}{\hbar}P_z & -\frac{1}{\sqrt{6}}\frac{P_0}{\hbar}P_- \\ \frac{1}{\sqrt{6}}\frac{P_0}{\hbar}P_+ & \sqrt{\frac{2}{3}}\frac{P_0}{\hbar}P_z \\ 0 & \frac{1}{\sqrt{2}}\frac{P_0}{\hbar}P_+ \\ -\frac{1}{\sqrt{3}}\frac{P_0}{\hbar}P_z & -\frac{1}{\sqrt{3}}\frac{P_0}{\hbar}P_- \\ -\frac{1}{\sqrt{3}}\frac{P_0}{\hbar}P_+ & \frac{1}{\sqrt{3}}\frac{P_0}{\hbar}P_z \end{pmatrix} \tag{III.70}$$

correspond to interactions between the conduction and valence bands.[52] Note that, in principle, the term $\mathbf{P}^2/(2m_0)$ should appear also in the main diagonal of $\mathrm{H}_v$. However, because of the large value of $m_0$ with respect to the electron effective mass in semiconductors (see Sec. E.1), $\mathbf{P}^2/(2m_0)$ is usually much smaller than $E_0$ and can be neglected in Eq. (III.68). In obtaining Eqs. (III.65)-(III.70) we have used the relations in Eq. (III.28). We have also introduced the shorthand notation

$$B_\pm = B_x \pm iB_y \;\; ; \;\; P_\pm = P_x \pm iP_y. \tag{III.71}$$

[52] Note that, taking the magnetic field as an argument, the relation $\mathrm{H}_{cv}(\mathbf{B}) = -\mathrm{H}^\dagger_{vc}(-\mathbf{B})$ holds.

By solving the system of eight coupled differential equations in Eq. (III.65) one can obtain the sets of envelope functions $(f_{n1}, f_{n2})$, $(f_{n3}, f_{n4}, f_{n5}, f_{n6})$, and $(f_{n7}, f_{n8})$ describing the eigenstates of the $\Gamma_{6c}$, $\Gamma_{8v}$, and $\Gamma_{7v}$, respectively.

We now focus on the case of conduction electrons (the generalization to the case of holes is straightforward). In order to obtain the envelope functions $f_{n1}$ and $f_{n2}$ corresponding to the $\Gamma_{6c}$ conduction band one has to eliminate all the other components of $\mathbf{f}$ from Eq. (III.65). Thus, the original problem is transformed into a reduced system of only two differential equations for $f_{n1}$ and $f_{n2}$. A general and systematic procedure for performing such a transformation is the so-called folding down method (Cohen and Heine, 1970). The main idea of this method is to introduce the matrix

$$\mathrm{U} = \begin{pmatrix} \mathbf{1}_{2\times 2} & -(\mathrm{H}_c - \mathrm{E})^{-1}\mathrm{H}_{cv} \\ -(\mathrm{H}_v - \mathrm{E})^{-1}\mathrm{H}_{vc} & \mathbf{1}_{6\times 6} \end{pmatrix}. \tag{III.72}$$

Inserting $\mathbf{1}_{8\times 8} = \mathrm{U}\mathrm{U}^{-1}$ in Eq. (III.65), one obtains

$$(\mathrm{H} - \mathrm{E})\mathrm{U}\mathrm{U}^{-1}\mathbf{f} = 0. \tag{III.73}$$

The equation above can be rewritten as follows

$$(\tilde{\mathrm{H}} - \mathrm{E})\tilde{\mathbf{f}} = 0, \tag{III.74}$$

where

$$\tilde{\mathbf{f}} = \mathrm{U}^{-1}\mathbf{f}, \tag{III.75}$$

$$\tilde{\mathrm{H}} = \mathrm{E} + (\mathrm{H} - \mathrm{E})\mathrm{U} = \begin{pmatrix} \tilde{\mathrm{H}}_c & 0 \\ 0 & \tilde{\mathrm{H}}_v \end{pmatrix}, \tag{III.76}$$

$$\tilde{\mathrm{H}}_c = \mathrm{H}_c - \mathrm{H}_{cv}(\mathrm{H}_v - \mathrm{E})^{-1}\mathrm{H}_{vc}, \tag{III.77}$$

and

$$\tilde{\mathrm{H}}_v = \mathrm{H}_v - \mathrm{H}_{vc}(\mathrm{H}_c - \mathrm{E})^{-1}\mathrm{H}_{cv}. \tag{III.78}$$

We are particularly interested in that eigenvector of Eq. (III.74) which has the form $\tilde{\mathbf{f}} = (\tilde{f}_1, \tilde{f}_2, 0, 0, 0, 0, 0, 0)^T$. For this eigenvector one has from Eqs. (III.72) and (III.75) that $\tilde{f}_1 = f_{n1}$ and $\tilde{f}_2 = f_{n2}$. One then obtains from Eq. (III.74) the following effective system of two coupled differential equations

$$\tilde{\mathrm{H}}_c\mathbf{f}_c = \mathrm{E}\mathbf{f}_c, \tag{III.79}$$

which contains only the envelope functions $\mathbf{f}_c = (f_{n1}, f_{n2})^T$ of the $\Gamma_{6c}$ conduction band.

Computing the matrix $\tilde{\mathrm{H}}_c$ requires the inversion of $(\mathrm{H}_v - \mathrm{E})$ [see Eq. (III.77)]. Such a matrix inversion is trivial for the case of zero magnetic field, since in this case $(\mathrm{H}_v - \mathrm{E})$ becomes diagonal [see Eq. (III.68)]. However, from the analytical point of view, the exact inversion of $(\mathrm{H}_v - \mathrm{E})$ can in general lead to cumbersome expressions in $\tilde{\mathrm{H}}_c$. A procedure for finding a simplified, approximate expression for $(\mathrm{H}_v - \mathrm{E})^{-1}$ consists in making the decomposition $(\mathrm{H}_v - \mathrm{E}) = \mathrm{A} - \mathrm{D}$, where A is a nonsingular, diagonal matrix. Thus, $\mathrm{H}_v - \mathrm{E} = \mathrm{A}(\mathbf{1}_{6\times 6} - \mathrm{A}^{-1}\mathrm{D})$ and, consequently,

$$(\mathrm{H}_v - \mathrm{E})^{-1} = (\mathbf{1}_{6\times 6} - \mathrm{A}^{-1}\mathrm{D})^{-1}\mathrm{A}^{-1} = [\mathbf{1}_{6\times 6} + \mathrm{A}^{-1}\mathrm{D} + (\mathrm{A}^{-1}\mathrm{D})^2 + (\mathrm{A}^{-1}\mathrm{D})^3 + ...]\mathrm{A}^{-1}. \quad \text{(III.80)}$$

The advantage of using Eq. (III.80) is that by appropriately choosing A, the series expansion may rapidly converge and few terms in the series suffice for a good estimation. Then, one can approximately compute $(\mathrm{H}_v - \mathrm{E})^{-1}$ by performing just a few matrix multiplications (since A is diagonal, it is trivial to compute $\mathrm{A}^{-1}$). A good choice for the case of Eq. (III.68) could be $\mathrm{A} = diag[V(\mathbf{r}) - E_0 - E, V(\mathbf{r}) - E_0 - E, V(\mathbf{r}) - E_0 - E, V(\mathbf{r}) - E_0 - E, V(\mathbf{r}) - E_0 - E - \triangle_0, V(\mathbf{r}) - E_0 - E - \triangle_0]$. Indeed, in such a case the non-vanishing elements of D are of the order of the Zeeman energy $g_0\mu_B B/2$. In many practical situations $g_0\mu_B B/2 \ll V(\mathbf{r}) - E_0 - E$ and the series in Eq. (III.80) rapidly converges.

## E. Bychkov-Rashba spin-orbit interaction

### E.1 Spin-orbit interaction in systems with structure inversion asymmetry

We have seen in Sec. D.2 that the SOI due to the lattice-periodic crystal potential can be diagonalized by using the *intelligent* basis set given in Tab. III.1. We show now that in systems with structure inversion asymmetry (SIA) there are other contributions to the SOI. Without loss of generality, we consider the case of SIA induced SOI in the conduction band for the case of an asymmetric, semiconductor quantum well (QW) in the presence of an external electric field oriented in the grown direction of the well. The grown direction $z$ is taken as the [001] crystallographic direction. The spatial dependence of the band edges is schematically represented in Fig. III.5. Because the materials in the left $(l)$, central $(c)$, and right $(r)$ regions are different, the parameters $E_0$ and $\triangle_0$ become piecewise constant, i.e.,

$$E_0(z) = E_0^{(l)}\Theta(-z - d/2) + E_0^{(c)}\Theta(d/2 - |z|) + E_0^{(r)}\Theta(z - d/2), \quad \text{(III.81)}$$

$$\triangle_0(z) = \triangle_0^{(l)}\Theta(-z - d/2) + \triangle_0^{(c)}\Theta(d/2 - |z|) + \triangle_0^{(r)}\Theta(z - d/2), \quad \text{(III.82)}$$

where $\Theta(x)$ and $d$ denote the Heaviside step function and the well width, respectively. Similarly, taking the zero of the energy scale at the bottom of the QW in the conduction band [i.e., $E_c^{(c)} = 0$], one obtains

$$V(\mathbf{r}) = V(z) = V_{ext}(z) + E_c(z), \quad \text{(III.83)}$$

where $V_{ext}(z)$ is the potential corresponding to the external electric field,[53] and

$$E_c(z) = E_c^{(l)}\Theta(-z - d/2) + E_c^{(r)}\Theta(z - d/2), \quad \text{(III.84)}$$

is the position-dependent band edge profile forming the QW in the conduction band (see Fig. III.5).

[53] Although we consider here an external electric field, the action of any other $z$-dependent, internal build-in electric potential $V_{int}$ (e.g., the depletion field) can be included in the same manner, i.e., by simple substitution of $V_{ext}$ by $V_{ext} + V_{int}$.

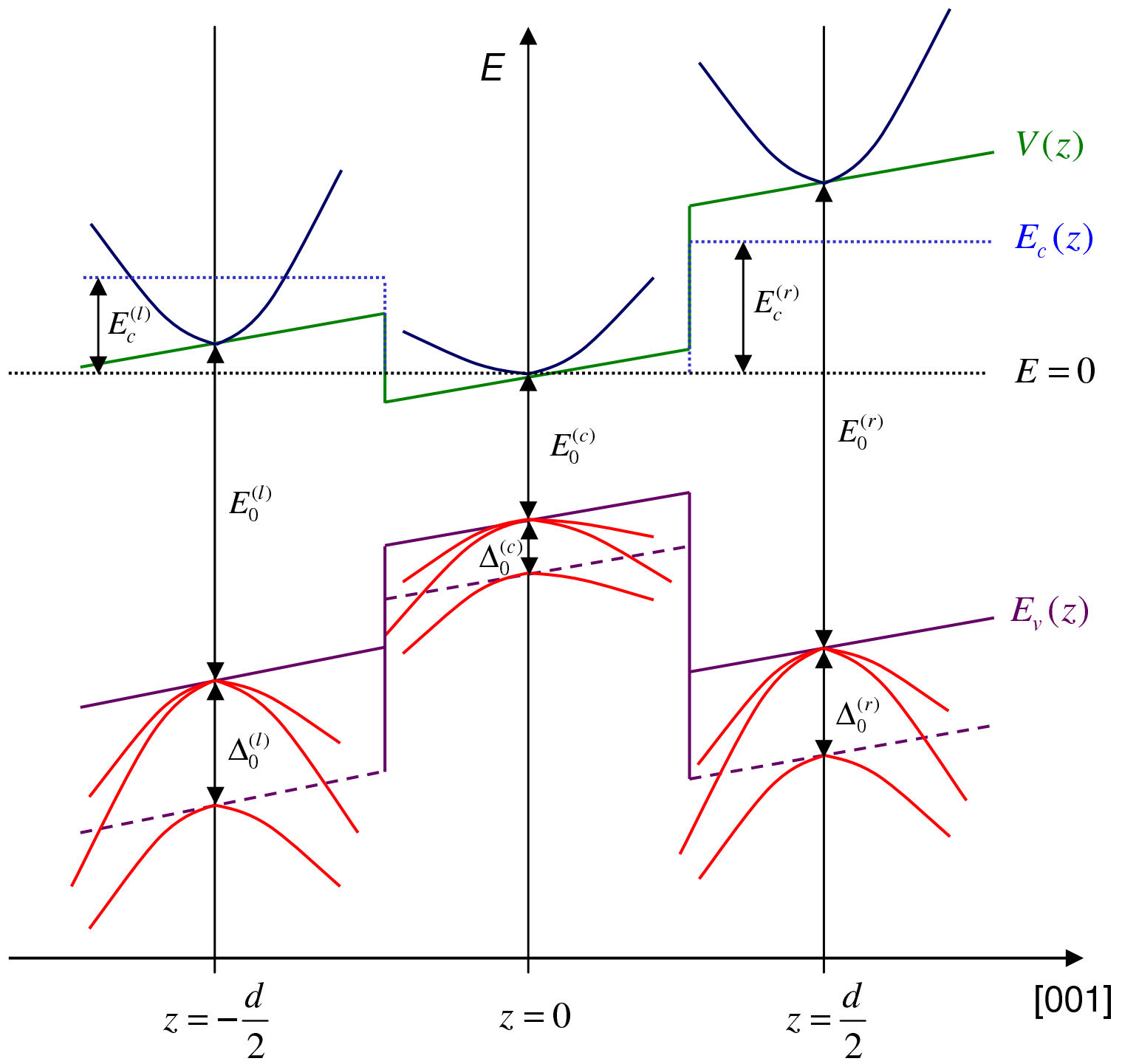

Fig. III.5. Kane model picture of the bulk-like band structure of an asymmetric quantum well in presence of an external electric field of potential $V(z)$.

The in-plane motion (i.e., in the directions parallel to the interfaces) is free, hence one can decouple the envelope function as

$$\mathbf{f} = e^{i\mathbf{k}_\parallel \cdot \mathbf{r}_\parallel} \mathbf{g}, \tag{III.85}$$

where $\mathbf{k}_\parallel = (k_x, k_y, 0)$ and $\mathbf{r}_\parallel = (x, y, 0)$ are the in-plane wave vector and in-plane position of the particle, respectively. Substituting this expression into Eq. (III.65) one obtains an eigenvalue problem for $\mathbf{g}$ with a Hamiltonian similar to Eq. (III.66) but with matrix blocks,

$$\mathrm{H}_c = \left[\frac{\hbar^2 k_\parallel^2}{2m_0} + \frac{p_z^2}{2m_0} + V_{ext}(z) + E_c(z)\right] \mathbf{1}_{2\times 2}, \tag{III.86}$$

$$\mathrm{H}_v = diag[E_v(z), E_v(z), E_v(z), E_v(z), E_v'(z), E_v'(z)], \tag{III.87}$$

where

$$E_v(z) = V_{ext}(z) + E_c(z) - E_0(z), \tag{III.88}$$

and

$$E'_v(z) = E_v(z) - \triangle_0(z), \tag{III.89}$$

represent the position-dependent band edge profile of the $\Gamma_{8v}$ and $\Gamma_{7v}$ valence bands, respectively (see Fig. III.5).

The matrices $\mathrm{H}_{cv}$ and $\mathrm{H}_{cv}$ have the same form as in Eqs. (III.69) and (III.70), respectively, but now with $P_z = p_z$ and $P_\pm = \hbar k_\pm = \hbar(k_x \pm ik_y)$.

Since $\mathrm{H}_v - E$ is diagonal, it is trivial to find its inverse. One then obtains for the folded-down Hamiltonian [see Eq. (III.77)],

$$\tilde{\mathrm{H}}_c = \left[ -\frac{\hbar^2}{2m_0}\frac{d^2}{dz^2} + \frac{\hbar^2 k_\parallel^2}{2m_0} + E_c(z) + V_{ext}(z) \right] \mathbf{1}_{2\times 2} - \begin{pmatrix} M & N \\ N^* & M \end{pmatrix}, \tag{III.90}$$

where

$$\begin{aligned} M &= \frac{P_0^2}{3\hbar^2}\left( \frac{2}{E_v(z) - E} + \frac{1}{E_v(z) - \triangle_0(z) - E} \right) \hbar^2 k_\parallel^2 \\ &- \frac{P_0^2}{3}\frac{d}{dz}\left[ \left( \frac{2}{E_v(z) - E} + \frac{1}{E_v(z) - \triangle_0(z) - E} \right) \frac{d}{dz} \right], \end{aligned} \tag{III.91}$$

and

$$N = \frac{iP_0^2 k_-}{3}\frac{d}{dz}\left( \frac{1}{E_v(z) - E} - \frac{1}{E_v(z) - \triangle_0(z) - E} \right). \tag{III.92}$$

In order to rewrite Eq. (III.90) in terms of the Pauli matrices we make the expansion,

$$\begin{pmatrix} M & N \\ N^* & M \end{pmatrix} = \lambda_0 \mathbf{1}_{2\times 2} + \boldsymbol{\lambda} \cdot \boldsymbol{\sigma}. \tag{III.93}$$

Taking the trace from both sides of Eq. (III.93) one obtains $\lambda_0 = M$. Furthermore, solving for the components of the vector $\boldsymbol{\lambda} = (\lambda_x, \lambda_y, \lambda_z)$ one finds

$$\lambda_0 = M \;\; ; \;\; \lambda_x = (N + N^*)/2 \;\; ; \;\; \lambda_y = i(N - N^*)/2 \;\; ; \;\; \lambda_z = 0. \tag{III.94}$$

By combining Eqs. (III.90)-(III.94) one can write

$$\tilde{\mathrm{H}}_c = -\frac{\hbar^2}{2}\frac{d}{dz}\left[ \frac{1}{m^*(z,E)}\frac{d}{dz} \right] + \frac{\hbar^2 k_\parallel^2}{2m^*(z,E)} + E_c(z) + V_{ext}(z) + \mathrm{H}_{BR}, \tag{III.95}$$

where (Lassnig, 1985; de Andrada e Silva *et al.*, 1997),

$$\frac{1}{m^*(z,E)} = \frac{1}{m_0} - \frac{2P_0^2}{3\hbar^2}\left( \frac{2}{E_v(z) - E} + \frac{1}{E_v(z) - \triangle_0(z) - E} \right) \tag{III.96}$$

is the inverse effective mass for the conduction band electrons, and

$$\mathrm{H}_{BR} = \alpha(z)(k_x\sigma_y - k_y\sigma_x), \tag{III.97}$$

represents the Bychkov-Rashba SOI. The Bychkov-Rashba parameter $\alpha(z)$ is given by

$$\alpha(z) = \frac{d\beta(z)}{dz} \quad ; \quad \beta(z) = \frac{P_{\rm o}^2}{3}\left(\frac{1}{E_v(z) - E} - \frac{1}{E_v(z) - \triangle_0(z) - E}\right). \tag{III.98}$$

The expression for the SOI contains only the position-dependent band edge profiles $E_v(z)$ and $E_v(z) - \triangle_0(z)$. Consequently, the SOI in the conduction band is modulated by the electric field in the valence band rather than by that in the conduction band, as already discussed in Sec. C.

An approximate expression for the effective mass in the $i$th region ($i = l, r, c$) can be obtained by noting that $E_0^{(i)}$ and $E_0^{(i)} + \triangle_0^{(i)}$ are the largest energy scales in that region. Taking into account Eq. (III.88), the two terms within the brackets in Eq. (III.96) can be expanded in powers of $[E_c^{(i)} + V_{ext}(z) - E]/E_0^{(i)}$ and $[E_c^{(i)} + V_{ext}(z) - E]/(E_0^{(i)} + \triangle_0^{(i)})$, respectively. One then obtains, up to the first order, the following relation,

$$\begin{aligned}\frac{1}{m^{*(i)}} \quad &\approx \quad \frac{1}{m_0} + \frac{2P_{\rm o}^2}{3\hbar^2}\left[\left(\frac{2}{E_0^{(i)}} + \frac{1}{E_0^{(i)} + \triangle_0^{(i)}}\right)\right. \\ &+ \quad \left.\left(\frac{2}{[E_0^{(i)}]^2} + \frac{1}{[E_0^{(i)} + \triangle_0^{(i)}]^2}\right)[E_c^{(i)} + V_{ext}(z) - E]\right].\end{aligned} \tag{III.99}$$

The energy dependence of the effective mass leads to the nonparabolicity of the dispersion relation. For analytical treatments one usually assumes the parabolic approximation, in which the second order term is neglected, i.e., (Lassnig, 1985; Bastard, 1998; de Andrada e Silva *et al.*, 1994; Winkler, 2003)

$$\frac{1}{m^{*(i)}} \approx \frac{1}{m_0} - \frac{2P_{\rm o}^2}{3\hbar^2}\left(\frac{2}{E_0^{(i)}} + \frac{1}{E_0^{(i)} + \triangle_0^{(i)}}\right). \tag{III.100}$$

Following the same procedure, one obtains from Eq. (III.98) the approximate expression,

$$\begin{aligned}\beta^{(i)}(z) \quad &= \quad \frac{P_{\rm o}^2}{3}\left[\left(\frac{1}{E_0^{(i)} + \triangle_0^{(i)}} - \frac{1}{E_0^{(i)}}\right)\right. \\ &+ \quad \left.\left(\frac{1}{[E_0^{(i)} + \triangle_0^{(i)}]^2} - \frac{1}{[E_0^{(i)}]^2}\right)[E_c^{(i)} + V_{ext}(z) - E]\right],\end{aligned} \tag{III.101}$$

which leads to the position dependence of $\beta(z)$,

$$\beta(z) = \beta^{(l)}(z)\Theta(-z - d/2) + \beta^{(c)}(z)\Theta(d/2 - |z|) + \beta^{(r)}(z)\Theta(z - d/2). \tag{III.102}$$

Considering Eqs. (III.98) and (III.102) one gets

$$\alpha(z) = \alpha_0(z) + \alpha_{int}(z). \tag{III.103}$$

The contribution

$$\alpha_0(z) = \alpha^{(l)}(z)\Theta(-z - d/2) + \alpha^{(c)}(z)\Theta(d/2 - |z|) + \alpha^{(r)}(z)\Theta(z - d/2), \tag{III.104}$$

with (de Andrada e Silva *et al.*, 1997; Ivchenko and Pikus, 1997; Winkler, 2003, 2004a)

$$\alpha^{(i)}(z) = \frac{d\beta^{(i)}(z)}{dz} = \frac{P_0^2}{3}\left(\frac{1}{[E_0^{(i)}+\triangle_0^{(i)}]^2} - \frac{1}{[E_0^{(i)}]^2}\right)\frac{dV_{ext}(z)}{dz}\ ;\ \ i=(l,c,r) \quad \text{(III.105)}$$

is related to the external field, while

$$\alpha_{int}(z) = [\beta^{(c)}(z)-\beta^{(l)}(z)]\delta(z+d/2) - [\beta^{(c)}(z)-\beta^{(r)}(z)]\delta(z-d/2), \quad \text{(III.106)}$$

with $\delta(x)$ being the Dirac delta function, contains the interface contributions only (de Andrada e Silva *et al.*, 1997).

The effective mass mismatch at the interfaces together with the interface Bychkov-Rashba SOI reflect on the boundary conditions. Integrating $\tilde{\text{H}}_c\mathbf{g} = E\mathbf{g}$ [with $\tilde{\text{H}}_c$ given by Eq. (III.95)] across the interface at $z = z_{ij}$ (i.e., between the $i$th and $j$th regions) one deduces the following boundary conditions (Sobkowicz, 1990; Bastard *et al.*, 1991; de Andrada e Silva *et al.*, 1997; Pfeffer, 1997; Zawadzki and Pfeffer, 2004; Voskoboynikov *et al.*, 1998, 1999)

$$\mathbf{g}^{(i)}(z_{ij}) = \mathbf{g}^{(j)}(z_{ij}), \quad \text{(III.107)}$$

$$\left.\frac{\hbar^2}{2m^{*(i)}}\frac{d\mathbf{g}^{(i)}}{dz}\right|_{z=z_{ij}} - \left.\frac{\hbar^2}{2m^{*(j)}}\frac{d\mathbf{g}^{(j)}}{dz}\right|_{z=z_{ij}} + [\beta^{(j)}(z_{ij})-\beta^{(i)}(z_{ij})](k_x\sigma_y - k_y\sigma_x)\mathbf{g}^{(i)}(z_{ij}) = 0. \quad \text{(III.108)}$$

The boundary condition in Eq. (III.108) depends on both the potential $V_{ext}(z_{ij})$ at the interfaces and the energy $E$. This dependence can be eliminated in a first approximation by neglecting the term proportional to $[E_c^{(i)} + V_{ext}(z) - E]$ in Eq. (III.101), i.e., by taking

$$\beta^{(i)} \approx \frac{P_0^2}{3}\left(\frac{1}{E_0^{(i)}+\triangle_0^{(i)}} - \frac{1}{E_0^{(i)}}\right). \quad \text{(III.109)}$$

In what follows, we assume such an approximation.

When the in-plane motion is of particular interest (as in the study of two-dimensional gases, for example) one can average out the $z$-dependent terms of the Hamiltonian $\tilde{\text{H}}_c$. Taking the average $\langle ...\rangle = \langle g_0|...|g_0\rangle$ with the state $g_0(z)$ corresponding to a given subband (without spin) of the QW one obtains for the Hamiltonian describing the in-plane motion,

$$\tilde{\text{H}}_{\parallel} = \langle\tilde{\text{H}}_c\rangle = \epsilon_0(\mathbf{k}) + \alpha_{BR}(k_x\sigma_y - k_y\sigma_x)\ ;\ \ \alpha_{BR} = \langle\alpha_0\rangle + \langle\alpha_{int}\rangle, \quad \text{(III.110)}$$

where $\epsilon_0(\mathbf{k})$ is the subband eigenenergy of the QW in the presence of the external field. We note that Eq. (III.110) is already an approximation. Generally speaking, when the structure itself lacks inversion symmetry, the SOI is coupled to the motion in the $z$-direction via the boundary condition in Eq. (III.108). In such a case it is not possible to exactly transfer the SOI to the Hamiltonian describing the in-plane motion. Equation (III.110) is, however, a good approximation for structures in which $\beta^{(j)}(z_{ij}) - \beta^{(i)}(z_{ij})$ is sufficiently small or in deep QWs, where Tab. III.2. Range of values of the Bychkov-Rashba parameter $\alpha_{BR}$ deduced from experiment.

| System | Spin splitting at the Fermi (meV) | $\alpha_{BR}$ ( meV Å) | Reference |
|---|---|---|---|
| AlSb/InAs/AlSb | 3.2 - 4.5 | 60 | (Heida *et al.*, 1998) |
| AlSb/InAs/AlSb | 0 | 0 | (Brosig *et al.*, 1999) |
| AlSb/InAs/AlSb | 0 | 0 | (Sasa *et al.*, 1999) |
| AlGaAs/GaAs/AlGaAs | - | $6.9 \pm 0.4$ | (Jusserand *et al.*, 1995) |
| 2DEG GaAs/AlGaAs | - | $5 \pm 1$ | (Miller *et al.*, 2003) |
| AlGaSb/InAs/AlSb | 5.6 - 13 | 120 - 280 | (Sasa *et al.*, 1999) |
| InAlAs/InGaAs/InAlAs | 1.5 | 40 | (Das *et al.*, 1989) |
| InAlAs/InGaAs/InAlAs | 4.9 - 5.9 | 63 - 93 | (Nitta *et al.*, 1997) |
| InAlAs/InGaAs/InAlAs | - | 50 - 100 | (Hu *et al.*, 1999) |
| InGaAs/InAs/InGaAs | 5.1 - 6.8 | 60 - 110 | (Nitta *et al.*, 1998) |
| InGaAs/InAs/InGaAs | 9 - 15 | 200 - 400 | (Grundler, 2000) |
| InGaAs/InP/InGaAs | - | 63 - 153 | (Engels *et al.*, 1997) |
| GaSb/InAs/GaSb | 3.7 | 90 | (Luo *et al.*, 1988) |
| Si/SiGe QW | - | 0.03 - 0.12 | (Malissa *et al.*, 2004) |
| SiO2/InAs/ | 5.5 - 23 | 100 - 300 | (Matsuyama *et al.*, 2000) |

the probability of finding the electron at the interfaces is small and the spin-dependent boundary condition becomes irrelevant.

For a system with inversion symmetry, the parameters $E_c(z)$, $E_0(z)$, $\triangle_0(z)$, and $V_{ext}(z)$ are even functions of $z$ and the corresponding eigenfunctions $\mathbf{g}(z)$ have well defined parity. One can see from Eqs. (III.104)-(III.106) that due to symmetry reasons the average parameters $\langle \alpha_0 \rangle$ and $\langle \alpha_{int} \rangle$ vanish in systems with inversion symmetry. Consider, for example, the simple case in which $V_{ext} = 0$. In such a case $\alpha^{(i)}(z) = 0$ and we need to consider only the interface contribution $\alpha_{int}$. If the QW is symmetric, the probabilities of finding the particle at the two interfaces $z = \pm d/2$ are, by symmetry, the same. In addition one has $\beta^{(l)}(-d/2) = \beta^{(r)}(d/2)$. One then concludes from Eq. (III.106) that $\alpha_{int}$ and, consequently, the Bychkov-Rashba SOI vanishes for the symmetric structure. On the contrary, when the inversion symmetry is broken by the asymmetric structure of the QW itself (even in absence of the external field) and/or by the presence of an external field for which $V_{ext}(z) \neq V_{ext}(-z)$ (this is the case of a constant electric field, for example), the SOI acquires a finite value. Thus, by tuning the spatial dependence and the strength of an external electric field (for example, by applying gate voltages), the spin of the carriers can be manipulated via the Bychkov-Rashba SOI. This interesting fact is at the heart of many investigations and proposals for spintronic devices (see Sec. V.) and has been experimentally observed. Values of the Bychkov-Rashba parameter $\alpha_{BR}$ inferred from experimental measurements are displayed in Tab. III.2. In the experimental situations, the parameter $\alpha_{BR}$ depends on both the external electric field corresponding to the applied gate voltage $V_g$ and the built in field determined by the electron density $n_s$. Thus, by changing $V_g$ and $n_s$, the Bychkov-Rashba SOI can be tuned within certain range of values.

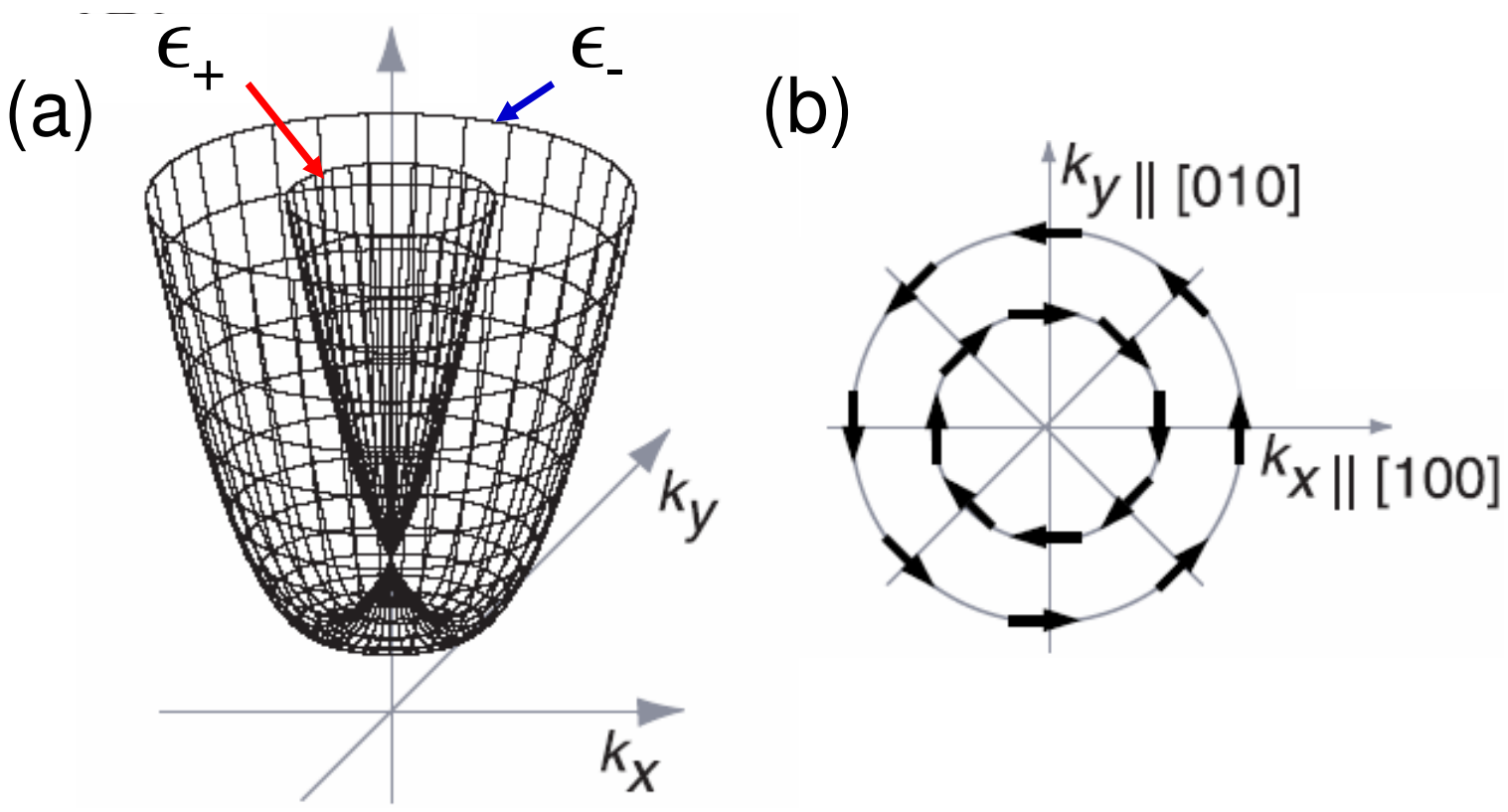

Fig. III.6. (a) Schematics of the $\mathbf{k}_{\parallel}$-dependence of the energy of the spin-splitted subbands $\epsilon_+$ and $\epsilon_-$. (b) Spin orientation at the Fermi surface for the $\sigma = 1$ (inner) and $\sigma = -1$ (outer) subbands. Reprinted with permission from S. D. Ganichev, V. V. Bel'kov, L. E. Golub, E. L. Ivchenko, P. Schneider, S. Giglberger, J. Eroms, J. De Boeck, G. Borghs, W. Wegscheider, D. Weiss, and W. Prettl, *Phys. Rev. Lett.* **92**, 256601 (2004). Copyright (2004) by the American Physical Society.

The Bychkov-Rashba SOI in Eq. (III.110) can be interpreted as the Zeeman-like interaction,

$$\mathrm{H}_{BR} = \mu_B \boldsymbol{\sigma} \cdot \mathbf{B}_{eff}(\mathbf{k}) \tag{III.111}$$

of the electron spin with the effective, the $\mathbf{k}$-dependent magnetic field (Ganichev *et al.*, 2004; Giglberger *et al.*, 2007),

$$\mathbf{B}_{eff}(\mathbf{k}) = \frac{1}{\mu_B} \alpha_{BR}(-k_y, k_x, 0), \tag{III.112}$$

which defines the spin quantization axis. Therefore, one can conclude that in presence of the Bychkov-Rashba SOI the spin orientation of an eigenstate is determined by the direction of $\mathbf{B}_{eff}$, i.e., $\sigma(-\sin\varphi, \cos\varphi, 0)$ for parallel ($\sigma = 1$) and antiparallel ($\sigma = -1$) spins.[54] Here we have used the in-plane polar coordinates $k_x = k_{\parallel} \cos\varphi$, $k_y = k_{\parallel} \sin\varphi$. Due to the SOI, the spin degenerate subband of the QW splits into two subbands. The Zeeman-like energy splitting is given by $\Delta\epsilon_0 = 2\mu_B |\mathbf{B}_{eff}| = 2\alpha_{BR} |\mathbf{k}_{\parallel}|$. The energies of the spin-splitted subbands are therefore given by (Winkler, 2003, 2004b)

$$\epsilon_\sigma(\mathbf{k}) = \epsilon_0(\mathbf{k}) + \sigma \alpha_{BR} k_{\parallel}; \; \sigma = \pm 1. \tag{III.113}$$

Schematics of the energy $\epsilon_\sigma(\mathbf{k})$ as a function of $k_x$ and $k_y$ is shown in Fig. (III.6) for the case of $\epsilon_0(\mathbf{k}) \sim k_{\parallel}^2$. In particular one can see that the energy splitting is isotropic.[55]

Estimations of the values of the Bychkov-Rashba parameter $\alpha_{BR}$ are given in Sec. F., where a more sophisticated variation of the Kane model is presented.

[54] In particular, since $\mathbf{B}_{eff} \cdot \mathbf{k}_{\parallel} = 0$, the spin is perpendicular to $\mathbf{k}_{\parallel}$ [see Fig. III.6(b)].

[55] Anisotropies may appear, however, when including higher order momentum terms in the SOI. To treat this problem one has to go beyond the Kane model.

### E.2 Spin-orbit related effects in symmetric structures

We have seen in Sec. E. that for structures with inversion symmetry the average value $\alpha_{_{BR}}$ of the Bychkov-Rashba parameter vanishes. However, even for inversion symmetric structures one may still observe spin-orbit related effects, as a consequence of intersubband coupling (Bernardes *et al.*, 2006). Unlike the average $\alpha_{_{BR}}$, the matrix elements of $\alpha(z)$ [see Eq. (III.103)] between different subbands are, in general, different from zero (even in inversion symmetric structures), having effects on the subband energies and the effective mass.

Consider a symmetric quantum well in the absence of external fields. In such a case, only the interface term $\alpha_{int}(z)$ contributes to the Bychkov-Rashba SOI and Eqs. (III.97) and (III.106) reduce to,

$$\mathrm{H}_{_{BR}} = \alpha_{int}(z)(k_x\sigma_y - k_y\sigma_x), \tag{III.114}$$

and

$$\alpha_{int}(z) = \Delta\beta[\delta(z + d/2) - \delta(z - d/2)], \tag{III.115}$$

respectively. Here $\Delta\beta = \beta^{(c)} - \beta^{(l)} = \beta^{(c)} - \beta^{(r)}$, in virtue of the symmetry of the QW.

Assume now that there are two subbands $\epsilon_0(\mathbf{k})$ and $\epsilon_1(\mathbf{k})$ in the QW (in absence of SOI) with $|g_0\sigma\rangle$ and $|g_1\sigma\rangle$ the spin-degenerate eigenvectors of the ground and excited states, respectively. With the help of Eq. (III.95), we can compute the matrix elements of $\mathrm{H}_c$ between the subbands of the QW. The result is

$$\langle g_i\sigma|\mathrm{H}_c|g_j\sigma'\rangle = \epsilon_i(\mathbf{k})\delta_{ij}\delta_{\sigma\sigma'} - i\sigma(1-\delta_{ij})(1-\delta_{\sigma\sigma'})\langle\alpha_{int}\rangle_{ij}(k_x - i\sigma k_y)\ ;\ \ i,j = 0,1. \tag{III.116}$$

Here $\delta_{ij}$ is the Kronecker delta function, $\langle\alpha_{int}\rangle_{ij} = \langle g_i|\alpha_{int}(z)|g_j\rangle$, $|\sigma\rangle = |\uparrow\rangle, |\downarrow\rangle$, and $\sigma = \pm 1$ ($\sigma_z|\sigma\rangle = \sigma|\sigma\rangle$). It is convenient to use the basis ordering $\{|g_0\uparrow\rangle, |g_1\downarrow\rangle, |g_1\uparrow\rangle, |g_0\downarrow\rangle\}$, for which the projected Hamiltonian $\mathrm{H}_{cp}$ becomes block diagonal [see Eq. (III.116)],

$$\mathrm{H}_{cp} = \begin{pmatrix} \epsilon_0(\mathbf{k}) & -i\langle\alpha_{int}\rangle_{01}k_- & 0 & 0 \\ i\langle\alpha_{int}\rangle_{10}k_+ & \epsilon_1(\mathbf{k}) & 0 & 0 \\ 0 & 0 & \epsilon_1(\mathbf{k}) & -i\langle\alpha_{int}\rangle_{10}k_- \\ 0 & 0 & i\langle\alpha_{int}\rangle_{01}k_+ & \epsilon_0(\mathbf{k}) \end{pmatrix}. \tag{III.117}$$

From the upper-left block one obtains the eigenvalues (Bernardes *et al.*, 2006),

$$\epsilon_\pm(\mathbf{k}) = \frac{1}{2}\left([\epsilon_0(\mathbf{k}) + \epsilon_1(\mathbf{k})] \pm \sqrt{[\epsilon_0(\mathbf{k}) - \epsilon_1(\mathbf{k})]^2 + |\langle\alpha_{int}\rangle_{01}|^2 k_\parallel^2}\right), \tag{III.118}$$

whose corresponding eigenvectors are (Bernardes *et al.*, 2006)

$$|\psi_u\rangle_+ = \sin(\xi/2)|g_0\uparrow\rangle + e^{i\vartheta}\cos(\xi/2)|g_1\downarrow\rangle, \tag{III.119}$$

$$|\psi_u\rangle_- = \cos(\xi/2)|g_0\uparrow\rangle - e^{i\vartheta}\sin(\xi/2)|g_1\downarrow\rangle. \tag{III.120}$$

Here we have used the notation, $\cos(\xi) = [\epsilon_0(\mathbf{k}) - \epsilon_1(\mathbf{k})]/\sqrt{[\epsilon_0(\mathbf{k}) - \epsilon_1(\mathbf{k})]^2 + |\langle\alpha_{int}\rangle_{01}|^2 k_\parallel^2}$, and, $e^{i\vartheta} = ik_+/k_\parallel$. For the lower-right block the energies are the same as in Eq. (III.118) but with eigenvectors (Bernardes *et al.*, 2006)

$$|\psi_l\rangle_+ = \cos(\xi/2)|g_1\uparrow\rangle + e^{i\vartheta}\sin(\xi/2)|g_0\downarrow\rangle, \tag{III.121}$$

$$|\psi_l\rangle_- = \sin(\xi/2)|g_1\uparrow\rangle - e^{i\vartheta}\cos(\xi/2)|g_0\downarrow\rangle. \tag{III.122}$$

In the limit case $[\epsilon_0(\mathbf{k}) - \epsilon_1(\mathbf{k})] \ll |\langle\alpha_{int}\rangle_{01}|k_\parallel$ one can rewrite Eq. (III.118) in the approximate form

$$\epsilon_\pm(\mathbf{k}) \approx \frac{1}{2}\left([\epsilon_0(\mathbf{k}) + \epsilon_1(\mathbf{k})] \pm |\langle\alpha_{int}\rangle_{01}|k_\parallel\right), \tag{III.123}$$

which resembles the energy spectrum in Eq. (III.113). On the other hand, when $|\langle\alpha_{int}\rangle_{01}|k_\parallel \ll [\epsilon_0(\mathbf{k}) - \epsilon_1(\mathbf{k})]$, one obtains

$$\epsilon_\pm(\mathbf{k}) \approx \epsilon_{0,1}(\mathbf{k}) \pm \frac{|\langle\alpha_{int}\rangle_{01}|^2 k_\parallel^2}{4[\epsilon_0(\mathbf{k}) - \epsilon_1(\mathbf{k})]}. \tag{III.124}$$

Denoting by $\epsilon_{iz}$ the $z$ component of the energy of the $i$th subband, we can rewrite Eq. (III.124) as

$$\epsilon_\pm(\mathbf{k}) \approx \epsilon_{(0,1)z} \pm \frac{\hbar^2 k_\parallel^2}{2m^*_{\parallel\pm}}, \tag{III.125}$$

where

$$\frac{1}{m^*_{\parallel\pm}} = \frac{1}{m^*}\left(1 \pm \frac{\varepsilon_{so}}{\epsilon_{0z} - \epsilon_{1z}}\right);\ \varepsilon_{so} = \frac{m^*|\langle\alpha_{int}\rangle_{01}|^2}{2\hbar^2}. \tag{III.126}$$

Thus, the action of the intersubband SOI produces an effective mass anisotropy in which the in-plane effective masses of the ground and first excited subbands are reduced and increased, respectively, with respect to their value in the $z$ direction. The matrix elements $\langle\alpha_{int}\rangle_{01}$ are, in general, different from zero, even if the well is perfectly symmetric. For the symmetric well the ground $g_0(z)$ and excited $g_1(z)$ wave functions are even and odd, respectively. Therefore, one obtains from Eq. (III.115) the relation

$$\langle\alpha_{int}\rangle_{01} = \Delta\beta[g_0(-d/2)g_1(-d/2) - g_0(d/2)g_1(d/2)] = -2\Delta\beta\ g_0(d/2)\ g_1(d/2) \neq 0. \tag{III.127}$$

For the case of an infinite QW, Bernardes *et al.* (2006) have found

$$\langle\alpha_{int}\rangle_{01} = \beta^{(c)}\frac{4\hbar^2}{d}\left(\frac{\pi}{d}\right)^2, \tag{III.128}$$

where $d$ is the well width, and $\beta^{(c)}$ is given by Eq. (III.109). For a well width $d = 12$nm, the parameter $\langle\alpha_{int}\rangle_{01}$ acquires the values of 14 meVÅ, 840 meVÅ, and 5600 meVÅfor the case of GaAs, InAs, and InSb QWs, respectively (Bernardes *et al.*, 2006).[56] Using these values one can see from Eq. (III.126) that the mass renormalization due to the inter-subband SOI is quite small for the case of GaAs. However, for InSb quantum wells, in which, typically, only the lower branch $\epsilon_-(\mathbf{k})$ is occupied, the in-plane effective mass is reduced by $20\%$ compared to the bulk value $m^*$ (Bernardes *et al.*, 2006). This reduction should produce sizable effects on physical quantities such as the cyclotron frequency and the mobility of the carriers.

[56] We note that the values of $\langle\alpha_{int}\rangle_{01}$ should not be directly compared to the Bychkov-Rashba parameter $\alpha_{BR}$ since they represent different quantities and influence the energy spectrum in different ways.

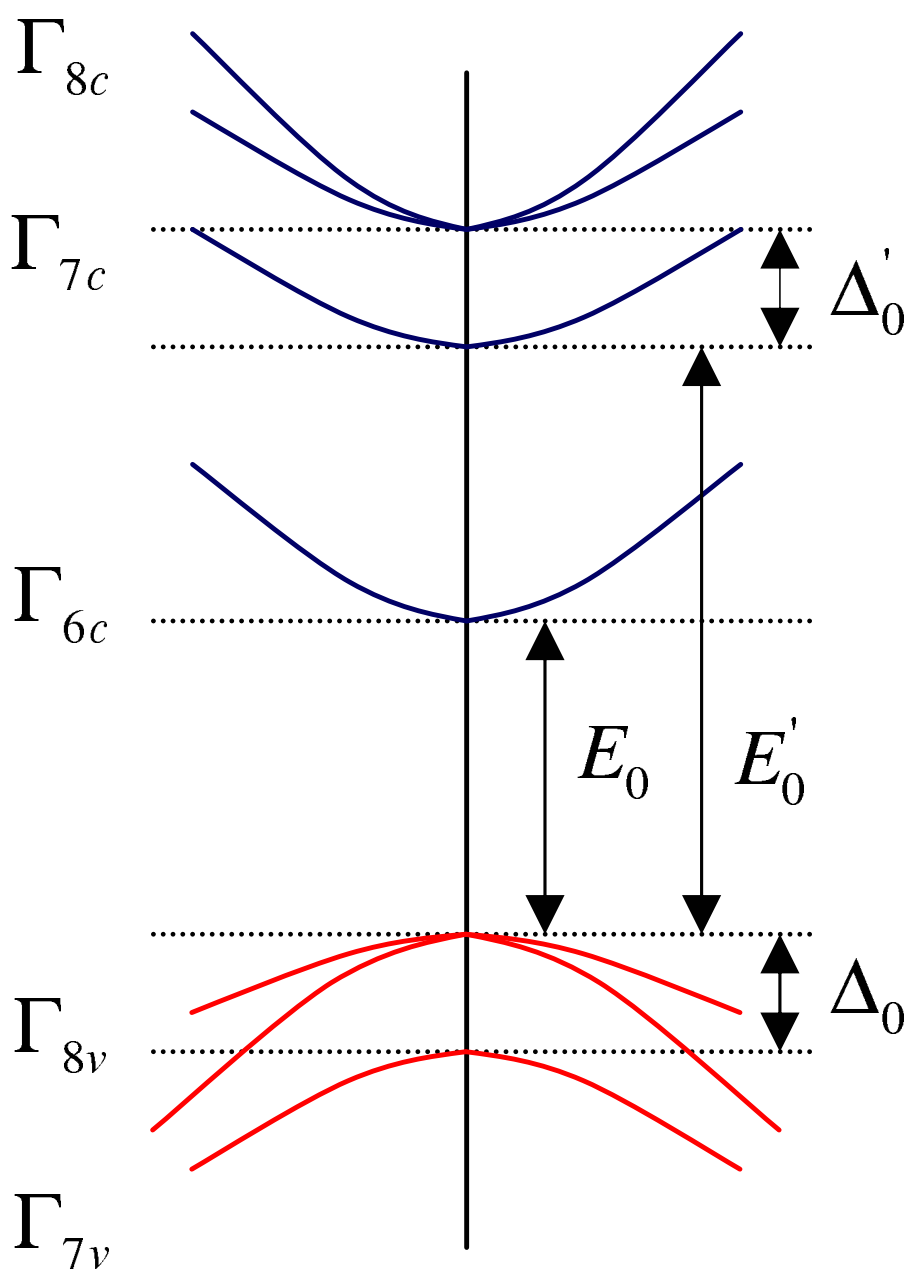

Fig. III.7. Schematics of the band structure corresponding to the extended Kane model.

## F. Dresselhaus spin-orbit interaction

### F.1 Spin-orbit interaction in systems with bulk inversion asymmetry

We have seen in Sec. E. that within the Kane model the Bychkov-Rashba SOI for the conduction electrons [see Eq. (III.97)] is linear in the momentum.[57] When the interaction with other bands is taken into account, new terms, non-linear in the momentum, appear in the SOI. A model that can capture this picture is the so-called extended Kane model (Rössler, 1984; Mayer and Rössler, 1991; Hermann and Weisbuch, 1977; Pfeffer and Zawadzki, 1990; Zawadzki and Pfeffer, 2004) in which in addition to the $\Gamma_{6c}$, $\Gamma_{8v}$, and $\Gamma_7$ bands included in the standard Kane model, the 6 $p$-like higher energy conduction bands are considered. The antibonding $p$-like conduction bands consist of a quadruplet $\Gamma_{8c}$ and a doublet $\Gamma_{7c}$ (see Fig. III.7). In order to account for these bands, the 8 dimensional *intelligent basis* of the standard Kane model has to be extended to a 14 dimensional basis following the same procedure discussed in Sec. D.2. The new basis of the extended Kane model is presented in Tab. III.3. Note that the basis functions of the valence and $p$-like conduction bands are similar. However, the conduction bands contain the new antibonding $p$-like orbitals $|X'\rangle$, $|Y'\rangle$, and $|Z'\rangle$.

For simplicity, we consider the case of zero magnetic field. Within the Kane model, the only non-vanishing momentum matrix elements are those given in Eq. (III.28), equal to $P_0$. In the extended Kane model new matrix elements appear. It turns out that, because of symmetry reasons, only a few of them are different from zero. Thus, in addition to $P_0$, we have now the new parameters,

$$P_1=\frac{i\hbar}{m_0}\langle X'|p_x|S\rangle=\frac{i\hbar}{m_0}\langle Y'|p_y|S\rangle=\frac{i\hbar}{m_0}\langle Z'|p_z|S\rangle, \tag{III.129}$$

$$Q=\frac{\hbar}{m_0}\langle X|p_y|Z'\rangle=-\frac{\hbar}{m_0}\langle X'|p_y|Z\rangle, \tag{III.130}$$

and[58]

$$\triangle^-=-\frac{3\hbar}{4m_0^2c^2}\langle X|[(\boldsymbol{\nabla}V_0)\times\mathbf{p}]_y|Z'\rangle=\frac{3\hbar}{4m_0^2c^2}\langle Z|[(\boldsymbol{\nabla}V_0)\times\mathbf{p}]_y|X'\rangle, \tag{III.131}$$

characterizing the only non-vanishing momentum matrix elements involving the antibonding, $p$-like conduction bands. In particular, $P_1\neq 0$ and $\triangle^-\neq 0$ in crystals whose structure does not have a center of inversion symmetry (e.g., zinc-blende structures) but they vanish in centrosymmetric structures such as diamond.

[57] Within the Kane model the Bychkov-Rashba SOI for the carriers in the valence band is already non-linear in the momentum. For more details see (Winkler, 2000, 2003).

Tab. III.3. Basis functions of the extended Kane model.

| Band | $\vert m\rangle$ | Basis functions | Energy |
|---|---|---|---|
| $\Gamma_{6c}$ | $\vert\frac{1}{2}\ \frac{1}{2}\rangle$ | $\vert S\uparrow\rangle$ | 0 |
| | $\vert\frac{1}{2}\ -\frac{1}{2}\rangle$ | $\vert S\downarrow\rangle$ | 0 |
| $\Gamma_{8v}$ | $\vert\frac{3}{2}\ \frac{3}{2}\rangle$ | $-\frac{1}{\sqrt{2}}\vert(X+iY)\uparrow\rangle$ | $-E_0$ |
| | $\vert\frac{3}{2}\ \frac{1}{2}\rangle$ | $\sqrt{\frac{2}{3}}\vert Z\uparrow\rangle-\frac{1}{\sqrt{6}}\vert(X+iY)\downarrow\rangle$ | $-E_0$ |
| | $\vert\frac{3}{2}\ -\frac{1}{2}\rangle$ | $\sqrt{\frac{2}{3}}\vert Z\downarrow\rangle+\frac{1}{\sqrt{6}}\vert(X-iY)\uparrow\rangle$ | $-E_0$ |
| | $\vert\frac{3}{2}\ -\frac{3}{2}\rangle$ | $\frac{1}{\sqrt{2}}\vert(X-iY)\downarrow\rangle$ | $-E_0$ |
| $\Gamma_{7v}$ | $\vert\frac{1}{2}\ \frac{1}{2}\rangle$ | $-\frac{1}{\sqrt{3}}\vert Z\uparrow\rangle-\frac{1}{\sqrt{3}}\vert(X+iY)\downarrow\rangle$ | $-(E_0+\triangle_0)$ |
| | $\vert\frac{1}{2}\ -\frac{1}{2}\rangle$ | $\frac{1}{\sqrt{3}}\vert Z\downarrow\rangle-\frac{1}{\sqrt{3}}\vert(X-iY)\uparrow\rangle$ | $-(E_0+\triangle_0)$ |
| $\Gamma_{8c}$ | $\vert\frac{3}{2}\ \frac{3}{2}\rangle'$ | $-\frac{1}{\sqrt{2}}\vert(X'+iY')\uparrow\rangle$ | $E_0'-E_0+\triangle_0$ |
| | $\vert\frac{3}{2}\ \frac{1}{2}\rangle'$ | $\sqrt{\frac{2}{3}}\vert Z'\uparrow\rangle-\frac{1}{\sqrt{6}}\vert(X'+iY')\downarrow\rangle$ | $E_0'-E_0+\triangle_0$ |
| | $\vert\frac{3}{2}\ -\frac{1}{2}\rangle'$ | $\sqrt{\frac{2}{3}}\vert Z'\downarrow\rangle+\frac{1}{\sqrt{6}}\vert(X'-iY')\uparrow\rangle$ | $E_0'-E_0+\triangle_0$ |
| | $\vert\frac{3}{2}\ -\frac{3}{2}\rangle'$ | $\frac{1}{\sqrt{2}}\vert(X'-iY')\downarrow\rangle$ | $E_0'-E_0+\triangle_0$ |
| $\Gamma_{7c}$ | $\vert\frac{1}{2}\ \frac{1}{2}\rangle'$ | $-\frac{1}{\sqrt{3}}\vert Z'\uparrow\rangle-\frac{1}{\sqrt{3}}\vert(X'+iY')\downarrow\rangle$ | $E_0'-E_0$ |
| | $\vert\frac{1}{2}\ -\frac{1}{2}\rangle'$ | $\frac{1}{\sqrt{3}}\vert Z'\downarrow\rangle-\frac{1}{\sqrt{3}}\vert(X'-iY')\uparrow\rangle$ | $E_0'-E_0$ |

[58] Eq. (III.130) also applies to the case of all the other equivalent combinations, say, $Q=\frac{\hbar}{m_0}\langle Y|p_x|Z'\rangle=-\frac{\hbar}{m_0}\langle Y'|p_x|Z\rangle$, etc. A similar situation occurs for the equivalent combinations of Eq. (III.131), say, $\triangle^-=-\frac{3\hbar}{4m_0^2c^2}\langle Y|[(\boldsymbol{\nabla}V_0)\times\mathbf{p}]_z|X'\rangle=\frac{3\hbar}{4m_0^2c^2}\langle X|[(\boldsymbol{\nabla}V_0)\times\mathbf{p}]_z|Y'\rangle$, etc.

Tab. III.4. Band structure parameters of the extended Kane model, obtained from a 40-band tight-binding model by Jancu *et al.* (2005). The parameter $C$ describes the influence of remote bands on the effective mass of conduction electrons (Cardona *et al.*, 1988; Hermann and Weisbuch, 1977; Pfeffer and Zawadzki, 1996). Available experimental values for $E_0$, $E_0'$, and $\triangle_0$ have been included for comparison.

| | AlAs | AlP | AlSb | GaAs | GaP | GaSb | InAs | InP | InSb |
|---|---|---|---|---|---|---|---|---|---|
| $E_0$ (eV) | 3.130 | 3.63 | 2.38 | 1.519 | 2.895 | 0.81 | 0.418 | 1.423 | 0.235 |
| (expt) | (3.099)[1] | (3.63)[1] | (2.386)[1] | (1.519)[1] | (2.886)[1] | (0.812)[1] | (0.417)[1] | (1.4236)[1] | (0.235)[1] |
| $E_0'$ (eV) | 4.55 | 4.78 | 3.53 | 4.54 | 4.38 | 3.11 | 4.48 | 4.78 | 3.18 |
| (expt) | - | - | - | (4.6)[2] | - | (3.19)[2] | - | (4.8)[2] | (3.15)[2] |
| $\triangle_0$ (eV) | 0.3 | 0.06 | 0.67 | 0.341 | 0.08 | 0.76 | 0.38 | 0.107 | 0.81 |
| (expt) | (0.28)[1] | (0.07)[1] | (0.676)[1] | (0.341)[1] | (0.08)[1] | (0.76)[1] | (0.39)[1] | (0.108)[1] | (0.81)[1] |
| $\triangle_0'$ (eV) | 0.15 | 0.04 | 0.24 | 0.2 | 0.09 | 0.33 | 0.31 | 0.19 | 0.46 |
| $\triangle^-$ (eV) | -0.19 | -0.03 | -0.41 | -0.17 | 0.04 | -0.4 | -0.05 | 0.11 | -0.26 |
| $P_0$ (eVÅ) | 8.88 | 9.51 | 8.57 | 9.88 | 9.53 | 9.69 | 9.01 | 8.45 | 9.63 |
| $P_1$ (eVÅ) | 0.34 | 0.19 | 0.51 | 0.41 | 0.36 | 1.34 | 0.66 | 0.34 | 1.2 |
| $Q$ (eVÅ) | 8.07 | 8.10 | 7.8 | 8.68 | 8.49 | 8.25 | 7.72 | 7.88 | 7.83 |
| $C$ | -1.07 | -1.36 | -0.72 | -1.76 | -1.77 | -1.7 | -0.85 | -1.33 | -1.19 |

[1] From (Vurgaftman *et al.*, 2001) [2] From (Madelung, 1996)

In addition to $E_0$ and $\triangle_0$, in the extended Kane model there is an energy gap $E_0'$ between the top valence band and the $\Gamma_{7c}$ bands at the $\Gamma$ point and a spin-orbit energy splitting $\triangle_0'$ between the $\Gamma_{7c}$ and $\Gamma_{8c}$ bands (see Fig. III.7).

Note that, unlike in the works of Winkler (2003); Pfeffer and Zawadzki (1990); Zawadzki and Pfeffer (2004), here we have adopted the definitions by Cardona *et al.* (1988) and Jancu *et al.* (2005), in which the parameters $P_0$, $P_1$, $Q$, $\triangle_0$, $\triangle_0'$, and $\triangle^-$ are real quantities. The values of the band structure parameters obtained by Jancu *et al.* (2005) from a 40-band tight-binding model are listed in Tab. III.4 for several typical semiconductors. The effects of $\triangle^-$ on the band effective mass and Bychkov-Rashba SOI are usually negligible (Cardona *et al.*, 1986b, 1988). However, the effects of $\triangle^-$ lead to a significant contribution to the Dresselhaus SOI and have to be considered for a quantitative description (Pfeffer and Zawadzki, 1999; Zawadzki and Pfeffer, 2004; Jancu *et al.*, 2005).

In a heterostructure the momentum matrix elements $P_0$, $P_1$, and $Q$ are position dependent, since they are material specific. Nevertheless, it is a usual practice to consider them as being constant along the heterostructure.[59] In what follows we adopt such an approximation (note that, although not mentioned, we have already made this assumption in Sec. E.). For the case of a QW one takes in both well and barrier regions, the values of $P_0$, $P_1$, and $Q$ corresponding to the material in the well.

Using the 14 dimensional *intelligent* basis and following the procedure discussed in Secs. D.2 and E. one obtains the Hamiltonian for the envelope functions. For the case of a quantum well

[59] The good agreement between theory and experiment seems to justify such an approximation.

grown in the $z$ direction, the result is

$$\mathrm{H} = \begin{pmatrix} \mathrm{H}_c & \mathrm{H}_{cv} & \mathrm{H}_{cc'} \\ \mathrm{H}_{vc} & \mathrm{H}_v & \mathrm{H}_{vc'} \\ \mathrm{H}_{c'c} & \mathrm{H}_{c'v} & \mathrm{H}_{c'} \end{pmatrix}, \tag{III.132}$$

where $\mathrm{H}_c$, $\mathrm{H}_{cv}$, $\mathrm{H}_{vc}$, and $\mathrm{H}_v$ are the same block matrices of the standard Kane model [see Sec. E. and Eqs. (III.67) - (III.70)]. The other block matrices are (Winkler, 2003; Rössler, 1984)

$$\mathrm{H}_{cc'} = \begin{pmatrix} -\frac{i}{\sqrt{2}}P_1 k_+ & i\sqrt{\frac{2}{3}}\frac{P_1}{\hbar}p_z & \frac{i}{\sqrt{6}}P_1 k_- & 0 & -\frac{i}{\sqrt{3}}\frac{P_1}{\hbar}p_z & -\frac{i}{\sqrt{3}}P_1 k_- \\ 0 & -\frac{i}{\sqrt{6}}P_1 k_+ & i\sqrt{\frac{2}{3}}\frac{P_1}{\hbar}p_z & \frac{i}{\sqrt{2}}P_1 k_- & -\frac{i}{\sqrt{3}}P_1 k_+ & +\frac{i}{\sqrt{3}}\frac{P_1}{\hbar}p_z \end{pmatrix}, \tag{III.133}$$

$$\mathrm{H}_{vc'} = \begin{pmatrix} \frac{i}{3}\triangle^- & \frac{i}{\sqrt{3}}Qk_+ & \frac{i}{\sqrt{3}}\frac{Q}{\hbar}p_z & 0 & -\frac{i}{\sqrt{6}}Qk_+ & -i\sqrt{\frac{2}{3}}\frac{Q}{\hbar}p_z \\ -\frac{i}{\sqrt{3}}Qk_- & \frac{i}{3}\triangle^- & 0 & \frac{i}{\sqrt{3}}\frac{Q}{\hbar}p_z & 0 & \frac{i}{\sqrt{2}}Qk_+ \\ -\frac{i}{\sqrt{3}}\frac{Q}{\hbar}p_z & 0 & \frac{i}{3}\triangle^- & -\frac{i}{\sqrt{3}}Qk_+ & -\frac{i}{\sqrt{2}}Qk_- & 0 \\ 0 & -\frac{i}{\sqrt{3}}\frac{Q}{\hbar}p_z & \frac{i}{\sqrt{3}}Qk_- & \frac{i}{3}\triangle^- & -i\sqrt{\frac{2}{3}}\frac{Q}{\hbar}p_z & \frac{i}{\sqrt{6}}Qk_- \\ \frac{i}{\sqrt{6}}Qk_- & 0 & \frac{i}{\sqrt{2}}Qk_+ & i\sqrt{\frac{2}{3}}\frac{Q}{\hbar}p_z & -\frac{2i}{3}\triangle^- & 0 \\ i\sqrt{\frac{2}{3}}\frac{Q}{\hbar}p_z & -\frac{i}{\sqrt{2}}Qk_- & 0 & -\frac{i}{\sqrt{6}}Qk_+ & 0 & -\frac{2i}{3}\triangle^- \end{pmatrix}, \tag{III.134}$$

and

$$\mathrm{H}_{c'} = diag[E_{c'}(z)+\triangle_0', E_{c'}(z)+\triangle_0', E_{c'}(z)+\triangle_0', E_{c'}(z)+\triangle_0', E_{c'}(z), E_{c'}(z)], \tag{III.135}$$

where

$$E_{c'}(z) = V_{ext}(z) + E_c(z) + [E_0'(z) - E_0(z)]. \tag{III.136}$$

From the Hermiticity of H one deduces that the matrix $\mathrm{H}_{c'c}$ can be obtained from $\mathrm{H}_{cc'}$ by transposition and the substitutions $k_\pm \to k_\mp$ and $P_1 \to -P_1$, while $\mathrm{H}_{c'v}$ results from $\mathrm{H}_{vc'}$ by the substitutions $k_\pm \to k_\mp$, $Q \to -Q$ and $\triangle^- \to -\triangle^-$.

Following the folding-down procedure described in Sec. D.2 we can obtain the Hamiltonian for the electrons in the lowest conduction subbands as

$$\tilde{\mathrm{H}}_c = \mathrm{H}_c - \mathrm{H}_{ur}(\mathrm{H}_{lr} - \mathrm{E})^{-1}\mathrm{H}_{ll}, \tag{III.137}$$

with the upper-right (ur) and lower-left (ll) block matrices, $\mathrm{H}_{ur} = (\mathrm{H}_{cv}\ \mathrm{H}_{cc'})$ and $\mathrm{H}_{ll} = (\mathrm{H}_{vc}^T\ \mathrm{H}_{c'c}^T)^T$, respectively. The lower right (lr) matrix is given by

$$\mathrm{H}_{lr} = \begin{pmatrix} \mathrm{H}_v & \mathrm{H}_{vc'} \\ \mathrm{H}_{c'v} & \mathrm{H}_{c'} \end{pmatrix}. \tag{III.138}$$

The inversion of the $(12 \times 12)$ matrix $(\mathrm{H}_{lr} - \mathrm{E})^{-1}$ can be quite cumbersome from the analytical point of view. Therefore it is convenient to find an approximate expression for $(\mathrm{H}_{lr} - \mathrm{E})^{-1}$ by using the scheme described at the end of Sec. D.2 [see, for example, Eq. (III.80)]. In the present case we can chose the matrix $A$ as a diagonal matrix whose matrix elements are those in the diagonal of $(\mathrm{H}_{lr} - E)$. Since these elements contain the largest energy scales in the system, the series expansion will converge fast. We keep only up to the second order in the expansion,[60] i.e., $(\mathrm{H}_{lr} - \mathrm{E})^{-1} \approx \mathrm{A}^{-1} + \mathrm{A}^{-1}\mathrm{D}\mathrm{A}{-}1 + (\mathrm{A}^{-1}\mathrm{D})^2\mathrm{A}^{-1}$, where

$$\mathrm{A} = \begin{pmatrix} \mathrm{H}_v - \mathrm{E} & 0 \\ 0 & \mathrm{H}_{c'} - \mathrm{E} \end{pmatrix}; \ \mathrm{D} = -\begin{pmatrix} 0 & \mathrm{H}_{vc'} \\ \mathrm{H}_{c'v} & 0 \end{pmatrix}. \tag{III.139}$$

Using Eqs. (III.137) - (III.139) one finds, after some algebra,[61] the following expression,

$$\tilde{\mathrm{H}}_c = -\frac{\hbar^2}{2}\frac{d}{dz}\left[\frac{1}{m^*(z,E)}\frac{d}{dz}\right] + \frac{\hbar^2 k_\parallel^2}{2m^*(z,E)} + E_c(z) + V_{ext}(z) + \mathrm{H}_{BR} + \mathrm{H}_D + \mathrm{H}'_D. \tag{III.140}$$

In addition to the corresponding value obtained within the standard Kane model [see Eq. (III.96)], the effective mass contains now the correction (Pfeffer and Zawadzki, 1999; Zawadzki and Pfeffer, 2004),

$$\delta\left(\frac{1}{m^*}\right) = -\frac{2{P_1}^2}{3\hbar^2}\left(\frac{1}{\tilde{E}_{c'}(z)} + \frac{2}{\tilde{F}_{c'}(z)}\right) - \frac{8P_1P_0\triangle^-}{9\hbar^2}\left(\frac{1}{\tilde{E}_{c'}(z)\tilde{F}_v(z)} - \frac{2}{\tilde{E}_v(z)\tilde{F}_{c'}(z)}\right), \tag{III.141}$$

Where we have introduced the notations

$$\tilde{E}_{c'}(z) = E_{c'}(z) - E, \ \tilde{E}_v(z) = E_v(z) - E, \tag{III.142}$$

and

$$\tilde{F}_{c'}(z) = \tilde{E}_{c'}(z) + \triangle'_0(z), \ \tilde{F}_v(z) = \tilde{E}_v(z) - \triangle_0(z). \tag{III.143}$$

The Bychkov-Rashba SOI has the same form as for the case of the standard Kane model [see Eq. (III.97)] but some corrections to $\alpha(z)$ appears due to the inclusion of the interaction with the $\Gamma_{7c}$ and $\Gamma_{8c}$ bands. The new expression for $\alpha(z)$ includes a correction $\delta\beta(z)$ to $\beta(z)$, which is given by

$$\delta\beta = \frac{P_1^2}{3}\left(\frac{1}{\tilde{F}_{c'}(z)} - \frac{1}{\tilde{E}_{c'}(z)}\right) - \frac{2P_0P_1\triangle^-}{9}\left(\frac{1}{\tilde{E}_v(z)\tilde{F}_{c'}(z)} + \frac{2}{\tilde{E}_{c'}(z)\tilde{F}_v(z)}\right). \tag{III.144}$$

The term $\mathrm{H}_D$ in Eq. (III.140) corresponds to the linear in $k_x$ and $k_y$ Dresselhaus SOI (Dresselhaus, 1955) and can be written as (Pfeffer and Zawadzki, 1999; Zawadzki and Pfeffer, 2004)

$$\mathrm{H}_D = (k_x\sigma_x - k_y\sigma_y)\frac{d}{dz}\left(\gamma(z)\frac{d}{dz}\right), \tag{III.145}$$

[60] In Sec. E. we considered only up to first order in the series expansion. Here we need to take the second order for incorporating the effects of the matrix elements $\triangle^-$.

[61] Fortunately, the tedious involved algebra can be efficiently performed by using softwares for analytical calculations such as Mathematica or Maple.

with the Dresselhaus spin-orbit parameter given by

$$\begin{aligned}\gamma(z) &= -\frac{4P_0 P_1 Q}{3}\left(\frac{1}{\tilde{E}_v(z)\tilde{E}_{c'}(z)} - \frac{1}{\tilde{F}_v(z)\tilde{F}_{c'}(z)}\right)\\ &+ \frac{4\triangle^- Q}{9}\left[\frac{P_1^2}{\tilde{E}_{c'}(z)\tilde{F}_{c'}(z)}\left(\frac{1}{\tilde{E}_v(z)} + \frac{2}{\tilde{F}_v(z)}\right)\right.\\ &- \left.\frac{P_0^2}{\tilde{E}_v(z)\tilde{F}_v(z)}\left(\frac{2}{\tilde{E}_{c'}(z)} + \frac{1}{\tilde{F}_{c'}(z)}\right)\right]. \end{aligned} \tag{III.146}$$

Higher order in $k_x$ and $k_y$ terms related to the bulk inversion asymmetry (BIA) of the system are included in

$$\mathrm{H}'_D = i(k_y^2 - k_x^2)\sigma_z \left[\frac{1}{2}\frac{d\gamma(z)}{dz} + \gamma(z)\frac{d}{dz}\right] + \gamma(z)(k_y\sigma_x - k_x\sigma_y)k_x k_y. \tag{III.147}$$

For the case of a two-dimensional gas confined in the $z$ direction, the $z$ degree of freedom is usually averaged out in a similar way as we did for obtaining Eq. (III.110). As a result one obtains a linearized (i.e., linear in the wave vectors) Dresselhaus SOI from Eq. (III.145), a quadratic in wave vectors contribution from the first term in Eq. (III.147) and the so-called cubic Dresselhaus SOI from the second term in Eq. (III.147). In what follows we consider the case in which the in-plane wave vector is small. We can then neglect the contribution of $\mathrm{H}'_D$ and concentrate in the analysis of $\mathrm{H}_D$.

It is worth remarking that $P_1 \neq 0$ and $\triangle^- \neq 0$ only in non-centrosymmetric crystals. For this reason bulk inversion asymmetry is required for the presence of the Dresselhaus SOI. The Dresselhaus SOI can be interpreted as the SOI induced by the electric field of the dipole resulting from the BIA (e.g., the local Ga-As dipole, in the case of bulk GaAs).

One can obtain simplified (energy-independent) expressions of Eqs. (III.141), (III.144), and (III.146) by following the same expansion technique discussed in Sec. E. The results for the $i$th region of the heterostructure are summarized in Tab. III.5. The expressions in Tab. III.5 are equivalent to the ones reported by Pfeffer and Zawadzki (1999) and Zawadzki and Pfeffer (2004).[62]

In the expression for the effective mass in Tab. III.5 we have included the effects of the parameter $C$, which characterizes the influence of the remote bands on the effective mass in the conduction band. We do not enter here in the details concerning this parameter but we have included it for a better accuracy when computing the effective mass. The origin of $C$ and the far band contribution to the effective mass has been extensively discussed by Cardona *et al.* (1988); Hermann and Weisbuch (1977); Pfeffer and Zawadzki (1996). We have computed the values of $m^*$, $\beta$, $\alpha$, and $\gamma$ from Tabs. III.4 and III.5. The results are shown in Tab. III.6. For the calculation of $\alpha$ we considered a triangular quantum well formed by a heterointerface and a uniform electric field $E = 10^5$ V/cm (de Andrada e Silva *et al.*, 1997; Miller *et al.*, 2003), which is proportional to the carrier concentration $n_s$. For the case of GaAs, such a field corresponds to a concentration $n_s \sim 10^{11}$ cm$^{-1}$ (de Andrada e Silva *et al.*, 1997). Theoretical results for $\gamma$ reported by Jancu *et al.* (2005) are also included in Tab. III.6. Values of the Dresselhaus parameter $\gamma$ deduced from experimental measurements are listed in Tab. III.7.

[62]Note, however, that these authors use different definitions for the band structure parameters than the ones assumed here.

Tab. III.5. Approximate expressions for the parameters $1/m^{*(i)}$, $\beta^{(i)}$, $\alpha^{(i)}(z)$, and $\gamma^{(i)}$, corresponding to the $i$th region of the structure. Here we have used the notation $\tilde{E}_0^{(i)} = E_0'^{(i)} - E_0^{(i)}$. The far band contribution to the effective mass has been included by introducing the parameter $C$.

| Parameter | Series expansion |
|---|---|
| $\frac{1}{m^{*(i)}}$ | $\frac{1+C}{m_0} + \frac{2P_0^2}{3\hbar^2}\left[\frac{2}{E_0^{(i)}} + \frac{1}{E_0^{(i)}+\triangle_0^{(i)}}\right] - \frac{2P_1^2}{3\hbar^2}\left[\frac{1}{\tilde{E}_0^{(i)}} + \frac{2}{\tilde{E}_0^{(i)}+\triangle_0'^{(i)}}\right] + \frac{8P_0P_1\triangle^-}{9\hbar^2}\left[\frac{1}{\tilde{E}_0^{(i)}(E_0^{(i)}+\triangle_0^{(i)})} + \frac{1}{E_0^{(i)}(\tilde{E}_0^{(i)}+\triangle_0'^{(i)})}\right]$ |
| $\beta^{(i)}$ | $\frac{P_0^2}{3}\left[\frac{1}{E_0^{(i)}+\triangle_0^{(i)}} - \frac{1}{E_0^{(i)}}\right] - \frac{P_1^2}{3}\left[\frac{1}{\tilde{E}_0^{(i)}} - \frac{1}{\tilde{E}_0^{(i)}+\triangle_0'^{(i)}}\right] + \frac{2P_1P_0\triangle^-}{9}\left[\frac{1}{E_0^{(i)}(\tilde{E}_0^{(i)}+\triangle_0'^{(i)})} + \frac{2}{\tilde{E}_0^{(i)}(E_0^{(i)}+\triangle_0^{(i)})}\right]$ |
| $\alpha^{(i)}(z)$ | $\frac{1}{3}\left[\frac{P_0^2}{(E_0^{(i)}+\triangle_0^{(i)})^2} - \frac{P_0^2}{(E_0^{(i)})^2} + \frac{P_1^2}{(\tilde{E}_0^{(i)})^2} - \frac{P_1^2}{(\tilde{E}_0^{(i)}+\triangle_0'^{(i)})^2}\right]\frac{dV_{ext}}{dz} - \frac{2P_1P_0\triangle^-}{9}\left[\frac{1}{E_0^{(i)}(\tilde{E}_0^{(i)}+\triangle_0'^{(i)})^2} - \frac{1}{(E_0^{(i)})^2(\tilde{E}_0^{(i)}+\triangle_0'^{(i)})} - \frac{2}{\tilde{E}_0^{(i)}(E_0^{(i)}+\triangle_0^{(i)})^2} + \frac{2}{(\tilde{E}_0^{(i)})^2(E_0^{(i)}+\triangle_0^{(i)})}\right]\frac{dV_{ext}}{dz}$ |
| $\gamma^{(i)}$ | $\frac{4P_0P_1Q}{3}\left[\frac{1}{E_0^{(i)}\tilde{E}_0^{(i)}} - \frac{1}{(E_0^{(i)}+\triangle_0^{(i)})(\tilde{E}_0^{(i)}+\triangle_0'^{(i)})}\right] - \frac{4P_0^2Q\triangle^-}{9E_0^{(i)}(E_0^{(i)}+\triangle_0^{(i)})}\left[\frac{2}{\tilde{E}_0^{(i)}} + \frac{1}{\tilde{E}_0^{(i)}+\triangle_0'^{(i)}}\right] - \frac{4P_1^2Q\triangle^-}{9\tilde{E}_0^{(i)}(\tilde{E}_0^{(i)}+\triangle_0'^{(i)})}\left[\frac{1}{E_0^{(i)}} + \frac{2}{E_0^{(i)}+\triangle_0^{(i)}}\right]$ |

Tab. III.6. Values of the parameters $m^*$, $\beta$, $\alpha$, $\gamma$, and $\gamma_D$ obtained from Tab. III.5 for different semiconductor materials. The calculations were performed by using the band structure parameters listed in Tab. III.4. Experimental values of $m^*$ (Vurgaftman *et al.*, 2001) and theoretical results for $\gamma$, obtained by Jancu *et al.* (2005) within a 40-band tight-binding (TB) model have been included for comparison.

| Parameter | AlAs | AlP | AlSb | GaAs | GaP | GaSb | InAs | InP | InSb |
|---|---|---|---|---|---|---|---|---|---|
| $m^*$ ($m_0$) | 0.159 | 0.163 | 0.131 | 0.067 | 0.136 | 0.041 | 0.023 | 0.080 | 0.0132 |
| $m^*$ ($m_0$) (expt) | 0.15 | 0.22 | 0.14 | 0.067 | 0.13 | 0.039 | 0.026 | 0.0795 | 0.0135 |
| $\beta$ (eV Å$^2$) | 0.81 | 0.14 | 2.62 | 4.01 | 0.26 | 19.92 | 30.91 | 1.13 | 103.25 |
| $\alpha$ (meV Å) | 0.43 | 0.07 | 1.56 | 4.72 | 0.19 | 35.52 | 112.49 | 1.57 | 534.21 |
| $\gamma$ (eV Å$^3$) | 11.55 | 2.11 | 41.50 | 24.45 | -2.42 | 178.51 | 48.63 | -10.34 | 473.61 |
| $\gamma$ (eV Å$^3$) (TB) | 10.6 | 1.9 | 39.3 | 23.6 | -1.4 | 168 | 42.3 | -8.6 | 389 |
| $\gamma_D$ (meV Å) | 7.91 | 1.45 | 28.44 | 16.75 | -1.66 | 122.35 | 33.33 | -7.09 | 324.60 |

Tab. III.7. Values of the Dresselhaus parameter $\gamma$ deduced from experimental measurements.

| System | $\gamma$ (eV Å$^3$) (exp) | Reference |
|---|---|---|
| GaAs | 24.5 | (Marushchak *et al.*, 1983) |
| GaAs | 17.4 - 26 | (Pikus *et al.*, 1988) |
| GaAs | $26.1 \pm 0.9$ | (Dresselhaus *et al.*, 1992) |
| GaAs | $16.5 \pm 3$ | (Jusserand *et al.*, 1995) |
| GaAs | 11 | (Richards *et al.*, 1996) |
| GaAs | 9 | (Krich and Halperin, 2007) |
| GaAs | $28 \pm 4$ | (Miller *et al.*, 2003) |
| GaSb | $\pm 185$ | (Pikus *et al.*, 1988) |
| InGaAs | 24 | (Knap *et al.*, 1996) |
| InP | $\pm(7.3 - 9.5)$ | (Gorelenko *et al.*, 1986) |
| InSb | 225 | (Cardona *et al.*, 1986a) |

Two key parameters in the description of zinc-blende compounds are the energy of the fundamental gap $E_0$ and the effective mass $m^*$.

In order to get a qualitative insight on the general trends in the changes of the SOI parameters when going from one material to another, we investigate how the Bychkov-Rashba $\alpha$ and Dresselhaus parameters $\gamma$ are correlated to the energy gap $E_0$ and effective mass $m^*$ of the different compounds related in Tab. III.6. The dependence of $\alpha$ and $\gamma$ on the energy gap $E_0$ and effective mass $m^*$ are shown in Figs. III.8(a) and (b), respectively. A clear trend of increasing the Bychkov-Rashba parameter when decreasing $E_0$ and/or $m^*$ is observed. The Dresselhaus parameter, on the other hand, exhibits an oscillating behavior. In particular, it has been experimentally observed that both the energy gap $E_0$ (Lautenschlager *et al.*, 1987) and the effective mass $m^*$ (Hazama *et al.*, 1986; Hopkins *et al.*, 1987) in systems such as GaAs decrease when the temperature increases. Consequently, we can expect from our qualitative analysis an increase of the Bychkov-Rashba parameter with the temperature. A more detailed discussion on the temperature dependence of $\alpha$ will be addressed later on in this section.

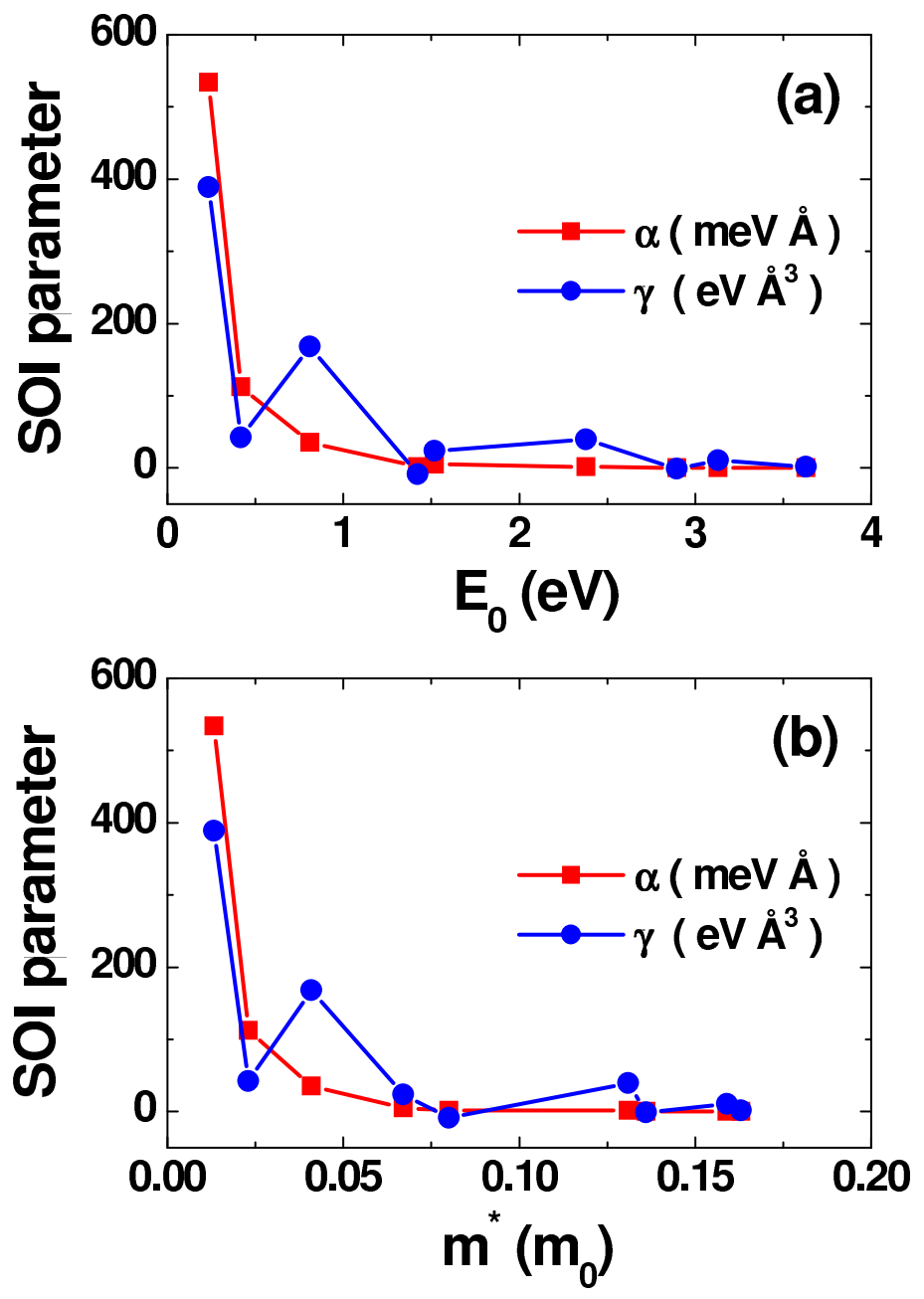

Fig. III.8. SOI parameters $\alpha$ and $\gamma$ as functions of the energy gap $E_0$ (a) and the effective mass $m^*$ (b). The data have been taken from Tab. III.6.

We remark that the values of $\gamma$ are indirectly extracted from the experiments, i.e., by confronting the experimental measurements with their corresponding theoretical estimation. Thus, the use of different theoretical approximations used for extracting the Dresselhaus parameter together with the different experimental techniques used by different authors can lead to rather different values of $\gamma$, as shown in Tab. III.7 [see also (Krich and Halperin, 2007)]. On the other hand different theoretical models predict distinct values of $\gamma$, varying from $8.5\,\text{eV}\text{Å}^3$ (Chantis *et al.*, 2006) to $30\,\text{eV}\text{Å}^3$ (Rössler, 1984) [for a detailed discussion see (Krich and Halperin, 2007)]. Consequently, although there is a qualitative consensus between all the different approximations, the question about the correct value of $\gamma$ is still a controversial issue (Krich and Halperin, 2007). In the case of heterostructures, there is an additional, interface contribution to the Dresselhaus SOI [see Eq. (III.149)] that is absent in bulk systems and is likely to play a role. However, in the theoretical approximations used for extracting $\gamma$ from experimental results as well as in many theoretical estimations of $\gamma$ in heterostructures, the interface contribution is usually neglected. Ignoring such a contribution is not always well justified and can be an additional source of uncertainties when determining the correct value of the Dresselhaus parameter $\gamma$. In fact, when the interface contribution becomes relevant, the energy splitting resulting from the Dresselhaus SOI in a given heterostructure can deviate from its value in the corresponding bulk

system.[63]

Since the Dresselhaus parameter $\gamma(z)$ is piecewise constant, i.e.,

$$\gamma(z) = \gamma^{(l)}\Theta(-z-d/2) + \gamma^{(c)}\Theta(d/2-|z|) + \gamma^{(r)}\Theta(z-d/2), \tag{III.148}$$

the Dresselhaus SOI can be rewritten as,

$$\mathrm{H}_D = (k_x\sigma_x - k_y\sigma_y)\left(\gamma(z)\frac{d^2}{dz^2} + \gamma_{int}(z)\frac{d}{dz}\right), \tag{III.149}$$

with an interface contribution determined by

$$\gamma_{int}(z) = \gamma^{(c)}[\delta(z+d^+/2) - \delta(z-d^-/2)] - \gamma^{(l)}\delta(z+d^-/2) + \gamma^{(r)}\delta(z-d^+/2). \tag{III.150}$$

Note that, unlike the interface Bychkov-Rashba SOI, the interface contribution to the Dresselhaus SOI does not vanish in systems with inversion symmetry. In fact, for a symmetric QW one has

$$\langle\gamma_{int}\rangle = 2g_0(d/2)\left[\gamma^{(l)}\left.\frac{\partial g_0^{(l)}}{\partial z}\right|_{z=-\frac{d}{2}} - \gamma^{(c)}\left.\frac{\partial g_0^{(c)}}{\partial z}\right|_{z=-\frac{d}{2}}\right], \tag{III.151}$$

which is, in general, different from zero. In obtaining Eq. (III.151) we have taken into account that for a symmetric QW, $\gamma^{(l)} = \gamma^{(r)}$ and the ground state $g_0(z)$ is an even function of $z$.

By requiring the probability flux conservation across the interfaces one can derive the generalization of the boundary conditions in Eqs. (III.107) and (III.108) to the case of the extended Kane model. The new boundary conditions are, (Pfeffer, 1997; Zawadzki and Pfeffer, 2004; Wang *et al.*, 2005)

$$\mathbf{g}^{(i)}(z_{ij}) = \mathbf{g}^{(j)}(z_{ij}), \tag{III.152}$$

$$\begin{aligned} &\frac{\hbar^2}{2m^{*(i)}}\left[1+\frac{2m^{*(i)}\gamma^{(i)}}{\hbar}(k_x\sigma_x - k_y\sigma_y)\right]\left.\frac{d\mathbf{g}^{(i)}}{dz}\right|_{z=z_{ij}} \\ -\;&\frac{\hbar^2}{2m^{*(j)}}\left[1+\frac{2m^{*(j)}\gamma^{(j)}}{\hbar}(k_x\sigma_x - k_y\sigma_y)\right]\left.\frac{d\mathbf{g}^{(j)}}{dz}\right|_{z=z_{ij}} \\ =\;&[\beta^{(i)}(z_{ij}) - \beta^{(j)}(z_{ij})](k_x\sigma_y - k_y\sigma_x)\mathbf{g}^{(i)}(z_{ij}). \end{aligned} \tag{III.153}$$

Here we have kept the same notation as for the standard Kane model but it is understood that all the parameters in the expression above contain already the corresponding corrections as listed in Tab. III.5.

[63]That the strength of the Dresselhaus SOI is different in heterostructures and bulk systems has already been noticed by Krich and Halperin (2007).

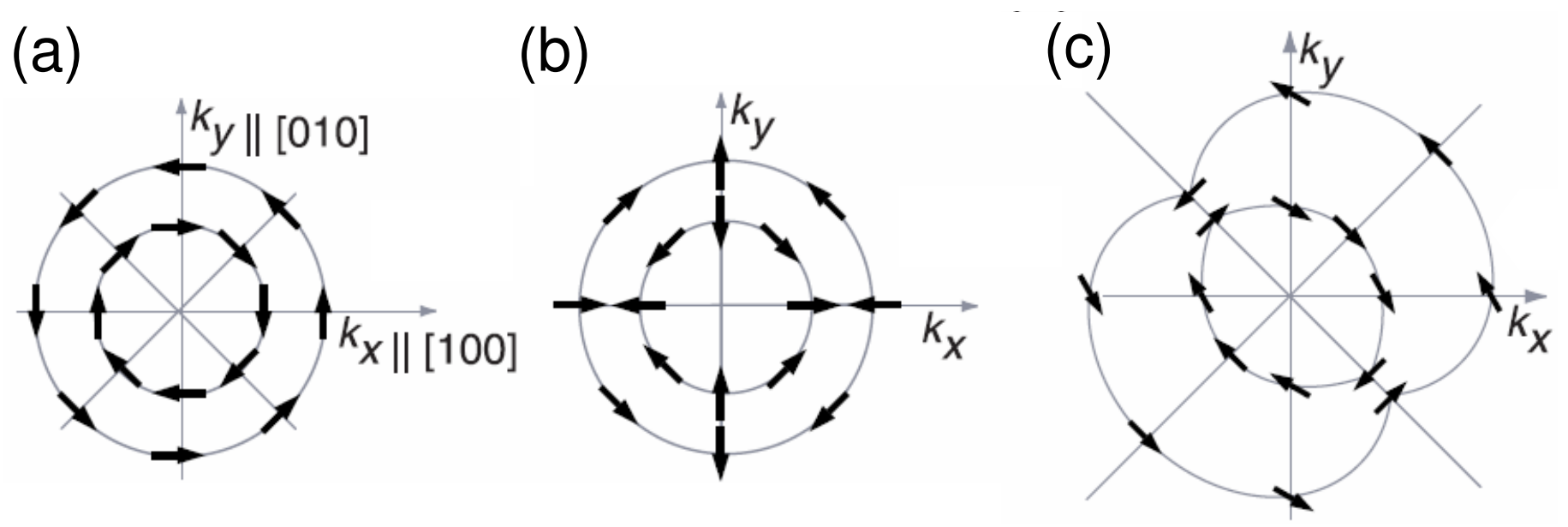

$\alpha_{BR} \neq 0;\ \gamma_D = 0$ $\alpha_{BR} = 0;\ \gamma_D \neq 0$ $\alpha_{BR} \neq 0;\ \gamma_D \neq 0$

Fig. III.9. Schematics of the spin orientation. (a) Only Bychkov-Rashba SOI is present. (b) Only Dresselhaus SOI is present. (c) Bychkov-Rashba and Dresselhaus SOIs are simultaneously present. Reprinted with permission from S. D. Ganichev, V. V. Bel'kov, L. E. Golub, E. L. Ivchenko, P. Schneider, S. Giglberger, J. Eroms, J. De Boeck, G. Borghs, W. Wegscheider, D. Weiss, and W. Prettl, *Phys. Rev. Lett.* **92**, 256601 (2004). Copyright (2004) by the American Physical Society.

### F.2 Interference between Bychkov-Rashba and Dresselhaus spin-orbit interactions

We consider now a QW in the presence of both the Bychkov-Rashba and Dresselhaus SOIs. After averaging out the $z$ degree of freedom, the SOI Hamiltonian reduces to

$$\mathrm{H}_{so} = \alpha_{BR}(k_x\sigma_y - k_y\sigma_x) + \gamma_D(k_x\sigma_x - k_y\sigma_y), \tag{III.154}$$

where $\gamma_D = \langle\partial_z[\gamma(z)\partial_z]\rangle$ is the linearized Dresselhaus coupling parameter. For the case of an infinite quantum well with well width $d$ one has $\gamma_D = \gamma^{(c)}\langle\partial_z^2\rangle$ and, therefore,

$$\gamma_D = \gamma^{(c)}\left(\frac{\pi}{d}\right)^2. \tag{III.155}$$

The values of the linearized Dresselhaus parameter $\gamma_D$ are listed in Tab. III.6 for the case of an infinite QW with $d = 12$nm.

The Zeeman-like effective magnetic field corresponding to the interaction in Eq. (III.154) is given by (Ganichev *et al.*, 2004; Giglberger *et al.*, 2007)

$$\mathbf{B}_{eff}(\mathbf{k}) = \frac{1}{\mu_B}(\gamma_D k_x - \alpha_{BR}k_y, \alpha_{BR}k_x - \gamma_D k_y). \tag{III.156}$$

This field determines the spin orientation when both SIA and BIA are present. The spin orientation at the Fermi surface is depicted in Fig. III.9 for the cases in which there is only Bychkov-Rashba (a) or Dresselhaus (b) SOIs. The situation corresponding to the presence of both Bychkov-Rashba and Dresselhaus SOIs is shown in Fig. III.9.

The SOI energy splitting is given by $\Delta\epsilon_0 = 2\mu_B|\mathbf{B}_{eff}|$, i.e.,

$$\Delta\epsilon_0 = 2k_\parallel\sqrt{\alpha_{BR}^2 + \gamma_D^2 + 2\alpha_{BR}\gamma_D\sin(2\varphi)}, \tag{III.157}$$

where we have taken into account that $k_x = k_\parallel\cos\varphi$ and $k_y = k_\parallel\sin\varphi$.

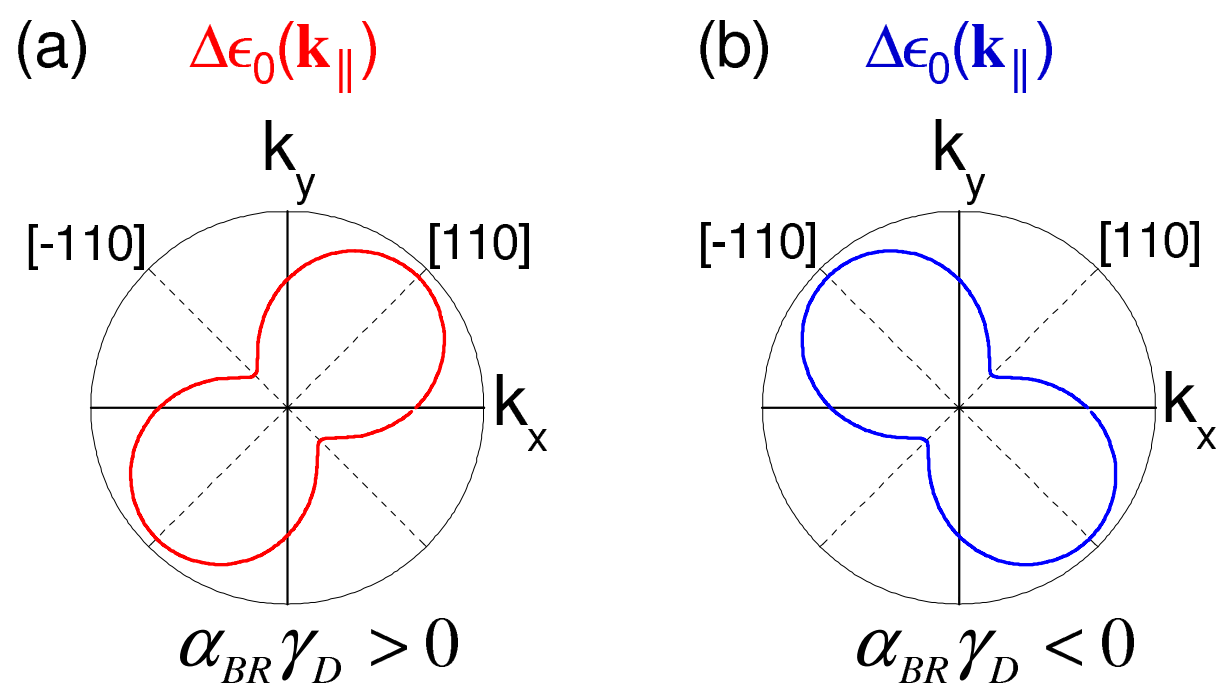

Fig. III.10. $\mathbf{k}_{\|}$-dependence of the energy splitting $\Delta\epsilon_0$ in presence of both Bychkov-Rashba and Dresselhaus SOIs. The anisotropy of $\Delta\epsilon_0$ manifests itself in a two-fold symmetry which symmetry axis depends on the sign of the product $\alpha_{BR}\gamma_D$.

A polar graph of the energy splitting is shown in Fig. (III.10), where the anisotropic nature of $\Delta\epsilon_0$ is apparent. This anisotropy originates from the interference of BIA and SIA. In fact, a symmetric QW with a diamond structure grown in the direction [001] has the point group $D_{4h}$. When BIA is present the symmetry is reduced to $D_{2d}$, whereas SIA reduces the point group to $C_{4v}$. Consequently, when both BIA and SIA are present, the symmetry is that of the point group $C_{2v}$. Even to leading order in $\mathbf{k}_{\|}$ (as we have assumed here) the $C_{2v}$ symmetry is already manifest, as shown in Fig. (III.10). Interestingly, the symmetry axis of the anisotropy can be flipped when the product $\alpha_{BR}\gamma_D$ inverts its sign. In practice, the parameter $\gamma_D$ constitutes a property of the constituents of the heterostructure and can not be externally tuned. The Bychkov-Rashba parameter can, however, be tuned by an external gate voltage or by changing the electron density. Thus, it is possible to invert the sign of $\alpha_{BR}$ by varying the voltage, resulting in the voltage-induced flipping of the symmetry axis of the anisotropy of the SOI. Such an effect has been experimentally observed in a GaAs 2DEG (Miller *et al.*, 2003) and in Fe/GaAs/Au heterojunctions (Moser *et al.*, 2007).

An interesting issue concerning systems in which the Bychkov-Rashba and Dresselhaus SOI are of the same order consists in the experimental separation of the relative contributions of each individual term to the spin-orbit coupling. To obtain the Bychkov-Rashba parameter, for example, the Dresselhaus contribution is usually neglected (Luo *et al.*, 1988; Nitta *et al.*, 1997; Engels *et al.*, 1997; Heida *et al.*, 1998; Hu *et al.*, 1999; Grundler, 2000). However, it has been noted that the interference between the Dresselhaus and Bychkov-Rashba terms can lead to the vanishing of microscopic effects even in the case when both terms are individually large (Averkiev *et al.*, 2002; Tarasenko and Averkiev, 2002). When both terms cancel each other, interesting effects such as a vanishing spin splitting in certain $\mathbf{k}$-space directions (Ganichev and Prettl, 2003), the disappearance of antilocalization (Knap *et al.*, 1996), the lack of Shubnikov-de Haas oscillations (Tarasenko and Averkiev, 2002), and the absence of spin relaxation in certain crystallographic direction (Averkiev *et al.*, 2002; Averkiev and Golub, 1999) occur. Furthermore, a nonballistic spin-field effect transistor operating in the regime $\alpha_{BR} = \gamma_D$ has been proposed (Schliemann *et al.*, 2003). It is therefore desirable to develop experimental methods for the direct measure of the relative contributions of the Bychkov-Rashba and Dresselhaus SOI. Novel techniques based in the spin-galvanic effect (Ganichev *et al.*, 2002; Ganichev and Prettl, 2006) and the circular photogalvanic effect (Ganichev *et al.*, 2001; Ganichev and Prettl, 2006) allow for the experimental separation of Bychkov-Rashba and Dresselhaus spin splittings in semiconductor QWs (Ganichev *et al.*, 2002, 2004; Giglberger *et al.*, 2007). The measured ratios are in the range $\alpha_{BR}/\gamma_D = -4.5 - 7.6$ for GaAs/AlGaAs Qws (Ganichev *et al.*, 2004; Giglberger *et al.*, 2007) and $\alpha_{BR}/\gamma_D = 1.8 - 2.3$ and $\alpha_{BR}/\gamma_D = 1.6$ for InAs/AlGaSb and InAs/InAlAs QWs, respectively (Giglberger *et al.*, 2007).

### G. Spin-orbit interaction in systems with interface inversion asymmetry

It is worth noting that in addition to the SIA and BIA induced SOI, there is a contribution to the spin-orbit coupling resulting from the interface inversion asymmetry (IIA), which is determined by the presence of different atoms at each side of the interface (Rössler and Kainz, 2002; Ivchenko *et al.*, 1996; Aleiner and Ivchenko, 1992). The evaluation of such a contribution requires the knowledge of the microscopic details at the interfaces.[64] For example, at an Fe/GaAs interface, one can have a situation in which GaAs is Ga-terminated and the Ga atoms lies at the interface with Fe atoms as nearest neighbors on one side and As atoms on the other side leading to a configuration of the type Fe-Ga-As [see Fig. III.11 (a)]. On the other hand, in the case of an As termination the configuration will be of type Fe-As-Ga [see Fig. III.11 (b)]. Obviously, these two situations are not equivalent and will, eventually, lead to different interface SOI.[65] Another aspect that has to be taken into account when considering the effects of IIA is the symmetry at the interface because it could be different to the symmetries present in other regions of the heterostructure. Consider, for example, a GaAs/AlAs interface along the [001] direction in absence of external electric fields (see Fig. III.12(a)). Away from the interface, the system has the symmetry of the point group $D_{2d}$ as a consequence of the BIA of the constituent materials. At the interface, however, the system has a $C_{2v}$ symmetry as sketched in Fig. III.12(b). For the case of an ideal symmetric QW grown in the [001] direction we have two such interfaces. We must then distinguish between common-atom and no-common atom interfaces (Magri and Zunger, 2000; Majewski and Vogl, 2003). In the presence of the BIA (say, for zinc-blende compounds), if the interfaces share a common atom [e.g., as in the case of the As layer in a GaAs/AlAs interface] the point group of the QW is $D_{2d}$.[66] If, on the contrary, the heterointerfaces do not share a common atom (e.g., as in an InAs/GaSb interface), then the microscopic atomic structure of a symmetric QW has the point group $C_{2v}$.

For interfaces between zinc-blende compounds both the interference of interface Bychkov-Rashba and Dresselhaus SOIs and the IIA lead to a $C_{2v}$ symmetry. However, these two contributions are, in general, different and may even lead to different symmetries of the SOI. This fact may manifest in interfaces between centrosymmetric materials. For such interfaces the interface Dresselhaus SOI vanishes and the interference contribution reduces to only the interface

[64]Note that the approach here discussed does not account for any microscopic detail at the interfaces.

[65]Note, however, that in both configurations the interface has the symmetry of the point group $C_{2v}$.

[66]This is consistent with our earlier results that did not consider microscopic details at the interfaces. Indeed, in our previous results we have Dresselhaus and Rashba terms at the interfaces, which can, in principle, lead to a $C_{2v}$ symmetry. However, since the QW is symmetric, the average of the Rashba terms vanishes and we end up with the $D_{2d}$ symmetry induced by the BIA.

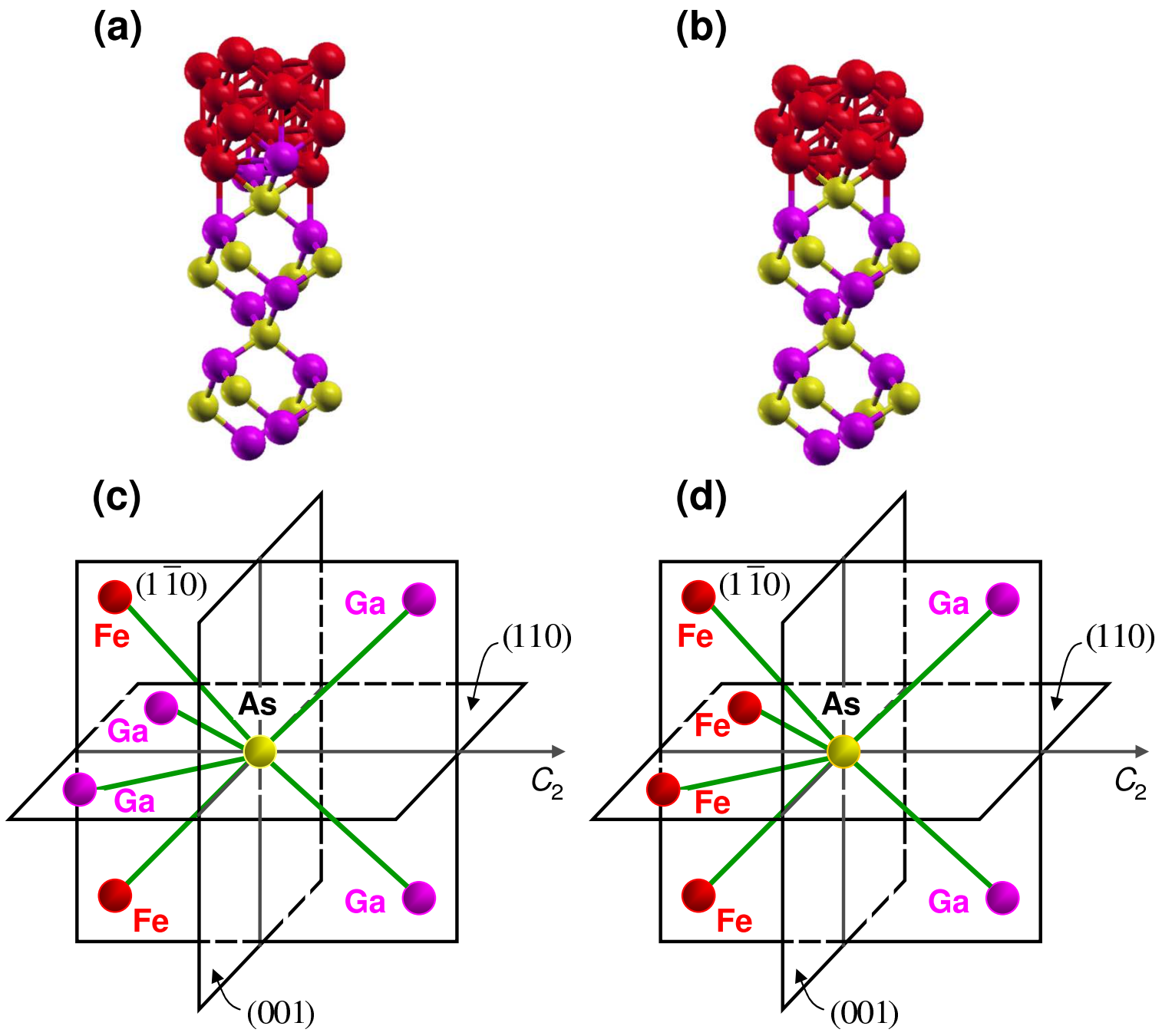

Fig. III.11. Atomic structure of an Fe/GaAs interface for Ga terminated (a) and As terminated (b) structures. Schematics of the nearest neighbors of an As atom in an Fe/GaAs interface. (c) For a Ga terminated structure. (d) For an As terminated structure. In both cases the symmetry of the interface is that of the point group $C_{2v}$, containing the twofold rotation axis $C_2$ parallel to the growth direction $[001]$ and two mirror planes $(110)$ and $(1\bar{1}0)$. Courtesy of M. Gmitra.

Bychkov-Rashba SOI which leads to a $C_{4v}$ symmetry. On the other hand, the IIA of the electronic structure at the interface may still exhibit a $C_{2v}$ symmetry, resulting in a two-fold symmetric SOI. Thus, even in the absence of BIA, the symmetry of the SOI may be reduced to $C_{2v}$ as a consequence of the IIA of centrosymmetric-centrosymmetric interfaces. To the best of our knowledge, no investigation exploring such a possibility has been reported.

A possible manifestation of the IIA is the experimentally observed in-plane anisotropy in the optical absorption of [001]-grown Gas/AlAs QWs (van Kesteren *et al.*, 1990), which exhibits a $C_{2v}$ symmetry. This anisotropy has been found to be particularly large in systems without common atoms in the well and barrier materials (Krebs *et al.*, 1997, 1998; Krebs and Voisin, 2000). Other experiments have explicitly shown the possibility of tuning the optical anisotropy by means of an external electric field oriented along the growth direction (Kwok *et al.*, 1992).

In order to explain the experimental findings of van Kesteren *et al.* (1990), Aleiner and Ivchenko (1992) and Ivchenko *et al.* (1996) introduced an additional, phenomenological interface term in the valence band block $\mathrm{H}_v$ of the Kane Hamiltonian. This additional term, compatible with the $C_{2v}$ symmetry of a GaAs/AlAs heterointerface, includes the mixing between the $|X\rangle$ and $|Y\rangle$ orbital states, which is possible due to the $C_{2v}$ symmetry of a (001) interface.

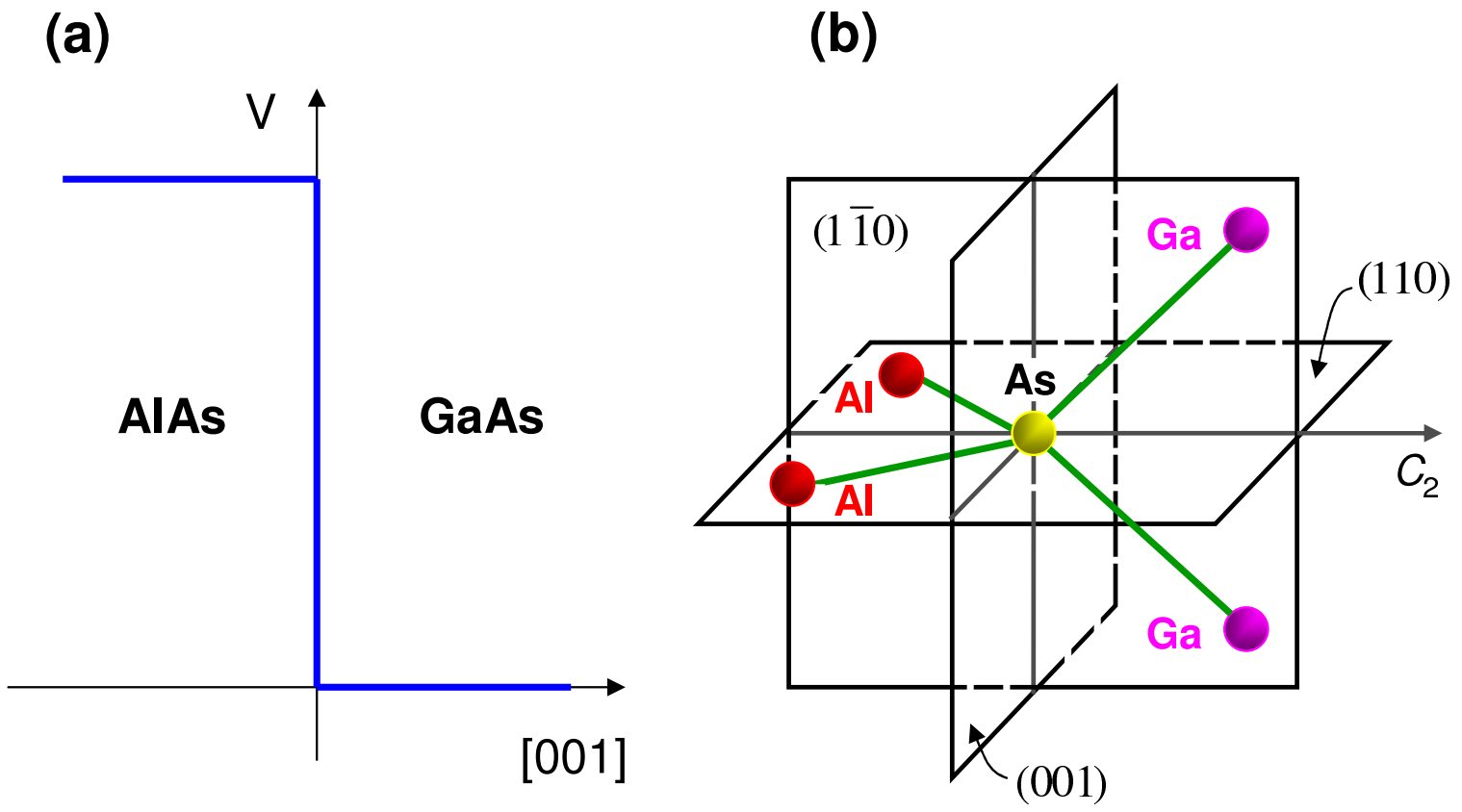

Fig. III.12. (a) Schematics of a GaAs/AlAs interface. (b) The nearest neighbors of an As interface atom. The point symmetry $C_{2v}$ of a single heterojunction contains the twofold rotation axis $C_2$ parallel to the growth direction $[001]$ and two mirror planes $(110)$ and $(1\bar{1}0)$ (Ivchenko *et al.*, 1996).

### H. Spin-orbit interaction: Temperature effects

An important but thus far ignored issue, relevant for any spintronic device application based on the Bychkov-Rashba and/or Dresselhaus SOIs, is the temperature dependence of the spin-orbit parameters determining the strength of the SOI. The $\mathbf{k}\cdot\mathbf{p}$ theory can describe quite well a wide range of low-temperature experimental results. When increasing temperature, however, discrepancies between the measured effective masses and Landé g-factor in GaAs and the corresponding predictions of the temperature-independent $\mathbf{k}\cdot\mathbf{p}$ theory have been observed (Hopkins *et al.*, 1987; Hazama *et al.*, 1986; Oestreich and Rühle, 1995; Hübner *et al.*, 2006).

The temperature dependence of the band-gap energies has been experimentally investigated by Viña *et al.* (1984) and Lautenschlager *et al.* (1987) who proposed the generic, semi-phenomenological functional relation

$$F(T) = F_{_B} - \alpha_{_B}\left(1 + \frac{2}{e^{\Theta/T} - 1}\right) \qquad \text{(III.158)}$$

for describing the temperature dependence of any of the band-gap energies $E_0$, $E_0'$, $\Delta_0$, and $\Delta_0'$ (see Fig. III.7). Note that in Eq. (III.158) the temperature dependence is included via an average Bose-Einstein statistical term for phonons with an average frequency $\Theta$ (Viña *et al.*, 1984; Lautenschlager *et al.*, 1987). The quantities $F_{_B}$, $a_{_B}$, and $\Theta$ are then considered as phenomenological parameters to be obtained by fitting the experimental results with the functional form given in Eq. (III.158). The resulting fitting parameters are shown in Tab. III.8 for the case of GaAs.

The temperature dependence can be introduced in the effective mass $m^*$, the Bychkov-Rashba and Dresselhaus SOI parameters, and the electron Landé g-factor by substituting the temperature -dependent band-gap energies into the expressions obtained within the $\mathbf{k}\cdot\mathbf{p}$ method (see Tab. III.5). The renormalized g-factor ($g^*$) obtained within the extended Kane model is Tab. III.8. Values of the parameters involved in Eq. (III.158). For the cases of $E_0$, $E_0 + \triangle_0$, $E'_0$, and $E'_0 + \triangle'_0$ the parameters $F_B$, $a_B$, and $\Theta$ are obtained by fitting the experimental data corresponding to the temperature dependence of the band-gap energies, while for the case of $2P_0^2/\hbar^2$ and $2P_1^2/\hbar^2$ fittings of the measured temperature dependence of the g-factor are used. Taken from Hübner *et al.* (2006).

| | $F_B$ (eV) | $a_B$ (meV) | $\Theta$ (K) |
|---|---|---|---|
| $E_0$ | 1.571 | 57 | 240 |
| $E_0 + \triangle_0$ | 1.907 | 58 | 240 |
| $E'_0$ | 4.563 | 59 | 323 |
| $E'_0 + \triangle'_0$ | 4.659 | 59 | 323 |
| $2P_0^2/\hbar^2$ | 30.58 | 1040 | 240 |
| $2P_1^2/\hbar^2$ | 8.84 | 1040 | 240 |

given by (Hermann and Weisbuch, 1977)

$$\frac{g^*}{g_0} = 1 - \frac{2P_0^2}{3\hbar^2}\left(\frac{2}{E_0} - \frac{1}{E_0 + \triangle_0}\right) - \frac{2P_1^2}{3\hbar^2}\left(\frac{1}{\tilde{E}_0} - \frac{1}{\tilde{E}_0 + \triangle_0}\right) + C_1, \tag{III.159}$$

where $g_0 = 2.0023$ is the free electron Landé g-factor and the constant $C_1$ accounts for the remote bands contribution. For the case of GaAs $C_1 = -0.02$ (Hübner *et al.*, 2006).

In addition to the temperature dependence of the energy band-gaps one has to consider also the changes in the strength of the different momentum matrix elements. The matrix elements $P_0$ and $P_1$ are both proportional to $a^{-1}$ (here $a$ is the lattice constant). Thus, a small temperature dependence is introduced in $P_0$ and $P_1$ due to the linear expansion of $a(T)$ with temperature as a result of the anharmonic lattice potential.[67] However, it has been shown that such a week dependence of $P_0$ and $P_1$ on the temperature is not enough for explaining the experimental results (Oestreich and Rühle, 1995; Hübner *et al.*, 2006). By using spin-quantum beat spectroscopy at low excitation densities, Oestreich and Rühle (1995) and Hübner *et al.* (2006) have performed high precision measurements of the temperature dependence of the electron g-factor in GaAs. They noticed that under the standard assumption that $P_i \sim 1/a(T)$ $(i = 0, 1)$, the $\mathbf{k} \cdot \mathbf{p}$ calculations of the electron g-factor in GaAs disagree with the experimental results (see Fig. III.13). A similar situation occurs for the case of the temperature dependence of the effective mass (see Fig. III.15). In order to solve these discrepancies between theory and experiment, Hübner *et al.* (2006) considered that not only the band-gap energies but also the momentum matrix elements $P_0$ and $P_1$ exhibit the functional temperature dependence given in Eq. (III.158). By fixing $\Theta$ to the value 240 K,[68] one is left with the linear prefactor in Eq. III.159 as the only fitting parameter. Then, by fitting the experimental results with the help of Eq. (III.159), the temperature dependence of the momentum matrix elements $P_0$ and $P_1$ can be completely determined. The fitting parameters defining the temperature dependence of $P_0^2$ and $P_1^2$ are given in Tab. III.8. In Fig. III.14 we show a comparison between the standard approximation $P_i \sim 1/a(T)$ $(i = 0, 1)$ and the results obtained when a temperature dependence of the form given in Eq. (III.158) is assumed. Unlike for the standard approximation, in the approximation proposed by Hübner *et al.* (2006) both $P_0^2$ and $P_1^2$ strongly decrease with increasing temperature. It has been proposed that such a strong temperature dependence originates from phonon induced fluctuations of the interatomic spacing and adiabatic following of the electrons (Hübner *et al.*, 2006). In such a model $P_i(T) \sim 1/a^*(T)$ $(i = 0, 1)$, where $a^*(T) = a(T) + \sqrt{\langle u^2(T)\rangle}$ now accounts for the phonon induced fluctuations of the interatomic spacing characterized by the mean squared displacement $\langle u^2(T)\rangle$ of the lattice atoms.

[67]For the case of GaAs one obtains $a(T) \approx 5.653251$ Å$+3.88 \cdot 10^{-5}$ ÅK$^{-1}(T - 300$ K) (Vurgaftman *et al.*, 2001).
[68]This is the value of $\Theta$ obtained from the fitting of the fundamental gap $E_0$ (see Tab. III.8).

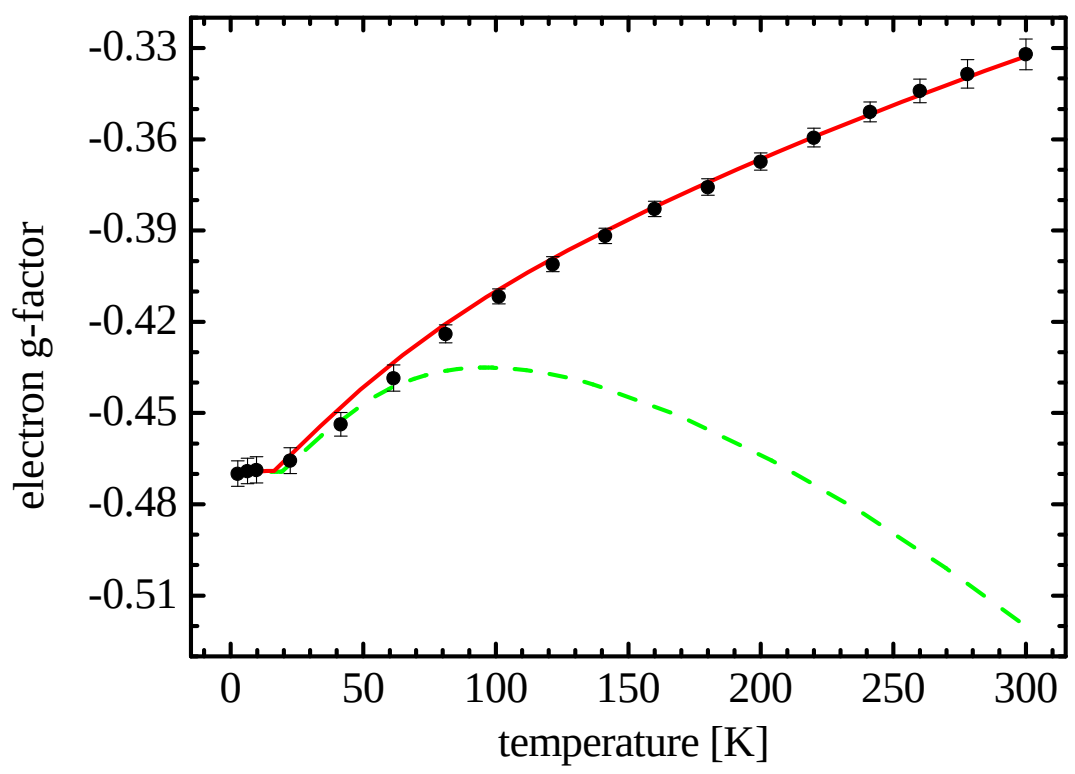

Fig. III.13. High precision measurements of the temperature dependence of the electron Landé g-factor in Bulk GaAs (filled circles). The red solid line is a fit of the experimental data by Eq. (III.159) with temperature dependent matrix elements $P_0$ and $P_1$ according to Eq. (III.158) [see also red solid lines in Figs. III.14(a) and (b)] and the fitting parameters listed in Tab. III.8. The green dashed line is the calculated $g^*$ with $P_i(T)$ $(i = 0, 1)$ depending only on anharmonic lattice expansion [see green dashed lines in Figs. III.14(a) and (b)]. In both cases the energy band-gaps were assumed to depend on the temperature according to Eq. (III.158). Reprinted with permission from Hübner *et al.* (2006).

Making use of the parameters displayed in Tab. III.8, which were obtained from the fitting of the $g^*$-data, one can compute the temperature dependence of the effective mass by using the expression of $m^*$ given in Tab. III.5.[69] The resulting temperature dependence of the effective mass is shown in Fig. III.15, where the good agreement between theory and experiment is apparent.

Following the same procedure as for the effective mass, we can use the expansion of the Bychkov-Rashba parameter $\alpha$ given in Tab. III.5 and the parameters in Tab. III.8 for calculating the temperature dependence of $\alpha$.[70]. The obtained temperature dependence of the Bychkov-Rashba parameter for the case of a GaAs 2D gas and an uniform electric field $E = 10^5$ V/cm is displayed in Fig. III.16. As already anticipated, the Bychkov-Rashba parameter increases with the temperature. This behavior originates from the decreasing of the band-gap energies when increasing the temperature.

[69]Note, however, that in the expansion of $m^*$ given in Tab. III.5 we have to neglect the term proportional to $\triangle^-$, since the temperature dependence of $\triangle^-$ will introduce new unknown parameters. It turns out that the contribution of such a term to the effective mass is quite small. Therefore, it is still a good approximation to consider only the first three terms in the expansion of $m^*$.

[70]Like in the case of the effective mass, we neglect the terms proportional to $\triangle^-$ in the expansion of $\alpha$.

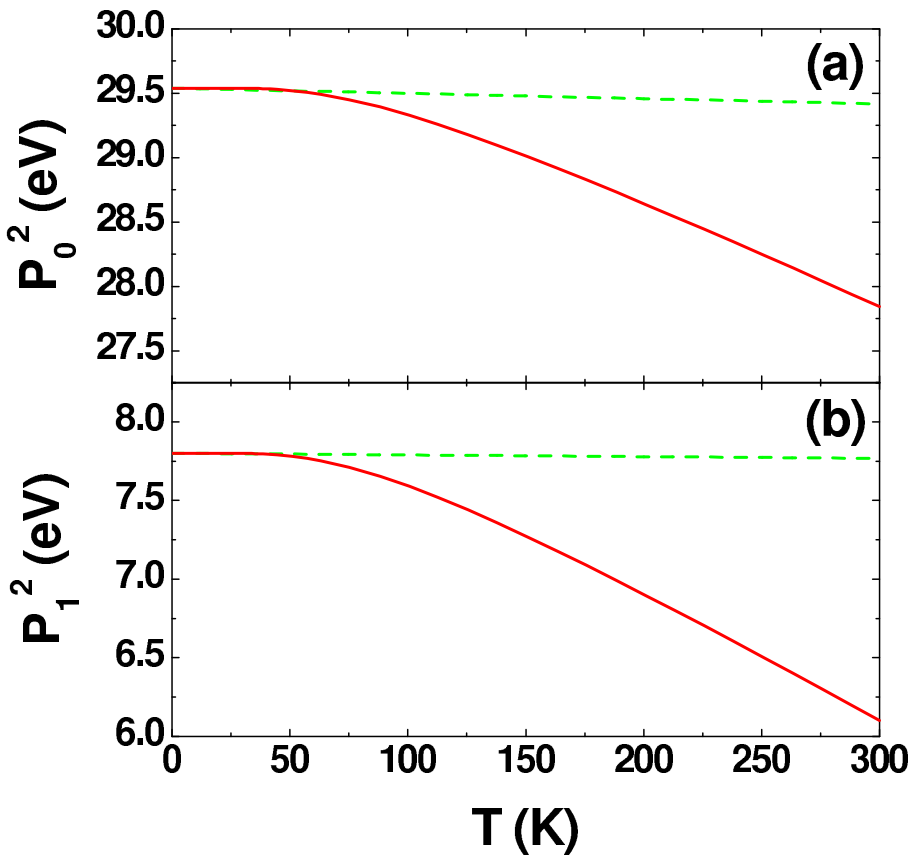

Fig. III.14. Temperature dependence of the $P_0^2(T)$ (a) and $P_1^2(T)$ (b). The red solid lines represent the results according to Eq. (III.158). The green dashed lines correspond to $P_i(T)$ $(i = 0, 1)$ depending only on anharmonic lattice expansion.

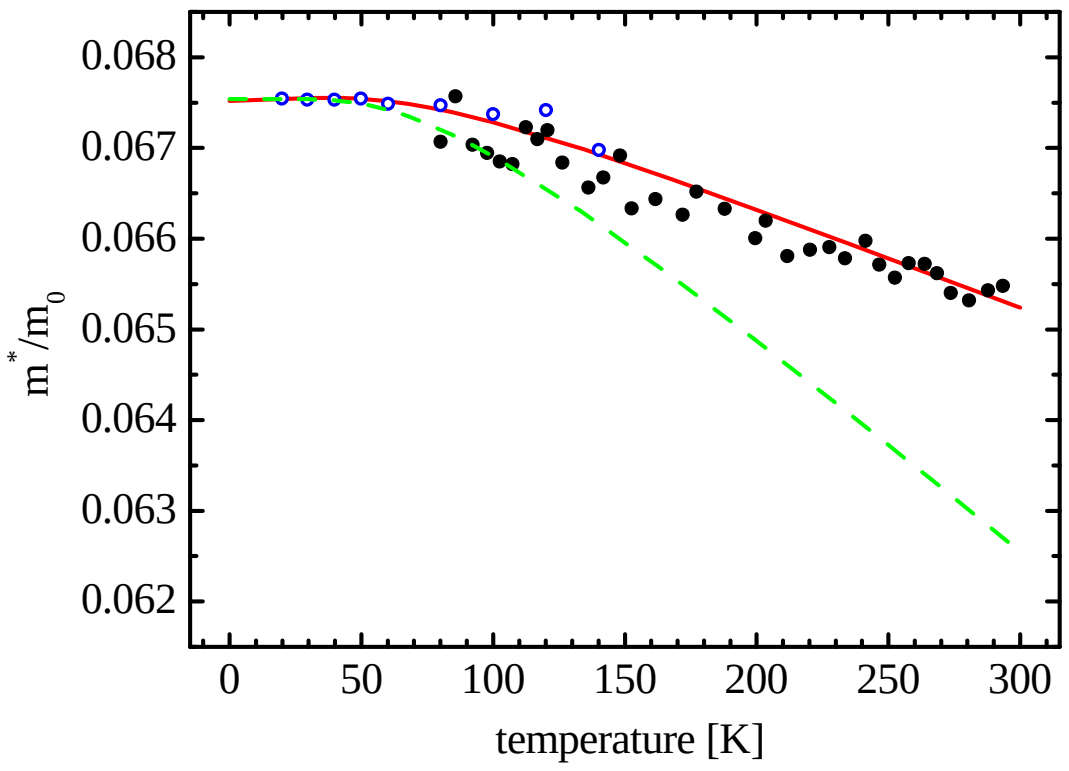

Fig. III.15. Temperature dependence of the effective conduction electron mass in Bulk GaAs. The experimental data points are taken from Hopkins *et al.* (1987) (hollow blue circles) and Hazama *et al.* (1986) (full black circles). The red solid line follows the effective mass expression (neglecting the term proportional to $\triangle^-$) given in Tab. III.5 with temperature dependent matrix elements $P_0$ and $P_1$ according to Eq. (III.158) [see also red solid lines in Figs. III.14(a) and (b)] and the fitting parameters listed in Tab. III.8. The green dashed line is the calculated $m^*$ with $P_i(T)$ $(i = 0, 1)$ depending only on anharmonic lattice expansion [see green dashed lines in Figs. III.14(a) and (b)]. In both cases the energy band-gaps were assumed to depend on the temperature according to Eq. (III.158). Reprinted with permission from Hübner *et al.* (2006).

Finally, we want to stress that although we have limited our analysis here to the case of conduction electrons, the methodology here explained is also applicable to the case of holes. An overview of the SOI for the case of holes is reviewed by Winkler, 2003. The magnetic field effects and the modifications of the electron g-factor are reviewed by Winkler (2003) and Zawadzki and Pfeffer (2004). Discussions on the SOI in presence of strain can be found in (Bir and Pikus,

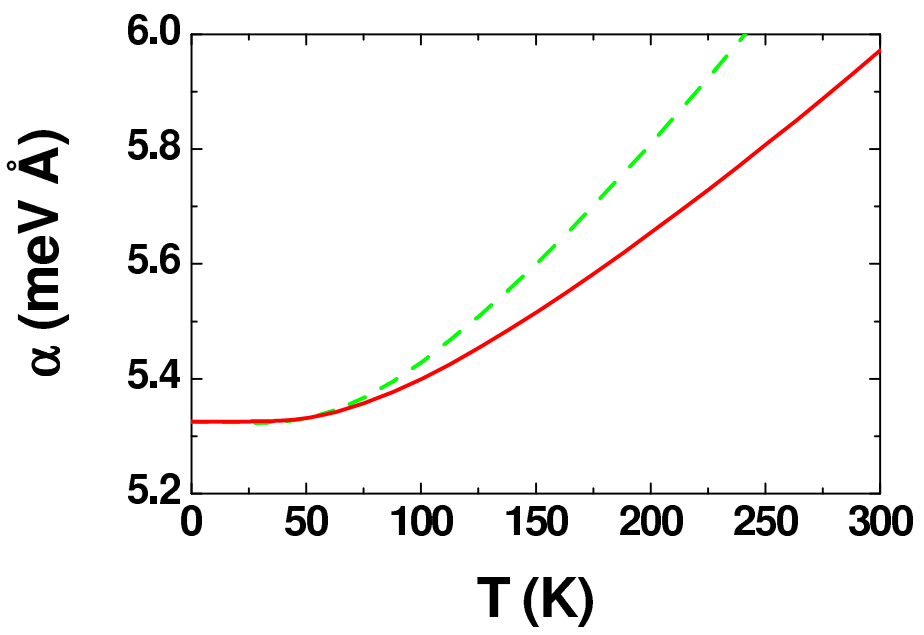

Fig. III.16. Temperature dependence of the Bychkov-Rashba parameter $\alpha$ in a triangular GaAs quantum well with a corresponding electric field $E = 10^5$ V/cm. The red solid line follows the expression for $\alpha$ (neglecting the term proportional to $\triangle^-$) given in Tab. III.5 with temperature dependent matrix elements $P_0$ and $P_1$ according to Eq. (III.158) [see also red solid lines in Figs. III.14(a) and (b)] and the fitting parameters listed in Tab. III.8. The green dashed line is the calculated $\alpha$ with $P_i(T)$ $(i = 0, 1)$ depending only on anharmonic lattice expansion [see green dashed lines in Figs. III.14(a) and (b)]. In both cases the energy band-gaps were assumed to depend on the temperature according to Eq. (III.158).

1974; Winkler, 2003). Complementarily to the method used here, the method of invariants is discussed by (Bir and Pikus, 1974; Trebin *et al.*, 1979; Winkler, 2003). This method is based on the fact that the Hamiltonian of the system must be invariant under all symmetry operations of the problem. From this, rather general, symmetry arguments, the theory of invariants can decide which terms may appear in the Hamiltonian and which terms must vanish. Thus, the method of invariants allows to obtain the general form[71] the different SOI contributions should have from symmetry considerations, without the necessity of explicitly performing all the $\mathbf{k}\cdot\mathbf{p}$ calculations.

[71]The price one pays for the elegance of this method is that it does not provide any estimation of the values of the different SOI parameters.

## IV. Spin relaxation, spin dephasing, and spin dynamics

### A. Bloch equations

The spin relaxation and dephasing times are usually defined with the help of Bloch equations. Suppose we have a spin $\mathbf{s}$, which is the total spin of an ensemble of electrons. We apply a static magnetic field $B_0$ in the z-direction, and a general oscillating field $\mathbf{B}_1(t)$. Let $s_{0z}$ be the equilibrium value of the spin in the static field $B_{0z}$. Then the time evolution of the three spin components in the total field $\mathbf{B} = B_0\hat{z} + \mathbf{B}_1$, is given by

$$\frac{\partial s_x}{\partial t} = \gamma\left(\mathbf{s}\times\mathbf{B}\right)_x - \frac{s_x}{T_2}, \tag{IV.1}$$

$$\frac{\partial s_y}{\partial t} = \gamma\left(\mathbf{s}\times\mathbf{B}\right)_y - \frac{s_y}{T_2}, \tag{IV.2}$$

$$\frac{\partial s_z}{\partial t} = \gamma\left(\mathbf{s}\times\mathbf{B}\right)_z - \frac{s_z - s_{0z}}{T_1}, \tag{IV.3}$$

also known as Bloch equations. The parameter $\gamma = g\mu_B/\hbar$ is the gyromagnetic ratio, with $g$ the effective electron g-factor. The time $T_1$ is called the *spin relaxation* time and the time $T_2$ the *spin dephasing* time.[72] The inverse of the spin relaxation time, $1/T_1$, gives the rate with which the spin along the static field decays to the equilibrium value. The spin system (electrons) need to exchange the energy during the spin relaxation process, if the static magnetic field is non-zero, due to the magnetic energy difference in the initial and final equilibrium state. In the absence of the magnetic field, the time $T_1$ describes the relaxation of a nonequilibrium spin population, or diagonal spin density matrix elements, towards equilibrium. The time $T_2$ describes the dephasing of the spin component transverse to the static field (here $x$ and $y$). In general, it describes the decay of coherent spin oscillations, or off-diagonal spin density matrix elements.

In experiments it is often that the so-called time $T_2^*$ is measured, instead of $T_2$. This “star $T_2$” time contains contributions from inhomogeneous broadening, such as due to g-factor inhomogeneities, in addition to the processes leading to $T_2$ (fluctuating effective magnetic fields in the lattice). Inhomogeneous g-factors cause different spin precession rate for different electrons, giving rise to dephasing. For electrons confined to impurities or quantum dots, in which the inhomogeneities are static, this dephasing contribution can be eliminated by what is called the spin echo, revealing the “homogeneous” dephasing time $T_2$ (Slichter, 1996). For itinerant electrons the inhomogeneous processes are typically washed out by motional narrowing: as electrons spread over the sample, they experience different g-factors, so that only the average contributes to the precession rate (essentially satisfying the central limit theorem).

How can we calculate $T_1$ and $T_2$ times? Usually we start with a Hamiltonian description of the electron system and try to obtain effective equations for the time evolution of the spin. In most cases of interest (but not in all cases!) we end up with Bloch-like equations so that we can directly identify $T_1$ and $T_2$. In electronic systems at relatively weak magnetic field usually $T_1 = T_2$ (up to anisotropy factors of order unity, if the system does not have cubic symmetry, or is not isotropic). Since it is often easier to calculate $T_1$, and measure $T_2$, such a relation is very useful.

[72] In the nuclear magnetic resonance literature $T_1$ is often called longitudinal, while $T_2$ the transverse time, in accord with their definition by the Bloch equations.

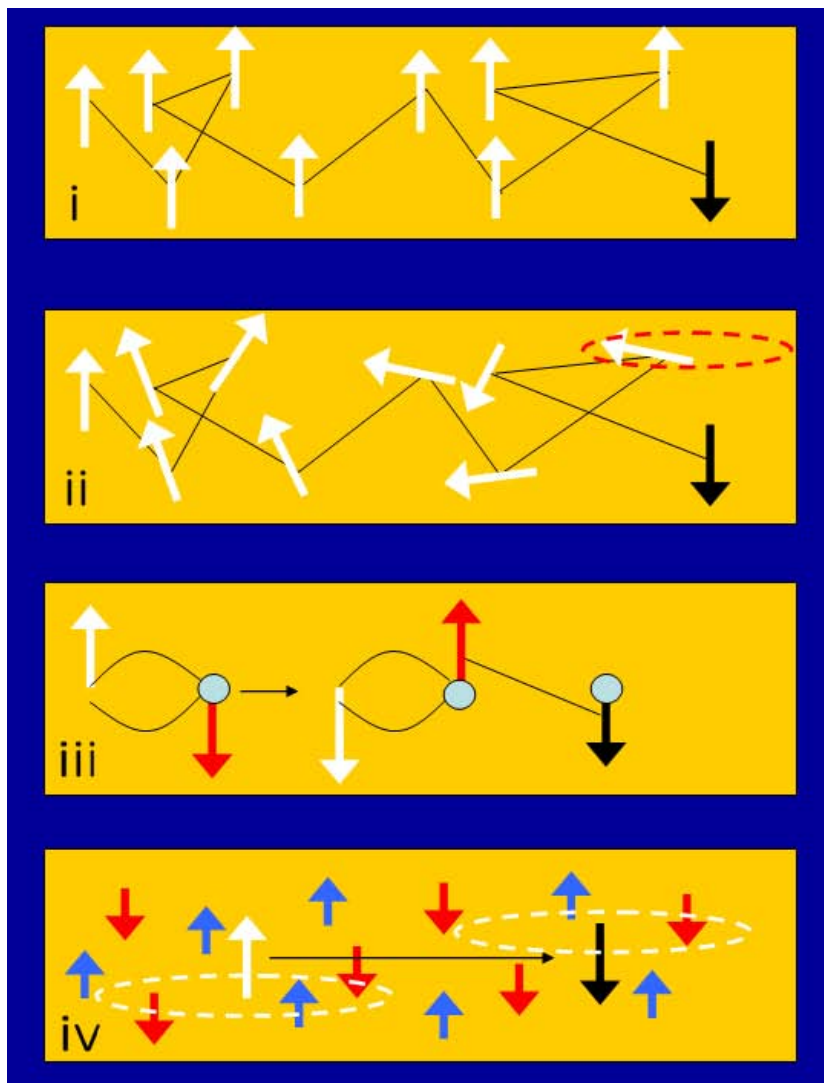

Fig. IV.1. Four important mechanisms of spin relaxation in semiconductors. From top to bottm: (i) The Elliott-Yafet mechanism, in which the electrons scattering off impurities or phonons has a tiny chance to flip its spin at each scattering. (ii) The Dyakonov-Perel mechanism in which electron spins precess along a magnetic field which depends on the momentum. At each scattering the direction and the frequency of the precession changes randomly. (iii) The Bir-Aronov-Pikus mechanism, in which electrons exchange spins with holes (circles), which then lose spins very fast due to the Elliott-Yafet mechanism. (iv) If electrons wave functions (dashed circles) are confined over a certain region with many nuclear spins, the hyperfine coupling causes spin relaxation and dephasing.

There are four important spin relaxation mechanisms of conduction electrons in semiconductors, see Fig. IV.1. (i) In the Elliott-Yafet mechanism, the spin relaxes by momentum scattering off impurities or phonons. Electron states are mixtures of spin up and down spinors, due to spin-orbit coupling. Since the coupling is weak, we can still label the states “up” and “‘down”, with respect to some quantization axis. Each momentum scattering gives a probability to flip the spin from “up ” to down, leading to spin relaxation. The spin relaxes during the scattering, which is why the faster is the momentum scattering, the faster is the spin relaxation. Typically an electron has to undergo $10^5$ scattering events before a spin flip occurs. The Elliott-Yafet mechanism operates in semiconductors with and without a center of inversion symmetry, although it is most prominent in the centrosymmetric ones (such as silicon).

(ii) The Dyakonov-Petel mechanism operates in semiconductors without a center of symmetry, such as GaAs (or zinc-blende in general). As we have seen in Sec. B. in such semiconductors the spin-orbit interaction manifests itself as an effective, momentum dependent magnetic field. An electron moving with one velocity feels one effective magnetic field along which the electron’s spin precesses. As the electron scatters by an impurity or a phonon, the electron changes its velocity and it feels a different (in both magnitude and direction) spin-orbit magnetic field. The precession axis and frequency changes randomly! We can view this situation as an electron’s spin in a fluctuating magnetic field, in line with the toy model to be studied in Sec. B.2. The spin performs a kind of random walk, with the spin flip occuring if the spin manages to walk as far as the opposite of its original direction. Unlike in the Elliott-Yafet mechanism, the spin relaxes in between the scattering events, and the more scattering events there are, the less is the spin relaxation.

In a p-doped semiconductor, the Bir-Aronov-Pikus mechanism (iii) dominates (Bir *et al.*, 1975). Usually, exchange interaction between electrons does not lead to spin relaxation as it preserves the total spin. In a p-doped semiconductor, there will be exchange coupling with holes. An electron with a spin up will exchange its spin with a hole with spin down. The total spin is preserved in the process. However, holes in most useful semiconductors (GaAs, say) lose their spins very fast, since the valence bands are strongly spin mixed due to spin-orbit coupling. The Elliott-Yafet mechanism then leads to very fast spin relaxation of holes; essentially any momentum scattering has a significant chance to give a spin flip. Holes then act as a reservoir for spin: spin-polarized electrons will dump their spin into this reservoir, in which the spins will get lost very fast.

Finally, when nuclear spins are present (they are in GaAs) the hyperfine interaction is capable of spin flips. For itinerant electrons this interaction will be motionally narrowed: electrons will move fast through nuclei with random spins, averaging their actions. However, for electrons confined on impurity levels or in quantum dots, the electron wave function will spread over a region containing, say $10^5$ nuclear spins (in GaAs), whose interaction with the electron will lead to a spin flip and, more significantly, spin dephasing.

We will first introduce a toy model of a spin in a fluctuating magnetic field, discussing the Markov approximation and give a method of deriving master equations. We will then discuss two most frequenly met spin relaxation methods: the Elliott-Yafet and the Dyakonov-Perel spin relaxation. More about spin relaxation in semiconductors can be found in (Meier and Zakharchenya (Eds.), 1984; Žutić *et al.*, 2004).

### B. Born-Markov approximation and a toy model of spin relaxation

This section gives first a general strategy of how to calculate the relaxation and dephasing rates of a system's degrees of freedom, due to an external random perturbation. This strategy is then applied to a toy model of spin relaxation in which Larmor precession is perturbed by a fluctuating magnetic field. Foundations and the methodology of the theory of dephasing and relaxation, resulting from the contact of the studied system with an environment, can be found in monographs (Breuer and Petruccione, 2002; Slichter, 1996; Blum, 1996).

#### B.1 General strategy

Suppose our Hamiltonian can be written as

$$H(t) = H_0 + V(t), \tag{IV.4}$$

where $H_0$ is the system Hamiltonian and $V(t)$ denotes a random fluctuating potential energy (both $H_0$ and $V$ are operators) coming from the interaction of the system with an environment. The equation of motion for the density matrix $\rho$ is,

$$\frac{d\rho}{dt} = \frac{1}{i\hbar}[H, \rho]. \tag{IV.5}$$

The density matrix fully specifies the state of our system. We will find it convenient to look for $\rho$ in the interaction picture. Both $\rho_I$ and $V_I(t)$ are defined by the usual expressions:

$$\rho_I(t) = e^{iH_0 t/\hbar}\rho\, e^{-iH_0 t/\hbar}, \tag{IV.6}$$

$$V_I(t) = e^{iH_0 t/\hbar} V(t) e^{-iH_0 t/\hbar}. \tag{IV.7}$$

The advantage of the interaction picture is that the time evolution of $\rho_I$ is determined explicitly by $V_I(t)$ only:

$$\frac{d\rho_I}{dt} = \frac{1}{i\hbar}\left[V_I(t), \rho_I\right]. \tag{IV.8}$$

The above equation has the following (exact) iterative solution:

$$\rho_I(t) = \rho_I(0) + \frac{1}{i\hbar}\int_0^t dt' \left[V_I(t'), \rho_I(0)\right] + \left(\frac{1}{i\hbar}\right)^2 \int_0^t \int_0^{t'} dt' dt'' \left[V_I(t'), \left[V_I(t''), \rho_I(t'')\right]\right]. \tag{IV.9}$$

Let us differentiate it and obtain an integro-differential evolution equation,

$$\frac{d\rho_I(t)}{dt} = \frac{1}{i\hbar}\left[V_I(t), \rho_I(0)\right] + \left(\frac{1}{i\hbar}\right)^2 \int_0^t dt' \left[V_I(t), \left[V_I(t'), \rho_I(t')\right]\right]. \tag{IV.10}$$

We will now find the evolution equation for the density operator averaged over the fluctuating fields. Assume for simplicity that the fields fluctuate around zero,[73] so that the ensemble average of $V$ vanishes:

$$\overline{V(t)} = \overline{V_I(t)} = 0. \tag{IV.11}$$

Upon ensemble averaging the evolution equation, Eq. (IV.10) becomes,

$$\frac{d\overline{\rho_I(t)}}{dt} = \left(\frac{1}{i\hbar}\right)^2 \int_0^t dt' \overline{\left[V_I(t), \left[V_I(t'), \rho_I(t')\right]\right]}. \tag{IV.12}$$

The term in Eq. (IV.10) linear in $V_I$ vanishes upon averaging (note that the initial condition $\rho_I(0)$ does not depend on $V$, so it is not affected by averaging).

Equation (IV.12) is an exact equation for the time evolution of the density operator averaged over random field realizations. Unfortunately, this equation is not possible to solve except for perhaps a few special cases. The reason is that while the left-hand side contains the time derivative of the average operator density, the right-hand side contains the averaged product of the density and the random field energies. If we consider the random field as a small perturbation, which is usually the case with spins in spintronics, this average of the product can be factorized into the product of two averages:

$$\frac{d\overline{\rho_I(t)}}{dt} = \left(\frac{1}{i\hbar}\right)^2 \int_0^t dt' \left[\overline{V_I(t),\ \left[V_I(t')\right.}, \overline{\rho_I(t')}\right]\Big]. \tag{IV.13}$$

[73]If this is not the case and the averaged field is not zero, we can rearrange the Hamiltonian such that $H = [H_0 + \overline{V(t)}] + [V(t) - \overline{V(t)}]$, and treat the first part as a regular term and the second as the random fluctuating energy with zero mean.

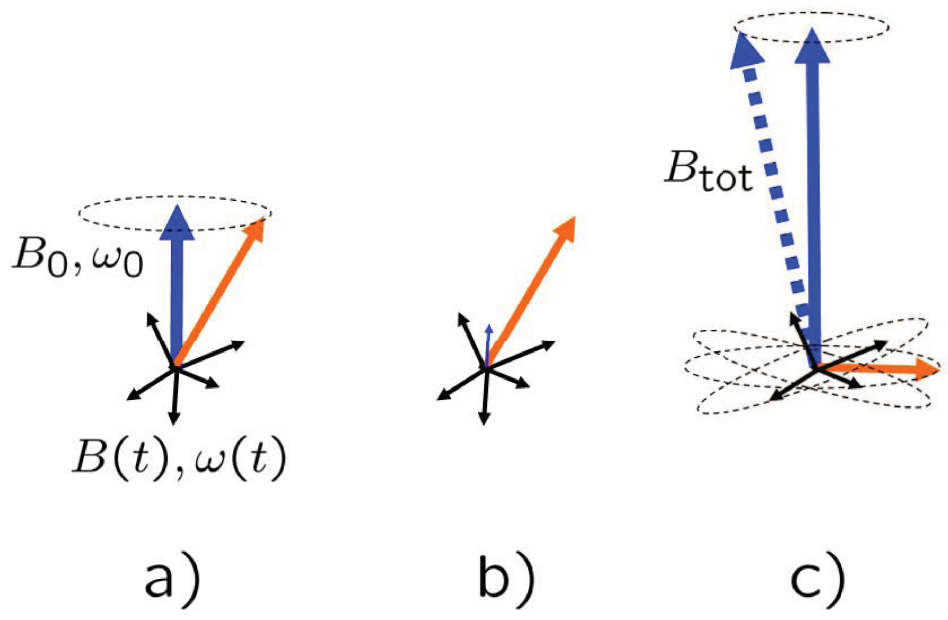

Fig. IV.2. Toy model of spin relaxation and dephasing. (a) A spin in the presence of a static magnetic field along the $z$ direction, $B_0$, giving rise to the Larmor precession frequency $\omega_0$. In addition, a randomly fluctuating magnetic field $\mathbf{B}(t)$, giving rise to the Larmor frequency $\omega(t)$, is applied. (b) If the static field is small, then all the spin components are equal, so that $T_1 = T_2$. (c) If the static field is large, transverse fluctuating fields are inefficient in flipping the spin (see the text).

The factoring, which is sometimes called the Born approximation, can be done since the density matrix $\overline{\rho(t)}$ is only weakly perturbed from its unperturbed value which does not depend on $V$. The above equation can be further simplified using the Markov approximation, which replaces $\rho(t')$ by $\rho(t)$ on the right-hand side. What is the physics behind this replacement? Typically the random fluctuating field has a small correlation time $\tau_c$, meaning that the values of the field separated in time by $\tau_c$ or more are not correlated. In the right-hand side of Eq. (IV.13) then only the terms for which $t'$ is within $\tau_c$ from $t$ contribute. We expect that the density operator varies only slowly over the time interval $\tau_c$, so we can assume that $\rho(t') \approx \rho(t)$ for the relevant times $t'$. The Markov approximation is often called course-graining procedure, since our time resolution in calculating the density operator is $\tau_c$. We give up on asking what is the behavior at finer time scales, which would result from the short memory of the fluctuating field. The final equation then reads,

$$\frac{d\overline{\rho_I(t)}}{dt} = \left(\frac{1}{i\hbar}\right)^2 \int_0^{t \gg \tau_c} dt' \left[\overline{V_I(t),\ [V_I(t'),} \overline{\rho_I(t)}]\right]. \tag{IV.14}$$

This is a Master equation for the density matrix, describing irreversible dynamics of a system which is in contact with environment.

### B.2 Electron spin in a fluctuating magnetic field

We now apply the analysis of the previous section to the simplest possible case, that of an electron spin in the presence of a static magnetic field as well as a randomly fluctuating magnetic field. Despite its simplicity, this toy model is representative for the spin relaxation and dephasing physics; for example, the Dyakonov-Perel mechanism of spin relaxation in semiconductors is of that kind. The model is illustrated in Fig. IV.2.

The Hamiltonian of the spin in the magnetic field is,

$$H = \frac{1}{2}\hbar\omega_0\sigma_z + \frac{1}{2}\hbar\boldsymbol{\omega}(t)\cdot\boldsymbol{\sigma}\,. \tag{IV.15}$$

Here $\omega_0$ is the time independent Larmor frequency–the frequency of the spin precession about the static magnetic field oriented along axis $z$, while $\boldsymbol{\omega}(t)$ is the time dependent, randomly fluctuating Larmor frequency vector: the magnitude is the frequency, while the direction is the precession axis, given by the direction of the fluctuating magnetic field. We can split the Hamiltonian as before, in Eq. (IV.4), with

$$\begin{aligned} H_0 &= \frac{1}{2}\hbar\omega_0\sigma_z, \\ V(t) &= \frac{1}{2}\hbar\boldsymbol{\omega}(t)\cdot\boldsymbol{\sigma}. \end{aligned} \tag{IV.16}$$

In the interaction picture the perturbation $V$ will be

$$V_I(t) = \frac{1}{2}\hbar\boldsymbol{\omega}(t)\cdot\boldsymbol{\sigma}_I(t), \tag{IV.17}$$

where the interaction picture of the Pauli matrices is

$$\boldsymbol{\sigma}_I(t) = e^{iH_0t/\hbar}\boldsymbol{\sigma}e^{-iH_0t/\hbar}. \tag{IV.18}$$

We will take a simple model for the correlation of the fluctuating field:

$$\overline{\omega_\alpha(t)\omega_\beta(t')} = \delta_{\alpha\beta}\overline{\omega_\alpha^2}e^{-|t-t'|/\tau_c}. \tag{IV.19}$$

The different cartesian coordinates $\alpha$ of the field are not correlated, while the fields of the same coordinates are correlated within the time scale of the correlation times $\tau_c$. Our starting point to derive the effective evolution equation for the spin in the presence of the fluctuating field is then Eq. (IV.14) (we omit the overline in the symbol for the density matrix):

$$\dot{\rho}_I = -\frac{1}{4}\sum_\alpha \overline{\omega^2}_\alpha \int_0^t dt' e^{-(t-t')/\tau_c}\left[\sigma_{I\alpha}(t), \left[\sigma_{I\alpha}(t'), \rho_I(t)\right]\right]. \tag{IV.20}$$

It is useful to work directly with the matrix elements rather than with operators. We choose for our basis set the eigenspinors of $\sigma_z$. We reserve the indexes $i$ and $j$ for the two spin states, called 1 for the lower energy and 2 for the upper energy) and denote as,

$$\omega_{ij} = (\varepsilon_i - \varepsilon_j)/\hbar, \tag{IV.21}$$

the frequency corresponding to the difference of the energies of states $i$ and $j$. We also introduce the integral spectral functions[74]

$$J_{ij} = \int_0^\infty dt' e^{-t'/\tau_c}e^{-i\omega_{ij}t'} = \frac{\tau_c - i\omega_{ij}\tau_c^2}{1+\omega_{ij}^2\tau_c^2}. \tag{IV.22}$$

[74]The upper limit on the integral is in fact $t \gg \tau_c$, which can be treated as infinity, since the integrand essentially vanishes at $t' \gg \tau_c$.

Since the two spin states of the Hamiltonian $H_0$ are separated by the Larmor energy, $\varepsilon_2 - \varepsilon_1 = \hbar\omega_0$, the diagonal matrix elements of $J$ are:

$$J_{11} = J_{22} = \tau_c, \tag{IV.23}$$

while the off-diagonal are

$$J_{12} = J_{21}^* = \tau_c \frac{1 + i\omega_0\tau_c}{1 + \omega_0^2\tau_c^2}. \tag{IV.24}$$

The spectral function $J_{ij}$ is proportional to the intensity of the fluctuating field at frequency $\omega_{ij}$, as follows from the definition of Eq. IV.22; for our ansatz, Eq. IV.19, $J$ vanishes at large $\omega$, while it stays constant at $\omega = 0$.

After some pages of algebra, which we leave as an exercise to the reader, we arrive at,

$$\begin{aligned} \dot{\rho}_{ij} &= i\omega_{ij}\rho_{ij} - \frac{1}{4}\overline{\omega_z^2}\left(J_{ii} + J_{jj}\right)\left(1 - \sigma_{z,ii}\sigma_{z,jj}\right)\rho_{ij} - \frac{1}{4}\left(\overline{\omega_x^2} + \overline{\omega_y^2}\right)\left(J_{-i,i} + J_{j,-j}\right)\rho_{ij} \\ &+ \frac{1}{4}\left(\overline{\omega_x^2}\sigma_{x,i,-i}\sigma_{x,-j,j} + \overline{\omega_y^2}\sigma_{y,i,-i}\sigma_{y,-j,j}\right)\left(J_{i,-i} + J_{-j,j}\right)\rho_{-i,-j}. \end{aligned} \tag{IV.25}$$

The above density matrix is already in the Schrödinger picture. We have also used the notation that $\bar{i}$ is the complementary state (opposite spin) to $i$. For example, if $i = 1$, then $\bar{i} = 2$.

Using that $J_{ii} = \tau_c$, we can obtain for the time derivative of the diagonal density matrix elements,

$$\dot{\rho}_{11} = -\frac{1}{4}\left(\overline{\omega_x^2} + \overline{\omega_y^2}\right)\left(J_{21} + J_{12}\right)\left(\rho_{11} - \rho_{22}\right), \tag{IV.26}$$

$$\dot{\rho}_{22} = -\frac{1}{4}\left(\overline{\omega_x^2} + \overline{\omega_y^2}\right)\left(J_{21} + J_{12}\right)\left(\rho_{22} - \rho_{11}\right), \tag{IV.27}$$

while the off-diagonal terms are given by

$$\dot{\rho}_{12} = i\omega_{12}\rho_{12} - \overline{\omega_z^2}\tau_c\rho_{12} - \frac{1}{2}\left(\overline{\omega_x^2} + \overline{\omega_y^2}\right)J_{21}\rho_{12} + \frac{1}{2}\left(\overline{\omega_x^2} - \overline{\omega_y^2}\right)J_{12}\rho_{21}, \tag{IV.28}$$

$$\dot{\rho}_{21} = i\omega_{21}\rho_{21} - \overline{\omega_z^2}\tau_c\rho_{21} - \frac{1}{2}\left(\overline{\omega_x^2} + \overline{\omega_y^2}\right)J_{12}\rho_{21} + \frac{1}{2}\left(\overline{\omega_x^2} - \overline{\omega_y^2}\right)J_{21}\rho_{12}. \tag{IV.29}$$

Since the expectation value of the average spin is,

$$\mathbf{s} = \frac{\hbar}{2}\mathrm{Tr}\left(\rho\boldsymbol{\sigma}\right), \tag{IV.30}$$

we get for the time evolution of the spin components

$$\dot{s}_x = -\omega_0\left[1 - \overline{\omega_y^2}\tau_c T_c\right]s_y - \left(\overline{\omega_z^2}\tau_c + \overline{\omega_y^2}T_c\right)s_x, \tag{IV.31}$$

$$\dot{s}_y = \omega_0\left[1 - \overline{\omega_x^2}\tau_c T_c\right]s_x - \left(\overline{\omega_z^2}\tau_c + \overline{\omega_x^2}T_c\right)s_y \tag{IV.32}$$

$$\dot{s}_z = -\left(\overline{\omega_x^2} + \overline{\omega_y^2}\right)T_c s_z, \tag{IV.33}$$

where we introduce the effective correlation time $T_c$ as,

$$T_c = \frac{\tau_c}{1 + \omega_0^2 \tau_c^2}, \tag{IV.34}$$

and use that $J_{12} + J_{21} = 2T_c$, and $J_{12} - J_{21} = 2iT_c\omega_0\tau_c$. The above equations for the spin dynamics show that the effects of the fluctuating field are twofold: first, the field lowers somewhat the frequency of the spin precession, and, second, the field induces spin relaxation. The change of the frequency does not qualitatively change the picture of the spin precession, and, in fact, is usually ignored (in most cases this is justified), while the second effect brings a qualitative change, allowing the spin to come into equilibrium.

Comparing Eqs. (IV.31), (IV.32), and (IV.33) with the Bloch equations, (IV.1), (IV.2), and (IV.3), we see the following problem: even though our static magnetic field points along the $z$ direction, the equations we have derived imply that also the spin $s_z$ component vanishes at large times, not settling at a finite equilibrium value. The reason is the absence of the heat bath in our consideration. We have effectively treated the system at infinite temperature in which the equilibrium spin is indeed zero. In reality, the fluctuating field will be in equilibrium with the thermal bath, having different spectral functions at different temperatures. Say, if the fluctuating field comes from phonons, there would be no spectral weight, $J$, at zero temperature for the transitions from a lower state 1 to the upper state 2, since no phonons are available at $T = 0$. The emission of phonons would still be allowed, for the transitions from 2 to 1. We can remedy our infinite temperature description by including the effect of the spin bath by changing Eq. (IV.33) to

$$\dot{s}_z = -\left(\overline{\omega_x^2} + \overline{\omega_y^2}\right) T_c (s_z - s_{0z}), \tag{IV.35}$$

where $s_{0z}$ is the temperature dependent equilibrium spin in the direction of the static field. Equation (IV.33) can then be considered as describing the time evolution of the nonequilibrium spin, $s_z - s_{0z}$.

We are now ready to extract the spin relaxation and the spin dephasing rates from Eqs. (IV.31), (IV.32), and (IV.35). The rates are

$$\frac{1}{T_1} = (\overline{\omega_x^2} + \overline{\omega_y^2}) \frac{\tau_c}{\omega_0^2 \tau_c^2 + 1} \tag{IV.36}$$

$$\frac{1}{T_{2x}} = \overline{\omega_z^2}\tau_c + \overline{\omega_y^2} \frac{\tau_c}{\omega_0^2 \tau_c^2 + 1}, \tag{IV.37}$$

$$\frac{1}{T_{2y}} = \overline{\omega_z^2}\tau_c + \overline{\omega_x^2} \frac{\tau_c}{\omega_0^2 \tau_c^2 + 1}. \tag{IV.38}$$

There are several important conclusions that can be drawn here. (i) The spin relaxation and spin dephasing rates increase with increasing of the correlation time. This counterintuitive effect (the more "random" is the fluctuating field, the less capable it is in dephasing spin) is known as motional or dynamical narrowing, to be considered in the following section. (ii) The spin dephasing of a given spin is due to the fluctuating fields perpendicular to the spin direction.[75] This is because only the perpendicular fields are capable of spin flipping. (iii) All the directions

[75] For an isotropic three-dimensional solid $\overline{\omega_x^2} = \overline{\omega_y^2} = \overline{\omega_z^2}$, which is why the Bloch equations contain only one

of the fluctuating field are treated on equal footing at $B_0 \approx 0$, see Fig. IV.2. However, the effect of the fluctuating field along the direction of the static field is not influenced by the static field itself, but the two transverse components of the fluctuating field are effectively diminished. The reason is, as will be discussed more below, that at large static fields the spin-flip abilities of the transverse fields are inhibited. This is best seen from the perspective of the coordinate frame rotated along $z$ with the Larmor frequency $\omega_0$. While the fluctuating field along $z$ is not affected, the $y$ component, say, rotates with $\omega_0$. This rotation diminishes the effective correlation time which is given by the integral over the correlation function (up to the overall fluctuating intensity):

$$\int_0^{t \gg \tau_c} dt' \overline{\omega_z(0)\omega_z(t')} = \overline{\omega_z^2}\tau_c \tag{IV.39}$$

$$\int_0^{t \gg \tau_c} dt' \overline{\omega_y(0)\omega_y(t')} \cos\omega_0 t' = \overline{\omega_y^2}\frac{\tau_c}{\omega_0^2\tau_c^2 + 1} \tag{IV.40}$$

The factor of cosine, due to the rotation of the transverse field, reduces the effect of the field on spin flips.

We can also explain the influence of the static field on spin relaxation and spin dephasing, using the notion of motional narrowing (see the next section if this concept is unfamiliar). As the static magnetic field increases, the transverse spin dephasing is diminished, while the longitudinal spin relaxation vanishes altogether. The higher is the $\omega_0$, the less effective are the transverse components ($x$ and $y$) of the fluctuating field in randomizing the spin. The efficiency of the longitudinal component is not affected. What is the cause? Suppose that initially we have a transverse spin, as in Fig. IV.2 c. Let us consider separately the two contributions to its dephasing. First, the transverse spin is dephased by the fluctuating magnetic field along the $z$ direction. This field causes random changes in the precession angles, leading to dephasing, equally with or without the static field. Second, the transverse spin is dephased by the transverse component of the fluctuating field, which is perpendicular to the spin; if the spin is along x, the relevant fluctuating field is along $y$. The total magnetic field, the static plus the fluctuating one, fills the cone of angle $\omega/\omega_0 \ll 1$, so that the spin precesses very close to the transverse plane. In other words, the transverse fluctuating field is very weak to rotate the spin over a full cycle. At most, the spin rotates out of the plane at an angle of magnitude $\omega/\omega_0$. Flipping the random transverse field after time $\tau_c$ changes the sign of the accumulated phase to $-\omega/\omega_0$. We thus end up with the picture of a random walk, with the step size of $\omega/\omega_0$, and the step time interval $\tau_c$.[76] After time $t$, the standard deviation of the accumulated angle is,

$$\varphi = \frac{\omega}{\omega_0}\sqrt{\frac{t}{\tau_c}}. \tag{IV.41}$$

We call time $t$ the spin relaxation time, $t = \tau_s$, if the standard deviation reaches one: $\varphi = 1$.

dephasing time $T_2$, equal for both transverse spin directions. In general the fluctuating field can be anisotropic in which case we need to distinguish spin dephasing in $x$ and $y$ directions; we do it here by introducing $T_{2x}$ and $T_{2y}$. In the most general case there is a tensor of the spin dephasing rates, though we can always diagonalize it to its three principal directions.

[76]The step size does not depend on $\tau_c$ since $\omega/\omega_0$ is the maximum magnitude we can get.

From the above equation we obtain,

$$\frac{1}{\tau_s} = \frac{1}{\tau_c}\frac{\omega^2}{\omega_0^2}, \tag{IV.42}$$

consistent with the large $\omega_0$ limit of Eqs. (IV.36), (IV.37), and (IV.38).

Consider now an isotropic case (or a cubic solid) in which $\overline{\omega^2} = \overline{\omega_x^2} = \overline{\omega_y^2} = \overline{\omega_z^2}$, that is, the fluctuating intensities in all directions are equal. We can write for the spin dephasing rate,

$$\frac{1}{T_2} = \frac{1}{T_2'} + \frac{1}{2T_1}. \tag{IV.43}$$

Here,

$$\frac{1}{T_2'} = \overline{\omega^2}\tau_c, \tag{IV.44}$$

is the contribution to the dephasing due to a random precession frequency modulation (motional narrowing, see the following section). This contribution is also termed *secular broadening*. The other contribution to the dephasing rate, $1/2T_2$, comes from the spin relaxation. This term is also called *lifetime broadening*. If the correlation time of the fluctuating field is small, such that $\omega_0\tau_c \ll 1$, the dephasing and the relaxation times are equal:

$$T_1 = T_2, \quad \omega_0\tau_c \ll 1. \tag{IV.45}$$

This result is not valid in general, although it appears to hold at high temperatures at which all the spectral components of the fluctuating fields (phonons, for example) are excited. At low temperatures usually the secular component is absent and the equality $T_2 = 2T_1$ holds.

In the opposite limit of large Larmor frequency, $\omega_0\tau_c \gg 1$, the spin relaxation rate vanishes,

$$T_1 \approx \frac{\overline{\omega^2}}{\omega_0^2}\frac{1}{\tau_c} \to \infty, \tag{IV.46}$$

while the spin dephasing time is given by the secular broadening only:

$$\frac{1}{T_2} \approx \overline{\omega_z^2}\tau_c. \tag{IV.47}$$

If secular broadening is absent, the leading term in the dephasing time will be, as in the relaxation,

$$\frac{1}{T_2} \approx \frac{\overline{\omega^2}}{\omega_0^2}\frac{1}{\tau_c}. \tag{IV.48}$$

In this limit the spin dephasing rate is proportional to the correlation rate, not to the correlation time.

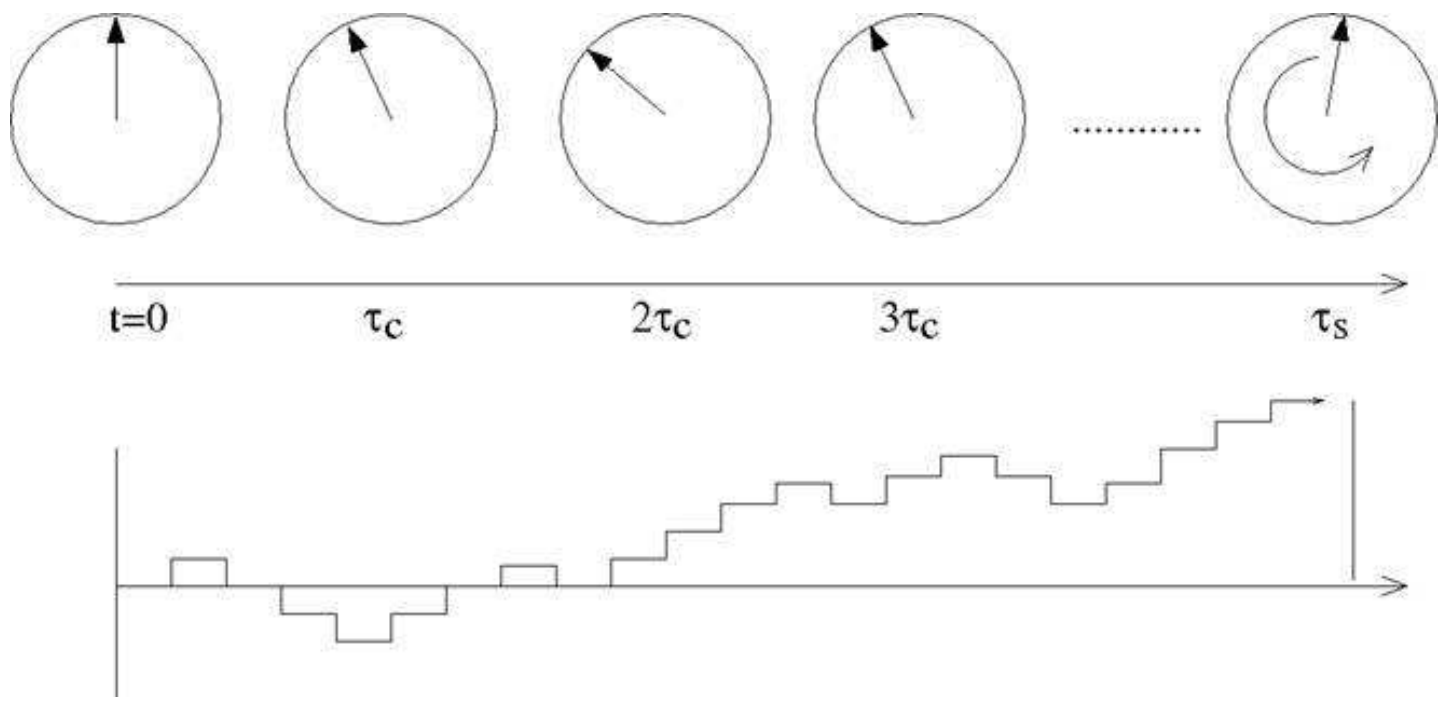

Fig. IV.3. Electron spin precesses along an axis randomly flipping its direction. After time $t$ (horizontal axis) the accumulated phase (vertical axis) will be proportional to $\sqrt{t}$, which is the signature of a random walk.

### B.3 Motional narrowing

The physics behind the spin dephasing due to the randomly fluctuating magnetic field is motional narrowing.[77] Consider a spin precessing about the $z$ axis with Larmor frequency $\Omega$. Let the frequency change randomly between $\Omega$ and $-\Omega$ with the correlation time $\tau_c$; see Fig. IV.3. That is, after time $\tau_c$, the spin has equal probability to continue precessing in the same direction or start turning backwards. The phase accumulated over $\tau_c$ is $\delta\varphi = \Omega\tau_c$. Looking at the spin precession as a random walk with step $\delta\varphi$, after $N$ steps the spread of the phase will be given by the standard deviation, see Sec. II.A.,

$$\varphi = \delta\varphi\sqrt{N}. \tag{IV.49}$$

We will call the spin dephasing time $\tau_s$ the time at which the standard deviation becomes $\varphi \approx 1$. Since $N = t/\tau_c$, we get,

$$1 = \delta\varphi^2 \frac{\tau_s}{\tau_c}, \tag{IV.50}$$

so the spin relaxation rate becomes,

$$\frac{1}{\tau_s} = \Omega^2\tau_c. \tag{IV.51}$$

The above equation is similar to the equation for the spin relaxation in our toy model of the previous section, but also in the Dyakonov-Perel mechanism, to be given later by Eq. (IV.96). We see that the correlation time $\tau_c$ corresponds to the time $\tau^*$. The Dyakonov-Perel mechanism can indeed be viewed as a random precession of electron spin due to fluctuations of precession frequencies and orientations. The greater is the correlation time, the greater is the spin relaxation rate.

[77]The term "motional narrowing" has its roots in nuclear spin resonance studies in liquids in which the absorption line is narrowed—the linewidth being proportional to the spin relaxation rate–if the spin experiences random kicks due to the motion of the spin carriers.

### C. Elliott-Yafet mechanism

The Elliott-Yafet mechanism of spin relaxation (Elliott, 1954; Yafet, 1963) is essentially a Fermi-golden rule mechanism of spin flipping due to the presence of impurities or phonons. There is, however, a subtlety, which was first recognized by Elliott (Elliott, 1954). There are two possible scenarios. First, electrons are described by Bloch states which are eigenstates of the Pauli $\sigma_z$ matrix. That is, the states look like,

$$\psi_{\mathbf{k}\sigma}(\mathbf{r}) = u_{\mathbf{k}}(\mathbf{r}) e^{i\mathbf{k}\cdot\mathbf{r}} \chi_\sigma, \tag{IV.52}$$

where $u_{\mathbf{k}}(\mathbf{r})$ is the modulation function, periodic with the lattice potential, and $\chi_\sigma$ are the Pauli spinors,

$$\chi_\uparrow = \begin{pmatrix} 1 \\ 0 \end{pmatrix}, \quad \chi_\downarrow = \begin{pmatrix} 0 \\ 1 \end{pmatrix}. \tag{IV.53}$$

If the impurity potential is not spin dependent, the scattering does not lead to a spin flip and, eventually, to spin relaxation. However, impurities induce spin-orbit coupling potential, of the form,

$$V_{so} = \frac{\hbar}{4m^2c^2} \nabla V_{\mathrm{imp}} \times \hat{\mathbf{p}} \cdot \boldsymbol{\sigma}, \tag{IV.54}$$

where $V_{\mathrm{imp}}$ is the spin-independent impurity potential and $\hat{\mathbf{p}}$ is the momentum operator. Spin-flip scattering due to the spin-orbit potential causes spin relaxation of conduction electrons. This spin relaxation is most pronounced for heavy impurities—the spin-orbit strength increases as $Z^2$ with the atomic number $Z$ of the impurity.

The second, more important case, is the spin relaxation due to the spin-orbit coupling induced by host ions. In the presence of spin orbit coupling, the Bloch states cannot be chosen as the Pauli spinors. If the conductor has a center of inversion symmetry, the Bloch states have the form of Eqs. (III.4) and (III.5). Since these two states are degenerate, we can make linear combinations of them such that the new eigenstates have the spin magnetic moment parallel or antiparallel to a given direction (say, $z$). Such states can then be called "spin up" and "spin down", even though they are an admixture of the Pauli spin up and spin down spinors. Any momentum relaxation process, such as impurity or phonon scattering, then couples the "spin up" with "spin down" states of a different momentum, giving spin relaxation. This spin relaxation depends on the $Z$ of the host atoms, typically again as $Z^2$. Naturally, semiconductors such as silicon, with low $Z$, have weaker spin relaxation than, say germanium whose spin-orbit coupling is larger.[78]

We will not present here the formalism to derive the Elliott-Yafet spin relaxation time. Instead, we give the formula for the spin relaxation by scattering off phonons, which is the ultimate limiting factor on spin relaxation at finite temperatures. For highly degenerate conductors, the spin relaxation rate is (Fabian and Das Sarma, 1999a),

$$1/T_1 = 8\pi T \int_0^\infty d\Omega \alpha_s^2 F(\Omega) \partial N(\Omega)/\partial T,$$

[78] Spin-orbit coupling of the host lattice itself, without momentum scattering, cannot lead to spin relaxation. Since the spin-orbit interaction of the host lattice is a periodic interaction, the Bloch theorem applies, allowing coupling between the states of the same momentum, but different energy (unless the bands cross)—vertical coupling in the band structure.

where

$$N(\Omega) \quad = \quad [\exp{(\hbar\Omega/k_BT)} - 1]^{-1}, \tag{IV.55}$$

is the distribution function for the phonons of frequency $\Omega$, and

$$\alpha_s^2 F(\Omega) = \frac{g_S}{2M\Omega} \sum_\nu \langle\langle g^\nu_{\mathbf{k}n\uparrow,\mathbf{k}'n'\downarrow} \delta(\omega_{\mathbf{q}\nu} - \Omega)\rangle_{\mathbf{k}n}\rangle_{\mathbf{k}'n'},$$

is the spin-flip Eliashberg function, measuring the efficiency of phonons of frequency $\Omega$ to flip the spin of electrons at the Fermi level. We have also denoted by $M$ the ion mass, $g_S$ the density of electronic states per spin, $\omega_{\mathbf{q}\nu}$ the frequency of a phonon of momentum $\mathbf{q}$ and polarization (displacement) label $\nu$, and $g^\nu_{\mathbf{k}n\uparrow,\mathbf{k}'n'\downarrow}$ is the matrix element of the electron-phonon interaction potential, corresponding to the phonon polarization $\nu$, in the Bloch states of "spin up" with momentum $\mathbf{k}$ and band index $n$, and states of "spin down", with momentum $\mathbf{k}' = \mathbf{k} + \mathbf{q}$, and band index $n'$. The double averaging is over the states at the Fermi level. The above formulas can be used, in combination with a realistic band structure as well as phonon displacement pattern and the electron-phonon coupling, to obtain realistic estimates of the spin relaxation time as a function of temperature.

A nice rule of thumb for the order of magnitude of the spin relaxation rate is the Elliott relation,

$$\frac{1}{T_1} \approx 10 \times \frac{(\Delta g)^2}{\tau_p}, \tag{IV.56}$$

in which $\Delta g = g - g_0$ is the deviation of the $g$-factor for a conduction electron from the free electron value, and $\tau_p$ is the momentum relaxation time. The factor of ten was suggested based on empirical observations of Beuneu and Monod (1978). The Elliott relation can be applied to metals and degenerate semiconductors.

## D. Dyakonov-Perel mechanism

We present below the full derivation of a spin relaxation mechanism proposed by D'yakonov and Perel' (1971b). This mechanism, which is essentially scattering induced motional narrowing of the spin precession due to the spin-orbit field of the host lattice, is effective in semiconductors without a center of inversion symmetry. The most prominent example is GaAs. We first derive the kinetic equation for the electron transport and spin dynamics, then solve the obtained equation in the motional narrowing limit.

### D.1 Kinetic equation for spin dynamics

The electron transport and dynamics in conductors can be, neglecting quantum orbital effects, described quasiclassically by considering electrons as wave packets of momentum $\mathbf{k}$, at space point $\mathbf{r}$ and time $t$; see, for example, (Ashcroft and Mermin, 1976). The orbital dynamics is given by Newton-like equations: In the presence of electric and magnetic fields $\mathbf{E}$ and $\mathbf{B}$, the time evolution of the position and momentum of the wave packet is given by :

$$\dot{\mathbf{r}} \quad = \quad \mathbf{v}_\mathbf{k} = \frac{\partial \varepsilon_\mathbf{k}}{\hbar \mathbf{k}}, \tag{IV.57}$$

$$\hbar\dot{\mathbf{k}} \quad = \quad \mathbf{F}_\mathbf{k} = -e\mathbf{E} - e\mathbf{v}_\mathbf{k} \times \mathbf{B}. \tag{IV.58}$$

Here $\varepsilon_{\mathbf{k}}$ is the energy of the electrons of momentum $\mathbf{k}$, while the momentum dependent force $\mathbf{F}_{\mathbf{k}}$ comprises the electric and Lorentz forces.

We now add the spin degrees of freedom, quantum mechanically. In the presence of a spin-dependent Hamiltonian, call it $H_{1\mathbf{k}}$, the spins are precessing in directions and frequencies specified for each momentum. To allow for the spin precession, as well as to describe a general spin state (pure or mixture) we introduce, for each electron wave packet, a $2 \times 2$ density matrix $\rho = \rho_{\mathbf{k}}(\mathbf{r}, t)$. The density matrix $\rho$ fully describes our, in principle nonequilibrium electronic system within the quasiclassical limit for the orbital degrees of freedom.[79] For example, the occupation number of the momentum state $\mathbf{k}$ at $\mathbf{r}$ is

$$n_{\mathbf{k}}(\mathbf{r}, t) = \mathrm{Tr}\, \rho_{\mathbf{k}}(\mathbf{r}, t), \tag{IV.59}$$

where trace $\mathrm{Tr}$ is over the spin space.[80] We can also calculate the total number of electrons $N$ and the spin $\mathbf{s}$, in the conductor of volume $V$:[81]

$$N = \frac{1}{V} \int d^3\mathbf{r} \sum_{\mathbf{k}} \mathrm{Tr}\, \rho_{\mathbf{k}}(\mathbf{r}, t) \tag{IV.60}$$

$$\mathbf{s}(t) = \frac{1}{V} \int d^3\mathbf{r} \sum_{\mathbf{k}} \mathrm{Tr}\, [\rho_{\mathbf{k}}(\mathbf{r}, t)\hat{\mathbf{s}}], \tag{IV.61}$$

where the spin operator is $\hat{\mathbf{s}} = \boldsymbol{\sigma}/2$. Equation (IV.60) is the normalization condition on our density matrix. The spin, or the associated magnetic moment $\mathbf{M} = \gamma \mathbf{s}$, can depend on time.

In contrast to the quasiclassical description of electron momenta and positions, we treat the spin evolution quantum mechanically. Due to the action of the spin-dependent term $H_1$, which we write as

$$H_{1\mathbf{k}} = \frac{\hbar}{2} \boldsymbol{\Omega}_{\mathbf{k}} \cdot \boldsymbol{\sigma}, \tag{IV.62}$$

the spin-dependent part of the density matrix at time $t - dt$ evolves to,

$$\rho(t) = e^{-iH_1 dt/\hbar} \rho(t - dt) e^{iH_1 dt/\hbar}, \tag{IV.63}$$

at time $t$; $dt$ is infinitesimally small. The kinetic equation for our density matrix can be obtained in a heuristic way, similarly to common derivations of the Boltzmann equation (Ashcroft and Mermin, 1976). Considering that $\rho(t) \equiv \rho[\mathbf{k}(t), \mathbf{r}(t), t]$, we can expand the above equation in the Taylor series in $dt$, to obtain,

$$\rho\,[\mathbf{k}(t), \mathbf{r}(t), t] = e^{-iH_1 dt/\hbar} \rho\,[\mathbf{k}(t - dt), \mathbf{r}(t - dt), t - dt]\, e^{iH_1 dt/\hbar} + \left(\frac{\partial \rho}{\partial t}\right)_{\mathrm{coll}} \tag{IV.64}$$

$$\approx \rho(\mathbf{k}, \mathbf{r}, t) - \frac{\partial \rho}{\partial \hbar \mathbf{k}} \cdot \mathbf{F}_{\mathbf{k}}\, dt - \frac{\partial \rho}{\partial \mathbf{r}} \cdot \mathbf{v}_{\mathbf{k}}\, dt - \frac{\partial \rho}{\partial t}\, dt - \frac{i}{\hbar}\, [H_{1\mathbf{k}}, \rho]\, dt + \left(\frac{\partial \rho}{\partial t}\right)_{\mathrm{coll}}. \tag{IV.65}$$

[79] One should view the single-electron density matrix $\rho_{\mathbf{k}}(\mathbf{r}, t)$ as a simplified notation for the more general density matrix, $\rho_{\mathbf{k}\mathbf{k}'}(\mathbf{r}, \mathbf{r}', t) = \delta_{\mathbf{k}\mathbf{k}'} \delta(\mathbf{r} - \mathbf{r}') \rho_{\mathbf{k}}(\mathbf{r}, t)$. In the quasiclassical approximation we allow only for probabilities (diagonal elements), no coherences (off-diagonal elements) in the momentum and position spaces, while keeping both probabilities and coherences for the spin.

[80] The occupation can be up to two electrons in a single momentum state $\mathbf{k}$.

[81] The division by $V$ appears since we treat the occupation number $n_{\mathbf{k}}$ as a dimensionless number, not as the occupation density. The number of electrons in a momentum state $\mathbf{k}$ in the space interval $[\mathbf{r}, \mathbf{r} + d\mathbf{r}]$ is $n_{\mathbf{k}}(\mathbf{r}) d^3\mathbf{r}/V$.

The term $(\partial\rho/\partial t)_{\rm coll}$, called collision integral, describes the net increase of the density matrix due to the collisions, typically with impurities, phonons, or other electrons. This term is added by hand. We have also used the quasiclassical equations of motion, Eqs. (IV.57) and (IV.58), to substitute for the time derivatives of the momenta and positions, neglecting possible spin dependence of the velocity. The spin Boltzmann kinetic equation follows as:

$$\frac{\partial\rho_{\mathbf{k}}}{\partial t}-\frac{1}{i\hbar}\left[H_{1\mathbf{k}},\rho_{\mathbf{k}}\right]+\frac{\partial\rho_{\mathbf{k}}}{\partial\hbar\mathbf{k}}\cdot\mathbf{F}_{\mathbf{k}}+\frac{\partial\rho_{\mathbf{k}}}{\partial\mathbf{r}}\cdot\mathbf{v}_{\mathbf{k}}=\left(\frac{\partial\rho_{\mathbf{k}}}{\partial t}\right)_{\rm coll}. \tag{IV.66}$$

The first term on the left hand side describes the change of the density matrix with time, at a fixed point in the momentum and position space. The second term, which is the von Neumann formula for the time evolution of the spin density matrix, describes spin precession; the third and fourth terms describe the time evolution of the momentum and position of the electronic wave packets. Taken as a whole, the left hand side is the total time derivative of the density matrix as attached to a moving point (state) in the momentum, position, and spin spaces.

In the absence of spin precession, if $H_1=0$, the spin Boltzmann equation, (IV.66), reduces to the Boltzmann equation for the spin diagonal components of the density matrix, $f_{\mathbf{k}\sigma}(\mathbf{r}t)=\rho_{\mathbf{k}\sigma\sigma}(\mathbf{r},t)$:

$$\frac{\partial f_{\mathbf{k}\sigma}}{\partial t}+\frac{\partial f_{\mathbf{k}\sigma}}{\partial\hbar\mathbf{k}}\cdot\mathbf{F}_{\mathbf{k}}+\frac{\partial f_{\mathbf{k}\sigma}}{\partial\mathbf{r}}\cdot\mathbf{v}_{\mathbf{k}}=\left(\frac{\partial f_{\mathbf{k}\sigma}}{\partial t}\right)_{\rm coll}. \tag{IV.67}$$

While the left hand side does not mix the spin components, the collision integral can include spin-flip processes (as is the case with the Elliott-Yafet mechanism), coupling the Boltzmann equations for spin up and spin down distribution functions $f_\sigma$.

It remains to decipher the collision integral. We will consider only elastic (that is, energy preserving) and spin-preserving scattering by impurities. An electron with momentum $\mathbf{k}$ scatters into another momentum state, $\mathbf{k}'$, keeping its spin unchanged. This is allowed as long as the other state, of the same spin, is empty, due to Pauli's principle. Let us denote the spin-preserving scattering rate, the collision probability per unit time, from $\mathbf{k}$ to $\mathbf{k}'$ by $W_{\mathbf{k}\mathbf{k}'}$. Then the net increase of the density matrix element $\rho_{\mathbf{k}\sigma\sigma'}$ is,

$$\left(\frac{\partial\rho_{\mathbf{k}\sigma\sigma'}}{\partial t}\right)_{\rm coll}=\sum_{\mathbf{k}'}\left[W_{\mathbf{k}'\mathbf{k}}\rho_{\mathbf{k}'\sigma\sigma'}\left(1-\rho_{\mathbf{k}\sigma\sigma'}\right)-W_{\mathbf{k}\mathbf{k}'}\rho_{\mathbf{k}\sigma\sigma'}\left(1-\rho_{\mathbf{k}'\sigma\sigma'}\right)\right]. \tag{IV.68}$$

The first term on the right hand side counts the number of collisions from all occupied states $\mathbf{k}'$ to an empty $\mathbf{k}$ state, increasing the density matrix, while the second term counts the collisions from the occupied $\mathbf{k}$ to all other states, on the energy shell, diminishing the density matrix. For an impurity potential the principle of detailed balance gives (Ashcroft and Mermin, 1976) $W_{\mathbf{k}\mathbf{k}'}=W_{\mathbf{k}\mathbf{k}'}$, which allows to write[82]

$$\left(\frac{\partial\rho_{\mathbf{k}\sigma\sigma'}}{\partial t}\right)_{\rm coll}=\sum_{\mathbf{k}'}W_{\mathbf{k}'\mathbf{k}}\left(\rho_{\mathbf{k}'\sigma\sigma'}-\rho_{\mathbf{k}\sigma\sigma'}\right). \tag{IV.70}$$

[82]It is tempting to write the collision integral in terms of density matrices (not their elements) as,

$$\sum_{\mathbf{k}\mathbf{k}'}W_{\mathbf{k}\mathbf{k}'}\left[\rho_{\mathbf{k}'}\left(1-\rho_{\mathbf{k}}\right)-\rho_{\mathbf{k}}\left(1-\rho_{\mathbf{k}'}\right)\right], \tag{IV.69}$$

and proceed to the final answer, Eq. (IV.70). This would be incorrect since $\rho_{\mathbf{k}}$ and $\rho_{\mathbf{k}'}$ do not, in general, commute.

We can now summarize that the kinetic description of quasiclassical orbital dynamics as well as quantum spin dynamics, in the presence of spin-preserving scattering, is given by the equation

$$\frac{\partial \rho_{\mathbf{k}}}{\partial t} - \frac{1}{i\hbar}\left[H_{1\mathbf{k}}, \rho_{\mathbf{k}}\right] + \frac{\partial f_{\mathbf{k}\sigma}}{\partial \hbar \mathbf{k}} \cdot \mathbf{F}_{\mathbf{k}} + \frac{\partial f_{\mathbf{k}\sigma}}{\partial \mathbf{r}} \cdot \mathbf{v}_{\mathbf{k}} = -\sum_{\mathbf{k}'} W_{\mathbf{k}\mathbf{k}'}\left(\rho_{\mathbf{k}} - \rho_{\mathbf{k}'}\right). \tag{IV.71}$$

We will solve this equation for a homogeneous case in the absence of external fields in the next section.

The spin Boltzmann equation, (IV.71), readily gives the spin continuity equation. Expressing the spin-dependent part of the Hamiltonian in the form of Eq. (IV.62), and calculating the time derivative of the spin of the momentum state $\mathbf{k}$ at point $\mathbf{r}$,

$$\mathbf{s}_{\mathbf{k}} = \frac{\hbar}{2}\mathrm{Tr}\left[\rho_{\mathbf{k}}(\mathbf{r}, t)\sigma\right], \tag{IV.72}$$

it is a simple exercise in commutator algebra to get

$$\frac{\partial \mathbf{s}_{\mathbf{k}}}{\partial t} - \left(\mathbf{\Omega}_{\mathbf{k}} \times \mathbf{s}_{\mathbf{k}}\right) + \frac{\partial \mathbf{s}_{\mathbf{k}}}{\partial \hbar \mathbf{k}} \cdot \mathbf{F}_{\mathbf{k}} + \frac{\partial \mathbf{s}_{\mathbf{k}}}{\partial \mathbf{r}} \cdot \mathbf{v}_{\mathbf{k}} = -\sum_{\mathbf{k}'} W_{\mathbf{k}\mathbf{k}'}\left(\mathbf{s}_{\mathbf{k}} - \mathbf{s}_{\mathbf{k}'}\right). \tag{IV.73}$$

The second term on the left describes the spin precession, originating from the commutator in Eq. (IV.71). The third term describes spin drift due to external fields, while the fourth term describes spin diffusion due to the presence of inhomogeneous spin distribution. The above equation is a generalization of the spin drift-diffusion transport equations discussed in Sec. II.B.

### D.2 Solution for spin relaxation

Since we consider elastic scattering only, the momentum scattering rate, $W_{\mathbf{k}\mathbf{k}'}$ connects momenta of the same magnitude (we also assume isotropic solid). We can then work on a single energy surface, $\varepsilon_k = \hbar^2 k^2/2m$, and the momentum variables are the azimuthal ($\vartheta$) and polar ($\varphi$) angles. This allows us to effectively use the decomposition of the density matrix, $\rho_{\mathbf{k}}$, of electrons with momentum $\mathbf{k}$, as,

$$\rho_{\mathbf{k}} = \overline{\rho} + \rho_{1\mathbf{k}}, \quad \overline{\rho}_{1\mathbf{k}} = 0. \tag{IV.74}$$

The bar denotes averaging over the directions of $\mathbf{k}$; $\overline{\rho}$ represents the isotropic part of the density matrix. This part results due to momentum relaxation which rounds up anisotropies—those are in turn represented by $\rho_{1\mathbf{k}}$. It is $\overline{\rho}$ which describes the spin relaxation process.

In the following we consider homogeneous case, $\partial \rho_{\mathbf{k}}/\partial \mathbf{r} = 0$, and no external fields present, $\mathbf{F}_{\mathbf{k}} = 0$. For such a situation the averaging over $\mathbf{k}$ of Eq. (IV.71) gives,

$$\frac{\partial \overline{\rho}}{\partial t} - \frac{1}{i\hbar}\overline{\left[H_{1\mathbf{k}}, \rho_{1\mathbf{k}}\right]} = 0, \tag{IV.75}$$

where we used that $\overline{H}_{1\mathbf{k}} = 0$, since $H_{1\mathbf{k}}$ is an odd function of $\mathbf{k}$; the right-hand side of Eq. (IV.71) vanishes identically upon averaging. Substitute now Eq. (IV.74) into Eq. (IV.75) to obtain the following equation for $\rho_{1\mathbf{k}}$:

$$\frac{\partial \rho_{1\mathbf{k}}}{\partial t} + \frac{1}{i\hbar}\overline{\left[H_{1\mathbf{k}}, \rho_{1\mathbf{k}}\right]} - \frac{1}{i\hbar}\left[H_{1\mathbf{k}}, \overline{\rho}\right] - \frac{1}{i\hbar}\left[H_{1\mathbf{k}}, \rho_{1\mathbf{k}}\right] = -\sum_{\mathbf{k}'} W_{\mathbf{k}\mathbf{k}'}\left(\rho_{1\mathbf{k}} - \rho_{1\mathbf{k}'}\right). \tag{IV.76}$$

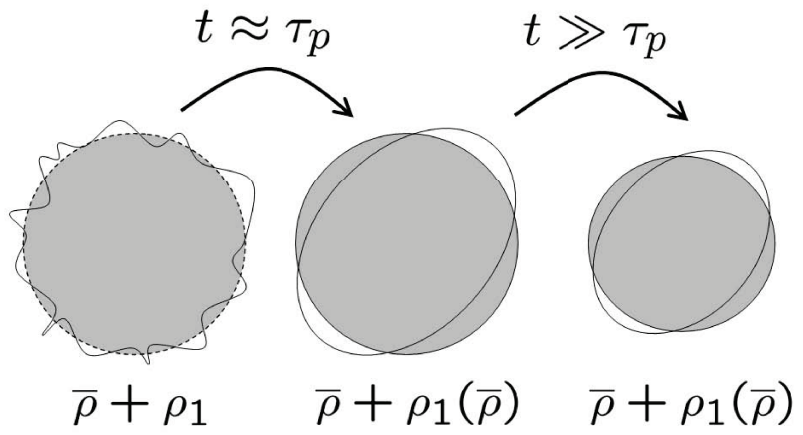

Fig. IV.4. The anisotropic part $\rho_1$ of the density matrix relaxes after the momentum relaxation time $\tau_p$ towards the quasistatic value $\rho_1 = \rho_1[\overline{\rho}]$. At longer times, $\rho_1$ evolves with time according to the time evolution of the uniform part $\overline{\rho}$, which, in turn, decays exponentially towards equilibrium within the spin relaxation time: $\rho_1(t) = \rho_1[\overline{\rho}(t)]$.

The anisotropic part $\rho_{1\mathbf{k}}$ decays towards instantaneous equilibrium defined by $\overline{\rho}$ very fast—on the order of the momentum relaxation time $\tau_p$ (the process is the equilibration of momenta, keeping spin intact). This scheme of solution is depicted in Fig. IV.4. Let us find this quasistatic value of $\rho_{1\mathbf{k}}$ by setting the time derivative in Eq. (IV.76) to zero and solve the resulting algebraic equation. We will see that $\rho_{1\mathbf{k}} \sim H_{1\mathbf{k}}$, and since $H_{1\mathbf{k}}$ is a small perturbation, we can neglect all terms containing the product $H_{1\mathbf{k}}\rho_{1\mathbf{k}}$ as negligible in the first approximation. We obtain

$$\frac{1}{i\hbar}\left[H_{1\mathbf{k}}, \overline{\rho}\right] = \sum_{\mathbf{k}\mathbf{k}'} W_{\mathbf{k}\mathbf{k}'}\left(\rho_{1\mathbf{k}} - \rho_{1\mathbf{k}'}\right). \tag{IV.77}$$

We look for the solution of Eq. (IV.77) in the form of the ansatz

$$\rho_{1\mathbf{k}} = \rho_{1\mathbf{k}}[\overline{\rho}] = \frac{\tau^*}{i\hbar}\left[H_{1\mathbf{k}}, \overline{\rho}\right]. \tag{IV.78}$$

Our task will be to determine the unknown parameter $\tau^*$, which has the dimension of time. To be specific, let us consider the spin Hamiltonian to be polynomial in the magnitude of the momentum (the angular dependence remains unspecified):

$$H_{1\mathbf{k}} \sim k^l, \tag{IV.79}$$

with integer $l$. The linear Bychkov-Rashba and Dresselhaus terms have $l = 1$, while the cubic Dresselhaus term has $l = 3$. We look here, as an example, at a three-dimensional (bulk) case for which $l = 3$, and expand the spin Hamiltonian in spherical harmonics:

$$H_{1\mathbf{k}} = \sum_m C_{lm} k^l Y_m^l(\vartheta, \varphi), \tag{IV.80}$$

where the angles $\vartheta$ and $\varphi$ are the angular coordinates of the momentum $\mathbf{k}$ in spherical coordinates, and the expansion parameters $C_{lm}$ are spin matrices of dimension $2 \times 2$.

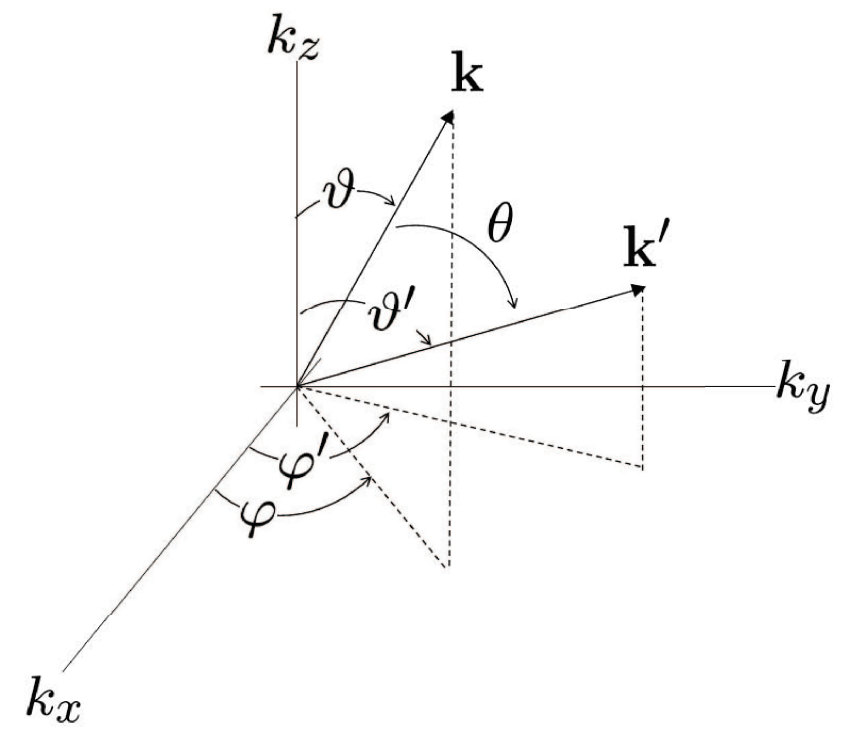

Fig. IV.5. Scheme of the scattering geometry, defining the spherical angles of the wave vectors $\mathbf{k}$ and $\mathbf{k}'$

The ansatz, Eq. (IV.78), leads to the following equations:

$$\rho_{1\mathbf{k}} = \frac{\tau^*}{i\hbar}\sum_m k^l Y_m^l(\vartheta,\varphi)\left[C_{lm},\overline{\rho}\right], \tag{IV.81}$$

$$\rho_{1\mathbf{k}'} = \frac{\tau^*}{i\hbar}\sum_m k^l Y_m^l(\vartheta',\varphi')\left[C_{lm},\overline{\rho}\right]. \tag{IV.82}$$

Since we consider elastic scattering, the magnitudes of the momenta are the same: $k = k'$; only the spherical angles differ. See Fig. IV.5 for the definition of spherical angles. Substituting the spherical expansions into Eq. (IV.77), we obtain the following formula for $\tau^*$ (which we further call $\tau_l^*$, to emphasize the dependence on the polynomial rank $l$):

$$\tau_l^* \sum_m \left[C_{lm},\overline{\rho}\right]\sum_{\mathbf{k}'} W_{\mathbf{kk}'}\left[Y_m^l(\vartheta,\varphi) - Y_m^l(\vartheta',\varphi')\right] = \sum_m \left[C_{lm},\overline{\rho}\right] Y_m^l(\vartheta,\varphi). \tag{IV.83}$$

We now use our earlier assumption that the scattering is isotropic, so that $W_{\mathbf{kk}'} \to W(\theta)$, where $\theta$ is the angle between $\mathbf{k}$ and $\mathbf{k}'$, see Fig. IV.5. Then

$$\sum_{\mathbf{k}'} W_{\mathbf{kk}'} \ldots \to \int \frac{d\Omega'}{4\pi} W(\theta)\ldots, \tag{IV.84}$$

where $d\Omega' = \sin\vartheta' d\vartheta' d\varphi'$ is the infinitesimal solid angle around $\mathbf{k}'$.

We can expand the function $W(\theta)$ in Legendre polynomials $P_l$,

$$W(\theta) = \sum_{l=0}^{\infty} W_l P_l(\cos\theta), \quad W_l = \frac{2l+1}{2}\int_{-1}^{1} W(\theta) P_l(\cos\theta), \tag{IV.85}$$

and use the addition theorem (Jackson, 1998) to factor out the individual angles $\vartheta$ and $\vartheta'$ from $\theta$:

$$P_l(\cos\theta) = \frac{4\pi}{2l+1}\sum_{m=-l}^{l} Y_{lm}^*(\vartheta',\varphi') Y_{lm}(\vartheta,\varphi). \tag{IV.86}$$

Using the orthonormality of spherical harmonics, $\int d\Omega Y^*_{lm}(\Omega)Y_{l'm'}(\Omega) = \delta_{ll'}\delta_{mm'}$, we readily obtain the following identity:

$$\int Y_{lm}(\vartheta', \varphi')W(\theta)\frac{d\Omega'}{2\pi} = Y_{lm}(\vartheta, \varphi)\int_0^\pi W(\theta)P_l(\cos\theta)\sin\theta d\theta. \tag{IV.87}$$

With this identity we can readily solve Eq. (IV.83), to give

$$\frac{1}{\tau_l^*} = \frac{1}{2}\int_{-1}^{1} d\cos\theta\, W(\theta)\left[1 - P_l(\cos\theta)\right]. \tag{IV.88}$$

If $l = 1$, we would get $\tau_l^* = \tau_p$, recovering the momentum relaxation time.

Knowledge of the parameter $\tau^*$ gives us the quasistatic solution for $\rho_{1\mathbf{k}}$, through Eq. (IV.78). We can now substitute this solution into Eq. (IV.75), describing the time evolution of the isotropic component of the density matrix. We obtain

$$\frac{\partial\overline{\rho}}{\partial t} = -\frac{\tau_l^*}{\hbar^2}\overline{[H_{1\mathbf{k}}, [H_{1\mathbf{k}}, \overline{\rho}]]}. \tag{IV.89}$$

This is the needed decay equation for the density matrix describing the slow, compared to momentum, spin relaxation. Let us see if we can transform the above into Bloch equations; compare also with the Born-Markov approximation, Eq. (IV.14).

For electrons with a given magnitude of the momentum $k$ (or energy $\varepsilon_k = \hbar^2k^2/2m$), it is relevant to consider only the averaged expectation value of the spin,

$$\mathbf{s} = \overline{\mathrm{Tr}\{\rho_{\mathbf{k}}\hat{\mathbf{s}}\}} = \mathrm{Tr}\{\overline{\rho}\hat{\mathbf{s}}\}, \tag{IV.90}$$

due to the fact that the momentum relaxation is much faster than spin relaxation. We can then write,

$$\overline{\rho} = \frac{1}{2} + \mathbf{s}\cdot\boldsymbol{\sigma}. \tag{IV.91}$$

Considering that our spin-dependent part of the Hamiltonian given by Eq. (IV.62), we obtain for the commutator in Eq. (IV.89),

$$[H_{1\mathbf{k}}, [H_{1\mathbf{k}}, \overline{\rho}]] = -\frac{1}{2}\hbar^2\left[(\boldsymbol{\Omega}_{\mathbf{k}}\cdot\mathbf{s})\,\Omega_{\mathbf{k}} - \Omega_{\mathbf{k}}^2\mathbf{s}\right]. \tag{IV.92}$$

We thus obtain the decay equations,

$$\frac{d\mathbf{s}}{dt} = \tau_l^*\left[(\boldsymbol{\Omega}_{\mathbf{k}}\cdot\mathbf{s})\,\Omega_{\mathbf{k}} - \Omega_{\mathbf{k}}^2\mathbf{s}\right]. \tag{IV.93}$$

To be specific, let us look at the decay equation for $s_z$:

$$\frac{ds_z}{dt} = -\tau^*\left[s_z\overline{\Omega_{\mathbf{k}}^2 - \Omega_{\mathbf{k}z}^2} - s_x\overline{\Omega_{\mathbf{k}x}\Omega_{\mathbf{k}z}} - s_y\overline{\Omega_{\mathbf{k}y}\Omega_{\mathbf{k}z}}\right]. \tag{IV.94}$$

Recall that the overline denotes angular averages, so

$$\overline{\Omega_{\mathbf{k}\alpha}\Omega_{\mathbf{k}\beta}} = \int\Omega_{\mathbf{k}\alpha}\Omega_{\mathbf{k}\beta}\frac{d\Omega}{4\pi}, \tag{IV.95}$$

and $\alpha$ and $\beta$ denote cartesian coordinates and $\Omega$ is the solid angle.

By comparing the decay equations, Eq. (IV.93), with Bloch equations given in Sec. IV.A we can write for the spin relaxation times,

$$\frac{1}{\tau_{s,\alpha\beta}} = \gamma_l^{-1}\tau_p\left(\overline{\Omega_{\mathbf{k}}^2} - \overline{\Omega_{\mathbf{k}\alpha}^2}\right), \tag{IV.96}$$

$$\frac{1}{\tau_{s,\alpha\neq\beta}} = -\gamma_l^{-1}\overline{\Omega_{\mathbf{k}\alpha}\Omega_{\mathbf{k}\beta}}. \tag{IV.97}$$

Above we denoted,

$$\gamma_l = \tau_p/\tau^*, \tag{IV.98}$$

a measure of the relative importance of the momentum relaxation and the effective randomization of the axis of $\Omega_{\mathbf{k}}$. Usually we say that the correlation time in the Dyakonov-Perel mechanism is given by the momentum relaxation. This is true only as an order of magnitude is concerned, since in general $\tau_l \neq \tau_p = \tau_1$. The correlation time is the time of randomization of the spin-orbit field. Different scattering events contribute differently to the spin randomization.

The physics comes in the choice of $\mathbf{\Omega_k}$. Note that $1/\tau_s$ is in general a tensor. For cubic systems (such as zinc-blende semiconductors like GaAs) it is a scalar, with the off-diagonal elements vanishing–as one can show explicitly with the Dresselhaus $\mathbf{\Omega_k}$. We remind that the Dresselhaus spin-dependent Hamiltonian for the conduction electrons in zinc-blende systems is,

$$\mathbf{\Omega_k} = \alpha_c\hbar^2(2m_cE_g)^{-1/2}\boldsymbol{\kappa}, \tag{IV.99}$$

where vector $\boldsymbol{\kappa}$ is given by,

$$\boldsymbol{\kappa} = \left[k_x(k_y^2 - k_z^2), k_y(k_z^2 - k_x^2), k_z(k_x^2 - k_y^2)\right]. \tag{IV.100}$$

It is a nice exercise to show that,

$$\overline{\kappa_i^2} = \frac{4}{105}k^6. \tag{IV.101}$$

With that we obtain for the spin relaxation rate,

$$\frac{1}{\tau_s} = \frac{32}{105}\gamma_3^{-1}\tau_p\alpha_c^2\frac{\varepsilon_{\mathbf{k}}^3}{\hbar^2E_g}. \tag{IV.102}$$

Recall that we are dealing with three-dimensional semiconductor such as GaAs for which the relevant spin-orbit field is cubic in momentum, that is, $l = 3$. The fact, that the spin relaxation rate is proportional to the momentum relaxation time (or correlation time, $\tau_3$), suggests that the origin of the Dyakonov-Perel mechanism is indeed motional narrowing. In this specific case, the spin relaxation strongly increases with increasing of the energy. For degenerate semiconductors, $\varepsilon_{\mathbf{k}} \approx \varepsilon_F$, and the temperature dependence of the spin relaxation time is given by the temperature dependence of $\tau_p$. For nondegenerate semiconductors one needs to perform ensemble averaging of the above result, essentially replacing $\varepsilon_{\mathbf{k}}$ with the thermal energy $k_BT$. Ramifications of the Dyakonov-Perel spin relaxation mechanism for specific cases of interests, as well as for the two-dimensional case, can be found in (Žutić *et al.*, 2004). The mechanism is studied in great detail in (Meier and Zakharchenya (Eds.), 1984).

Finally, we will show that the spin diffusion length in the Dyakonov-Perel mechanism does not depend on $\tau_p$, that is, on the degree of disorder. Indeed, writing schematically $1/\tau_s = \overline{\Omega^2}\tau_p$, gives

$$L_s = \sqrt{D\tau_s} = \sqrt{\frac{v_F^2 \tau_p}{\overline{\Omega^2}\tau_p}} = \frac{v_F}{\sqrt{\overline{\Omega^2}}}. \tag{IV.103}$$

Since the diffusion parameter $D \sim \tau_p$ and the spin relaxation time, $\tau_s \sim 1/\tau_p$, their product is independent on the momentum relaxation rate.

## E. Spin relaxation in semiconductors

The most studied semiconductor for spin relaxation is GaAs. This case study was given in detail in our reference review (Žutić *et al.*, 2004). There have been since several important developments. One is the measurement of spin relaxation in GaAs by analyzing Faraday-rotation noise spectroscopy (Oestreich *et al.*, 2005); for the theoretical treatment see (Braun and König, 2007). Another is the influence of electron-electron interactions on spin relaxation. This is discussed in more detail below. With respect to the recent demonstration of electrical spin injection into silicon by Appelbaum *et al.* (2007), we also include a compilation of experimental results of spin relaxation in silicon, in Sec. E.2.

### E.1 Electron-electron interaction effects in spin relaxation in GaAs

Most studies of spin relaxation in semiconductors have focused on impurity (somewhat less phonon) mediated spin flips. Recently it has been observed that electron-electron interactions as well play important role in spin relaxation and dephasing in semiconductor quantum wells. The coulomb interaction influences spin relaxation in two major ways: (i) The electron-electron scattering leads to additional momentum relaxation and, via the motional narrowing of the Dyakonov-Perel type, to inhibited spin dephasing (Glazov and Ivchenko, 2002) as measured recently in n-GaAs/AlGaAs quantum wells (Stich *et al.*, 2007; Leyland *et al.*, 2007). In a magnetic field the electron-electron scattering effect on the spin dephasing has been considered by (Wu and Ning, 2002). (ii) While the electron-electron exchange (Hartree-Fock) interaction preserves spin, in electron systems of high spin polarization the exchange leads to a momentum-dependent effective magnetic field.[83] This field, while containing a random component, points in one direction—of the spin polarization—and reduces spin dephasing, not unlike the external magnetic field reduces spin dephasing in the toy model of Sec. B.2. This effect, predicted in (Weng and Wu, 2003; Weng *et al.*, 2004) and elaborated by (Glazov and Ivchenko, 2004), has been recently observed (Stich *et al.*, 2007).

The experiment of Stich *et al.* (2007) measured spin relaxation time of electrons in GaAs/ $Al_{0.3}Ga_{0.7}As$ modulation doped quantum wells, oriented along the [001] direction. The width of the well was 20 nm, while the conduction electron density was $n = 2.1 \times 10^{11}$ cm$^{-2}$. Shining circularly polarized light perpendicular on the quantum well creates spin-polarized electrons and holes, by the process known as optical orientation or optical pumping (Meier and Zakharchenya (Eds.), 1984; Žutić *et al.*, 2004). Roughly, the photon angular momentum is transferred to electrons and holes, due to the presence of spin-orbit coupling in GaAs.[84] In optical spin pumping, which is optical spin orientation of n-doped samples, the resulting spin polarization depends on the excitation rate: the more electrons we pump into the conduction band, the higher is the spin polarization (Žutić *et al.*, 2004). The densities of the photoexcited spin-polarized electron-hole pairs in the experiment of Stich *et al.* (2007) ranged from $9 \times 10^9$ to $6 \times 10^{11}$ cm$^{-2}$; the highest density is higher than the equilibrium density. At the highest excitation density the initial electron spin polarization was about 30%.

[83] Unlike the effective spin-orbit fields, the exchange field breaks time reversal symmetry, as it results from nonequilibrium spin polarization, which, by itself, breaks the time symmetry.

How is the spin detected? A nice tool to observe spin decay, or spin precession in real time, is the so-called Faraday rotation. This technique has proven extremely useful in spintronics, leading to fundamental observations[85] of spin coherent transport (Kikkawa and Awschalom, 1999; Awschalom and Kikkawa, 1999), spin relaxation and dephasing (Kikkawa and Awschalom, 1998; Beschoten *et al.*, 2001), and, recently, the so-called spin Hall effect (Kato *et al.*, 2004; Sih *et al.*, 2006; Sih and Awschalom, 2007). We have already seen nice examples of the related Kerr rotation spectroscopy of electrical spin injection in Sec. II.F.1.

The scheme of a Faraday rotation experiment is shown in Fig. IV.6. The pump pulse, incident close to normal, is circularly polarized, generating spin-polarized electrons and holes, manifested as the magnetization of the sample. At a delay time $\Delta t$ (which can be as small as a hundred femtoseconds or a picosecond) a probe light pulse, of much smaller intensity, is applied. The probe, which is also incident close to the normal direction, is linearly polarized. As the probe pulse transmits through the sample, the polarization axis of the pulse (the direction of the electric field of the photons) rotates in the polarization plane, proportional to the magnetization of the sample. Detecting this Faraday rotation angle at different time delays gives direct information about the time dependence of the magnetization.

The results of the Faraday rotation experiment of Stich *et al.* (2007) are shown in Fig. IV.7. At small time delays the magnetization decays rather fast, presumably due to the fast spin relaxation of photoexcited holes (Žutić *et al.*, 2004). At longer times the exponential decay can be fitted to give information about spin relaxation time, here denoted as $T_2^*$, although the figure shows a plain decay of the spin, not decay of coherent oscillations (which were also detected in (Stich *et al.*, 2007)). The oscillations seen at small times and for small electron polarizations $P$ are due to the ensemble electron spin oscillations about the average spin-orbit field $\Omega$, as first seen by Brand *et al.* (2002). These oscillations are in the opposite limit to the limit of the Dyakonov-Perel mechanism: whereas in the latter the spin precession during momentum relaxation is slow, the coherent oscillations require that the spin of an electron in a given momentum state rotates by the spin-orbit field $\Omega_{\mathbf{k}}$ at least a full circle before being scattered.

[84] Electrons in the conduction band have spin 1/2, while in the valence (so-called heavy hole) band of the quantum well they have spin or rather angular momentum, of 3/2. A circularly polarized light to the right carries angular momentum 1, able to excite an electron with spin -3/2 from the valence band to the spin -1/2 state of the conduction band. A hole (missing electron) of spin +3/2 and an electron of spin -1/2 are created, resulting in the spin-polarization of charge carriers. The spin is oriented along the direction (in fact, opposite) of the photon momentum.

[85] Often (magnetoptic) Kerr rotation is used instead of Faraday rotation. The only difference between the two is that while in Faraday rotation one detects the rotation of the polarization plane of the light transmitted through a magnetized sample, in Kerr rotation one looks at the reflected light. One just needs to be careful to distinguish spin injection from the magneto-optic effects which could be intrinsic to the ferromagnetic films used as spin injectors (Tanner *et al.*, 2006).

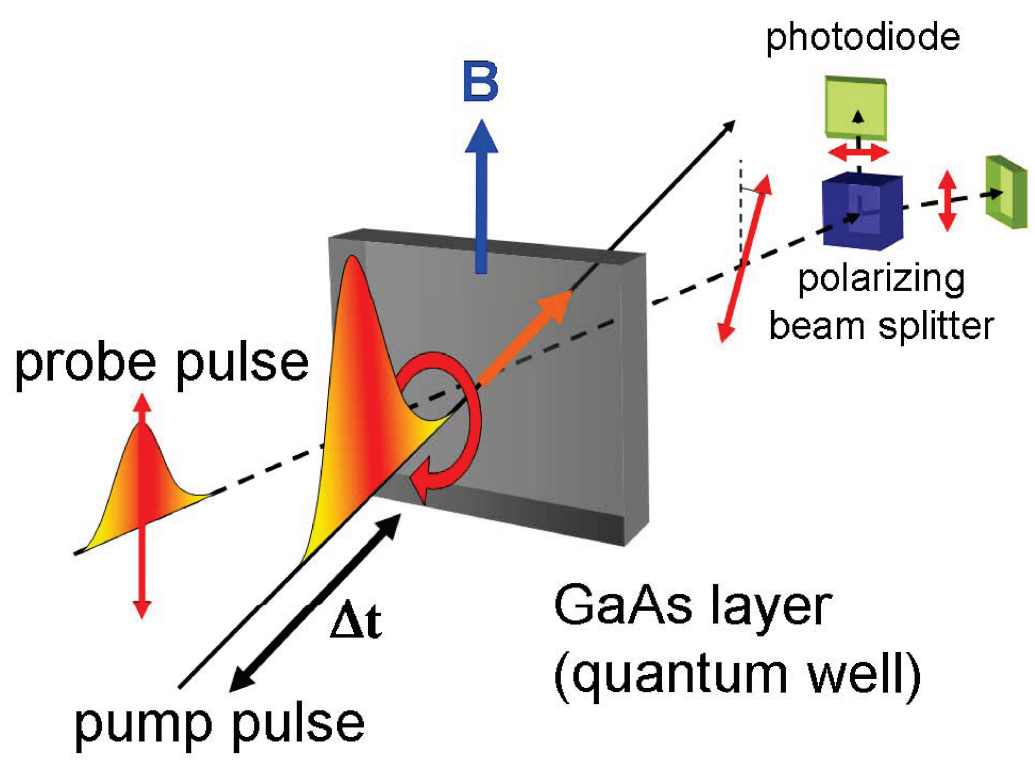

Fig. IV.6. Scheme of the pump and probe Faraday rotation experiment. The pump pulse of a circularly-polarized light generates spin-polarized carriers in the quantum well; the spins are oriented perpendicular to the plane of the well, along the light propagation direction. The probe pulse of linearly polarized light is applied after a time delay. The rotation angle of the polarization angle of the transmitted probe light through the wall is proportional to the magnetic moment (spin) of the quantum well. If, in addition, an external magnetic field $B$ is applied, the magnetization precesses, giving an oscillating Faraday angle. The decay of the oscillation gives the spin dephasing time $T_2^*$. Courtesy of Christian Schüller.
.

The most important result of Fig. IV.7 is the strong dependence of the electron spin relaxation time on the initial spin polarization $P$. The higher the $P$ is, the weaker is the spin relaxation. A microscopic calculation confirms that this behavior is quantitatively consistent with processes (ii) of the electron-electron exchange coupling induced effective magnetic field inhibiting spin relaxation, (Stich *et al.*, 2007).

### E.2 Spin relaxation in silicon

Silicon is not the best studied semiconductor for spin relaxation. The best case study is GaAs. But because GaAs has been extensively covered in Žutić *et al.* (2004), and due to the relatively unknown experimental data on spin relaxation of conduction electrons in silicon, as well as the expanding interest in silicon spin relaxation stirred by the recent reports of electrical spin injection into silicon (Appelbaum *et al.*, 2007), we present its case here.

Spin relaxation in silicon is less efficient than in GaAs. The silicon g-factor is about 0.001 below the free electron value, Eq. (I.2),[86] indicating weak spin-orbit coupling (compare to $g \approx -0.44$ for GaAs). The Elliott relation, Eq. (IV.56), then suggests that the spin relaxation time in silicon is about $10^5$ greater than the momentum relaxation time. For a picosecond momentum relaxation one should expect a 100 nanosecond spin relaxation time, which is what is observed, as described below.

[86] For example, in (Young *et al.*, 1997) the value of $g = 1.9995$ ($\Delta g = -2.8 \times 10^{-3}$) was found for conduction electrons at 3.5 K and 125 K. For comparison, $g = 1.99875$ ($\Delta g = -3.6 \times 10^{-3}$) was found for electrons in the impurity band of silicon at 1.3 K (Feher, 1959), independent of the type of donor. It had long been believed that the g-factors of conduction electrons and electrons in the impurity band of shallow donors are the same, since the shallow impurity levels forming the impurity band are derived from the conduction band states. Finally, the g-factors of electrons on shallow donor levels and in the impurity band differ very little.

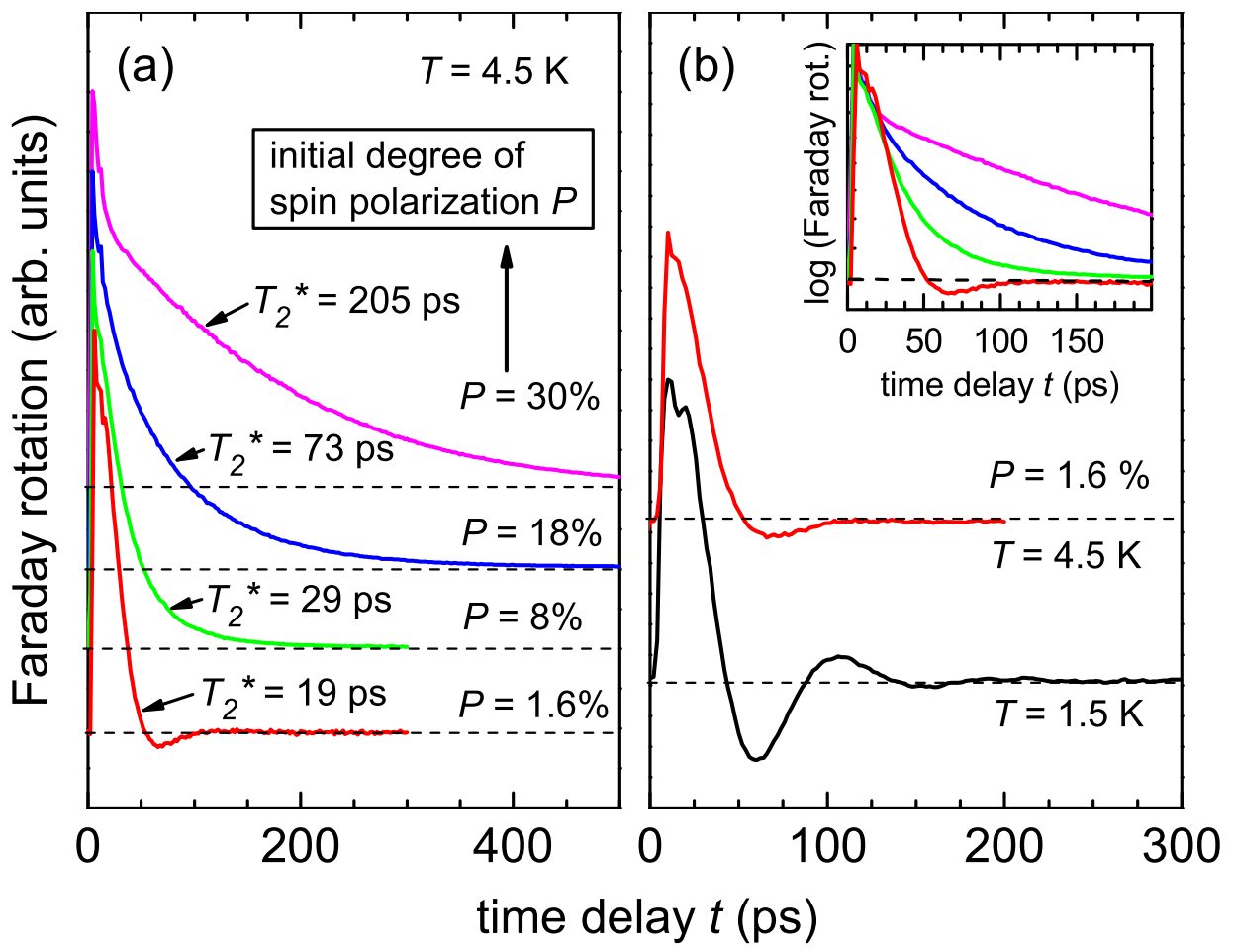

Fig. IV.7. Left figure (a) shows the decay of the magnetization (detected as the Faraday rotation angle) of the photoexcited GaAs quantum well at 4.5 K for different initial spin polarizations $P$. The inhomogeneous dephasing times $T_2^*$ are extracted from the exponential decay at longer times. In the right figure the time evolution of magnetization for the $P = 1.6\%$ polarization and two different temperatures is shown. The magnetization (spin) precession of the electron ensemble due to spin-orbit fields $\Omega$ are visible at times below 200 ps. The inset is the logarithmic plot of the left figure, showing the exponential decay at long times. Reprinted figure with permission from D. Stich *et al.*, *Physical Review Letters* **98** 176401 (2007). Copyright 2007 by the American Physical Society.

The silicon samples used for conduction electron spin resonance (CESR)[87] experiments below, were doped with phosphorus (P) donors. Each donor contributes electronic states within the silicon band gap; the ground state for the P donor in silicon is 45 meV below the conduction band edge (Kittel, 1996). For low doping densities, say, $10^{16}$ cm$^{-3}$, and temperatures below about 150 K, a sizable fraction of electrons reside in the donor states (ground or excited). For high doping levels, the donor levels broaden to form an impurity band. This happens, for P donors in silicon, at the critical concentration of $n_c = 3.7 \times 10^{18}$ cm$^{-3}$ (Kittel, 1996). For doping densities below $n_c$, electrons are confined on the donor sites; for doping densities above $n_c$ the electrons become itinerant (albeit with heavy effective masses) with electron states described by extended wave functions in the impurity band. The behavior at doping densities less than $n_c$ is termed insulating, while the behavior at densities greater than $n_c$ is termed metallic. The transition at $n_c$ is then referred to as the metal-to-insulator transition (MIT). At temperatures higher than about 150 K, most electrons are in the conduction band.

[87] In a spin resonance experiment an external magnetic field casues Zeeman splitting of the electron energy states. Electromagnetic radiation of the frequency tuned to the energy splitting shows a pronounced absorption peak, as a function of frequency (in actual experiments usually the magnetic field is varied, keeping the frequency locked). The peak coincides with the splitting, giving us information about the electron g-factors, while the width of the peak is proportional to the spin relaxation rate. For conduction electrons the splitting occurs at microwave energies. Spin resonance experiments with nuclei employ radiofrequency waves.

Thus far the most comprehensive experimental study of spin relaxation of conduction electrons in silicon has been reported by Lepine (1970), for low doping densities in the ranges of $10^{14}$ to almost $10^{17}$ cm$^{-3}$. At low temperatures such densities correspond to the insulating regime. The results are shown in Fig. IV.8, where we plot the spin relaxation time as converted from the measured CESR linewidths. To convert a CESR linewidth into the spin relaxation time, one can use the formula,

$$\alpha \frac{1}{T_1} = \gamma \Delta B = 18 \times \Delta B[\text{gauss}]\ \text{MHz}, \tag{IV.104}$$

where $\gamma = g\mu_B/\hbar$ is the electron gyromagnetic ratio and $\alpha$ is a parameter, of order one (typically varies between one and two), reflecting the conditions of the spin resonance experiment and the definition of the linewidth.[88] For the linewidth of 1 gauss (typical order of magnitude observed in silicon), the lifetime can then be 50 to 100 ns, depending on $\alpha$. Since Lepine (1970) reports observing a single Lorentzian line of the resonance spectrum, we have used $\alpha = 1$ in Eq. (IV.104) to convert the absorption linewidth to $T_1$; the reported linewidth is the half width at half maximum (Lepine, 2007).

As seen from Fig. IV.8, the spin relaxation time decreases with increasing doping density, for temperatures below 200 K. At higher temperatures $T_1$ appears rather insensitive to doping. The maximum, $T_1 \approx 100$ ns, is observed for the most dilute sample, at around 100 K. This maximum spin relaxation time occurs at roughly the same temperature for all studied doping densities. The region above 150 K is the region dominated by electrons in the conduction band, as all the donor levels are thermally excited. This temperature range is dominated by phonon scattering, with the spin relaxation described by the Elliott-Yafet mechanism, as suggested by Lepine (1970). At room temperature, the observed spin relaxation time is about 10 ns, large enough for spintronics applications.

At lower temperatures the spin relaxation physics is different. For the lowest temperatures, below about 50 K, the observed $T_1$ is the spin relaxation of electrons in the ground state of the donor levels. The mechanism is the hyperfine interaction with nuclei of the phosphorus dopants, which have spin of one half. At the temperatures of 50 to 75 K, it is again the hyperfine interaction which is responsible for $T_1$ of the donor states, but now the interaction is motionally narrowed (this is why $T_1$ increases with increasing temperature) by exchange coupling with conduction electrons. According to Lepine (1970), the region around the maximum $T_1$, of temperatures between 75 and 150 K, corresponds to the spin relaxation of electrons in the first excited donor state. The mechanism is the spin-orbit interaction of that state, assisted by thermal excitation to the conduction band and by the decay to the donor ground state. The opposite temperature trends for the motional narrowing of the hyperfine-interaction at low temperatures ($T_1$ increasing with $T$) and the phonon-induced Elliott-Yafet spin relaxation a high temperatures ($T_1$ decreases with $T$ as the phonons become more populated), results in the peak observed at about 100 K.

[88] The connection between the linewidth and the spin relaxation rate is quite subtle, as it depends on the actual observed shape of the microwave absorption power. The shape, in turn, depends on the relative magnitudes of the skin depth, the spin diffusion length, and the sample thickness. A sample thinner than the skin depth gives a symmetric Lorentzian absorption line for which a full linewidth gives $\alpha = 2$. If the sample is thick, it is best to make a fit to the theoretical asymmetric (Dysonian) shape, as this depends on other parameters, such as the ratio of the skin depth and the spin relaxation length. If the spin relaxation length dominates, the frequency derivative of the Dysonian shape has the full linewidth of $\alpha \approx 1$. The CESR theory has been worked out by Dyson (1955) and a nice introduction to the variety of possible line shapes can be found in Feher and Kip (1955).

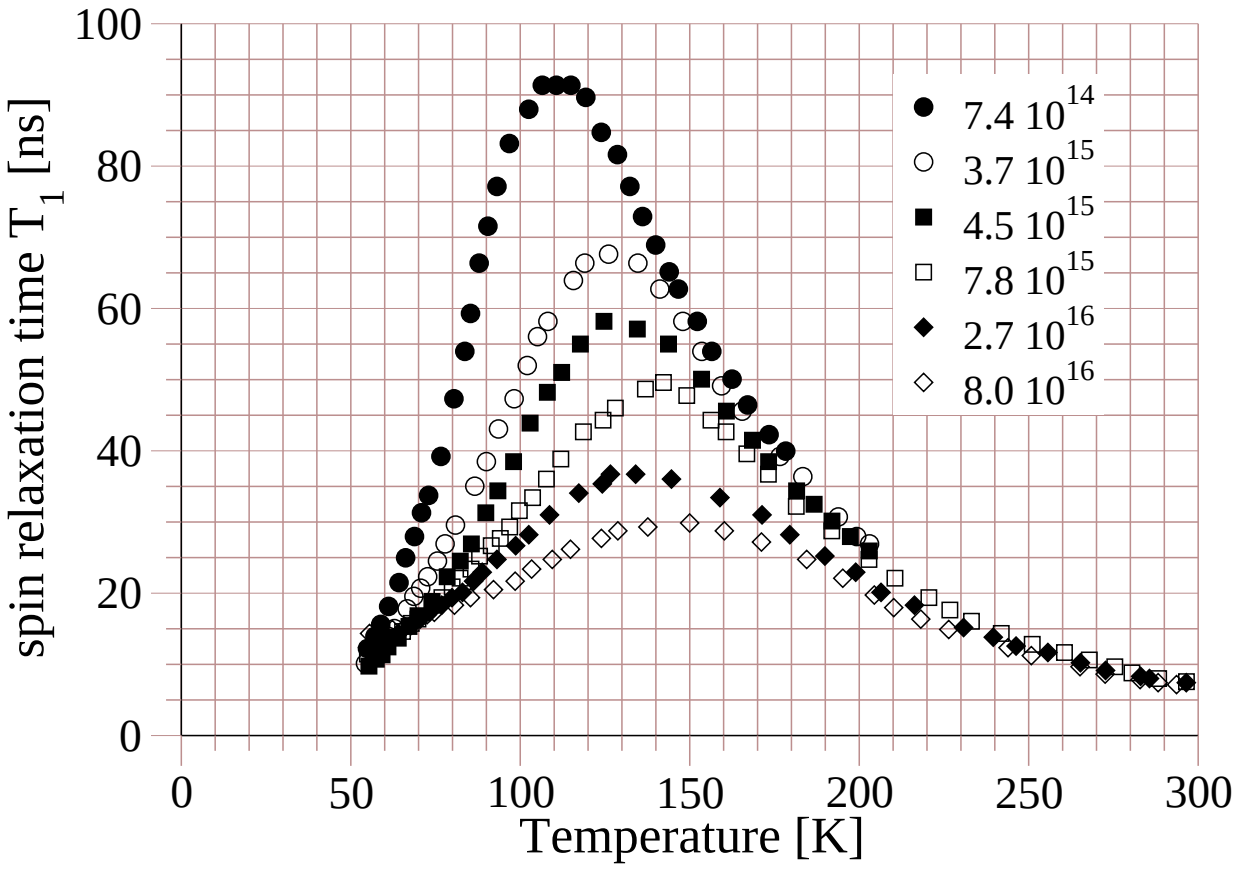

Fig. IV.8. Temperature dependence of the spin relaxation time in lightly doped silicon (doping density indicated), extracted using Eq. (IV.104) by taking $\alpha = 1$, from the conduction electron spin relaxation linewidths (half widths at full maximum) reported in (Lepine, 1970). While individual experimental data are not shown in (Lepine, 1970), the symbols in the graph above are sampled points from the reported continuous curves.

Let us look closer at the high temperature region which describes conduction electrons spin relaxation useful for spintronics applications. Yafet predicted that the Elliott-Yafet spin relaxation mechanism in semiconductors such as Si, in which the minimum of the conduction band is not at $\mathbf{k} = 0$, should change with temperature as (Yafet, 1963)

$$\frac{1}{T_1(\Delta g)^2} \sim T^{5/2}. \tag{IV.105}$$

This temperature dependence is due to phonons and the symmetry of the electron-phonon spin-flip matrix elements. Figure IV.9 shows CESR extracted spin lifetimes for a P doped silicon, in which a constant-temperature background is removed (Lancaster *et al.*, 1964). Spin relaxation at high temperatures is due to phonons. Yafet's dependence, Eq. (IV.105), seems to work reasonably well (Lancaster *et al.*, 1964), although the spin relaxation time appears to decrease somewhat faster with increasing temperature than the $T_1 \sim T^{-5/2}$ law. The comparison between the experiment and Yafet's relation is even more striking after taking into account the temperature variation of $\Delta g$ and plotting $1/T_1\Delta(g)^2$, as was done in (Ochiai and Matsuura, 1978).

The CESR measurements of Lancaster *et al.* (1964) are for dopings an order of magnitude higher than those of Lepine (1970). Nevertheless, $T_1$ in Fig. IV.8 agrees with $T_1$ in Fig. IV.9; see the inset of this figure. This suggests that the spin relaxation is not sensitive to the doping density, up to densities of $10^{17}$ cm$^{-3}$. Another CESR study, of heavily phosphorus doped silicon, shows the experimental linewidths at high temperatures increasing with increasing doping (Ochiai and Matsuura, 1978). There appears to be a trend emerging from these works, that the linewidth (and $1/T_1$) does not vary much with doping at dopant concentrations below the critical density of the metal-to-insulating transition, $n_c$, while the linewidth is increasing with increasing doping density above $n_c$.

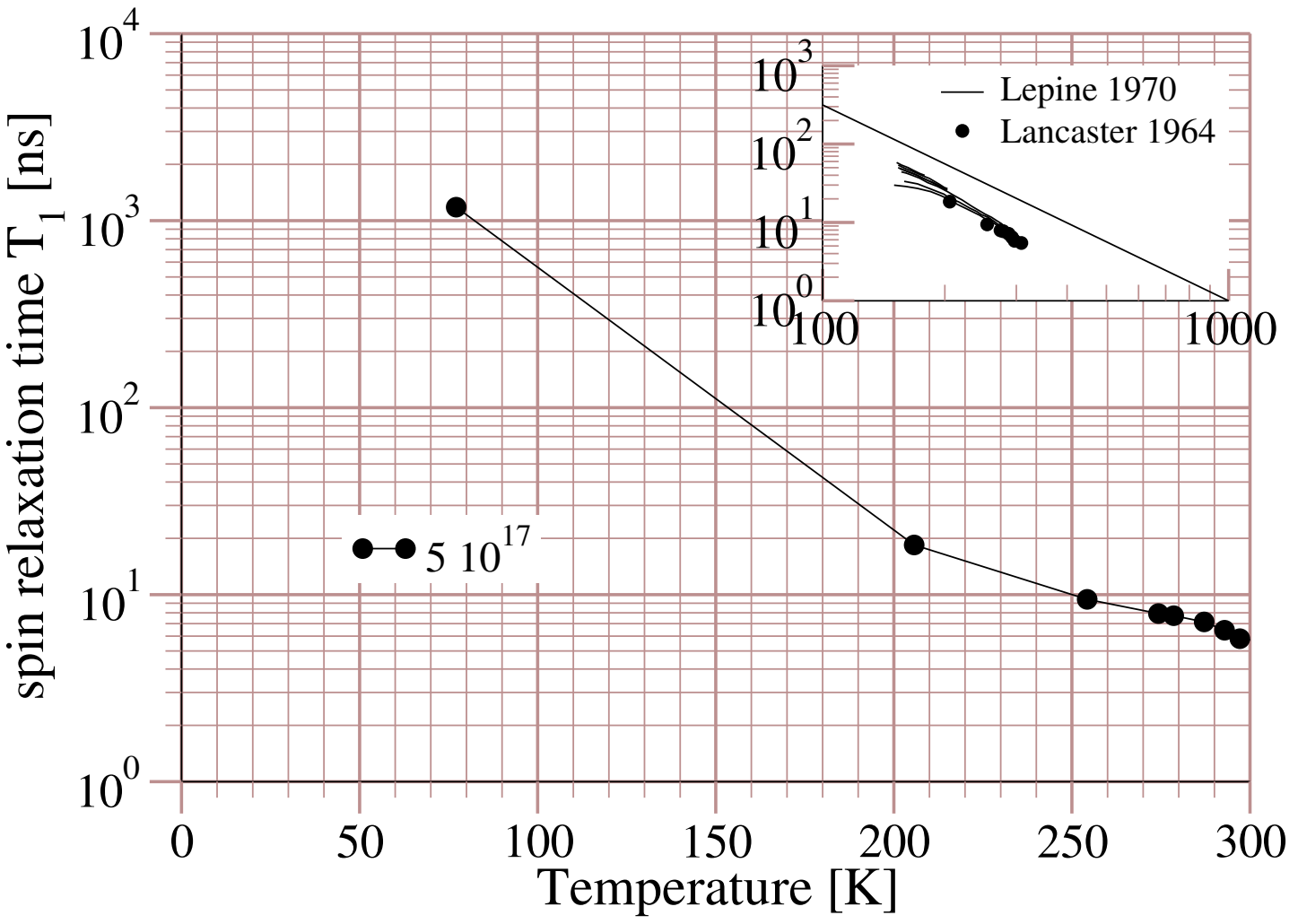

Fig. IV.9. Temperature dependence of the spin relaxation time in phosphorus doped silicon with the doping density of $5 \times 10^{17}$. The data are extracted from the CESR linewidths of Lancaster *et al.* (1964), taking $\alpha = 2/\sqrt{3}$ in Eq. (IV.104), since the reported linewidth is the distance between the maximum slopes of the Lorentzian line shape. The inset is the log-log plot of the high-temperature region (the datum at 75 K is absent), along with a straight line indicating Yafet's power law, $T_1 \sim T^{-5/2}$. Data from Fig. IV.8, for temperatures above 150 K, are also included for reference.

The spin relaxation of conduction electrons in silicon has been investigated by electron spin resonance also in the region around the metal-to-insulator transition at temperatures below 4.2 K for As (Zarifis and Castner, 1987) and Sb (Zarifis and Castner, 1998) dopants. At these critical doping concentrations the spin relaxation appears to be mainly due to the motional narrowing of the exchange interaction, as well as due to the impurity spin-orbit interaction. CESR studies were also performed at higher temperatures, up to 300 K, for densities near the MIT (Ochiai and Matsuura, 1976). As far as theoretical understanding goes, the spin relaxation around the MIT has received little attention. An earlier study looked at spin relaxation in a strongly disordered conductor, treating nuclear and spin-orbit induced spin flips, concluding that spin relaxation in strongly disordered Si:P can be described by the latter mechanism, the inter-valley[89] spin-orbit scattering, which is the Elliott-Yafet mechanism of spin relaxation. One recent study has shown that the Dyakonov-Perel mechanism, operating in semiconductors without the center of symmetry (such as GaAs), with hopping conductivity instead of conduction band transport, persists to the MIT region (Shklovskii, 2006), while a spin relaxation mechanism on the theme of Elliott-Yafet, with hopping-assisted spin-orbit induced spin flips, has been proposed for electrons in the impurity band, close to the MIT (Tamborenea *et al.*, 2007).

[89]The term valley refers to one of the six ellipsoids forming the Fermi surface of silicon (Kittel, 1996). The scattering between the valleys is caused by impurities in dirty samples. The spin-flip is allowed by the silicon-atom induced spin-orbit coupling.

Spin relaxation was also measured for heavily doped silicon (more than $10^{19}$ P donors in $cm^3$), at temperatures below 77 K, finding that the relaxation rate increases in proportion to the donor density (Quirt and Marko, 1972), indicating the impurity-dominated (not host-dominated) Elliott-Yafet spin relaxation processes. Similar findings were reported for P and As doped Si (Pifer, 1975). The effects of Fe and Mn impurities on Si:P spin relaxation have been studied by CESR (Kennedy and Pifer, 1975), extracting the spin-flip scattering cross section of silicon conduction electrons on iron impurities. An early CESR study of heavily doped Si:P, below 100 K, (Ue and Maekawa, 1971) found a linear increase of the spin relaxation rate with increasing temperature. Such an increase could be normally attributed to phonons (although with a different power, see Eq. (IV.105)), but since it was observed at low temperatures, the authors concluded it must be due to the presence of localized magnetic moments in their samples.

In addition to spin resonance studies, spin relaxation in bulk silicon has also been extracted from spin-valve and Hanle effect signals, in an electrical spin-injection scheme (Appelbaum *et al.*, 2007; Huang *et al.*, 2007c), as discussed in Sec. II.(F.2). The largest value (better, the lower bound on the spin relaxation time) reported thus far is 202 ns at 85 K, and 65 ns at 150 K, corresponding to coherent spin transport through 350 $\mu$m of a pure (nominally undoped) silicon sample (Huang *et al.*, 2007c), more than enough for spintronics applications. These long spin relaxation times are consistent with the above discussed CESR results on lightly doped silicon, considering that the spin injection studies employ very clean, float zone samples, in which one expects, based on the trends exhibited in Fig. (IV.8), highest lifetimes. In comparison, bulk GaAs has the largest spin relaxation times of about 200 ns at low doping densities close to the MIT (which is about $2 \times 10^{16}$ $cm^{-3}$ for GaAs) and low temperatures, around 4 K; the spin relaxation times decrease to about 1 ns at 100 K (Kikkawa and Awschalom, 1998), which is two orders of magnitudes smaller than in silicon. Similarly for metals. Say, aluminum at 100 K has the spin relaxation time of about 1 ns (Fabian and Das Sarma, 1999a), and 85 ps at room temperature (Fabian and Das Sarma, 1999a; Jedema *et al.*, 2002, 2003). Although, as shown by the CESR experiments, the spin relaxation time decreases with increasing dopant concentration, silicon appears far superior to GaAs, or any other studied conductor, in terms of the longevity of the conduction electron spin. A comprehensive discussion of spin lifetimes in metals and semiconductors can be found in the review (Žutić *et al.*, 2004).

Much more systematic investigations of spin relaxation in silicon have been performed for silicon quantum wells, such as formed by Si/Ge heterostructures (Jantsch *et al.*, 1998; Sanderfeld *et al.*, 2000; Wilamowski and Jantsch, 2002; Tyryshkin *et al.*, 2005). The main spin-relaxation mechanism is found to be the Dyakonov-Perel one, due to the effective Bychkov-Rashba spin-orbit coupling[90] induced by the loss of inversion symmetry at the heterostructure interfaces (Wilamowski *et al.*, 2002). Spin dephasing times $T_2$ as large as 3 $\mu$s, and spin relaxation times $T_1$ of 2.3 $\mu$s have been reported at 4.2 K (Tyryshkin *et al.*, 2005). The fact that $T_2 > T_1$ arises apparently from the anisotropy of the two-dimensional electron system. The Bychkov-Rashba spin-orbit field, which gives the fluctuating magnetic field in our toy model (see Sec. B.2), gives vanishing fluctuating fields in the direction perpendicular (call it $z$) to the quantum well plane, $\overline{\omega_z^2} \approx 0$, so that one can expect $T_2 = 2T_1$ (see Eqs. (IV.36), (IV.37), and (IV.38)) if no perpendic-ular fluctuating fields are present, indicating that the anisotropy is not perfect. At present it is not clear why the spin relaxation in two-dimensional Si/Ge electron gases is an order of magnitude greater than in the bulk, discussed above.

[90]The extracted Bychkov-Rashba parameter for the studied silicon quantum wells was $0.55 \times 10^{-12}$ eV cm, two to four orders of magnitude smaller than found in III-V heterostructures.

### F. Spin relaxation of an electron confined in a quantum dot

An electron confined in a semiconductor quantum dot (Reimann and Manninen, 2002) is a promising system for potential applications in quantum information processing (Loss and DiVincenzo, 1998; Nielsen and Chuang, 2000). Especially the spin degree of freedom in quantum dots (Das Sarma *et al.*, 2001; Cerletti *et al.*, 2005; Hanson *et al.*, 2007; Fabian and Hohenester, 2005) has attracted attention for two reasons. First, the electron spin provides a natural quantum two level system, suitable for encoding the information bit. Second, the spin is less strongly coupled to the environment compared to electron orbital degrees of freedom, thus providing longer coherence time. This is crucial for quantum computation – the time needed for an elementary operation, such as a controlled spin flip, has to be much smaller than a time after which the information initially encoded in the spin is lost. Disturbances due to environment thus introduce stringent limits on any practical realization of a quantum bit.

Bearing in mind this importance, we will study in this Section the influence of a dissipative environment on an electron confined in a semiconductor quantum dot. We first define spin and orbital relaxation, discuss their main mechanisms in quantum dots, and why they differ by orders of magnitude. Since we want to assess the possibility for exploitation of a confined quantum dot electron spin as a qubit, we will focus on the spin relaxation, which sets an ultimate timescale limit for using the spin in quantum computation. We will learn that at large magnetic fields (above tesla) the relaxation is caused by coaction of phonons and spin-orbit interactions. In magnetic fields below tesla, this channel is not effective and the magnetic moments of nuclei of the material are dominant in disturbing the electron spin.

Then we review the recent experiments in measuring the single electron spin relaxation time and check our theoretical understanding. First such single particle measurement was done in 2004 and since then quite a few results were obtained manifesting rapid experimental progress. From the results follows that relevant spin computation setups, being one, or two electrons in single or double quantum dot, with relaxation times of milliseconds are reproducible with current technology. Even more importantly, coherence times of microseconds were demonstrated here, supporting the initial hopes that using spin instead of charge offers the advantage of a strong isolation of the qubit from the environment. We will also discuss in detail several spin-to-charge conversion schemes, which solve the problem of readout of such isolated spin entity.

Finally, we present a theoretical description, using the density matrix formalism, which allows us to analytically study the role of dissipation in the interesting phenomena of Rabi oscillations and photon assisted tunneling, which play an important role in the manipulation and probing of a spin qubit. Bursts of resonant fields lead to Rabi oscillations allowing controlled coherent spin flips, what was also experimentally demonstrated in quantum dot spin qubits. This is another important achievement showing that lateral quantum dots are serious candidates for physical qubit realization.

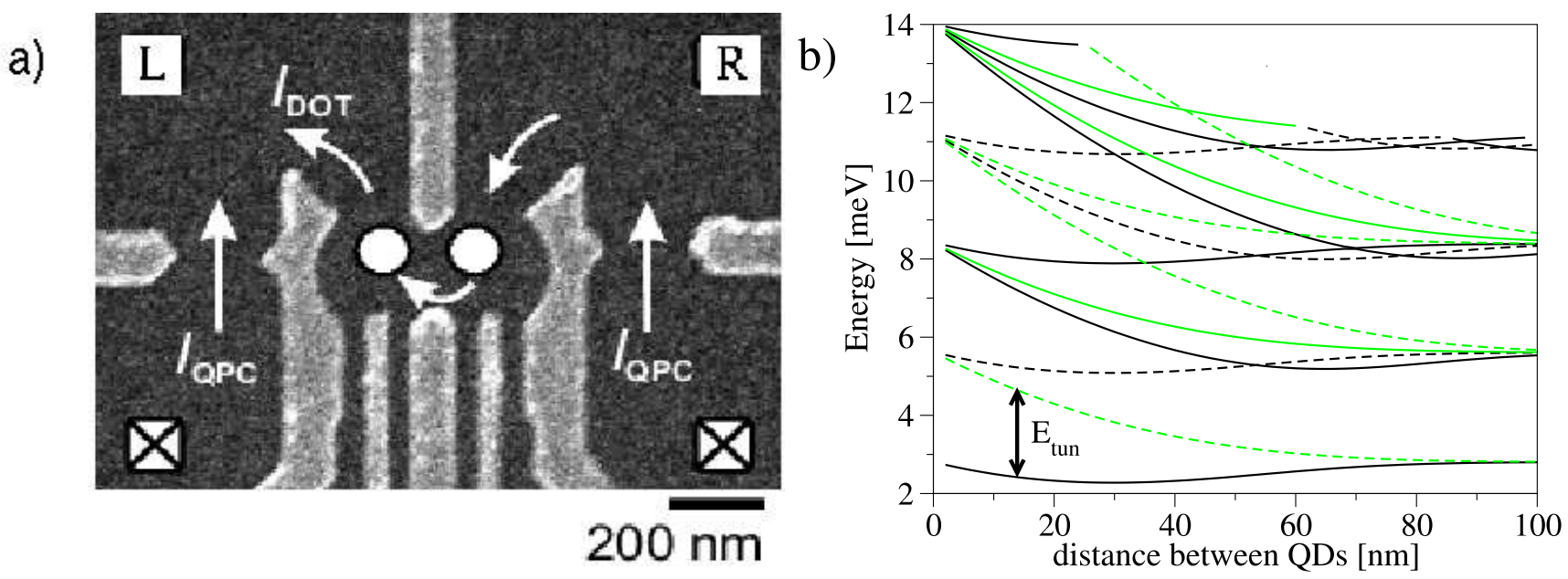

Fig. IV.10. a, Double lateral quantum dot in GaAs/AlGaAs semiconductor heterostructure. The gates (lighter color) provide electrostatic potential with minima (white dots), where electrons can be localized. Electrons can enter the dot from leads (L and R – reservoirs of electrons kept at fixed chemical potential) through narrow contacts (white arrows entering/leaving the dots). Thus, if a voltage across the dot is applied, current $I_{\rm DOT}$ flows through. The quantum point contact (QCP) currents $I_{\rm QPC}$ are used as meters of the charge on the dots. Scanning electron micrograph picture, from Ref. (Hanson *et al.*, 2007), courtesy of R. Hanson. b, Energy spectrum of a symmetric double dot as a function of the interdot distance. Different types of lines denotes states with different symmetry. See Ref. (Stano and Fabian, 2005) for details.

### F.1 Mechanisms of spin relaxation in quantum dots

A semiconductor quantum dot is an electrostatically created potential minimum, where electrons can be trapped. A typical experimental setup is in Fig. IV.10a. Electrons confined to a two dimensional plane of the GaAs/AlGaAs heterostructure are further bound to a small region, typically tens of nanometers, by electrostatic top gates. Electrons can enter the dots one by one, their number being monitored by nearby quantum point contact (QPC): the more charge is on the dot, the smaller is the current through a QPC, due to the Coulomb repulsion between the electrons on the dot and ones flowing through the QPC. The great advantage of lateral quantum dots is in their versatility – the shape of the confining potential, the coupling between individual dots and between the dots and nearby leads can be controlled electrically to large extent by applying voltages on the gates. This enables quick changes between different dot configurations allowing wide control over electron states.

To illustrate the control on a particular example, we give in Fig. IV.10b the theoretically obtained energy spectrum of a double dot such as in Fig. IV.10a, as a function of the interdot distance. The interdot distance can be changed applying voltage on the middle gate in Fig. IV.10a – more negative voltage pushes the two minima further apart. At zero interdot distance, in the left on in Fig. IV.10b, the degeneracies in the spectrum reveal the circular symmetry of the potential addopted in the theoretical model. On the other hand, for very large interdot distances the two dots are isolated and the spectrum is that at the zero distance doubly degenerate. For finite distances the dot is in the true double dot regime. Here an electron placed initially in one of the dot coherently oscillates between the two dots, with the frequency proportional to the energy difference of the two lowest levels (tunneling energy $E_{\rm tun}$). It can be seen in Fig. IV.10b how the tunneling energy can be controlled by changing the interdot distance.

In the first approximation, if we suppose there is one electron on the dot, the electron Hamil-tonian $H_e$ consists of the confining potential, defined by the gates, and the kinetic energy, defined by the crystal. Compared to a free particle, in addition to the leading term quadratic in momenta, the kinetic energy contains also other terms polynomial in momenta, as discussed in Sec. III.B. for GaAs. The electron time evolution can be written independently on the details of the Hamiltonian as

$$\Psi(t) = \sum_{j=1}^{\infty} \alpha_j e^{-\mathrm{i}E_j t/\hbar} \Psi_j. \tag{IV.106}$$

Here, at time $t = 0$ the wavefunction is a certain superposition, given by the coefficients $\alpha_j$, of eigenstates $\Psi_j$ of the electron Hamiltonian, $H_e \Psi_j = E_j \Psi_j$. The whole time dependence is a phase linearly growing with time for each eigenstate in the expansion. The probability to find the electron in a certain eigenstate will be time independent – there will be no transitions between states.

The electron is, however, immersed in a condensed matter environment with its own dynamics. Phonons, impurity charge fluctuations, or nuclei are few examples of entities interacting with the electron whose influence is not included in $H_e$. Such interactions will change the electron wavefunction non-trivially. A change is called inelastic, if it involves a transfer of energy. The corresponding timescale is called relaxation time and denoted $T_1$. On the other hand, processes preserving the electron energy are called elastic and happen on the timescale of the decoherence time $T_2$. For a nondegenerate spectrum, the relaxation is a process by which moduli of coefficients $\alpha$ in Eq. (IV.106) are changed, while the decoherence time quantifies time after which the phases start to differ considerably from the linear evolution in Eq. (IV.106). Such a definition of transition times is not unambiguous and more time scales expressing the environment influence on the electron are used (Žutić *et al.*, 2004). An important example is the dephasing time $T_2^*$. It appears if the decoherence is measured on an ensemble of dots (or if the result is obtained by averaging over many different measurements on a single dot). In such a case the measured signal decays if the individual dots feel different local environments (for example different nuclear magnetic fields), even if each of the dots evolves coherently. In principle, the decay of the signal due to the dephasing can be removed,[91] for example, by the spin echo technique (Petta *et al.*, 2005).

The relaxation and decoherence also reflect the difference between classical and quantum information bit. If classical information is encoded in the qubit, it is contained in the states occupations and is lost after the relaxation time. Quantum information processing is much more demanding – here the information is encoded also in the phase of the states. Therefore, the information lifetime is the decoherence time, which can easily be orders of magnitude smaller than the relaxation time. To assess the possibility to use quantum dot qubits in information processing, it is thus a crucial task to quantify the two timescales and find conditions maximizing them.

### *F.1.1 Spin relaxation*

We will now focus on the relaxation and decoherence of spin. We will consider the two lowest electron eigenstates, denoted by $\Psi_1$ and $\Psi_2$, where the first is the ground state having, say, spin up, while the second is the same orbital state with the opposite spin. These two states are energetically split by the Zeeman energy due to an applied magnetic field. The two logical states of the qubit (0 and 1) can be then encoded into the electron states $\Psi_1$ and $\Psi_2$, realizing a spin qubit. Starting initially with, say, $\Psi_2$, looking at the system at a later time, we could find it in state $\Psi_1$. Such transition happens due to interactions with the environment, with a rate given by the inverse of the relaxation time. Note that if no magnetic field is applied and the two states $\Psi_1$ and $\Psi_2$ are degenerate, the timescale of transitions between them is the decoherence time.

[91] The possibility (at least in principle) of the removal can be taken as the definition for the dephasing, distinguishing it from the decoherence.

One can similarly encode the information using different orbital states, using instead of $\Psi_2$ an excited orbital state $\Psi_3$ with the same spin as the ground state. This would correspond to a charge qubit and the orbital relaxation time would quantify timescales for transitions between $\Psi_1$ and $\Psi_3$. The advantage of using spin is in much longer relaxation time, with the explanation given in next paragraphs, where we list the main channels for the spin relaxation. We start with phonons since, compared to other environment fluctuations, phonons can not be get rid off in the crystal and define a fundamental upper limit for the relaxation time.

**Phonons.** Phonons and electron can interact in several ways. A phonon is an oscillating wave-like deformation of the crystal lattice. On the other hand, compression of a crystal changes the band structure (bands are shifted). If the amount of the crystal compression depends on the position, as depicted in Fig. IV.11a, it results in space dependent bands shifts, or, in another words, electric field. The interaction of a charged particle with this field is called the deformation potential. As follows from the explanation, as the phonon wavevector goes to zero, the induced electric field vanishes, since the lattice deformation becomes homogeneous. Similarly, if the crystal atoms are charged, an inhomogeneous displacement induces dipole moments in the material. This (Fröhlich) interaction is relevant only for optical phonons, for which the neighboring atoms move out of phase, see Fig. IV.11c, whereby their displacement is substantially larger than for acoustic phonons. A homogeneous deformation can induce electric field too, by asymmetrically distorting positive and negative charge distributions in a polar material, such as GaAs, see Fig. IV.11b. This is denoted as piezoelectric effect, and finishes our list of the electron-phonon interactions (Grodecka *et al.*, unpublished; Mahan, 2000), which can all be viewed as the electron interacting with an electric field induced by phonons. However, the electric field does not couple directly to the electron spin – for that a further spin-dependent mechanism is needed. Such mechanisms can be divided into two main groups (Khaetskii and Nazarov, 2000, 2001; Khaetskii, 2001) for a more comprehensive discussion of phonon-induced spin dephasing and relaxation see, for example, (Semenov and Kim, 2007).

First, suppose there is a spin-dependent term in the electron Hamiltonian, such as the spin-orbit interaction, that does not allow to define a common spin quantization axis. Then the electron eigenstates are not Pauli spin like. Nevertheless, one can always attach labels to a state, such as "spin up" or "spin down", according to the state spin expectation value along the applied magnetic field direction. If the spin-orbit interaction is small, what is usually the case, such "spin" will be well defined since the spin expectation values will be close to $\pm\hbar/2$. A "spin down" state can be then written as a Pauli spin down state plus a small amount of the Pauli spin up state. This small *admixture* of Pauli spin opposite state allows the transition to a "spin up" through spin preserving phonons and gives the name for this *admixture mechanism* of the spin relaxation (Khaetskii and Nazarov, 2001). For free electrons, this corresponds to Elliott-Yafet mechanism, see Sec. C.

Second, say one neglects the influence of the spin mixing terms in the electron Hamiltonian and considers the electron eigenstates to be Pauli spins. Phonons can, under certain conditions, induce spin-dependent coupling to the electron (Frenkel, 1991) called direct spin-phonon coupling. This happens if the environment where the electron is localized is anisotropic in such a way that the phonon induces fluctuations of the parameters of the spin-dependent part in the electron Hamiltonian. The examples are an anisotropic g-factor (that is, the Zeeman energy depends on the magnetic field orientation) (Calero *et al.*, 2005), the phonon induced spin-orbit interaction (Alcalde *et al.*, 2004), ripple coupling originating from the different materials forming the heterostructure (Knipp and Reinecke, 1995; Woods *et al.*, 2002; Alcalde *et al.*, 2005), or phonon modulation of the hyperfine coupling (Semenov and Kim, 2004).

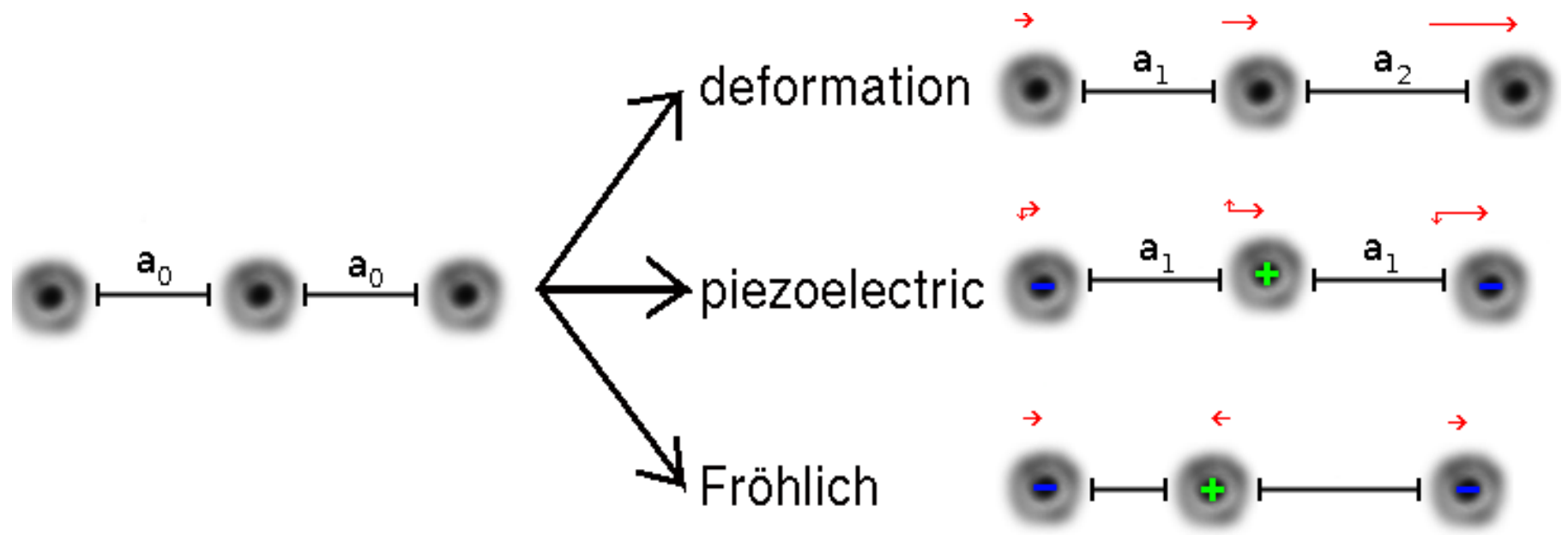

Fig. IV.11. Three ways how phonon induces electric field in the crystal. (i) Deformation potential: a position dependent crystal compression leads to position dependent bands shift. (ii) Piezoelectric interaction: A compression of crystal distorts positive and negative charges asymmetrically leading to non zero dipole field even if the deformation is homogenous. (iii) Fröhlich coupling to the optical phonon: Out of phase displacement of the oppositely charged neighbouring atoms produces a dipole field. Electron-phonon Hamiltonians of the three interactions are given in Sec. G.2.

The role of phonons can in principle be taken over by fluctuating electric fields of any other origin, for example fluctuations of the electric potential of the gates (Marquardt and Abalmassov, 2005; Borhani *et al.*, 2006) or the background charge fluctuations (Galperin *et al.*, 2006). While the direct coupling of the electron spin and those fields is absent in the leading order, the phonon and electric fluctuations have different spectral density. Phonon density drops with smaller phonon energy as a power law; the gates' potential fluctuations are mostly described as a white noise with a constant spectral density. Other baths can play role, such as ohmic fluctuations from a nearby current and 1/f background charge fluctuations (San-Jose *et al.*, 2006). In analogy with the phonons, also the fluctuating fields can couple to the electron spin directly: time dependent currents in the leads induce magnetic field, the spins of the electrons in the leads interact with the electrons through the exchange interaction (which requires the overlap of the confined and lead electrons) or dipole interaction (Onoda and Nagaosa, 2006) (no overlap is needed).

**Atomic nuclei.** The second important spin relaxation source are magnetic moments, if present in the region of the quantum dot. They couple directly to the electron spin through the dipole interaction. In magnetic diluted semiconductors, electrons confined at some crystal atoms have unpaired magnetic moments (Yang and Chang, 2005). However, even in non-magnetic materials an important spin thermal bath is present if the constituent atoms of the material have nuclear magnetic moments. All III-V semiconductors are such, while both most important IV group elements, silicon and carbon, have main isotopes with zero nuclear spin. The electron-hyperfine interaction is (Merkulov *et al.*, 2002) [compare also with Eq. (V.54)]

$$V^{\mathrm{hf}} = \sum_{j \in \mathrm{ions}} A_j \mathbf{S}_j . \boldsymbol{\sigma} \delta(\mathbf{r} - \mathbf{R}_j), \tag{IV.107}$$

where $j$ labels ions with a nuclear magnetic moment, $A_j$ is a material constant, $\mathbf{S}_j$ is the vector of operators for the nuclear spin, $\mathbf{r}$ is the electron position operator, and $\mathbf{R}_j$ is the position of the ion $j$. Due to the delta function, the coupling of the electron to a specific nucleus is proportional to the electron wavefunction at the nucleus position. Plugging in numbers for GaAs, it would seem that nuclear spins are much more efficient in spin relaxation than second-order processes including spin-preserving phonons. Fortunately from the perspective of long spin relaxation time, the energy conservation blocks a direct process where the electron and a nucleus flip their spins, since the electron and nuclear magneton differ by a factor of $\sim 2000$. Another co-acting interaction, for example, electron-phonon, is thus needed to provide the energy conservation leading to a similar second order process, as discussed in the previous part dedicated to phonons (Erlingsson *et al.*, 2001; Abalmassov and Marquardt, 2004).

Important difference compared to phonons is that now the thermal bath elements (the nuclear spins) have much longer lifetime, leading to long time correlation effects. Three main timescales (Merkulov *et al.*, 2002) can be identified in the mutual interaction between the electron and an ensemble of nuclear spins – a typical lateral quantum dot in GaAs contains $\sim 10^5$ of nuclear spins. (i) The shortest timescale, being $\sim 1$ ns, is the precession time of an electron spin in the magnetic field of the nuclear spins. On this timescale, the nuclear spins can be considered frozen and described by an effective magnetic field (Erlingsson and Nazarov, 2002; Khaetskii *et al.*, 2002; Coish and Loss, 2005). (ii) Second, $\sim 1\mu$s, is the time of precession of a nuclear spin in the magnetic field of the electron, being three orders of magnitude smaller due to electron wavefunction being delocalized over many nuclei. A simultaneous flip of the spin of both the electron and a nucleus can be used for a dynamical polarization of the nuclear spins (Eble *et al.*, 2006; Lai *et al.*, 2006). Due to the large number of the nuclear spins, some sort of a cut-off scheme is inevitable for analytical description of such mutually influencing nuclei-electron dynamics. Taking into account only pair wise interactions of the electron with a picked nucleus (Deng and Hu, 2006), instead of considering the whole set of nuclear spins, seems a possible way. A systematic formulation of such approach was recently done by a cluster expansion (Witzel *et al.*, 2005). (iii) The slowest is a nuclear spin precession in the dipole magnetic field of neighbor nuclei (Hüttel *et al.*, 2004), being $\sim 100\ \mu$s. It effectively leads to spin diffusion, by which the nulear spin bath thermalizes. The electron in the quantum dot strongly influences this diffusion (Lyanda-Geller *et al.*, 2002; Deng and Hu, 2005), but for the back action of the diffusion on the electron, no theoretical work exists, apart from an exact numerical simulation that can encompass only up to 20 nuclear spins (Dobrovitski *et al.*, 2006). In experiments (Ono and Tarucha, 2004) one has observed complicated electron behavior over long-times (seconds), reflecting the mutual influence between the three discussed processes.

It is natural to expect that having electron with certain spin, the nuclei will be ineffective in relaxing the electron spin, if they all are polarized in the same direction as the electron. The same applies to the electron decoherence which is dominated by non-uniform hyperfine couplings induced by space dependence of the electron wavefunction (Khaetskii *et al.*, 2002). Based on this finding, several schemes were proposed to suppress this decoherence channel by polarization or narrowing the quantum state[92] of the nuclear spins (Stepanenko *et al.*, 2006; Klauser *et al.*, 2006).

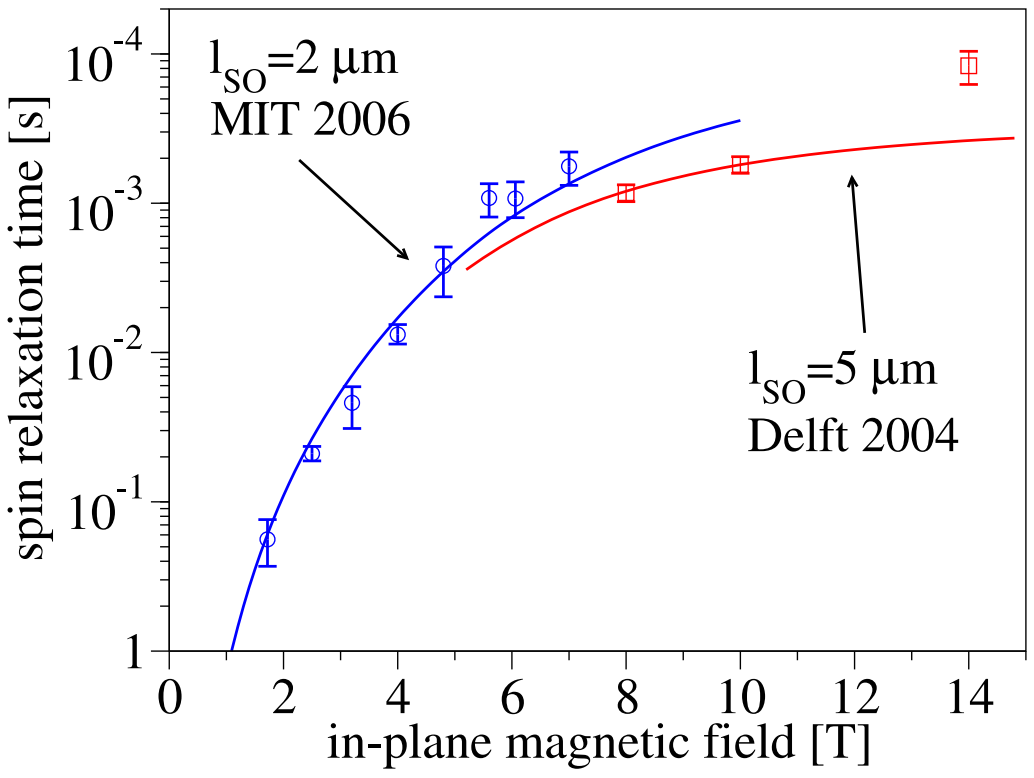

Fig. IV.12. Spin relaxation time for a single electron in a single lateral GaAs quantum dot. The two data sets (red and blue squares with error bars) were obtained in two different experiments, Refs. (Elzerman *et al.*, 2004), and (Amasha *et al.*, unpublished, cond-mat/0607110), respectively. The solid lines are theoretical results obtained by fitting one parameter, effective spin-orbit length $l_{SO}$, in a model where the spin relaxation is due to acoustic phonons and admixture mechanism (Stano and Fabian, 2006a,b).

**Dominant channels of spin relaxation.** Concluding from the previous part, both main sources of spin relaxation, that is phonons and nuclear spins (the other mentioned spin relaxation mechanisms turn out to be less important), are blocked in the leading order. This is the reason for the large discrepancy of spin and orbital relaxation times, which can be many (typically 6) orders of magnitude.

As for the dominating source in lateral dots in III-V semiconductors, such as GaAs, it seems that in magnetic fields above $\sim 1$ tesla, the acoustic phonons combined with the admixture mechanism due to the spin-orbit coupling is the dominant channel for the spin relaxation. The spin relaxation time is of order of 0.1 ms and analytical results fit experiments neatly (Amasha *et al.*, unpublished, cond-mat/0607110) – see Fig. IV.12. A peculiarity of this mechanism is an enhancement of the spin relaxation nearby a spin hot-spot (anti-crossing) (Bulaev and Loss, 2005), where the spin-orbit influence on the electron spin is much more profound. This results in strong anisotropy of the spin relaxation with respect to the orientation of the magnetic field (Golovach *et al.*, 2004) or electron momentum (Averkiev and Golub, 1999) (the latter is meaningful only for free electrons). This anisotropy could be used to control the spin relaxation, measure the strength of the spin-orbit interactions, and would be a definite proof of the spin relaxation origin.

Less clear situation is for sub-tesla magnetic fields, where nuclear spins combined with phonons are believed to dominate the spin relaxation. The ground for this belief comes from the decoherence measurements, where spin echo (Petta *et al.*, 2005) and a suppression of the decoherence by magnetic fields (Johnson *et al.*, 2005) are indications for the electron-nuclei interactions to play the dominant role.

[92]Narrowing means in this context to go from a statistical mixture towards a pure state.

*Spin relaxation in two electron QD:* From the point of view of the quantum computation a very important case is the spin relaxation in a two electron quantum dot – a transition between singlet and triplet states (Golovach *et al.*, unpublished). Surprisingly, such transitions in experimentally relevant double dot setup still lacks a comprehensive quantitative analysis. In a parabolic quantum dot the linear spin-orbit terms couple the ground state to different states according to selection rules (Climente *et al.*, 2007), which are loosen if the circular symmetry is lowered (Florescu and Hawrylak, 2006; Florescu *et al.*, 2004). The recent experiment (Meunier *et al.*, 2007) suggests that the double electron case is not qualitatively different from the single electron case. Further analytical work is needed to clarify the role of cubic Dresselhaus spin-orbit term, influence of the higher excited states (Climente *et al.*, 2007), appearance of a new spin-orbit interaction originating in the Coulomb electric field (Badescu and Reinecke, unpublished) or the possible non-spin-orbit origin of the spin relaxation (Hu and Das Sarma, 2006).

### *F.1.2 Orbital relaxation*

To complete the discussion, we now shortly comment on transitions between spin-like electron states which are not blocked by the spin conservation. The above discussed spin-dependent mechanisms can be neglected and the transition is induced by any fluctuation producing electric field. The main possible sources of such fluctuations are phonons, potential of the circuitry (confinement gates, measuring units like a quantum point contact), heterostructure background charge fluctuations (Fedorov *et al.*, 2003), and interactions with the leads. Apart from phonons, the previous mechanisms can be suppressed: improving the circuit, putting the dot farther away from the doped region, lowering the coupling to the leads, respectively, for the three listed sources. Phonons are always present and can be regarded as principally the dominant source of the orbital relaxation. The single electron relaxation time due to phonons is mainly given by the energy difference between the initial and final states. To compare with the spin relaxation, the orbital relaxation time is of order of $0.1$ ns for the same dot as considered to obtain data in Fig. IV.12. Other details, such as the shape of the potential and the magnetic field, have only minor influence. Apart from the single dot confining energy, in the double dot there is an additional handle to influence the relaxation – the distance between the two potential minima (Fedichkin and Fedorov, 2004; Stavrou and Hu, 2005). If the dot is populated by more electrons, the transition rates tend to decrease comparing to single electron case, since the Coulomb interaction mixes the lowest states with higher single electron orbitals (Bertoni *et al.*, 2005; Climente *et al.*, 2006).

Concerning the orbital decoherence, it is not dominated by phonons (Vorojtsov *et al.*, 2005; Liang, 2005), but the true source of the orbital decoherence is not yet clear, the most probable candidate are circuit potential fluctuations, making the decoherence rate strongly dependent on sample details.

## **F.2 Experiments on single electron spin relaxation**

After reviewing the sources, we discuss now experimental techniques for measuring the spin relaxation time. For that it is necessary to measure the state of the electron spin. This is, however, a nontrivial task since the electron magnetic moment is very small. To illustrate, the method of nuclear magnetic resonance (Vandersypen and Chuang, 2005), a state of art of magnetic moment detection, resolves only magnetic moments larger than those corresponding to roughly $10^7$ electrons. There are methods capable of single spin detection, for example using scanning tunneling microscope (Manassen *et al.*, 1989), or a refined version of the nuclear magnetic resonance (Rugar *et al.*, 2004), but we discuss here a different system, where the spin is observed optically, due to certain selection rules. The reason is that it introduces a useful idea of conversion of the spin state into other property of the system, allowing to detect the spin indirectly.

### *F.2.1 Example of an indirect spin observation: nitrogen vacancy defect*

The nitrogen vacancy defect is a charged defect in a diamond sample. An information of the spin of the defect state can be deduced from the detection of the luminescence connected with a transition between internal states of the defect (Epstein *et al.*, 2005). The ground state, $^3$A, includes singlet and triplet states split by a small energy difference, while $^3$E is an excited state, see Fig. IV.13. Important is that only the ground singlet state is active in the laser induced $^3$A-$^3$E transitions, while the triplet is a dark state. Then, if a laser is on, the presence/absence of the luminescence means the system is in singlet/triplet state. The observed luminescence signal switches spontaneously between being on and off reflecting intrinsic singlet-triplet transitions. The average time between the switches reveals the spin relaxation time, measured (Jelezko *et al.*, 2002) between milliseconds at room temperature and seconds at two Kelvins. If an additional oscillating field is applied, resonant with the ground state singlet to triplet transition, it induces coherent oscillations between these two states (we will discuss these so called Rabi oscillation in detail later), which are damped by the decoherence. From the time dependence of the damped signal, the decoherence time of 1 $\mu$s at 2K was obtained (Jelezko *et al.*, 2004; Hanson *et al.*, 2006). More importantly, we learn that it is possible to detect the spin state not by measuring its magnetic moment, but indirectly, by transforming the spin state into an optical signal. This is in a close analogy with spin-to-charge conversions used to detect spin states in quantum dots, what we discuss next.

### *F.2.2 Measuring spin relaxation in quantum dots*

Detection of the spin state in a quantum dot suffers the same problem of a very tiny magnetic moment, which was solved using a spin-to-charge conversion (Hanson *et al.*, 2007). Namely, the spin state is transformed into occupation (that is charge) of the dot, that is being measured after a specially designed sequence of voltage pulses applied on the gates. In fact, it is used that the spin is just a part of the label (quantum numbers) for the electron wave function – different states have all kinds of different properties not connected directly to the magnetic moment (energy, angular moment, spatial extent of the wave function, etc.). By measuring these properties, one can learn about the electron spin. We review now three spin-to-charge conversion schemes that have been realized in quantum dots.

**Transient current method: energy resolved readout (ERO).** The first method is the energy resolved readout, proposed and demonstrated in Ref. (Fujisawa *et al.*, 2001). A voltage is applied across the dot, which is connected to leads such that only the two lowest states are relevant for the current through the dot. A two step pulse (low and high negative gate voltage shifting the dot energy levels with respect to the chemical potential of the leads) is applied repeatedly, while the current is measured by averaging over many cycles. Changing the overall chemical potential of the leads, as depicted in Fig. IV.14, there are three possible configurations when a current can flow. First, the current flows through the ground state during the low voltage pulse. Second, for a higher leads potential, the current flows through the ground state during the high voltage pulse. In between these two there is the transient regime where the current can flow through the excited state during the low voltage pulse. However, now the current flows only until the ground state is populated. If this happens, the current is blocked until the next high voltage pulse, where the dot is emptied and the cycle is repeated. By prolonging the time interval of the low voltage pulse, a decay of the transient current allows to deduce the relaxation time for the transition from the excited to the ground state.

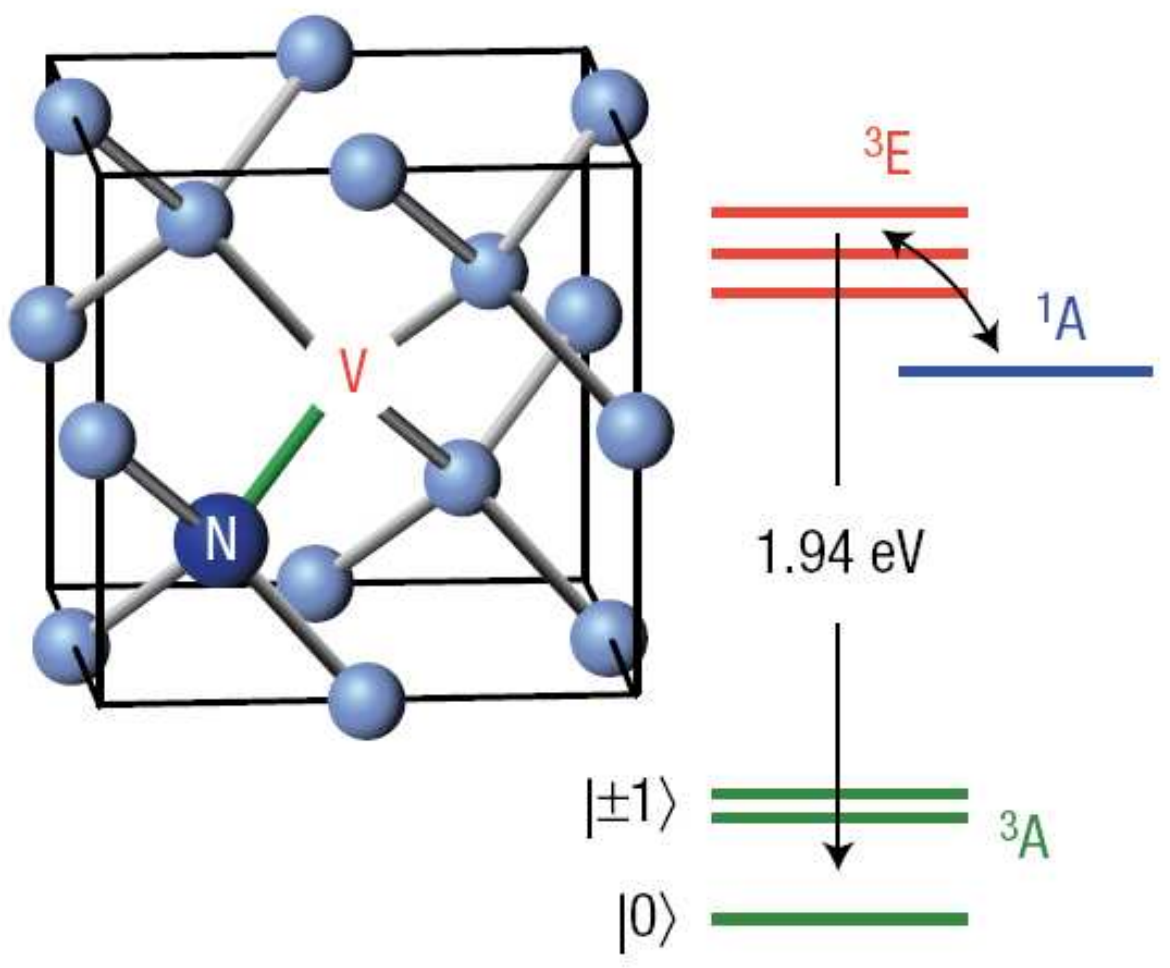

Fig. IV.13. Nitrogen vacancy defect. Atomic structure of the defect and energy levels. Figure from Ref. (Epstein *et al.*, 2005), courtesy of D. D. Awschalom. Reprinted by permission from Macmillan Publishers Ltd: *Nature Physics* **1**, *94 copyright (2005).*

The method is based on the fact that in the transient regime the ground and excited states are discriminated by their energies. In between these energies the chemical potential of the right lead is placed. Only if the electron is in the excited state, it can leave the dot into the right lead and contribute to the measured current. Exploitation of the energy difference gives the method its name. If the excited and ground states have different spins, the method realizes a spin-to-charge conversion.

The high voltage pulse in the transient regime is further denoted as the empty step, since the dot, possibly initially occupied, is emptied. Similarly, the low voltage pulse is denoted as the probe step, since the state of the electron is measured. Figure IV.15a summarizes the energy positions for the two steps (the fill&wait step will be discussed later). Figure IV.15b introduces definitions of tunneling rates needed for a quantitative description. The relaxation rate is denoted as $W$, while the tunneling rates to/from the leads are denoted by $\Gamma$ with indexes $L$ and $R$ standing for the left and right leads, and $E$ and $G$ for the excited and ground states. It is assumed that the tunneling rates are independent on the chemical potential of the leads and that the left lead is much more strongly coupled to the dot than the right dot, and also compared to the relaxation rate, $\Gamma_L \gg \Gamma_R, W$. The leads' chemical potentials define whether the electron can tunnel to/from the dot. For example, in the probe step the electron can tunnel to the excited state only from the left lead, while from the excited state it can tunnel only to the right lead or to the ground state. The last assumption is that there can be only one electron in the dot, meaning the charging energy (the energy needed to add a second electron into the dot) is much larger than the chemical potentials of the leads. In the probe configuration, populations of the ground $g$ and excited $e$ states are described by the following set of equations (for the comments on the derivation, see page 190; here we give the motivation for Eqs. (IV.108) by Fig. IV.15b):

$$\dot{e} = \Gamma_{LE}(1 - e - g) - (\Gamma_{RE} + W)e, \tag{IV.108a}$$

$$\dot{g} = (\Gamma_{LG} + \Gamma_{RG})(1 - e - g) + We. \tag{IV.108b}$$

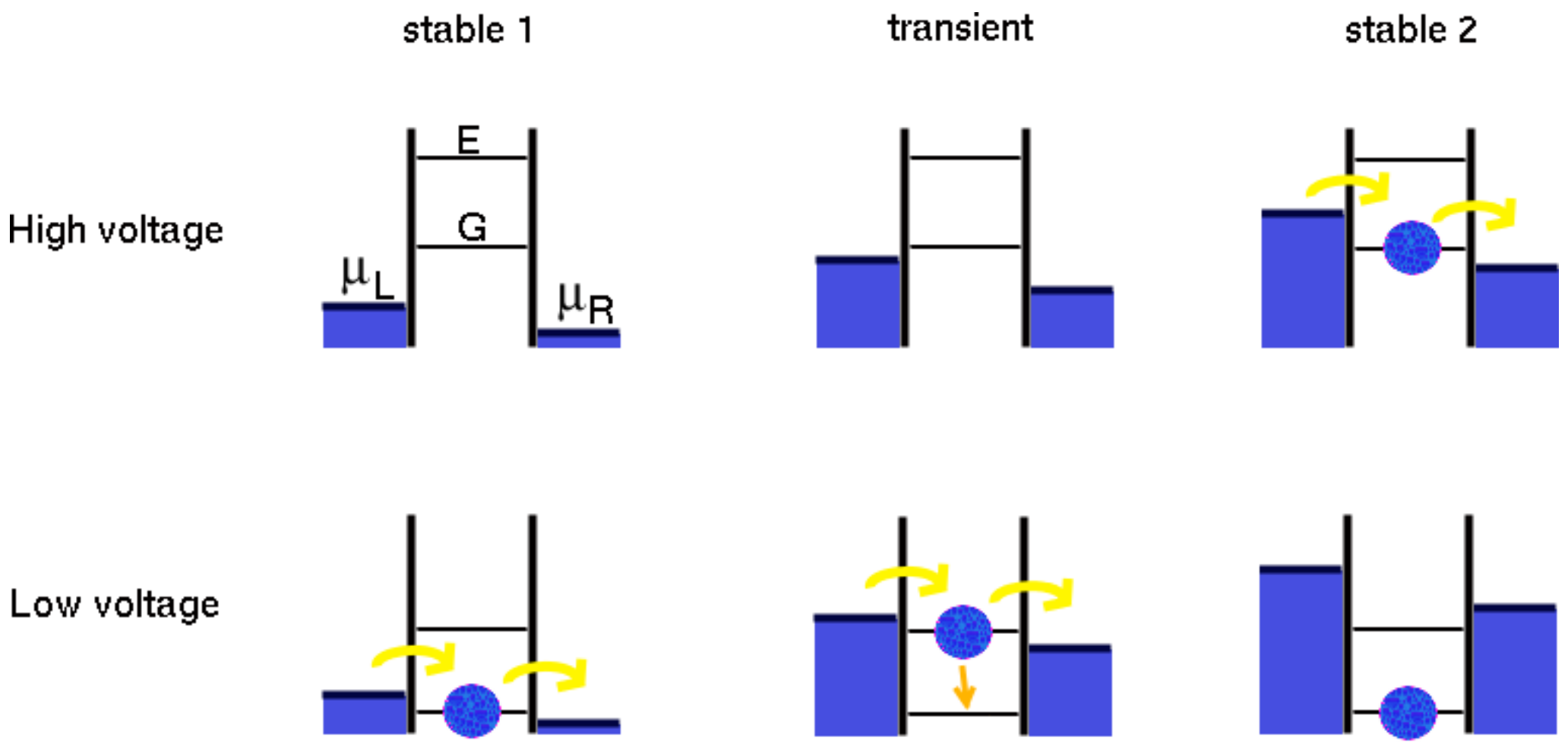

Fig. IV.14. Two stable and one transient current dot configurations for low and high gate voltages. G and E denote the ground and excited states (more precise, their energies), while $\mu_L$ and $\mu_R$ are the chemical potential of the left and right leads. In the stable current 1 and transient current configuration the current flows during the low voltage pulse, while in stable current 2 configuration during the high voltage pulse.

An initial condition of an empty dot, $g(0) = e(0) = 0$, leads to the following solutions (see Sec. G.1 for the derivation)

$$e(t) \approx \frac{\Gamma_{LE}}{\Gamma}\left(1 - e^{-\Gamma t}\right) e^{-Dt}, \tag{IV.109a}$$

$$g(t) \approx \frac{\Gamma_G}{\Gamma}\left(1 - e^{-\Gamma t}\right) + \frac{\Gamma_E}{\Gamma}\left(1 - e^{-Dt}\right). \tag{IV.109b}$$

Here tunneling rates without the index for a lead are total tunneling rates for the corresponding

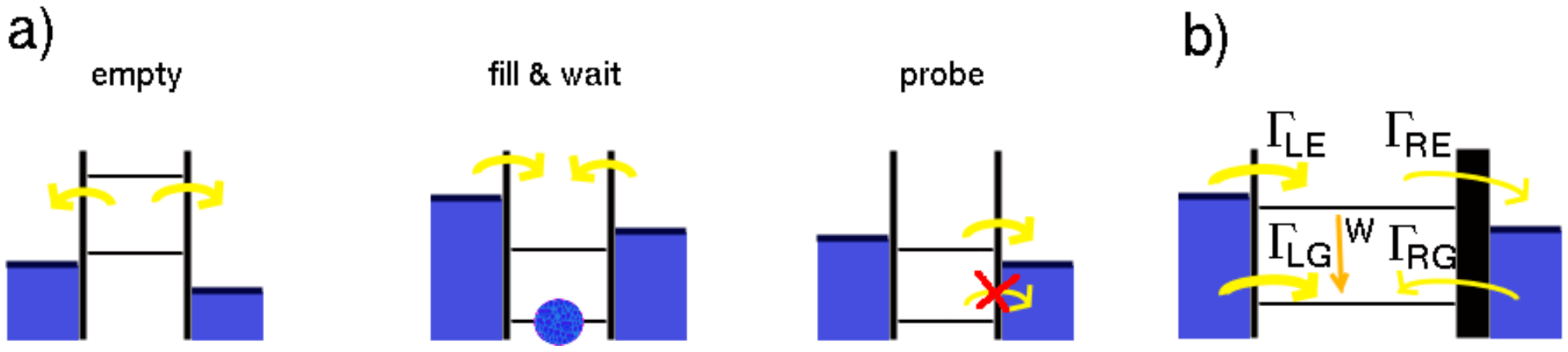

Fig. IV.15. a, Relative positions of the energy levels and chemical potentials of the leads for the three steps in the energy resolved readout. b, Tunneling rates between the leads and the dot. The first index denotes the lead (left or right), while the second denotes the state of the dot (ground or exited). The relaxation rate from the excited to the ground state is denoted as $W$. The right barrier is much thicker than the left one.

state,

$$\Gamma_G = \Gamma_{LG} + \Gamma_{RG}, \tag{IV.110a}$$

$$\Gamma_E = \Gamma_{LE} + \Gamma_{RE}, \tag{IV.110b}$$

$$\Gamma = \Gamma_G + \Gamma_E, \tag{IV.110c}$$

and $D$ is an effective relaxation rate

$$D = W + \frac{\Gamma_G}{\Gamma}\Gamma_{RE}. \tag{IV.111}$$

The approximate solutions are written in a form allowing straightforward physical interpretation. Considering an empty dot at time zero, there is an initial filling of the dot due to the electron tunneling from the leads. This is a fast process and leaves the states of the dot occupied according to the "filling efficiency" – the average excited state population equals $\Gamma_{LE}/\Gamma$, while the ground state is populated with a probability of $\Gamma_G/\Gamma$. After the initial filling, the population of the excited state decays in favor of the ground state on a longer time scale, given by the effective relaxation rate $D$. This rate reveals two ways how an electron can get from the excited state to the ground state. It relaxes directly, with a rate $W$, or it leaves the dot going to the right lead (rate $\Gamma_{RE}$) and another electron tunnels into the ground state, with a probability given by the ground state filling efficiency. The second process is called direct injection. In experiments the duration of a particular voltage step is much longer than $\Gamma^{-1}$. Then the terms decaying with the rate $\Gamma$ (the initial filling) are not resolved and only the trade-off between the excited and the ground state is observed.

*Introduction of the third step:* From Eq. (IV.111) it follows that the relaxation rate $W$ can be extracted only if it is at least comparable with the direct injection rate. To overcome this restriction, there was an intermediate step introduced in Ref. (Fujisawa *et al.*, 2002a). It is denoted as "fill&wait" in Fig. IV.15a, and is such that both dot states are below the chemical potential of the right lead. The dot, if empty, is filled by an electron from one of the leads. One can solve for the time evolution of the populations analogously to the probe configuration (see Sec. G.1). The behavior of the system is the same – there is an initial filling followed by the exchange of the excited and ground state populations. However, now the electron, once captured in the dot, cannot escape. Therefore the effective (that is measured) relaxation rate is not renormalized and equals the intrinsic relaxation rate $W$.

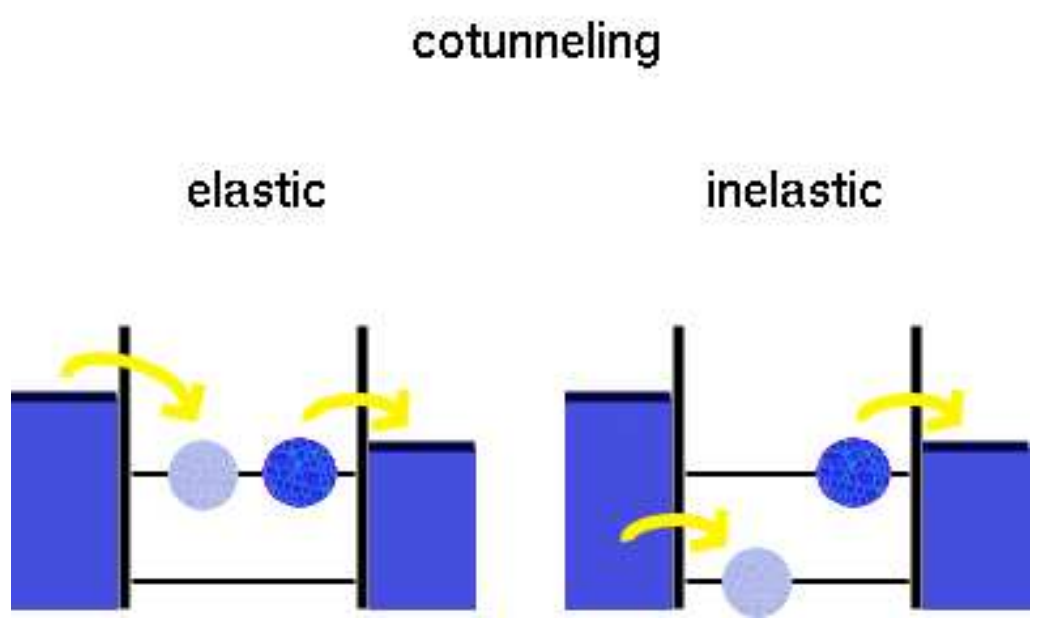

Fig. IV.16. Elastic and inelastic cotunneling as a second order tunneling process. In the elastic cotunneling the electron initially in the dot tunnels out while another electron tunnels simultaneously into the same state. In inelastic cotunneling the incoming electron enters into a different state of the dot.

To complete the picture, the effective relaxation in early experiments using ERO was dominated by cotunneling (Averin and Nazarov, 1990). This is a quantum mechanical process, thus not included in our classical description in Eq. (IV.108), illustrated in Fig. IV.16. Here an electron being in the dot can tunnel out while another electron from a lead tunnels in *simultaneously*, to the same (elastic cotunneling) or to a different state of the dot (inelastic cotunneling). This process is of the second order in the tunneling rates. For our purposes here the case of interest is when the initial electron is in the excited state, while the replacing electron tunnels into the ground state, contributing to the relaxation. (The elastic cotunneling contributes to the decoherence.)

In the first experiment using ERO (Fujisawa *et al.*, 2001) a lateral quantum dot contained $\sim 50$ electrons and the only successful relaxation time measured was 3 ns, attributed to an orbital relaxation process. It took some time to obtain results for the spin relaxation, since the relaxation rates were dominated by direct injection and cotunneling (Fujisawa *et al.*, 2002b,a). In a vertical dot, measuring the spin relaxation was successful both for singly and doubly occupied dot, but it turned out to be problematic to use this method in a lateral dot. The reason is that forcing the dot to be occupied by smaller number of electrons by applying larger negative gate voltage makes also the tunneling rates smaller, resulting in smaller current. The minimum number of electrons where the current signal was still measurable has gone from 50 in 2001 (Fujisawa *et al.*, 2001) to 8 in 2005 (Sasaki *et al.*, 2005). Even though this technical problem was solved later by proper gate design, in between a different technique proved to be very useful.

*Using QPC:* The idea is to use a quantum point contact (QPC) to measure the charge instead of measuring the current (Hanson *et al.*, 2007). QPC placed nearby the dot is able to measure the charge of the dot with a resolution of $\sim 0.1$e. It is thus possible to resolve each single electron tunneling event. This works the better, the smaller the tunneling rates, since the longer is the dot charge constant – the shortest time over which the charge can be deduced is $10\ \mu$s, setting the QPC resolution.

In Fig. IV.17 a typical time trace of a QPC current during the three steps of ERO is sketched.

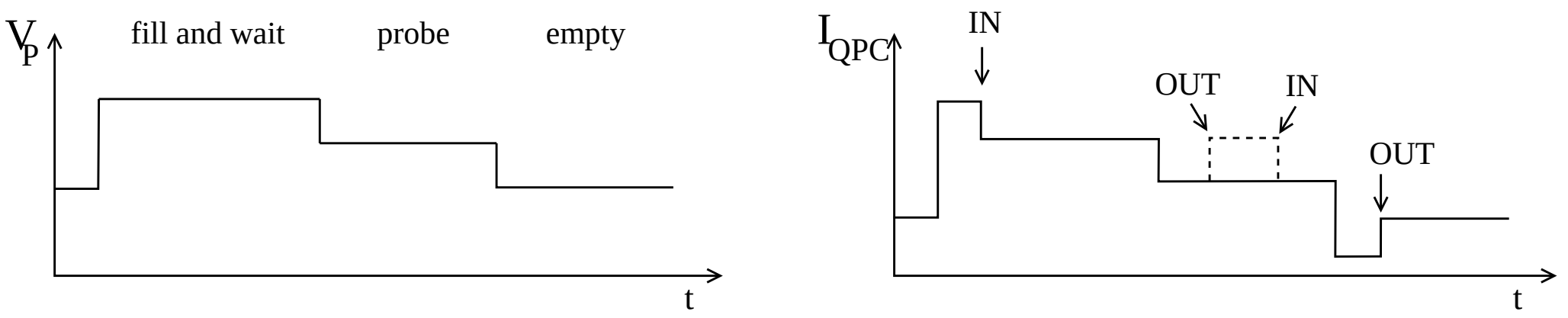

Fig. IV.17. Time dependence of the gate voltage (left) and QPC current (right) during the three step ERO sequence. See text for the explanation.

An electron enters the dot, which is being initially empty, during the fill&wait step, what is observed as a decrease of the QPC current. If in the excited state, the electron tunnels to the right lead during the probe step and another electron tunnels into the ground state, what is observed as a temporary enhancement of the QPC current. If the electron had relaxed during fill&wait step, or entered the dot into the ground state, no enhancement is observed during the probe step. Finally, the dot is emptied during the empty step, observed as an increase of the QPC current and the cycle is restarted. The current through the dot can be obtained by counting the number of tunneling electrons. In addition, tunneling rates can be deduced by averaging the corresponding time for a particular tunneling to occur.

Using ERO with QPC as the charge detector, the Delft group succeeded in measuring the spin relaxation of a single electron in a single lateral quantum dot (Elzerman *et al.*, 2004) (the first successful measurement of this kind). The results of this measurement, together with results obtained more recently by the MIT group using the same method for a larger range of the magnetic field (Amasha *et al.*, unpublished, cond-mat/0607110), are presented in Fig. IV.12.

A limitation of the ERO method is that for the readout the energy difference has to be large enough to overcome the blurring of the levels due to a finite temperature and shifts of the levels due to background charge fluctuations. Since the Zeeman energy is small, a high magnetic field is required for ERO. (In the recent MIT spin relaxation measurement the minimal magnetic field was pushed down to 1.7 T.) For smaller magnetic field, a different spin-to-charge conversion scheme was proposed.

**Transient current in small magnetic field: tunneling resolved readout (TRRO).** In ERO the read out of the electron state is possible since the tunneling rate out of the ground state during the probe step is strictly zero. Now, if we are not able to energetically resolve the ground and excited states, the readout will be still possible, if both states are above the right dot chemical potential, but the tunneling rates are different, as depicted in Fig. IV.18. If the dot is occupied by two electrons, the difference in the rates originates in the fact that the triplet state is spatially more extended (being antisymmetric) than singlet, leading to a larger overlap with the lead. Measurement using TRRO for two electron states was done (Hanson *et al.*, 2005), giving the triplet to singlet relaxation time also at zero energy difference. For single electron states the spatial extent is the same for both spin states, therefore another mechanism for discriminating tunneling rates has to be used. Some were proposed (Engel *et al.*, 2004), up to now without a successful experimental realization.

Due to the fact that the "unwanted" tunneling rate out of the ground state during the probe step is not exactly zero, the occurrence of the tunneling does not exactly correspond to the electron state. The probability of the correct assessment of the electron state is called visibility. (A wrong measurement would be an observed tunneling even if the electron was in the ground state, or no tunneling if the electron was in the excited state.) The visibility is a function of the tunneling rates and the duration of the probe step, see Sec. G.2. To get a notion, in the TRRO experiment the visibility was 80%, while in previously discussed ERO experiments it was of a comparable value, where the reduction below 100% originated in thermal fluctuations and cotunnelings.

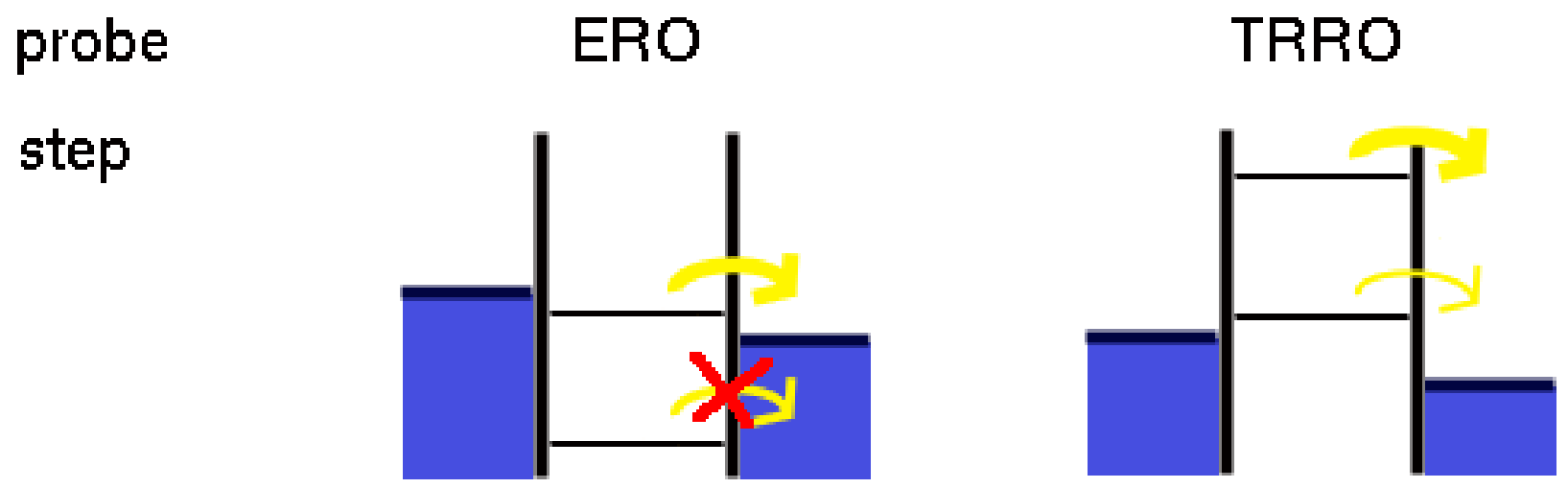

Fig. IV.18. Comparison of probe step in the energy and tunneling resolved readouts. In ERO the tunneling out of the ground state is strictly zero. In TRRO, a tunneling out of the ground states happens, but with smaller rate than the tunneling out of the excited state.

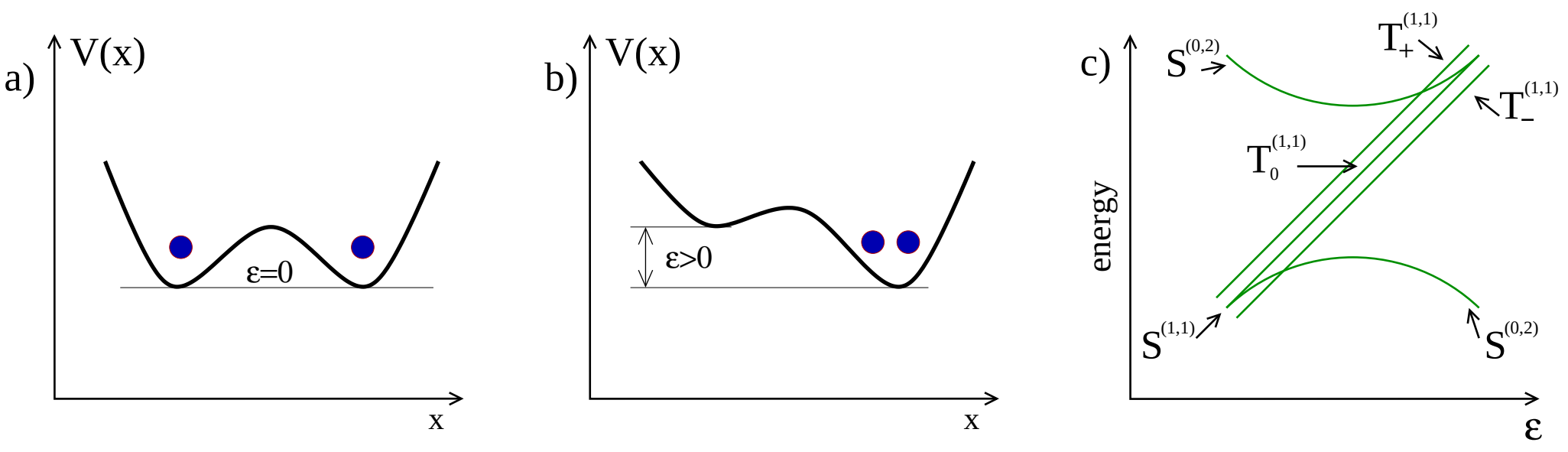

Fig. IV.19. Two electron double dot system. a, Sketch of the dots potential profile for zero detuning, when in the ground state each dot is occupied by one electron. b, Potential for a non-zero detuning. In the ground state both electrons are in the right dot. c, Energy of the states as a function of the detuning $\epsilon$, at finite magnetic field. Singlet (triplet) states are denoted by $S(T)$. The two upper indexes denote the occupation of the left and right dots. The lower index denotes the projection of the total spin along the magnetic field direction.

**Singlet to singlet spin-to-charge conversion.** We finish this section by describing yet another spin-to-charge conversion scheme due to its great potential and proven suitability for studying electron spin dynamics. The setup consists of a double dot occupied by two electrons with controllable asymmetry $\epsilon$ (detuning of the ground state energy of the left and right single dots when considering them to be isolated). If the detuning is small, as in Fig. IV.19a, the preferable occupation is one electron per dot. In this case the exchange energy is small and the ground state is four times degenerate, comprising one singlet $S^{(1,1)}$ and three triplet states $T_0^{(1,1)}$, $T_+^{(1,1)}$, and $T_{-}^{(1,1)}$ (here and further the two numbers in the superscript denote the population of the left and the right dot, respectively). If a magnetic field is applied, the triplet states are split. On the other hand, if the asymmetry is large, as in Fig. IV.19b, the dot lower in energy is preferably occupied and the ground state is a single dot singlet $S^{(0,2)}$. Since both electrons occupy one dot, the exchange energy is high and the single dot triplet states are far above the ground state and can be neglected. The spectrum as a function of the detuning is in Fig. IV.19c. Suppose the detuning is small and the qubit is encoded into the singlet $S^{(1,1)}$ and triplet $T_0^{(1,1)}$ states. If the detuning is adiabatically enlarged, the singlet $S^{(1,1)}$ evolves into $S^{(0,2)}$, where two electron occupy the right dot, while the triplet $T_0^{(1,1)}$ stays delocalized over both dots. The afterward measurement of the occupation of one of the dots then discriminates between the two states and realizes a spin-to-charge conversion.

At first, a two electron singlet to triplet relaxation time was measured at tesla (Petta *et al.*, unpublished) and sub-tesla (Johnson *et al.*, 2005) fields. The results at small magnetic fields, such as suppression of the relaxation by magnetic field, indicate that the nuclei are the source of the spin relaxation here (Merkulov *et al.*, 2002). It was also possible to measure the spin dephasing time by observing the decay of coherent oscillations between the degenerate $S^{(1,1)}$ and $T_0^{(1,1)}$ states. In a remarkable experiment (Petta *et al.*, 2005) using the spin echo technique, a two orders of magnitude difference between the dephasing and decoherence time was found.

To conclude this part, we have discussed several experiments studying spin relaxation in lateral quantum dots. Table IV.1 summarizes main results together with used methods. The problem of detection of the spin state is solved using the spin-to-charge conversion. Comparision of the first measurement in 2001 where just an upper limit for the spin relaxation was obtained in a dot containing 50 electrons, with the demonstrated coherent single spin evolution over 1 microsecond in 2005, illustrates the great amount of the experimental progress achieved on the road towards a spin qubit realization.

One remaining outstanding issue is a direct detection of spin relaxation of one electron in a double dot. In this setup the relaxation should be much more sensitive to the spin hot spot influence which can reveal the anisotropy of the spin-orbit interaction. Figure IV.20 shows the theoretically obtained spin relaxation rate as a function of the tunneling energy (y axis) and orientation of the inplane magnetic field (x axis). Taking a horizontal slice at maximal tunneling, on the bottom of the figure, which corresponds to the single dot regime, the relaxation rate varies with the magnetic field orientation. However, the variation is much more pronounced in the double dot regime (smaller tunnelings) and reaches six orders of magnitude at the hot spot (at the tunneling energy of $\sim$0.1 meV). Experimental observation of the anisotropy would be a strong proof of the spin-orbit origin of the relaxation. The strengths of the spin-orbit interactions can be also deduced from the position of the relaxation rate minimum (which is 39° in the Fig. IV.20). Most importantly, if the relaxation rate is so strongly anisotropic, placing the dot intentionally into the configuration with minimal relaxation could substantially prolong the qubit relaxation time.

### F.3 Relaxation and decoherence in the density matrix formalism

In this section we present a theoretical description of a dissipative phonon environment of the quantum dot. We use the density matrix formalism (Blum, 1996; Slichter, 1996) which is suit-Tab. IV.1. Towards a measurement of the relaxation time of the spin of a single confined electron – experimental results: used method, physical system, reference publication, temperature, and result – we give only the maximal relaxation/decoherence time measured. Short hand notations: N-V defect means nitrogen vacancy defect, MRFM means magnetic resonant force microscopy, QPC means quantum point contact, SD means single dot, DD means double dot, S-T means singlet-triplet, $T_1$ denotes relaxation time, and $T_2$ denotes decoherence time.

| method | system | Ref. | T [K] | result | | |
|---|---|---|---|---|---|---|
| scanning tunneling microscope | single spin | (Manassen *et al.*, 1989) | 300 | spin detection | | |
| fluorescence | N-V defect | (Gruber *et al.*, 1997) | 300 | spin detection | | |
| | | (Jelezko *et al.*, 2002) | 2 | $T_1$ | = | 1 s |
| | | (Jelezko *et al.*, 2004) | 300 | $T_2$ | = | 1 $\mu$s |
| MRFM | single spin | (Rugar *et al.*, 2004) | 1.6 | $T_1$ | = | 0.76 s |
| optical pump&probe | $10^4$ dots (self assembled) | (Kroutvar *et al.*, 2004) | 1 | $T_1$ | = | 20 ms |
| energy resolved readout | lateral SD, 50 e | (Fujisawa *et al.*, 2001) | 0.15 | $T_1^{\mathrm{orbital}}$ | = | 3 ns |
| (ERO) | lateral SD, 50 e | (Fujisawa *et al.*, 2002b) | 0.15 | $T_1$ | $>$ | 2 $\mu$s |
| | vertical SD, 2 e | (Fujisawa *et al.*, 2002a) | 0.1 | $T_1^{\mathrm{S-T}}$ | $>$ | 0.2 ms |
| | lateral SD, 8 e | (Sasaki *et al.*, 2005) | 0.09 | $T_1^{\mathrm{S-T}}$ | = | 0.2 ms |
| resonance assisted current | lateral DD, 2 e | (Koppens *et al.*, 2006) | 0.1 | $T_2^{S-T}$ | $\lesssim$ | 1 $\mu$s |
| ERO+QPC | lateral SD, 1 e | (Hanson *et al.*, 2003) | 0.02 | $T_1$ | $>$ | 50 $\mu$s |
| (B$\geq$ 8 T) | lateral SD, 1 e | (Elzerman *et al.*, 2004) | 0.03 | $T_1$ | = | 0.85 ms |
| (B$\geq$ 1.7 T) | lateral SD, 1 e | Amasha *et al.* unpublished | 0.12 | $T_1$ | = | 0.15 s |
| | lateral SD, 2e | (Meunier *et al.*, 2007) | 0.18 | $T_1^{\mathrm{S-T}}$ | = | 1 ms |
| tunneling resolved readout (TRRO) | lateral SD, 2 e | (Hanson *et al.*, 2005) | 0.02 | $T_1^{\mathrm{S-T}}$ | = | 2.6 ms |
| singlet to singlet | lateral DD, 2 e | (Johnson *et al.*, 2005) | 0.16 | $T_1^{\mathrm{S-T}}$ | $\lesssim$ | 1 ms |
| spin-to-charge conversion | | (Petta *et al.*, 2005) | 0.14 | $T_2^{*\mathrm{S-T}}$ | = | 10 ns |
| | | | | $T_2^{\mathrm{S-T}}$ | = | 1 $\mu$s |

able due to the statistical nature of the relaxation and decoherence. We first derive the Fermi's Golden Rule in this formalism. After that we show how the decoherence leads to the decay of the Rabi oscillations, which occur if a resonant microwave field is present. We finish by showing that if electrons are allowed to flow into/out of the dot, the decoherence can be deduced also in a steady state measurement, from the width of the resonantly induced current peak. Similar considerations, using randomly fluctuating fields to describe dissipative environment, were already presented on a more elementary level in Sec. B. There we saw that the fluctuation averaging does not provide for the temperature dependence of the average spin. This is remedied by consdering the more general form of a heat bath described by its own density matrix. This more advanced derivation, presenting a different viewpoint, will be presented below, although the physics involved in indeed very similar.

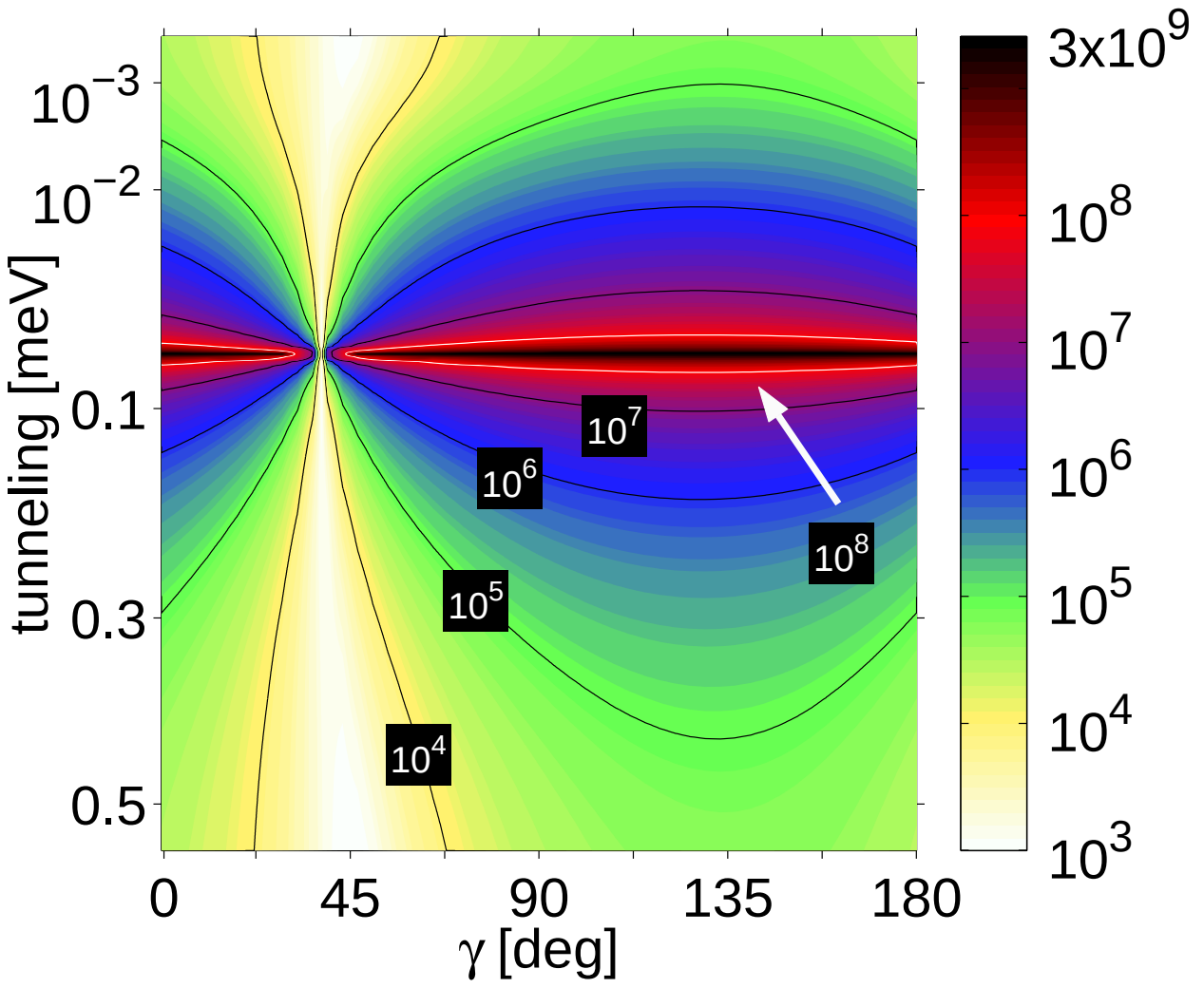

Fig. IV.20. Single spin relaxation rate in a logarithmic scale (right) in inverse seconds for a single electron in a double dot. The double dot is in the plane of a [001] grown GaAs quantum well. The axis of the dots is along [100]. The orientation of the in-plane magnetic field (of 5 tesla) relative to [100] is varied on the horizontal axis. The tunneling energy is varied on the y axis. After (Stano and Fabian, 2006b).

### *F.3.1 Electron in a phonon bath*

We begin by considering the electron and phonons as two interacting subsystems of a composite system with Hamiltonian $H_T = H_e + H_p + V$, where $H_e$ refers to the isolated electron, $H_p$ describes phonons if no electron is present and $V$ is an electron-phonon interaction. The Hilbert space is spanned over basis vectors $\{|i,\alpha,n\rangle\}$, where $i$ labels the electron state, $\alpha$ phonon state, and $n$ is the occupation number for the phonon state $\alpha$. We are interested in the time evolution defined by the Schrödinger equation with the total Hamiltonian $H_T$ and an initial state. However, since we do not have detailed information about the state of the phonons, we use the density matrix (the superscript $S$ is for the Schrödinger picture),

$$\rho_T^S(0) = \sum_{i,\alpha,n} W_{i,\alpha,n} |i,\alpha,n\rangle\langle i,\alpha,n|, \tag{IV.112}$$

which reflects that we know only statistical probabilities $W_{i,\alpha,n}$ for certain initial condition to occur. If each function in the expansion obeys the Schrödinger equation, the time evolution of the density matrix is described by the Liouville equation

$$\mathrm{i}\hbar\, \partial_t \rho_T^S(t) = [H_T(t), \rho_T^S(t)]. \tag{IV.113}$$

To get rid of the unobserved phonons we trace out the redundant degrees of freedom of the system, defining the reduced electron density matrix (denoted by dropping the subscript $T$)

$$\rho^S(t) = \mathrm{tr}_p\left(\rho_T^S(t)\right) = \sum_{\alpha,n} \langle \alpha,n|\rho_T^S(t)|\alpha,n\rangle. \tag{IV.114}$$

In the interaction picture (denoted by dropping the superscript $S$) electron relaxation and decoherence rates appear in the following form (see Sec. G.2 for the derivation)

$$\partial_t^R \rho_{ii}(t) = -\sum_{k\neq i} 2\Gamma_{ik}\rho_{ii}(t) + \sum_{k\neq i} 2\Gamma_{ki}\rho_{kk}(t), \tag{IV.115a}$$

$$\partial_t^R \rho_{ij}(t) = -\left(\sum_{k\neq i}\Gamma_{ik} + \sum_{k\neq j}\Gamma_{jk}\right)\rho_{ij}(t) = -\gamma_{ij}\rho_{ij}(t). \tag{IV.115b}$$

We have added superscript $R$ on the time derivative to denote that the relaxation and decoherence is the source for these terms. The time evolution of the diagonal and off-diagonal terms is decoupled, and all decay exponentially into their steady state values. The off-diagonal terms decay to zero with the decoherence rates $\gamma_{ij}$. The diagonal terms equilibrium value is defined by the temperature through the relaxation rates, such as $\Gamma_{ij}$ which gives the transition rate from state $i$ to $j$. At temperature low enough such that spontaneous excitations are negligible ($\Gamma_{ij} = 0$ for $i < j$), for a two level system we get $\gamma_{21} = \Gamma_{21}$. From that we obtain the decoherence time $T_2 = 2T_1$, which says that the relaxation contributes to the decoherence and which can be thought of as the upper limit for the decoherence time. Taking into account other environment fluctuation (for example, nuclei) often leads to $T_2 \ll T_1$. While Eqs. (IV.115) do not look to contain much more than the Fermi's Golden Rule, the density matrix formalism allows to treat dissipation, which is of statistical nature, together with a coherent manipulation of the electron spin by resonant fields and in running a current through the dot, as we will show next.

### *F.3.2 Rabi oscillations*

Starting with a localized electron subject to dissipation, described by Eqs. (IV.115), we are interested now in the possibility of spin and charge manipulation by resonant microwave field. For this purpose, we consider that at time zero a monochromatic microwave field is turned on. The field couples only to the electron, hence it can be expressed in the basis of the electron states, defining the field matrix elements $\Omega_{ij}$,

$$\begin{aligned} V^M(\mathbf{r},t) = V^M(\mathbf{r})\cos\omega t &= \sum_{i,j} |\Psi_i\rangle\langle\Psi_i|V^M(\mathbf{r})|\Psi_j\rangle\langle\Psi_j|\cos\omega t \\ &= \sum_{i,j} |\Psi_i\rangle\hbar\Omega_{ij}\langle\Psi_j|\cos\omega t. \end{aligned} \tag{IV.116}$$

We could include this term into the electron Hamiltonian obtaining an analogue of eigenstates (which would be time dependent) and eigenenergies (Shirley, 1965), and use these states as a basis for a generalized "Fermi's Golden Rule" (Jiang *et al.*, 2006). Such approach is needed if the oscillating field is strong such that it substantially changes the electron wavefunction and energy. If the field is weak, which is our case, we can treat the time dependent field as a perturbation in the same sense as we treated the electron-phonon interaction. Including $V^M$ into the Liouville equation (IV.113) results in the leading order in additional contributions to the time derivative (denoted by superscript M for the microwave) of the reduced electron density matrix. The microwave influences only the resonant states, denoted by indexes $a$ and $b$, which are the states whose energy difference is close to the frequency of the oscillating field, $E_b - E_a = \hbar\omega_{ba} \approx \hbar\omega$, what is also expressible as (the absolute value of) the detuning

$$\Delta = \omega_{ba} - \omega \tag{IV.117}$$

being much smaller than the frequency of the field itself. Without a loss of generality, we suppose $\omega_{ba}, \omega > 0$. The microwave influence is described by (see Sec. G.3 for the derivation)

$$\begin{aligned}
\partial_t^M \rho_{aa}(t) &= -(\mathrm{i}/2)\Omega_{ab}\rho_{ba}(t)\exp[\mathrm{i}(\omega_{ab}+\omega)t] \\
&+ (\mathrm{i}/2)\Omega_{ba}\rho_{ab}(t)\exp[-\mathrm{i}(\omega_{ab}+\omega)t], && \text{(IV.118a)}\\
\partial_t^M \rho_{bb}(t) &= -(\mathrm{i}/2)\Omega_{ba}\rho_{ab}(t)\exp[-\mathrm{i}(\omega_{ab}+\omega)t] \\
&+ (\mathrm{i}/2)\Omega_{ab}\rho_{ba}(t)\exp[\mathrm{i}(\omega_{ab}+\omega)t], && \text{(IV.118b)}\\
\partial_t^M \rho_{ab}(t) &= -(\mathrm{i}/2)\Omega_{ab}[\rho_{bb}(t)-\rho_{aa}(t)]\exp[\mathrm{i}(\omega_{ab}+\omega)t]. && \text{(IV.118c)}
\end{aligned}$$

One can readily solve for the time evolution of the electron if the field is resonant with the two lowest states, neglecting the presence of the higher lying states. If the two lowest states are the spin opposite states, we are describing the spin resonance. The total time derivative of the density matrix, $\partial_t = \partial_t^R + \partial_t^M$, in the two level basis is

$$\begin{aligned}
\partial_t\rho_{11}(t) &= 2\Gamma_{21}\rho_{22}(t) - 2\Gamma_{12}\rho_{11}(t) - (\mathrm{i}/2)\Omega_{12}\rho_{21}(t)\exp(-\mathrm{i}\Delta t) \\
&+ (\mathrm{i}/2)\Omega_{21}\rho_{12}(t)\exp(\mathrm{i}\Delta t), && \text{(IV.119a)}\\
\partial_t\rho_{22}(t) &= 2\Gamma_{12}\rho_{11}(t) - 2\Gamma_{21}\rho_{22}(t) - (\mathrm{i}/2)\Omega_{21}\rho_{12}(t)\exp(\mathrm{i}\Delta t) \\
&+ (\mathrm{i}/2)\Omega_{12}\rho_{21}(t)\exp(-\mathrm{i}\Delta t), && \text{(IV.119b)}\\
\partial_t\rho_{12}(t) &= -\gamma_{12}\rho_{12}(t) - (\mathrm{i}/2)\Omega_{12}[\rho_{22}(t)-\rho_{11}(t)]\exp(-\mathrm{i}\Delta t). && \text{(IV.119c)}
\end{aligned}$$

The explicitly time dependent factors can be removed by expressing the off diagonal density matrix element by

$$\rho_{12}(t) = \overline{\rho}_{12}(t)e^{-\mathrm{i}\Delta t} \tag{IV.120}$$

If $\overline{\rho}_{12}(t)$ does not depend on time, the off diagonal density matrix elements in the Schrödinger picture rotate with the frequency of the field – this is what characterizes the steady state, as we will see later; by steady state we mean a situation in which the diagonal elements of the density matrix (that is, states occupations) are time independent. To describe the steady state we therefore use the ansatz,

$$\rho_{12}(t) := \overline{\rho}_{12}e^{-\mathrm{i}\Delta t}, \tag{IV.121}$$

giving from Eqs. (IV.119),

$$\overline{\rho}_{12} = \frac{\Omega_{12}}{2}\frac{\rho_{22}(t)-\rho_{11}(t)}{\Delta - \mathrm{i}\gamma_{12}}, \tag{IV.122}$$

and,

$$\begin{aligned}
\partial_t\rho_{11}(t) = (\partial_t^R + \partial_t^M)\rho_{11}(t) &= \Gamma_{21}\rho_{22}(t) - \Gamma_{12}\rho_{11}(t) \\
&+ 2[\rho_{22}(t)-\rho_{11}(t)]J, && \text{(IV.123a)}\\
\partial_t\rho_{22}(t) = (\partial_t^R + \partial_t^M)\rho_{22}(t) &= \Gamma_{12}\rho_{11}(t) - \Gamma_{21}\rho_{22}(t) \\
&+ 2[\rho_{11}(t)-\rho_{22}(t)]J, && \text{(IV.123b)}
\end{aligned}$$

where we introduced the induced rate,

$$J = \frac{|\Omega_{21}|^2}{4} \frac{\gamma_{21}}{\Delta^2 + \gamma_{21}^2}. \tag{IV.124}$$

The zero time derivative of the occupations in the steady state can be viewed as a balance between two competing processes: relaxation, which drives the system towards the thermodynamical equilibrium ($\rho_{22}/\rho_{11} = \Gamma_{12}/\Gamma_{21}$), and microwave induced transition, which equilibrates occupations of the resonant states ($\rho_{22} = \rho_{11}$). The effectiveness of the oscillating field in driving the system out of thermal equilibrium is characterized by the induced rate $J$. Enlarging the detuning $\Delta$ (that is going away from the resonance), the induced rate decays; the microwave is less effective in influencing the system.

The interpretation of $J$ as a rate of occupation transition is, however, valid only in the steady state. Equations (IV.123) do not describe the microwave influence on the resonant states before the steady state is reached. This temporal regime can be obtained by solving Eqs. (IV.119). For that we introduce the following variables

$$R(t) = \Omega_{21}\rho_{12}(t)e^{\mathrm{i}\Delta t} + \Omega_{12}\rho_{21}(t)e^{-\mathrm{i}\Delta t}, \tag{IV.125a}$$
$$\mathrm{i}\, I(t) = \Omega_{21}\rho_{12}(t)e^{\mathrm{i}\Delta t} - \Omega_{12}\rho_{21}(t)e^{-\mathrm{i}\Delta t}, \tag{IV.125b}$$
$$D(t) = \rho_{22}(t) - \rho_{11}(t). \tag{IV.125c}$$

If in the steady state and for real field matrix elements, $R$ and $I$ are proportional to the real and imaginary part of the off diagonal density matrix element $\overline{\rho}_{12}(t)$, while $D$ is the difference in the occupations of the resonant states. The time dependence of the new variables follows from Eqs. (IV.119):

$$[\partial_t + 2(\Gamma_{12} + \Gamma_{21})]D(t) = I(t) - 2(\Gamma_{21} - \Gamma_{12}), \tag{IV.126a}$$
$$(\partial_t + \gamma_{12})I(t) = \Delta\, R(t) - D(t)|\Omega_{21}|^2, \tag{IV.126b}$$
$$(\partial_t + \gamma_{12})R(t) = -\Delta\, I(t). \tag{IV.126c}$$

One can see that a steady state solution [that is, constant $D(t)$] requires the time independence of $R(t)$ and $I(t)$, validating the ansatz in Eq. (IV.121). We rewrite Eqs. (IV.126) as a third order differential equation for $D(t)$:

$$\{[(\partial_t + \gamma_{12})^2 + \Delta^2][\partial_t + 2(\Gamma_{12} + \Gamma_{21})] + |\Omega_{12}|^2(\partial_t + \gamma_{12})\}[D(t) + \overline{D}] = 0, \tag{IV.127}$$

where we denote by $\overline{D}$ the steady state occupation difference,

$$\overline{D} = \frac{2(\Gamma_{21} - \Gamma_{12})(\gamma_{12}^2 + \Delta^2)}{2(\gamma_{12}^2 + \Delta^2)(\Gamma_{12} + \Gamma_{21}) + \gamma_{12}|\Omega_{21}|^2}. \tag{IV.128}$$

Equation (IV.127) has three linearly independent solutions

$$D(t) = -\overline{D} + \sum_{i=1}^{3} A_i e^{(-\gamma_{12} + a_i)t}. \tag{IV.129}$$

Here $A_i$ are given by the initial conditions and $a_i$ are the roots of the following algebraic equation:

$$(a^2 + \Delta^2)(a + \gamma) + |\Omega_{21}|^2 a = 0, \tag{IV.130}$$

where $\gamma = 2(\Gamma_{12} + \Gamma_{21}) - \gamma_{12}$ (for phonons and at small temperatures $\gamma = \gamma_{12}$). The cubic equation has three real solutions, or one real and two complex conjugated solutions. If the first case occurs, the occupation difference decays towards the steady state exponentially. The second case corresponds to Rabi oscillations, where the Rabi frequency is the imaginary part of the complex solution to Eq. (IV.130). Instead of algebraically complicated general expressions for $a_i$, we give two examples of interest.[93]

First, close to the resonance, $\Delta \ll |\Omega_{12}|, \gamma$, up to the first order in $\Delta$, we have

$$a_1 = 0, \qquad a_{2,3} = -\frac{\gamma}{2} \pm \mathrm{i}\sqrt{|\Omega_{12}|^2 - \gamma^2/4}. \tag{IV.131}$$

Second, for small damping, $\gamma \ll |\Omega_{12}|, \Delta$, up to the first order in $\gamma$, we obtain

$$a_1 = -\gamma \frac{\Delta^2}{\Delta^2 + |\Omega_{12}|^2}, \qquad a_{2,3} = -\gamma \frac{|\Omega_{12}|^2}{2(\Delta^2 + |\Omega_{12}|^2)} \pm \mathrm{i}\sqrt{|\Omega_{12}|^2 + \Delta^2}. \tag{IV.132}$$

The importance of the Rabi oscillation can be illustrated by a simple example considering an exactly resonant driving without dissipation, $\Delta = 0, \gamma_{12} = 0$. Starting initially in state $\Psi_1$ [that is $D(0) = -1$], the evolution of the system written in the basis of $\{\Psi_1, \Psi_2\}$ is

$$\Psi(t) = \Psi_1 \cos(|\Omega_{12}|t) + \Psi_2 \sin(|\Omega_{12}|t), \tag{IV.133}$$

meaning, the populations of the two states oscillate harmonically. Letting the system evolve for time $\pi/2|\Omega_{12}|$ when the final state is $\Psi_2$ realizes a coherent spin flip – one of the basic operations needed for quantum computation. The demonstration of Rabi spin oscillations in a quantum dot induced by magnetic field is an important experimental breakthrough (Koppens *et al.*, 2006).

Nonzero damping and detuning make the solution more complicated than Eq. (IV.133), but qualitatively they cause the following: (i) decoherence causes the oscillations to decay, and reduces the Rabi frequency. (ii) The detuning makes the oscillation amplitude smaller and enlarges the Rabi frequency. (iii) The existence of the third non oscillatory solution expresses the fact that the amplitude of oscillations depends on initial conditions.

### *F.3.3 Current*

Having studied a driven localized electron with dissipation, we widen our model further, allowing, in addition, for the electron to tunnel out/into the dot by connecting the dot to the leads. This is relevant for the experiments since many properties of the system are probed by current measurements – most profound example is the dot spectroscopy (that is measuring the energy levels), which can be done by the resonant tunneling technique (van der Wiel *et al.*, 2003). As a further example, we show here how the spin relaxation and decoherence can be deduced from a current measurement in a resonantly driven dot (Engel and Loss, 2001), if in the probe step configuration (discussed in Sec. F.2), as depicted in Fig. IV.21.

[93] If none of these two applies, then the field matrix element $|\Omega_{12}|$ is not dominant, and either $\gamma_{12}$ is large and the damping is too strong to observe Rabi oscillations or $\Delta$ is large and the amplitude of the oscillations is small – both cases are not of interest here.

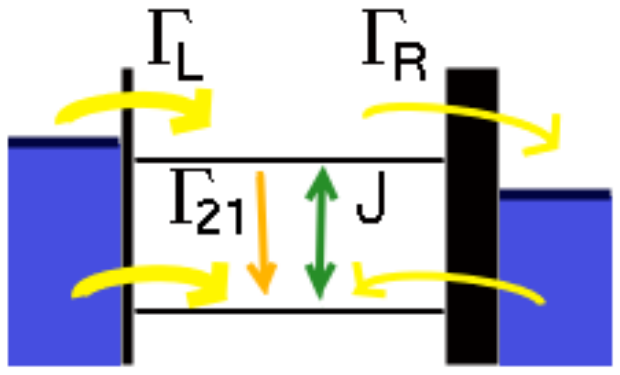

Fig. IV.21. A two level model of a quantum dot with dissipation (described by relaxation rate $\Gamma_{21}$). The applied oscillating field results in induced rate $J$. The dot is connected to the left and right leads, each characterized by a single tunneling rate ($\Gamma_L$ and $\Gamma_R$). The electron can enter the ground state of an empty dot from any lead, while the excited state can be filled only from the left lead. Once in the dot, the electron can leave only from the excited state to the right lead.

The presence of the leads appears as an additional contribution to the density matrix time derivative ($l$ stands for leads),

$$\partial_t^l \rho_{22}(t) = -\Gamma_R \rho_{22} + \Gamma_L [1 - \rho_{22}(t) - \rho_{11}(t)], \quad \text{(IV.134a)}$$

$$\partial_t^l \rho_{11}(t) = (\Gamma_R + \Gamma_L)[1 - \rho_{22}(t) - \rho_{11}(t)]. \quad \text{(IV.134b)}$$

Here the leads are characterized by tunneling rates $\Gamma_L$ and $\Gamma_R$ which describe the possibility of the electron to tunnel out of/into the dot. For simplicity, we suppose no dependence of the these tunnel rates, on the electron state. Also, compared to the case of an isolated electron considered in the previous part of Rabi oscillations, now the dot can be also empty, denoted by $\rho_{00}(t)$. Due to the normalization of the density matrix $\rho_{00}(t) = 1 - \rho_{22}(t) - \rho_{11}(t)$.

Equations such as Eqs. (IV.134), and the rates appearing therein, can be derived in the lowest order of the interaction of the dot and the Fermi electron sea in the leads. For the derivation of the rates, see Refs. (Gurvitz and Kalbermann, 1987; Gurvitz and Prager, 1996) and Sec. V.A.4. We mention here only two important properties when considering a rate between the dot state $i$ and the lead $l$: First, the rate is proportional to the tunneling amplitude between $i$ and $l$, reflecting the possibility of tuning the rate by changing the barrier between the dot and the lead. Second, the incoming (outgoing) rate is proportional to the density of the occupied (empty) states in the lead $l$ at the energy $E_i$. Thus, at small temperatures the electron can enter the dot state $i$ only from a lead with the chemical potential higher than $E_i$ and leave the dot into a lead with a smaller chemical potential – the rules that we have used for the description of the energy resolved readout in Sec. F.2.

We consider now only the steady state measurement, which is appropriate if the current is measured over time much longer compared to the time of transient regime of the decaying Rabi oscillations. We know that the steady state solution for the off-diagonal terms in the density matrix appears as an effective induced rate $J$, so the steady state can be described by putting $\partial_t \rho(t) = 0$ in Eqs. (IV.123). This, together with the normalization, allows us to find the steady state occupations for the three concerned states (0,1,2). For example, the steady state occupation

of the excited state is

$$\overline{\rho}_{22} = \frac{J + \Gamma_{12}}{2J + \Gamma_{12} + \Gamma_{21}}. \tag{IV.135}$$

The current through the dot can be obtained as the sum of all contributions filling the dot from the left lead. Plugging in the steady state result for the density matrix leads to

$$I = \frac{\Gamma_L \Gamma_R (J + \Gamma_{12})}{\Gamma_R[\Gamma_R + 3J + \Gamma_{21} + 2\Gamma_{12}] + \Gamma_L\{\Gamma_R + 2[2J + \Gamma_{21} + \Gamma_{12}]\}}, \tag{IV.136}$$

which simplifies in several limiting cases, with straightforward physical explanation:

$$I = 2e\Gamma_L \overline{\rho}_{22}, \qquad \text{if} \quad \Gamma_L \ll \Gamma_R \ll J + \Gamma_{12}, \tag{IV.137a}$$

$$I = e\Gamma_R \overline{\rho}_{22}, \qquad \text{if} \quad \Gamma_R \ll \Gamma_L \ll J + \Gamma_{12}, \tag{IV.137b}$$

$$I = 2e\frac{\Gamma_L}{\Gamma_L + \Gamma_R}(J + \Gamma_{12}), \qquad \text{if} \quad \Gamma_L, \Gamma_R \gg J + \Gamma_{12}. \tag{IV.137c}$$

For small tunneling rates, the current is proportional to the excited state population and the smaller tunneling rate. For large tunneling rates, the current is proportional to the filling efficiency of the dot from the left lead and the effective rate of excitation from the ground to the excited state. By the current measurement in the transient regime configuration it is thus possible to measure the excited state population, and the induced rate, by changing the coupling to the leads. From these one can deduce the spin decoherence and relaxation rate [as can be expected, since $\overline{\rho}_{22}$ and $J$ depend on the relaxation and decoherence rate – see Ref. (Stano and Fabian, unpublished) for details], interestingly indeed, since the measurement is done in a steady state. This is another way, alternative to the decay of Rabi oscillations, how the spin decoherence can be measured.

## G. Appendix

### G.1 Time evolution of the state occupations

#### *G.1.1 ERO: Probe pulse*

Here we derive Eqs. (IV.109) from Eqs. (IV.108). We introduce the missing charge in the ground state, $x = 1 - g$, which transforms Eq. (IV.108) into a homogeneous system:

$$\dot{e} = \Gamma_{LE}(x - e) - (\Gamma_{RE} + W)e, \tag{IV.138a}$$

$$\dot{x} = -(\Gamma_{LG} + \Gamma_{RG})(x - e) - We. \tag{IV.138b}$$

The first equation says that an empty excited state is populated from the left lead (first term), while an occupied excited state decays into the right lead or into the ground state (second term). Similarly, the second equation describes the fact that the ground state can be populated from both leads if the dot is empty, and from the excited state if an electron is in the excited state. Using the total tunneling rates, Eqs. (IV.110), we express $e$ from Eq. (IV.138b),

$$e(\Gamma_G - W) = \dot{x} + \Gamma_G x, \tag{IV.139}$$

and insert it into Eq. (IV.138a) getting

$$\ddot{x} + (\Gamma_T + W)\dot{x} + [\Gamma_G \Gamma_{RE} + W(\Gamma - \Gamma_{RE})]x = 0. \quad \text{(IV.140)}$$

In further we use that $W$ and $\Gamma_{RE}$ are much smaller than the total (or left) tunneling rate and express all quantities in the lowest order in small frequencies. The solution of Eq. (IV.140) is

$$x(t) = Ae^{-\omega_1 t} + Be^{-\omega_2 t}, \quad \text{(IV.141)}$$

where the frequencies $\omega_{1,2}$ are in the leading order

$$\omega_1 \approx \Gamma, \qquad \omega_2 \approx \frac{\Gamma_G}{\Gamma}\Gamma_{RE} + W = D. \quad \text{(IV.142a)}$$

(IV.142b)

The initial condition of an empty dot, $e(0) = 0$, $x(0) = 1$ gives through Eqs. (IV.138) and Eq. (IV.141) the coefficients $A$ and $B$,

$$\left.\begin{array}{l} A + B = 1, \\ A\omega_1 + B\omega_2 = \Gamma_G, \end{array}\right\} => \begin{array}{l} A = \frac{\Gamma_G - \omega_2}{\omega_1 - \omega_2} \approx \frac{\Gamma_G}{\Gamma}, \\ B = 1 - A, \end{array} \quad \text{(IV.143)}$$

so that

$$x(t) \approx \frac{\Gamma_G}{\Gamma}e^{-\Gamma t} + \frac{\Gamma_E}{\Gamma}e^{-Dt}, \quad \text{(IV.144)}$$

from where Eq. (IV.109b) for the ground state population $g = 1 - x$ follows.

As seen from Eq. (IV.139), the excited state will have the same functional form as Eq. (IV.141) with different coefficients, which are again given by the initial condition of an empty dot:

$$\left.\begin{array}{l} A' + B' = 0, \\ A'\omega_1 + B'\omega_2 = -\Gamma_{LE}, \end{array}\right\} => \begin{array}{l} A' = \frac{\Gamma_{LE}}{\omega_2 - \omega_1} \approx -\frac{\Gamma_{LE}}{\Gamma}, \\ B' = -A', \end{array} \quad \text{(IV.145)}$$

From where

$$e(t) \approx \frac{\Gamma_{LE}}{\Gamma}(e^{-Dt} - e^{-\Gamma t}), \quad \text{(IV.146)}$$

which is, within the leading order in $\Gamma^{-1}$, equivalent to the solution in Eq. (IV.109a).

### *G.1.2 ERO: Fill&wait pulse*

In an analogous way to the probe configuration, the equations for the time evolution of the population during the fill&wait step are

$$\dot{e} = (\Gamma_{LE} + \Gamma_{RE})(1 - e - g) - We, \quad \text{(IV.147a)}$$

$$\dot{g} = (\Gamma_{LG} + \Gamma_{RG})(1 - e - g) + We. \quad \text{(IV.147b)}$$

This means that if the dot is empty, both states can be filled from both leads, with corresponding rates. Once the dot is occupied, the only evolution will be the occupied state decaying into the ground state. The solution, again for the initial condition of an empty dot, is

$$e(t) \approx \frac{\Gamma_E}{\Gamma}\left(1-e^{-\Gamma t}\right)e^{-Wt}, \tag{IV.148a}$$

$$g(t) \approx \frac{\Gamma_G}{\Gamma}\left(1-e^{-\Gamma t}\right)+\frac{\Gamma_E}{\Gamma}\left(1-e^{-Wt}\right). \tag{IV.148b}$$

To derive Eqs. (IV.148), we would introduce the total missing charge on the dot $c = 1 - g - e$, whereby Eqs. (IV.147) are transformed into a homogeneous system, and continue as in the previous.

### *G.1.3 TRRO: Probe pulse*

Here we derive the visibility for the TRRO measurement. We suppose the dot is in the probe step, as depicted in Fig. IV.18, and described by the following equations for the populations:

$$\dot{e} = -(\Gamma_T + W)e, \tag{IV.149a}$$

$$\dot{g} = -\Gamma_S g + We. \tag{IV.149b}$$

The electron from the singlet (triplet) states tunnels out of the dot into the lead with the rate $\Gamma_S$ ($\Gamma_T$). In addition to that the triplet can decay also into the singlet with rate $W$. Solving these equation in the same way as before we get

$$e(t) = e(0)e^{-(\Gamma_T+W)t}, \tag{IV.150a}$$

$$g(t) = Ae^{-(\Gamma_T+W)t} + Be^{-\Gamma_S t}. \tag{IV.150b}$$

We consider the dot initially occupied, $e(0) + g(0) = 1$, giving two equations for the three unknown coefficients $e(0)$, $A$, and $B$. The third equation follows by specifying where the electron initially is: The first part of the possible error is the probability that a tunneling has occurred, although the electron was in the ground state,

$$\alpha(t) = 1 - g(t) - e(t) = [\text{for } \ g(0) = 1] = 1 - e^{-\Gamma_S t}. \tag{IV.151}$$

The second part is the probability that a tunneling has not occurred, with an electron in the excited state initially,

$$\begin{aligned}\beta(t) &= e(t) + g(t) = [\text{for } \ e(0) = 1] \\ &= \frac{1}{\Gamma_T + W - \Gamma_S}\left(We^{-\Gamma_S t} + (\Gamma_T - \Gamma_S)e^{-(\Gamma_T+W)t}\right).\end{aligned} \tag{IV.152}$$

The visibility is defined (Hanson *et al.*, 2005) as $v(t) = 1 - \alpha(t) - \beta(t)$, and for given tunneling rates can be optimized (maximized) as a function of the duration of the probe step $t$.

### G.2 Liouville equation for an electron in a phonon bath

We derive here Eqs. (IV.115) describing the relaxation and decoherence of a localized electron due to interactions with phonons. Our derivation closely follows Ref. (Blum, 1996), where broader discussion and further details can be found. The Hamiltonian $H_T$ of a composite electron-phonon system consists of

$$H_e = \sum_i E_i |\Psi_i\rangle\langle\Psi_i|, \tag{IV.153}$$

$$H_p = \sum_\alpha \hbar_\alpha \omega_\alpha (n_\alpha + 1/2), \tag{IV.154}$$

$$V = \sum_\alpha D_\alpha (b_\alpha + b^\dagger_{-\alpha}) e^{\mathrm{i}\mathbf{k}\cdot\mathbf{r}}. \tag{IV.155}$$

The basis vectors of the Hilbert space of the total system are tensor products of electron and phonon states, $\{|\Psi_i\rangle \otimes |\alpha, n\rangle\}$, denoted as $|i, \alpha, n\rangle$ in Eq. (IV.112). Further we use $n$ to denote the phonon states occupation, other Roman letters to index the electron states, and Greek letters for phonon states.

The eigenfunctions $\Psi_i$ of the electron Hamiltonian $H_e$ are localized – an illustrative example is a ground state in a two dimensional harmonic potential, $\langle \mathbf{r}|\Psi_1\rangle = (1/r_0\sqrt{\pi})\exp(-r^2/2r_0^2)$. For an example of the explicit form of $H_e$ in GaAs lateral quantum dot, see Ref. (Stano and Fabian, 2005).

Concerning phonons, the phonon index $\alpha$ comprises the wavevector $\mathbf{k}$, the polarization (one longitudinal and two transversal), and the branch (acoustic or optical), the last two denoted together by $\lambda$, $\alpha \equiv (\mathbf{k}, \lambda)$, [where further $-\alpha \equiv (-\mathbf{k}, \lambda)$], $b^\dagger_\alpha$, $b_\alpha$ are the creation and annihilation operators for a phonon in state $\alpha$, and $n_\alpha = b^\dagger_\alpha b_\alpha$ is the number operator. The phonon Hamiltonian $H_p$ describes non-interacting phonons, meaning our model does not contain processes equilibrating phonons. We, however, suppose that phonons are always in thermal equilibrium – we will insert this assumption by hand at the appropriate place. The electron-phonon interaction is encoded into the coupling constants $D_\alpha$. For the illustration and reference, we list here the electron-phonon interactions depicted in Fig. IV.11, thereby giving the explicit form of $D_\alpha$.

The deformation potential interaction is

$$V^{\mathrm{df}} = \sigma_e \sum_{\mathbf{k},\lambda=\{l,a\}} \sqrt{\frac{\hbar k}{2\rho V c_\lambda}} (b_{\mathbf{k},\lambda} + b^\dagger_{-\mathbf{k},\lambda}) e^{\mathrm{i}\mathbf{k}\cdot\mathbf{r}}. \tag{IV.156}$$

Only longitudinal acoustic phonons play role, $\lambda = \{l, a\}$, $\rho$ is the material density, $c_\lambda$ is the phonon velocity, $V$ is the material volume, and $\sigma_e$ is the deformation potential constant.

The piezoelectric interaction is

$$V^{\mathrm{pz}} = -\mathrm{i}h_{14} \sum_{\mathbf{k},\lambda=\{\{l,t_1,t_2\},a\}} \sqrt{\frac{\hbar}{2\rho V c_\lambda}} M_\lambda(\mathbf{k}) (b_{\mathbf{k},\lambda} + b^\dagger_{-\mathbf{k},\lambda}) e^{\mathrm{i}\mathbf{k}\cdot\mathbf{r}}, \tag{IV.157}$$

where the $h_{14}$ is the piezoelectric constant and polarization-dependent geometrical factors are

$$M_\lambda(\mathbf{k}) = \frac{2}{k^2}[k_x k_y (\vec{e}_{\mathbf{k},\lambda})_z + k_y k_z (\vec{e}_{\mathbf{k},\lambda})_x + k_z k_x (\vec{e}_{\mathbf{k},\lambda})_y]. \tag{IV.158}$$

One possible choice of the phonon polarizations is

$$\vec{e}_{\mathbf{k},l,a} = (k_x, k_y, k_z)/k, \tag{IV.159a}$$

$$\vec{e}_{\mathbf{k},t_1,a} = (-k_y, k_x, 0)/\sqrt{k_x^2+k_y^2}, \tag{IV.159b}$$

$$\vec{e}_{\mathbf{k},t_2,a} = (k_z k_x, k_z k_y, -k_x^2-k_y^2)/\left(k\sqrt{k_x^2+k_y^2}\right). \tag{IV.159c}$$

Finally, the Hamiltonian for the Fröhlich coupling is

$$H_{e-ph}^{(Fr)} = \sum_{\mathbf{k},\lambda=\{l,o\}} \sqrt{\frac{2\pi\hbar\Omega_\lambda}{V}\left(\frac{1}{\epsilon_\infty}-\frac{1}{\epsilon_s}\right)}\frac{e}{k}(b_{\mathbf{k},\lambda}+b^\dagger_{-\mathbf{k},\lambda})e^{\mathrm{i}\mathbf{k}\cdot\mathbf{r}}, \tag{IV.160}$$

where $\epsilon_s$, and $\epsilon_\infty$ is the static and high frequency dielectric function, respectively, and $\hbar\Omega_\lambda$ is the optical phonon energy. For the derivation of these interactions, see Refs. (Grodecka *et al.*, unpublished; Mahan, 2000). In the following we derive the master equation in the Born-Markov approximation, properly taking into account the environment. This derivation is a generalization of the master equation with the classical oscillating field, presented in Sec. B.

#### *G.2.1 Liouville equation in the interaction picture*

We will work in the interaction picture, which is suitable if the Hamiltonian can be divided into two parts, where for the first, representing an unperturbed system $H_0$, the solution is known and the second represents the perturbation, influence of which we study. For us, the unperturbed part are electron and phonons noninteracting with each other,

$$H_0 = H_e + H_p, \tag{IV.161}$$

and the perturbation is the electron-phonon interaction. The interaction picture for a general operator is defined as

$$O^I = \exp(\mathrm{i}H_0 t/\hbar)O\exp(-\mathrm{i}H_0 t/\hbar). \tag{IV.162}$$

If we express the Liouville equation using the interaction picture operators (this is the only place where we write the superscript $I$ for the density matrix explicitly, we omit it in further and in the main text)

$$\mathrm{i}\hbar\,\partial_t\rho_T^I(t) = [V^I(t), \rho_T^I(t)], \tag{IV.163}$$

we see that in the interaction picture it is only the perturbation that is responsible for a non-trivial evolution of the system. We rewrite the previous equation as an integral one

$$\rho_T(t) = \rho_T(0) + \frac{1}{\mathrm{i}\hbar}\int_0^t \mathrm{d}\tau\,[V^I(\tau), \rho_T(\tau)], \tag{IV.164}$$

which allows us to obtain an equivalent form for Eq. (IV.163)

$$\partial_t\rho_T(t) = \frac{1}{\mathrm{i}\hbar}[V^I(t), \rho_T(0)] - \frac{1}{\hbar^2}\int_0^t \mathrm{d}\tau\,[V^I(t), [V^I(\tau), \rho_T(\tau)]]. \tag{IV.165}$$

Here the hidden phonon dynamics shows up as the dependence of the evolution of the electron at time $t$ [$\rho(t)$ on the left hand side] on the previous electron state at times $\tau < t$ [$\rho(\tau)$ on the right hand side]. It is reasonable to neglect the time memory of the system in our model, since the model does not encompass the phonon dynamics, crucially influencing such cross time correlations. We thus adopt Markov approximation, in which the density matrix derivative depends only on the present state of the system, replacing $\rho_T(\tau)$ by $\rho_T(t)$ on the right hand side of Eq. (IV.165). Also related to the phonon dynamics being absent in our model is the Born approximation, in which the missing phonon equilibration is inserted by hand – we assume the phonon system is always in equilibrium, irrespective on the interaction with and state of the electron, $\rho_T(t) = \rho(t) \otimes \rho^p$. Here $\rho^p = Z^{-1}\exp(-H_p/k_B T)$, where $k_B$ is the Boltzmann constant, $T$ temperature, and $Z$ the normalization constant (partition function) assuring $\mathrm{tr}_p(\rho^p) = 1$. We note that such $\rho^p$ is diagonal in the chosen phonon basis $|\alpha, n\rangle$. If we apply these approximations and trace out the unobserved phonons defining the electron reduced density matrix, see Eq. (IV.114), we arrive at

$$\partial_t \rho(t) = \frac{1}{\mathrm{i}\hbar}\mathrm{tr}_p\Big([V^I(t), \rho_T(0)]\Big) - \frac{1}{\hbar^2}\int_0^t \mathrm{d}\tau\, \mathrm{tr}_p\Big([V^I(t), [V^I(\tau), \rho_T(t)]]\Big), \qquad \text{(IV.166)}$$

the starting equation for the further analysis. In the remaining sections of this Appendix we discuss some important ramifications of the above formalism.

### G.2.2 Statistical average of phonon operators

To proceed, we rewrite the electron-phonon Hamiltonian Eq. (IV.155) in the interaction picture using the definition (IV.162). For the boson phonon operators, it is straightforward to obtain the standard result, $b^I_\alpha(t) = b_\alpha \exp(-\mathrm{i}\omega_\alpha t)$. For the oscillating wave in the interaction picture we use the following short hand notation:

$$\begin{aligned}[\exp(\mathrm{i}\mathbf{k}_\alpha.\mathbf{r})]^I(t) = R_\alpha(t) &= \sum_{i,j} |\Phi_i\rangle\langle\Phi_i| \exp(\mathrm{i}\mathbf{k}_\alpha.\mathbf{r})|\Phi_j\rangle\langle\Phi_j| \exp(\mathrm{i}\omega_{ij} t) \\ &= \sum_{i,j} |\Phi_i\rangle R^{ij}_\alpha \langle\Phi_j| \exp(\mathrm{i}\omega_{ij} t), \end{aligned} \qquad \text{(IV.167)}$$

where $\hbar\omega_{ij} = E_i - E_j$ is the difference of energies of electron states $i$ and $j$.

We introduce $B_\alpha = b_\alpha + b^\dagger_{-\alpha}$ for the particular combination appearing in the electron-phonon interaction and get zero for the statistical average of a single phonon operators due to the diagonality of $\rho^p$:

$$\overline{B_\alpha(t)} = \mathrm{tr}_p\Big(B_\alpha(t)\rho^p\Big) = \sum_{\beta,n}\langle\beta, n|B_\alpha(t)\rho^p|\beta, n\rangle = \sum_{\beta,n}\langle\beta, n|B_\alpha(t)|\beta, n\rangle \rho^p_{\beta,n,\beta,n} = 0. \qquad \text{(IV.168)}$$

The statistical average of a pair of phonon operators is non zero

$$
\begin{aligned}
\overline{B_\alpha(t)B_\beta(\tau)} &= \sum_{\gamma,n}\langle\gamma,n|B_\alpha(t)B_\beta(\tau)\rho^p|\gamma,n\rangle \\
&= \sum_{\gamma,n}\langle\gamma,n|[b^I_\alpha(t)+b^{I\dagger}_{-\alpha}(t)][b^I_\beta(\tau)+b^{I\dagger}_{-\beta}(\tau)]|\gamma,n\rangle\rho^p_{\gamma,n,\gamma,n} \\
&= \delta_{\alpha,-\beta}\sum_{\gamma,n}\langle\gamma,n|b^I_\alpha(t)b^{I\dagger}_\alpha(\tau)+b^{I\dagger}_{-\alpha}(t)b^I_{-\alpha}(\tau)|\gamma,n\rangle\rho^p_{\gamma,n,\gamma,n} \\
&= \delta_{\alpha,-\beta}\sum_{\gamma,n}\langle\gamma,n|(1+n_\alpha)\exp[\mathrm{i}\omega_\alpha(\tau-t)]+n_{-\alpha}\exp[\mathrm{i}\omega_{-\alpha}(t-\tau)]|\gamma,n\rangle\rho^p_{\gamma,n,\gamma,n} \\
&= \delta_{\alpha,-\beta}\mathrm{tr}_p\Big((n_\alpha+1)\rho^p\Big)\exp[\mathrm{i}\omega_\alpha(\tau-t)]+\mathrm{tr}_p\Big(n_{-\alpha}\rho^p\Big)\exp[\mathrm{i}\omega_{-\alpha}(t-\tau)] \\
&= \delta_{\alpha,-\beta}\Big\{(\overline{n}_\alpha+1)\exp[\mathrm{i}\omega_\alpha(\tau-t)]+\overline{n}_{-\alpha}\exp[\mathrm{i}\omega_{-\alpha}(t-\tau)]\Big\} \qquad \text{(IV.169)}
\end{aligned}
$$

Here we have introduced the average occupation of the phonon state, $\overline{n}_\alpha = \mathrm{tr}_p(n_\alpha\rho^p) = 1/[\exp(E_\alpha/k_BT)-1]$. Note the difference to Sec. B. – there the dissipation was caused by a fluctuating classical field, with a correlation function that decays in time with a characteristic time $\tau_c$, Eq. (IV.19). Contrary to this, the statistical average in Eq. (IV.169) does not decay. It reflects that in our model phonons are always in equilibrium, there are no fluctuations. We will see that the two terms in Eq. (IV.169) lead to transitions including the emission (the first term) and absorption (second) of a phonon. The rates of the two processes depend on the mean phonon occupation number as $\overline{n}+1$ and $\overline{n}$, respectively, introducing temperature, which was effectively infinite in Sec. B. due to the classical description of the fluctuating field.

### *G.2.3 Relaxation and decoherence rates*

The first term on the right hand side of Eq. (IV.166) is zero, due to Eq. (IV.168):

$$
\begin{aligned}
\mathrm{tr}_p\Big([V^I(t),\rho(0)\rho^p]\Big) &= \sum_\alpha D_\alpha R_\alpha(t)\rho(0)\,\mathrm{tr}_p\Big(B_\alpha(t)\rho^p\Big) \\
&- D_\alpha\rho(0)R_\alpha(t)\,\mathrm{tr}_p\Big(\rho^p B_\alpha(t)\Big) = 0. \qquad \text{(IV.170)}
\end{aligned}
$$

The integrand in Eq. (IV.166) is

$$
\begin{aligned}
\mathrm{tr}_p\Big([V^I(t),[V^I(\tau),\rho(t)\rho^p]]\Big) &= \mathrm{tr}_p\Big(V^I(t)V^I(\tau)\rho(t)\rho^p - V^I(t)\rho(t)\rho^pV^I(\tau) \\
&- V^I(\tau)\rho(t)\rho^pV^I(t) + \rho(t)\rho^pV^I(\tau)V^I(t)\Big) \\
&= \sum_{\alpha,\beta} D_\alpha D_\beta\Big\{R_\alpha(t)R_\beta(\tau)\rho(t)\,\mathrm{tr}_p\Big(B_\alpha(t)B_\beta(\tau)\rho^p\Big)
\end{aligned}
$$

$$
\begin{aligned}
&- R_\alpha(t)\rho(t)R_\beta(\tau)\,\mathrm{tr}_p\Big(B_\alpha(t)\rho^p B_\beta(\tau)\Big) \\
&- R_\beta(\tau)\rho(t)R_\alpha(t)\,\mathrm{tr}_p\Big(B_\beta(\tau)\rho^p B_\alpha(t)\Big) + \rho(t)R_\beta(\tau)R_\alpha(t)\,\mathrm{tr}_p\Big(\rho^p B_\beta(\tau)B_\alpha(t)\Big)\Big\} \\
&= \sum_\alpha |D_\alpha|^2\Big\{R_\alpha(t)R_{-\alpha}(\tau)\rho(t)\overline{B_\alpha(t)B_{-\alpha}(\tau)} - R_\alpha(t)\rho(t)R_{-\alpha}(\tau)\overline{B_{-\alpha}(\tau)B_\alpha(t)} \\
&- R_{-\alpha}(\tau)\rho(t)R_\alpha(t)\overline{B_\alpha(t)B_{-\alpha}(\tau)} + \rho(t)R_{-\alpha}(\tau)R_\alpha(t)\overline{B_{-\alpha}(\tau)B_\alpha(t)}\Big\}.
\end{aligned}
\tag{IV.171}
$$

We have used that the Hermitivity of the electron-phonon Hamiltonian requires $D_\alpha = D^\dagger_{-\alpha}$. Inserting the statistical average of the pair of phonon operators we get for a particular element of the density matrix the following:

$$
\begin{aligned}
\partial_t\rho_{ij}(t) &= -\frac{1}{\hbar^2}\sum_{\alpha,k,l}|D_\alpha|^2 \\
&\times \int_0^t \mathrm{d}\tau\Big\{R^{ik}_\alpha R^{kl}_{-\alpha}\rho_{lj}(t)\exp(\mathrm{i}\omega_{ik}t+\mathrm{i}\omega_{kl}\tau)\{\overline{n}_{-\alpha}\exp[\mathrm{i}\omega_{-\alpha}(t-\tau)] \\
&+ (\overline{n}_\alpha+1)\exp[-\mathrm{i}\omega_\alpha(t-\tau)]\} \\
&- R^{ik}_\alpha\rho_{kl}(t)R^{lj}_{-\alpha}\exp(\mathrm{i}\omega_{ik}t+\mathrm{i}\omega_{lj}\tau)\{\overline{n}_\alpha\exp[-\mathrm{i}\omega_\alpha(t-\tau)] \\
&+ (\overline{n}_{-\alpha}+1)\exp[\mathrm{i}\omega_{-\alpha}(t-\tau)]\} \\
&- R^{ik}_{-\alpha}\rho_{kl}(t)R^{lj}_\alpha\exp(\mathrm{i}\omega_{ik}\tau+\mathrm{i}\omega_{lj}t)\{\overline{n}_{-\alpha}\exp[\mathrm{i}\omega_{-\alpha}(t-\tau)] \\
&+ (\overline{n}_\alpha+1)\exp[-\mathrm{i}\omega_\alpha(t-\tau)]\} \\
&+ \rho_{ik}(t)R^{kl}_{-\alpha}R^{lj}_\alpha\exp(\mathrm{i}\omega_{kl}\tau+\mathrm{i}\omega_{lj}t)\{\overline{n}_\alpha\exp[-\mathrm{i}\omega_\alpha(t-\tau)] \\
&+ (\overline{n}_{-\alpha}+1)\exp[\mathrm{i}\omega_{-\alpha}(t-\tau)]\}\Big\}.
\end{aligned}
\tag{IV.172}
$$

We will now make several approximations based on the same reasoning. From the final result of this computation, listed in Eqs. (IV.115), one can see that the relaxation and decoherence rates define timescales for the changes of the density matrix, meaning that over much shorter times the density matrix is essentially constant. Also, we are not interested in such short time (or equivalently large frequency) effects. To give an example, a typical orbital relaxation in the quantum dot is $\sim 0.1$ ns, which is to be compared with the time corresponding to a typical energy difference $\hbar\omega_{ik}$ appearing in Eq. (IV.172), being for two orbital states typically $\hbar/1$ meV$\sim$ ps. For two states split by the Zeeman energy the energy difference is smaller, corresponding in 1 tesla field to 0.1 ns. However, here the relaxation is much slower ($> 1\mu$s), resulting in similarly large difference between $\omega_{ik}$ and the relaxation rate for the two states.

The first place where we use the difference of the timescales is in the time integration in Eq. (IV.172) – since $t \gg 1/\omega$, we can make the limit $t \to \infty$, getting the delta function $\int_0^t \mathrm{d}\tau\exp(\mathrm{i}\omega\tau) \approx \int_0^\infty \mathrm{d}\tau\exp(\mathrm{i}\omega\tau) = \pi\delta(\omega)$ [we neglected the principal value part of the integral, which leads to small imaginary terms which only slightly renormalize the electron energy

(Blum, 1996)]. The equation for the density matrix simplifies to

$$\begin{aligned}
\partial_t \rho_{ij}(t) &= -\frac{\pi}{\hbar^2} \sum_{\alpha,k,l} |D_\alpha|^2 \\
&\times \Big\{ R_\alpha^{ik} R_{-\alpha}^{kl} \rho_{lj}(t) \exp(\mathrm{i}\omega_{ik} t) \{\overline{n}_{-\alpha} \delta(\omega_{kl} - \omega_{-\alpha}) \exp(\mathrm{i}\omega_{-\alpha} t) \\
&+ \delta(\omega_{kl} + \omega_\alpha)(\overline{n}_\alpha + 1) \exp(-\mathrm{i}\omega_\alpha t)\} \\
&- R_\alpha^{ik} \rho_{kl}(t) R_{-\alpha}^{lj} \exp(\mathrm{i}\omega_{ik} t) \{\overline{n}_\alpha \delta(\omega_{lj} + \omega_\alpha) \exp(-\mathrm{i}\omega_\alpha t) \\
&+ (\overline{n}_{-\alpha} + 1) \delta(\omega_{lj} - \omega_{-\alpha}) \exp(\mathrm{i}\omega_{-\alpha} t)\} \\
&- R_{-\alpha}^{ik} \rho_{kl}(t) R_\alpha^{lj} \exp(\mathrm{i}\omega_{lj} t) \{\overline{n}_{-\alpha} \delta(\omega_{ik} - \omega_{-\alpha}) \exp(\mathrm{i}\omega_{-\alpha} t) \\
&+ (\overline{n}_\alpha + 1) \delta(\omega_{ik} + \omega_\alpha) \exp(-\mathrm{i}\omega_\alpha t)\} \\
&+ \rho_{ik}(t) R_{-\alpha}^{kl} R_\alpha^{lj} \exp(\mathrm{i}\omega_{lj} t) \{\overline{n}_\alpha \delta(\omega_{kl} + \omega_\alpha) \exp(-\mathrm{i}\omega_\alpha t) \\
&+ (\overline{n}_{-\alpha} + 1) \delta(\omega_{kl} - \omega_{-\alpha}) \exp(\mathrm{i}\omega_{-\alpha} t)\} \Big\}. \qquad \text{(IV.173)}
\end{aligned}$$

We use the delta functions to express all the oscillating time factors by the electron energy differences

$$\begin{aligned}
\partial_t \rho_{ij}(t) &= -\frac{\pi}{\hbar^2} \sum_{\alpha,k,l} |D_\alpha|^2 \\
&\times \Big\{ R_\alpha^{ik} R_{-\alpha}^{kl} \rho_{lj}(t) \exp(\mathrm{i}\omega_{il} t) \{\overline{n}_{-\alpha} \delta(\omega_{kl} - \omega_{-\alpha}) \\
&+ \delta(\omega_{kl} + \omega_\alpha)(\overline{n}_\alpha + 1)\} \\
&- R_\alpha^{ik} \rho_{kl}(t) R_{-\alpha}^{lj} \exp(\mathrm{i}\omega_{ik} t + \mathrm{i}\omega_{lj} t) \{\overline{n}_\alpha \delta(\omega_{lj} + \omega_\alpha) \\
&+ (\overline{n}_{-\alpha} + 1) \delta(\omega_{lj} - \omega_{-\alpha})\} \\
&- R_{-\alpha}^{ik} \rho_{kl}(t) R_\alpha^{lj} \exp(\mathrm{i}\omega_{ik} t + \mathrm{i}\omega_{lj} t) \{\overline{n}_{-\alpha} \delta(\omega_{ik} - \omega_{-\alpha}) \\
&+ (\overline{n}_\alpha + 1) \delta(\omega_{ik} + \omega_\alpha)\} \\
&+ \rho_{ik}(t) R_{-\alpha}^{kl} R_\alpha^{lj} \exp(\mathrm{i}\omega_{kj} t) \{\overline{n}_\alpha \delta(\omega_{kl} + \omega_\alpha) \\
&+ (\overline{n}_{-\alpha} + 1) \delta(\omega_{kl} - \omega_{-\alpha})\} \Big\} \qquad \text{(IV.174)}
\end{aligned}$$

Now we make another approximation in the same spirit by keeping only the terms without oscillating time factors. This is so called on-shell (or secular) approximation and can be understood from the following – let us integrate both sides of Eq. (IV.174) over time $T$ much shorter compared to the timescale for a change of the density matrix but much larger compared to a particular $1/\omega_{ij}$. Then on the right hand side only the exponential function is time dependent and picking just one term from the sum we get

$$\left.\frac{\triangle\rho}{\triangle T}\right|_\alpha = \left.\frac{\rho_{ij}(T) - \rho_{ij}(0)}{T}\right|_\alpha \propto \frac{1}{T}\int_0^T \exp(\mathrm{i}\omega_{il} t)\} \begin{cases} \sim 1/\omega_{il} T \sim 0, & \text{if } \omega_{il} \neq 0, \\ = 1, & \text{if } \omega_{il} = 0. \end{cases} \qquad \text{(IV.175)}$$

Therefore only the terms without explicit time dependent factors contribute to the derivative. In the first term in Eq. (IV.174) only terms with $l = i$ contribute, so we can replace the exponential by the Kronecker delta $\delta_{il}$. In the second term, unless there is regularity in the spectrum, a case we do not consider, there are two possibilities how the frequency in the exponential can be zero: $i = k$ and $l = j$, or $i = j$ and $l = k$. Applying these results gives

$$\begin{aligned}
\partial_t \rho_{ij}(t) &= -\frac{\pi}{\hbar^2} \sum_{\alpha,k,l} |D_\alpha|^2 \\
&\times \Big\{ R_\alpha^{ik} R_{-\alpha}^{kl} \rho_{lj}(t) \delta_{il} \{ \overline{n}_{-\alpha} \delta(\omega_{kl} - \omega_{-\alpha}) \\
&+ \delta(\omega_{kl} + \omega_\alpha)(\overline{n}_\alpha + 1) \} \\
&- R_\alpha^{ik} \rho_{kl}(t) R_{-\alpha}^{lj} (\delta_{ik}\delta_{lj} + \delta_{ij}\delta_{kl}) \{ \overline{n}_\alpha \delta(\omega_{lj} + \omega_\alpha) \\
&+ (\overline{n}_{-\alpha} + 1) \delta(\omega_{lj} - \omega_{-\alpha}) \} \\
&- R_{-\alpha}^{ik} \rho_{kl}(t) R_\alpha^{lj} (\delta_{ik}\delta_{lj} + \delta_{ij}\delta_{kl}) \{ \overline{n}_{-\alpha} \delta(\omega_{ik} - \omega_{-\alpha}) \\
&+ (\overline{n}_\alpha + 1) \delta(\omega_{ik} + \omega_\alpha) \} \\
&+ \rho_{ik}(t) R_{-\alpha}^{kl} R_\alpha^{lj} \delta_{kj} \{ \overline{n}_\alpha \delta(\omega_{kl} + \omega_\alpha) \\
&+ (\overline{n}_{-\alpha} + 1) \delta(\omega_{kl} - \omega_{-\alpha}) \} \Big\}. \qquad \text{(IV.176)}
\end{aligned}$$

We consider now the diagonal and non-diagonal terms separately. For the diagonal the result can be written in the form

$$\partial_t \rho_{ii}(t) = -\sum_{k \neq i} 2\Gamma_{ik} \rho_{ii}(t) + \sum_{k \neq i} 2\Gamma_{ki} \rho_{kk}(t), \qquad \text{(IV.177)}$$

where

$$2\Gamma_{ij} = \frac{2\pi}{\hbar^2} \sum_\alpha |D_\alpha|^2 |R_\alpha^{ij}|^2 \{ \overline{n}_{-\alpha} \delta(\omega_{ij} + \omega_{-\alpha}) + (\overline{n}_\alpha + 1) \delta(\omega_{ij} - \omega_\alpha) \} \qquad \text{(IV.178)}$$

is the transition rate from state $i$ to state $j$ due to the perturbation of the system by phonons, computed according to Fermi's Golden rule: The rate is a sum of individual phonon contributions, where each one is proportional to the overlap between the two electron states through the electron-phonon interaction potential. The energy is conserved – if $\omega_{ij} < 0$ a phonon is absorbed, providing the needed energy, while if $\omega_{ij} > 0$, a phonon is emitted taking away the energy. Thus Eq. (IV.177) says that a particular state is (de)populated by (out)in scatterings (to)from all other states of the system. We note that terms with $k = i$ canceled themselves.

For better understanding, we write the result for a non-diagonal density matrix element assuming $\omega_\alpha = \omega_{-\alpha}$ which leads to $n_\alpha = n_{-\alpha}$:

$$\partial_t \rho_{ij}(t) = -\gamma_{ij} \rho_{ij}(t) - \left( \frac{\pi}{\hbar^2} \sum_\alpha |D_\alpha|^2 |R_\alpha^{ii} - R_\alpha^{jj}|^2 (2\overline{n}_\alpha + 1) \delta(\omega_\alpha) \right) \rho_{ij}(t). \qquad \text{(IV.179)}$$

The first term describes the fact that the relaxation always contributes to the decoherence with the rate

$$\gamma_{ij} = \sum_{k \neq i} \Gamma_{ik} + \sum_{k \neq j} \Gamma_{jk}. \qquad \text{(IV.180)}$$

The second term in Eq. (IV.179) is specific only to the decoherence – in our model it is zero due to the zero density of phonons at zero phonon energy [which is required by the term $\delta(\omega_\alpha)$]. However, if there would be a way for phonons to exchange energy (for example, with each other, or with some energy reservoir) and phonon occupations would fluctuate, each phonon that changes the energy of the states $i$ and $j$ differently (that is $R_\alpha^{ii} - R_\alpha^{jj} \neq 0$) causes decoherence. This is because such energy changes due to phonons appear as random shifts of the phases of the two states.

Finally we note that from Eqs. (IV.178) and (IV.180) the results for high and low temperature stated in Sec. B., Eq. (IV.45), follow. Indeed, considering two states, at temperature high enough with respect to the energy difference of the states, the mean phonon occupation is high and approximately $\overline{n} \approx \overline{n} + 1$. Then the emission and absorption rates are equal $\Gamma_{12} = \Gamma_{21}$. We get $\gamma_{12} = 2\Gamma_{21}$, meaning $T_1 = T_2$. On the other hand, if the temperature is much lower than the states energy difference, there are no phonons available, $\overline{n} \approx 0$ and the absorption process is not possible, $\Gamma_{12} = 0$. From that we get $T_2 = 2T_1$.

### G.3 Oscillating field in the rotating wave approximation

Derivation of the oscillating resonant field contribution to the time derivative of the electron reduced density matrix, Eqs. (IV.118), due to the electron-field interaction, Eq. (IV.116), can be done in the same way as the way we treated the electron-phonon interaction in the previous section: We include the interaction $V^M$, which in the interaction picture is

$$\begin{aligned} (V^M)^I &= \sum_{i,j} \hbar\Omega_{ij} |\Phi_i\rangle\langle\Phi_j| \exp(\mathrm{i}\omega_{ij}t)\cos(\omega t) \\ &= \sum_{i,j} \hbar\Omega_{ij} |\Phi_i\rangle\langle\Phi_j| \exp(\mathrm{i}\omega_{ij}t)[\exp(\mathrm{i}\omega t) + \exp(-\mathrm{i}\omega t)]/2, \end{aligned} \tag{IV.181}$$

into the interaction Hamiltonian $V$ in the Liouville equation Eq. (IV.166). If we neglect terms representing mixed phonon-microwave contributions, the phonons and microwave contribute separately and it is more suitable to go back to Eq. (IV.163) to get the part due to the microwave. Inserting here Eq. (IV.181) and tracing over phonons, what now just changes $\rho_T$ into $\rho$ without other consequences, we get

$$\begin{aligned} \partial_t^M \rho_{ij}(t) &= \sum_k \frac{1}{2} \left[ -\mathrm{i}\Omega_{ik}\rho_{kj}(t)\exp(\mathrm{i}\omega_{ik}t) + \mathrm{i}\Omega_{kj}\rho_{ik}(t)\exp(\mathrm{i}\omega_{kj}t) \right] \\ &\times \left[ \exp(\mathrm{i}\omega t) + \exp(-\mathrm{i}\omega t) \right]. \end{aligned} \tag{IV.182}$$

We now once again use the approximation of keeping only the terms with the lowest frequency, which is in this case called the rotating wave approximation, but is of the same nature as the approximation described under Eq. (IV.172). Supposing the frequency of the oscillating field is close to the energy difference of a particular pair of states $a$ and $b$, we find that the off resonant terms are not influenced by the field, $\partial_t^M \rho_{ij} = 0$, if $i, j \notin \{a, b\}$ and the only non zero contributions following from Eq. (IV.182) are those listed in Eq. (IV.118).

## V. Spintronics devices and materials

In a narrow sense spintronics refers to spin control of electronics. Say, flip a spin or turn on magnetic field and the current stops flowing, ideally. Similarly, we would like a device which would orient spin by passing a current or applying a gate voltage. In this way the spin would be fully integrated with electronics and we could write, store and manipulate, as well as read the information based on spin.

The goal is to make useful electronic devices that would enhance functionalities of the existing semiconductor technology. Thus far this goal has been elusive, although the field has gone through immense progress keeping us optimistic about its potential. The case at hand is metal spintronics, which has already revolutionized computer industry with a device based on giant magnetoresistance. Earlier, hard-disk information stored in the magnetization of the grains of the disk was read by the so-called anisotropic magnetoresistance effect: in a bulk ferromagnetic conductor, due to the crystal anisotropy, the resistance of the conductor depends on the orientation of the magnetization of the ferromagnet. Resistance variations are typically rather small, of a percent or so. As the magnetization beneath a read head changed, a small change in the current was sensed.

The discovery of the giant magnetoresistance effect (GMR) by Binasch *et al.* (1989) and Baibich *et al.* (1988) allowed to increase the density of the information stored in hard disks, leading to more than a hundredfold increase in their capacity. The idea is to increase the sensitivity of the electrical current due to magnetization changes. The GMR effect occurs in ferromagnetic layered nanostructures, such as the one shown in Fig. V.1. Two ferromagnetic layers sandwich over a nonmagnetic conductor. If the magnetizations of the two layers are parallel, the resistance is small, if they are antiparallel, the resistance is large. The relative change of the resistance is called giant magnetoresistance.[94] At room temperature the changes are typically about 10–50%, with the upper values obtained in multilayer systems (Grünberg, 2001). Why is the resistance different for the different relative orientations of magnetization? Take parallel magnetizations. The spacer layer between the ferromagnets is a few nanometers thick. Electrons injected there from one layer keep their spin orientation and can relatively easily continue to the second ferromagnetic layer. If the magnetizations are antiparallel, the injected electrons will be more likely reflected from the second interface, due to the reduced density of states for that spin in the second ferromagnet. This interface scattering increases the resistance of the antiparallel orientation.

[94]The word giant refers to the magnitude of the effect: it is giant if the magnetoresistance change is more than about 10%. Nowadays the term refers to the specific magnetoresistance due to the difference in the parallel and antiparallel magnetizations of the layers. Such a structure is also called a spin valve.

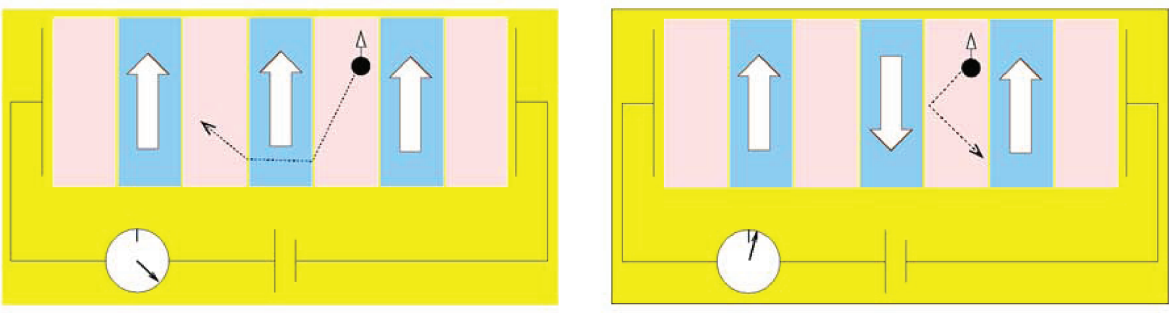

Fig. V.1. Scheme of a giant magnetoresistive (GMR) multilayer system. If the magnetizations of the ferromagnetic layers are parallel (left), the current in the circuit is larger than in the case of antiparallel magnetizations (right).

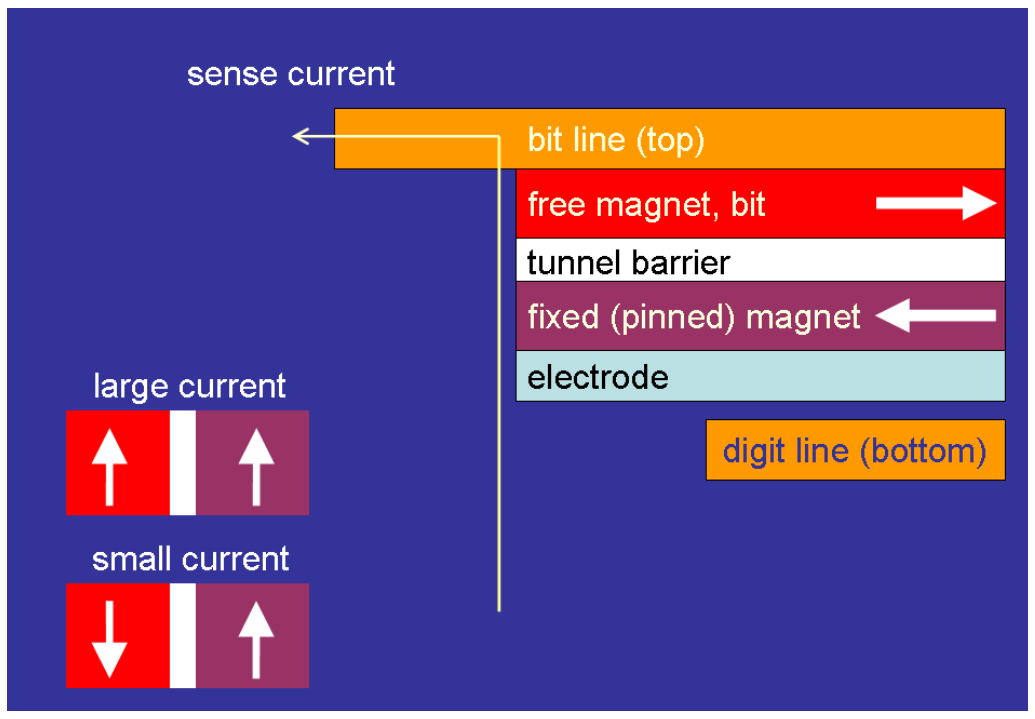

Fig. V.2. Scheme of a tunnel magnetoresistive (TMR) element of the magnetic random access memory (MRAM). The information is stored as the magnetization direction, left or right, in the free ferromagnetic layer. The magnetization can be flipped by current induced magnetic field if electric current flows through both the bit and the digit lines simultaneously. The information about the magnetization (bit) is read by passing current through the three-layer tunnel junction formed by the two magnetic layers and the insulator in between.

Another important spin-valve-like effect is the tunneling magnetoresistance, discussed in Sec. II.H.1. The physics is similar in description to GMR, although the transport is by tunneling through a nonmagnetic insulating layer, not ballistic transport through a metallic nano-region. A metallic TMR is being employed as a non-volatile magnetic random access memory (MRAM), whose operating principle is shown in Fig. V.2. Non-volatility is crucial here: the information about a memory element is stored in the magnetization configuration of the ferromagnetic layers; this information need not be refreshed, nor does it disappear after the power is switched off.

Where does the semiconductor spintronics stand in terms of device applications? The most sought for semiconductor spintronic device is spin transistor. Since there are many spin transistors proposed, and only few convincingly demonstrated, it is too early to say in which direction the field develops. Various designs have various advantages and disadvantages, but without experimental demonstrations theoretical proposals are hard to judge. We should also mention that the word transistor is often liberally used to describe any three terminal device, without a prospect for current amplification. Such devices can be useful for electrical injection or as spin valves, but not for logic elements which require voltage controlled ON and OFF states with the ratio of the electrical currents in these states of at least 1000 to 1.

Selected spin transistor schemes are shown in Fig. V.3. The most straightforward scheme is a spin metal-oxide-silicon field-effect transistor (spin MOSFET), see (Sugahara and Tanaka, 2004, 2005). This device would act as a spin valve in the setting of a conventional field-effect transistor. If the magnetizations of the ferromagnetic source and drain (also called emitter and collector) are parallel, the transport channel is open (ON); if the magnetizations are antiparallel, the channel is closed (OFF). This structure is yet to be demonstrated, in particular now that spin injection into silicon has been demonstrated, see Sec. II.F.2.

The so-called Datta-Das spin transistor (Datta and Das, 1990), in Fig. V.3 b, uses spin-orbit coupling of the Bychkov-Rashba type for its operational principle. The source and drain are ferromagnetic. The transport channel is a two-dimensional electron gas. The top gate, which in conventional field-effect transistors controls the channel conductance, now controls the spin-orbit coupling strength. Electrons injected with momentum parallel to the transport feel an effective magnetic spin-orbit field transverse to that direction. The electrons' spins then precess with a single precession frequency, assuming a ballistic one dimensional transport. Depending on the precession speed, the spin either precesses very little (or even multiples of $\pi$), in which case the electron will enter into the drain (ON), or by $\pi$ (or its odd multiples), in which case the electron bounces back, increasing the channel's resistance (OFF). The Datta-Das transistor has not been demonstrated, although its conceptual simplicity and originality has been source of much inspiration in the field. Different variants of the Datta-Das scheme have been proposed (Schliemann *et al.*, 2003; Bandyopadhyay and Cahay, 2004a).

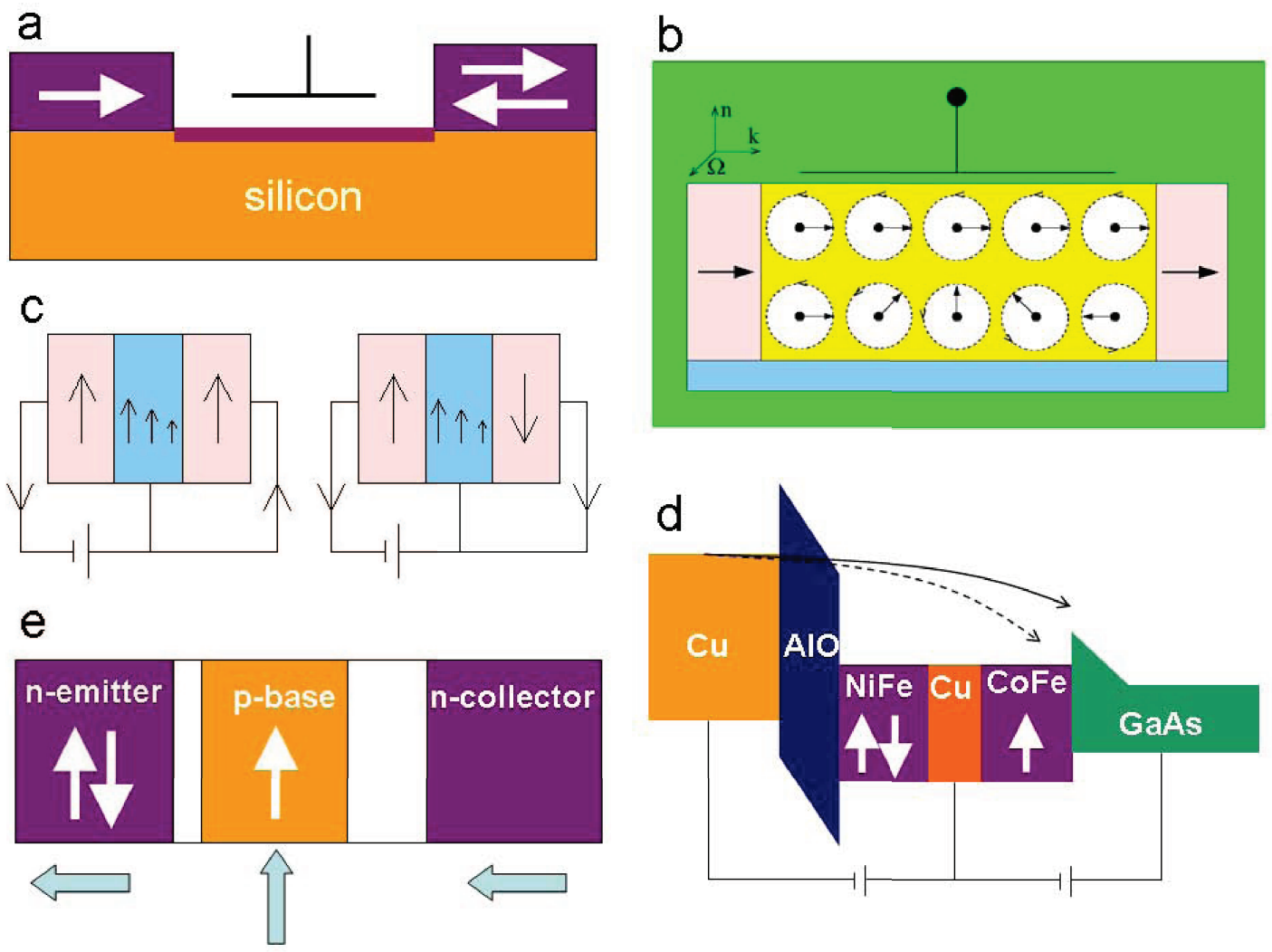

Fig. V.3. (a) spin metal-oxide-silicon field effect transistor (spin MOSFET), (b) Datta-Das spin FET, (c) Johnson's spin-valve transistor, (d) a hot electron spin-valve transistor, and (e) magnetic bipolar junction transistor.

The Johnson spin switch, shown in Fig. V.3 c, originally proposed as an all-metal Ohmic transistor (but the physics is equally applicable to highly degenerate semiconductors) is based on a spin-valve geometry in which the nonmagnetic layer offers an additional contact. This structure offers little amplification, if any, since the base current would be similar in magnitude to the collector current.[95] The transistor works as follows: if the magnetizations of the two ferromagnetic layers are parallel, the collector current flows as indicated, into the collector. If the magnetizations are antiparallel, the collector current is opposite.

Hot electron spin transistors form a large class of devices, see (Žutić *et al.*, 2004). These are usually hybrid metal/semiconductor structures, offering again little potential for current amplification due to the large base current. Nevertheless, these devices offer huge magnetocurrents, as well as practical ways to inject spin into semiconductors such as silicon, see Sec. II.F.2. In the structure shown in Fig. V.3 d, the emitter is a nonmagnetic metal (Cu), while the collector is a nonmagnetic semiconductor (GaAs). The base is formed by a ferromagnetic spin-valve: NiFe/Cu/CoFe, after (van Dijken *et al.*, 2003a). The base forms a Schottky barrier (Kittel, 1996) with the collector. Electrons that tunnel from the emitter to the base are hot electrons (not thermalized to the Fermi level of the base). These electrons lose energy depending on the relative orientation of the magnetizations in the spin valve. If the magnetizations are parallel, the energy loss is greater than for an antiparallel orientation. The loss of energy is directly reflected in the collector current, since only the electrons of energy high enough to overcome the potential Schottky barrier contribute to the collector current. This large spin filtering effect has been demonstrated to give magnetocurrent (relative ratio of the collector current for parallel and antiparallel orientation of the spin valve) exceeding 3000% (van Dijken *et al.*, 2003b). An all-semiconductor version of a hot electron spin transistor has also been proposed (Mizuno *et al.*, 2007).

[95] As we will see later, the current amplification is defined as the ratio of the collector to the base current. In semiconductor transistors this ratio is on the order of one hundred.

Finally, the magnetic bipolar transistor (Fabian *et al.*, 2002a; Flatté *et al.*, 2003; Lebedeva and Kuivalainen, 2003), depicted in Fig. V.3 e, is based on the conventional junction transistor design, substituting ferromagnetic semiconductors in the active regions, say the base. This transistor allows for spin-control of current amplification (Fabian *et al.*, 2002a; Fabian and Žutić, 2004a) due to spin-dependent tunneling across the emitter/base contact (called the depletion layer). Although the diode version of the transistor (a single magnetic p-n junction) has been demonstrated, the transistor is still a theoretical concept. We will discuss the physics of the magnetic bipolar transistor in more detail below.

Other spintronic devices include a proposed scheme for reconfigurable logic (Dery *et al.*, 2007), a room temperature spin-transference device (Dery *et al.*, 2006), electron spin resonance transistor (Vrijen *et al.*, 2000), a 2d channel spin valve controlled by ferromagnetic gates (Ciuti *et al.*, 2002), spin capacitor (Žutić *et al.*, 2001a; Datta, 2005), spin Esaki diodes (Kohda *et al.*, 2001; Johnston-Halperin *et al.*, 2002), spin lasers (Rudolph *et al.*, 2003), unipolar magnetic diodes (Flatté and Vignale, 2001), spin light-emitting diodes (Fiederling *et al.*, 1999, 2003; Ohno *et al.*, 1999), field-effect magnetic switch (Ohno *et al.*, 2000; Matsukura *et al.*, 2002a), or spin-flip or single electron spin-valve transistors (Brataas *et al.*, 2006; Wetzels *et al.*, 2005). In the following section we describe in detail spintronic devices based on magnetic resonant tunneling.

In order to provide a balanced view, we also note that not all agree on the potential of certain classes of spintronics devices, see (Bandyopadhyay and Cahay, 2005, 2004b). Since at the moment most devices are theoretical concepts, it is simply too early to say which schemes will be practical, as well as which role and which functionalities will be played and taken over by spintronic devices from the future semiconductor technologies.

Figure V.4 summarizes prospects for spin transistors as emerging electronic devices. The chart was proposed by the influential International Technology Roadmap for Semiconductors, produced by technology experts. Spin transistors appear at the tail, behind molecular transistors, whose at least short-term prospects appear far less certain than those of spin transistors.[96] It is still gratifying to know that the spin transistor got into the chart at all.

[96]Figure out the risk for a combined molecular spin transistor! Risks should multiply.

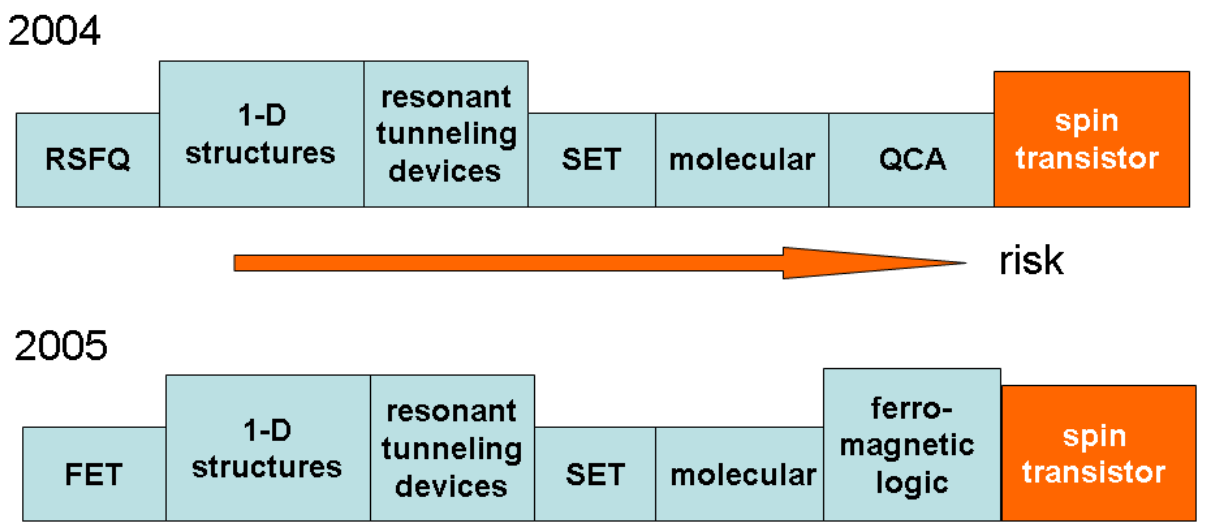

Fig. V.4. International Technology Roadmap for Semiconductors 2004 and 2005 evaluated prospects of various emerging devices for electronics applications (Semiconductor Industry Association, 2004, 2005). Spin transistors got into the chart, albeit at the last position, doomed most risky. The 2006 edition brought no changes to the 2005 one (Semiconductor Industry Association, 2006). More info can be found on the Roadmap web site, www.itrs.net.

In the following we describe two classes of semiconductor spintronic devices: magnetic resonant tunneling diodes and magnetic bipolar diodes and transistors. The presentation of these two classes reflects more the authors' own interest than an attempt to pick the most perspective devices. Nevertheless, both classes have been studied theoretically as well as experimentally, allowing to nicely illustrate the nature and complexities of spintronic devices. We also include a section on diluted magnetic semiconductors reviewing their properties as well as presenting a mean field model to explain the occurance of ferromagnetism in certain classes of these materials (such as GaMnAs).

## A. Resonant tunneling diodes

### A.1 Introduction

The realization of the anticipated advantages of spintronics needs an electrically controllable creation, manipulation and detection of spin-polarized currents (Prinz, 1999; Wolf *et al.*, 2001; Gregg *et al.*, 2002; Žutić *et al.*, 2004; Ivanov *et al.*, 2004). This requires in general that spin-dependent transport processes can be modulated by external factors as applied gate voltages or magnetic fields. A promising approach to tackle these tasks is to use stacked ultrathin layers of the order of a few nanometers of both magnetic and nonmagnetic semiconductors. In such heterostructures the material properties are changing on the length scales comparable to the phase coherence length of the carriers, i.e., on distances on which the carriers preserve the memory of their initial wave function phase. Hence, quantum interference effects become important for describing the transport properties in the vertical growth direction of the structure. Typically, some of the layers constitute energetic barriers for the incident carriers, which can only be overcome by tunneling. In particular for double or multi-barrier structures very high transmission probabilities up to unity can occur for some resonant energies; an effect which is known as *resonant tunneling*. The magnetic layers in the structure cause a strongly spin-dependent transmission, since in magnetic semiconductors the spin degeneracy of the valance and conduction bands are distinctly broken by an exchange or giant Zeeman field. [97] This spin splitting gives rise to an energy and *spin selective* resonant tunneling of the carriers through the structure, which effectively allows for the realization of very efficient spin filters and spin detectors.

In view of possible device applications one can think of several desired properties, which ideal magnetic layers should possess: the ferromagnetic order should remain even at high temperatures well above room temperature; the exchange splitting of the bands should well exceed the thermal energy $k_B T$, where $k_B$ denotes the Boltzmann constant and $T$ is the temperature; for flexibility different magnetic materials with either a particle density dependent or a particle density independent ferromagnetic state should be available; the growing of high quality structures with clean interfaces should be feasible; for the purpose of integration in existing technologies close lattice-matching to common semiconductors is favorable; and in terms of mobility and spin lifetime $n$-type conductivity usually appears advantageous. Luckily, nowadays magnetic semiconductors materials can already meet at least several of these criterions thanks to the tremendously growing research interest and the achieved enormous progress in developing magnetic materials in the last decade.

The current state of the art in spin-dependent transport in heterostructures shows that spintronic devices based on resonant tunneling can be used to accomplish several different important aims and functionalities necessary for future spintronic applications. They can be utilized as (i) tunneling magnetoresistance (TMR)- or spin-valve devices, which exhibit high TMR-effects even at higher applied biases, (ii) highly efficient spin injectors, which can produce a high degree of spin-polarization, following different routes, e.g., all magnetic heterostructures, magnetic interband or Zener tunneling resonant tunnel diodes (RTDs), coupled quantum wells, which combine the resonant tunneling with the spin-blockade effect, or nonmagnetic RTDs exploiting the spin-orbit coupling, (iii) electrically controllable spin switching devices by using a magnetic quantum well, (iv) as spin-detectors, exploiting the energy- and spin-resolving spectroscopic nature of resonant tunneling, and (v) as ultimate magnetoelectronic devices, in which the ferromagnetic order can be controlled by external voltages. In view of the progress of growing high-quality structures and the invention and development of novel high $T_c$ magnetic semiconductors such band-engineered heterostructures appear to be a perfect playground for exploiting the possibilities to utilize the carriers spin degree of freedom in semiconductor electronics. A better understanding of the interplay of the transport and magnetic properties in low-dimensional systems will enrich the functionalities of already known device concepts and inspire the invention of novel magnetoelectronic devices.

In what follows we briefly discuss at first the basic physics of resonant tunneling in nonmagnetic structures for both fully coherent and sequential tunneling. Transport in such conventional nonmagnetic systems has been thoroughly investigated in the past. A detailed introduction to resonant tunneling can be found for instance in the textbooks Ferry and Goodnick (1997); Datta (1995) and a good collection of articles about the underlying physics and its possible applications is given in Chang *et al.* (1991); Mizuta and Tanoue (1995); Capasso (1990) and references therein. The resonant tunneling diode (RTD) was actually one of the first realized quantum devices (Tsu and Esaki, 1973) and its theoretical investigation considerably initiated the afterwards rapidly growing interest in quantum transport. Quantum devices based on resonant tunneling are technologically interesting due to their extreme high-speed (the intrinsic tunneling time is typically of the order of 100 femtoseconds) and low-power performance. The industrial large-scale production, however, is a challenging task, since room temperature operation needs the growing of high quality structures with reproducible device characteristics.

[97] The effective electron $g$-factor can exceed 100 in magnetic semiconductors, which is giant compared to the free electron value of $g \approx 2$. This causes a giant Zeeman splitting $\Delta E = g\mu_B B$ in external magnetic fields $B$ with $\mu_B$ is the Bohr magneton.

After the introductory discussion of conventional RTDs we describe how magnetic layers can lead to a spin-dependent transmission. For this purpose we apply a mean field description of the magnetism in thin dilute magnetic semiconductors (DMS) layers. Endowed with this basic knowledge, we finally review the research on magnetic resonant tunneling structures done so far and discuss possible spintronic device applications, describing in more detail the concept of a magnetic monostable bistable logic element (m-MOBILE). Readers familiar with conventional resonant tunneling can directly skip to Sec. B.

### A.2 Theory of resonant tunneling

The striking quantum phenomena of tunneling refers to the possibility that quantum particles can traverse regions, which are from a classical point of view energetically forbidden. Tunneling is an intimate consequence of the wave properties of matter and the probabilistic interpretation of the wave function. Quantum tunneling was already considered from the early days of quantum mechanics in connection with the problem of field ionization of atoms and the nuclear decay of alpha particles. Shortly thereafter, the concept of tunneling was firstly applied in solid state physics to explain the field emission of electrons from metals into vacuum. A brief overview of the history of tunneling can be found in Ferry and Goodnick (1997); Garcia-Calderon (1993). Single barrier tunneling has found widespread applications. One of the most prominent is the invention of the scanning tunneling microscope (STM), in which particles tunnel through a controllable vacuum barrier and which made it possible to make images on an atomic scale.

In the case of tunneling through a single barrier of height $V_0$, the energy-dependent transmission probability $T(E)$, which is defined as ratio of the transmitted to the incident flux, decreases exponentially with the barrier width $W$:

$$T(E) \propto e^{-2W\sqrt{2m(V_0-E)}/\hbar}, \tag{V.1}$$

where $m$ denotes the particle mass. When a second barrier of same width is added one might intuitively suggest that, following Ohm's law, the total resistance of the structure is just doubled. This is indeed true if the region between the barriers is much larger than the de Broglie wavelength of the electrons, which in semiconductors is typically of the order of 10-100 nanometers.[98] However, if the middle region is only a few nanometers in width the carrier transport remains phase-coherent and for some incident energies $E_m$ within a small energy range of width $\gamma$, the particle is transmitted with a high probability, eventually up to one. This extraordinary enhancement of the transmission probability is known as *resonant tunneling*. The physical explanation is that the resonant energies correspond to the energies of the quasibound eigenstates of the quantum well formed by the double-barrier system. These states are not truly bound, because electrons, which are localized in such a state, can leak out through the barriers with a finite probability. Due to the uncertainty principle, the finite lifetime $\tau$ of the electron causes an uncertainty in the energy $\Delta E\tau \approx \hbar$, which effectively leads to the broadening of the resonance $\gamma = \Delta E \approx \hbar/\tau$. The whole process of resonant tunneling can be understood as a constructive interference between the waves leaking through the first barrier and the reflected waves of the second barrier, similar to what happens to electromagnetic waves in a Fabry-Perot etalon. In a more particle-like picture corresponding to wave packets an incident electron at resonant energy tunnels through the first barrier, bounces then several times back and forth in the quantum well in a way that adds up coherently, and finally tunnels out through the second barrier. The single versus double-barrier tunneling process is schematically sketched in Fig. V.5.

[98] It is rather the dephasing or decoherence length that determines the tunneling character. Typically an inellastic scattering, in which energy is exchange with phonons or with other electrons, contribute to decoherence. Elastic scattering off static impurities does not cause irreversible loss of phase, though it can change the tunneling properties.

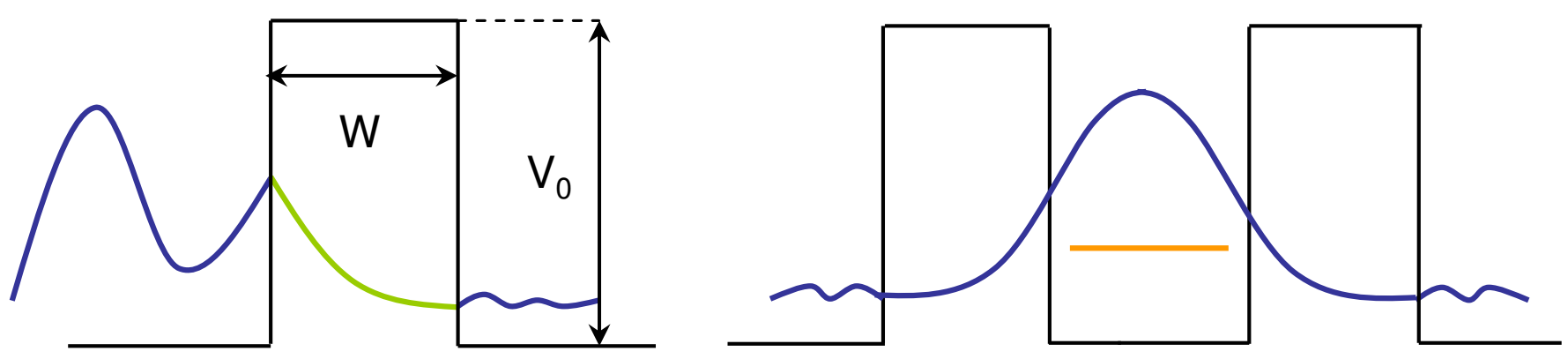

Fig. V.5. Left: Tunneling through a single barrier; the wave function amplitude decreases exponentially with the barrier thickness $W$ and height $V_0$. Right: Occurrence of resonant tunneling through a double barrier structure when the energy of the incident electrons coincides with one of the discrete well state energies.

In the pioneering work of Tsu and Esaki (1973) such double barrier structures were realized by an epitaxial growth of alternating ultrathin films of two semiconductor materials with different band gaps. Using GaAs as smaller band gap material and $Ga_{1-x}Al_xAs$ as barrier with the barrier height controlled by the molecular fraction $x$ of Al, the conduction band profile of the layered structure exhibits sharp discontinuities at the heterointerfaces, effectively realizing a double barrier structure as shown in Fig. V.6. Nowadays it is possible to grow well-controlled semiconductor heterostructure layers with atomic precision due to the impressive advances in epitaxial growth techniques such as molecular beam epitaxy (MBE) (Yu and Cardona, 2001). The double barrier structure is usually surrounded by heavily doped layers, which provide low-resistance emitter and collector contacts. To prevent diffusion of the dopants from the high doped regions into the inner double-barrier structure usually also thin undoped buffer layers are included in experiments. By attaching ohmic contacts to the whole structure an external bias can be applied to the *resonant tunneling diode* (RTD). The $N$-shaped current-voltage (IV) characteristics exhibits a region of negative differential resistance (NDR), which allows for interesting technological applications such as high frequency oscillators, switching devices, or realization of simplified digital logic circuits (Capasso, 1990; Chang *et al.*, 1991; Mizuta and Tanoue, 1995).

This NDR-behavior of a RTD can be qualitatively easily understood if we recognize that the electrons which are trapped between the two barriers exhibit a discrete energy spectrum whose spacing increases if the confinement gets stronger, i.e., the quantum well width becomes smaller. Let us assume, for simplicity, that the quantum well is thin enough that there is only one quasibound state in the energy range of interest as shown in Fig. V.7. With a positive bias $V_a$ applied to the right (collector) lead the resonant energy level is lowered relative to the energy of the incident electrons from the left (emitter) lead. In a first approximation one can assume that the voltage drops linearly from the emitter to the collector side. The electrons in the left (emitter) and right (collector) contact are considered to be always in thermal equilibrium which allows to introduce chemical potentials $\mu_L, \mu_R$ for both reservoirs and to describe the electrons distribution by the Fermi-Dirac function. This means that at low temperatures incident electrons from the emitter with energies reaching from the bottom of the conduction band up to the Fermi energy are available. However, since the RTD effectively acts as an energy filter only electrons with the resonant energy $E_0$ can transmit to the collector side if there are unoccupied states at that energy; otherwise the electrons are blocked by the Pauli principle. By applying positive bias to the collector the resonant level passes through the emitter's Fermi energy and current starts flowing (see point A in Fig. V.7). Increasing the bias leads to higher and higher current magnitudes. However, at a certain voltage (the peak voltage) the resonant level becomes energetically aligned with the bottom of the emitter's conduction band as illustrated by point B in Fig. V.7. Further bias pushes the resonant level $E_0$ below this edge (depicted by point C in Fig. V.7), which suddenly cuts off the supply of emitter's electrons causing a sharp drop in the current and thereby leading to the phenomenon of NDR.

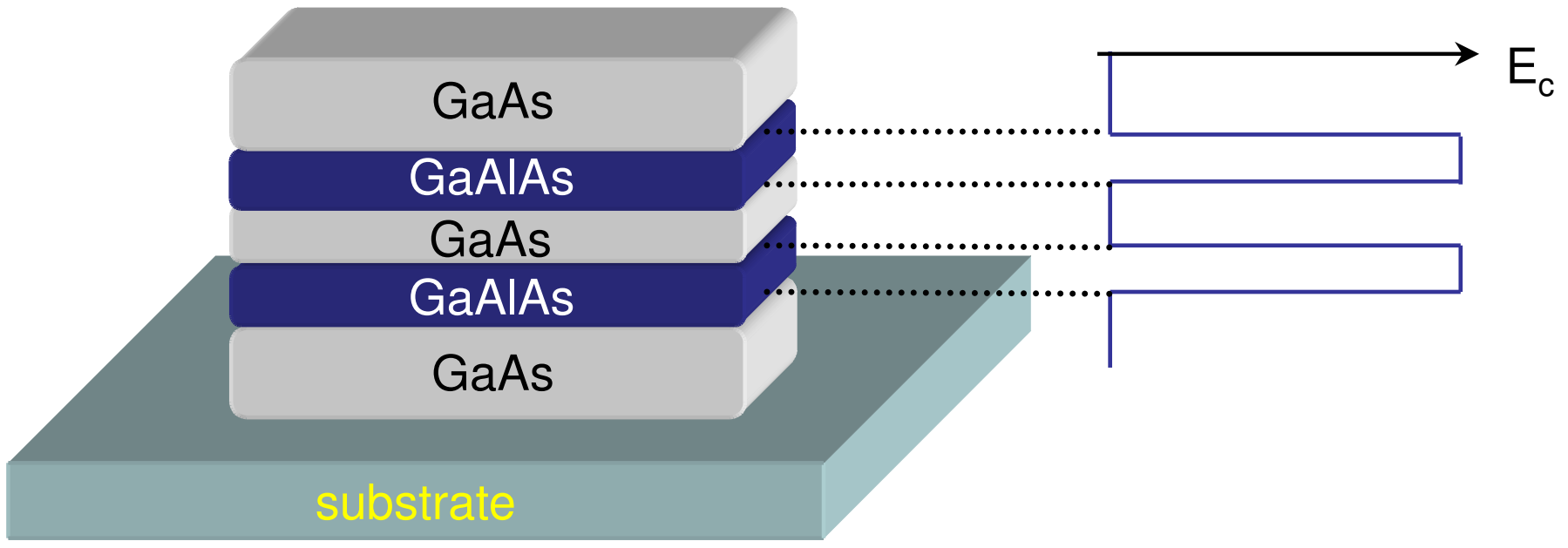

Fig. V.6. A double barrier structure realized by the epitaxial growth of a few nanometers thick planar layers of AlGaAs and GaAs. By sandwiching the middle GaAs layer between two AlGaAs barriers a quantum well is formed, in which conduction electrons become confined.

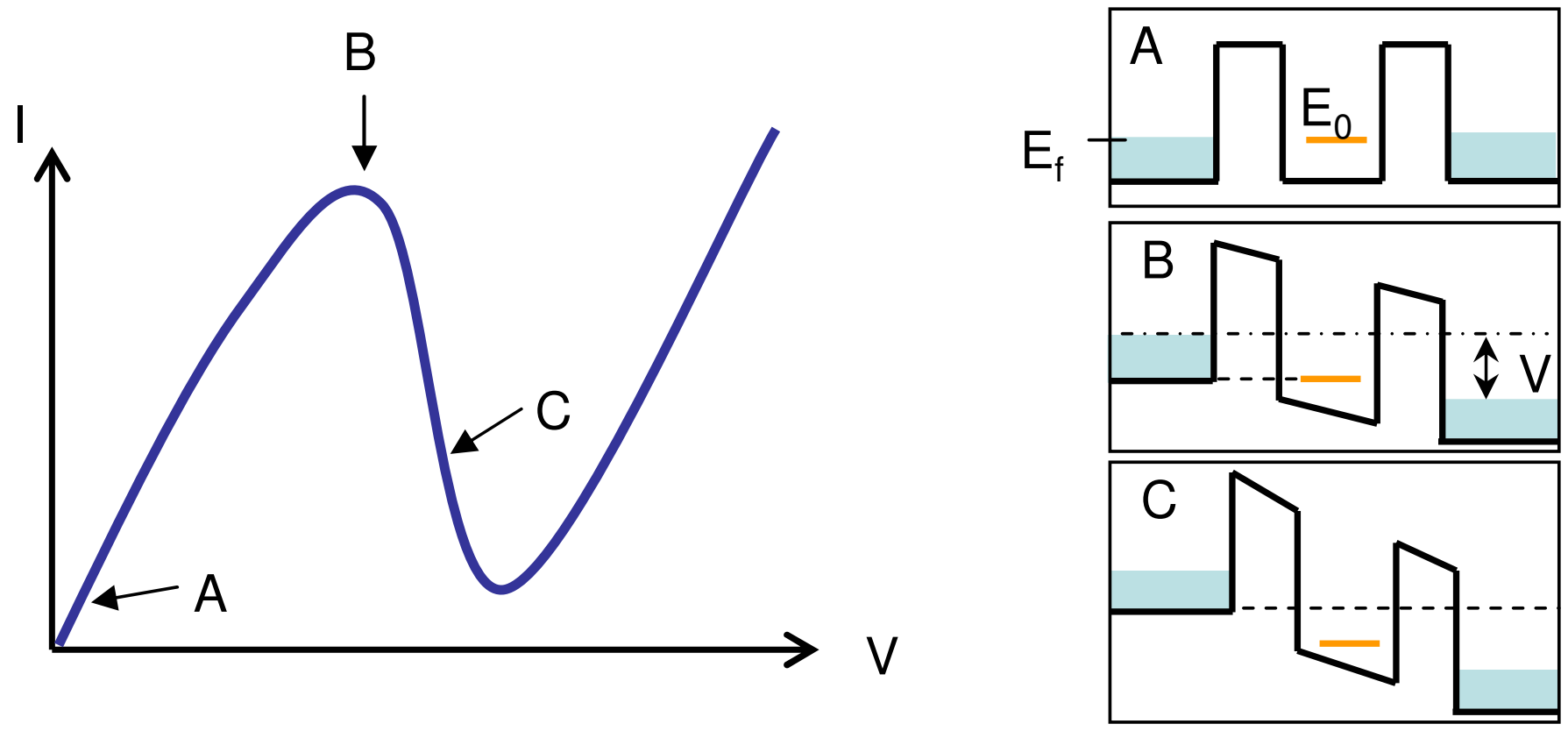

Fig. V.7. Left: Typical current-voltage ($IV$) characteristic of a resonant tunneling diode exhibiting a negative differential resistance (NDR) region. Right: Conduction band profile for different applied biases. Point A (small applied voltages): current starts to flow through the resonant level when the resonant well level $E_0$ is pulled down to the emitter's Fermi energy $E_f$. Point B (peak voltage): the resonant well level coincides with the emitter conduction band edge $E_0 = 0$. Point C (NDR-regime): $E_0 < 0$.

### A.3 Coherent tunneling

For the purpose of obtaining a more quantitative understanding of resonant tunneling and the related NDR-effect in semiconductor heterostructures we assume at first that the transport through the structure is fully phase-coherent.[99] This assumption allows to apply a wave function treatment of the transport similar to what is done in the description of electromagnetic wave propagation in planar layers of different permittivity. To emphasize the basic conceptual issues we restrict our discussion here primarily to electrons in a parabolic conduction band, e.g., one can think of the Γ-valley ($\mathbf{k} \approx 0$) electrons in GaAs. In the case of coherent transport between two contacts the flowing current density can be obtained in general from the Landauer-Büttiker formula (Landauer, 1957, 1970; Büttiker *et al.*, 1985; Datta, 1995; Ferry and Goodnick, 1997)

$$j = \frac{2e}{h} \int \hat{T}(E)[f_L(E) - f_R(E)]\mathrm{d}E, \tag{V.2}$$

where the factor 2 takes into account the spin degeneracy, $e$ is the elementary charge, $h = 2\pi\hbar$ is Planck's constant, and $f_{L,R}$ are the electrons distribution function in the left and right reservoir, which are usually assumed to be given by Fermi-Dirac functions. The single particle transmission function $\hat{T}(E)$ describes physically how likely a single electron of energy $E$ can transmit through the structure and is more rigorously defined as the sum over all transmission probabilities $T_{n\leftarrow m}(E)$ of an electron starting in the input mode $m$ and ending up in the output mode $n$ of the left and right leads, respectively, which connect the reservoirs with the structure (Datta, 1995). As we will see below, in the specific case of planar heterostructures these lead modes are easily identified with the plain wave electron states of fixed in-plane momentum $\mathbf{q}$, i.e., of a certain momentum component perpendicular to the growth direction. If we assume that the in-plane momentum is conserved during the transport, which means that there is no scattering from one lead mode to another, the transmission function can be written as

$$\hat{T}(E) = \sum_{\mathbf{q}',\mathbf{q}} T_{\mathbf{q}'\leftarrow\mathbf{q}} = \sum_{\mathbf{q}',\mathbf{q}} \delta_{\mathbf{q}',\mathbf{q}}\, T_{\mathbf{q}}(E) = \sum_{\mathbf{q}} T_{\mathbf{q}}(E). \tag{V.3}$$

This assumption is reasonable for elastic scatterers, which do not change the electron's momentum considerably, and as long as inelastic scattering processes are not important (which should be actually the case to allow for a phase-coherent propagation).

[99] In the coherent transport regime the phase of the wave function is preserved, which excludes inelastic scattering processes, which randomize the particles' phase.

The transmission function $T_{\mathbf{q}}(E)$ can be determined from the solution of the single-particle Schrödinger equation if the electrons can be treated as independent coherently propagating quasi-particles. This demands that the effect of electron-electron interactions is describable by an effective single-particle potential, which, by following the approach of local density functional theory, depends only on the local electron density. For simplicity we will include here only the selfconsistent Hartree terms and neglect the exchange potentials or other electron-electron correlations. The influence of the periodic lattice potential of the crystal on the electrons is treated in the effective mass approximation. Under these assumptions the steady-state envelope function $\psi(\mathbf{r}, z)$ of an single electron in the heterostructure can be determined from the Schrödinger-like equation

$$\left(\frac{\hbar^2}{2}\frac{\partial}{\partial z}\frac{1}{m_l(z)}\frac{\partial}{\partial z} + \frac{\hbar^2}{2m_t}\nabla_{\mathbf{r}}^2 + V_{\text{eff}}(z)\right)\psi(\mathbf{r}, z) = E\psi(\mathbf{r}, z). \tag{V.4}$$

Here, $\mathbf{r}$ is the in-plane or transversal position vector and $z$ denotes the growth direction or what we call the longitudinal direction, $m_l(z)$ is the longitudinal effective mass perpendicular to the heterointerface, $m_t$ is the in-plane effective mass and $E$ denotes the total energy. The kinetic energy operator for the longitudinal motion takes into account the $z$-dependence of the longitudinal effective mass and satisfies the requirement of being Hermitian. The effective potential $V_{\text{eff}}(z) = U_i(z) + U_{\text{el}}(z)$ contains the intrinsic conduction band discontinuities $U_i$ and the electrostatic potential $U_{\text{el}}(z)$, which depends on the fixed ionized impurity density and the electron density profile in the structure. Since the effective potential varies only in the longitudinal direction the in-plane motion of the electrons, which is of free electrons plane-wave type, can be separated from the growth-direction dynamics, justifying a product ansatz for the envelope function: $\psi(\mathbf{r}, z) \propto e^{i\mathbf{qr}}\varphi(z)$. With this the lead input and output modes can be characterized by the plane wave states $e^{i\mathbf{qr}}$ and Eq. (V.4) can be reduced to an effective one-dimensional Schrödinger equation for the growth direction motion

$$\left(\frac{\hbar^2}{2}\frac{\partial}{\partial z}\frac{1}{m_l(z)}\frac{\partial}{\partial z} + V_{\text{eff}}(z)\right)\varphi(z) = E_l\varphi(z), \tag{V.5}$$

where we introduce the longitudinal energy $E_l = E - \hbar^2 q^2/2m_t = E - E_t$, which we always measure in the following from the bottom of the emitter's conduction band. From the definition of the longitudinal energy it is evident that $E_l$ is conserved during the transport if we assume that the total energy and the in-plane momentum $\mathbf{q}$ are conserved and that $m_t$ is independent of $z$. As we will see, these assumptions considerably simplify all further calculations, since the transmission function $T_{\mathbf{q}}(E) = T(E_l)$ will only depend on the longitudinal energy, having no explicit dependence on the in-plane momenta $\mathbf{q}$.

According to our mean field approach for the electron-electron interaction the electrostatic potential $U_{\text{el}} = -e\phi$ can be obtained from the Poisson equation

$$\frac{\partial}{\partial z}\epsilon(z)\frac{\partial}{\partial z}\phi(z) = \frac{1}{\epsilon_0}[en(z) - \rho_{\text{imp}}(z)], \tag{V.6}$$

where $\epsilon$ denotes the, in general, $z$-dependent static dielectric constant, $\epsilon_0$ is the permeability of the vacuum, $\rho_{\text{imp}}(z)$ is the fixed impurity charge density of the structure, and $n(z)$ is the electron density. The Poisson equation (V.6) is nonlinearly coupled to the envelope function equation (V.5) via the particle density $n[\varphi(z)]$, since the electron density profile of the structure is established by occupying the energy-dependent scattering states $\varphi(z)$ according to the distribution functions of the electron reservoirs in the emitter and collector leads. Hence, the coupled Schrödinger-Poisson system has to be solved in a selfconsistent way, which can be done iteratively by alternately solving both equations and using the solution of one equation as input for the other, until convergence is reached.

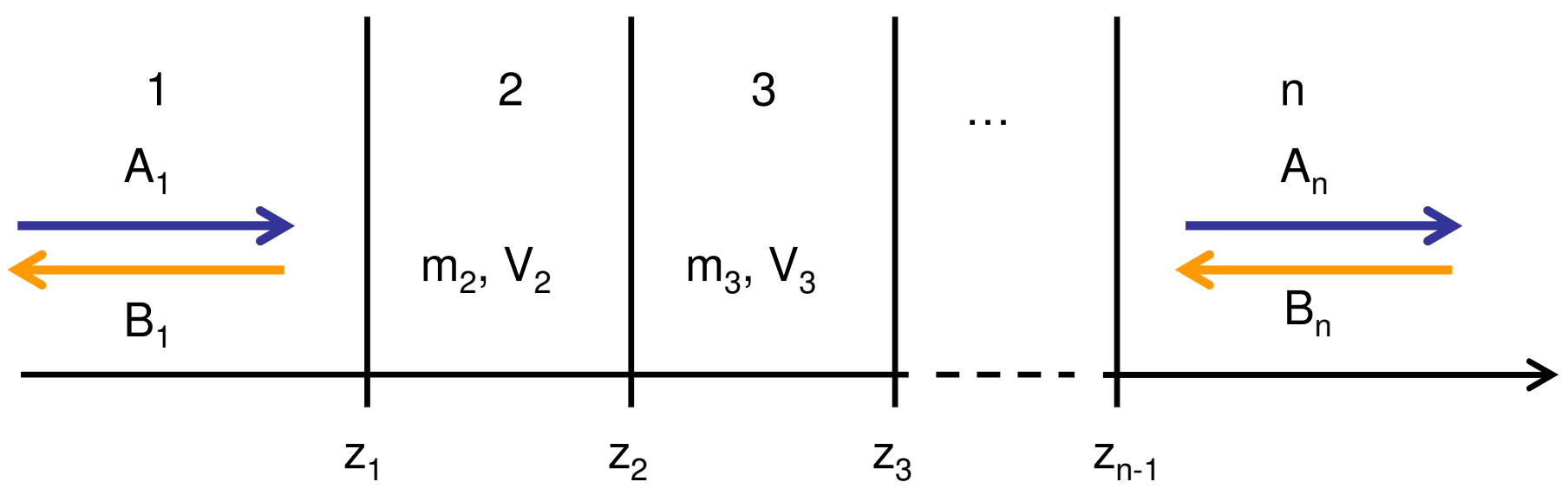

Fig. V.8. Illustration of the transfer matrix technique. The structure is divided along the growth direction $z$ into a sequence of $n$ different layers $[z_{i-1}, z_i]$ with constant effective masses $m_i$ and potentials $V_i, i = 1, \dots, n$. The total transfer matrix $M$ connects the wave amplitudes of the incoming and reflected electrons from the first layer $(A_1, B_1)$ with those of the last layer $(A_n, B_n)$.

In order to solve the Schrödinger equation (V.5) and to find the transmission function $T(E_l)$ we introduce here the *transfer matrix technique*, which is widely used in literature (Tsu and Esaki, 1973; Vassell *et al.*, 1983; Ricco and Azbel, 1984; Ferry and Goodnick, 1997). The basic idea of the method is to divide the $z$-axis into a sequence of regions where the solution can be obtained analytically. These local solutions are then composed to a global one by using the continuity conditions of the wave function between the different regions. To illustrate the method let us assume that we have $n$ different layers with different effective masses $m_i$ and constant effective potentials $V_i$ in each layer as illustrated in Fig. V.8. The solution for each individual layer $z_{i-1} \leq z \leq z_i$ can then generally be written as the sum of left and right moving plane wave states

$$\varphi_i = A_i \varphi_i^+ + B_i \varphi_i^- = A_i e^{ik_i z} + B_i e^{-ik_i z} \tag{V.7}$$

with $k_i = \sqrt{2m_i(E_l - V_i)}/\hbar$ and $A_i$, $B_i$ denoting the amplitudes of right and left moving waves, respectively. The continuity of the wave function demands,

$$\varphi_i(z_i) = \varphi_{i+1}(z_i) \tag{V.8}$$

and the conservation of the probability current leads to,

$$\frac{1}{m_i}\frac{d}{dz}\varphi_i(z_i) = \frac{1}{m_{i+1}}\frac{d}{dz}\varphi_{i+1}(z_i). \tag{V.9}$$

These relations between neighboring layers can be rewritten in matrix form,

$$U_i(z_i)\begin{pmatrix} A_i \\ B_i \end{pmatrix} = U_{i+1}(z_i)\begin{pmatrix} A_{i+1} \\ B_{i+1} \end{pmatrix}, \quad i = 1, \ldots, n-1 \tag{V.10}$$

with the matrix

$$U_i(z) = \begin{pmatrix} \varphi_i^+ & \varphi_i^- \\ \frac{1}{m_i}(\varphi_i^+)' & \frac{1}{m_i}(\varphi_i^-)' \end{pmatrix} \tag{V.11}$$

where the prime denotes the derivative with respect to $z$. Starting with $i = 1$, Eq. (V.10) allows to express the transition amplitudes of the second layer as a function of the amplitudes of the first one, $C_2 = U_2^{-1}(z_1)U_1(z_1)C_1$, using the vector notation $C_i = (A_i, B_i)$. The matrix $M_1 = U_2^{-1}(z_1)U_1(z_1)$ is called a transfer matrix between the first and second region since it connects the corresponding amplitudes. Repeating successively this procedure for $i = 2, \ldots, n-1$ finally allows to correlate the amplitudes of the last layer with those of the first one:

$$\begin{pmatrix} A_n \\ B_n \end{pmatrix} = M\begin{pmatrix} A_1 \\ B_1 \end{pmatrix}, \tag{V.12}$$

where we have introduced the composed transfer matrix,

$$M = U_n^{-1}(z_{n-1})U_{n-1}(z_{n-1})U_{n-1}^{-1}(z_{n-2})\ldots U_2(z_2)U_2^{-1}(z_1)U_1(z_1) = \prod_{i=1,n-1} M_i. \tag{V.13}$$

Hence, the total transfer matrix can be composed by the individual transfer matrices $M_i$ just by using conventional matrix multiplications.

The amplitudes $C_1$ are determined by the boundary conditions of the Schrödinger equation. For instance, if we assume only impinging electrons from the left we can set $A_1 = 1$ and $B_n = 0$. Using the relation $C_n = MC_1$ leads to $B_1 = -M_{21}/M_{22}$ with $M_{ij}$ denoting the matrix elements of $M$. The knowledge of the first layer amplitudes $C_1$ allows to successively calculate all other layer amplitudes $(C_2 = M_1C_1, C_3 = M_2C_2, \ldots)$, constructing in this way the envelope function throughout the whole structure.

The transfer matrix connects the left and right amplitude coefficients of the structure. This representation is not unique and it is often more convenient to connect the incoming and outgoing amplitudes by the *scattering matrix*

$$\begin{pmatrix} B_1 \\ A_n \end{pmatrix} = S\begin{pmatrix} A_1 \\ B_n \end{pmatrix} = \begin{pmatrix} r & t' \\ t & r' \end{pmatrix}\begin{pmatrix} A_1 \\ B_n \end{pmatrix}. \tag{V.14}$$

The S-matrix is a natural representation for scattering problems, since the diagonal elements are given by the reflection amplitudes $r$ and $r'$ for waves coming from the left and right hand side of the sample, respectively, and the off-diagonal elements are related to the wave transmission amplitudes $t$ and $t'$. This physical interpretation of $S_{ij}$ becomes immediately evident by recognizing that the outgoing amplitudes can be always composed by reflected and transmitted parts of incoming wave amplitudes of electrons impinging from the same and opposite side, e.g., ($B_1 = S_{11}A_1 + S_{12}B_n = rA_1 + t'B_n$). By using Eqs. (V.14) and (V.12) the transfer matrix can be also expressed in terms of these wave amplitudes

$$M = \begin{pmatrix} t - r'(t')^{-1}r & r'(t')^{-1} \\ -(t')^{-1}r & (t')^{-1} \end{pmatrix}. \tag{V.15}$$

It should be noted that for the general case of $N$ incoming channels the amplitudes $A_1, B_1, A_n$ and $B_n$ become complex vectors of length $N$ and the transmission and reflection amplitudes are replaced by $N \times N$ matrices. The elements of the transfer matrix are not independent due to the flux conservation and other physical symmetries. For instance, for symmetric structures time reversal symmetry leads to the relation $t = (t')^T$, where the superscript $T$ denotes transposition of the matrix. In the simple one-dimensional case, as considered here, this simplifies to $t = t'$ confirming the intuitive expectation that the transmission amplitude is the same for left and right incident electrons of equal energy, since the left-moving electron follows the time-reserved trajectory of the right moving one.

If the transmission matrix is known, the single particle transmission function $T(E_l)$ can be easily obtained as follows. Physically the transmission function is defined as the ratio of the transmitted to the incident probability flux of a particle: $T = f_{\text{trans}}/f_{\text{inc}}$. Similarly, the reflection coefficient is defined by $R = f_{\text{refl}}/f_{\text{inc}}$ with $f_{\text{refl}}$ denoting the reflected probability flux. Conservation of the total particles flux demands that $T + R = 1$. The incident probability flux is given by the squared wave amplitude times the group velocity of the incident electron, which we assume here to impinge from the left, $f_{\text{inc}} = |A_1|^2\hbar k_1/m_1$, and the reflected and transmitted fluxes are accordingly determined by $f_{\text{refl}} = |B_1|^2\hbar k_1/m_1$ and $f_{\text{trans}} = |A_n|^2\hbar k_n/m_n$. With these definitions the transmission function reads as

$$T(E_l) = \frac{k_n m_1 |A_n|^2}{k_1 m_n |A_1|^2}. \tag{V.16}$$

By applying the corresponding boundary conditions of left incident electrons ($A_1 = 1, B_n = 0$) and by using the relation $C_n = MC_1$ we obtain

$$A_n = \frac{\det M}{M_{22}} A_1. \tag{V.17}$$

The determinant of the transfer matrix results in $\det M = k_1 m_n / k_n m_1$, since one easily finds $\det[U_i(z_i)U_i^{-1}(z_{i-1})] = 1$ and $\det[U_n^{-1}(z_{n-1})U_1(z_1)] = k_1 m_n / k_n m_1$, which can be easily verified by using the explicit expressions for $M$ and $U_i$ stated in Eq. (V.13) and (V.11), respectively. With this the transmission function can finally be written as

$$T(E_l) = \frac{k_1 m_n}{k_n m_1} \frac{1}{|M_{22}|^2}. \tag{V.18}$$

An important point to note here is that the transmission function can also be defined as the squared *current* transmission amplitude $T = |\tilde{t}|^2$. The current transmission amplitude $\tilde{t}$ is related to the wave transmission amplitude $t$, which we have introduced in the definition of the S-matrix in Eq. (V.14), by $\tilde{t} = t_{L\to R}\sqrt{v_R/v_L}$, where $v_L, v_R$ are the left and right side group velocities. The renormalized scattering matrix based on current amplitudes, $\tilde{S}_{ij} = S_{ij}\sqrt{v_i/v_j}$, has the advantage of being unitary due to current flux conservation (Ferry and Goodnick, 1997). With this it follows that $T(E) = (v_R/v_L)|t|^2$, which is consistent with our previous results Eq. (V.18) and Eq. (V.15) by taking into account that $\tilde{t} = \tilde{t'}$ and, hence, $t'\sqrt{v_L/v_R} = t\sqrt{v_R/v_L}$ according to time reversal symmetry.

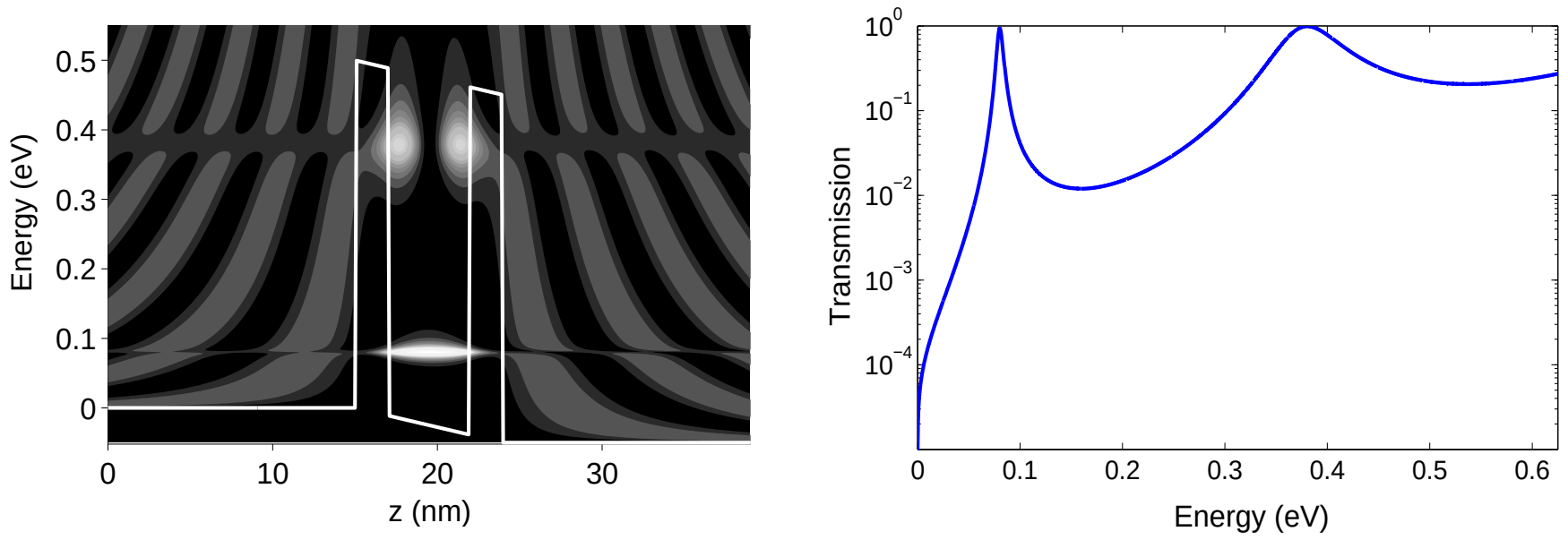

Fig. V.9. Energetics of a model tunnel diode at applied bias. Left: Contour plot of the local density of states of electrons versus energy and growth direction $z$. Bright regions correspond to high densities and dark to small ones. The first two quasibound states in the quantum well and their energetic broadening is clearly visible. The conduction band profile is indicated by the white solid line. Right: Transmission probability of the electrons as a function of the incident electrons energy. Resonant tunneling $T(E) \approx 1$ occurs at the well state energies. Near resonance the transmission function can be described in good approximation by a Lorentzian profile according to the Breit-Wigner formula.

In order to investigate the basic physics of resonant tunneling we apply these general results to the special case of a double barrier structure. The typical appearance of the transmission function versus the electron's incident energy is illustrated in the right plot of Fig. V.9 showing its strongly "spiky" characteristic, whereas the left plot of Fig. V.9 displays the local density of states of the conduction electrons, in which the forming of quantum well states and their energetic broadening become clearly apparent. Since such a double-barrier structure consists of two single barriers in series we can calculate the total transmission matrix $M$ by using the composition law $M = M_2 M_1$, where $M_1$ and $M_2$ are the transfer matrices of the first and second single barrier. Using the general expression given in Eq. (V.15) the composed transmission amplitude $t$ results in

$$t = \frac{t_1 t_2}{1 - r_1' r_2}, \tag{V.19}$$

where $t_i, r_i$ and $r_i'$ denote the amplitudes of the single barriers $i = 1, 2$. The transmission function is given by the squared current transition amplitude

$$T(E_l) = \frac{v_R}{v_L}|t|^2 = \frac{T_1 T_2}{1 - 2\sqrt{R_1 R_2}\cos(\theta) + R_1 R_2} \tag{V.20}$$

with $T_1 = v_w/v_L|t_1|^2, T_2 = v_R/v_w|t_2|^2$ with $v_w$ denoting the group velocity in the well, $R_i = |r_i|^2 = |r_i'|^2, i = 1, 2$ and $\theta$ is the phase of $r_1' + r_2$. The phase shift $\theta$ corresponds to the phase acquired by the electron when it makes one round-trip between the two barriers, which means that the electron is reflected once from each barrier before transmitting the structure. The

analytical form of the transmission function $T_1, T_2$ for the single barriers is easily obtained from the transfer matrix technique (Ferry and Goodnick, 1997) showing an exponential dependence on the barrier width $W$, as stated in Eq. (V.1), in the limit of thick and/or high barriers $Wk_b \gg 1$, where $k_b = \sqrt{2m(V_0 - E)}/\hbar$.

The expression, Eq. (V.20), for the composed transmission function of the double barrier structure can be further simplified if we assume, as is normally the case, that $T_1, T_2 \ll 1$, and consequently the reflection coefficients are of the order of unity, $R_1, R_2 \sim 1$:

$$\begin{aligned} T(E_l) &= \frac{T_1 T_2}{(1 - \sqrt{R_1 R_2})^2 + 2\sqrt{R_1 R_2}(1 - \cos\theta(E_l))} \\ &\approx \frac{T_1 T_2}{[(T_1 + T_2)/2]^2 + 2[1 - \cos\theta(E_l)]}. \end{aligned} \tag{V.21}$$

Resonance occurs when the denominator becomes very small, which means that $\theta$ is a multiple of $2\pi$. At resonance $T_{\text{res}} = 4T_1T_2/(T_1 + T_2)^2$, which approaches unity for the case of symmetric barriers $T_1 = T_2$. In the off-resonant case, $T \approx T_1T_2/4$, indicating that the double barrier behaves as two independent barriers. Close to the resonance, $E_l = E_0$, we can further simplify Eq. (V.21) by performing a Taylor series expansion of the cosine function

$$1 - \cos[\theta(E_l)] \approx \frac{1}{2}\left[\theta(E_l) - 2n\pi\right]^2 \approx \frac{1}{2}\left(\left.\frac{\mathrm{d}\theta}{\mathrm{d}E_l}\right|_{E_0}\right)^2 (E_l - E_0)^2. \tag{V.22}$$

This yields the well-known *Breit-Wigner-formula* for the transmission function near the resonance,

$$T(E_l) \approx \frac{\gamma_1\gamma_2}{(E_l - E_0)^2 + [(\gamma_1 + \gamma_2)/2]^2} \tag{V.23}$$

where

$$\gamma_i = \left.\frac{\mathrm{d}E_l}{\mathrm{d}\theta}\right|_{E_0} T_i, \quad i = 1, 2. \tag{V.24}$$

Historically, the Breit-Wigner-formula was first derived in studying the decay of resonant states in nuclear problems and is often also written in the form,

$$T(E_l) = \frac{\gamma_1\gamma_2}{\gamma_1 + \gamma_2} A(E_l - E_0) \tag{V.25}$$

with the Lorentzian function

$$A(\xi) = \frac{\gamma}{\xi^2 + (\gamma/2)^2}, \quad \gamma = \gamma_1 + \gamma_2. \tag{V.26}$$

This analytical expression shows that the transmission function is sharply peaked around the resonant energy $E_0$ and that its broadening is determined by $\gamma$, which corresponds to the full width at half maximum (FWHM) of $A(\xi)$. Physically $\gamma_1/\hbar$ and $\gamma_2/\hbar$ represents the rate at which an electron leaks out of the quantum well through barrier 1 and 2, respectively. To make this more plausible one can roughly approximate the acquired round-trip phase by $\theta \approx 2k_w a$, where $k_w$ is the longitudinal momentum of the electron in the quantum well and $a$ is the width of the well. Hence,

$$\frac{\gamma_i}{\hbar} = \frac{1}{\hbar}\left.\frac{\mathrm{d}E_l}{\mathrm{d}\theta}\right|_{E_0} T_i \approx \frac{1}{2a\hbar}\left.\frac{dE_l}{\mathrm{d}k_w}\right|_{E_r} T_i = \frac{v_w}{2a}T_i, \tag{V.27}$$

where $v_w$ is the group velocity of the electron in the quantum well at the resonant energy level. The attempt frequency $v_w/2a$ tells us the number of escape attempts of the electron per second through a single barrier when the electron bounces forth and back in the quantum well. Multiplying the attempt frequency by the transmission probability of the single barrier gives us the rate of successful escapes of the electron per second. Hence, the lifetime of the electron is given by the inverse of the total escape rate $\gamma/\hbar = (\gamma_1 + \gamma_2)/\hbar$. Since $\gamma$ is the FWHM of the resonant transmission peak this again leads to the general result that the energetic broadening of the quasibound state $E_0$ is inversely proportional to the lifetime of the electron in this state.

If the transmission function is known the current density can be calculated by using the Landauer-Büttiker formula, Eq. (V.2). In the case that the transmission depends only on the longitudinal energy, as considered here, Eq. (V.2) results in

$$j = \frac{4\pi e m}{(2\pi)^3\hbar^3}\int_0^\infty \mathrm{d}E_l T(E_l)\int_0^\infty \mathrm{d}E_t[f_L(E_l,E_t) - f_R(E_l,E_t)], \tag{V.28}$$

where we have rewritten the summation over the in-plane momentum $\mathbf{q}$ in the usual integral form,

$$\hat{T}(E) = \sum_{\mathbf{q}} T_{\mathbf{q}} = \frac{S}{(2\pi)^2}\int \mathrm{d}q_x \mathrm{d}q_y T(E_l) \tag{V.29}$$

with $S$ denoting the cross sectional area of the structure, and transforming to the longitudinal and transversal energy as integration variables. Assuming Fermi-Dirac distributions in the leads

$$f_{L,R} = \frac{1}{1 + \exp[(E_l + E_t - \mu_{L,R})/k_B\Theta]}, \tag{V.30}$$

where $\mu_{L,R}$ are the chemical potentials in the left and right lead ($\mu_R = \mu_L - eV_a$), $\Theta$ denotes the reservoir temperature to avoid confusion with the transmission $T$, and $k_B$ is the Boltzmann constant, the integration over the transversal energy is easily evaluated to give the *Tsu-Esaki formula* (Tsu and Esaki, 1973):

$$j = \frac{emk_B\Theta}{2\pi^2\hbar^3}\int_0^\infty \mathrm{d}E_l T(E_l)\ln\left(\frac{1 + \exp[(\mu_L - E_l)/k_B\Theta]}{1 + \exp[(\mu_R - E_l)/k_B\Theta]}\right). \tag{V.31}$$

The logarithmic term is the so-called supply function which determines the energy interval of interest. The range of electron energies, which can contribute to the total current, is restricted to the energy window between the left and right chemical potentials $[\mu_R, \mu_L]$ plus/minus several $k_B\Theta$ due to the thermal smearing of the Fermi-Dirac functions in the leads. The dominant contributions to the current integral are given by the resonant peaks of the transmission function.

If we assume that there is only one single transmission peak in the energy range of interest and that $T(E_l)$ is very sharply peaked around $E_0$ due to thick and/or high barriers we can approximate its Lorentzian form by a Dirac-Delta function by using the asymptotic limit

$$\delta(E_l - E_0) = \frac{1}{2\pi} \lim_{\gamma \to 0} A(E_l - E_0). \tag{V.32}$$

With this approximation the current density results in

$$j = \frac{e}{\hbar} \frac{\gamma_1 \gamma_2}{\gamma} k_B \Theta D_0 \ln \left( \frac{1 + \exp[(\mu_L - E_0)/k_B \Theta]}{1 + \exp[(\mu_R - E_0)/k_B \Theta]} \right), \tag{V.33}$$

where we have introduced the constant density of states of a two-dimensional (2D) electron gas $D_0 = m/\pi\hbar^2$. In the special case of zero temperature, $\Theta = 0$, this further simplifies to the expression

$$j = \frac{e}{\hbar} D_0 \frac{\gamma_1 \gamma_2}{\gamma} [\mu_L - E_0(V_a)], \quad 0 < E_0 < \mu_L, \tag{V.34}$$

where the voltage dependence of the current is "hidden" in the voltage-dependent resonant energy level $E_0(V_a)$, which is shifted energetically downwards by the applied bias $V_a$. If we assume, in a first approximation, that the voltage is equally divided between the barriers the voltage dependence of the resonant level can be explicitly written as $E_0(V_a) = E_{00} - eV_a/2$ with $E_{00} = E_0(V_a = 0)$ denoting the resonant level position when no bias is applied. Equation (V.34) shows that the current initially increases linearly with the applied voltage, reaching its peak value

$$j_p = \frac{e}{\hbar} D_0 \frac{\gamma_1 \gamma_2}{\gamma} \mu_L \tag{V.35}$$

when the resonant level approaches the bottom of the emitters conduction band ($E_0 = 0$) at the corresponding peak voltage of $eV_p = 2E_0$. At higher voltages the quasibound state becomes off-resonant causing a sudden cutoff of the current, as long as no other higher lying resonant level is pulled down into the energy window of interest.

### A.4 Sequential Tunneling

In our discussion of resonant tunneling so far we assumed that inelastic, phase-breaking scattering processes are negligible, which enables us to apply a wave function treatment of the underlying electron transport. However, if scattering is important the electrons will lose their phase memory during propagation and the transport becomes incoherent. In this case one can use the *sequential tunneling* model introduced by Luryi (1985). In a sequential tunneling process electrons tunnel through the first barrier, reside some time in the quantum well where they lose coherence by phase-randomizing scattering processes and, finally, tunnel out through the second barrier by a second uncorrelated tunneling process. The regime of sequential tunneling can be characterized by the condition $\tau_{\rm ph}\gamma \ll \hbar$ saying that the lifetime $\hbar/\gamma$ of the electrons in the quantum well is much greater than the phase breaking time $\tau_{\rm ph}$.

As argued by Luryi, NDR generally follows from the reduction of the dimensionality as the electrons tunnel from a three dimensional Fermi sea in the emitter to a 2D electron gas in the quantum well. Assuming an energy and in-plane momentum conserving tunneling process leads to the constraining condition $E_l = \hbar k_z^2/2m = E_0$, where $E_0$ is the energy of the resonant level in the well, measured from the bottom of the emitter conduction band. Therefore, only electrons with the fixed longitudinal momentum $k_z = k_0 = \sqrt{2mE_0}/\hbar$ can tunnel from the emitter Fermi sea into the quantum well, as illustrated by the shaded disk in Fig. V.10. As the bias is increased the shaded disk moves downwards, leading to an linear increase of the disk area and correspondingly of the current proportional to $\mu_L - E_0$. The maximum current is reached at the equatorial plane $k_0 = 0$. If $E_0 < 0$ no resonant tunneling from the emitter into the well is possible anymore, which leads to an abrupt drop of the current giving rise to NDR. This explanation shows that for the occurrence of NDR it does not matter if the electrons propagation is coherent or not.

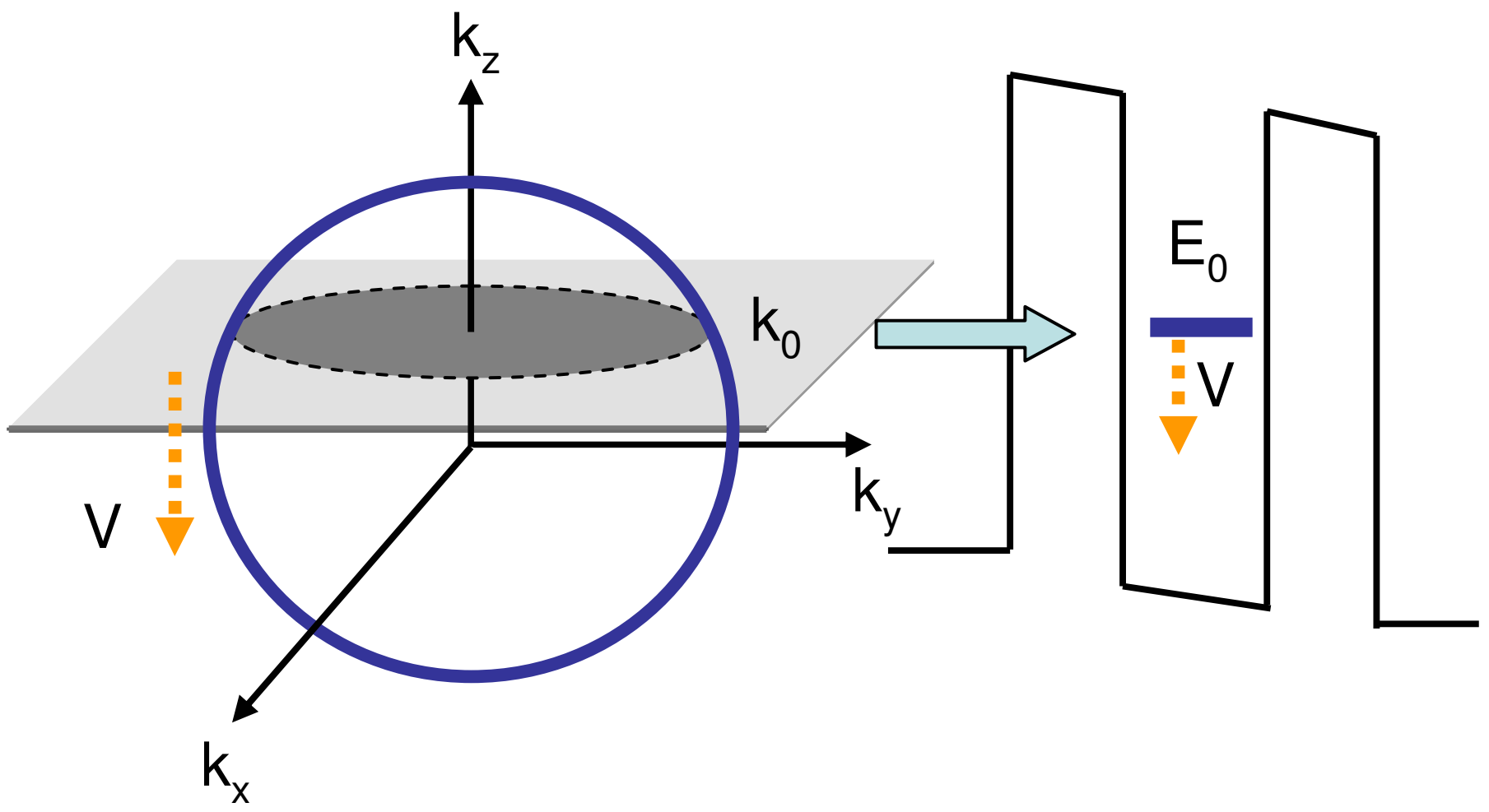

Fig. V.10. Schematic explanation of the occurrence of negative differential resistance (NDR) caused by the tunneling from a 3D Fermi sea to a 2D electron gas in the quantum well according to Luryi (1985). Only those electrons from the 3D emitter Fermi sea with conserved longitudinal momentum ($k_z = k_0$), which are indicated by the dark disk, can tunnel through the resonant level $E_0$. As the applied bias $V$ increases the resonant level and the disk move downwards. The current and the corresponding disk area increases linearly until the equatorial plane is reached (or equally $E_0$ is pushed down to the emitter's conduction band edge). Further bias cuts off the electrons supply from the Fermi sea giving rise to a sharp drop in the $IV$-characteristics.

To calculate the current in the sequential tunneling regime we can use a master equation approach, since the in- and out-tunneling processes become uncorrelated. For this purpose, we introduce a single particle distribution function $f_\alpha$ for the electron states $|\alpha\rangle$ in the quantum well. The states $|\alpha\rangle = |m, \mathbf{q}\rangle$ are characterized by the in-plane momentum $\mathbf{q}$ of the electrons and the subband index $m$, which enumerates the well quasibound states starting from the ground state $m = 0$. In real-space representation the state $|\alpha\rangle$ reads $\langle \mathbf{r}, z|\alpha\rangle \sim e^{i\mathbf{qr}}\phi_m(z)$, where $\phi_m(z)$ is the quasibound wave function. In the leads the electrons occupy plane-wave Bloch

states, shortly denoted by $|k\rangle$. With these definitions the master equation for the quantum well distribution function reads as

$$\frac{\partial f_\alpha}{\partial t} = \sum_{jk} \Gamma^j_{\alpha k} f^j_k (1 - f_\alpha) - \Gamma^j_{k\alpha} f_\alpha (1 - f^j_k), \tag{V.36}$$

where $f^j_k$ denotes the electron distribution function in the left and right lead $(j = 1, 2 = L, R)$ and $\Gamma^j_{\alpha \leftarrow k}$ denotes the transition rate from state $|k\rangle$ in the lead $j$ to the state $|\alpha\rangle$ in the quantum well. The physical meaning of the two terms on the right hand side of Eq. (V.36) is easily understood. The first term is the gain term which describes the tunneling of the electrons from the leads into the quantum well state $|\alpha\rangle$ by taking into account the Pauli blocking factor $(1 - f_\alpha)$, whereas the second term describes all loss processes due to tunneling out of the state $|\alpha\rangle$. The transition rate $\Gamma^j_{k\alpha}$ can be calculated by using the *transfer Hamiltonian* approach (Bardeen, 1961; Harrison, 1961; Duke, 1969), which was first developed for describing single barrier tunneling and has been extensively used in the context of transport in superconducting tunnel junctions. In the case of single barrier systems the basic idea of the method is to represent the total Hamiltonian of the system by $H = H_L + H_R + H_T$, where $H_L$ and $H_R$ describes the Hamiltonian of the left and right subsystem and $H_T$ is the tunneling Hamiltonian describing the transport between the two subsystems. The main advantage of the method is that if the coupling between the two subsystems is weak, $H_T$ can be treated as a perturbation term, which allows to use perturbative techniques developed in many-body theory. In our case of a double barrier structure the total Hamiltonian consist of three subsystems: the emitter $H_L$, the well $H_w$, and the collector $H_R$ Hamiltonian, which are connected by two tunneling Hamiltonians $H^j_T$ for the left and right barrier:

$$H = H_L + H_R + H_w + H^L_T + H^R_T. \tag{V.37}$$

Assuming a free electron gas in the emitter and collector the corresponding Hamiltonians read

$$H_{L,R} = \sum_k E^{L,R}_k c^+_k c_k, \tag{V.38}$$

and the well Hamiltonian is given by

$$H_w = \sum_\alpha E_\alpha c^+_\alpha c_\alpha, \tag{V.39}$$

with $c_k, c_\alpha$ and $c^+_k, c^+_\alpha$ denoting the annihilation and creation operators of the leads and well states, respectively. The energies, $E^{L,R}_k = (\hbar k_t)^2/2m + (\hbar k_z)^2/2m + U_{L,R}$, and $E_\alpha = E_m + (\hbar q)^2/2m + U_w$, include the electrostatic energies of the reservoirs $U_{L,R}$ and the well $U_w$. By measuring the energy from the conduction band edge of the emitter it follows that $U_L = 0$ and $U_R = -eV_a$. The electrostatic potential of the well $U_w$ depends on the space charge density in the structure, and has to be calculated in general in a selfconsistent way. The tunneling Hamiltonians are formulated in the standard form,

$$H^j_T = \sum_{\alpha,k} t^j_{\alpha k} c^+_\alpha c_k + \text{h.c.}, \tag{V.40}$$

where h.c. abbreviates the hermitian conjugate of the first term and $t^j_{\alpha k}$ are the tunneling matrix elements. If we assume that the leads are weakly coupled to the well, the tunneling Hamiltonian can be treated as a perturbation term and the transition rates between the well and the lead states follow from Fermi's golden rule, which gives in first order

$$\Gamma^j_{k\alpha} = \frac{2\pi}{\hbar}\left|\langle k|H^j_t|\alpha\rangle\right|^2 \delta(E_\alpha - E_k) = \frac{2\pi}{\hbar}\left|t^j_{k\alpha}\right|^2 \delta(E_\alpha - E_k). \tag{V.41}$$

By assuming that the in-plane momentum is conserved during the tunneling process this becomes

$$\Gamma^j_{k\alpha} = \frac{2\pi}{\hbar}|t^j_m(k_z)|^2 \delta_{k_t,q}\delta(E_\alpha - E_k). \tag{V.42}$$

with the $k_z$-dependent tunneling matrix element $t^j_m(k_z)$, which physically corresponds to the overlap of the lead and well wave function in the barriers and is given by Bardeen's formula (Bardeen, 1961):

$$t^j_m(k_z) = \frac{\hbar^2}{2m}\left[\psi^j_{k_z}(z)\frac{\mathrm{d}}{\mathrm{d}z}\phi^*_m(z) - \phi^*_m(z)\frac{\mathrm{d}}{\mathrm{d}z}\psi^j_{k_z}(z)\right]_{z=z_0}. \tag{V.43}$$

Here, $\psi^j_{k_z}$ is the longitudinal part of the lead wave function, which is exponentially decaying in the barrier regions, the superscript $^*$ denotes complex conjugation, and the expression has to be evaluated at some point $z_0$ inside the $j$th barrier-region.

The total leaking rates from a certain quantum well state $|\alpha\rangle = |m, \mathbf{q}\rangle$ through the left and right barriers into the leads are defined by

$$\frac{\gamma^m_1}{\hbar} = \sum_{k,k_z<0} \Gamma^1_{k\alpha}, \quad \frac{\gamma^m_2}{\hbar} = \sum_{k,k_z>0} \Gamma^2_{k\alpha}, \tag{V.44}$$

which by using Eq. (V.42) is readily simplified to

$$\gamma^m_j = \frac{m^{1/2}L}{\sqrt{2}\hbar^2\sqrt{E^j_m}}|t^j_m(k^j_m)|^2, \quad E^j_m = \frac{\hbar(k^j_m)^2}{2m} = E_m + U_w - U_j. \tag{V.45}$$

with $L$ denoting the length of the leads. With these definitions and by exploiting the microscopic reversibility of the tunneling processes $\Gamma^j_{k\alpha} = \Gamma^j_{\alpha k}$, the master equation Eq. (V.36) can be written in the form

$$\frac{\partial f_\alpha}{\partial t} = \sum_j \frac{\gamma^m_j}{\hbar}\left[f^j(E_\alpha) - f_\alpha\right], \tag{V.46}$$

If we now assume, as before, that $f_j$ are given by Fermi-Dirac distributions and that there is only one resonant level $E_0$ in the energy range of interest we can obtain a simple rate equation for the quantum well particle density, which is defined by

$$n = \sum_\alpha f_\alpha. \tag{V.47}$$

Summation of the master equation over all well states $\alpha = |0, \mathbf{q}\rangle$ yields the rate equation

$$\frac{\mathrm{d}n}{\mathrm{d}t} = \frac{\gamma_1}{\hbar} n_1 + \frac{\gamma_2}{\hbar} n_2 - \frac{\gamma}{\hbar} n \tag{V.48}$$

with

$$n_j = \sum_\alpha f_j(E_\alpha) = D_0 k_B \Theta \ln\{1 + \exp[(\mu_j - E_0)/k_B\Theta]\}, \tag{V.49}$$

and $\gamma_j = \gamma_j^0$, $\gamma = \gamma_1 + \gamma_2$. The steady state particle density $n_0$ follows from the condition $\mathrm{d}n/\mathrm{d}t = 0$, which by using Eq. (V.48) results in

$$n_0 = \frac{\gamma_1 n_1 + \gamma_2 n_2}{\gamma}. \tag{V.50}$$

This expression confirms the naive expectation that at steady state the quantum well has to establish a "compromise" between the opposing efforts of equilibrating with both leads at the same time and, hence, the particle density becomes a balanced sum of the lead particle densities weighted according to the coupling strengths to the particle reservoirs. At steady state the current is constant throughout the whole structure, since from the particle continuity equation it follows that $\mathrm{d}n/\mathrm{d}t = \nabla \cdot \mathbf{j} = \mathrm{d}j/\mathrm{d}z = 0$. Therefore, it does not matter at which $z$-point the current density is evaluated, and calculating the current at the first barrier yields

$$\begin{aligned} j_0 &= e\frac{\gamma_1}{\hbar}(n_1 - n_0) = \frac{e}{\hbar}\frac{\gamma_1\gamma_2}{\gamma}(n_1 - n_2) \\ &= \frac{e}{\hbar}\frac{\gamma_1\gamma_2}{\gamma} k_B \Theta D_0 \ln\left(\frac{1 + \exp[(\mu_L - E_0)/k_B\Theta]}{1 + \exp[(\mu_R - E_0)/k_B\Theta]}\right), \end{aligned} \tag{V.51}$$

which is exactly the same result as we get for the coherent model in Eq. (V.33) in the limit of a delta-like resonant level. This limit is physically reasonable, since in order to apply the transfer Hamiltonian formalism we had to assume that the well is only very weakly coupled to the reservoirs and accordingly the electrons can stay a long time in the well before they tunnel out. A long lifetime in the well corresponds to only a very narrow energetic broadening of the quasibound states resulting in a delta-like resonance.

At first glance it appears surprising that the sequential and coherent tunneling models give essentially the same values for resonant currents, although the underlying physical pictures are very different. In particular the peak current of the IV-characteristic has been shown to be insensitive to scattering (Weil and Vinter, 1987). This conclusion can be justified by using a more general model that includes both a coherent and sequential part of the tunnel current (Jonson and Grincwaijg, 1987; Stone and Lee, 1985; Datta, 1995), showing that scattering processes effectively lead to an additional broadening of the resonant level. This broadening hardly influences the total current density, bearing in mind that the current, as shown in Eq. (V.31), is proportional to the folding integral of the transmission function with the supply function. In contrast to the peak current, the off-resonant valley current depends strongly on the presence of inelastic scattering processes. In the coherent model such processes are completely ignored and there is no simple way to include scattering terms into the Schrödinger equation. This is actually the reason why the coherent model predicts valley currents, which are usually much smaller than what is observed in experiments, and the theoretical results overestimate the peak to valley ratio (PVR), which is an important figure of merit for technical applications of RTDs. Although it is possible to include inelastic scattering processes in the sequential tunneling model by adding additional scattering terms on the right hand side of the master equation (V.36),[100] a proper treatment of incoherent quantum transport needs the introduction of single-particle quantum distribution functions as formulated in the density matrix, Wigner function or non-equilibrium Greens functions approach (Frensley, 1990; Datta, 1995; Ferry and Goodnick, 1997). The latter has become most popular nowadays, where scattering is included by self-energy terms (Lake *et al.*, 1997), which can be evaluated perturbatively by diagrammatic techniques. In the framework of these advanced formulations, the coherent and sequential model can be understood as two extreme limit cases of completely phase conserving and phase randomizing transport regimes in the well.

### A.5 Space charge effects and bistability

The fixed ionized impurities and the free carrier charge density stored in the well give rise to an electrostatic potential profile in the structure, which might considerably differ from the linear voltage drop considered so far. The free electron density due to the electron flux from the left and right contacts is determined by occupying the scattering states $\phi_{L,R}(k_z)$ according to the electron distributions in the contacts,

$$n_{L,R}(z) = \frac{2}{(2\pi)^3}\int \mathrm{d}\mathbf{k}_t \int_0^\infty \mathrm{d}k_z |\phi_{L,R}(k_z,z)|^2 f_{L,R}(\mathbf{k}_t,k_z). \tag{V.52}$$

The scattering states are determined from the Schrödinger equation (V.5) by using the boundary conditions of left and right incident plane waves, respectively. If we again assume that $f_{L,R}$ are given by Fermi-Dirac functions we can easily perform the integration over the transverse motion, which yields

$$n_{L,R} = \frac{D_0 k_B\Theta}{2\pi}\int_0^\infty \mathrm{d}k_z |\phi_{L,R}(k_z)|^2 \ln\{1+\exp[(\mu_{L,R}-E_z)/k_B\Theta]\}. \tag{V.53}$$

As already discussed, using a mean-field approach for the electron-electron interaction allows to calculate the electrostatic potential profile from the Poisson equation (V.6), which couples via Eq.(V.53) to the Schrödinger equation (V.5). Hence, one needs to find a selfconsistent solution of the Schrödinger-Poisson system, which is usually obtained iteratively by solving both equations alternatively until convergence is reached (Ohnishi *et al.*, 1986; Cahay *et al.*, 1987; Brennan, 1987; Pötz, 1989). For the boundary conditions local charge neutrality is assumed in the asymptotic contact regions far away from the tunneling region, which leads to flat conduction bands beyond certain points in the lead. These points can be used to define the boundaries of the active system, where the interior solution is matched to the plane-wave states of the contacts.

The space charge effects lead to several important changes in the potential profile and consequently in the IV-characteristic of RTD. A characteristic result for the conduction band profile obtained from a selfconsistent calculation is shown in Fig. V.11. Typically a substantial fraction of the bias already drops outside of the double barrier structure, which resembles in some sense a series resistance effect. Therefore, the peak current is reached at a higher applied voltage as found in non-selfconsistent simulations. Most interestingly, at higher voltages a potential notch is formed on the emitter side of the double barrier structure as shown in Fig. V.11. The quantum confinement in triangular-like well profile of the notch leads to the occurrence of quasibound notch states, where electrons can become trapped. This leads to the forming of an accumulation layer adjacent to the tunnel barrier. The contribution to the total current due to tunneling out of the notch states is not taken into account in the above considered transport model, since the trapping of electrons in these states requires some sort of inelastic scattering processes. Again, a quantitative treatment of describing the injection from the emitter quasibound states needs advanced formalisms based on quantum distribution functions (Klimeck *et al.*, 1995).

[100] For instance one can introduce in a first approach a simple relaxation time approximation for the scattering term: $(f_\alpha - f_0)/\tau_{\mathrm{rel}}$, where $f_0$ is an equilibrium distribution and $\tau_{\mathrm{rel}}$ denotes the scattering relaxation time.

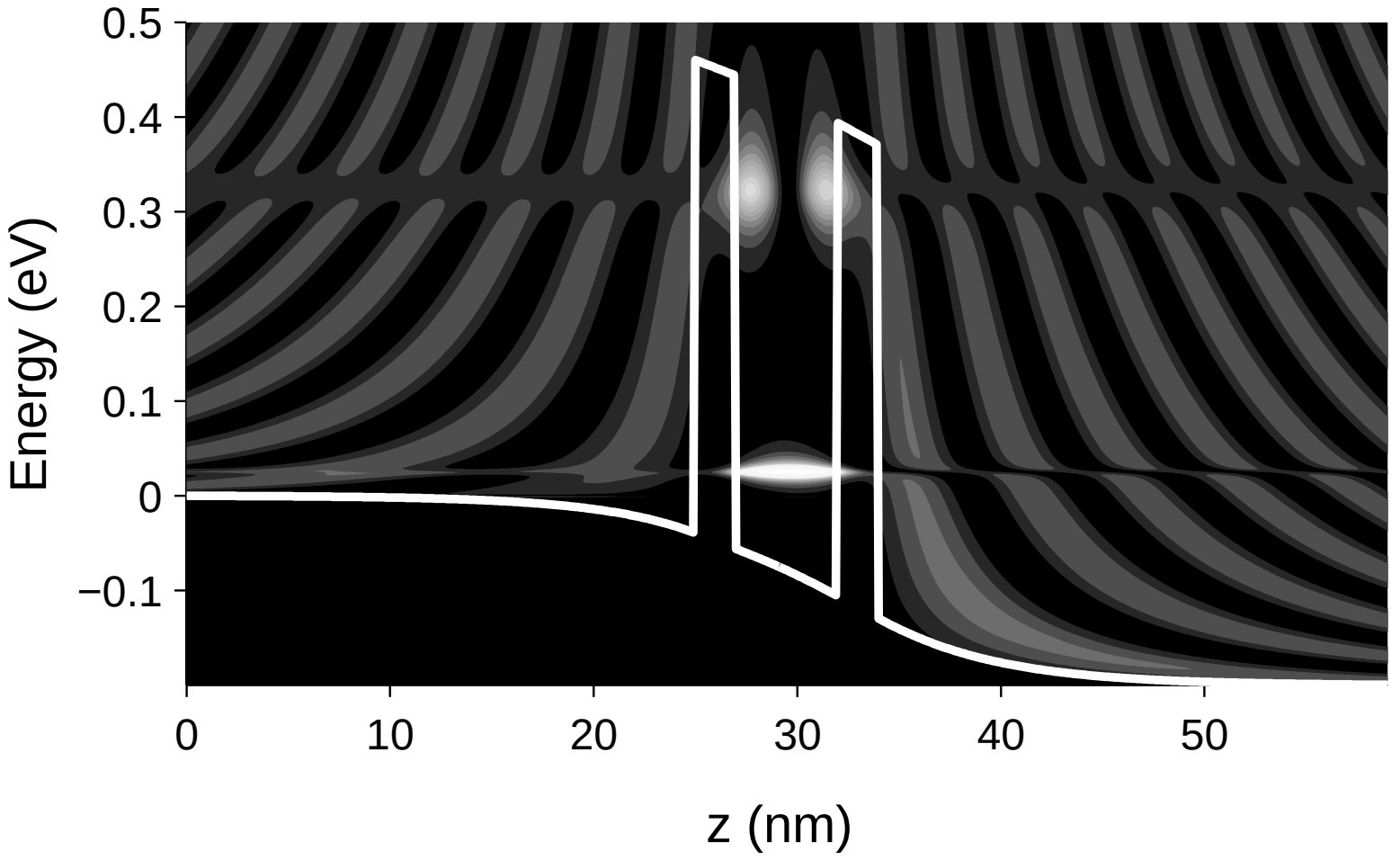

Fig. V.11. Contour plot of the local density of states of the conduction electrons for a model tunnel diode at applied bias. Bright (dark) areas correspond to high (low) densities. The solid white line indicates the selfconsistent conduction band profile, illustrating the forming of a potential notch at the emitters side (starting at around $\approx 18$ nm) at higher applied voltages. As can be seen a substantial fraction of the applied voltage already drops in the emitter and collector lead regions.

Another striking consequence of including space charge effects is the occurrence of intrinsic bistability in the IV-characteristic due to the charge storage of free electrons in the quantum well (Goldman *et al.*, 1987; Mains *et al.*, 1989), as illustrated in Fig. V.12. For increasing voltages free charges pile up in the well, causing a negative electrostatic well potential which pushes the resonant level to higher energies as compared to the case when no electrons would be stored in the well. This causes different peak voltages for increasing and decreasing voltages. The well is charged in the case of an up-sweep of the applied voltage shortly before the resonant state becomes off-resonant, i.e., when it drops below the emitters conduction band edge. Whereas for the down-sweep the well is almost uncharged before the quasibound state becomes resonant. This causes a higher peak voltage for increasing than for decreasing voltages leading to a hysteresis loop in the IV-characteristics in the NDR-region. In practical experiments, however, there is a great difficulty to separate this predicted intrinsic bistability due to the charge storage in the well from external measurements circuits effects when the RTD is driven into the unstable NDR-regime. Nevertheless, convincing observations are possible by using asymmetric barriers which emphasizes the effect (Zaslavsky *et al.*, 1988).

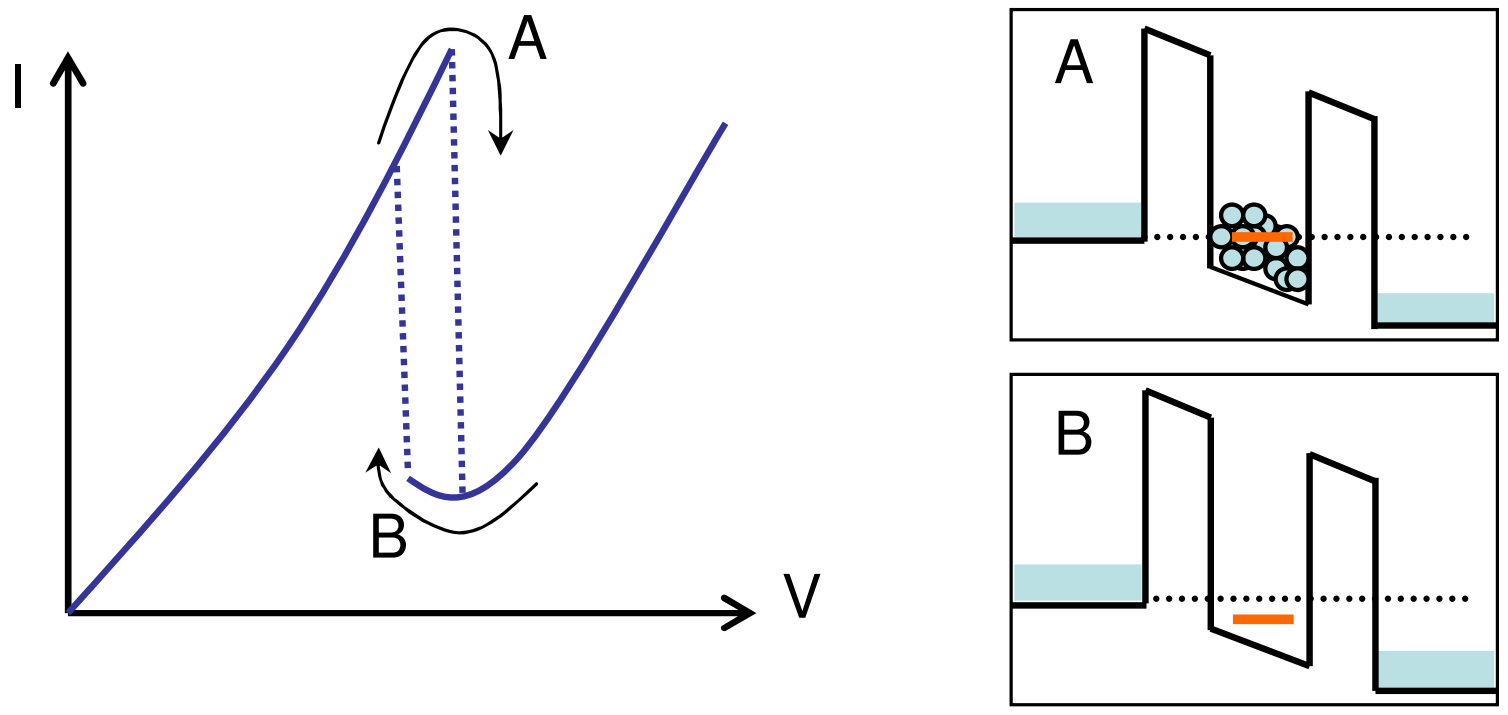

Fig. V.12. Left: schematic illustration of the occurrence of an intrinsic bistability in the $IV$ curve of a RTD. The peak voltage in the case of an up sweep of the applied voltage (case A) is higher than for the down-sweep (case B), since in case A charges are piled up in the quantum well, which pushes the resonant level to higher energy values as compared to an almost empty well (case B). Right: The conduction band profile and the relative position of the resonant well level for both cases (A and B) at the peak voltage of the voltage up-sweep.

### B. Diluted magnetic semiconductor heterostructures

So far we discussed resonant tunneling in double barrier structure made of planar layers of non-magnetic semiconductors only. In order to attain a spin-dependent transmission it is a fairly straightforward idea to use heterostructures, in which one or several layers are made of magnetic semiconductors as illustrated in Fig. V.13. In ferromagnetic materials the spin up and down states of the valence and/or conduction band are appreciably split in energy. This makes, e.g., tunneling through a single magnetic barrier already spin selective, because electrons with different spin components experience unequal tunnel barrier heights and, hence, the tunneling probability of one spin type is exponentially suppressed compared to the other. Spin filtering becomes even more efficient in a double barrier structure with a magnetic quantum well, where spin split quasi-bound states are formed. For instance, pulling one spin-resolved state down to off-resonance, it means below the emitter's conduction band edge, by an applied voltage, while the other spin state is still resonant, would allow to create highly spin-polarized currents. The phenomenology of spin-dependent transport becomes further enriched in triple barriers structures with two magnetic quantum wells. In such structures, resonant tunneling only occurs if the spin-split resonant levels of both wells are energetically aligned as well as they correspond to the same spin state. In the case that the aligned adjacent well levels belong to different spin states, the current becomes blocked; a phenomenon which is known as the spin-blockade effect. These simple examples of magnetic barrier systems give a first impression of how band-structure engineered magnetic heterostructures can open up a whole plethora of novel opportunities for controlling and tuning the spin-dependent magneto-transport properties in low dimensional semiconductor systems.

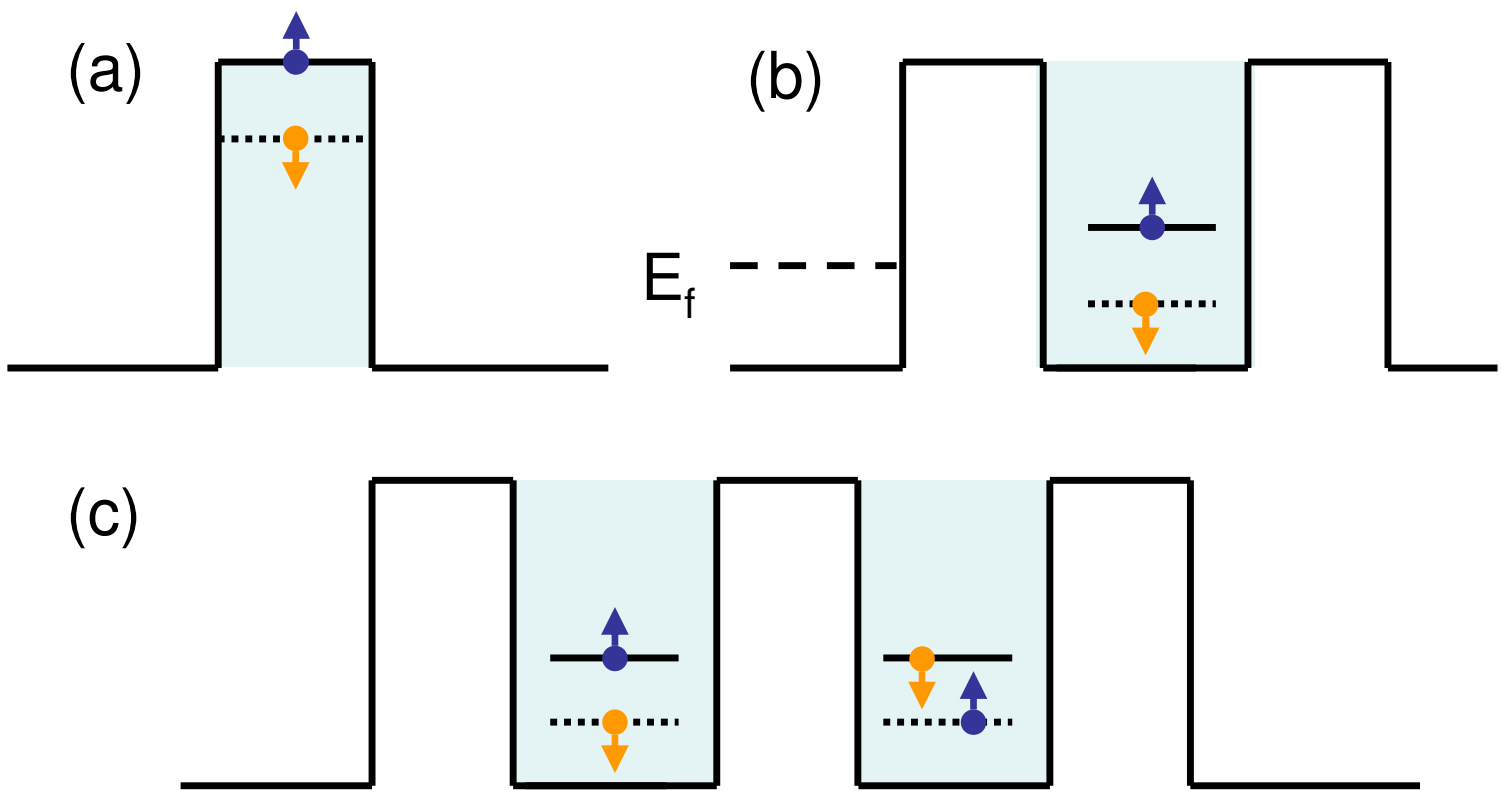

Fig. V.13. Spin filtering in heterostructures made of magnetic semiconductors. The magnetic layers have colored background. (a) Single magnetic barrier with different barrier heights for spin up and down electrons. Here, the tunneling of the spin up electrons is exponentially suppressed. (b) Double barrier structure with magnetic quantum well, in which the well states become spin split. Spin filtering is achieved by bringing one spin level to off-resonance, while the other remains in resonance. (c) Spin blockade effect in coupled magnetic quantum wells. Resonant tunneling occurs only if the adjacent well states are energetically aligned and are of the same spin type.

### B.1 Diluted magnetic semiconductors

For a long time few magnetic semiconductors have been known, e.g., europium based chalcogenides (e.g. EuO) (Kasuya and Yanase, 1968), or Cr-based spinels (e.g. $CdCr_2Se_4$) (Baltzer *et al.*, 1965; Park *et al.*, 2002b). This changed substantially with the discovery of diluted magnetic semiconductors (DMSs) in the 1980s (Furdyna, 1988; Dietl, 1994). DMSs are made magnetic by doping with transition metal elements. The dopants are substituted more or less randomly on the host crystal sites where they introduce local magnetic moments. It turns out that manganese dopants are especially appropriate for this purpose, since they provide well defined high spin $S = 5/2$, $Mn^{2+}$ local moments and appear to alter the band structure of the host crystal only very weakly. First mainly Mn-doped II-VI semiconductor compounds, e.g., CdMnSe, CdMnTe, ZnMnSe, or ZnMnTe, made of group II and VI elementary semiconductors have been investigated. Since Mn exhibits the same valence ($s^2$) as the cations of the host they are easily incorporated on the cation sites. The Mn moments can be aligned by relatively weak magnetic fields, which causes a giant Zeeman splitting of the carriers bands of the order of tens of meV leading to a rich variety of magneto-optical and magneto-electrical effects. However, most of the II-VI compounds remain in a paramagnetic state; long range ferromagnetic order, if any, usually only occurs at very low temperatures [here, ZnCrTe constitutes an interesting exception being ferromagnetic up to room temperature (MacDonald *et al.*, 2005)]. A breakthrough in the research for DMSs was the discovery of ferromagnetism in Mn-doped III-V semiconductors in the 1990s, first in InMnAs (Ohno *et al.*, 1992; Munekata *et al.*, 1989) and then in GaMnAs (Ohno *et al.*, 1996; Ohno, 1999; Matsukura *et al.*, 2002b). In the nowadays prototypical $Ga_{1-x}Mn_xAs$ compound the transition temperature reaches values well above 100 K, when heavily doped with Mn. [The current reported record lies at 173 K (Jungwirth *et al.*, 2005)]. The key difference to II-VI DMSs is that in III-V hosts the Mn acts at the same time as an acceptor introducing holes in valence band. The ferromagnetic order only appears if the doping and the corresponding hole density is high enough. The low equilibrium solubility of Mn in GaAs is overcome by non-equilibrium, low temperature MBE, allowing for doping densities of about $x \approx 5 - 10\%$ and hole densities of about $p \approx 1 - 3 \times 10^{20}$ cm$^{-3}$. Together with theoretical predictions of possible room temperature ferromagnetic transition temperatures in other doped III-V compounds (Dietl *et al.*, 2000) these findings initiated an enormously growing research interest in ferromagnetic DMSs. The theoretical predictions and the experimental properties of bulk ferromagnetic III-V compounds have been reviewed by several authors (Jungwirth *et al.*, 2006; Dietl, 2002; MacDonald *et al.*, 2005; König *et al.*, 2003; Dietl, 2007).

For a significant practical impact, however, magnetic semiconductors are required which exhibit transition temperatures well above room temperature and can be simply incorporated into semiconductor electronics. Due to the considerable research efforts in the last years there have already been several experimental reports on above room temperature ferromagnetism in different classes of semiconductors when doped with transition metals, e.g., in wide band gap III-V semiconductors (GaN, GaP), in III-IV-$V_2$ chalcopyrites ($CdGeP_2$, $ZnSnAs_2$), in the group IV semiconductors (Ge, Si) or in some oxide semiconductors (ZnO, $TiO_2$). A discussion of promising spintronic materials and criterions for an ideal ferromagnetic semiconductor are given in the review articles (MacDonald *et al.*, 2005; Pearton *et al.*, 2003b,c; Ivanov *et al.*, 2004; Felser *et al.*, 2007). In Tab. V.1 we present a selection of important candidates with their reported Curie temperatures and transport properties, including references to more detailed review papers. The ongoing progress has substantially expanded the list of possible magnetic semiconductors, but the current understanding of the ferromagnetism in these new materials is far from complete. The challenge of finding the optimal magnetic semiconductor is an exciting, still open question in current material research.[101]

### B.2 Mean field model of ferromagnetism in heterostructures

In order to understand how the transport and magnetic properties become closely intertwined in ferromagnetic heterostructures, we discuss here a minimal mean field approximation of the Zener model of the ferromagnetism in the most prominent DMS GaMnAs, following the treatments of Refs. (Das Sarma *et al.*, 2003b) and (Lee *et al.*, 2000) for the bulk and spatially inhomogeneous case, respectively. In GaMnAs the Mn ions substitutionally replace Ga at the cation sites and act simultaneously as acceptors in the ideal case, as shown in Fig. V.14(a). The provided Mn moments couple to holes by an on-site exchange interaction due to the overlap of the hole wave functions with the d-orbitals of the local Mn electrons. If the concentration of dopants is high enough a long range ferromagnetic state among the local moments is established by the itinerant holes; a mechanism which is usually denoted as *carrier mediated ferromagnetism* and which is schematically sketched in Fig. V.14(b). Such a ferromagnetic state, which is tuneable by the carrier density, is especially interesting for magneto-electronic applications, since the carrier density in semiconductors can be simply modulated by external electric fields. The electrical control of ferromagnetism has already been demonstrated for high gate voltages in a field effect transistor structure comprising a conduction channel made of InMnAs (Ohno *et al.*, 2000). Moreover, the phase coherence length in GaMnAs nanowires and rings was experimentally demonstrated to be of the order of 100 nm at low temperatures, although the typical mean free path of holes in metallic GaMnAs is only of the order of a few lattice constants (Wagner *et al.*, 2006). This remarkable finding is of particular importance for resonant tunneling diodes or other interference device applications, whose functioning rely on a coherent particle propagation. The analysis of the observed universal conductance fluctuations as shown in Fig. V.15 for a nanowire of 400 nm diameter revealed a $T^{-1}$ dependence of the dephasing time.

[101] There are encouraging results that DMS nanostructures, such as Mn-doped quantum dots have desirable materials properties (Mackowski *et al.*, 2004; Leger *et al.*, 2006; Holub *et al.*, 2004; Govorov, 2005; Abolfath *et al.*, 2007a). The interplay of strong Coulomb interactions and quantum confinement can lead to the onset of magnetization at temperatures much higher than in their bulk counterparts (Fernandez-Rossier and Brey, 2004; Holub *et al.*, 2004; Abolfath *et al.*, 2007b). Such quantum dots could provide a versatile control of magnetism including switching the magnetization *on* and *off* by the gate voltage at fixed number of carriers in the absence of applied magnetic field (Abolfath *et al.*, 2007c).

Tab. V.1. Selection of important semiconductor spintronic materials (FM abbreviates ferromagnetism).

| Material | $T_c$ [K] | Comments | References |
|---|---|---|---|
| **II-VI Diluted Magnetic Semiconductors** | | | |
| ZnCrTe | $\sim 300$ | recent findings show that FM is caused by the formation of Cr-rich metallic nanocrystalls embedded in a Cr-poor matrix | (Saito *et al.*, 2003; Kuroda *et al.*, 2007; MacDonald *et al.*, 2005) |
| CdTe, ZnSe, CdSe, CdS | $\sim 1$ | $n$-doping at most $10^{19}$ cm$^{-3}$, $p$-doping difficult; paramagnetic phase with giant Zeeman splitting of bands | (Furdyna, 1988; Dietl, 1994) |
| **III-V Diluted Magnetic Semiconductors** | | | |
| InMnAs | 35 | $p$-type, first reported III-V DMS | (Ohno *et al.*, 1992; Pearton *et al.*, 2003b; Schallenberg and Munekata, 2006) |
| GaMnAs | $< 170$ | $p$-type, $p \approx 10^{20} - 10^{21}$ cm$^{-3}$, prototype of carrier mediated ferromagnet, Mn acts as an acceptor | (Ohno *et al.*, 1996; Jungwirth *et al.*, 2006; Dietl, 2002; MacDonald *et al.*, 2005; König *et al.*, 2003; Dietl, 2007) |
| InGaMnAs | 110 | $p$-type | (Pearton *et al.*, 2003b) |
| GaMnN | 10-940 | mostly $n$-type, $n \approx 10^{19}$ cm$^{-3}$ at 300 K, $p$-doping with most common Mg acceptor is limited up to $10^{18}$ cm$^{-3}$ at room temperature due to deep acceptor levels; mean field theory not very accurate, role of electrons in stabilization of ferromagnetism is not clear | (Pearton *et al.*, 2003b,a; Liu *et al.*, 2005; Sasaki *et al.*, 2002; Overberg *et al.*, 2001; Thaler *et al.*, 2002) |
| GaMnP | $> 300$ | $p$-type, $p \approx 10^{20}$ cm$^{-3}$, $T_c$ suppressed for $n$-type GaP substrates, nearly lattice matched with Si | (Pearton *et al.*, 2003b,d; Theodoropoulou *et al.*, 2002) |
| **Group IV Ferromagnetic Semiconductors** | | | |
| $Mn_xGe_{1-x}$ | $\approx 100$ | $T_c$ depends linearly on Mn concentration, hole mediated exchange, $p$-type with $p \approx 10^{19} - 10^{20}$ cm$^{-3}$ | (Park *et al.*, 2002a) |
| Si:Mn | $> 400$ | ion implantation of Mn, both $n$- and $p$-doping, saturation magnetization is decreased for $n$-type samples, MnSi exhibits a quantum phase transition at low temperatures resulting in a non-Fermi liquid behavior | (Bolduc *et al.*, 2005; Pfleiderer *et al.*, 1997) |

Table V.1 continues on the next page.

Table V.1

| Material | $T_c$ [K] | Comments | References |
|---|---|---|---|
| **Magnetic Oxide Semiconductors** | | | |
| ZnMnO | 30-425 | transparent films, mainly $p$-type but $n$-type doping possible | (Pearton *et al.*, 2003b; Liu *et al.*, 2005; Pearton *et al.*, 2003c; Fukumura *et al.*, 2005; Sharma *et al.*, 2003) |
| Zn(Co)O | $> 300$ | $n$-type; films are transparent, FM might not be carrier induced and is predicted without need of additional charge carriers | (Ueda *et al.*, 2001; Pearton *et al.*, 2003b; MacDonald *et al.*, 2005; Sluiter *et al.*, 2005; Xu *et al.*, 2006b) |
| (Co)$TiO_2$ | $> 300$ | can be made $n$-type ($n \approx 10^{19}$ cm$^{-3}$), experimental reports of forming of Co nanocluster | (Pearton *et al.*, 2003b; Matsumoto *et al.*, 2001; Fukumura *et al.*, 2005; Champers *et al.*, 2003) |
| **Chalcopyrites (II-IV-$V_2$)** | | | |
| $CdMnGeP_2$, $ZnMnGeP_2$, $ZnMnSiGeN_2$, $ZnMnSnAs_2$ | $> 300$ | exhibit interesting non-linear optical properties useful for applications, both $n$- and $p$-type conductivity were reported, first principle calculations identified a small set of promising chalcopyrites exhibiting lattice-matching with common semiconductors | (Pearton *et al.*, 2003b; Medvedkin *et al.*, 2000; Erwin and Žutić, 2004; Picozzi, 2004; Cho *et al.*, 2002) |
| **Concentrated Magnetic Semiconductors** | | | |
| Cr-chalcogenide spinels ($CdCr_2Se_4$,$CdCr_2S_4$, etc.) | | | |
| $CdCr_2Se_4$ | $< 130$ | $n$-type, reasonably lattice matched with Si and GaP ($\approx 1.7\%$ tensile mismatch) | (Baltzer *et al.*, 1965; Kioseoglou *et al.*, 2004; Park *et al.*, 2002b) |
| Eu-chalcogenides (EuSe, EuS, EuO) | | | |
| EuO | $< 100$ | thoroughly investigated in the 1960s, used as FM barriers in tunneling structures | (Kasuya and Yanase, 1968; Moodera *et al.*, 2007) |

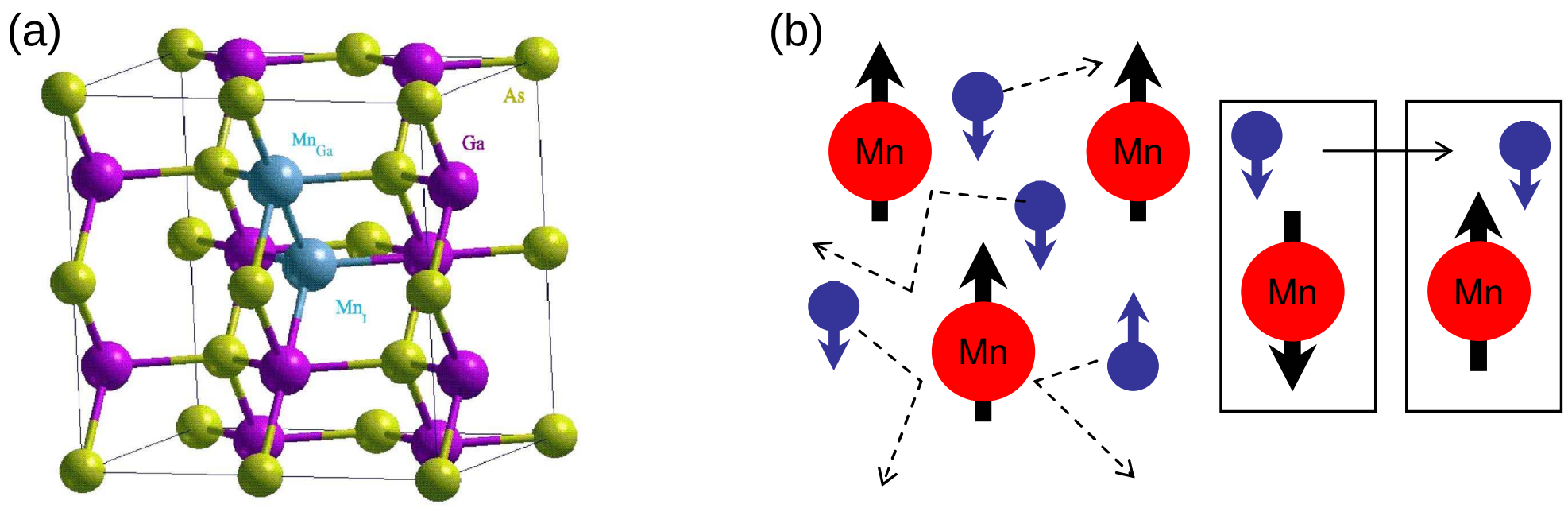

Fig. V.14. (a) Zincblende lattice structure of GaMnAs. In the ideal case Mn substitutes for Ga at the lattice sites ($Mn_{Ga}$) acting as an acceptor and providing a local magnetic moment $S = 5/2$. The hole density is usually heavily compensated by antisite and interstitial defects ($Mn_I$). Reprinted figure with permission from T. Jungwirth, J. Sinova, J. Mašek, J. Kučera, and A. H. MacDonald, *Rev. Mod. Phys.* **78**, 809 (2006). Copyright (2006) by the American Physical Society. (b) Schematic illustration of carrier mediated ferromagnetism. The itinerant holes (blue circles) couple antiferromagnetically to the local Mn-moments via the exchange interaction due to the overlap of the hole wave function with the d-orbitals of the local Mn electrons. For high enough hole densities a long range ferromagnetic order is established. The bordered boxes illustrate the flipping of a local Mn-moment caused by the kinetic exchange interaction with a passing-by hole.

The strong kinetic exchange coupling between the holes and the Mn spins, which proves to be antiferromagnetic in $Ga_{1-x}Mn_x$ As, is the basic physical mechanism underlying the occurrence of ferromagnetic order. Usually the hole density $p$ is a small fraction (of the order of 10%) of the magnetic dopant density, since the holes become heavily compensated by both antisite defects, which means that As is bound at a Ga site, and interstitial defects (i.e. Mn settles down on an interstitial site) (Bouzerar *et al.*, 2005). These defects act as double donors, delivering two free electrons. The local Mn moment density $n_i$ is usually also smaller than the total Mn concentration in the crystal host, because the presence of the unavoidable interstitial defects lowers the density of magnetically active Mn ions. In the spirit of presenting here a minimal model we neglect all band structure effects and assume a single parabolic band of the holes with a constant effective mass. Actually, it has been shown (Dietl *et al.*, 2000) that including the band-structure allows to understand the anisotropy effects in GaMnAs, e.g., the orientation of magnetic easy axis or the influence of strain. Therefore it has been argued that in relatively disorder-free metallic systems ($x > 5\%$) the ferromagnetism is mediated by valence band holes. However, it has been also suggested, particulary in the metal-insulator-transition regime, that the extended hole states of the impurity band are responsible for the establishment of the ferromagnetic order.

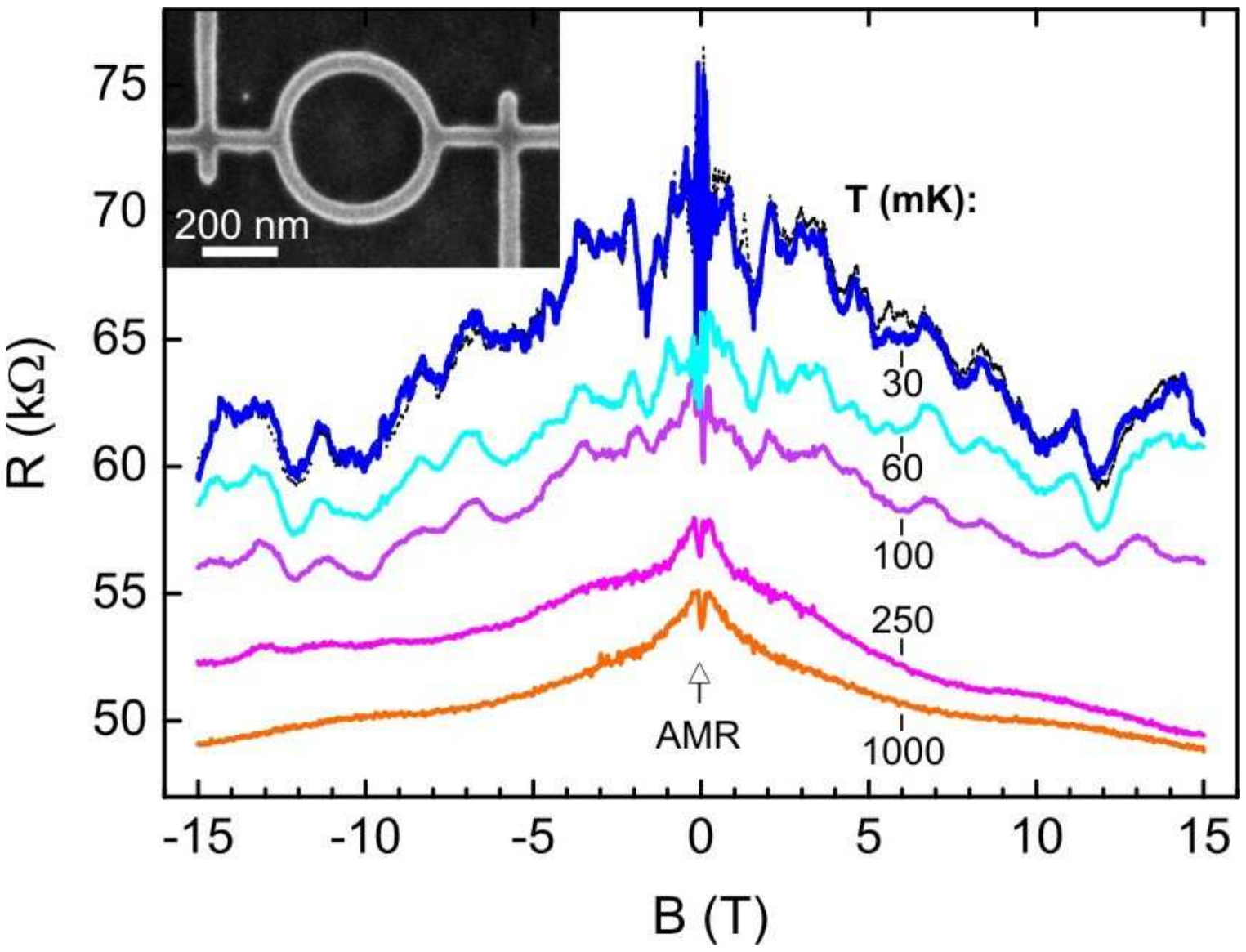

Fig. V.15. Magnetoresistance of a GaMnAs ring with a diameter of 400 nm and a ring width of 40 nm. The inset displays a top view of the ring. At temperatures below 200 mK reproducible resistance fluctuations emerge due to long hole phase coherence lengths of the order of 100 nm. To demonstrate the reproducibility of the observed resistance oscillations the 30 mK trace is shown for an up (blue line) and down sweep (dashed black line) of the magnetic field B. Reprinted figure with permission from K. Wagner, D. Neumaier, M. Reinwald, W. Wegscheider, and D. Weiss, *Phys. Rev. Lett.* **97**, 56803 (2006). Copyright (2006) by the American Physical Society.

Indeed, very recent experiments have shown strong evidence that impurity band holes play a major role, challenging the standard model for GaMnAs based on valence band holes (Burch *et al.*, 2006). This newly flared up debate about which type of holes mediate the ferromagnetic order points out again that our current understanding of DMSs, even in the most prominent GaMnAs, is not complete. With these open questions in mind the simple parabolic effective mass approximation for the holes is a good starting point to understand at least the typical features of DMSs.[102]

In the Zener model of ferromagnetism the exchange interaction between the impurities and the holes is described by the contact kinetic exchange Hamiltonian

$$H_{\mathrm{KE}} = \int \mathrm{d}^3 r \sum_j J_{pd} \mathbf{S}_j \cdot \mathbf{s}(\mathbf{r}) \delta(\mathbf{r} - \mathbf{R_j}) = \sum_j J_{pd} \mathbf{S}_j \mathbf{s}(\mathbf{R_j}), \tag{V.54}$$

where $J_{pd}$ is the $p$-$d$ coupling strength between the impurity spin $\mathbf{S}_j$ located at $\mathbf{R}_j$ and the local hole spin density $\mathbf{s}(\mathbf{r})$. In the simplest mean field approach both the action of the spin hole density on a single magnetic impurity and the action of the magnetic impurities on the local spin density can be expressed in terms of effective mean fields.

[102]Here we do not consider the temperature dependence of carrier density which can significantly modify the magnetic phase diagram (Petukhov *et al.*, 2007).

We first discuss the effect of the holes on a single local impurity. Following mean field theory (Ashcroft and Mermin, 1976), the effective Hamiltonian for the local manganese impurity reads as

$$H^i_{KE,\text{eff}} = J_{pd}\langle s_z\rangle S_z, \tag{V.55}$$

where $\langle s_z\rangle$ is the mean hole spin density with $z$ labeling the direction of the spontaneous magnetization of the spins. In planar heterostructures $\langle s_z\rangle$ depends in general on the growth direction, which we denote in the following by $\xi$ to avoid confusion with the spin quantization axis $z$. The relative simplicity of the mean-field model also allows to account for the direct Mn-Mn antiferromagnetic (AF) exchange interaction among the impurities themselves, which can be described in general by a Heisenberg spin Hamiltonian of the form,

$$H^{AF} = \sum_{ij} J^{AF}_{ij}\mathbf{S}_i \cdot \mathbf{S}_j. \tag{V.56}$$

Usually the contribution of this AF coupling is much smaller than that of kinetic exchange coupling because the AF coupling strength $J^{AF}_{ij}$ decays rapidly with the distance between the ions and for the typical doping densities $x \ll 1$ the magnetic ions are separated from each other by several nonmagnetic atoms. According to the relative weakness of the direct Mn-Mn exchange coupling, we will in the following consider its effect only for the bulk case, but neglecting it for inhomogeneous problems. By singling out one of the ions and replacing the polarizations of all others by a mean value $\langle S_z\rangle$, we get the additional effective Hamiltonian,

$$H^{AF}_{\text{eff}} = z_{AF}J^{AF}S_z\langle S_z\rangle, \tag{V.57}$$

where we take into account only the nearest neighbor interactions described by a single coupling constant $J^{\text{AF}}$ and an effective number of surrounding magnetic impurities $z_{\text{AF}}$. If we introduce an effective magnetic field $B^i_{\text{eff}}$ acting upon a single magnetic impurity given by,

$$B^i_{\text{eff}} = \frac{J_{pd}}{g_i\mu_B}\langle s_z\rangle + \frac{J^{AF}z_{AF}}{g_i\mu_B}\langle S_z\rangle, \tag{V.58}$$

the total effective Hamiltonian can be now rewritten in the form

$$H^i_{\text{eff}} = g_i\mu_B S B^i_{\text{eff}}, \tag{V.59}$$

with $S = 5/2$ being the Mn spin, $g_i$ the Mn $g$-factor, and $\mu_B$ the Bohr magneton. The paramagnetic response of the impurity spin to this effective field at the temperature $T$ is then given by the expression (Ashcroft and Mermin, 1976)

$$\langle S_z\rangle = SB_S\left(\frac{Sg_i\mu_B B^i_{\text{eff}}}{k_BT}\right), \tag{V.60}$$

where the Brillouin function $B_S(x)$ is defined by

$$B_S(x) = \frac{2S+1}{2S}\coth\frac{2S+1}{2S}x - \frac{1}{2S}\coth\frac{1}{2S}x. \tag{V.61}$$

The expression, Eq. (V.60), together with the effective magnetic field given by Eq. (V.58) couples the mean fields of both the impurities $\langle S_z \rangle$ and of the holes $\langle s_z \rangle$.

In order to calculate the effect of the magnetic ions on the itinerant carriers (in our case holes), we proceed in an analogous way by rewriting the kinetic exchange Hamiltonian (V.54) in a mean field form,

$$H_{\text{eff}}^c = J_{pd} \int \mathrm{d}^3 r \sum_j \langle S_z \rangle s_z(\mathbf{r}) \delta(\mathbf{r} - \mathbf{R}_j). \tag{V.62}$$

Comparing this mean field hole Hamiltonian with

$$H_{\text{eff}}^c = g_c \mu_B \int \mathrm{d}r^3 s_z(\mathbf{r}) B_{\text{eff}}^c(\mathbf{r}), \tag{V.63}$$

where $g_c$ denotes the $g$-factor of the holes, the action of the impurities can again be expressed in terms of an effective magnetic field given by

$$B_{\text{eff}}^c(\mathbf{r}) = \frac{J_{pd} \langle S_z \rangle}{g_c \mu_B} \sum_j \delta(\mathbf{r} - \mathbf{R}_j). \tag{V.64}$$

If we furthermore assume that the randomly distributed magnetic ions are dense in the in-plane (i.e. in the plane perpendicular to $\xi$) of the heterostructure on a scale given by Fermi wave vectors of the free holes, we can apply a continuum limit,

$$\sum_j \delta(\mathbf{r} - \mathbf{R_j}) \approx n_i(\xi). \tag{V.65}$$

where the volume density of magnetically active ions $n_i$ can be experimentally controlled during the growth of the structure. In this way we get rid of the necessity to take care of the concrete random impurity distribution, greatly simplifying the theoretical treatment. The continuum limit corresponds to a virtual-crystal approximation for the positional disorder of the Mn-atoms.

In the effective magnetic field, $B_{\text{eff}}^c$, caused by the polarized impurities the holes experience a spin-dependent kinetic exchange potential,

$$H_{pd} = \frac{\sigma}{2} h_{pd}(\xi) = \frac{\sigma}{2} g_c \mu_B B_{\text{eff}}^c = \frac{\sigma}{2} J_{pd} \langle S_z \rangle (\xi) n_i(\xi) \tag{V.66}$$

with $\sigma = \pm 1$ corresponding to the spin up $(s_z = 1/2, \uparrow)$ and down state $(s_z = -1/2, \downarrow)$. Following spin density functional theory, the dynamics of the holes can be described by an envelope function equation (Jungwirth *et al.*, 1999) completely analogous to the previously discussed case of electrons in Sec. A.3, Eq. (V.4). But since the holes experience now different potentials depending on their spin state, we have to consider two equations, one for each component of the hole spinor wave function $\psi_\sigma(\xi)$. Assuming plane wave motion of the holes in the in-plane of the structure and parabolic bands, the problem can be again reduced to an effective one-dimensional Schrödinger equation similar to Eq. (V.5),

$$\left( \frac{\hbar^2}{2} \frac{\partial}{\partial \xi} \frac{1}{m_l(\xi)} \frac{\partial}{\partial \xi} + V_{\text{eff}}(\xi) - \frac{\sigma}{2} h_{pd}(\xi) \right) \psi_\sigma(\xi) = E_{l,\sigma} \psi_\sigma(\xi), \tag{V.67}$$

with $V_{\text{eff}}$ including the band offsets, the electrostatic potential due to the hole and ionized impurity density, and eventually the spin-dependent hole exchange correlation potential. By inserting the mean field results for $\langle S_z \rangle$ given by Eq. (V.60) and for the effective magnetic field acting on the impurities $B^i_{\text{eff}}$, stated in Eq. (V.58), into Eq. (V.66), the kinetic exchange potential becomes

$$h_{pd}(\xi) = J_{pd} n_i(\xi) S B_S [J_{pd} S (n_\uparrow - n_\downarrow) s / k_B T], \tag{V.68}$$

where we neglected the weak direct AF coupling between the Mn impurities and we used the relation $\langle s_z \rangle = s(n_\uparrow - n_\downarrow)$ with $s = 1/2$ denoting the hole spin. Since the spin densities $n_\sigma(\xi)$ have to be calculated by a summation over the hole spinor envelope functions $\psi_\sigma(\xi)$ analogous to Eq. (V.52), both the kinetic exchange potential and the Poisson equation for the electrostatic potential are coupled to the Schrödinger equation (V.67).

In the framework of the spinor envelope function description also the effect of the giant Zeeman splitting, which appears especially in paramagnetic DMSs layers when an external magnetic field $B$ is applied, is easily included. The magnetic field aligns the manganese spins leading to a mean polarization $\langle S_z \rangle$ of the ions, where the $z$-axis now corresponds to the external magnetic field direction. Due to the Zener exchange interaction the Mn-polarization induces again an effective magnetic field acting on the carriers $B^c_{\text{eff}} = J_{pd} \langle S_z \rangle n_i / g_c \mu_B$ as given by Eq. (V.66). In paramagnetic DMSs both the $s$-like conduction and $p$-like valence bands can exhibit a giant Zeeman splitting depending on the relative strength of the corresponding coupling coefficients $J_{sd}$ and $J_{pd}$ for electrons and holes, respectively. Typically the splitting is several times larger for holes than for electrons (Brandt and Moshchalkov, 1984). In literature the effective magnetic field is usually written in the phenomenological form

$$B^c_{\text{eff}} = N_0 \alpha x_{\text{eff}} \langle S_z \rangle / g_c \mu_B = N_0 \alpha x_{\text{eff}} S B_S \left[ \frac{S g_i \mu_B B}{k_B (T + T_{\text{eff}})} \right], \tag{V.69}$$

where $N_0\alpha$ is the so-called $sp$-$d$ exchange integral, [103] $x_{\text{eff}}$ denotes the effective manganese concentration, and the effective temperature $T_{\text{eff}}$ takes into account that the direct antiferromagnetic interaction of the Mn-impurities counteracts a parallel Mn-spin alignment. The Zeeman field leads to a spin-splitting of the carriers bands introducing an additional spin-dependent potential in the hole Hamiltonian of Eq. (V.67)

$$H_{\text{Zeeman}} = \frac{\sigma}{2} g_c \mu_B B_{\text{eff}} = \frac{\sigma}{2} g_{\text{eff}} \mu_B B, \tag{V.70}$$

where we introduced an effective giant $g$-factor given by

$$g_{\text{eff}} = \frac{g_c N_0 \alpha x_{\text{eff}}}{B} S B_S \left[ \frac{S g_i \mu_B B}{k_B (T + T_{\text{eff}})} \right]. \tag{V.71}$$

This discussion shows that mean field theory allows to describe the transport properties of magnetic heterostructures made of both ferromagnetic and paramagnetic DMSs by a closed, coupled set of a few physically transparent equations: (i) the spinor Schrödinger equation (V.67),

[103] It is believed that the $s$-$d$ coupling between $s$-like conduction band electrons and electrons localized in the $d$-shell of the Mn impurities is ferromagnetic in the bulk but there are experimental reports of finding antiferromagnetic coupling in heterostructures (Myers *et al.*, 2005).

(ii) the Poisson equation (V.6), and (iii) the spin-dependent potentials given by Eqs. (V.68) and (V.70). It is straightforward to solve these equations numerically in a selfconsistent way (Lee *et al.*, 2000; Ganguly *et al.*, 2005). However, one should be aware that in such a mean field model the dynamic correlations between the spins are completely neglected and its application is mostly feasible due to the relative simplicity of the model.

### B.3 Curie temperature in the bulk and in a magnetic quantum well

The mean field description allows a qualitative insight into the anticipated phenomena occurring in magnetic heterosystems. In particular, it allows to derive an explicit analytic expressions for the collective Curie temperature of a magnetic quantum well. However, before discussing this spatially inhomogeneous system it is elucidating to derive at first the Curie temperature for the simpler bulk case. In the bulk all mean fields are homogenous, which means that we can ignore any $\xi$-dependence in the above derived mean field expressions. For temperatures closely below the Curie temperature the effective fields are weak and we can linearize Eq. (V.60) using the first order expansion of the Brillouin function, $B_S(x) \approx (S+1)x/3S$, which yields

$$\langle S_z \rangle = \frac{S(S+1)}{3} \frac{g_i \mu_B B^i_{\text{eff}}}{k_B T}. \tag{V.72}$$

Analogously we can linearize the relationship between the spin hole density and the effective magnetic field acting on the holes,

$$\langle s_z \rangle = \frac{\partial \langle s_z \rangle}{\partial B^c_{\text{eff}}} B^c_{\text{eff}} = \chi_s B^c_{\text{eff}}, \tag{V.73}$$

by introducing the spin susceptibility $\chi_s = \partial \langle s_z \rangle / \partial B^c_{\text{eff}}$. Inserting the mean field results for both effective magnetic fields $B^i_{\text{eff}}$ and $B^c_{\text{eff}}$, given in Eqs. (V.58) and (V.64), into Eqs. (V.73) and (V.72), and combining the results, yields the selfconsistent equation for the mean ion magnetization, defining the Curie temperature $T_c$:

$$\langle S_z \rangle = \langle S_z \rangle \frac{S(S+1)}{3} \frac{1}{k_B T_c} (J^2_{pd} \frac{\chi_s}{g_c \mu_B} n_i + J_{AF} z_{AF}). \tag{V.74}$$

The Curie temperature in the bulk immediately follows as

$$k_B T_c = \frac{S(S+1)}{3} (J^2_{pd} \frac{\chi_s}{g_c \mu_B} n_i + J_{AF} z_{AF}). \tag{V.75}$$

The hole spin susceptibility can be expressed in terms of the magnetic Pauli susceptibility $\chi_m = \partial M / \partial B = g_c \mu_B \chi_s$, with $M$ denoting the hole magnetization. In the case of a degenerate hole gas at thermal equilibrium, $\chi_m$ is proportional to the hole density of states at the Fermi energy $D(E_f)$ (Ashcroft and Mermin, 1976):

$$\chi_m = (g_c \mu_B s)^2 D(E_f). \tag{V.76}$$

Hence, the equilibrium bulk Curie temperature results in

$$k_B T_c = \frac{S(S+1)}{3} \left[ J^2_{pd} n_i s^2 D(E_f) + J_{AF} z_{AF} \right], \tag{V.77}$$

leading to the important conclusion that $T_c$ is controllable by the hole density $p$ since $D(E_f) \propto p^{1/3}$. Moreover, $T_c$ depends quadratically on the coupling constat, which indicates that high transitions temperatures are most likely in $p$-type materials, since the $p$-$d$ coupling for holes is usually much stronger than the $s$-$d$ coupling of the electrons.

In the inhomogeneous case of a narrow magnetic quantum well the above derivation for the bulk can be generalized as follows (Lee *et al.*, 2000). Let us assume that the quantum well is narrow enough that only the first hole subband of the quantum well is occupied by holes. The kinetic exchange potential given by Eq. (V.66) leads to a rigid spin splitting of the subband. Since the effective field $B^c_{\text{eff}}$ or, equivalently, the Mn polarization $\langle S_z \rangle$ can be considered to be weak just below the Curie temperature, we can calculate the spin-dependent subband energies $\varepsilon_\sigma$ by applying first order perturbation theory:

$$\varepsilon_\sigma = \langle \phi_0 | H_{pd} | \phi_0 \rangle = \frac{\sigma}{2} \int \mathrm{d}\xi J_{pd} n_i(\xi) \langle S_z \rangle(\xi) |\phi_0(\xi)|^2, \tag{V.78}$$

denoting by $|\phi_0\rangle$ the unperturbed lowest subband envelope function which depends on the applied bias. Hence, the subband splitting $\Delta$ results in

$$\Delta = \varepsilon_\downarrow - \varepsilon_\uparrow = J_{pd} \int \mathrm{d}\xi \, n_i \langle S_z \rangle |\phi_0|^2. \tag{V.79}$$

Since we assumed that only the first well subband is occupied by holes, it is reasonable that the profile of the spin hole density $n_\sigma(\xi)$ is proportional to the normalized envelope functions $\psi_\sigma(\xi)$ of the spin-split subband,

$$n_\sigma(\xi) = N_\sigma |\psi_\sigma(\xi)|^2, \quad \int |\psi_\sigma(\xi)|^2 \mathrm{d}\xi = 1, \tag{V.80}$$

where we introduced the two-dimensional quantum well densities $N_\sigma$,

$$N_\sigma = \int n_\sigma(\xi) \mathrm{d}\xi. \tag{V.81}$$

If we approximate the spin up and down wave functions in zeroth order by the unperturbed ground state, $\psi_\uparrow(\xi) = \psi_\downarrow(\xi) = \phi_0(\xi)$, the $\xi$-dependent spin density can be written as

$$\langle s_z \rangle(\xi) = s(N_\uparrow - N_\downarrow) |\phi_0(\xi)|^2 = \langle s_{2D} \rangle |\phi_0(\xi)|^2, \tag{V.82}$$

where we have defined the 2D spin density of the quantum well, $\langle s_{2D} \rangle = s(N_\uparrow - N_\downarrow)$. Close to $T_c$ we can assume a linear relationship between $\langle s_{2D} \rangle$ and the band splitting $\Delta$, analogous to Eq. (V.73) in the 3D bulk case:

$$\langle s_{2D} \rangle = \frac{\partial \langle s_{2D} \rangle}{\partial \Delta} \Delta = \chi_{2D} \Delta, \tag{V.83}$$

introducing the 2D-spin susceptibility, $\chi_{2D} = \partial \langle s_{2D} \rangle / \partial \Delta$. Hence, by using Eq. (V.79), the spin density profile results in

$$\langle s_z \rangle(\xi) = \chi_{2D} |\phi_0(\xi)|^2 \Delta = \chi_{2D} |\phi_0(\xi)|^2 J_{pd} \int \mathrm{d}\xi' \, n_i(\xi') \langle S \rangle(\xi') |\phi_0(\xi')|^2. \tag{V.84}$$

In order to obtain a selfconsistent equation from which one can obtain the Curie temperature, we use the linearized mean field expression for the Mn-polarization $\langle S_z \rangle$ given by Eq. (V.72), and insert the explicit form of the effective field $B_{\text{eff}}^i$ given by Eq. (V.58), which yields

$$\langle S_z \rangle = \frac{S(S+1)}{3} \frac{1}{k_B T} (J_{pd} \langle s_z \rangle + J_{AF} z_{AF} \langle S_z \rangle). \tag{V.85}$$

By combining this result with Eq. (V.84), and again neglecting the contribution of the weak direct antiferromagnetic coupling, we obtain the desired selfconsistent relation,

$$\langle S_z \rangle(\xi) = \frac{S(S+1)}{3k_B T_c} J_{pd}^2 \chi_{2D} |\phi_0(\xi)|^2 \int \mathrm{d}\xi' n_i(\xi') \langle S_z \rangle(\xi') |\phi_0(\xi')|^2 \tag{V.86}$$

which represents a linear integral equation for $\langle S_z \rangle(\xi)$. This relation can be also understood as an eigenvalue problem $\hat{I} \langle S_z \rangle = k_B T_c \langle S_z \rangle$ of the integral operator,

$$\hat{I}[f(\xi)] = \frac{S(S+1)}{3} J_{pd}^2 \chi_{2D} |\phi_0(\xi)|^2 \int \mathrm{d}\xi' n_i(\xi') |\phi_0(\xi')|^2 f(\xi'). \tag{V.87}$$

The critical temperature is given by the highest temperature at which $\hat{I} \langle S_z \rangle = k_B T_c \langle S_z \rangle$ is fulfilled, corresponding to the largest eigenvalue with the eigenfunction $f(\xi) \propto |\phi_0(\xi)|^2$. Simply inserting the relation Eq. (V.86) for $\langle S_z \rangle$ iteratively, yields the desired analytical expression for the collective Curie temperature in a magnetic quantum well

$$k_B T_c = \frac{S(S+1)}{3} J_{pd}^2 \chi_{2D} \int \mathrm{d}\xi n_i(\xi) |\phi_0(\xi)|^4. \tag{V.88}$$

In the case of a 2D-hole gas, which is in thermal equilibrium, the 2D-spin susceptibility is again proportional to the 2D density of states per spin, $\chi_{2D} = sD_0/2 = sm_*/2\pi\hbar^2$, yielding the equilibrium critical temperature

$$k_B T_c = \frac{S(S+1)}{12} D_0 J_{pd}^2 \int \mathrm{d}\xi n_i(\xi) |\phi_0(\xi)|^4. \tag{V.89}$$

The latter two expressions for $T_c$ show that we can define a single Curie temperature for the whole quantum well instead of a $\xi$-dependent $T_c$, which would follow from a naive adaption of the bulk results to quasi-2D systems. Most important, in contrast to the bulk case, there is no obvious carrier density dependence of $T_c$ at equilibrium, since the 2D density of states $D_0$ is a constant.

However, there is an "hidden" indirect dependence, because the coupling constant becomes generally a function of the carrier density in quasi-2D systems (Priour, Jr. *et al.*, 2005) and for nonequilibrium cases the spin susceptibility can change considerably when an external bias is applied (Ganguly *et al.*, 2005). Moreover, Eq. (V.89) points out the intriguing possibility of changing $T_c$ by modulating the overlap of the impurity density profile $n_i(\xi)$ with the quantum well wave function $\phi_0(\xi)$. Since the wave function profile can be influenced by changing the confinement potential due to external electric fields, this would allow to control the appearance of ferromagnetism in confined systems electrically.

### C. Resonant tunneling in magnetic double and multi-barrier systems: a review

In the last years there have been several theoretical and experimental studies on magnetic double and multibarrier heterosystems demonstrating the enriched possibilities of such band-engineered structures. The work can be loosely grouped into three different categories: (i) structures in which only the emitter lead is magnetic and the nonmagnetic quantum well structure is used to detect the spin polarization of the injected carriers either optically, e.g., in spin light emitting diodes (spin-LEDs) (Fiederling *et al.*, 1999; Jonker *et al.*, 2000; Hanbicki *et al.*, 2002; Ohno *et al.*, 1999; Holub and Bhattacharya, 2007) or electrically by resonant tunneling spectroscopy (Ohno *et al.*, 1998; Nonoyama and Inoue, 2001; Slobodskyy *et al.*, 2007), (ii) tunneling magneto resistance (TMR)-structures or spin-valve structures in which a nonmagnetic quantum well is sandwiched between magnetic emitter and collector leads, and (iii) magnetic or spin-RTDs, which are double- and multibarrier structures where also the quantum wells and/or barriers are made of magnetic semiconductors, and which allow to realize highly efficient spin filtering and spin switching devices or to electrically control the ferromagnetic order in the quantum wells. In the following we will mainly focus on the latter two types of structures, (ii) and (iii), where resonant tunneling and quantum size effects are expected to be crucial for the understanding of the observed transport properties. A review of spin-LEDs and of the related more general issues of all-semiconductor spin injection can be found in the Refs. (Žutić *et al.*, 2004; Jonker *et al.*, 2003; Schmidt, 2005).

#### C.1 Double-barrier TMR-structures

In TMR-structures of kind (ii) the emitter serves as a spin polarizer whereas the collector acts as a spin analyzer, realizing in this way the concept of a spin valve. The resistance of a spin-valve structure depends on the relative alignment of the magnetization directions of the two magnetic layers, which is illustrated in Fig. V.16. The TMR ratio is usually defined as $(R_{\rm AP} - R_{\rm P})/R_{\rm P}$, where $R_{\rm P}$ and $R_{\rm AP}$ are the resistances for parallel and antiparallel alignments of the lead magnetizations, respectively. In the last years metallic magnetic tunnel junctions, which consist of two metallic ferromagnetic layers separated by a single thin insulator barrier, e.g. Fe/MgO/Fe, have been studied exhaustively (see Sec. II.H.1). Nowadays such metallic structures have already become the key building blocks in the latest generation of magnetoelectronic devices, such as magnetic random access memories or magnetic read heads in hard disks. However, for integration into existing semiconductor processing techniques an all-semiconductor magnetic tunnel junction would be a significant advance. All-semiconductor single barrier structures such as GaMnAs/AlAs/GaMnAs or GaMnAs/GaAs/GaMnAs have been grown and demonstrated to exhibit large TMR-ratios up to 290 % at low temperatures (Tanaka and Higo, 2001; Chiba *et al.*, 2004; Elsen *et al.*, 2006; Mattana *et al.*, 2005) and they have been studied theoretically in several papers (Krstajić and Peeters, 2005; Brey *et al.*, 2004; Sankowski *et al.*, 2007, 2006). However, in single-barrier structures the TMR and the spin polarization decreases rapidly with the increase of the applied bias. It is believed that this problem can be overcome by using a double barrier structure sandwiched between the magnetic leads (Petukhov *et al.*, 2002; Yuasa *et al.*, 2002; Kalitsov *et al.*, 2004). Such a resonant tunneling spin-valve structure was proposed by (Bruno and Wunderlich, 1998). A theoretical investigation of resonant tunneling through the double barrier system GaMnAs/AlAs/GaAs/AlAs/GaMnAs predicts a tremendous enhancement of the

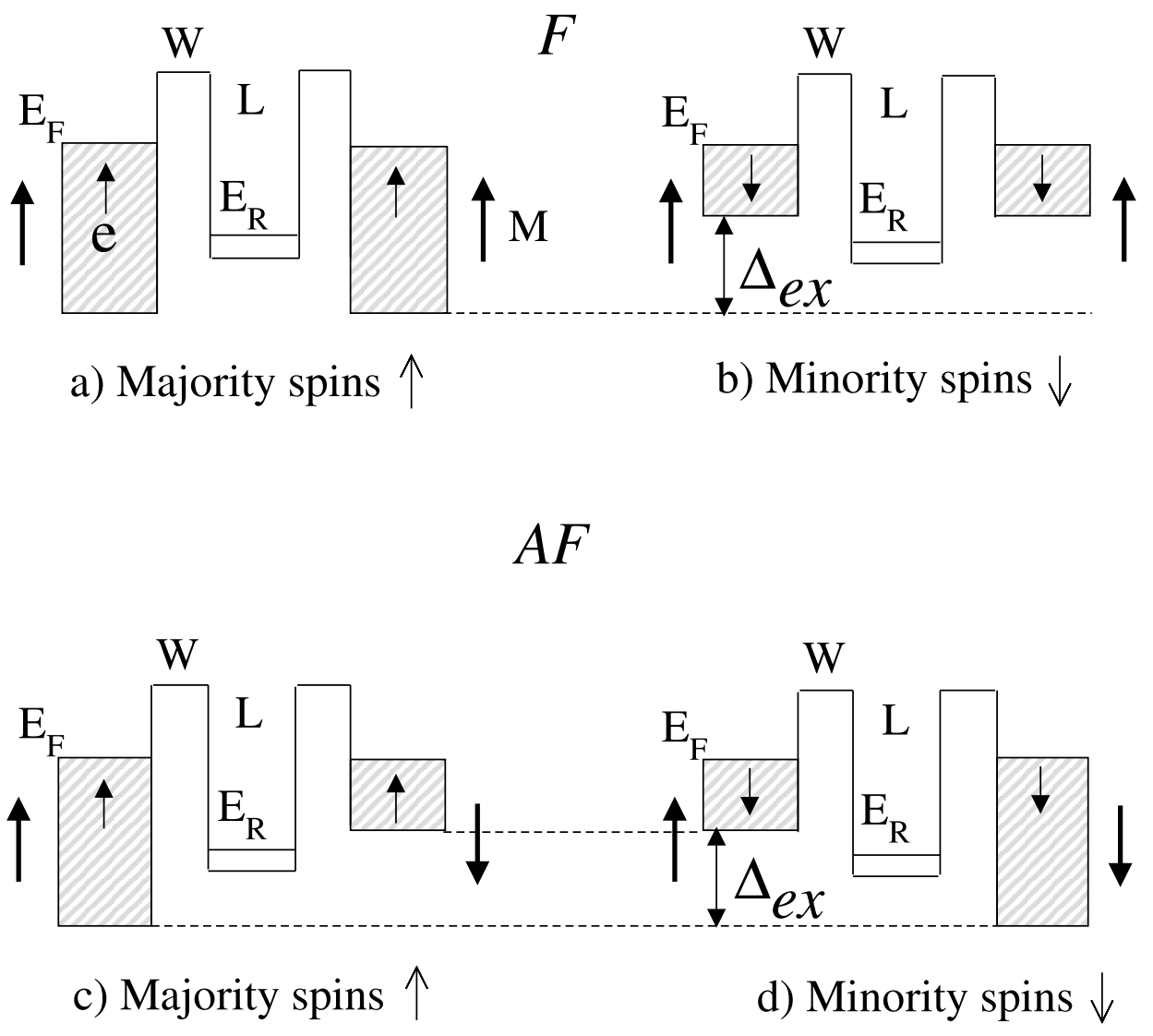

Fig. V.16. Schematic flat band diagrams illustrating the spin-valve effect in a double-barrier TMR structure with barriers of width $W$ and a quantum well of width $L$. The upper plots show the band profile in the case of a parallel (here denoted as ferromagnetic, F) alignment of the lead magnetizations $M$ for (a) the majority and (b) the minority spins. The band diagrams in the case of an antiparallel (here denoted as antiferromagnetic, AF) orientation are sketched in the lower plots (c) and (d) for majority and minority spins, respectively. The band edge in the leads is spin-split by the exchange field $\Delta_{ex}$ with the energy measured here from the *unsplit* band edge. If the first resonant level $E_R$ falls into the energy interval $0 < E_R < \Delta_{ex}/2$, as assumed in the diagrams and which can be achieved by properly tuning the geometric parameters of the structure, only the majority channel of the F configuration contributes to the conductance; all other channels are blocked. Reprinted figure with permission from A. Petukhov, A. Chantis, and D. Demchenko, *Phys. Rev. Lett.* **89**, 107205 (2002). Copyright (2002) by the American Physical Society.

TMR if the thickness of the quantum well is properly tuned (Petukhov *et al.*, 2002). The effect has been calculated to be as high as 10000% for generic parabolic bands and calculations based on a more realistic $k \cdot p$ band structure model still reveal high TMRs of about 800%, as shown in Fig. V.17. By replacing the GaMnAs leads with digitally doped Mn monolayers adjacent to the RTD-barriers, TMR values in excess of 1000% were predicted (Stewart and van Schilfgaarde, 2003). However, in transport measurements of the first realized GaMnAs-based double barrier structures resonant tunneling of holes could not be clearly observed (Mattana *et al.*, 2003; Hayashi *et al.*, 2000), although magneto-optical measurements show blueshifts of the magnetic circular dichroism (MCD) spectra, which strongly suggest the existence of quantum size-effects (Shimizu and Tanaka, 2002; Oiwa *et al.*, 2004). One possible reason for the difficulty to observe resonant tunneling is due to the band profile in the structure, since it has been shown that in terms of the hole energy the valance band bottom of GaAs is about 87-140 meV higher than the Fermi level in GaMnAs (Ohno *et al.*, 2002). Hence, one needs to apply high biases of at least 200 meV to pull down the resonant levels to the emitter Fermi energy. Unfortunately, at such high biases the TMR ratio becomes already very small and, hence, it is difficult to observe TMR associated with resonant tunneling in such GaMnAs-based structures. The valence band offset of the quantum well can be decreased by using InGaAs instead of GaAs as the well material. The observed oscillating behavior of the TMR in GaMnAs/AlAs/InGaAs/AlAs/GaMnAs structures between positive and negative values with increasing the AlAs barrier thickness has been explained by the resonant tunneling effect (Ohya *et al.*, 2005).

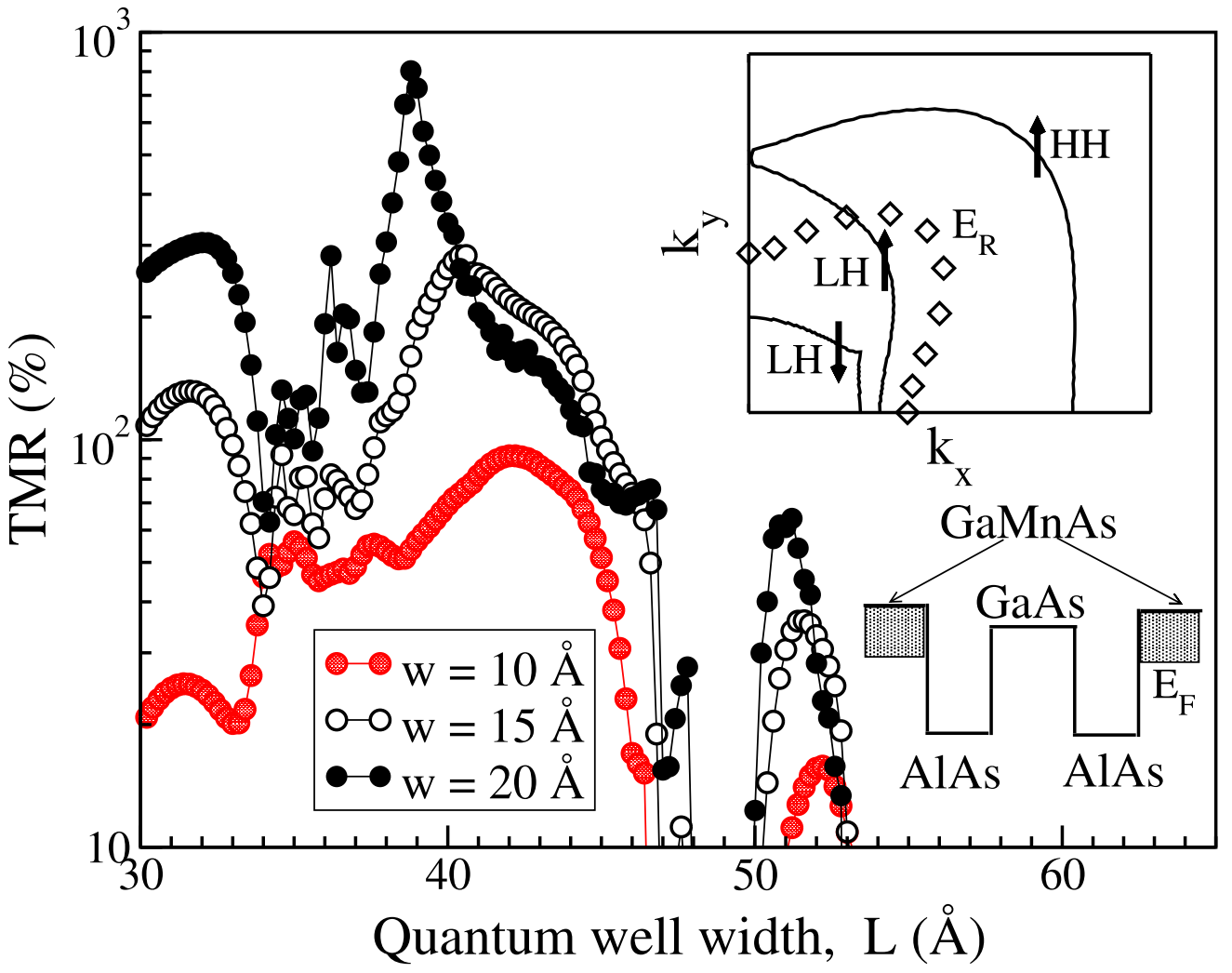

Fig. V.17. Tunneling magneto resistance (TMR) of GaMnAs/AlAs/GaAs/AlAs/GaMnAs double barrier structure as a function of the quantum-well width $L$ for different barrier widths $w$ calculated by using the Kohn-Luttinger hole Hamiltonian model. The upper inset shows the Fermi surface in the emitter (solid line) and the quantum well (diamonds) for heavy (HH) and light holes (LH), respectively, corresponding to the maximum TMR. Lower inset: schematic band diagram. Reprinted figure with permission from A. Petukhov, A. Chantis, and D. Demchenko, *Phys. Rev. Lett.* **89**, 107205 (2002). Copyright (2002) by the American Physical Society.

Further enhancement of the TMR-ratio up to $10^6\%$ is expected by introducing a ferromagnetic quantum well, e.g., GaMnAs/AlAs/GaMnAs/AlAs/GaMnAs, according to the resonant tunneling effect through the spin-split resonant levels in the GaMnAs quantum well [104](Hayashi *et al.*, 2000). Clear experimental evidence of resonant tunneling and the correlated increase of TMR has been found in GaMnAs/AlGaAs/GaMnAs/AlAs/Be-doped GaAs heterostructures in which thin GaAs spacer layers between the magnetic and nonmagnetic layers were inserted to prevent Mn diffusion into the barrier layers and to smooth the interfaces (Ohya *et al.*, 2007, 2006).

[104]Strictly following our classification scheme this structure falls into category (iii) but we discuss it also here, since it promises high TMR-ratios.

## C.2 Magnetic RTDs

Magnetic RTDs of type (iii) with a magnetic quantum well open up the interesting opportunity of realizing efficient voltage-controlled spin injectors or spin switching devices, i.e., structures in which the spin character of the carriers ending up at the collector side can be controlled electrically. Such a controllable spin-selective injection by magnetic RTDs would be very desirable, since the usual spin injection based on magnetic contacts can only be utilized to transfer the majority spin from the magnetic into nonmagnetic layer, making it necessary to apply an external magnetic field to flip the injected spin polarization by switching the magnetization of the emitter contact. Furthermore, magnetic RTDs also allow for new routes of determining the spin polarization of a injected current employing the double barrier structure as a spin detector (Giazotto *et al.*, 2003). The basic idea of a magnetic RTD used for spin injection, switching or detection is fairly straightforward. Since the quasibound states in magnetic quantum wells are spin-split, the collector current becomes spin polarized by bringing either the spin up or spin down state of the quantum well into resonance.

As another interesting field of applications, magnetic RTDs comprising a ferromagnetic DMS quantum well afford the possibility of realizing a voltage-controlled phase transition between the para- and ferromagnetic order, which would offer prospects of new functionalities in magnetoelectronic device applications (Lee *et al.*, 2002).

In the following we will group the work done so far on magnetic RTDs into five categories: (i) RTDs with a quantum well made of a paramagnetic semiconductor material, (ii) RTDs with a ferromagnetic or delta-doped quantum wells, (iii) heterostructures, in which only the barriers are magnetic, (iv) structures, which employ interband tunneling, and (v) nonmagnetic heterosystems, which utilize the spin-splitting due to spin-orbit coupling.

## C.3 Paramagnetic spin-RTDs

In RTDs with a paramagnetic quantum well a giant Zeeman splitting of the well states of the order of tens of meV is induced by applying moderate external magnetic fields of a few tesla. This is due to the fact that an applied magnetic field aligns the local impurity spins leading to an impurity spin polarization, which via the Zener exchange coupling gives rise to an strong effective magnetic field acting on the carriers. Such paramagnetic structures provide a perfect framework for investigating the potential of magnetic resonant tunneling structures, since the growing of high quality structures is possible and the magnetoelectronic effects are tuneable by the experimentally adjustable magnetic field strengths. Most of the structures employ transport of electrons, which in terms of mobility and spin life time is advantageous compared to $p$-type conductivity, usually encountered in III-V ferromagnetic DMS.

The forming of spin superlattices by a periodical doping with paramagnetic ions were proposed by Ortenberg (von Ortenberg, 1982) and later realized for Mn and Fe-based multilayer systems (Dai *et al.*, 1991; Chou *et al.*, 1991; Guo *et al.*, 2001b). Magnetic RTDs, in which the leads are nonmagnetic and only the quantum well is made of a magnetic material were realized by Brehmer *et al.* (1995), who fabricated RTDs consisting of semimetallic ErAs quantum well and AlAs barriers sandwiched between GaAs layers and observed a large spin splitting of the electron states in the quantum well in magnetic field. The unusual behavior of resonant tunneling through ErAs layers in magnetic fields of different orientation was theoretically explained in terms of tunneling of electrons into the valence band states of ErAs in the vicinity of the $\Gamma$ point (Petukhov *et al.*, 1996). Further realizations of paramagnetic quantum wells became possible with the discovery of the II-VI DMSs coming along with the technical progress of growing high-quality magnetic heterostructures. Haury *et al.* (1997) realized a modulation doped quantum well made of CdMnTe with a very low Curie temperature of $T_c = 2$ K evidenced by photoluminescence measurements. However, the electrical properties were not studied. The diffusive in-plane transport in magnetic two dimensional electron gases and its spin dependence has been observed and investigated in single quantum wells utilizing ZnSe/$Zn_{1-x-y}Cd_xMn_ySe$ (Smorchkova *et al.*, 1997, 1998), ZnTe/$Cd_{1-x}Mn_xSe$ (Knobel *et al.*, 1999), or $Cd_{1-y}Mg_yTe/Cd_{1-x}Mn_xTe$ heterojunctions (Jaroszyński *et al.*, 2000). An overview of the interesting work done at that time concerning the spin dynamics and quantum transport properties in II-VI DMS quantum structures can be found in the review paper of (Awschalom and Samarth, 1999).

The vertical transport properties of a double barrier structure doped with magnetic impurities under the influence of parallel electric fields were firstly examined theoretically by (Sugakov and Yatskevich, 1992). The occurrence of a spin filtering effect in ZnSe/ZnMnSe heterostructures with a single paramagnetic layer was proposed by Egues (1998), finding a strong suppression of one spin-component of the current density with increasing magnetic fields. Similar results were also predicted by (Chang and Peeters, 2001) revealing conductivity oscillations with increasing magnetic field and the thickness of the single DMS-layer. These considerations were generalized by Guo *et al.* (2000) by taking into account an external electric field, the conduction band offset of ZnSe and ZnMnSe (Zhai *et al.*, 2001) and studying systems which comprise symmetric and asymmetric paramagnetic double layers (Guo *et al.*, 2001a). Their results showed that the spin-dependent transmission can become either enhanced or suppressed depending on the applied bias as well as on the structural asymmetry. By employing an additional semimagnetic emitter contact it has been shown theoretically (Egues *et al.*, 2001) that the voltage-dependent magnetoresistance should exhibit robust features like spin split kinks and beating patterns in the case of single barrier and double barrier structures, respectively. The theoretical investigation of asymmetric tunnel structures with differently doped paramagnetic layers reveal high spin filtering effects of up to nearly 100% for suitable magnetic and electric fields in the case of conduction electrons in ZnSe-based structures (Zhai *et al.*, 2003; Zhu and Su, 2004; Papp *et al.*, 2005, 2006; Saffarzadeh *et al.*, 2005) as well as for holes and electrons in CdTe/$Cd_{1-x}Mn_xTe$ heterosystems (Malkova and Ekenberg, 2002; Gnanasekar and Navaneethakrishnan, 2006; Lev *et al.*, 2006). Recently, Borza *et al.* (2007) investigated the interesting possibility of an electric field manipulation of the electronic states in a two-partitioned quantum well, which consists of a magnetic and nonmagnetic layer, resulting in the forming of a potential step in the quantum well for one spin component, while the other experiences a deeper well in the magnetic layer region. Also all-magnetic RTD, in which the whole structure including the barriers, the emitter and collector lead consists of II-VI DMSs layers of different magnetic ion concentrations $x$, were investigated, e.g., using $Zn_{1-x}Mn_xS$ (Beletskii *et al.*, 2005), or $Cd_{1-x}Zn_xTe$ (Chitta *et al.*, 1999). In the latter work also the influence of spin flip scattering caused by thermal fluctuations of the magnetic ion moments was taken into account, showing that it is inefficient in depolarizing the current.

The prospects of very high spin filtering effects also inspired several experimental realizations of II-VI paramagnetic RTDs. The experimental observation of resonant tunneling in semimagnetic (Zn,Mn)Se/BeTe double barrier structures were reported by Keim et al. (Keim *et al.*, 1999) although no spin splitting of subbands could be detected in electrical measurements. A suc-cessful efficient injection of spin-polarized current into GaAs using BeTe/ZnMnSe/BeTe RTDs were later demonstrated by the same group (Gruber *et al.*, 2001). Clear evidence of spin splitting of the transmission resonance into two separate peaks was found in the IV-characteristics of ZnSe/BeZnSe/ZnMnSe/BeZnSe/ZnSe RTDs (Slobodskyy *et al.*, 2003; Gould *et al.*, 2004b). The experimental setup of Ref. (Slobodskyy *et al.*, 2003) and the measured IV-curves for a Mn concentration of 4% and 8%, respectively, in the quantum well for different applied magnetic fields at low temperatures $T = 1.3$ K are shown in Fig. V.18. For appropriate high magnetic fields the giant Zeeman splitting of the well states becomes sufficient to be also observable in the IV-curves. The experimental results agree very well with modeled values (indicated by circles in Fig. V.18), using the simplifying assumptions that both spin-split levels exhibit approximately the same conductivity and that the level conductivity depends merely on the relative alignment of the well level to the emitter Fermi energy. The electrical demonstration of the well splitting constitutes a first important step for realizing a voltage controlled spin switching device, which works by selectively bringing either the spin-up or down level into resonance. Such a device operation was also theoretically investigated (Havu *et al.*, 2005) simulating the experimental setup of Ref. (Slobodskyy *et al.*, 2003) and comparing the numerical results obtained from both the Wigner and Greens function approach. Very recently another successful realization of a magnetic RTD based on paramagnetic $Zn_{1-x-y}Mn_yCd_xSe$ with a pillar diameter down to 6 $\mu$m have been reported (Fang *et al.*, 2007), again demonstrating the current peak splitting in moderate external magnetic fields. At high fields, when the giant Zeeman splitting in the well is already saturated, the Zeeman splitting in the contacts starts to play a role, leading to a spin-polarization of the injected carriers. As a result the peak current becomes modified dependent on the injected spin polarization; an effect which can be utilized for spin detection (Sánchez *et al.*, 2007).

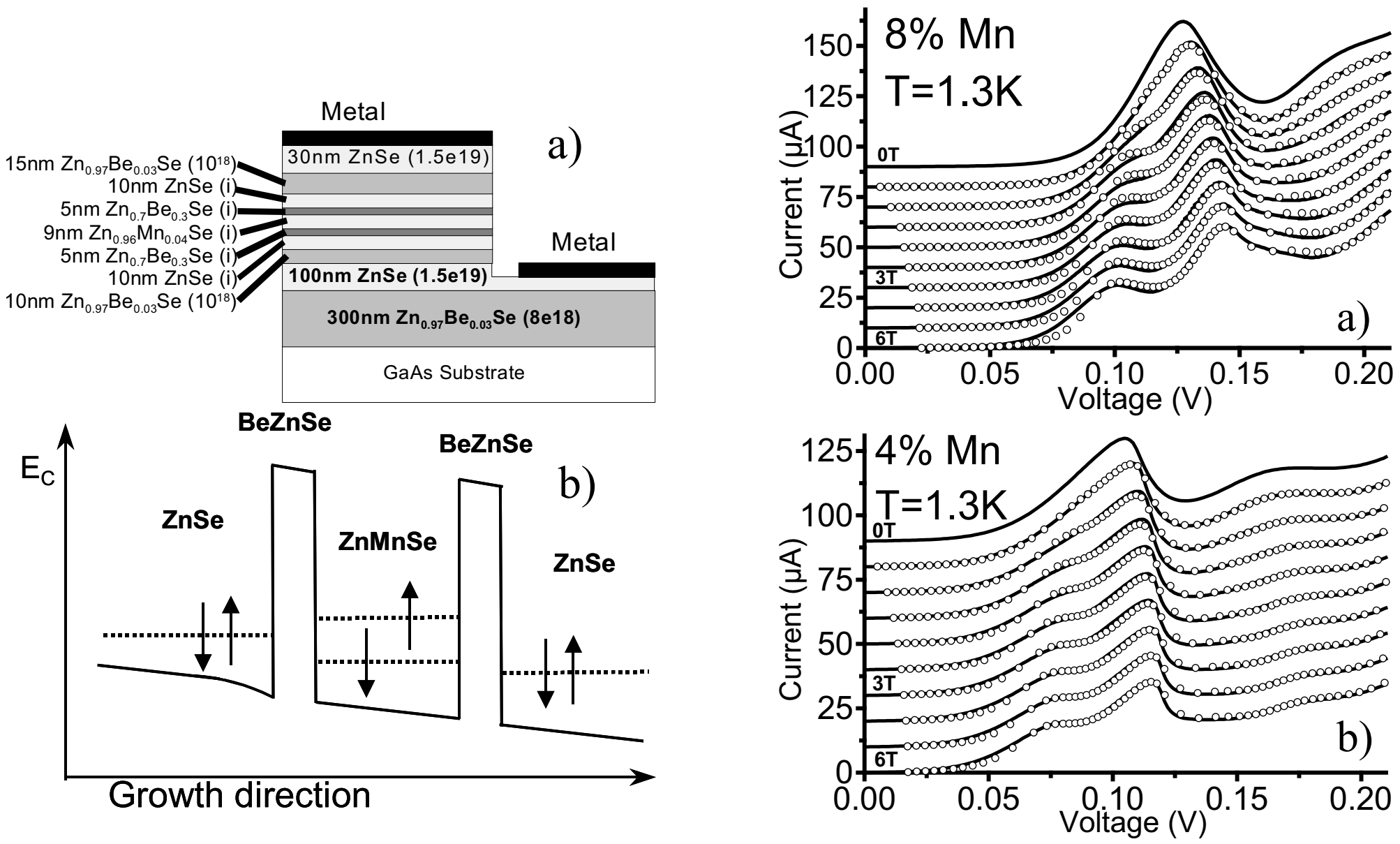

Fig. V.18. Left: (a) Layered structure of the spin-RTD with a 9 nm thick paramagnetic quantum well made of ZnMnSe. (b) Schematic profile of the conduction band of the structure under applied bias. Right: Experimental (lines) and modeled (circles) $IV$ curves of the magnetic RTD with a ZnMnSe quantum well exhibiting (a) 8% Mn concentration and (b) 4% Mn concentration. For clarity curves are offset by 10 $\mu$A and are taken in 0.5 T intervals from 0 to 3 T and in 1 T intervals form 3 to 6 T. The spin-splitting of the quantum well levels leads to a splitting of the transmission resonance into two separate peaks, which becomes clearly observable for higher magnetic fields. Reprinted figures with permission from A. Slobodskyy, C. Gould, T. Slobodskyy, C. Becker, G. Schmidt, and L. W. Molenkamp, *Phys. Rev. Lett.* **90**, 246601 (2003). Copyright (2003) by the American Physical Society.

Moreover, structures with several quantum wells have been realized, e.g., modulation-doped ZnSe/(Zn,Cd,Mn)Se systems (Berry *et al.*, 2000). The theoretical investigation of such multiple quantum well systems reveals that the nonlinear transport phenomenology become considerably enriched by an interplay between the charge accumulation in the wells and the resonant interwell tunneling, resulting in a formation of electric field domains (Sánchez *et al.*, 2001, 2002).

## C.4 Ferromagnetic spin-RTDs

The exhaustive work done so far on II-VI DMSs based RTDs resulted in a better understanding of the magnetic properties of thin transition metal doped paramagnetic layers as well as revealed important findings for possible high efficient all-semiconductor spin injection and detection schemes. However, in such structures always external magnetic fields of a few tesla are necessary to induce the giant Zeeman splitting. By using ferromagnetic DMSs layers *nonvolatile* devices could be realized, which can operate without the need of auxiliary, external magnetic fields. The successful growing of short-period GaMnAs/GaAs superlattices showed that the ferromagnetic order can sustain in layers of a few nm width (Shimizu and Tanaka, 2002; Mathieu *et al.*, 2002; Kolovos-Vellianitis *et al.*, 2006) and the formation of spin-resolved quantized states in III-V-based ferromagnetic quantum wells has been confirmed by optical measurements (Oiwa *et al.*, 2004). The appearance of a ferromagnetic, paramagnetic, or a canted spin phase in AlAs/GaMnAs quantum wells, depending on the carrier concentration and magnetic layer width has been studied with the aid of Monte Carlo simulations (Boselli *et al.*, 2000). Theoretical considerations of the ferromagnetism in inhomogeneous layer systems based on mean field theory and local spin density functional theory (Dietl *et al.*, 1997; Jungwirth *et al.*, 1999; Lee *et al.*, 2000; Fernández-Rossier and Sham, 2001; Brey and Guinea, 2000; Ghazali *et al.*, 2001) revealed that the magnetic properties of a Mn-doped quantum well considerably depends on the overlap of the subband wave function with the spatial profile of the magnetic impurity density, as already discussed in Sec. B.3. The wave function can be modulated in situ by an external bias voltage, thereby strongly influencing the collective Curie temperature of the well (Lee *et al.*, 2000). The idea of a voltage-controlled magnetization switch in a RTD with a ferromagnetic quantum well made of GaMnAs has already been proposed (Ganguly *et al.*, 2005) and it has been shown that the effect should be also observable if moderate scattering is present in the well (Ganguly *et al.*, 2006b). In such magnetic RTDs the voltage-dependent Curie temperature can become bistable (Ganguly *et al.*, 2006a) due to the hysteretic behavior of the IV-characteristic of the diode (see Sec. A.5). The forming of a bistable state through hole redistribution between a magnetic and nonmagnetic quantum well has also been predicted in double quantum well structures, which was proposed to be utilized as a stable memory element, down-scalable to a few-hole regime (Semenov *et al.*, 2005). Theoretical investigations of the ferromagnetic properties of quantum wells made of InMnP, GaMnN, or ZnO (Kim *et al.*, 2006, 2005) based on the mean field approximation have shown that by taking into account the carrier-carrier exchange correlations the Curie temperature of the wells can be substantially enhanced to values well above room temperature. Recently, magnetic RTDs comprising two coupled magnetic quantum wells made of GaMnN were proposed as a spintronic device offering large magnetocurrents, which are defined as relative current magnitudes for parallel and antiparallel orientations of the quantum well magnetizations (Ertler and Fabian, 2006b).

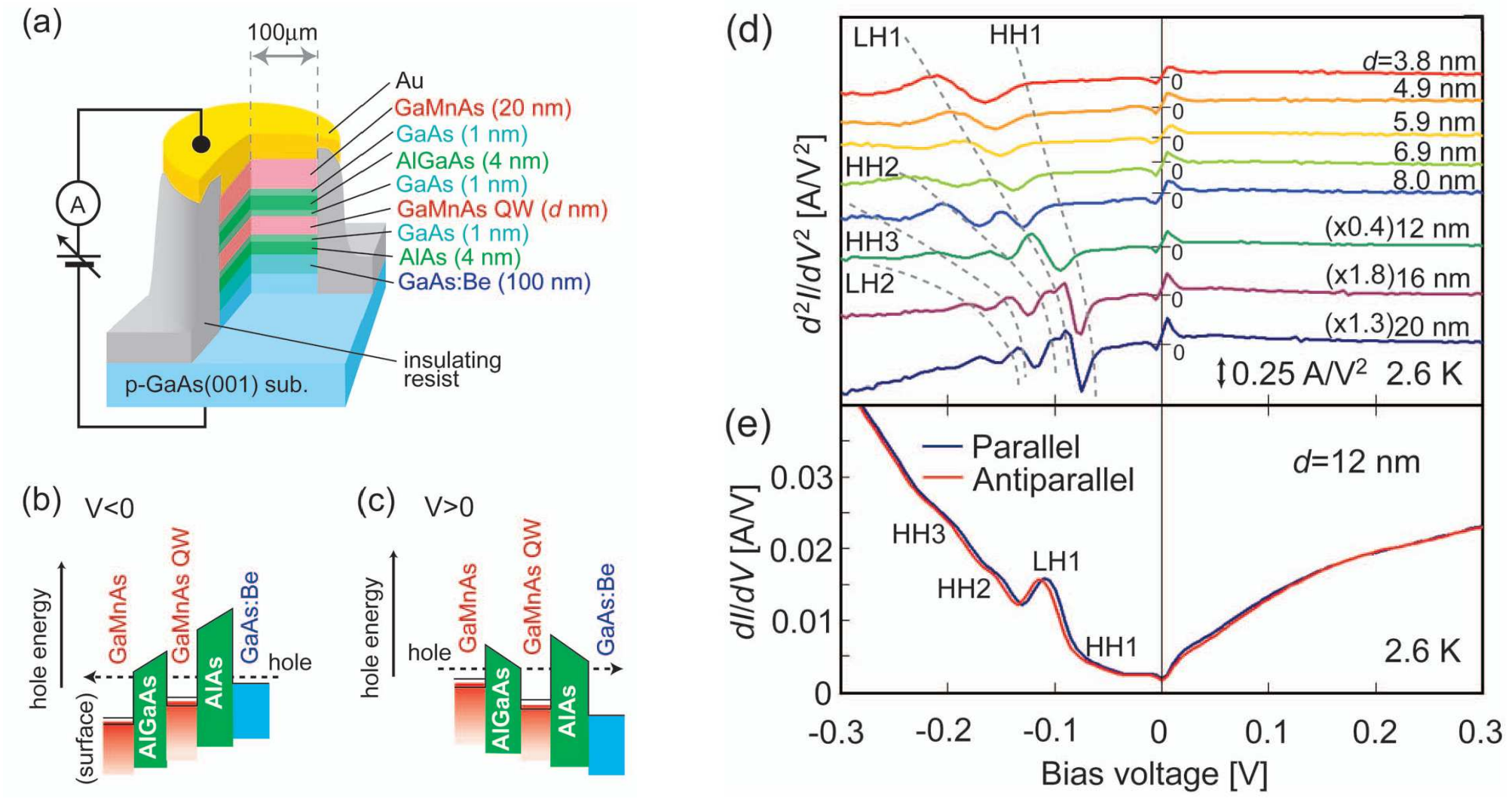

Fig. V.19. (a) Schematic device structure of a GaMnAs/AlGaAs/GaMnAs/AlAs/GaAs:Be RTD junction with imbedded thin GaAs spacer layers for smoothing the interfaces and preventing Mn diffusion. Schematic band diagrams for negative (b) and positive (c) applied biases, respectively. (d) Measured $d^2I/dV^2 - V$ characteristic of the ferromagnetic RTD for various quantum well thicknesses $d$ for parallel magnetization of the emitter and well at 2.6 K. The peak voltages and the period of the oscillations become smaller with increasing $d$, which indicates that the oscillating features are induced by resonant tunneling of light and heave holes (LH,HH). (e) Measured $dI/dV - V$ characteristics of the junction with $d = 12$ nm for parallel (blue curve) and antiparallel (red curve) magnetization. Reprinted figures with permission from S. Ohya, P. N. Hai, Y. Mizuno, and M. Tanaka, *Phys. Rev. B* **75**, 155328 (2007). Copyright (2007) by the American Physical Society.

The quantum transport of holes and the corresponding spin filtering effect through magnetic RTDs with a GaMnAs quantum well have been investigated in several theoretical studies. In the simplest approach simple parabolic or tight-binding band models were used to calculate the spin-polarized current by solving either the Schrödinger-Poisson system (Makler *et al.*, 2002) or by applying the Keldysh non-equilibrium Greens function technique (Lebedeva and Kuivalainen, 2005; Ganguly *et al.*, 2005). For more quantitative predictions the complex band structure of the holes has to be considered by using either the hole Kohn-Luttinger Hamiltonian (Petukhov *et al.*, 2000; Wu *et al.*, 2003a) derived from the $k \cdot p$ approximation of the band structure or by applying an empirical multiorbital tight-binding description (Sankowski *et al.*, 2007). Also experimentally the resonant tunneling effect of holes through a double barrier structure with a GaMnAs quantum well has been observed (Ohya *et al.*, 2007, 2006). The characteristic oscillatory features become eminently apparent in the voltage dependent $\mathrm{d}^2I/\mathrm{d}V^2$ curves, i.e., in the second derivative of the measured $IV$ characteristics, as shown in Fig. V.19. The growing of high quality structures with clean interfaces appears difficult, since the necessary usage of low temperature MBE for the magnetic layers leads to suboptimal grow conditions for the whole structure. Nevertheless, the successful growth of highly periodic and homogenous heterostructures without structural and compositional fluctuations has been reported recently (Kolovos-Vellianitis *et al.*, 2006).

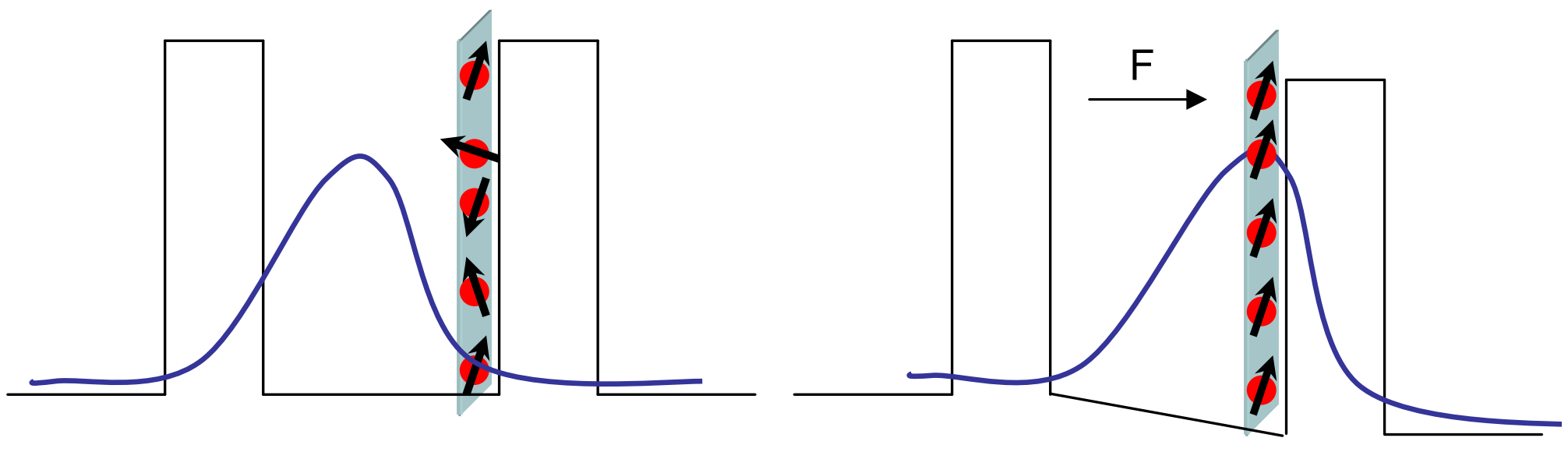

Fig. V.20. Schematic illustration of a $\delta$-doped ferromagnetic quantum well with a single layer grown of MnAs. The ferromagnetic order can be controlled by an external electric field, which modulates the overlap of the subband wave function with the Mn-atoms. Left: paramagnetic state. Right: ferromagnetic state.

Another interesting approach to realize high $T_c$ ferromagnetic quantum wells is the inclusion of so-called magnetic $\delta$-doped layers (Kawakami *et al.*, 2000; Nazmul *et al.*, 2003, 2004), in which a single or several monolayers are grown of MnAs. According to the delta function-like doping profile, the Mn atoms are expected to substitute for all Ga sites in a single layer, which can exhibit high transition temperatures close to room temperature due to local high dopant and carrier concentration (Nazmul *et al.*, 2005). Recent theoretical considerations show that ferromagnetism can be stabilized in such a single layer of magnetic ions (Melko *et al.*, 2007), although the Mermin-Wagner theorem implies that in the absence of magnetic anisotropy gapless spin excitations always destroy the ferromagnetic order in two dimensions. The stabilized ferromagnetic order in 2D layers however comes along with quantum fluctuations, which suppresses the total magnetic moment from its fully saturated value. Investigation based on mean field models reveal that $p$-$d$ exchange interaction in $\delta$-doped layers tends to be further enhanced by an additional confinement of the carriers, e.g., in a quantum well, leading accordingly to higher Curie temperatures (Fernández-Rossier and Sham, 2001; Kim and Yi, 2002; Lee *et al.*, 2002). In such digital ferromagnetic heterostructures the occurrence of spin separation (the majority and minority-spin carriers reside in different spatial regions) has been predicted (Fernández-Rossier and Sham, 2002). Furthermore, it has been shown that the Curie temperature can be controlled over a broad range by external electric fields, which modulate the overlap of the subband wave functions with the Mn $\delta$-doping profile (Nazmul *et al.*, 2004; Lee *et al.*, 2000), as schematically sketched in Fig. V.20. A strong enhancement of $T_c$ in a asymmetric double quantum wells structure, which consist of a Mn $\delta$-doped and a $p$-type doped well, was found by Kim *et al.* (2004) under applying moderate external electric fields of about $\approx 1.5$ meV/nm. Such an enhancement was also proposed very recently in a single quantum well of about 40 nm width for low electric fields ($\approx 0.3$ meV/nm) (Lv *et al.*, 2007).

## C.5 Spin-RTDs with magnetic barriers

Spin-filtering devices can also be realized by structures, which consist of a nonmagnetic quantum well but utilize magnetic barriers to achieve spin-dependent tunneling. Since the barrier height is spin-dependent the transmission of one spin component is strongly reduced compared to the other one leading to a spin filtering effect. High spin polarizations of about 80% were achieved experimentally in samples with a single barrier made of EuS (Moodera *et al.*, 1988; Hao *et al.*, 1990; Moodera *et al.*, 2007); a spin injection device based on two EuS magnetic tunnel barriers have been proposed by Filip *et al.* (2002). The TMR of double barrier structures was investigated in both the coherent and sequential tunneling limit and predicted to reach values of about 100% (Wilczyński *et al.*, 2003; Saffarzadeh, 2003). Spin dependent resonant tunneling has been also found in a parabolic quantum well, which is sandwiched between two magnetic barriers and subjected to external electric and magnetic fields, where in the case of resonance the effective potential becomes similar to a double quantum well profile (Santiago and Guimarães, 2003). In order to achieve room temperature operation without external magnetic fields, a ferromagnetic RTD with a InGaN quantum well and magnetic barriers made of GaMnN has been proposed recently (Li *et al.*, 2006a), confirming the intuitive result that the spin splitting of the barrier height has to exceed at least $k_B T$ to observe the spin splitting also in the IV-characteristics. The temperature evolution of the spin splitting of the barrier height directly affects these IV characteristics (Lebedeva and Kuivalainen, 2003; Oliveira *et al.*, 2007).

## C.6 Magnetic interband RTDs

Another interesting approach for injecting highly spin-polarized electrons is to employ interband or Zener tunneling. It has been proven experimentally that electrons have remarkably long spin life times in III-V semiconductors, whereas the hole spin relaxes much faster (Žutić *et al.*, 2004; Kikkawa and Awschalom, 1999). Hence, the injection of spin polarized electrons rather than holes appears favorable for further spin manipulations. By using Zener-Esaki-diodes, successful electrical injection of electrons by interband tunneling from the valence band of $p$-GaMnAs into the conduction band of an adjacent $n$-GaAs has been demonstrated (Kohda *et al.*, 2001; Johnston-Halperin *et al.*, 2002; Kohda *et al.*, 2006b). Recently, high spin polarizations of the injected electron current of about 80% were obtained in such devices (Dorpe *et al.*, 2004; Kohda *et al.*, 2006a). However, the polarization decreases rapidly with increasing applied bias (Sankowski *et al.*, 2007). Again theoretical considerations suggest that double barrier structures, which utilize spin-dependent resonant tunneling in magnetic heterostructures with type-II broken-gap band alignment, allow for very efficient electron spin injection even at higher biases (Petukhov *et al.*, 2003; Vurgaftman and Meyer, 2003a,b). The system investigated in these studies was an interband RTDs based on a InAs/AlSb/GaMnSb/AlSb/InAs heterostructure with its schematic band structure shown in Fig. V.21. Since the bottom of the conduction band of InAs is energetically below the top of the valence band of GaMnSb, electrons from the InAs emitter can tunnel through the hole states of GaMnSb to the InAs collector side. According to the spin-split quantized hole states the emerging electrons become spin polarized, resulting in a highly efficient spin injection of the electrons into the collector semiconductor material. The theoretical results for the calculated transmission coefficient and the spin polarization are shown in Fig. V.22, in which a $8\times 8\,\mathbf{k}\cdot\mathbf{p}$ Kane Hamiltonian was used to describe the carriers dynamics (Petukhov *et al.*, 2003).

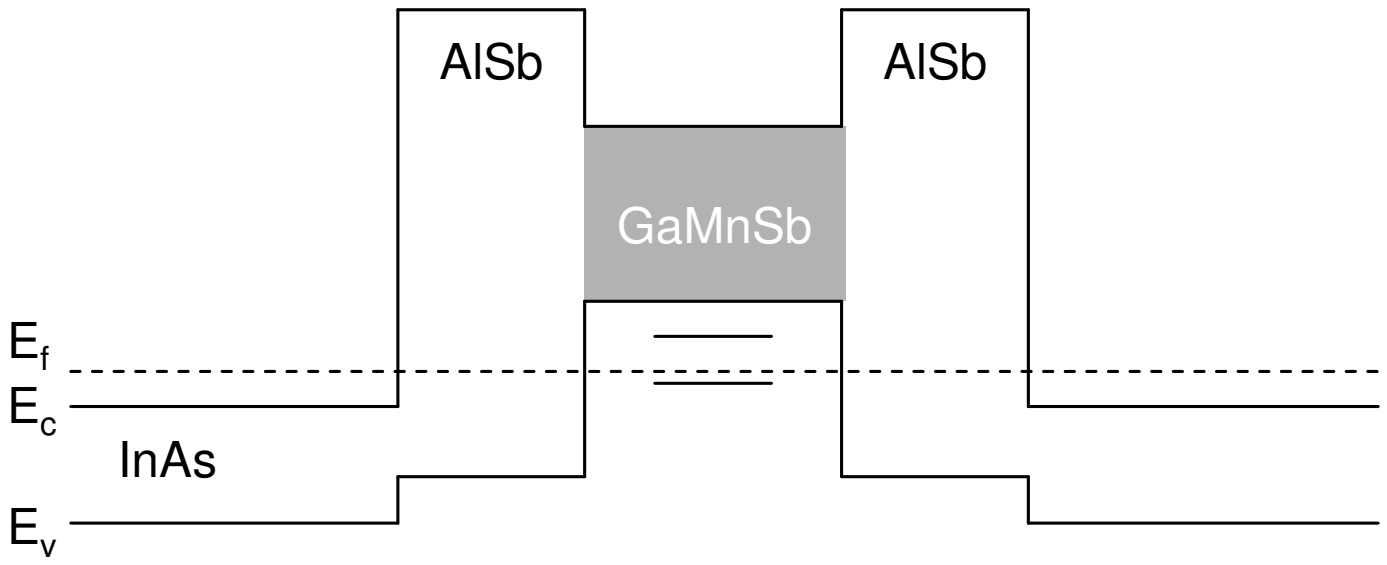

Fig. V.21. Schematic flat band diagram of the conduction, $E_c$, and valance, $E_v$, bands for a resonant interband tunneling diode. Electrons from the InAs emitter tunnel through the confined spin-split hole states of GaMnSb to the collector side, resulting in an efficient spin injection.

### C.7 Nonmagnetic spin-RTDs based on spin-orbit coupling

So far all discussed structures included some magnetic layers, giving rise to a spin-dependent vertical transport. However, it has been shown that the creation of spin-polarized current is also possible in nonmagnetic structures, which consist only of nonmagnetic semiconductors, by utilizing the spin-orbit coupling effects to induce spin-split resonant levels. The spin-orbit interaction in III-V semiconductors is usually described by two contributions; one is referred to as the Bychkov-Rashba term, which is induced by the inversion asymmetry of the confining potential profile and the other is known as Dresselhaus term caused by the bulk inversion asymmetry of the zinc-blende lattice structure. For narrow-gap semiconductors the Rashba coupling has been shown to be the dominant effect (de Andrada e Silva, 1992; de Andrada e Silva *et al.*, 1994) compared to the Dresselhaus term, which is therefore often neglected in a first account. Nonmagnetic spin filters are interesting from the viewpoints of both the absence of magnetic stray fields, which can cause undesirable effects, and the possible growth of high-quality structures.

In asymmetric heterostructures with built-in or external electric fields, the spin-orbit Rashba interaction provides a coupling of the spin to the in-plane motion of the electrons, which is controllable by electric fields. A Rashba spin filter based on resonant tunneling in double barrier structures has been proposed by Voskoboynikov *et al.* (1999, 2000), showing that the structure can provide some degree of spin polarization ($\approx 40\%$). The spin splitting in such structures in the absence of external magnetic fields was also confirmed experimentally (Yamada *et al.*, 2002; de Carvalho *et al.*, 2006). To achieve even higher spin polarizations Koga *et al.* (2002) suggested to use a nonmagnetic triple barrier RTD, as shown in Fig. V.23, which combines the resonant tunneling with the spin blockade effect. The current spin polarization exceeds almost 100% at the peak positions of the IV-curve. High spin filter efficiencies were also predicted in asymmetric resonant interband RTDs (Ting and Cartoixà, 2002), exploiting the strong spin-orbit interaction in the valence band. In a first realized side-gated prototype of such a Rashba interband-RTD based on a AlSb/InAs/GaSb/AlSb heterosystem the tunneling current show modulations with the applied side gate bias (Moon *et al.*, 2004). Ting and Cartoixà (2003) showed that a further enhancement of the current spin polarization in such interband structures is possible if the effect of the Dresselhaus term is included in the theoretical description. A nonmagnetic spin transistor based on the seminal Datta-Das proposal (Datta and Das, 1990), in which nonmagnetic interband RTDs are used as spin injector and detectors and the spin precession in the lateral transport channel is controlled by the gate voltage exploiting the Rashba coupling, has been proposed by Hall et al. (Hall *et al.*, 2003).

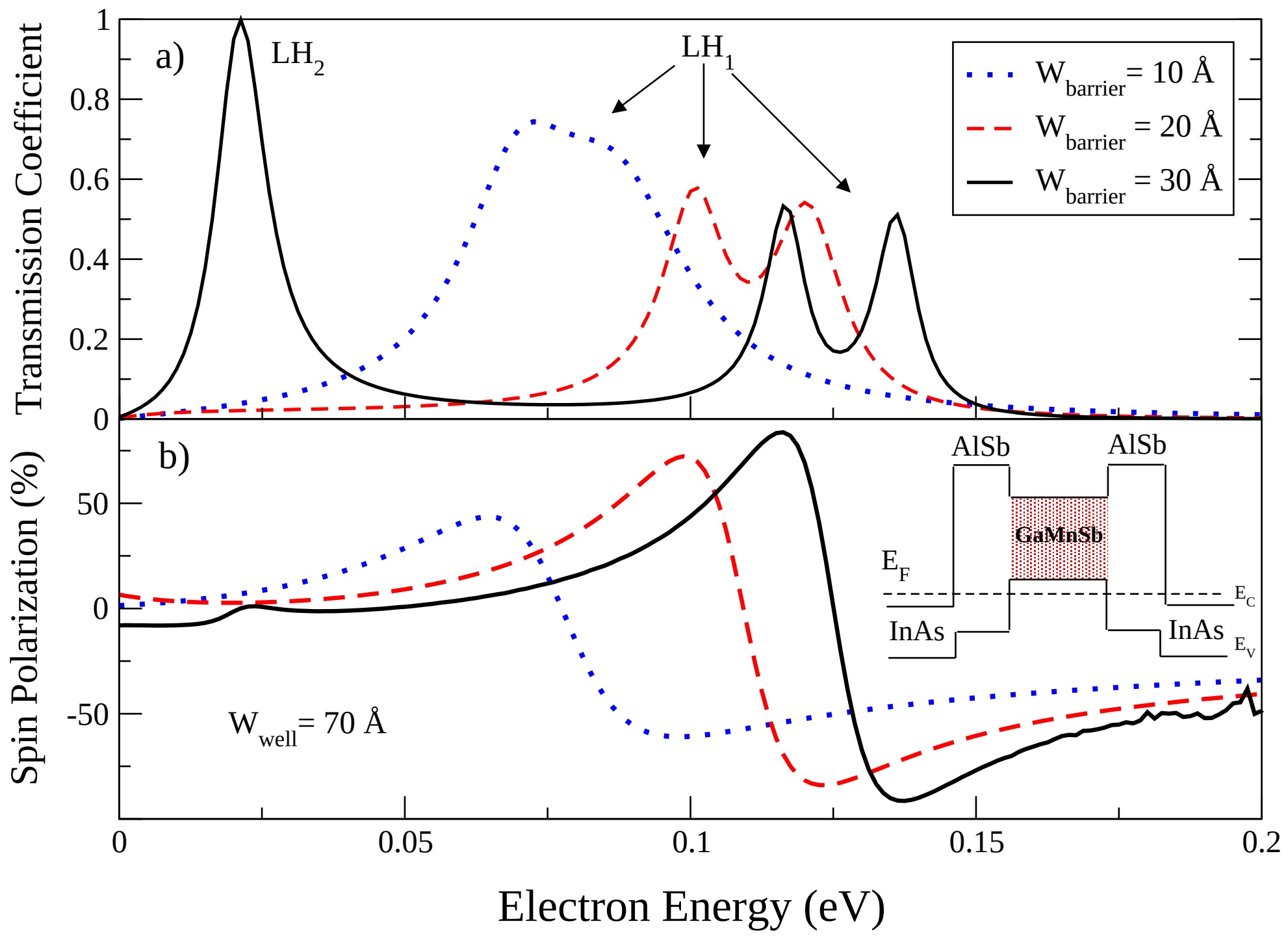

Fig. V.22. (a) Transmission coefficients versus electron energy of a InAs/AlSb/GaMnSb heterostructure with a 70 Å quantum well and various barrier widths $w$. (b) Calculated spin polarization for carriers with vanishing in-plane momentum $k_t = 0$. Reprinted figure with permission from A. Petukhov, D. Demchenko, and A. N. Chantis, *Phys. Rev. B* **68**, 125332 (2003). Copyright (2003) by the American Physical Society.

Due to the Dresselhaus coupling spin-polarized tunneling can also occur through a single *symmetric barrier*, where Rashba-splitting is ineffective owing to the structural symmetry (Perel' *et al.*, 2003). The Dresselhaus term can also induce an observable spin splitting of the trans-mission resonances of a symmetrical double-barrier structure (Glazov *et al.*, 2005). Dynamical investigations of the resonant tunneling process revealed that the tunneling time of electrons of opposite spin orientations can vary over a few order of magnitudes (Yu and Voskoboynikov, 2005; Wang *et al.*, 2002; Guo *et al.*, 2005; Wu *et al.*, 2003b). Such a spin-dependent tunneling time can actually serve as the basis for a dynamical or time-resolved spin filtering device (Romo and Ulloa, 2005; Yu and Voskoboynikov, 2005; Li and Guo, 2006). A recent more exhaustive discussion of possible device concepts based on Rashba and Dresselhaus spin splitting can be found in Ting and Cartoixà (2005).

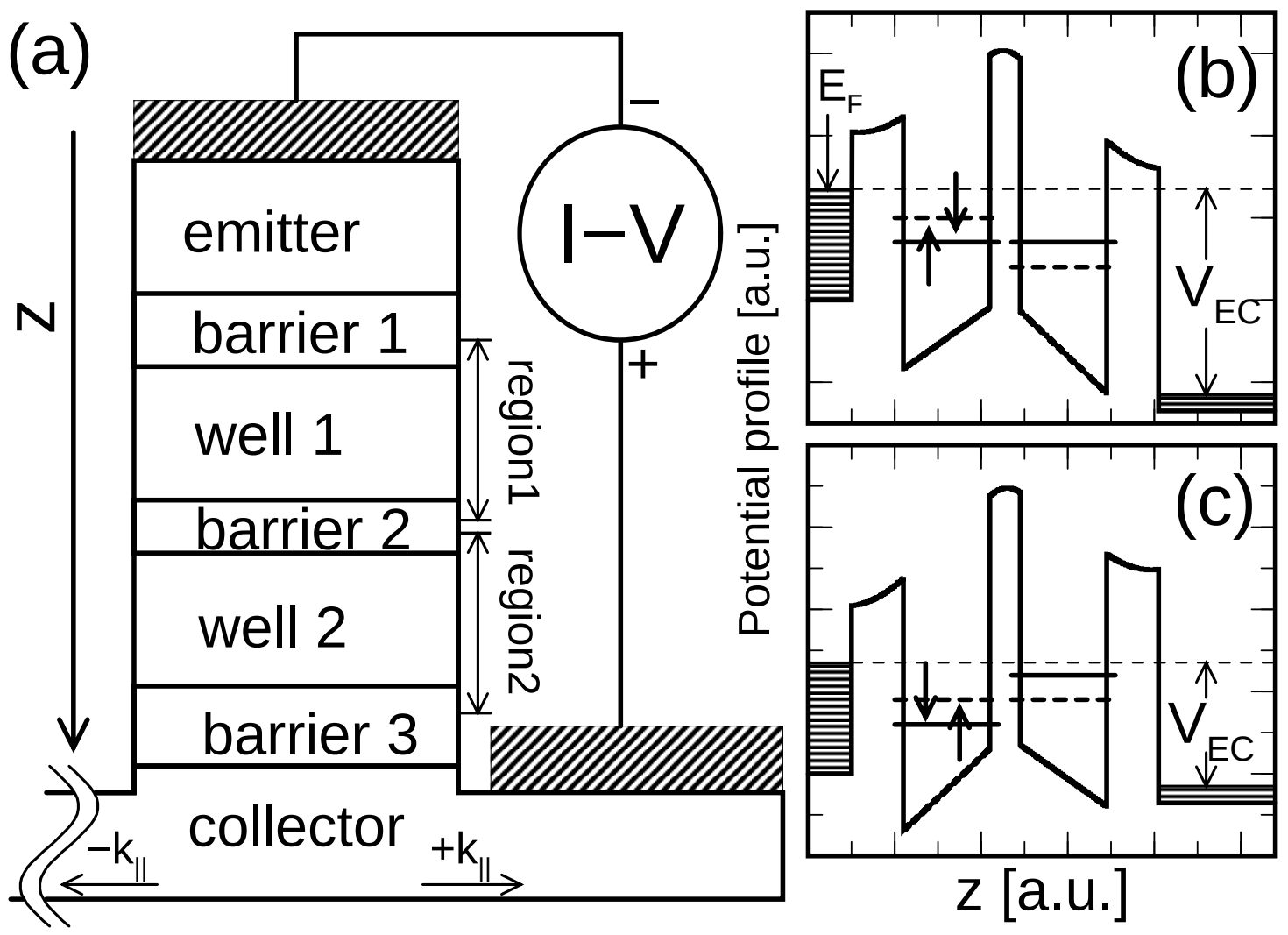

Fig. V.23. (a) Schematic illustration of the spin filter device based on the Rashba effect in a nonmagnetic triple barrier structure. (b) and (c) Conduction band profile of the device illustrating how the matching of the spin-split well levels is performed by controlling the collector bias $V_{\mathrm{EC}}$. Reprinted figure with permission from T. Koga, J. Nitta, H. Takayanagi and S. Datta, *Phys. Rev. Lett.* **88**, 126601 (2002). Copyright (2002) by the American Physical Society.

## D. Digital magneto resistance in magnetic MOBILEs

As an example for a magnetoelectronic device exploiting resonant tunneling we will discuss here in detail the concept of a magnetic monostable bistable logic element, shortly denoted as m-MOBILE. The principle of the device operation, which rests on the nonlinear $N$-shaped IV-characteristics of RTDs, has been proposed and realized by (Maezawa and Mizutani, 1993; Maezawa and Förster, 2003). The conventional MOBILE consists of two nonmagnetic RTDs, a load and a driver, which are connected in series, as schematically illustrated in Fig. V.24. The device is driven by an oscillating input voltage $V_{\mathrm{in}}$, performing a transition between the mono- and bistable working point regimes of the circuit. The occurrence of a bistable state at higher input voltages can be readily understood by drawing a circuit load line diagram. From Kirchoff's laws it follows that $V_{\text{load}} + V_{\text{out}} = V_{\text{in}}$ and $I_{\text{load}}(V_{\text{load}}) = I_{\text{driver}}(V_{\text{out}})$, where $V_{\text{out}}$ is the output voltage, $V_{\text{load}}$ denotes the voltage drop at the load-RTD, and $I_{\text{load}}$ and $I_{\text{driver}}$ are the currents flowing through the load and driver RTD, respectively. Combining both expressions yields the condition

$$I_{\text{load}}(V_{\text{in}} - V_{\text{out}}) = I_{\text{driver}}(V_{\text{out}}), \tag{V.90}$$

which determines the dc-working point of the circuit. The possible solutions to this working point equation can be found in a graphical way. For this purpose one plots both the driver and the load IV-curves, $I_{\text{driver}}(V_{\text{out}})$ and $I_{\text{load}}(V_{\text{in}} - V_{\text{out}})$, as functions of the output voltage, resulting in a mirrored load IV starting at $V_{\text{out}} = V_{\text{in}}$ as illustrated in the left plot of Fig. V.24. The crossing points of both curves correspond to the possible dc-working points of the circuit fulfilling the condition Eq. (V.90). As we can see from Fig. V.24, for low input voltages only one crossing point appears; the circuit is said to be in the monostable regime. However, by increasing the input voltage the mirrored load IV-curve is shifted to the right and at appropriate high input voltages we end up with three crossing points and, hence, three possible dc working points of the circuit. The middle crossing point in the NDR region is proven to be unstable as shown in detailed stability analysis including the dynamic behavior of the circuit (Kidner *et al.*, 1991; Chow, 1964). For a stable working point the sum of load and driver conductivity $G_{\text{load}}, G_{\text{driver}}$ has to be greater than zero, giving the stability criterion $G_{\text{load}} + G_{\text{driver}} > 0$, which is clearly violated in the NDR region. Therefore, at high input voltages the circuit resides in a bistable regime with a stable working point either at low or high output voltage.

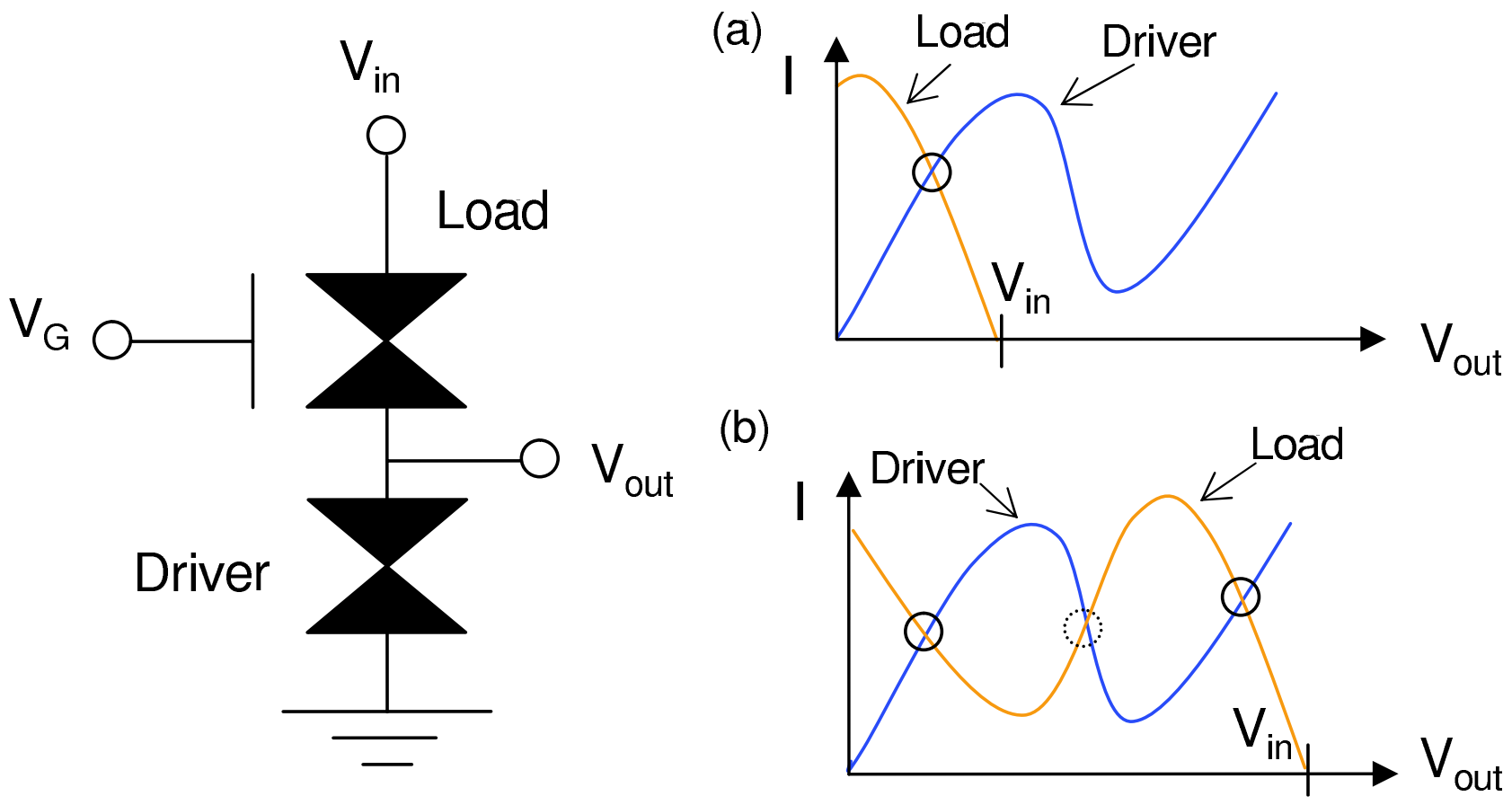

Fig. V.24. Left: Circuit configuration of the nonmagnetic MOBILE proposed by Maezawa and Mizutani (Maezawa and Mizutani, 1993). The peak current of the load RTD can be modified by an external gate voltage $V_G$. Right: Schematic load line diagrams for (a) low input voltage and (b) high input voltage $V_{\text{in}}$. For small input voltages only one stable dc-working point is possible (monostable regime), whereas for high input voltages two stable working points become feasible (bistable regime).

Which of the two possible working points is actually realized depends on the difference of the load and driver peak currents. As illustrated in Fig. V.25(a), if the load peak current is smaller than the driver peak value, the working point is unable to “overcome” the driver’s current peak when the input voltage is increased and, hence, the working point remains always at low output voltages. However, in the opposite case of a higher load peak current, the working point can "climb" over the driver peak voltage, resulting in a high output voltage for high input voltages, as displayed in Fig. V.25(b). In short, the whole device acts as a comparator of the load and driver's peak currents, giving either low or high output in the bistable regime. For performing the switching between the low and high output voltage states the difference of the peak currents can be actually very small, since the transition is in some sense analogous to a second order phase transition as discussed in Maezawa and Mizutani (1993).

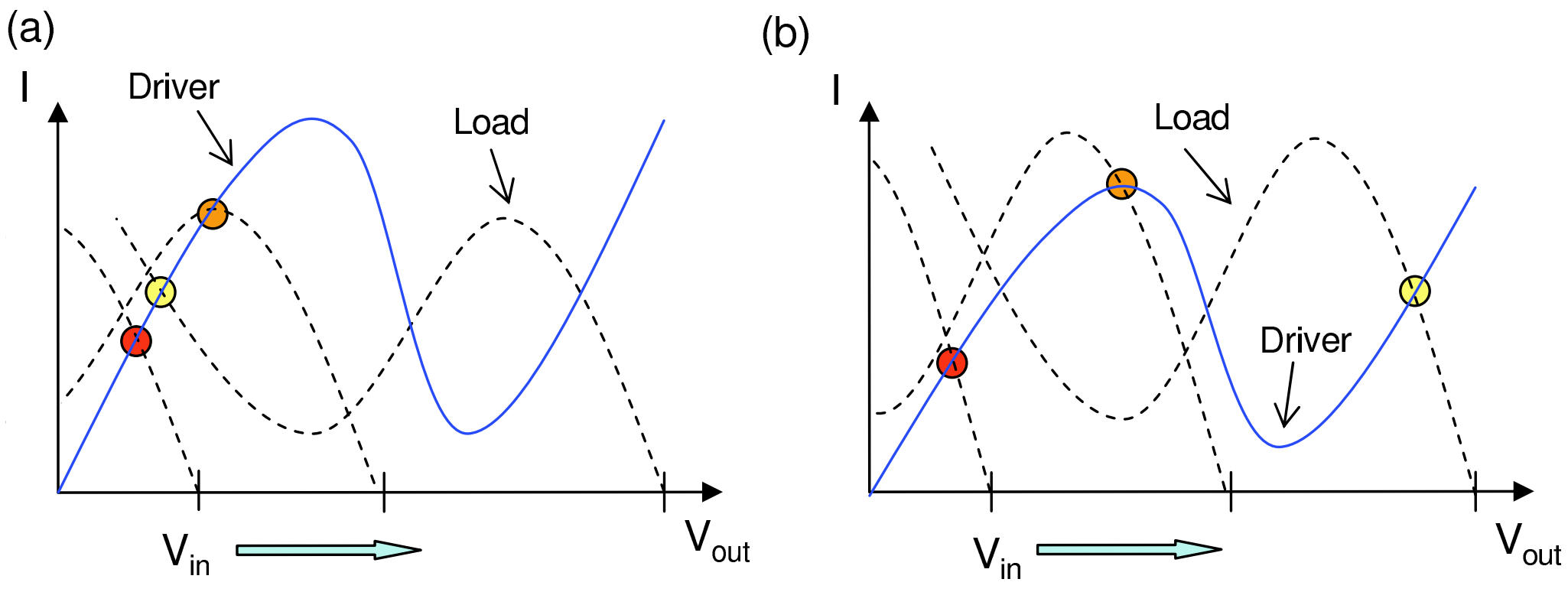

Fig. V.25. Schematic load line diagrams for the case of a lower (a) and higher (b) load peak current compared to the driver peak value. The mirrored load IV-characteristic is shifted from the left to the right when the input voltage $V_{\rm in}$ is increased and the actual (stable) working point of the circuit is marked by a circle. Only in case (b) the working point can "overcome" the driver's current peak.

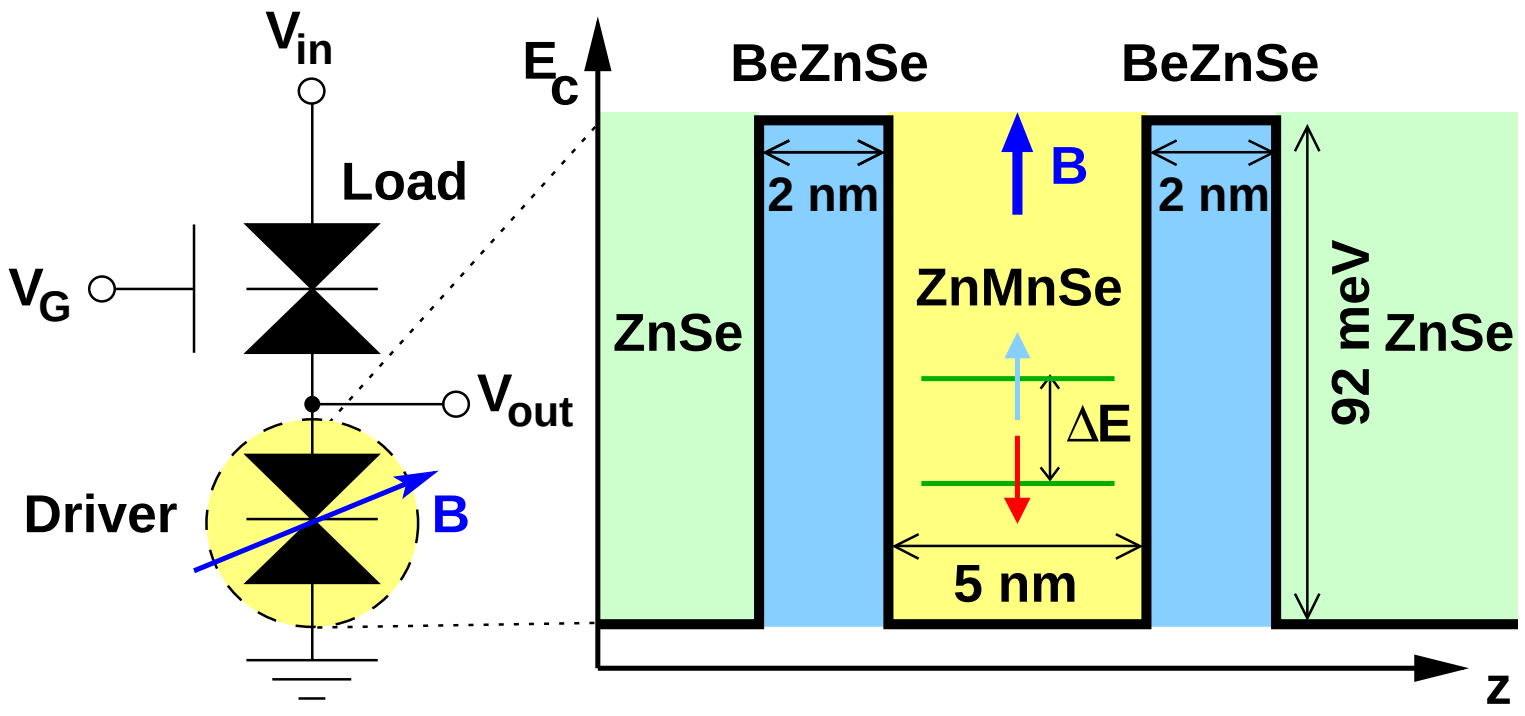

Fig. V.26. Left: The circuit configuration of the paramagnetic-MOBILE. The load is a conventional RTD, whose peak current can be modified by an external gate voltage $V_G$. The driver device consists of a magnetic RTD [here made of a Zn(Be,Mn)Se material system]. The peak current of the driver is controlled by the magnetic field. Right: Schematic conduction band profile of the magnetic RTD with giant Zeeman splitting $\Delta E$ of the first well state.

A magnetic variant of the MOBILE can be realized by replacing the conventional driver

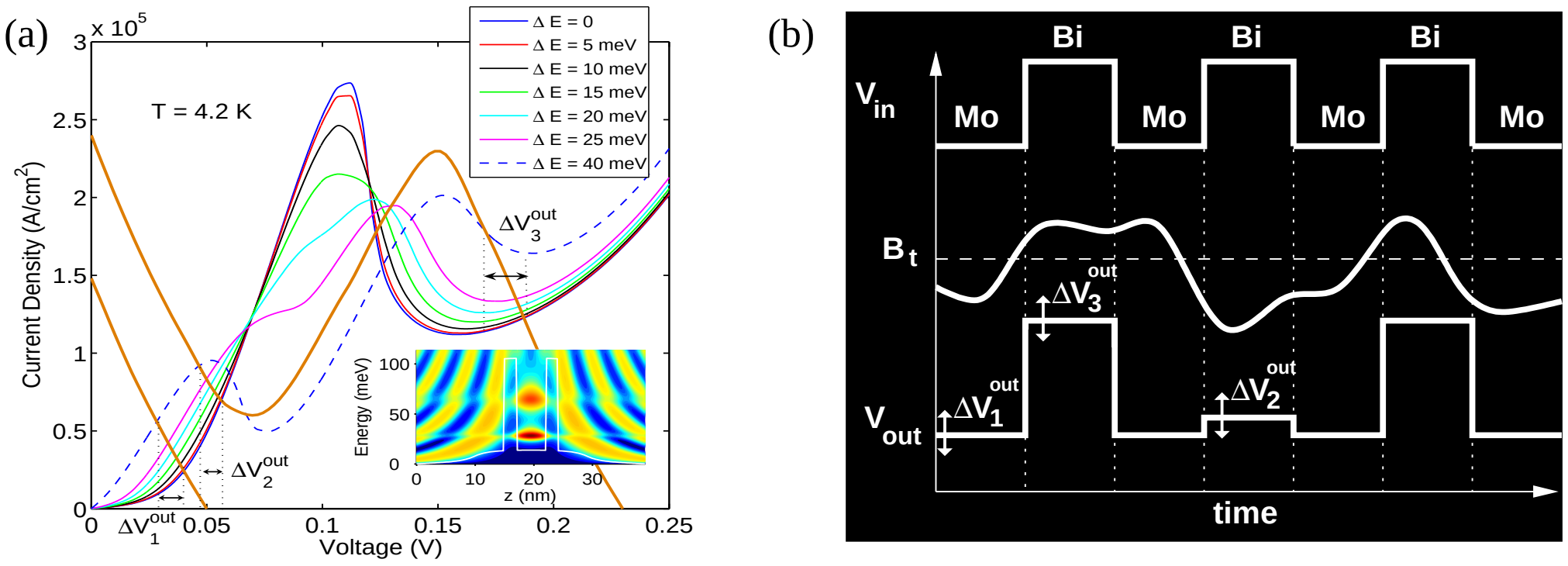

Fig. V.27. (a) Current-Voltage characteristics of the magnetic RTDs at the temperature of $T = 4.2$ K for different Zeeman energy splittings $\Delta E$. The thick solid lines indicate the mirrored IV-curve of the load RTD for low and high input voltages. The inset shows a contour plot of the local density of states versus energy and growth direction $z$ at zero bias and $\Delta E = 40$ meV. The solid line in the inset indicates the conduction band profile. The two spin resonances are visible by the large density (dark) in the well. (b) Scheme of the operation principle of the digital magneto resistance MOBILE. The vertical dotted lines indicate the points of time when a mono-to-bistable transition is performed. The output voltages fluctuate within the intervals $\Delta V_i^{\text{out}}, i = 1, 2, 3$.

RTD by a magnetic RTD, which comprises either a para- or ferromagnetic quantum well (Ertler and Fabian, 2006a, 2007). Figure V.26 shows the schematic circuit diagram for a paramagnetic MOBILE. The magnetoelectronic device operation is based on the possibility of modulating the peak current of the magnetic driver RTD by applying an external magnetic field, which controls the spin splitting of the quantum well states. For instance, in a paramagnetic RTD with a quantum well made of ZnMnSe, selfconsistent numerical calculations have shown that the peak current becomes appreciably decreased by increasing the applied external magnetic field (Ertler and Fabian, 2006a) as shown in Fig. V.27(a). For high magnetic fields the Zeeman splitting of the well states becomes also observable in the IV-characteristics, leading to two separated current peaks; an effect which has already been observed experimentally (Slobodskyy *et al.*, 2003; Fang *et al.*, 2007; Sánchez *et al.*, 2007). The influence of the magnetic field on the driver peak current can be utilized to realize the effect of *digital magneto resistance* (DMR) introduced by Ertler and Fabian (2006a, 2007): the output voltage makes a discrete jump from low to high after performing the mono-to-bistable transition if the external magnetic field is higher than some electrically controllable threshold value, or to define it more generally, if a particular device characteristic, which can be influenced by an external magnetic field, exceeds a certain threshold.

The DMR effect relies on the following operation principle. Let us assume that at zero magnetic field the driver peak current is higher than the load peak current (the load peak current can for instance be modulated by an external gate voltage $V_G$). By applying an external magnetic field the driver peak current is reduced, and at some threshold value $B_t$ both load and drive peak currents become equal. The threshold field depends on the initial difference (at $B = 0$) of load and driver peak currents, which can be easily controlled by $V_G$. By applying an input voltage, which oscillates between low and high voltage levels, the circuit constantly performs a transition between the mono- and bistable working point regime. If the magnetic field is higher (lower) than the threshold value during the transition the output voltage in the bistable regime ends up at a high (low) level. As illustrated in Fig. V.27(a) the spreading of the possible working points $\Delta V_i^{\mathrm{out}}, i = 1, 2, 3$ in the mono- and bistable regime according to different magnetic field values is much smaller than the separation of the low and high voltage levels, which allows for a direct digital interpretation of the output voltage. The operation principle is schematically illustrated in Fig. V.27(b), where for simplicity a rectangular input signal is assumed. In the monostable regime ($V_{\mathrm{in}}$ is low) the output voltage is always low. However, if the magnetic field is higher than the threshold $B_t$, which is indicated by the horizontal dashed line in Fig. V.27(b), at the moment of transition from a low to a high input signal, i.e., from monostable to bistable regime, as marked by the vertical dashed lines in Fig. V.27(b), the output voltage in the bistable regime is high; otherwise low. Ultrahigh frequency operations of nonmagnetic MOBILEs of about 100 GHz have been demonstrated by employing a symmetric clock configuration (Maezawa *et al.*, 2006). As an application this proposed paramagnetic MOBILE might be potentially used as a very fast read head of conventional hard disks performing a direct conversion of the magnetically stored information into a binary electrical signal.

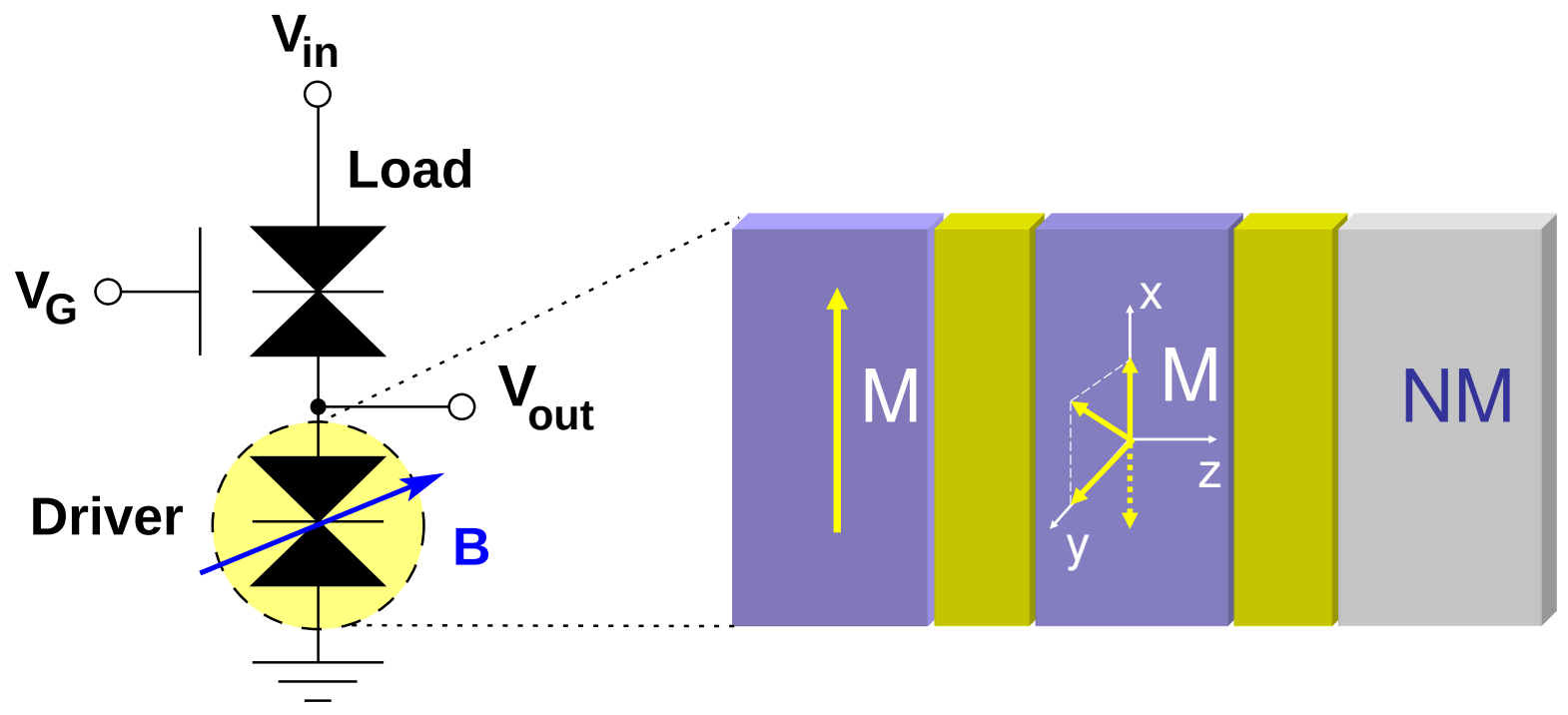

Fig. V.28. Left: Sketch of the circuit configuration of the ferromagnetic MOBILE. The load is a conventional RTD, whereas the driver device consists of a RTD with a ferromagnetic emitter and quantum well. The peak current of the driver is controlled by twisting the well magnetization, which is assumed to be soft.

The effect of DMR has, in fact, already been experimentally demonstrated in a somewhat different circuit setup, in which a metallic giant magneto resistance (GMR) element is shunted to a nonmagnetic driver RTD (Hanbicki *et al.*, 2001). Depending on the relative orientation of the magnetization directions of the ferromagnetic layers (parallel or antiparallel) in the GMR element either a high or low output voltage is observed in the bistable regime. In this way the circuit becomes nonvolatile upon the loss of power, since the state of the device is actually stored in the alignment of the ferromagnetic layers. Nonvolatile devices are very attractive for a fast and reliable data storage, e.g, in random access memory applications (Daughton, 1999), or for reprogrammable logic (Dery *et al.*, 2007), in which the mode of logical operation can be modified by changing the magnetic state of the device.

Such a nonvolatile ferromagnetic MOBILE might be also realized in a simplified setup as shown in Fig. V.28, where instead of a GMR-element a ferromagnetic driver RTD is used (Ertler and Fabian, 2007). One conceivable way to modulate the driver's peak current would be to change the magnitude of the spin-splitting of the ferromagnetic quantum well, by varying the value of the well magnetization using external magnetic fields similar to what is done in the above discussed paramagnetic MOBILE. However, in ferromagnetic layers it is usually much easier to change the orientation of the magnetization than its absolute value. Indeed, in the case of a ferromagnetic well, in which the exchange splitting is strongly anisotropic, a simple twisting of the well magnetization would be sufficient to observe DMR. For the isotropic case, however, DMR becomes only possible if an additional ferromagnetic layer is used. It appears advantageous to employ a ferromagnetic emitter instead of a ferromagnetic collector lead in order to obtain a noticeable effect on the peak current when the well magnetization direction is changed. This is due to the fact that the exchange splitting in the collector lead, which can be assumed to be typically of the order of a few tens of meV, becomes ineffective in influencing the current magnitude at high voltages, since the collector's band edge is then already shifted far below the resonant well states, which determine the carriers transmission. Hence, using an additional ferromagnetic collector lead would result only in a very small magnetocurrent at the peak voltage, making the MOBILE operation unfeasible. This reasoning also suggests that hardly any DMR can be observed by utilizing double barrier TMR-structures with a nonmagnetic quantum well sandwiched between two ferromagnetic leads.

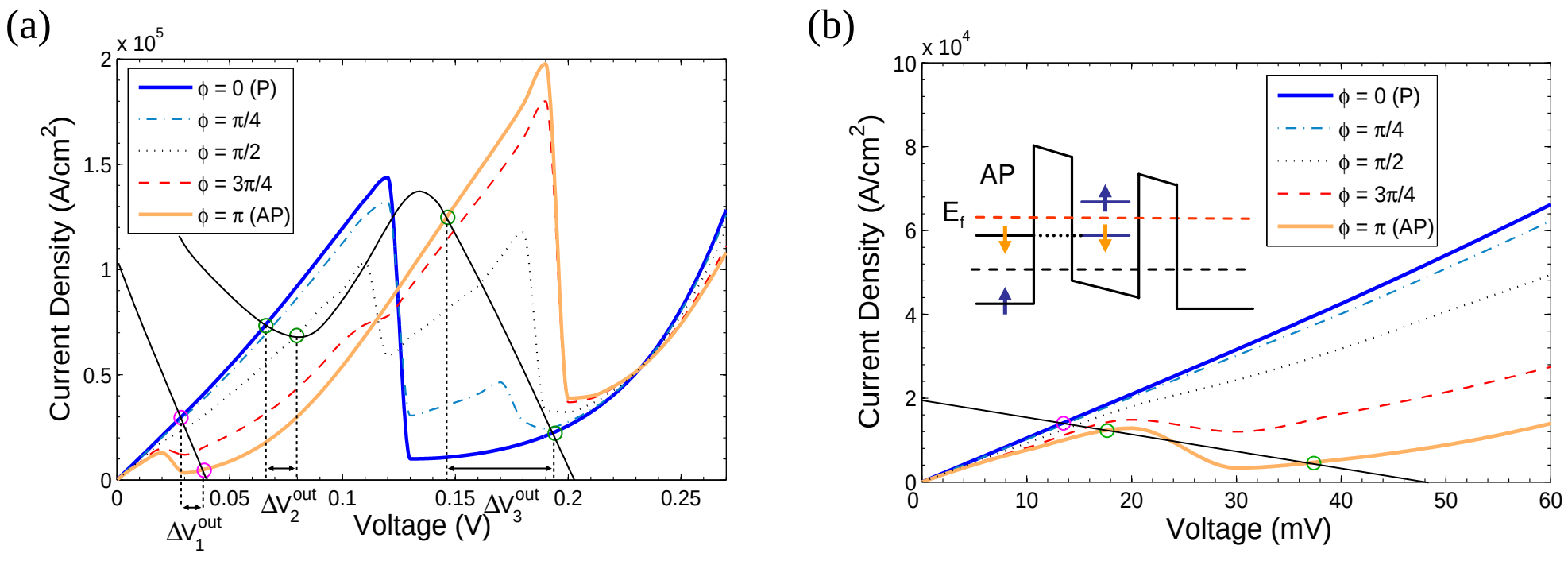

Fig. V.29. (a) Selfconsistent current-voltage characteristics of the ferromagnetic driver-RTD for several relative orientations of the quantum well magnetization (indicated by the angle $\varphi$) at the temperature of $T = 100$ K. The solid black lines show the mirrored IV curve of the nonmagnetic load-RTD for low and high input voltages, respectively; working points are indicated by circles. For these fixed low and high input voltages the output voltages are restricted to the intervals $\Delta V_i^{\rm out}, i = 1, 2, 3$. (b) Blow up of the selfconsistent $IV$ curves displayed in (a) for low voltages. By replacing the load-RTD ba a linear resistance (indicated by the black solid line) allows to perform the monostable-to-bistable transition by flipping the well magnetization from parallel to antiparallel configuration. The working points are indicated by circles. The inset shows schematically the arrangement of the spin-split well levels at the peak voltage in the antiparallel case.

Selfconsistent numerical simulations of the IV-characteristics of a RTD with a ferromagnetic emitter and quantum well made of a GaMnN-like system suggest a strong dependence of the peak current on the relative magnetization orientations (Ertler and Fabian, 2007), as shown in Fig. V.29(a). The magnetization of the emitter lead is assumed to be fixed, whereas the well is "soft", which means that its magnetization direction indicated by the angle $\phi$ can be altered by an external magnetic field. For room temperature operation of the ferromagnetic MOBILE a band splitting of the order of several tens of meV (well exceeding $k_B T$) is needed, regardless by which underlying microscopic mechanism the exchange splitting is actually induced. From the viewpoint of observing the DMR-effect the most prominent feature of the IV-curves in Fig. V.29(a) is that the driver peak current is remarkably reduced when the well magnetization is distorted a little bit, let's say by an angle of a few tens of degrees, from its parallel alignment $\phi = 0$. If we assume that the load peak current is smaller than the driver's one in the initial case of parallel magnetization alignment, we can define a threshold angle $\phi_{\mathrm{th}}$ as the angle of distortion where the load and driver peak currents become equal. The threshold angle depends on the difference of load and driver peak currents at $\phi = 0$, which can be controlled by applying a gate voltage to the nonmagnetic load RTD, influencing thereby its peak current value. When the ferromagnetic MOBILE is driven into the bistable regime by applying a high input voltage the output voltage is low for $\phi < \phi_{\mathrm{th}}$ but suddenly jumps to a high value if $\phi > \phi_{\mathrm{th}}$, effectively realizing DMR. In this way the distortion of the well magnetization above a electrical controllable angle can be converted directly into an digital electrical signal; an effect which again might be utilized for a very fast "readout" of magnetically stored information.

Moreover, by properly tuning the emitter's Fermi energy of such a ferromagnetic RTD one can modify the IV-characteristic in the low voltage regime from ohmic to NDR behavior just by changing the relative magnetization orientations. This is illustrated in Fig. V.29(b) showing a blow-up of the IV-curves of Fig. V.29(a) at low voltages. For the antiparallel case the spin down quasibound state becomes off-resonant before carriers from the emitter can flow via the spin up level, as sketched in the inset of Fig. V.29(b), effectively leading to the appearance of a NDR region in the IV-curve. This interesting behavior might be utilized, for instance, to perform a mono-to-bistable working point transition merely by flipping the well magnetization instead of increasing the input voltage from low to high as done usually. For this purpose, we assume a modified circuit setup, in which the load RTD of Fig. V.28 is replaced by a simple linear load resistance. By drawing the load line diagram (see. Fig. V.29)(b) it becomes evident that we always obtain only one single working point for the parallel alignment but in the case of antiparallel orientation we can acquire two stable points for appropriate high input voltages. This allows again to realize DMR as follows. Let us assume that in the beginning the magnetic layers are antiparallel aligned and that the circuit operates at the high voltage level of the bistable regime. By increasingly distorting the well magnetization out of the antiparallel configuration at some threshold angle the circuit is switched from the bistable to the monostable regime. This results in a sudden swing of the working point to a low voltage state, which allows to detect electrically any disturbance of the antiparallel orientation above a certain threshold. After recovering the antiparallel orientation the circuit will end up in the low voltage state of the bistable regime. In order to reset the device to the initial state a small current pulse can be applied, which induces a voltage swing back to the high voltage state.

As another application one can also think of using this circuit as a memory element, in which the parallel and antiparallel alignments are utilized to store a binary information. The state of the device can then be simply read out by applying a small current pulse. Such a pulse would provoke a voltage swing to the high voltage state if the circuit operates initially in the low voltage state of the antiparallel configuration, but it would remain ineffective for the other cases of parallel magnetization alignment or that the circuit is already in the high voltage state of the antiparallel setup. Hence, after the short current pulse one would always end up with a low or high voltage state corresponding to the parallel or antiparallel magnetization orientation.

This example of a spintronic device with its operation relying on resonant tunneling through paramagnetic or ferromagnetic layers nicely illustrates that the current characteristics can be drastically modulated and engineered just by changing the magnetic characteristics of the device. The intriguing effects rest largely on the physical fact that resonant tunneling is a strongly energy filtering process and, hence, even small energy splittings for the different spin states can have substantial implications on the carriers transmission. This makes resonant tunneling devices very promising for realizing all-semiconductor spintronic concepts, in which the charge current is strongly modulated by the carriers spin state.

## E. Bipolar spintronic devices

Bipolar spintronic devices refer to semiconductor devices in which the spin-polarized transport of both electrons and holes determine the device workings. We will discuss magnetic p-n junctions as magnetic diodes and magnetic bipolar transistors. In order to explain how the magnetic devices operate, we also introduce relavant terminology and scheme of the conventional counterparts, p-n junction and bipolar junction transistors. Although not everything will be derived from the underlying drift-diffusion models, we present qualitative arguments as well as computational scheme of calculating the I-V characteristics of magnetic diodes and magnetic transistors. Spin transport theory of magnetic p-n junctions has been worked out in (Fabian *et al.*, 2002b); see also (Lebedeva and Kuivalainen, 2003).

### E.1 Conventional p-n junctions

We will briefly remind the reader of the physical principles of conventional, spin-unpolarized p-n junctions. Take a semiconductor material whose left side is p-type, doped with $N_a$ acceptors per unit volume, and whose right side is n-type, doped with $N_d$ donors per unit volume. We consider the whole system to be made of the same semiconductor[105] with the intrinsic carrier density $n_i$. At high enough temperatures (usually above 100 K) most of the donors and acceptors are thermally excited, with carriers residing in the semiconductor bands. Below we consider low doping densities for which the Bolztmann statistics limit of the Fermi-Dirac distribution applies, see Sec. II.C. In the p-side there are the $p = N_a$ holes and,

$$n_{0p} = n_i^2/N_a, \tag{V.91}$$

electrons at equilibrium. Holes are the majority carriers, electrons form the minority carriers.[106] The electron equilibrium distribution follows from statistical considerations (Ashcroft and Mermin, 1976; Tiwari, 1992). In the n-region, we have $N_d$ electrons and,

$$p_{0n} = n_i^2/N_d, \tag{V.92}$$

[105] Such a system is called a homojunction. If the two regions were of different underlying material, we speak of a heterojunction.

[106] Not to be confused with the majority and minority carriers in ferromagnetic conductors, in which these terms refer to the carriers of higher and lower spin density, respectively.

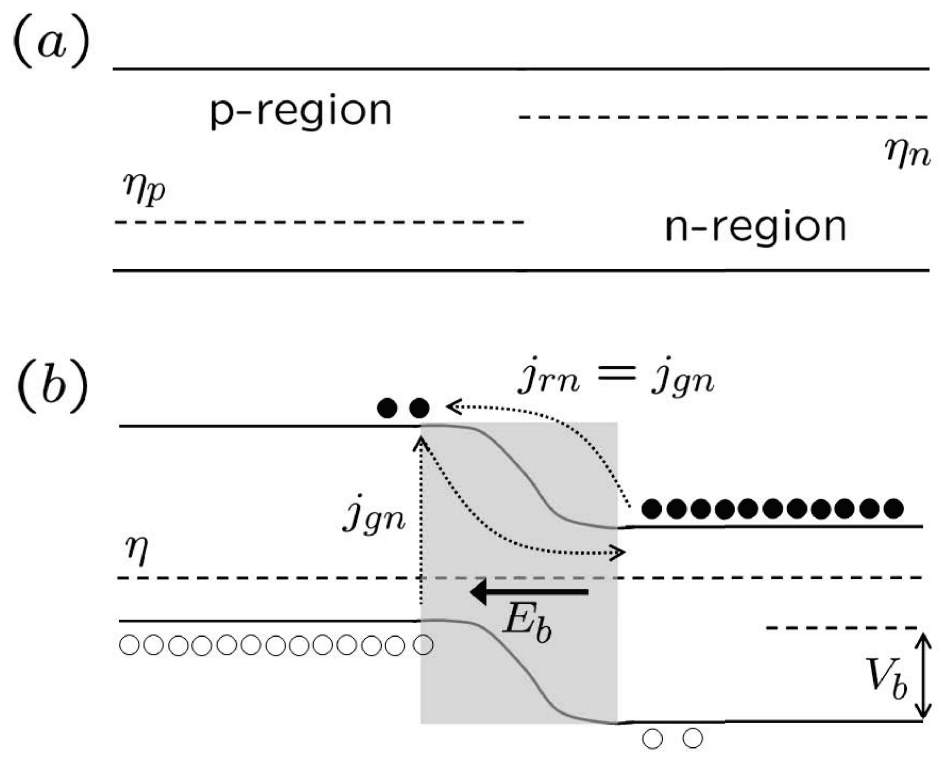

Fig. V.30. (a) Two separate regions, one of p-type, the other of n-type, of the same semiconductor material have different chemical potentials $\eta_p$ and $\eta_n$, respectively. The chemical potentials are shown relative to the conduction and valence bands. (b) If electrons are allowed to flow between the two regions, a common thermodynamic equilibrium is established with a uniform chemical potential $\eta$. As a result of the charge rearrangement, a space-charge region (shaded) is formed, in which a macroscopic, so-called built-in field $E_b$ is present. The initial difference of the chemical potentials is now reflected in the built-in voltage drop, $V_b$, across the depletion region. The electron generation current, $j_{gn}$, equals the electron recombination current, $j_{rn}$ that flows in the opposite direction. Filled circles indicate electrons, empty one holes.

holes. Electrons are the majority, holes are the minority carriers here. Considered separately, the two regions would have different chemical potentials, $\eta_p$ and $\eta_n$, as seen in Fig. V.30 a. If the regions are connected and electrons can flow through the contact, a single chemical potential, $\eta$ is established, as in Fig. V.30 b. To establish the uniform chemical potential, electrons flow from the n-region to the p-region. This leaves a positive, uncompensated charge in the $n$-region, close to the contact, as well as a negative charge in the $p$-region. As a result, a built-in electric field is induced in the contact. The contact region, in which the field exists, is called the depletion layer, or the space-charge region, since the built-in field drives the electrons (holes) into the n-region (p-region). Since the carrier densities are small in the depletion layer, there is a space charge due to the uncompensated donors and electrons. It is the physics of the space-charge region that gives rise to the interesting properties of bipolar junction diodes and transistors.

The built-in electric field is an equilibrium field, driving no macroscopic current. The corresponding electrostatic potential drop, called the built-in potential, is not an electromotive force (emf). No current would flow if we closed the circuit. In order for a current to flow in a p-n junction, we need to create a difference in the chemical potentials at the two ends of the junctions. This could be done by attaching battery contacts onto the two regions, or by shining light on the junction. The latter method, which is the principle behind solar cells, results from the existence of the built-in field. Excess electron-hole pairs from the optical excitation, arriving or generated at the space-charge region, are separated by the built-in field: electrons go to the n-side, holes to the p-side, resulting in an electrical current flowing in the direction of holes, in a closed circuit.

Here we consider the usual current generation by attaching battery contacts on the junction. How does electrical current flow in the junction? We will first give a qualitative estimate of the current, then solve the problem using a drift-diffusion model for p-n junctions. Let us start with a qualitative account. In order to understand the currents that flow under applied bias $V$, we need to understand first the currents that flow at $V = 0$. There are two such currents. Since the physics is identical for electrons and holes, we only look at electrons. If there is a thermal excitation of an electron-hole pair in or near the space-charge region, the electrons will be pushed to the n-region and holes to the p-region, by the built-in field. This gives the so-called generation current $j_{gn}$, the name indicating that the current originates from thermal generation of carriers. As there should be no macroscopic equilibrium current, there needs to be an opposite flowing current. Indeed, there is the so-called recombination current, $j_{rn}$, due to the thermal activation of the electrons from the n-region to the p-region. The terminology comes from the ultimate fate of the thermally activated electrons: they recombine in the p-region with holes. In equilibrium, $j_{gn} = j_{rn}$. This equilibrium currents are indicated in Fig. V.30 b.

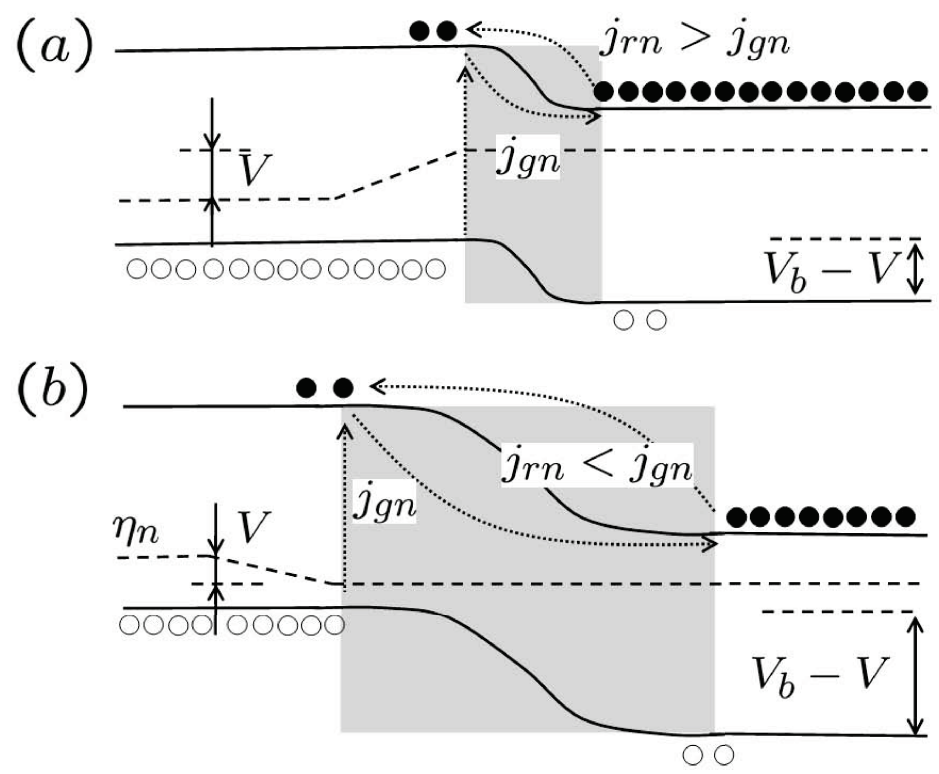

Fig. V.31. (a) Forward bias. If positive voltage, $V > 0$, is applied to the p-region (say, the n-region is grounded), the depletion layer gets narrower and the band bending smaller. The electron recombination current is greater than the electron generation current (similarly for holes), resulting in a flow of electrical current. The electron chemical potential, $\eta_n$, is indicated. The potential drops mainly in the p-region. (b) Reverse bias. If negative voltage, $V < 0$, is applied to the p-region, the depletion layer widens and the band bending increases. The electron recombination current decreases well below the generation current. Only a small electrical current, of the magnitude essentially of the generation current flows.

**Qualitative picture of the currents in a p-n junction.** If there is a bias, $V$, applied across the junction, a net macroscopic current will flow. As the depletion layer is the region with the highest resistance, all the bias drops there. If $V > 0$, in the convention that the higher voltage is applied to the p-region, electrons flow from the n- to the p-region, and the band bending in the depletion layer will decrease, as shown in Fig. V.31 a. This case is called the *forward bias*, since it results in a relatively large current. If $V < 0$, the band bending increases, as shown in Fig. V.31. This case is called the *reverse bias*, since a very small current flows through the junction. The asymmetry between the forward and reverse bias gives the p-n junction its current rectification property.

If the band bending is modified from its equilibrium value, the recombination current has to change. Indeed, the recombination current depends on the thermal activation over the barrier, since the electrons have to be activated from the n- to the p-region:

$$j_{rn} = KN_d e^{q(-V_b+V)/k_BT}. \tag{V.93}$$

Here we indicate that the current is proportional to both the available density of electrons, $N_d$, as well as to the rate of thermal transfer, $\exp[q(-V_b + V)/k_BT]$, where $q$ is the magnitude of the electron charge, $q = |e|$. The higher is the built-in field $V_b$, the smaller is the transfer; opposite holds for the applied bias, in the adapted sign convention.

On the contrary, the generation current does *not* depend on the applied bias. This is crucial to the rectification property of the junction. The reason why $j_{gn}$ does not depend on $V$ is that the generation current does not depend on the band bending; it depends on the energy gap,[107] which does not change with the bias. The generation current also depends on the density of available electrons close to the depletion layer (as we will see later from the drift-diffusion model). The rate of the electron transfer from the p-region to the n-region due to the generation depends only on the rate of the electron generation. The rate does not depend on how fast the electron moves in the depletion layer due to the presence of the built-in field, since the field is so large, that every electron that enters the layer is swept to the n-region. For one electron thermally excited in the p-region close to the depletion layer, one electron enters the n-region, per unit time. We thus require that in equilibrium, in which no net currents flow,

$$j_{gn} = -j_{rn}(V = 0) = -KN_d e^{-qV_b/k_BT}. \tag{V.94}$$

The electron current flowing as a result of an applied bias is,

$$j_n = j_{gn} + j_{rn} = j_{gn}\left(e^{qV/k_BT} - 1\right). \tag{V.95}$$

An analogous expression can be written for holes:

$$j_p = j_{gp} + j_{rp} = j_{gp}\left(e^{qV/k_BT} - 1\right). \tag{V.96}$$

Putting electrons and holes together, we finally obtain the I-V characteristic of a p-n junction in the form,

$$j = j_g\left(e^{qV/k_BT} - 1\right), \tag{V.97}$$

where $j_g = j_{gn} + j_{gp}$. This equation describes the rectification of a p-n junction diode. In the forward bias, when $qV \gtrsim k_BT$, the current exponentially increases; in the reverse bias, when $V < 0$, the current drops to $j \approx -j_g$.

**Diffusion model of Shockley for a p-n junction.** In order to calculate the generation current, namely, the value of the unknown parameter $K$, we need to consider a more quantitative model. An intuitive model, which turned out extremely practical, was proposed by Shockley (Tiwari, 1992). The model is based on the following assumptions: (i) The transport of the minority carriers away from the depletion layer is due to diffusion. The drift can be neglected. (ii)

[107] The generation current is limited by the thermal excitations of the carriers, electron-hole pairs, in one region only. Thermal forces do not have to act further to bring, say, the electrons across the depletion layer. All the electrons generated thermally at the depletion layer are swept by the electric built-in field.

The chemical potential (or, rather, the quasichemical potential since we are not in equilibrium) is constant across the space-charge region (depletion layer). The chemical potential, which is different for electrons and holes in the depletion layer, drops in the region in which the carriers are in minority; see Fig. V.31. This assumption suggests that the carriers are in a sort of quasiequilibrium inside the depletion layer. As we will see, assumption (ii) gives boundary conditions for the diffusion model (i). Related to (ii) is the additional assumption (iii), that the electron-hole recombination inside the depletion layer is inhibited. Electrons and holes are kept at different chemical potentials throughout the layer. This assumption gives the continuity of the electron and hole currents across the depletion layer, allowing us to connect the currents at the two regions. This assumption, unlike (i) and (ii), may be relaxed while still allowing analytical solutions. We stress that the Shockley model works only at low biases (say, below the voltages corresponding to the energy band gap), at which we are in the regime of low carrier injection. This means that the density of the injected carriers, say electrons in the p-region, is much below the corresponding equilibrium density of electrons there. Fortunately, most useful properties of bipolar junction devices (diodes and transistors) are in this regime.

Let us formulate the transport model based on Shockley's conditions for our p-n junction. We will calculate the electron current and present the expression for the hole current by analogy. The electron transport in the p-region is governed, due to assumption (i), by the diffusion equation:

$$\frac{d^2\delta n}{dx^2} = \frac{\delta n}{L_{np}^2}, \tag{V.98}$$

where $\delta n = n - n_{0p}$ is the nonequilibrium electron density and $L_{np}$ is the electron diffusion length in the p-region. The diffusion length is limited by electron-hole recombination (typically nanoseconds in GaAs) in the region, and can be micrometers long.[108] There are two boundary conditions: one, at $x \to -\infty$, at the far left edge of the p-region, the other one at $x = 0$, which we take to be the interface of the depletion layer and the bulk p-region. Usually one assumes ohmic contacts to the bulk regions, which means,

$$\delta n(-\infty) = 0, \tag{V.99}$$

expressing the fact that nonequilibrium carrier densities are absent in ohmic contacts.

The second boundary condition can be obtained from assumption (ii). If the chemical potential is constant across the depletion layer, the Boltzmann statistics gives

$$\delta n(0) = n_{0p}(e^{qV/k_BT} - 1), \tag{V.100}$$

since the separation of the chemical potential from the conduction band at $x = 0$ changes by $qV$ upon application of bias. The higher is the forward bias, the exponentially more minority carriers appear at the contact to the depletion layer. These nonequilibrium carriers diffuse towards the bulk, disappearing at the distance of $L_{np}$. On the other hand, at reverse biases, the number of minority carriers decreases below the equilibrium value.

The solution to the diffusion equation with the above specified boundary conditions is,

$$\delta n(x) = \delta n(0)e^{x/L_{np}}. \tag{V.101}$$

[108] In indirect band-gap semiconductors such as silicon the recombination is less effective, since phonons must assist to conserve momentum. The carrier diffusion length is then much longer.

From this density profile we can calculate the electric current due to the diffusion of electrons in the p-region:

$$j_n(x) = qD_{np}\frac{\delta n(x)}{dx}, \tag{V.102}$$

where $D_{np}$ is the diffusion coefficient of electrons in the p-region. We obtain,

$$j_n(x) = \frac{qD_{np}}{L_{np}}\delta n(0)e^{x/L_{np}}. \tag{V.103}$$

In particular, the electron current at the contact with the depletion layer is,

$$j_n(0) = \frac{qD_{np}}{L_{np}}n_{0p}\left(e^{qV/k_BT} - 1\right). \tag{V.104}$$

Due to the assumption (iii), this is also the electron current that appears at the contact of the depletion layer with the n-region. This current would be more difficult to calculate directly since it is carried by the majority carriers for which both diffusion and drift are important. By analogy, the hole current across the depletion layer is,

$$j_p(0) = \frac{qD_{pn}}{L_{pn}}p_{0n}\left(e^{qV/k_BT} - 1\right), \tag{V.105}$$

where the notation is symmetric to the electron case; for example, $L_{pn}$ is the diffusion length of holes in the n-region.

The total electrical current at any point in the junction is the sum of the electron and hole currents at that point. The current continuity guarantees that we can indeed calculate the current at any point, at our convenience. If we choose, for example, point $x = 0$, the Shockley assumptions give us,

$$j = j_n(0) + j_p(0), \tag{V.106}$$

which leads to

$$j = \left(\frac{qD_{np}}{L_{np}}n_{0p} + \frac{qD_{pn}}{L_{pn}}p_{0n}\right)\left(e^{qV/k_BT} - 1\right). \tag{V.107}$$

We see that the Shockley model gives us the same rectification characteristic as that obtained from our qualitative model. We can now identify the generation currents as,

$$j_{gn} = \frac{qD_{np}}{L_{np}}n_{0p}, \tag{V.108}$$

$$j_{gp} = \frac{qD_{pn}}{L_{pn}}p_{0n}. \tag{V.109}$$

The rectification behavior of the diode is then due to the nonequilibrium densities of the minority carriers. This is an important message which will carry through the rest of this chapter when we discuss magnetic diodes and transistors.

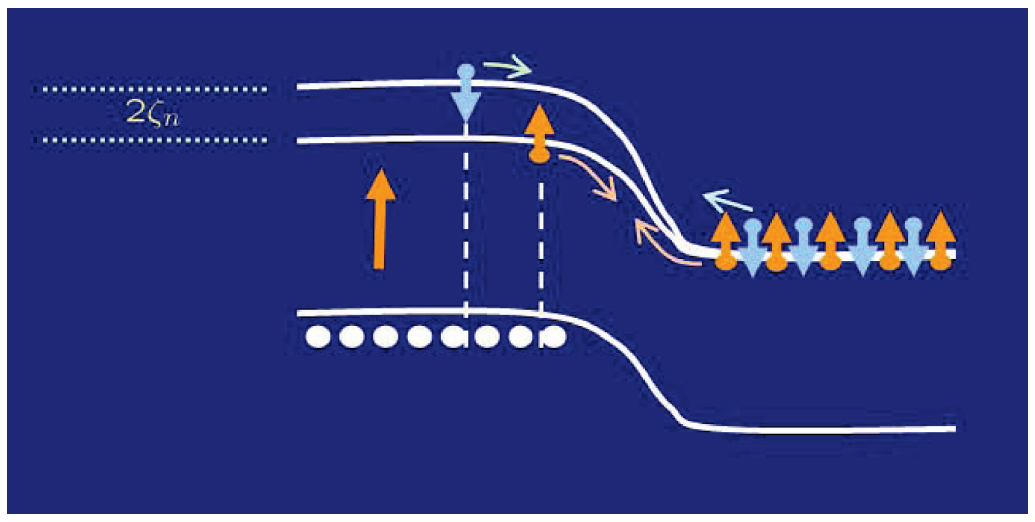

Fig. V.32. Scheme of a magnetic diode with the magnetic p-region. The n-region is nonmagnetic. The spin splitting of the magnetic region is $2\zeta_n$. Generation and recombination currents are indicated.

### E.2 Magnetic diode

Consider now a straightforward extension of the above picture of bipolar junction diodes including magnetic semiconductors (Žutić *et al.*, 2002; Fabian *et al.*, 2002b); the resulting magnetic diode is a generalization of the concept of spin-polarized diodes (Žutić *et al.*, 2001b,a, 2003). We can have either one, or both regions magnetic,[109] while there is a source of nonequilibrium spin in the junction. The source can be either electrical or optical. For simplicity we consider only one type of magnetic diodes—those with magnetic p-region. We further assume that holes are spin unpolarized, which is a reasonable approximation since holes, due to their strong spin-orbit coupling, usually lose their nonequilibrium spins very fast in comparison to electrons. However, if needed, the model can be extended to include hole spin polarization (Fabian *et al.*, 2002b; Žutić *et al.*, 2006b). The description of bipolar junction diodes (Žutić *et al.*, 2002; Fabian *et al.*, 2002b) could also be relevant to various manganite-based junctions (Li *et al.*, 2006c,b; Cai and LI, 2005; Nakagawa *et al.*, 2005) and novel class of ferromagnetic semiconductors in which carrier polarity can be changed by impurity doping (Motomitsu *et al.*, 2005).

The scheme of our magnetic diode is shown in Fig. V.32. The p-region is magnetic with the spin-splitting of the conduction band to be $2\zeta_n$. The generation current is different for spin up and for spin down electrons, similarly for the recombination current. This current can be controlled by magnetic field, modifying the splitting of the electron bands in the magnetic region, or by introducing nonequilibrium spins in the n-region. We will use a qualitative model, developed in the previous section, to calculate the current through a magnetic diode and discuss its ramifications; general approach to calculate spin-polarized currents through arrays of magnetic p-n junctions will be given later in the section on magnetic bipolar transistors.

The electron recombination current in the magnetic diode of Fig. V.32 is $j_{rn\uparrow}+j_{rn\downarrow}$. Suppose the equilibrium spin polarization of electrons in the p-region is $P_0$, and the nonequilibrium spin polarization in the n-region, due to a spin source, is $\delta P$. Then the spin-up and spin-down electron

[109] We do not address the origin of the magnetism: the semiconductor can be either ferromagnetic, in which case the carrier band spin splitting is due to exchange coupling, while in case of a general dilute magnetic semiconductor the spin splitting is due to giant g-factors and an applied magnetic field, giving giant Zeeman splitting.

currents are,

$$j_{rn\uparrow} = \frac{1}{2}K(1+P_0)(1+\delta P)N_d e^{q(-V_b+V)/k_BT}, \tag{V.110}$$

$$j_{rn\downarrow} = \frac{1}{2}K(1-P_0)(1-\delta P)N_d e^{q(-V_b+V)/k_BT}. \tag{V.111}$$

The information about $\zeta_n$ is encoded in $P_0$. The above equations have simple interpretations. The number of spin up electrons available for thermal excitation from the n- to the p-region is proportional to $1+\delta P$. On the other hand, the spin up electrons have a lower barrier to cross, giving a factor of $\exp(q\zeta_n/k_BT)$ for the thermal excitation. The Boltzmann statistics says that this factor is proportional to $1+P_0$.

The generation current does not depend on nonequilibrium conditions. This means that,

$$j_{gn\uparrow} = -j_{rn\uparrow}(V=0,\delta P=0) = -\frac{1}{2}K(1+P_0)N_d e^{qV_b/k_BT}, \tag{V.112}$$

$$j_{gn\uparrow} = -j_{rn\uparrow}(V=0,\delta P=0) = -\frac{1}{2}K(1+P_0)N_d e^{qV_b/k_BT}. \tag{V.113}$$

Summing up all the contributions to the electron current we obtain,

$$j_e = j_{gn}\left[e^{qV/k_BT}(1+P_0\delta P)-1\right]. \tag{V.114}$$

The above equation expresses spin-charge coupling in magnetic p-n junctions. The proximity of an equilibrium and nonequilibrium spins gives rise to modifications of the I-V characteristis of the junction. For a parallel orientation of the spins, the current is enhanced; for an antiparallel orientation, the current is reduced. The relative change of the current with respect to the orientation of the equilibrium and nonequilibrium spin gives rise to a giant magnetoresistive effect in magnetic diodes. Interestingly, the effect is also present at zero bias. Even for $V=0$ electric current flows due to the spin-charge coupling. The current is either positive or negative, depending on the sign of the product of $P_0\delta P$. This phenomenon is called spin-voltaic effect (Žutić *et al.*, 2002; Fabian *et al.*, 2002b; Žutić *et al.*, 2003; Žutić and Fabian, 2003), producing an emf from nonequilibrium spin, similarly to the Silsbee-Johnson spin-charge coupling of Sec. II.D.9. [110]

The spin-voltaic effect is illustrated in Fig. V.33. In simple terms, we can pose the question of how can we make the current in the magnetic diode larger or smaller. Looking at the potential barrier for going from the n- to the p-region, we immediately see that by making the spin in the n-region nonequilibrium, pointing up, electrons will have a lower barrier to cross, increasing the recombination current. On the other hand, introducing more spin down electrons in the n-region, electrons have to climb a higher barrier, reducing the recombination current. Since the generation current is not influenced by the nonequilibrium properties, the current for the parallel orientation will be larger than thar for an antiparallel orientation of the equilibrium and nonequilibrium spins.

The spin-voltaic effect have been observed in magnetic p-n junction diodes based on GaMnAs (Chen *et al.*, 2006). Fig. V.34 shows the experimental setup. The magnetic diode is formed by n-GaAs and the ferromagnetic p-GaMnAs. The spin source is the iron electrode connected via the Schottky barrier to the n-region. Spin-charge coupling results from the existence of nonequilibrium spin, due to the spin injection from the iron electrode, at the depletion layer with the p-region.

[110] There are various proposals (Žutić *et al.*, 2004) which give emf from nonequilibrium spin, often referred to as spin(-polarized) pumps, cells, or batteries. They even do not need to have a magnetic element and can span a range of structures from nonmagnetic p-n junctions (Žutić *et al.*, 2001a) and semiconductors sandwiched between the two conducting leads (Wu and Ahn, 2006) to double quantum dots (Chia *et al.*, 2006).

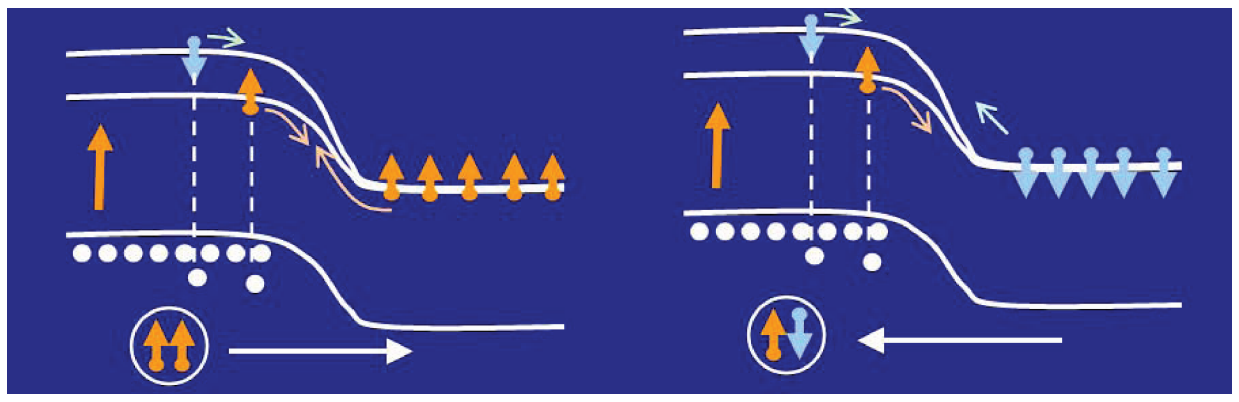

Fig. V.33. Spin-voltaic effect in a magnetic diode. The p-region is magnetic, while the nonmagnetic $n$-region contains nonequilibrium spin. In the case that the equilibrium spin is parallel to the nonequilibrium one, electrons tend to flow left. If the two spins are antiparallel, electrons flow right. Electrical current flows even in the absence of applied bias, for $V = 0$.

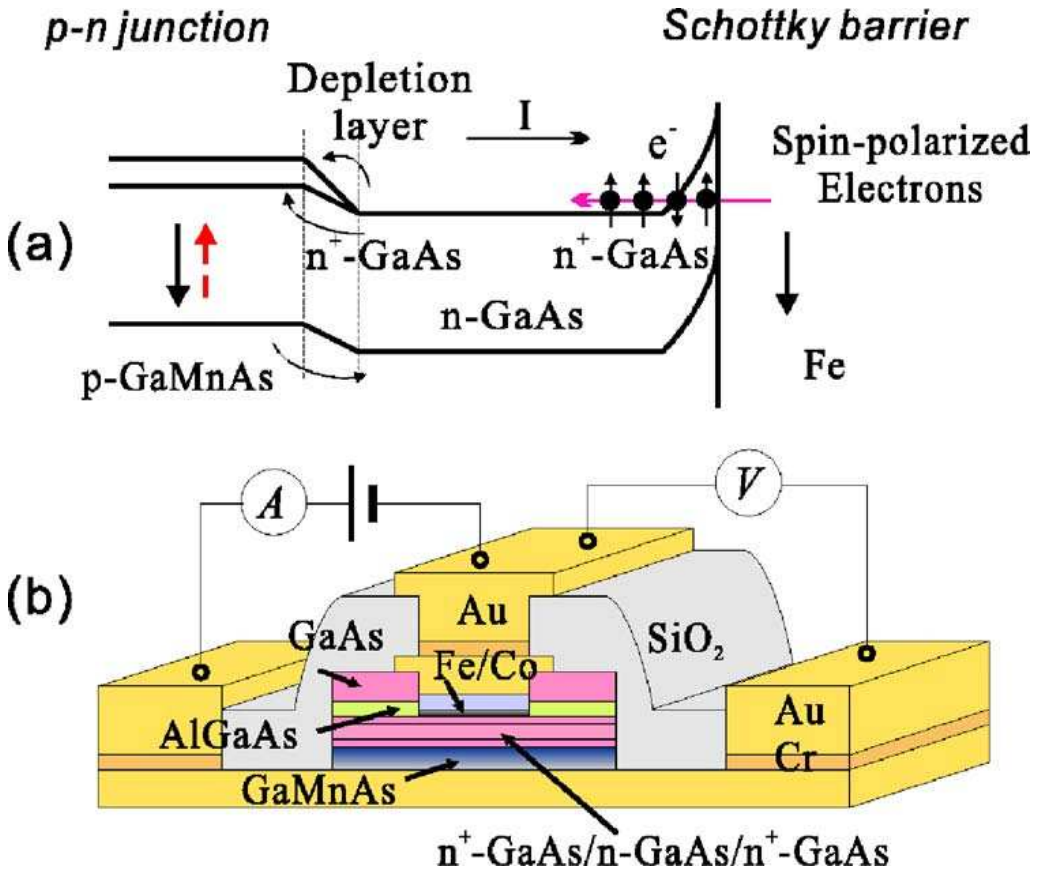

Fig. V.34. (a) Schematic band diagram of the device structure. The p-region is ferromagnetic, based on GaMnAs. The n-region is formed by GaAs, which makes a Schottky barrier contact with an iron spin injecting electrode. (b) The actual experimental setup. The Co/Fe layer is included to magnetically bias the Fe layer. Reprinted figure with permission from P. Chen *et al.*, *Physical Review B* **74** 241302 (R) (2006). Copyright 2006 by the American Physical Society.

The results of the experiment are shown in Fig. V.35. The iron electrode has higher coercive field, due to the magnetic biasing with the Fe/Co layer, see Fig. V.34, allowing both parallel and antiparallel orientations of the two ferromagnets at a large window of external magnetic fields. As the traces of the resistance versus magnetic field, in different directions of the field indicate, there is spin-charge coupling due to the injected nonequilibrium spin in GaAs and the equilibrium magnetization in GaMnAs.[111]

[111]Fig. V.35 shows low resistance for antiparallel and high resistance for parallel magnetizations. The actual magneti-

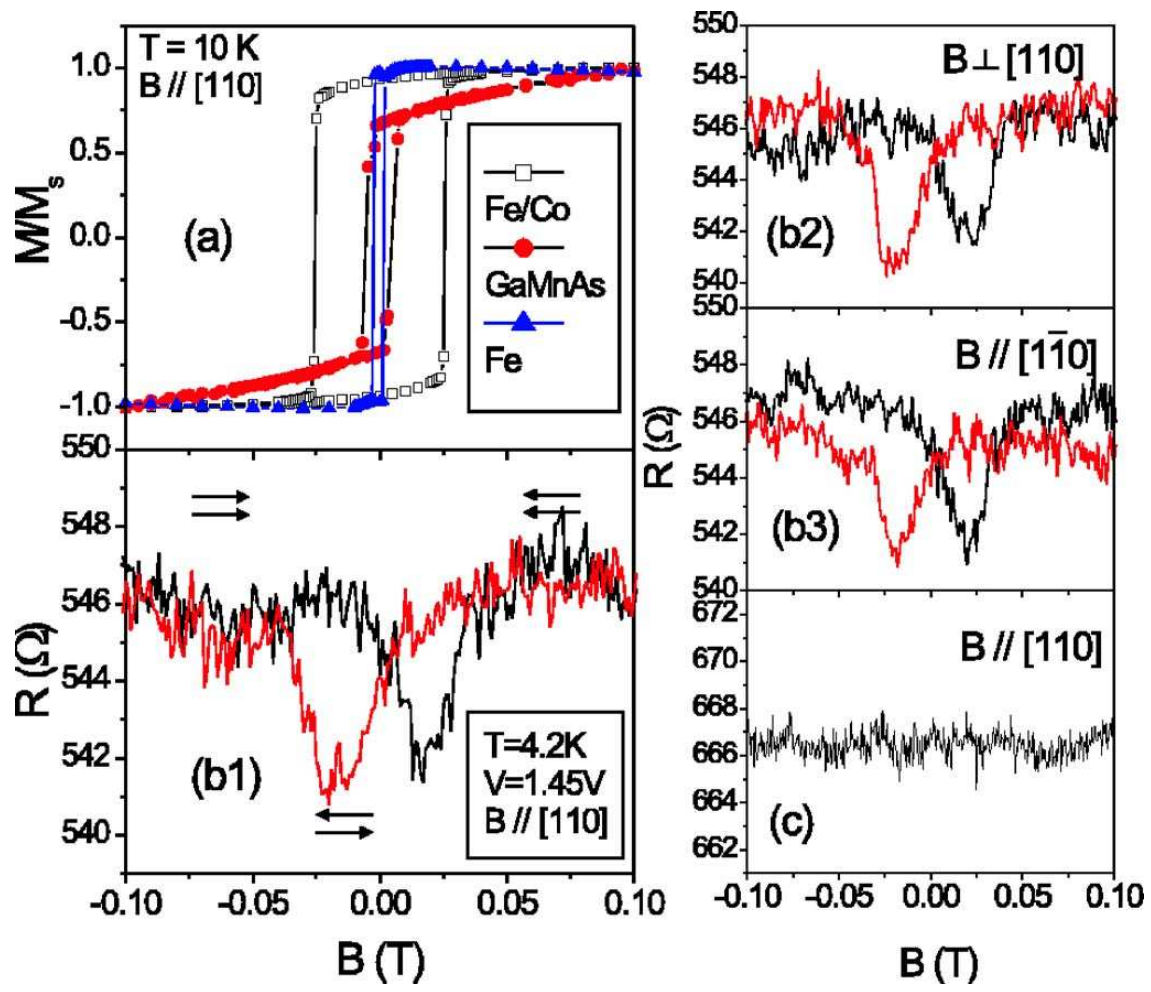

Fig. V.35. (a) SQUID measurements of the magnetizations of various ferromagnetic layers in the device. (b1) Magnetoresistance for the [110], an in-plane, orientation of the magnetic field. Arrows indicate orientations of the magnetizations of the GaMnAs and Fe electrode. (b1) and (b2) show the same, for the indicated orientation of the magnetic field. (c) Reference data, with no magnetically biasing Co layer. Without this layer, GaMnAs and Fe in the structure have similar coercivities, so antiparallel orientation is not realized. The resistance is then constant. Reprinted figure with permission from P. Chen *et al.*, *Physical Review B* **74** 241302 (R) (2006). Copyright 2006 by the American Physical Society.

Similar findings of the spin-voltaic effect were observed in p-InGaAs/n-AlGaAs spin-polarized p-n junctions (Kondo *et al.*, 2006) in applied magnetic fields, using the g-factor differences of the n- and p-regions: AlGaAs composition was selected to have $g \approx 0$, while InGaAs had $g \approx -1.9$. In effect the p-region had a finite spin splitting, while the n-region could be considered to have zero equilibrium spin. The source spin was induced by optical spin orientation. Spin-polarized transort in ferromagnetic p-n junctions based on p-InMnAs and n-InAs has been investigated experimentally in (May and Wessels, 2005), in manganite based junctions in (Cai and LI, 2005).

Finally we address briefly the question of spin injection across a magnetic p-n junction. Suppose we have a magnetic n-side and wish to inject the spin-polarized electrons into the p-side. It turns out that at low biases there will be no spin accumulation—spin injection is inefficient in transforming spin-polarized majority electrons into spin-polarized minority electrons. The reason is rather simple: although there are more, say, spin up electrons than spin down in the n-region, due to the equilibrium magnetization, the activation barrier for spin up is greater than for spin down electrons in crossing over to the p-region. Both effect cancel each other, so that the current flowing across the depletion layer is spin unpolarized (Fabian *et al.*, 2002b). Spin injection, as well as spin extraction, is possible only in a high injection limit, in which first a zation arrangement depends on whether the majority or minority spins are injected across the Schottky barrier, as well as on the effective g-factor of electrons in GaMnAs.

nonequilibrium spin accumulation in the magnetic side is set up (Žutić *et al.*, 2002).

### E.3 Magnetic bipolar transistor

The structure of the magnetic bipolar junction transistor has been proposed by Fabian *et al.* (2002a, 2004); Fabian and Žutić (2004a,b, 2005), as well as by Flatté *et al.* (2003) and Lebedeva and Kuivalainen (2003). The transistor is identical to the conventional (nonmagnetic, spin-unpolarized) bipolar junction transistor of Shockley (Tiwari, 1992). Two magnetic p-n junctions are connected in series and a third contact is added to the middle region called the base. The transistor can have equilibrium spin, coming from the magnetization or the spin splitting of the carrier bands in any region. In addition, there can be a source of a nonequilibrium spin, such as coming from electrical spin injection or optical spin orientation (Žutić *et al.*, 2004). Although the equilibrium spin is enough to allow for a magnetic control of amplification—what is termed *magnetoamplification*—see also Ref. (Lebedeva and Kuivalainen, 2003; Fabian and Žutić, 2004a), fascinating new effects appear from the interaction of the equilibrium and nonequilibrium spin, when they are in an electrical contact. This interaction, which is a realization of the Silsbee-Johnson spin-charge coupling, see Sec. II.D.9, gives rise to *giant magnetoamplification*, the control of current amplification by the relative orientation of the equilibrium and nonequilibrium spins (Fabian *et al.*, 2002a; Fabian and Žutić, 2004a).

Important for spintronics application is transfer of spin within the transistor. There are two possibilities. One is the spin injection from a source or a nonequilibrium spin in the emitter, another is the spin injection from the equilibrium spin in a magnetic base. The former is similar to spin injection through a magnetic diode (Žutić *et al.*, 2002), except that we now have two diodes in series, with opposite polarities (in the useful configuration of forward active regime). The latter, less trivial spin injection process, has no counterpart in the diode physics and appears due to a nonequilibrium spin accumulation in the base, resulting from the electrical injection of initially unpolarized carriers, into the base. The carriers polarize in the base by spin relaxation processes. We call this process *intrinsic* spin injection.

Conventional bipolar junction transistors have many applications in information technology. The junction transistors are faster than field effect transistors; the disadvantage is the vertical design of junction transistors. They are used for high-speed digital circuits in mobile communication systems, for example, in small signal amplification devices, in high frequency analog circuits, as well as for bipolar complementary metal-oxide-semiconductor (CMOS) technologies. Bipolar transistors form about 20% of the integrated circuit market (the rest is MOSFETs). Their proposed magnetic variants have chance to greatly enhance functionalities of the present technologies, with applications such as reprogrammable logic (Black and Das, 2000).[112]

### E.4 Conventional bipolar junction transistor

Bipolar junction transistor is a three terminal semiconductor structure comprising two p-n junctions in series. While p-n junctions rectify current, junction transistors amplify current. In addition, they are used for fast logic devices, as they allow ON and OFF operations, depending on the configuration of the bias drops across the two p-n junctions. To be specific, consider an npn transistor sketched in Fig. V.36. The n-doped emitter serves to emit electrons; usually it is highly doped. The collector, which collects electrons, is usually weakly doped. Finally, the base makes a third contact (bipolar junction transistor is a three-terminal device) which controls the current between the emitter and the collector. In the usual operating mode, called *forward active*, the base-emitter junction is forward biased, $V_{be} > 0$, while the base-collector junction is reverse biased, $V_{bc} < 0$. This is the situation shown in Fig. V.36. The base-emitter barrier is lower than the equilibrium intrinsic built-in barrier of the junction, pushing electrons from the emitter to the base. The base-collector barrier is higher than the intrinsic value, allowing the electrons that reach the barrier from the base region to be swept, without the need for thermal activation, to the collector.

[112]Reprogrammable or reconfigurable logic is a term describing logic devices which can be roconfigured by, say, magnetic field. One set of magnetic transistors can work as an AND logic element for one specific configuration of the magnetizations, and as an OR element for another configuration.

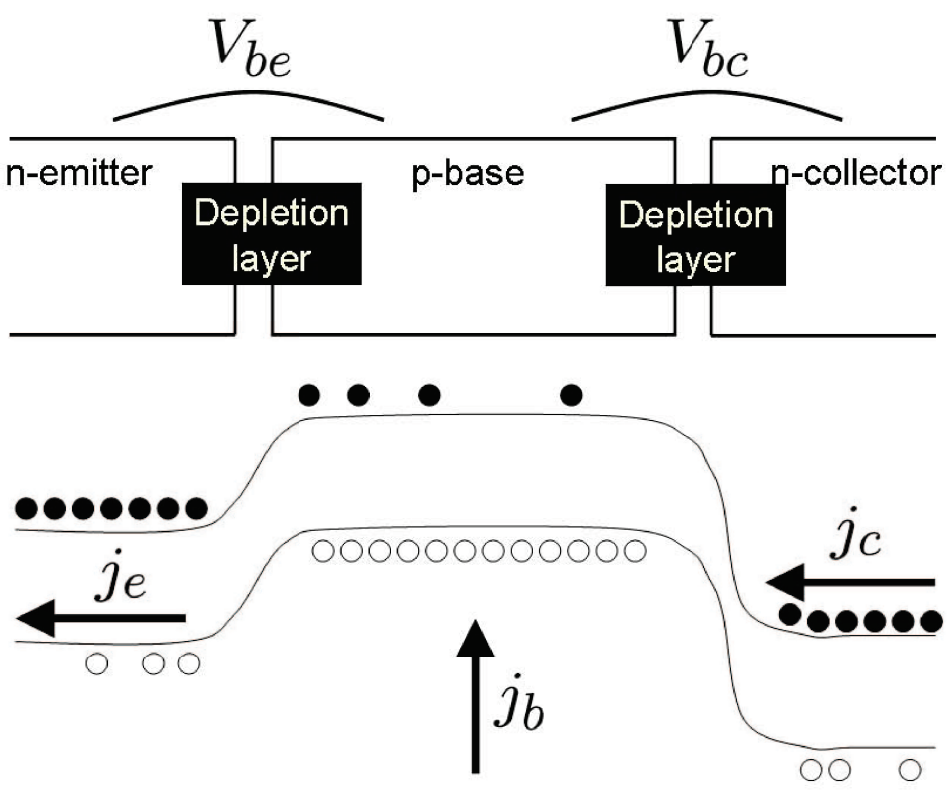

Fig. V.36. Scheme of an npn bipolar junction transistor in the forward active regime. The lower figure shows the conduction and valence bands, as well as the direction of the electric current in the three regions.

The electric current flowing in the three regions is indicated in Fig. V.37. The current in the emitter, $j_e$, flows to the left, reflecting the flow of electrons to the right. Similarly for the collector current, $j_c$. In the base the electrons which recombine with holes leave the region, so the base electric current, $j_b$, flows into the base. The current gain, or amplification factor, of the transistor is the ratio of the collector and emitter currents:

$$\beta = \frac{j_c}{j_b}. \tag{V.115}$$

Typically, this ratio is several hundreds. This means that for each electron that recombines in the base and forms the base current, there are hundreds flowing into the collector. Without the base electrode there would be no current flowing through the transistor in the forward active region. Indeed, the electrons recombining in the base would stay there and induce electric force opposing further injection of electrons from the emitter.

In addition to the forward active regime, there are three more modes in which the bipolar junction transistor can operate. All four modes, summarized in Tab. V.2, are: (i) forward active, defined by the forward emitter-base, $V_{be} > 0$ and reverse collector base, $V_{bc} < 0$, junctions, as discussed above; (ii) reverse active, with reverse polarities: $V_{be} < 0$ and $V_{bc} > 0$; (iii) saturation, with both junctions forward biased, $V_{be} > 0$ and $V_{bc} > 0$, and, finally, (iv) cutoff, with both junctions reverse biased, $V_{be} < 0$ and $V_{bc} < 0$. In the forward active mode the transistor acts as an amplifier. Due to its design, with high emitter and low collector doping, the reverse active mode has a much smaller amplification factor than the forward active mode. The saturation and cutoff modes are used in logic operations as ON and OFF states, respectively. We will see that magnetic bipolar transistors, while providing additional functionalities in these conventional modes, offer another regime due to spin-charge coupling.

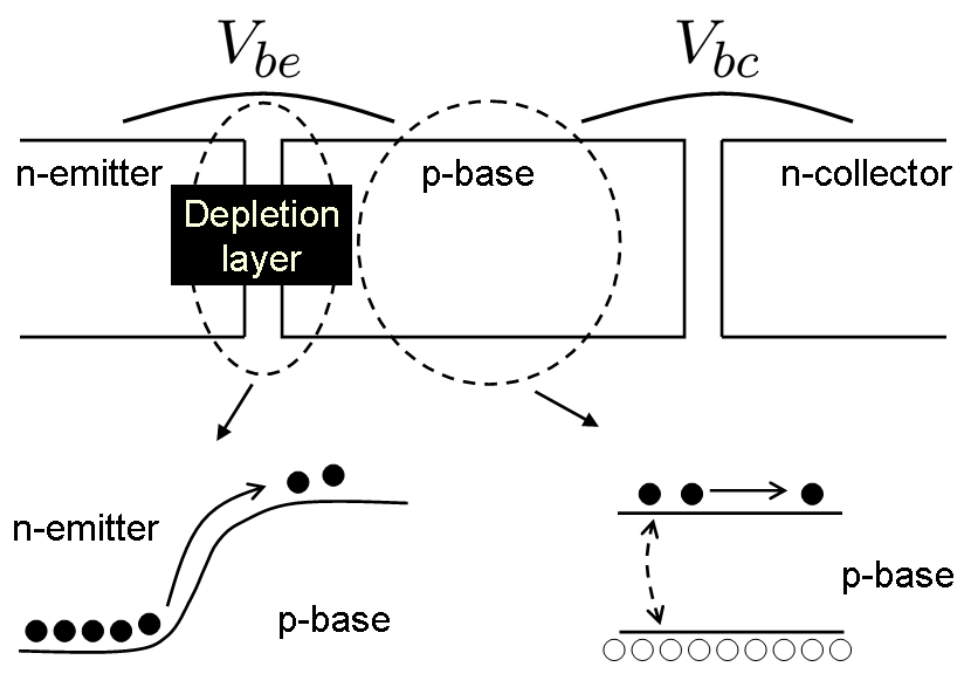

Fig. V.37. Currents in a conventional npn bipolar junction transistor. Electrons from the emitter easily overcome the activation barrier to the base in the forward biased base-emitter junction. Once in the base, the electrons diffuse towards the collector; some stay in the base because of recombination with holes.

Tab. V.2. Operational modes of bipolar junction transistors (BJT) and magnetic bipolar transistors (MBT). Forward (F) and reverse (R) bias means positive and negative voltage over the indicated region, respectively. Symbols MA and GMA stand for magnetoamplification and giant magnetoamplification, while ON and OFF are modes of small and large resistance, respectively; SPSW stands for spin switch.

| mode | $V_{be}$ | $V_{bc}$ | BJT | MBT |
|---|---|---|---|---|
| forward active | F | R | amplification | MA, GMA |
| reverse active | R | F | amplification | MA, GMA |
| saturation | F | F | ON | ON, GMA, SPSW |
| cutoff | R | R | OFF | OFF |
| spin-voltaic | 0 | 0 | OFF | SPSW |

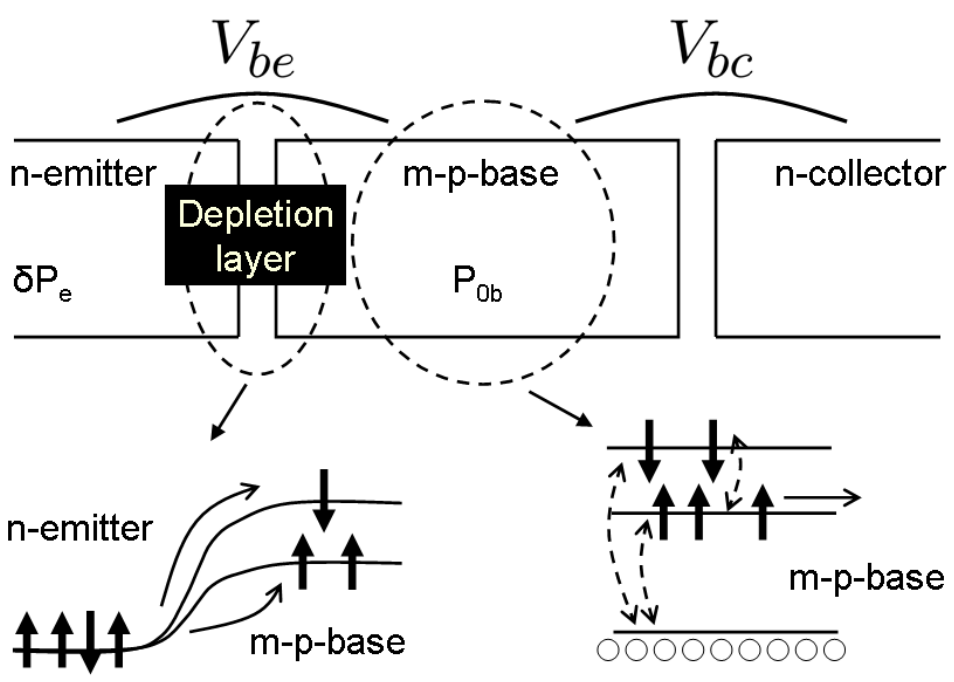

Fig. V.38. Physical processes leading to the spin control of amplification of magnetic bipolar transistors. There is a spin-charge coupling in the base-emitter junction, which controls the injection of the electrons to the base. In the base the electrons diffuse towards the collectors, recombine with holes, or flip their spin in an attempt to reach the spin equilibrium.

### E.5 Magnetic bipolar transistor with magnetic base

What happens if we replace one of the regions of the transistor, say base, with a magnetic semiconductor? For simplicity we assume that the conduction band of the base is exchange or Zeeman spin split, leading to electron spin polarization $P_{0b}$, while the holes are spin unpolarized. The equilibrium spin in the base influences the equilibrium electron density there, through the shift of the chemical potential. Since the current in the base-emitter junction depends on the minority electron density in the base, this current depends also on the equilibrium spin polarization in the base. This dependence allows some control over the amplification factor.

Nontrivial effects appear if we allow for the possibility of nonequilibrium spin in the emitter. Such a spin can appear as a result of a spin injection, electrical or optical. The origin is not essential for our discussion. We only assume that there is a nonequilibrium spin polarization $\delta P_e$ in the emitter, see Fig. V.38. What happens if the spin polarizations are parallel, that is, if $\delta P_e P_{0b} > 0$? More electrons in the emitter have spin up than spin down. At the same time there is a smaller energy barrier for spin up than for spin down, to cross the base-emitter barrier. The injection of the electrons from the emitter increases, giving a larger collector current and with it a larger current amplification $\beta$. On the other hand, if the two polarizations are antiparallel, that is, if $\delta P_e P_{0b} < 0$, the injection as well as the current gain are reduced. This is the basic principle behind the spin control of the current gain proposed for magnetic bipolar transistors.

In the following we introduce a formalism allowing to calculate the I-V characteristics of magnetic transistors in the low injection limit. In fact, the formalism applies to a whole array of magnetic p-n junctions connected in series.

### E.6 Magnetic p-n junctions in series

A magnetic transistor is the simplest nontrivial example of an array of magnetic p-n junctions serially connected. Indeed, the transistor comprises two such junctions in series. The next example would be a magnetic thyristor, comprising three junctions. In general, we impose boundary conditions on the carrier densities as well as on the spin at the two boundaries of the array. What needs to be determined are the carrier densities and the spin inside, especially at the depletion layers. In the following we present a general theory of serial magnetic p-n junctions in which the spin is in the conduction band. The valence band will be taken unpolarized for simplicity. As we wish this to be a reference section, we introduce all the relevant notation.

Each magnetic p-n junction is characterized by its equilibrium properties: acceptor doping $N_a$ and donor doping $N_d$, giving the majority hole and electron densities; equilibrium minority electron density $n_{0p}$ in the $p$ region and the equilibrium minority hole density $n_{0n}$ in the $n$ region; equilibrium spin density $s_{0p}$ in the $p$ region and equilibrium spin density $s_{0n}$ in the $n$ region. The corresponding equilibrium spin polarizations are $P_{0p} = s_{0p}/n_{0p}$ and $P_{0n} = s_{0n}/N_d$. The widths of the $p$ and $n$ regions are $w_p$ and $w_n$, respectively. The minority electron and hole carrier diffusion lengths are $L_{np}$ and $L_{pn}$, respectively, while the electron spin diffusion length in the $n$ and $p$ regions are $L_{sn}$ and $L_{sp}$, respectively. Similarly, the electron diffusion length in the $n$ and $p$ regions are denoted by $D_{nn}$ and $D_{np}$. Finally, we let $V$ denote the voltage drop across the junction, positive for a forward bias.

The currents across a particular junction are determined by the nonequilibrium minority carrier densities, $\delta n = n - n_0$, and $\delta p = p - p_0$, as well as by the nonequilibrium spin densities, $\delta s = s - s_0$, at both sides of the depletion layer. For a single p-n junction we define the scalar,

$$u = \delta s_n, \tag{V.116}$$

and the column vector

$$\mathbf{v} = \begin{pmatrix} \delta n_p \\ \delta s_p \end{pmatrix}. \tag{V.117}$$

The scalar $u$ describes the nonequilibrium electron spin density in the $n$ region of the junction. The majority electron density has the equilibrium value of $N_d$ in the low injection limit considered here. On the other hand, both the minority electron and spin densities can have nonequilibrium values in the $p$ region. These quantities are described by the vector $\mathbf{v}$. All the relevant physics of spin injection across magnetic p-n junctions is contained in $u$ and $\mathbf{v}$.

In spin-unpolarized p-n junctions the amount of the nonequilibrium minority carrier density at the depletion layer depends solely on the voltage drop, $V$, across the layer, see Sec. E.2. This follows from the Shockley condition of the constant chemical potential across the layer. Spin introduces two complications: (i) The nonequilibrium carrier and spin densities at one side of the depletion layer depend on the densities on the other side. This coupling, referred to as *intrajunction* spin-charge coupling, appears due to the generalized Shockley conditions of constant spin-resolved chemical potentials as well as uniform spin current across the depletion layer, see (Fabian *et al.*, 2002b). In addition, (ii) the nonequilibrium carrier and spin densities at one depletion layer depend on the densities at the adjacent layer (a single $p$ or $n$ region is sandwiched between the two layers). The reason for this coupling, called *interjunction* spin-charge coupling, lies in the carrier and spin diffusion across the region between the two layers.

As a result of the intra and interlayer spin-charge couplings, the minority carrier and spin densities at each internal magnetic p-n junction depend on the densities at all other junctions. Since in the low injection limit the couplings are linear, we can write down algebraic expressions describing the coupling and solve for all the relevant densities in a self-consistent manner. The apparent simplicity of the algebraic structure is very appealing as the algebraic solution can be readily found. We will see the technique below for the npn magnetic transistor case.

Let us first define the nonequilibrium carrier and spin densities in the absence of spin-charge couplings. We start with the more familiar,

$$\mathbf{v}^0 = \left(e^{qV/k_BT} - 1\right) \begin{pmatrix} n_{0p} \\ s_{0p} \end{pmatrix}. \tag{V.118}$$

If no nonequilibrium spin is present in the junction, the minority electron density in the $p$ region as well as the spin in the $p$ region would be given by the above formula. Only the voltage drop across the layer determines the nonequilibrium values which increase exponentially with $V$. While the nonequilibrium majority electron density in the $n$ region is negligible compared to $N_d$, the nonequilibrium spin $\delta s$ needs to be considered, as in general the equilibrium spin density can have any value smaller or equal $N_d$. As a result of electron extraction not taking into account spin-charge coupling, the nonequilibrium spin density is (Fabian *et al.*, 2002b),

$$u^0 = -\gamma_2 \cosh(w_p/L_{np}) s_{0p} \left(e^{qV/k_BT} - 1\right). \tag{V.119}$$

The electron extraction to the $p$ region with equilibrium spin $s_{0p}$ is spin-selective resulting in the spin imbalance (with the sign opposite to $s_{0p}$) in the $n$ region, described by the above formula; $\gamma_2$ is a structure factor to be specified below.

Before we proceed and describe the effects of spin-charge couplings on the carrier and spin densities in an array of magnetic p-n junctions, we introduce two column vectors describing the strengths of the coupling and the spin injection efficiency across the depletion layer. The vector $\mathbf{C}$, which is dimensionless, characterizes the intrajunction coupling:

$$\mathbf{C} = \begin{pmatrix} P_{0p}(\gamma_2 - \gamma_1) \\ \gamma_1 \end{pmatrix}. \tag{V.120}$$

The vector $\mathbf{D}$, again dimensionless, characterizes the strength of the interjunction coupling,

$$\mathbf{D} = \frac{n_{0p}}{N_d} \frac{e^{qV/k_BT}}{1 - P_{0n}^2} \begin{pmatrix} P_{0p} - P_{0n} \\ 1 - P_{0p}P_{0n} \end{pmatrix}. \tag{V.121}$$

**A single magnetic p-n junction.** We first recall the magnetic *p-n* junction equations (Fabian *et al.*, 2002b; Fabian and Žutić, 2004a), which we will later generalize using the notation introduced above. The reader should refer to Fig. V.39 where the geometry of the p-n junction is introduced. The p-region is on the left, the n-region on the right. The effective widths of the regions are $w_p$ and $w_n$, the corresponding dopings $N_a$ (acceptor density) and $N_d$ (donor density). We assume, quite generally, that both regions can have equilibrium electron spin, with the spin polarizations $P_{0p}$ and $P_{0n}$. The boundary conditions are fixed densities: $n_p$ and $s_p$, for the electron and spin density at the far left of the $p$ region, and $s_n$, the spin density at the far right of the $n$ region. In the array these boundary densities will no longer be fixed, as they will appear at the depletion layers to other regions. Going back to our single junction: we need to find the electron and spin densities at the depletion layer: $n_L$ and $s_L$, at the *left* of the layer, and $s_R$ at the *right* of the layer; the electron density at $R$ is just $N_d$ (see Fig. V.39).

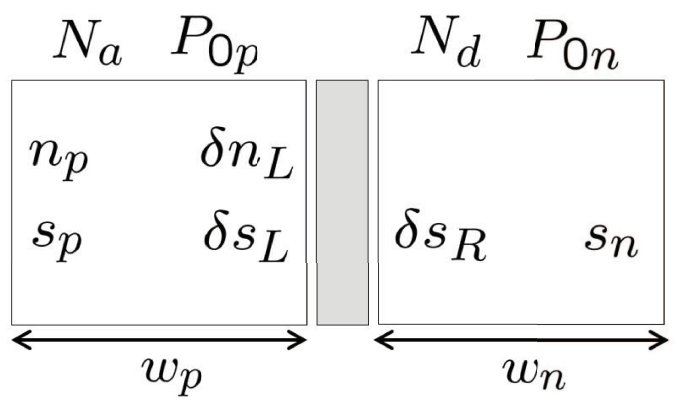

Fig. V.39. Scheme of a single p-n junction with identified boundary conditions (far left and right regions) and unknown carrier and spin densities at the left ($L$) and right ($R$) of the depletion layer.

Here we only quote the result, which for the nonequilibrium spin density, $\delta s_R = s_n - s_{0n}$, at the right side of the depletion layer gives (Fabian *et al.*, 2002b; Fabian and Žutić, 2004a):

$$\begin{aligned} \delta s_R &= \gamma_0 \delta s_n + \gamma_1(\delta s_p - P_{0p}\delta n_p) + \gamma_2 P_{0p}\delta n_p \\ &- \gamma_2 \cosh(w_p/L_{np})\, s_{0p}\left(e^{qV/k_BT} - 1\right). \end{aligned} \quad \text{(V.122)}$$

The structure factors $\gamma$ are,

$$\gamma_0 = \frac{1}{\cosh(w_n/L_{sn})}, \quad \text{(V.123)}$$

$$\gamma_1 = \left(\frac{D_{np}L_{sn}}{D_{nn}L_{sp}}\right)\frac{\tanh(w_n/L_{sn})}{\sinh(w_p/L_{sp})}, \quad \text{(V.124)}$$

$$\gamma_2 = \left(\frac{D_{np}L_{sn}}{D_{nn}L_{np}}\right)\frac{\tanh(w_n/L_{sn})}{\sinh(w_p/L_{np})}, \quad \text{(V.125)}$$

$$\gamma_3 = \gamma_2 \cosh(w_p/L_{np}). \quad \text{(V.126)}$$

Equation (V.122) is accurate up to the terms of the relative order of $n_0 \exp(qV/k_BT)/N_d$. While such terms can be safely neglected when dealing with the spin and carrier densities, they must be included when calculating the spin current in the $n$ region. The correct formula for the injected spin density at the low injection limit, $\delta s_R$, can be cast in the form of Eq. (V.122), with the coefficients $\gamma$ divided by the factor $1 + \nu$:

$$\gamma \to \gamma/(1+\nu), \quad \text{(V.127)}$$

where,

$$\nu = \frac{n_{0p}e^{qV/k_BT}}{N_d}[\gamma_1 \cosh(w_p/L_{sp})\frac{1-P_{0p}^2}{1-P_{0n}^2} + \gamma_3 P_{0p}\frac{P_{0p}-P_{0n}}{1-P_{0n}^2}]. \quad \text{(V.128)}$$

Typically $\nu$ is a number smaller than 0.1. The equation (V.122) gives the spin injection efficiency across the p-n junction.

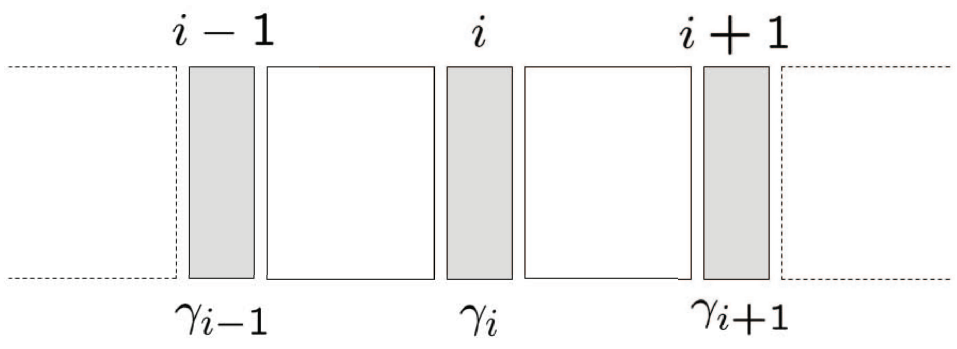

Fig. V.40. Array of (in general magnetic) p-n junctions. The top indexes refer to the number of the depletion layer, characterizing the corresponding junction. The bottom indexes show the spin injection parameters $\gamma$ corresponding to the junctions.

If $\delta P_R = \delta s_R/N_d$ is known, we can also calculate the injected minority densities $\delta n_L$ and $\delta s_L$:

$$\delta n_L = n_{0p}\left[e^{qV/k_BT}\left(1+\delta P_R\frac{P_{0p}-P_{0n}}{1-P_{0n}^2}\right)-1\right], \tag{V.129}$$

$$\delta s_L = s_{0p}\left[e^{qV/k_BT}\left(1+\frac{\delta P_R}{P_{0p}}\frac{1-P_{0p}P_{0n}}{1-P_{0n}^2}\right)-1\right]. \tag{V.130}$$

The following relation connects the spin polarization across the depletion layer:

$$P_L = \frac{P_{0p}\left(1-P_{0n}^2\right)+\delta P_R\left(1-P_{0p}P_{0n}\right)}{1-P_{0n}^2+\delta P_R\left(P_{0p}-P_{0n}\right)}. \tag{V.131}$$

Equations (V.122), (V.129)-(V.131) form what we call the magnetic p-n junction equations.

**An array of magnetic p-n junction** The magnetic p-n junction equations can be cast in a very compact form using the vector notation. Consider an array of p-n junctions depicted in Fig. V.40.

If the junction is the $i$th in the series, that is, the depletion layer of the junction has label $i$, Eqs. (V.122) and (V.129) for an oriented p-n junction are,

$$u_i = u_i^0 + \gamma_{0,i}u_{i+1} + \mathbf{C_i}\cdot\mathbf{v_{i-1}}, \tag{V.132}$$

$$\mathbf{v}_i = \mathbf{v}_i^0 + \mathbf{D}_i u_i. \tag{V.133}$$

In case the $i$th junction is of the directed n-p type (that is, the n-region is on the left and the p-region is on the right), the equations are,

$$u_i = u_i^0 + \gamma_{0,i}u_{i-1} + \mathbf{C_i}\cdot\mathbf{v_{i+1}}, \tag{V.134}$$

$$\mathbf{v}_i = \mathbf{v}_i^0 + \mathbf{D}_i u_i. \tag{V.135}$$

The label $i$ goes from 1 to the number of junctions (two for a transistors). The boundary conditions are the spin and carrier densities at the left ($i = 0$) and right ($i$ is the number of junctions plus 1) of the array. The resulting algebraic set of equations can be readily solved. Figures V.41 and V.42 show the stencils for the calculation of the electron and spin densities in the two oriented magnetic junctions.

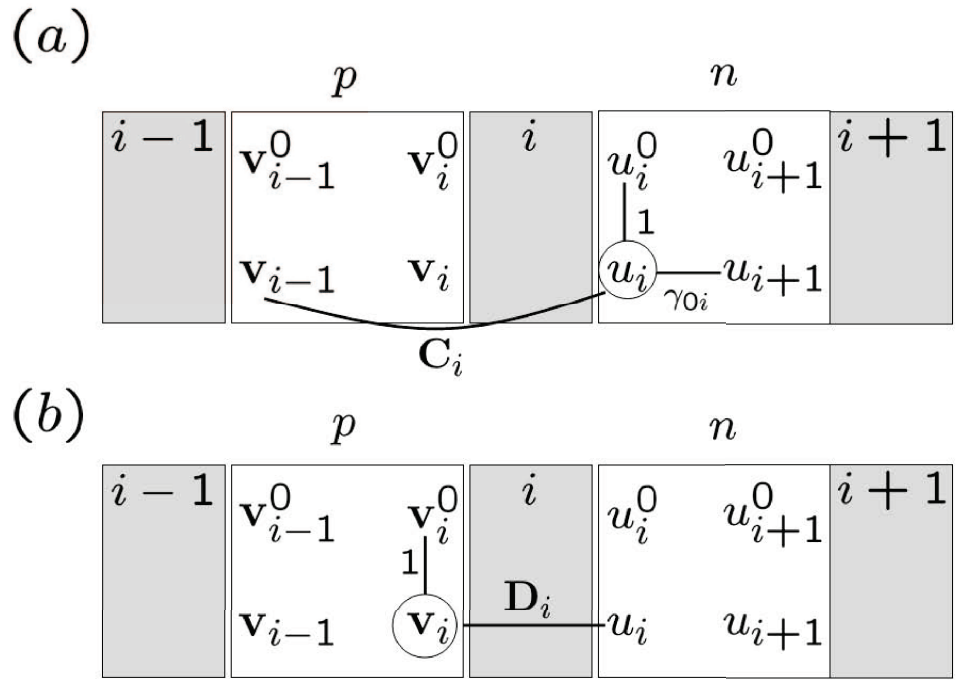

Fig. V.41. Stencil for an oriented magnetic p-n junction.

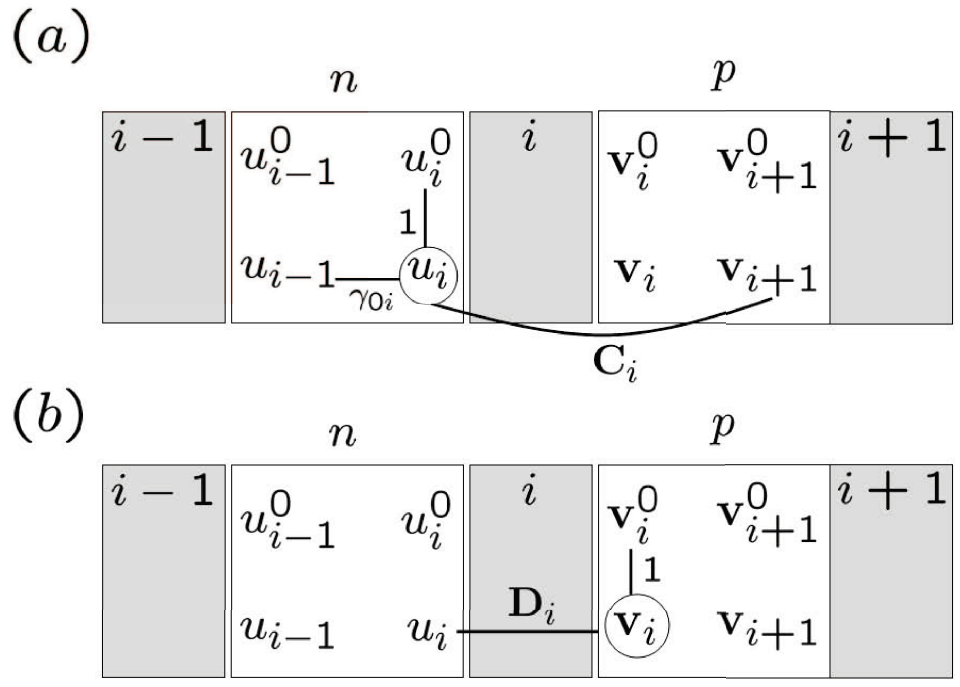

Fig. V.42. Stencil for an oriented magnetic n-p junctions.

**Example: npn magnetic transistor.** We will demonstrate the solution for the case of a npn magnetic transistor. We have the following equations for the first junction, the n-p emitter-base one, from Eqs. (V.134) and (V.135),

$$u_1 = u_1^0 + \gamma_{0,1} u_0 + \mathbf{C}_1 \cdot \mathbf{v}_2, \tag{V.136}$$

$$\mathbf{v}_1 = \mathbf{v}_1^0 + \mathbf{D}_1 u_1. \tag{V.137}$$

The second junction, the p-n base-collector one, described by Eqs. (V.132) and (V.133), is given by,

$$u_2 = u_2^0 + \gamma_{0,2} u_3 + \mathbf{C}_2 \cdot \mathbf{v}_1, \tag{V.138}$$

$$\mathbf{v}_2 = \mathbf{v}_2^0 + \mathbf{D}_2 u_2. \tag{V.139}$$

These equations can be solved analytically giving,

$$u_2 = (\mathbf{C}_2) \cdot (\mathbf{D}_1)(\gamma_{0,1} u_0 + u_1^0) + \mathbf{C}_2 \cdot \mathbf{v}_1^0 + u_2^0 + \gamma_{0,2} u_3. \tag{V.140}$$

Terms of order $[n_{0p}\exp(eV/k_BT)/N_d]^2$ have been neglected as small in the low-injection limit; the magnetic p-n junction equations are themselves valid in this limit only. Since $u_2$ is the nonequilibrium spin density in the collector (at the junction with the base), Eq. (V.140) describes the spin injection through the magnetic transistor. The knowledge of $u_2$ gives us also the expressions for $u_1$, $\mathbf{v}_1$, and $\mathbf{v}_2$, from Eqs. (V.136), (V.137), and (V.139).

The first term in Eq. (V.140) describes the spin injection through the transistor, from a source of nonequilibrium spin ($u_0$) in the emitter; the term containing $u_1^0$ can be neglected in the low injection limit, see below. The second term, proportional to $\mathbf{v}_1^0$, describes the process of the intrinsic spin injection resulting from the equilibrium spin polarization in the base. No source spin is needed for spin injection here! This term vanishes if the base is nonmagnetic. The third term would describe spin extraction from the collector due to the magnetic base, if the base-collector junction were forward biased. If the bias is reverse, as in the forward active regime, this term can be neglected. Finally, the last term describes diffusion of the spin from a possible source of spin in the collector.

### E.7 Spin injection through magnetic bipolar transistor

We will now look at ramifications of Eq. (V.140) and consider two apparently distinct physical conditions. The first is a nonmagnetic transistor, but with a spin-polarized emitter in which there is a source (such as electrical or optical spin injection) of nonequilibrium spin. The second is a magnetic-base transistor, but with no source of nonequilibrium spin in the emitter.

**Spin injection due to source spin in the emitter.** Consider a conventional bipolar junction transistor with a source of nonequilibrium spin in the emitter. In the forward active regime, what is the nonequilibrium spin in the collector? The nonequilibrium spin there appears as a result of the electric spin injection through *two* depletion regions! First, the emitter spin is transmitted to the base. There the accumulated spin diffuses towards to collector, as well as relaxes to equilibrium in the base itself. The spin that moves towards to the collector is swept by the electric field in the base-collector depletion layer and accumulates in the collector. Looking at Eq. (V.140), for the case at hand we have the following conditions: (i) $u_3 = 0$, meaning there is no nonequilibrium spin in the far boundary of the collector. (ii) $u_1^0 = u_2^0 = \mathbf{v}_2^0 = 0$, reflecting our conventional transistor with no equilibrium spin as well as with little nonequilibrium electrons at the reverse biased base-collector junction. Using these conditions we obtain for the spin injection density in the collector:

$$u_2 = \gamma_{0,1}\mathbf{C}_2 \cdot \mathbf{D}_1 u_0, \tag{V.141}$$

and

$$\mathbf{D}_1 = \frac{n_{0,1}}{N_{d,1}} e^{qV_{12}/k_bT} \begin{pmatrix} 0 \\ 1 \end{pmatrix}, \tag{V.142}$$

$$\mathbf{C}_2 = \begin{pmatrix} 0 \\ \gamma_{1,2} \end{pmatrix}. \tag{V.143}$$

This translates directly to the emitter-base-collector language:

$$\delta s_c = \gamma_{0,be} \frac{n_{0b}}{N_{d,e}} e^{qV_{be}/k_BT} \gamma_{1,bc} \delta s_e. \tag{V.144}$$

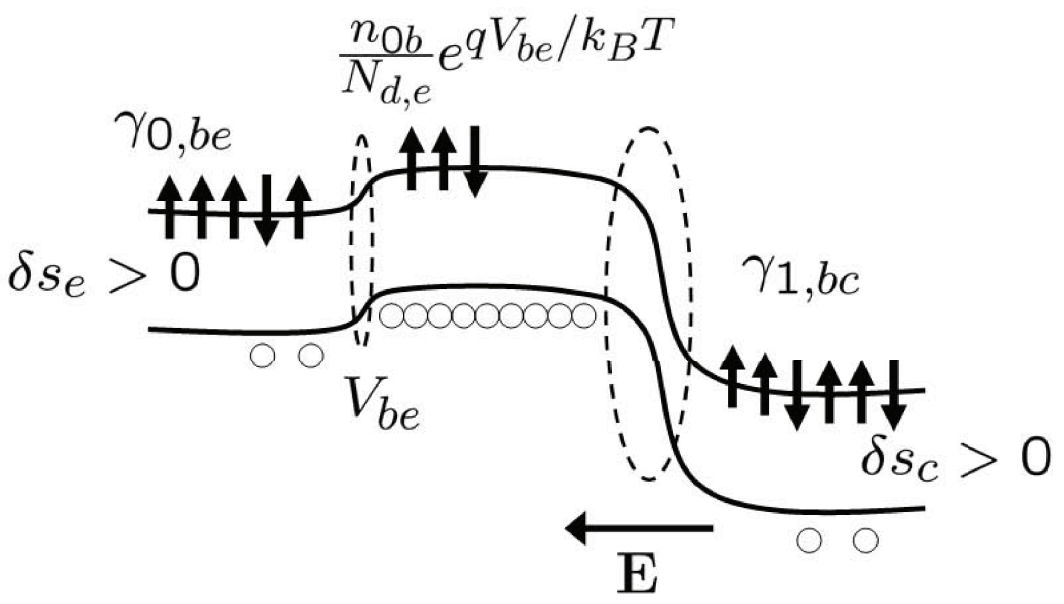

Fig. V.43. Spin injection through a conventional bipolar junction transistor, from a nonequilibrium spin, $\delta s_e$, in the emitter. First the spin is injected into the base. In the base the spin density is lowered due to the low density of the injected spins. The spin that reaches the depletion layer to the collector is swept by the large electric field $\mathbf{E}$ to the collector where it accumulates. The corresponding factors lowering the initial spin density in the process are shown at the top.

The factor of $\gamma_{0,be}$ describes the diffusion of the nonequilibrium spin within the emitter. This factor is close to one if the emitter's width is comparable to the spin diffusion length, which would be a practical case of submicron transistors. The spin density that arrives at the emitter-base depletion layer can then be as large as the source spin itself.

The spin then proceeds through the depletion layer. Since the spin in our model is attached to electrons, once it overcomes the barrier and comes to the base, it becomes part of the minority carrier density, since the base is p-doped. This is why the spin density drops by the factor of,

$$\frac{n_{0b}}{N_{de}} e^{qV_{be}/k_BT}, \tag{V.145}$$

which defines the electron density in the base. This lowers the spin density to less than perhaps 10% of the initial value in the emitter. In the base the spin relaxes as well as diffuses towards the collector. The spin injection through the base-collector depletion layer proceeds through the process of the minority-carrier spin pumping, akin to the processes that appear in spin-polarized p-n junctions or solar cells (Žutić *et al.*, 2001b,a). This process is described by $\gamma_{1,bc}$. The whole spin injection process is illustrated in Fig. V.43.

The injected value of the spin density in the collector is much smaller than the density of the source spin, in large part due to the low injection intensity from the emitter to the base. What about the spin *polarization* in the collector? It is,

$$\delta P_c = \gamma_{0,be} \frac{n_{0b}}{N_{dc}} e^{qV_{be}/k_BT} \gamma_{1,bc} \delta P_e. \tag{V.146}$$

The denominator is now $N_{dc}$, which is typically much less than the emitter donor density $N_{de}$, in practical transistors. Next, the factor $\gamma_{1,bc}$ can be quite large in the limit of a narrow base and a wide collector (relative to the spin diffusion lengths), $\gamma_{1,bc} \approx L_{sc}/w_b$, which can be a hundred or so. As a result, even in the low injection limit, the injected spin polarization through the conventional bipolar junction transistor can be a significant portion of the source spin polarization in the emitter.

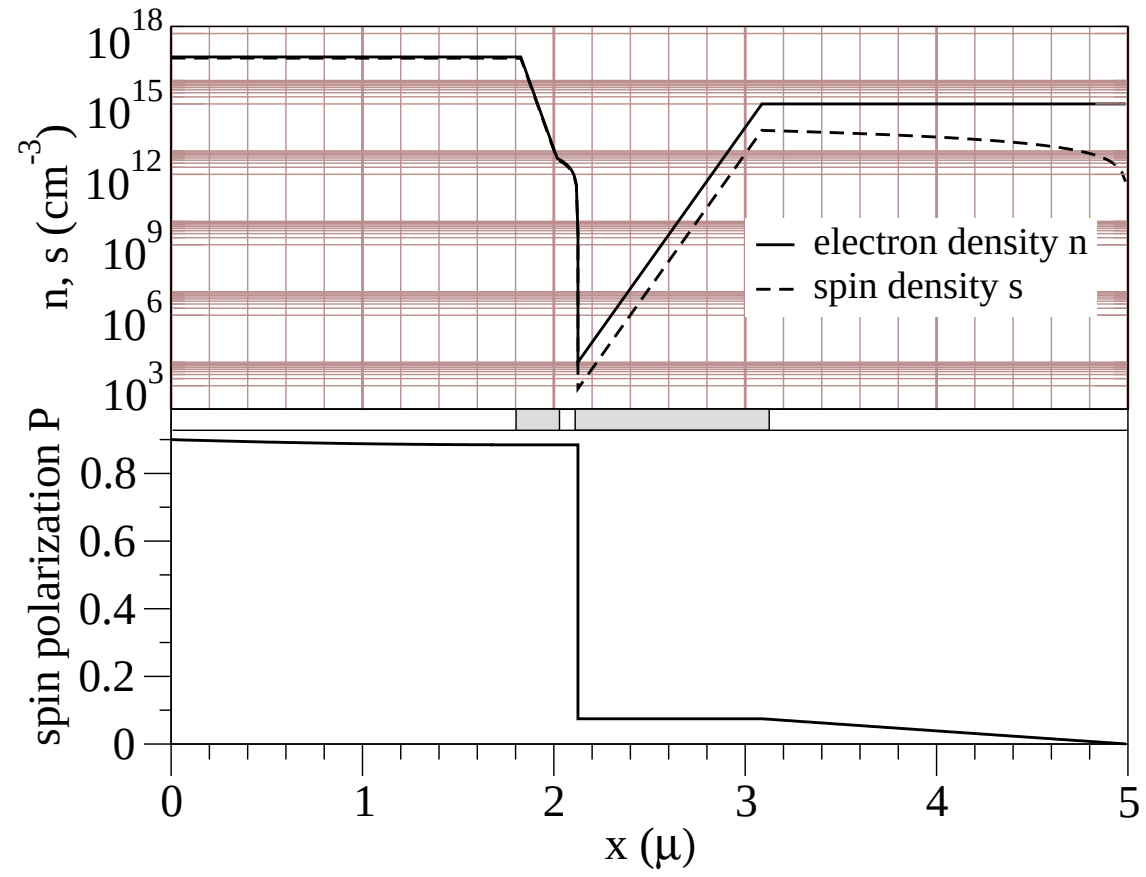

Fig. V.44. Calculated spin injection of a source spin through a conventional bipolar junction transistor. A source spin is injected into the emitter. The spin is injected by a forward current to the base, diffuses towards the collector-base junction, where it is swept into the collector. In the figure the source spin is $\delta P_e = 0.9$. The calculated electron and spin densities (top) as well as the spin polarization $P$ profile (bottom) are shown. The geometry of the transistor is indicated between the graphs, with the shaded areas corresponding to the depletion layers. After (Fabian *et al.*, 2002b).

A realistic calculation, for a GaAs based npn transistor, is shown in Fig. V.44. The electron as well as the spin densities are shown as they change within the transistor. The source spin polarization in the emitter is $P_e = 0.9$. The spin density as well as the spin polarization drop when reaching the base. While the spin density still increases as going from the base to the collector, due to the minority electron spin pumping, the spin polarization is the same (it eventually vanishes at the right edge of the collector, due to our boundary conditions). The spin injection results in $P_c \approx 0.1$. The forward voltage drop across the base-emitter junction in the calculation is $V_{be} = 0.5$ V (the low injection limit holds for values up to about 0.8 V). No voltage drop is applied for the base-collector junction, keeping it effectively in the reverse bias. Details of the calculations as well as the geometric and materials parameters of the simulated structure can be found in (Fabian and Žutić, 2004a).

**Spin injection due to equilibrium spin in the base.** Suppose now that there is only an equilibrium spin in the transistor, in the base. The equilibrium spin polarization is $P_{0b}$. No source of nonequilibrium spin exists in the emitter. Will there be any spin injected across the transistor? Based on the impossibility of spin injection due to equilibrium spin in magnetic p-n junctions (Žutić *et al.*, 2002; Fabian *et al.*, 2002b) we would be tempted to answer this question negative. However, transistors are not simple extensions of diode and the answer is actually positive, due to simultaneous build up of charge and spin in the base in the forward active region.

The situation we are describing has the following conditions for Eq. (V.140): (i) $u_0 = u_3 = 0$, since there are no source spins in the emitter or collector and (ii) $u_2^0 \approx 0$, since we are in the forward active region and can set $V_{23} = 0$. As the only equilibrium spin polarization is $P_{0b}$ we have that,

$$u_2 = \mathbf{D}_1 \cdot \mathbf{C}_2 u_1^0 + \mathbf{C}_2 \cdot \mathbf{v}_1^0, \tag{V.147}$$

where now,

$$u_1^0 = -\gamma_{2,1} \cosh(w_{p,1}/L_{np,1}) s_{0p,1} e^{qV_{12}/k_BT}, \tag{V.148}$$

$$\mathbf{v}_1^0 = e^{qV_{12}/k_BT} \begin{pmatrix} n_{0p,1} \\ s_{0p,1} \end{pmatrix}, \tag{V.149}$$

and

$$\mathbf{D}_1 = \frac{n_{0p,1}}{N_{d,1}} e^{qV_{12}/k_BT} \begin{pmatrix} P_{0p,1} \\ 1 \end{pmatrix}, \tag{V.150}$$

$$\mathbf{C}_2 = \begin{pmatrix} P_{0p,2}(\gamma_{2,2} - \gamma_{1,2}) \\ \gamma_{1,2} \end{pmatrix}. \tag{V.151}$$

In the expressions for $u_1^0$ and $\mathbf{v}_1^0$ we have used the fact that since in the forward active regime $qV_{12} \gg k_BT$, the exponent is much larger than one. We see that,

$$\mathbf{D}_1 \cdot \mathbf{C}_2 u_1^0 \approx s_{0p,1} e^{qV_{12}/k_BT} \left( \frac{n_{0p,1}}{N_{d,1}} e^{qV_{12}/k_BT} \right), \tag{V.152}$$

is much smaller than,

$$\mathbf{v}_1^0 \cdot \mathbf{C}_2 \approx s_{0p,1} e^{qV_{12}/k_BT}, \tag{V.153}$$

due to the low injection limit in which $\frac{n_{0p,1}}{N_{d,1}} e^{qV_{12}/k_BT} \ll 1$. We are thus left with

$$u_2 \approx \mathbf{C}_2 \cdot \mathbf{v}_1^0, \tag{V.154}$$

which translates into the npn transistor language to,

$$\delta s_c \approx \gamma_{0,be} \gamma_{1,bc} e^{qV_{be}/k_BT} s_{0b}. \tag{V.155}$$

The spin polarization at the collector then is,

$$\delta P_c \approx \gamma_{0,be} \gamma_{2,bc} \frac{n_{0b}}{N_{dc}} e^{qV_{be}/k_BT} P_{0b}. \tag{V.156}$$

In the limit of a thin base, that is, for $w_b \ll L_{nb}, L_{sb}$, the factor $\gamma_2 \approx \gamma_1$. Then the above equation for the spin injection is exactly the same as the spin injection due to the source spin, Eq. (V.146), if we replace the source spin $\delta P_e$ with the equilibrium spin in the base, $P_{0b}$. The spin injection has the same intensity in both cases!

Since there is no analogue of the spin injection of the equilibrium spin in the diode, we call the spin injection *intrinsic*. The physical reason for the intrinsic spin injection is as follows. The electrons injected from the emitter into the base accumulate there and form a nonequilibrium electron density. These electrons, initially spin unpolarized, become polarized in the base with the same spin polarization as that of the base, by both spin-selective transfer across the depletion layer as well as by spin relaxation. While the base spin *polarization* is kept at equilibrium, the spin *density* is not, as it is tied with the nonequilibrium electron density. It is this nonequilibrium spin density in the base, at the depletion layer with the emitter, that serves as the spin source for the minority electrons which is injected into the collector by the process of minority electron spin pumping. This is in line with our statements that first a nonequilibrium spin density has to be build in order to have a spin injection across a depletion layer. The intrinsic spin injection is illustrated in Fig. V.45.

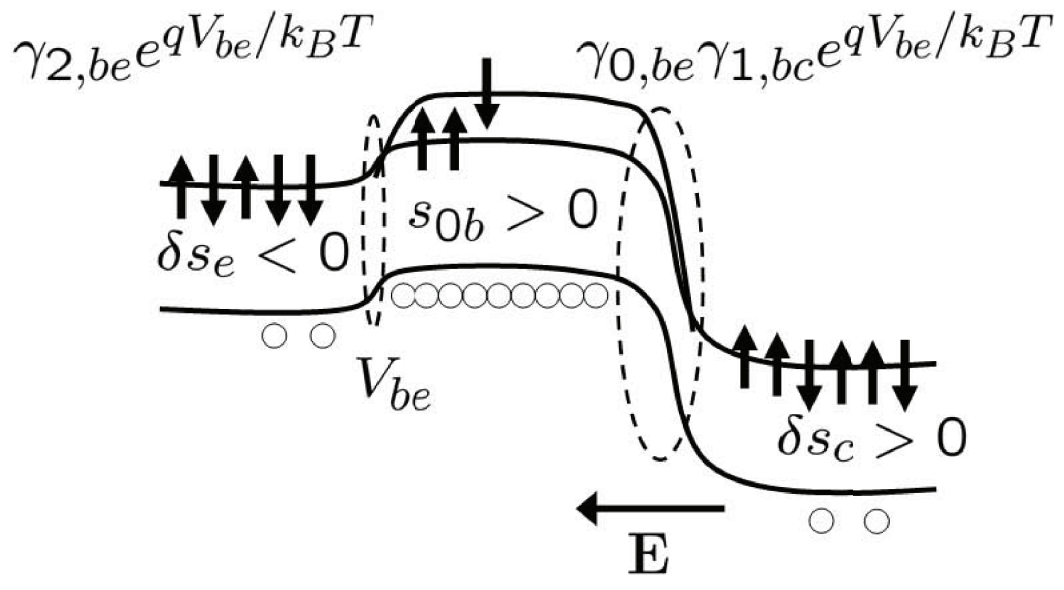

Fig. V.45. Intrinsic spin injection and spin extraction in a magnetic-base magnetic bipolar junction transistor. The extracted spin density in the emitter is negative, the corresponding spin polarization being significantly lower than the equilibrium spin polarization in the base. The intrinsic spin injection into the collector is positive; the injected spin polarization can be a large fraction of the equilibrium polarization. The corresponding factors changing the initial equilibrium spin density in the process are shown at the top.

As we drive electrons from the emitter to the magnetic base, we also extract spin from the emitter. The reason is that electrons with spins parallel to the preferred spin in the base have larger probability to cross—their barrier is lower than the opposite spin, due to the spin splitting of the conduction band. Is this behavior predicted by our model above? From Eq. (V.136) we have ($u_0 = 0$):

$$u_1 = u_1^0 + \mathbf{C}_1 \cdot \mathbf{v}_2. \tag{V.157}$$

Now, Eq. (V.139) gives ($\mathbf{v}_2^0 \approx 0$ due to the reversed biased junction 2):

$$\mathbf{v}_2 = \mathbf{D}_2 u_2, \tag{V.158}$$

where, finally, $u_2$ is given by Eq. (V.156). In the low injection limit we can neglect the second term in Eq. (V.157), as it contains the small factor $n_{0b}/N_{dc}$. We are left with

$$u_1 \approx u_1^0, \tag{V.159}$$

which in the language of the npn transistor reads

$$\delta s_e \approx -\gamma_{2,be} \cosh(w_b/L_{nb}) e^{qV_{be}/k_BT} s_{0b}. \tag{V.160}$$

The negative sign reflects the spin extraction. The extracted spin polarization,

$$\delta P_c \approx -\gamma_{2,be} \cosh(w_b/L_{nb}) \frac{n_{0b} e^{qV_{be}/k_BT}}{N_{de}} P_{0b}, \tag{V.161}$$

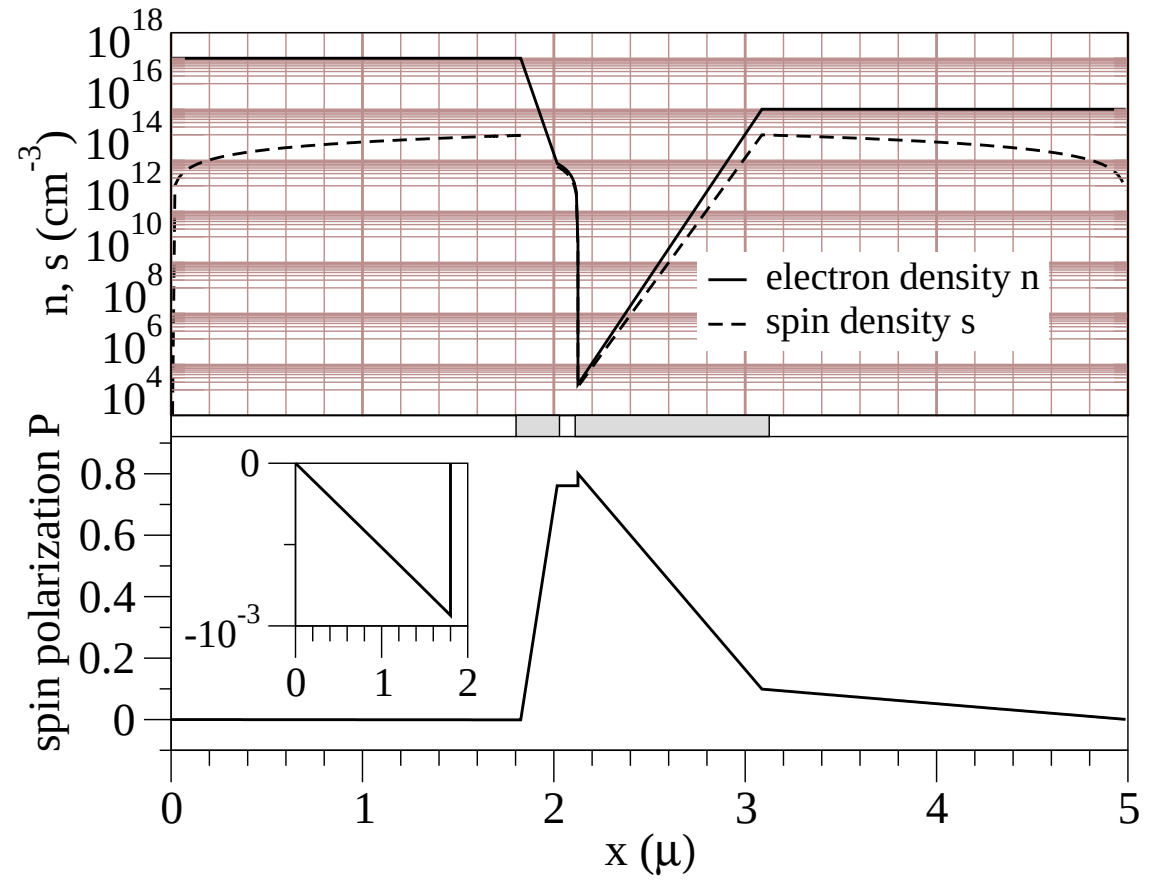

Fig. V.46. Calculated spin extraction and intrinsic spin injection in a magnetic bipolar transistor with a magnetic base. No source spin is present. Calculated electron- and spin-density profiles (top) and the spin density polarization $P$ (bottom). In the emitter region the spin density is negative, here plotted as positive in the log scale. The intrinsic spin injection results in a spin polarization in the collector of $\delta P_c \approx 0.1$. After (Fabian and Žutić, 2004a).

is rather small due to the small value of $n_{0b}/N_{de}$. Note that the intrinsic spin injection into the collector results in a large spin polarization there since the collector donor doping is by design smaller than in the emitter. The spin extraction is illustrated in Fig. V.45.

A realistic calculation of the intrinsic spin injection across a magnetic bipolar transistor is shown in Fig. V.46. The equilibrium spin polarization in the base is set to $P_{0b} = 0.762$, which corresponds to the conduction band splitting of $k_B T$. The bias voltages are $V_{be} = 0.5$ V and $V_{bc} = 0$ V, as in Fig. V.44. The spin density in the emitter is actually negative, as expected for spin extraction. The injected spin polarization in the emitter is about 10%. Details of the calculations as well as the geometric and materials parameters of the simulated structure can be found in Fabian and Žutić (2004a).

### E.8 Magnetoamplification effects

Magnetic bipolar transistors allow for magnetic and spin control of current amplification. In addition, they can operate as a spin switch, even in the absence of bias, only due to spin-charge coupling. Both phenomena enhance functionalities of conventional transistors.

**Currents in a transistor.** The primary application of bipolar junction transistors is current amplification or gain. In order to evaluate the current gain in magnetic bipolar transistors, we need to look at the distribution of electron and hole currents in all three regions of the transistor. This distribution is shown in Fig. V.47. There is a zoo of relevant currents. We first calculate the electron currents in the base. Electrons in the base are minority carriers whose transport, in the low injection limit, is given solely by diffusion. The electron (charge) current then is

$$j_b^n(x) = -qD_{nb}\frac{d\delta n_b(x)}{dx}, \tag{V.162}$$

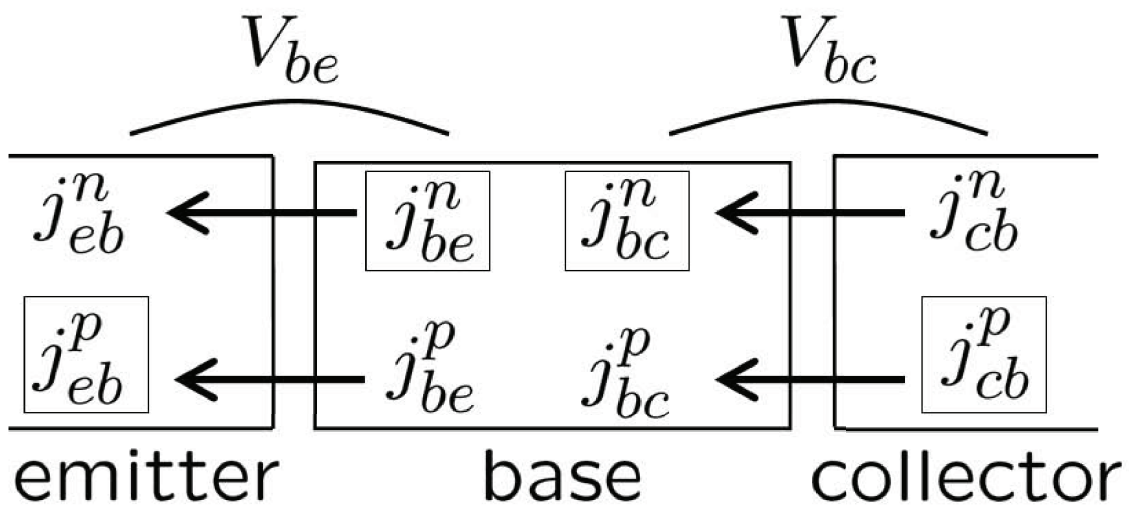

Fig. V.47. The distribution of currents flowing in a npn bipolar junction (magnetic or not) transistor. The currents in a box result from the diffusion of the minority carriers. All other come from both diffusion and drift of the majority carriers.

where $x$ measures the distance from the base-emitter junction (see Fig. V.47). We have included the minus sign in order to keep with the convention that in the forward active regime the positive currents are opposite to the x direction, see Fig. V.36. In order to find the current, we need to calculate first the nonequilibrium electron density profile $\delta n_b$ in the base.

The electron density in the base is given by the diffusion equation,

$$\frac{d^2\delta n_b}{dx^2} = \frac{\delta n_b}{L_{nb}^2}, \tag{V.163}$$

with the boundary conditions fixed by the electron densities at the depletion layer with the emitter, $\delta n_{be} \equiv \delta n_b(0)$, and at the depletion layer with the collector, $\delta n_{bc} \equiv \delta n_b(w_b)$. With these boundary conditions the solution to the above diffusion equation is,

$$\delta n_b = \delta n_{be}\frac{\sinh[(w_b - x)/L_{nb}]}{\sinh(w_b/L_{nb})} + \delta n_{bc}\frac{\sinh(x/L_{nb})}{\sinh(w_b/L_{nb})}. \tag{V.164}$$

The electric current can now be readily obtained from Eq. (V.162). We do not need to know the full electron current profile inside the base, only the two boundary values: $j_{n,be} \equiv j_{nb}(0)$ and $j_{nbc} \equiv j_{nb}(w_b)$. These two values are,

$$j^n_{be} = \frac{eD_{nb}}{L_{nb}}\left[\frac{\delta n_{be}}{\tanh(w_b/L_{nb})} - \frac{\delta n_{bc}}{\sinh(w_b/L_{nb})}\right], \tag{V.165}$$

$$j^n_{bc} = \frac{qD_{nb}}{L_{nb}}\left[\frac{\delta n_{be}}{\sinh(w_b/L_{nb})} - \frac{\delta n_{bc}}{\tanh(w_b/L_{nb})}\right]. \tag{V.166}$$

We will also need the hole currents flowing in the emitter and collector, where holes are minority carriers. Again, in these two regions holes are transported by diffusion only. The boundary conditions are now simpler since we can assume that the contacts with electrodes are ohmic, in which case the nonequilibrium density at the contact vanishes (no charges build up in ohmic contacts). For example, if now $x$ runs within the emitter from the contact to an external electrode, to the depletion layer with the base, the hole density profile is,

$$\delta p_e = \delta p_{eb}\frac{\sinh(x/L_{pe})}{\sinh(w_e/L_{pe})}. \tag{V.167}$$

At $x = 0$, the place of the Ohmic contact, the nonequilibrium density vanishes. Similarly, if now $x$ measures the distance in the collector from the depletion layer to the base all the way to the contact with an external electrode, the hole profile is,

$$\delta p_c = \delta p_{cb} \frac{\sinh[(w_c - x)/L_{pc})]}{\sinh(w_c/L_{pc})}. \tag{V.168}$$

Again at the contact, $x = w_c$, the density vanishes. The hole current can be obtained from (see the sign convention for the current in Fig. V.36),

$$j^p = qD_p \frac{d\delta p}{dx}, \tag{V.169}$$

which gives for the hole current in the emitter, at the depletion layer with the base, the value of,

$$j^p_{eb} \equiv j_{pe}(w_e) = \frac{qD_{pe}}{L_{pe}} \coth(w_e/L_{pe}) \delta p_{eb}, \tag{V.170}$$

and for the hole current in the collector, at the depletion layer with the base,

$$j^p_{cb} \equiv j_{pc}(0) = -\frac{qD_{pc}}{L_{pc}} \coth(w_c/L_{pc}) \delta p_{cb}. \tag{V.171}$$

The signs of the currents are such that in the forward active regime all the currents point left (positive direction for the currents), as shown in Fig. V.47.

Unlike the individual electron and hole currents, the full charge current in any region must be uniform. Take the emitter. The current anywhere within the region must be equal to $j^n_{eb} + j^p_{eb}$, the current flowing at the depletion layer with the base. We have calculated already $j^p_{eb}$, from the diffusion transport model of minority holes. A direct calculation of $j^n_{eb}$ would be more difficult since both diffusion and drift contribute to the majority carrier transport. Fortunately, we can, to a good accuracy, assume that as the electrons are transported through the depletion layer, electron-hole recombination does not severely reduce the electron current, so we can write that $j^n_{eb} = j^n_{be}$, which we already calculated. Similarly for other majority currents. Here is the summary:

$$j^n_{eb} = j^n_{be}, \tag{V.172}$$
$$j^p_{be} = j^p_{eb}, \tag{V.173}$$
$$j^n_{cb} = j^n_{bc}, \tag{V.174}$$
$$j^p_{bc} = j^p_{cb}. \tag{V.175}$$

We can thus use the simple expressions obtained from the diffusion currents and use them for diffusion-drift currents!

The electric current flowing in the emitter then is,

$$j_e = j^n_{be} + j^p_{eb}. \tag{V.176}$$

Substituting Eqs. (V.165) and (V.170) for the electron and hole currents, we obtain,

$$j_e = j^n_{gb} \left[ \frac{\delta n_{be}}{n_{0b}} - \frac{1}{\cosh(w_b/L_{nb})} \frac{\delta n_{bc}}{n_{0b}} \right] + j^p_{ge} \frac{\delta p_{eb}}{p_{0e}}. \tag{V.177}$$

Here we simplified the notation by introducing the generation currents in the base and the emitter,

$$j_{gb}^{n} = \frac{qD_{nb}}{L_{nb}} n_{0b} \coth(w_b/L_{nb}), \tag{V.178}$$

$$j_{ge}^{p} = \frac{qD_{pe}}{L_{pe}} p_{0e} \coth(w_e/L_{pe}). \tag{V.179}$$

Similarly, the collector current is given by,

$$j_c = j_{bc}^{n} + j_{cb}^{p}. \tag{V.180}$$

Substituting Eqs. (V.166) and (V.171) we get,

$$j_c = j_{gb}^{n} \left[ -\frac{\delta n_{bc}}{n_{0b}} + \frac{1}{\cosh(w_b/L_{nb})} \frac{\delta n_{be}}{n_{0b}} \right] - j_{gc}^{p} \frac{\delta p_{cb}}{p_{0c}}. \tag{V.181}$$

Here,

$$j_{gc}^{p} = \frac{qD_{pc}}{L_{pc}} p_{0c} \coth(w_c/L_{pc}). \tag{V.182}$$

Finally, the current in the base,

$$j_b = j_e - j_c, \tag{V.183}$$

follows from the current conservation (see Fig. V.36).

Equations (V.177), (V.181), and (V.183), allow us to calculate the gain factor $\beta$, given by Eq. (V.115). It is remarkable that the equations are valid for both conventional as well as for magnetic bipolar magnetic transistors. Indeed, except for the materials parameters like diffusivities and mean free paths, the only parameters that specify the type of the transistor are the nonequilibrium carrier densities. These, in turn, can be obtain from our vector model solution, Eq. (V.140) and the defining equations Eqs. (V.136), (V.137), and (V.139).

**Magnetoamplification.** To explain the phenomenon of magnetoamplification, we will try to stay within the formalism already developed for conventional transistors since this is how the contrast is best illustrated. Let us first rewrite the gain factor $\beta$ as (Tiwari, 1992),

$$\beta = \frac{1}{\alpha_T' + \gamma'}, \tag{V.184}$$

where $\alpha_T'$, a base transport parameter, is a measure of how inefficient is electron recombination with holes in the base, and $\gamma'$, an emitter injection parameter, is a measure of how inefficient is the electron injection from the emitter to the base. In electronics literature one usually introduces the efficiency factors $\alpha_T$ and $\gamma$ by,

$$\alpha_T = \frac{1}{1 + \alpha_T'}, \tag{V.185}$$

$$\gamma = \frac{1}{1 + \gamma'}. \tag{V.186}$$

We need to specify what $\alpha_T'$ and $\gamma'$ are. For a general magnetic transistor Eq. (V.115), as well as Eqs. (V.177), (V.181), (V.183) lead to, in the forward active regime,

$$\alpha_T' = \cosh(w_b/L_{nb}) - 1, \tag{V.187}$$
$$\gamma' = \gamma_0'/\eta, \tag{V.188}$$

where the *magnetoamplification coefficient* $\eta$ equals one for a conventional transistor, for which the emitter injection parameter is

$$\gamma_0' = \frac{N_{ab}}{N_{de}}\frac{D_{pe}}{D_{nb}}\frac{n_{ie}^2}{n_{ib}^2}\frac{L_{nb}}{L_{pe}}\frac{\sinh(w_b/L_{nb})}{\tanh(w_e/L_{pe})}, \tag{V.189}$$

and $n_i$ denote the intrinsic carrier concentrations.

The expression for the base transport parameter is the same in conventional and magnetic transistors. In the thin base limit, $w_b \ll L_{nb}$, it becomes,

$$\alpha_T' \approx \frac{w_b^2}{2L_{nb}^2}. \tag{V.190}$$

The base transport parameter decreases quadratically with decreasing of the width of the base. The thinner is the base, the less chance the electrons have to recombine with holes and more electrons arrive at the depletion layer with the collector. The thinner is the base, the greater is the current gain $\beta$.

What distinguishes magnetic from conventional transistors is the emitter injection parameter $\gamma'$. The factor $\eta$ depends both on the equilibrium spin, as well as on spin-charge coupling. We will see that the former dependence leads to magnetoamplification, while the latter to giant magnetoamplification. In the thin base limit the emitter injection parameter behaves as,

$$\gamma_0 \sim \frac{w_b}{L_{nb}}. \tag{V.191}$$

The parameter decreases linearly with decreasing of the width. In essence the smaller is the width of the base, the faster is the diffusion of electrons as the electron density gradient increases, so that more electrons can be injected from the emitter to accommodate in the base (recall that $\gamma'$ is inefficiency, not efficiency of emitter injection). Important for gain consideration is that for practical thin-base transistors, one can neglect the base transport factor as small compared to the emitter efficiency, due to the quadratic versus linear dependence on $w_b/L_{nb}$. In this case, which we will use further, the current gain is determined by,

$$\beta = \eta\beta_0, \tag{V.192}$$

where $\beta_0$ is the gain of the conventional transistor with a thin base.

In the following we will obtain $\eta$ in two important cases: (i) magnetic transistor without a source of a nonequilibrium spin, and (ii) magnetic transistor with a source of a nonequilibrium spin. In the case (i), the magnetoamplification coefficient is readily found to be,

$$\eta = \sqrt{\frac{1 - P_{0e}}{1 - P_{0b}^2}}, \tag{V.193}$$

so that the current gain of the transistor is,

$$\beta = \beta_0 \sqrt{\frac{1 - P_{0e}}{1 - P_{0b}^2}}. \tag{V.194}$$

The current gain can be simply controlled by the equilibrium magnetizations either in the emitter or in the collector. The dependence is opposite. Increasing the emitter magnetization leads to an increase of the minority hole density. Indeed, in the presence of a spin-split bands the Boltzmann statistics gives for holes,

$$p_{0e} = \frac{n_i^2}{N_{dc}} \frac{1}{\sqrt{1 - P_{0e}^2}}. \tag{V.195}$$

More minority holes means a large hole current (since the generation current is proportional to the equilibrium minority density) relative to the electron current, so the emitter injection parameter $\gamma'$ must increase (efficiency decreases). The current gain decreases. On the other hand, the equilibrium electron density in the base is,

$$n_{0b} = \frac{n_i^2}{N_{ab}} \frac{1}{\sqrt{1 - P_{0b}^2}}. \tag{V.196}$$

The electron density increases with increasing of the equilibrium spin polarization in the base, leading to the increase of the electron current from the emitter and thus decrease of $\gamma'$. The current gain increases. We call *magnetoamplification* the change of the current gain due to the change in the equilibrium spin polarization in the transistor. Magnetoamplification was described also in pnp transistors by Lebedeva and Kuivalainen (2003).

The above described behavior is illustrated in Fig. V.48, using a realistic calculation for GaAs-based hypothetical magnetic transistor. As expected from the analytical formula, see Eq. (V.194), the current gain increases with increasing the magnitude of the polarization (the sign is not relevant for the statistics) if the base is magnetic. If the emitter is magnetic, the gain, as predicted, decreases.

Let us move to the less trivial case (ii). We have an equilibrium spin polarization as well as a source of a nonequilibrium spin. From our analytical model we obtain the following expression for the magnetoamplification parameter:

$$\eta = \sqrt{\frac{1 - P_{0e}^2}{1 - P_{0b}^2}} [1 + \delta P_e \frac{P_{0b} - P_{0e}}{1 - P_{0e}^2}]. \tag{V.197}$$

If the nonequilibrium spin vanishes, that is, $\delta P_e = 0$, we recover the magnetoamplification factor for the equilibrium spin, Eq. (V.193). In order to illustrate the spin-charge coupling effects in current amplification, let us first consider the case of a magnetic base: $P_{0b} \neq 0, P_{0e} = 0$. The current gain becomes,

$$\beta = \beta_0 \frac{1 + \delta P_e P_{0b}}{\sqrt{1 - P_{0b}^2}}. \tag{V.198}$$

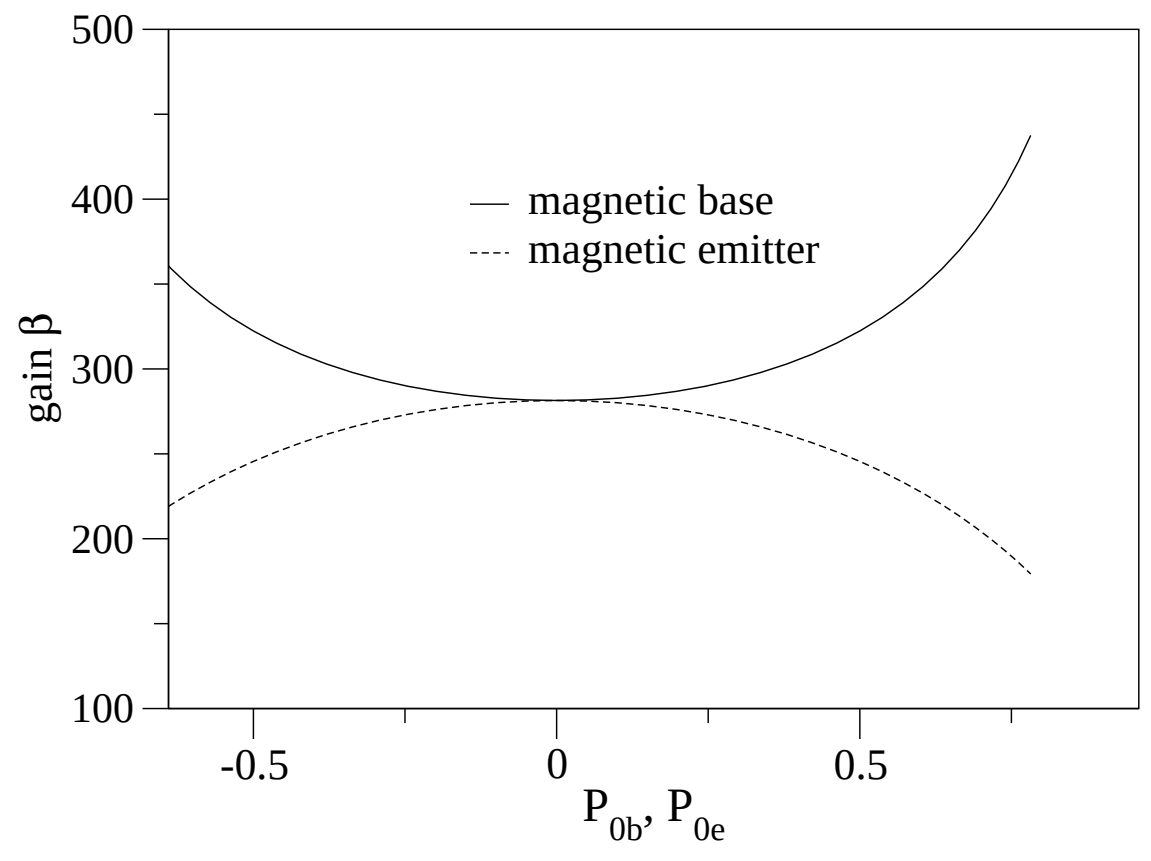

Fig. V.48. Calculated magnetoamplification of a magnetic-base and a magnetic-emitter npn magnetic bipolar transistor. Gain $\beta$ is plotted as a function of the equilibrium spin polarization $P_{0b}$ in the base (solid) and of the equilibrium spin polarization $P_{0e}$ in the emitter (dashed). No source spin is present. After (Fabian and Žutić, 2004a).

The spin-charge coupling is described by the product $\delta P_e P_{0b}$. If the product is positive, the gain increases. If the product is negative, the gain decreases. Where does this behavior come from? We have already encountered similar physics Sec. E.2. If the equilibrium and nonequilibrium spins point in the same direction, the current across the depletion layer increases, since there are more electrons with a lower barrier to overcome. In the context of magnetic transistors, this spin-charge coupling operates at the emitter-base junction. If the nonequilibrium spin in the emitter is aligned with the equilibrium spin in the base, the injection efficiency from the emitter increases, $\gamma'$ decreases, and the gain goes up. If the two spins are antiparallel, the things go in the other direction. To put it briefly, spin-charge coupling controls the emitter injection parameter which, in turn, controls the current gain in the thin base limit.

The effect is opposite if we have a magnetic emitter. Indeed, now the current gain is,

$$\beta = \beta_0(1 - \delta P_e P_{0e})\sqrt{1 - P_{0b}^2}. \tag{V.199}$$

The spin-charge coupling is described by $-\delta P_e P_{0b}$. If the equilibrium and nonequilibrium spins are aligned, now we have a large barrier for a large number of electrons to cross, so emitter injection decreases. If the spins are antiparallel, the situation is reversed. Nevertheless, the control of the gain by spin is equally effective as in the case of the magnetic base.

In order to formalize the above analysis, we call magnetoamplification the effects of spin-charge coupling on the gain. The corresponding giant magnetoamplification factor, defined as,

$$\text{GMA} = \frac{\beta_{\max} - \beta_{\min}}{\beta_{\min}}, \tag{V.200}$$

depends on whether we deal with a magnetic base or emitter. In the above definition $\beta_{\max}$ is the maximum gain, while $\beta_{\min}$ is the minimum gain, as determined by the relative orientation of the equilibrium and nonequilibrium spins. To be specific, for a magnetic base we have,

$$\text{GMA} = \frac{2|\delta P_e P_{0b}|}{1 - |\delta P_e P_{0b}|}. \tag{V.201}$$

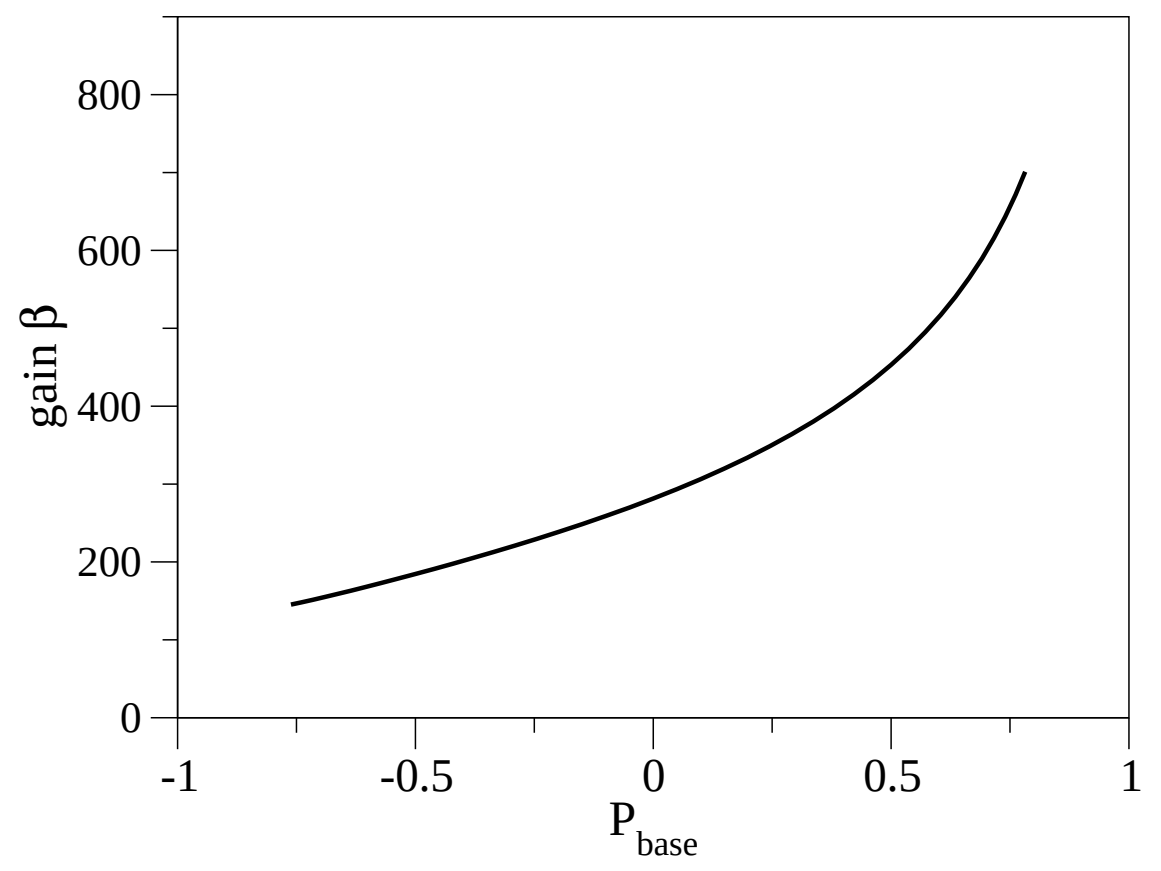

Fig. V.49. Calculated gain $\beta$ for a magnetic-base magnetic bipolar transistor, as a function of the equilibrium spin polarization in the base, $P_{0b}$. The emitter spin polarization, $\delta P_e$, is kept at $\delta P_e = 90\%$. From Ref. (Fabian and Žutić, 2004a).

If the emitter is magnetic, the coefficient becomes,

$$\text{GMA} = \frac{2|\delta P_e P_{0e}|}{1 - P_{0e}^2 + |\delta P_e P_{0be}|}. \tag{V.202}$$

For a typical spin polarization of 50%, one would get a GMA of about 67% for the case of a magnetic base, and 50% for the case of a magnetic emitter.

Figure V.49 illustrates giant magnetoamplification in a realistic calculation using a GaAs-based hypothetical magnetic transistor with a magnetic base.

### E.9 Spin switching

Magnetic bipolar transistors can amplify currents even in the saturation regime, in which both the emitter-base and the collector-base junctions are forward biased, see Tab. V.2. Both the collector current, $j_c$, as well as the current gain, $\beta$, change the sign with the change of the relative orientation of the equilibrium spin in the base and the nonequilibrium spin in the emitter and collector (Fabian and Žutić, 2004b).

Let us first assume that $V_{be} = V_{bc} \gg k_B T$. We also assume that the emitter and the collector doping is much larger than the doping in the base so that we can neglect the hole current (typically in conventional transistors the collector doping is rather small). In the thin base limit (the width of the base is smaller than the electron diffusion length), it can be shown (Fabian and Žutić, 2004b) that the collector current obeys the following relation:

$$j_c \sim P_{0b}(\delta P_e - \delta P_c). \tag{V.203}$$

The sign of the collector current depends on the relative orientation of the equilibrium spin in the base and the nonequilibrium spins in the emitter or collector. This switching behavior can detect the presence of a nonequilibrium spin. In effect, it is the semiconductor counterpart of the Johnson spin switch (Johnson, 1993).

Even more remarkable is the amplification behavior of the transistor in the saturation regime. Conventional transistors do not amplify signal when both junctions are forward biased. Magnetic transistors, on the other hand, do amplify signals. Moreover, the amplification can be negative, meaning, that the amplified collector current can change sign. Indeed, using the above assumptions the current gain is (Fabian and Žutić, 2004b),

$$\beta = \frac{P_{0b}\,(\delta P_e - \delta P_c)}{w_b^2/L_{nb}^2 + j_{ge}^p/j_{gb}^n + j_{gc}^p/j_{gb}^n}. \tag{V.204}$$

The gain is solely due to spin-charge coupling: changing the orientation of either the equilibrium or excess spins the sign of the amplification changes (of course the amplification factor itself is the amplitude of $\beta$). We note that spin switching effects as well as spin-induced amplification should be observable even when the bias voltages across the two junctions are not identical but differ by a factor smaller than the thermal energy (Fabian and Žutić, 2004b).

**Acknowledgement:** We are grateful to I. Adagideli, I. Appelbaum, F. Baruffa, G. Bayreuther, S. C. Erwin, S. Ganichev, S. Giglberger, M. Gmitra, U. Hohenester, E. L. Ivchenko, B. T. Jonker, M. Johnson, F. Koppens, D. Lepine, K. T. Maezawa, J. Moser, S. J. Pearton, A. G. Petukhov, E. I. Rashba, U. Rössler, C. Schüller, L. Vandersypen, T. Waho, and D. Weiss for useful hints and discussions. We thank I. Appelbaum, D. Awschalom, S. Crooker, M. Gmitra, R. Hanson, T. Jungwirth, T. Koga, L. Molenkamp, M. Oestreich, C. Schüller, and M. Tanaka for providing us with illustrative figures of their works. This work was supported by the Deutsche Forschungsgemeinschaft (DFG) through the Sonderforschungsbereich (SFB) 689, Schwerpunktprogramm (SPP) 1285, Graduiertenkolleg (GK) 638, as well as by the US ONR, NSF-ECCS CAREER, CNMS at ORNL and the CCR at SUNY Buffalo.

## References

Abalmassov, V. A., and F. Marquardt, Electron-nuclei spin relaxation through phonon-assisted hyperfine interaction in a quantum dot, 2004, Phys. Rev. B **70**, 75313.

Abolfath, R. M., P. Hawrylak, and I. Žutić, Electronic states of magnetic quantum dots, 2007a, preprint.

Abolfath, R. M., P. Hawrylak, and I. Žutić, Tailoring magnetism in quantum dots, 2007b, Phys. Rev. Lett. **98**, 207203.

Abolfath, R. M., A. G. Petukhov, and I. Žutić, Piezomagnetic quantum dots, 2007c, cond-mat/07072805.

Adagideli, I., G. E. W. Bauer, and B. I. Halperin, Detection of current-induced spins by ferromagnetic contacts, 2006, Phys. Rev. Lett. **97**, 256601.

Affleck, I., J.-S. Caux, and A. M. Zagoskin, Andreev scattering and Josephson current in a one-dimensional electron liquid, 2000, Phys. Rev. B **62**, 1433.

Alcalde, A. M., O. O. Diniz Neto, and G. E. Marques, Spin-flip relaxation due to phonon macroscopic deformation potential in quantum dots, 2005, Microel. J. **36**, 1034.

Alcalde, A. M., Q. Fanayo, and G. E. Marques, Electron-phonon induced spin relaxation in InAs quantum dots, 2004, Physica E **20**, 228.

Aleiner, I. L., and E. L. Ivchenko, Anisotropic exchange splitting in type-II GaAs/AlAs superlattices, 1992, JETP Lett. **55**, 692.

Amasha, S., K. MacLean, I. Radu, D. M. Zumbuhl, M. A. Kastner, M. P. Hanson, and A. C. Gossard, Measurements of the spin relaxation rate at low magnetic fields in a quantum dot, unpublished, cond-mat/0607110.

Anderson, P. W., Localized magnetic states and Fermi-surface anomalies in tunneling, 1966, Phys. Rev. Lett. **17**, 95.

Ando, T., A. B. Fowler, and F. Stern, Electronic properties of two-dimensional systems, 1982, Rev. Mod. Phys. **54**, 437.

Andreev, A. F., The thermal conductivity of the intermediate state in superconductors, 1964, Zh. Eksp. Teor. Fiz. **46**, 1823, [Sov. Phys. JETP **19**, 1228-1231 (1964)].

Appelbaum, I., B. Huang, and J. Monsma, Electronic measurement and control of spin transport in silicon, 2007, Nature **447**, 295.

Appelbaum, J., 's-d' exchange model of zero-bias tunneling anomalies, 1966, Phys. Rev. Lett. **17**, 91.

Aronov, A. G., Spin injection in metals and polarization of nuclei, 1976, Zh. Eksp. Teor. Fiz. Pisma Red. **24**, 37, [JETP Lett. **24**, 32-34 (1976)].

Ashcroft, N. W., and N. D. Mermin, 1976, *Solid State Physics* (Saunders, Philadelphia).

Auth, N., G. Jakob, T. Block, and C. Felser, Spin polarization of magnetoresistive materials by point contact spectroscopy, 2003, Phys. Rev. B **68**, 024403.

Averin, D. V., and Y. V. Nazarov, Virtual electron diffusion during quantum tunneling of the electric charge, 1990, Phys. Rev. Lett. **65**, 2446.

Averkiev, N. S., and L. E. Golub, Giant spin relaxation anisotropy in zinc-blende heterostructures, 1999, Phys. Rev. B **60**, 15582.

Averkiev, N. S., L. E. Golub, and M. Willander, Spin relaxation anisotropy in two-dimensional semiconductor systems, 2002, J. Phys.: Condens. Matter **14**, R271.

Awschalom, D. D., and M. Flatté, Challenges for semiconductor spintronics, 2007, Nature Physics **3**, 153.

Awschalom, D. D., M. E. Flatté, and N. Samarth, Spintronics, 2002, Scientific American, June , 66.

Awschalom, D. D., and J. M. Kikkawa, Electron spin and optical coherence in semiconductors, 1999, Phys. Today **52** (**6**), 33.

Awschalom, D. D., and N. Samarth, Spin dynamics and quantum transport in magnetic semiconductor quantum structures, 1999, J. Magn. Magn. Mater. **200**, 130.

Badalyan, S. M., G. Vignale, and C. S. Kim, Spin coulomb drag beyond the random phase approximation, 2007, cond-mat/07061403.

Badescu, S. C., and T. L. Reinecke, Mixing of two-electron spin states in a semiconductor quantum dot, unpublished, cond-mat/0610405.

Baibich, M. N., J. M. Broto, A. Fert, F. Nguyen Van Dau, F. Petroff, P. Etienne, G. Creuzet, A. Friederich, and J. Chazelas, Giant magnetoresistance of (001)Fe/(001)Cr magnetic superlattices, 1988, Phys. Rev. Lett. **61**, 2472.

Baltzer, P. K., H. W. Lehmann, and M. Robbins, Insulating ferromagnetic spinels, 1965, Phys. Rev. Lett. **15**, 493.

Bandyopadhyay, S., and M. Cahay, Alternate spintronic analog of the electro-optic modulator, 2004a, Appl. Phys. Lett. **85**, 1814.

Bandyopadhyay, S., and M. Cahay, Reexamination of some spintronic field-effect device concepts, 2004b, Appl. Phys. Lett. **85**, 1433.

Bandyopadhyay, S., and M. Cahay, Are spin junction transistors suitable for signal processing?, 2005, Appl. Phys. Lett. **86**, 133502.

Bardarson, J. H., I. Adagideli, and P. Jacquod, Mesoscopic spin hall effect, 2007, Phys. Rev. Lett. **98**, 196601.

Bardeen, J., Tunneling from a many-particle point of view, 1961, Phys. Rev. Lett. **6**, 57.

Bass, J., and W. P. Prat Jr., Spin-diffusion lengths in metals and alloys, and spin-flipping at metal/metal interfaces: an experimentalist's critical review, 2007, J. Phys.: Condens. Matter. **19**, 183201.

Bastard, G., Superlattice band structure in the envelope-function approximation, 1981, Phys. Rev. B **24**, 5693.

Bastard, G., 1998, *Wave Mechanics Applied to Semiconductor Heterostructures* (Les Editions de Physique, Les Ulis).

Bastard, G., J. A. Brum, and R. Ferreira, Electronic states in semiconductor heterostructures, 1991, Solid State Phys. **44**, 229.

Batelaan, H., T. J. Gay, and J. J. Schwendiman, Stern-gerlach effect for electron beams, 1997, Phys. Rev. Lett. **79**, 4517.

Bauer, G. E. W., A. Brataas, Y. Tserkovnyak, B. I. Halperin, M. Zwierzycki, and P. J. Kelly, Dynamic ferromagnetic proximity effect in photoexcited semiconductors, 2004, Phys. Rev. Lett. **92**, 126601.

Bauer, G. E. W., Y. Tserkovnyak, A. Brataas, J. Ren, K. Xia, M. Zwierzycki, and P. J. Kelly, Spin accumulation and decay in magnetic schottky barriers, 2005, Phys. Rev. B **72**, 155304.

Beletskii, N. N., G. P. Berman, and S. A. Borysenko, Controlling the spin polarization of the electron current in a semimagnetic resonant-tunneling diode, 2005, Phys. Rev. B **71**, 125325.

Bergeret, F. S., A. F. V. AF, and K. B. Efetov, Odd triplet superconductivity and related phenomena in superconductor-ferromagnet structures, 2005, Rev. Mod. Phys. **77**, 1321.

Bernardes, E. S., J. Schliemann, J. C. Egues, and D. Loss, Spin-orbit interaction in symmetric wells and cycloidal orbits without magnetic fields, 2006, unpublished, cond-mat/0607218.

Berry, J. J., R. Knobel, O. Ray, W. Peoples, and N. Samarth, Modulation-doped ZnSe/(Zn,Cd,Mn)Se quantum wells and superlattices, 2000, J. Vac. Sci. Tech. B **18**, 1692.

Bertoni, A., M. Rontani, G. Goldoni, and E. Molinari, Reduced electron relaxation rate in multi-electron quantum dots, 2005, Phys. Rev. Lett. **95**, 66806.

Beschoten, B., E. Johnston-Halperin, D. K. Young, M. Poggio, J. E. Grimaldi, S. Keller, S. P. DenBaars, U. K. Mishra, E. L. Hu, and D. D. Awschalom, Spin coherence and dephasing in GaN, 2001, Phys. Rev. B **63**, 121202(R).

Beuneu, F., and P. Monod, The Elliott relation in pure metals, 1978, Phys. Rev. B **18**, 2422.

Binasch, G., P. Grünberg, F. Saurenbach, and W. Zinn, Enhanced magnetoresistance in layered magnetic structures with antiferromagnetic interlayer exchange, 1989, Phys. Rev. B **39**, 4828.

Bir, G. L., A. G. Aronov, and G. E. Pikus, Spin relaxation of electrons due to scattering by holes, 1975, Zh. Eksp. Teor. Fiz. **69**, 1382, [Sov. Phys. JETP **42**, 705-712 (1976)].

Bir, G. L., and G. E. Pikus, 1974, *Symmetry and Strain-Induced Effects in Semiconductors* (Wiley, New York).

Black, W. C., and B. Das, Programmable logic using giant-magnetoresistance and spin-dependent tunneling devices, 2000, J. Appl. Phys. **87**, 6674.

Blonder, G. E., and M. Tinkham, Metallic to tunneling transition in Cu-Nb point contacts, 1983, Phys. Rev. B **27**, 112.

Blonder, G. E., M. Tinkham, and T. M. Klapwijk, Transition from metallic to tunneling regimes in superconducting microconstrictions: Excess current, charge imbalance, and supercurrent conversion, 1982, Phys. Rev. B **25**, 4515.

Blum, K., 1996, *Density matrix theory and applications,* 3rd Ed. (Plenum press, New York).

Bolduc, M., C. Awo-Affouda, A. Stollenwerk, M. B. Huang, F. G. Ramos, G. Agnello, and V. P. LaBella, Above room temperature ferromagnetism in Mn-ion implanted Si, 2005, Phys. Rev. B **71**, 33302.

Bolotin, K. I., F. Kuemmeth, and D. C. Ralph, Anisotropic magnetoresistance and anisotropic tunneling magnetoresistance due to quantum interference in ferromagnetic metal break junctions, 2006, Phys. Rev. Lett. **97**, 127202.

Borhani, M., V. N. Golovach, and D. Loss, Spin decay in a quantum dot coupled to a quantum point contact, 2006, Phys. Rev. B **73**, 155311.

Borza, S., F. M. Peeters, P. Vasilopoulos, and G. Papp, Electric field manipulation of spin states in confined non-magnetic/magnetic heterostructures, 2007, J. Phys.: Condens. Matter **19**, 176221.

Boselli, M. A., A. Ghazali, and I. C. da Cunha Lima, Ferromagnetism and canted spin sphase in AlAs/$Ga_{1-x}Mn_x$As single quantum wells: Monte carlo simulation, 2000, Phys. Rev. B **62**, 8895.

Bourgeois, O., P. Gandit, A. Sulpice, J. Chaussy, J. Lesueur, and X. Grison, Transport in superconductor/ferromagnet/superconductor junctions dominated by interface resistance, 2001, Phys. Rev. B **63**, 064517.

Bouzerar, G., T. Ziman, and J. Kudrnovský, Compensation, interstitial defects, and ferromagnetism in diluted ferromagnetic semiconductors, 2005, Phys. Rev. B. **72**, 125207.

Božović, M., and Z. Radović, Coherent effects in double-barrier ferromagnet/superconductor/ferromagnet junctions, 2002, Phys. Rev. B **66**, 134524.

Braden, J. G., J. S. Parker, P. Xiong, S. H. Chun, and N. Samarth, Direct measurement of the spin polarization of the magnetic semiconductor (Ga,Mn)As, 2003, Phys. Rev. Lett. **91**, 056602.

Brand, M. A., A. Malinowski, O. Z. Karimov, P. A. Marsden, R. T. Harley, A. J. Shields, D. Sanvitto, D. A. Ritchie, and M. Y. Simmons, Precession and motional slowing of spin evolution in a high mobility two-dimensional electron gas, 2002, Phys. Rev. Lett. **89**, 236601.

Brandt, N. B., and V. V. Moshchalkov, Semimagnetic semiconductors, 1984, Adv. Phys. **33**, 193.

Brataas, A., G. E. W. Bauer, and P. J. Kelly, Non-collinear magnetoelectronics, 2006, Phys. Rep. **427**, 157.

Bratkovsky, A. M., Tunneling of electrons in conventional and half-metallic systems: Towards very large magnetoresistance, 1997, Phys. Rev. B **56**, 2344.

Bratkovsky, A. M., and V. V. Osipov, Efficient spin extraction from nonmagnetic semiconductors near forward-biased ferromagnetic-semiconductor modified junctions at low spin polarization of current, 2004, J. Appl. Phys. **96**, 4525.

Braude, V., and Y. V. Nazarov, Fully developed triplet proximity effect, 2007, Phys. Rev. Lett. **98**, 077003.

Braun, M., and J. König, Faraday-rotation fluctuation spectroscopy with static and oscillating magnetic fields, 2007, Phys. Rev. B **75**, 085310.

Brehmer, D. E., K. Zhang, C. J. Schwarz, S. Chau, S. J. Allen, J. P. Ibbetson, J. P. Zhang, C. J. Palmstrøm, and B. Wilkens, Resonant tunneling through ErAs semimetal quantum wells, 1995, Appl. Phys. Lett. **67**, 1268.

Brennan, K. F., Self-consistent analysis of resonant tunneling in a two-barrier-one-well microstructure, 1987, J. Appl. Phys. **62**, 2392.

Breuer, H. P., and F. Petruccione, 2002, *The theory of open quantum systems* (Oxford University Press, New York).

Brey, L., and F. Guinea, Phase diagram of diluted magnetic semiconductor quantum wells, 2000, Phys. Rev. Lett. **85**, 2384.

Brey, L., C. Tejedor, and J. Fernández-Rossier, Tunnel magnetoresistance in GaMnAs: Going beyond Jullière formula, 2004, Appl. Phys. Lett. **85**, 1996.

Brosig, S., K. Ensslin, R. J. Warburton, C. Nguyen, B. Brar, M. Thomas, and H. Kroemer, Zero-field spin splitting in InAs-AlSb quantum wells revisited, 1999, Phys. Rev. B **60**, R13989.

Bruno, P., and J. Wunderlich, Resonant tunneling spin valve: A novel magnetoelectronics device, 1998, J. Appl. Phys. **84**, 978.

Bugoslavsky, Y., Y. Miyoshi, S. K. Clowes, W. R. Branford, M. Lake, I. Brown, A. D. Caplin, and L. F. Cohen, Possibilities and limitations of point-contact spectroscopy for measurements of spin polarization, 2005, Phys. Rev. B **71**, 104523.

Bulaev, D. V., and D. Loss, Spin relaxation and anticrossing in quantum dots: Rashba versus Dresselhaus spin orbit coupling, 2005, Phys. Rev. B **71**, 205324.

Burch, K. S., D. B. Shrekenhamer, E. J. Singley, J. Stephens, B. L. Sheu, R. K. Kawakami, P. Schiffer, N. Samarth, D. D. Awschalom, and D. N. Basov, Impurity band conduction in a high temperature ferromagnetic semiconductor, 2006, Phys. Rev. Lett. **97**, 87208.

Burt, M. G., The justification for applying the effective-mass approximation to microstructures, 1992, J. Phys.: Condens. Matter **4**, 6651.

Butler, W. H., X.-G. Zhang, T. C. Schulthess, and J. M. MacLaren, Spin-dependent tunneling conductance of Fe/MgO/Fe sandwiches, 2001, Phys. Rev. B **63**, 054416.

Büttiker, M., Y. Imry, R. Landauer, and S. Pinhas, Generalized many-channel conductance formula with application to small rings, 1985, Phys. Rev. B **31**, 6207.

Buzdin, A. I., Proximity effects in superconductor-ferromagnet heterostructures, 2005, Rev. Mod. Phys. **77**, 935.

Bychkov, Y. A., and E. I. Rashba, Oscillatory effects and the magnetic susceptibility of carriers in inversion layers, 1984a, J. Phys. C: Solid State Phys. **17**, 6039.

Bychkov, Y. A., and E. I. Rashba, Properties of a 2D electron gas with lifted spectral degeneracy, 1984b, JETP Lett. **39**, 78.

Cahay, M., M. McLennan, S. Datta, and M. S. Lundstrom, Importance of space-charge effects in resonant tunneling devices, 1987, Appl. Phys. Lett. **50**, 612.

Cai, T. Y., and Z. Y. LI, Effects of temperature on good rectifying characteristic of manganite-based p-n junction, 2005, Appl. Phys. Lett. **86**, 3405.

Calero, C., E. M. Chudnovsky, and D. A. Garanin, Field dependence of the electron spin relaxation in quantum dots, 2005, Phys. Rev. Lett. **95**, 166603.

Capasso, F. (ed.), 1990, *Physics of Quantum Electron Devices* (Springer-Verlag, Heidelberg).

Cardona, M., N. E. Christensen, M. Dobrowolska, J. K. Furdyna, and S. Rodríguez, Spin splitting of the conduction band of InSb along [110], 1986a, Solid State Commun. **60**, 17.

Cardona, M., N. E. Christensen, and G. Fasol, Terms linear in k in the band structure of zinc-blende-type semiconductors, 1986b, Phys. Rev. Lett. **56**, 2831.

Cardona, M., N. E. Christensen, and G. Fasol, Relativistic band structure and spin-orbit splitting of zinc-blende-type semiconductors, 1988, Phys. Rev. B **38**, 1806.

Cayssol, J., and G. Montambaux, Exchange-induced ordinary reflection in a single-channel superconductor-ferromagnet-superconductor junction, 2004, Phys. Rev. B **70**, 224520.

Cayssol, J., and G. Montambaux, Incomplete Andreev reflection in a clean superconductor-ferromagnet-superconductor junction, 2005, Phys. Rev. B **71**, 012507.

Cerletti, V., W. A. Coish, O. Gywat, and D. Loss, Recipes for spin-based quantum computing, 2005, Nanotechnology **16**, R27.

Champers, S. A., T. Droubay, C. M. Wang, A. S. Lea, R. F. C. Farrow, L. Folks, V. Deline, and S. Anders, Clusters and magnetism in epitaxial Co-doped $TiO_2$ anatese, 2003, Appl. Phys. Lett. **82**, 1257.

Chang, K., and F. M. Peeters, Spin polarized tunneling through diluted magnetic semiconductor barriers, 2001, Solid State Comm. **120**, 181.

Chang, L. L., E. E. Mendez, and C. Tejedor (eds.), 1991, *Resonant Tunneling in Semiconductors: Physics and Applications* (Plenum Press, New York).

Chantis, A. N., K. D. Belashchenko, E. Y. Tsymbal, and M. van Schilfgaarde, Tunneling anisotropic magnetoresistance driven by resonant surface states: First-principles calculations on an Fe(001) surface, 2007, Phys. Rev. Lett. **98**, 046601.

Chantis, A. N., M. van Schilfgaarde, and T. Kotani, *Ab initio* prediction of conduction band spin splitting in zinc blende semiconductors, 2006, Phys. Rev. Lett. **96**, 086405.

Chen, P., J. Moser, P. Kotissek, J. Sadowski, M. Zenger, D. Weiss, and W. Wegscheider, All electrical measurement of spin injection in a magnetic p-n junction diode, 2006, Phys. Rev. B **74**, 241302(R).

Chen, Z. Y., A. Biswas, J. C. Read, S. B. Ogale, R. L. Greene, T. Venkatesan, A. Gupta, A. Anguelouch, and G. Xiao, $CrO_2$/Ag/YBCO interface study with a flip-chip configuration, 2005, J. Supercond. **18**, 499.

Chen, Z. Y., A. Biswas, I. Žutić, T. Wu, S. B. Ogale, R. L. Greene, and T. Venkatesan, Spin-polarized transport across a $La_{0.7}Sr_{0.3}MnO_3/YBa_2Cu_3O_{7-x}$ interface: Role of Andreev bound states, 2001, Phys. Rev. B **63**, 212508.

Chia, F., J.-L. Liub, L.-L. Sunc, and Y.-J. Gao, Spin-polarized current through a lateral double quantum dot with spin-orbit interaction, 2006, Phys. Lett. A **363**, 302.

Chiba, D., F. Matsukura, and H. Ohno, Tunneling magnetoresistance in (Ga,Mn)As-based heterostructures with a GaAs barrier, 2004, Physica E **21**, 966.

Chitta, V. A., M. Z. Maialle, S. A. L. ao, and M. H. Degani, Spin-polarized current in semimagnetic semiconductor heterostructures, 1999, Appl. Phys. Lett. **74**, 2845.

Cho, S., S. Choi, G.-B. Cha, S. C. Hong, Y. Kim, Y.-J. Zhao, A. J. Freeman, J. B. Ketterson, B. J. Kim, Y. C. Kim, and B.-C. Choi, Room-temperature ferromagnetism in $Zn_{1-x}Mn_xGeP_2$ semiconductors, 2002, Phys. Rev. Lett. **88**, 257203.

Chou, W. C., A. Petrou, J. Warnock, and B. T. Jonker, Spin superlattice behavior in $ZnSe/Zn_{0.99}Fe_{0.01}Se$ quantum wells, 1991, Phys. Rev. Lett. **67**, 3820.

Chow, W. F., 1964, *Principles of Tunnel Diode Circuits* (Wiley, New York).

Ciuti, C., J. P. McGuire, and L. J. Sham, Spin-dependent properties of a two-dimensional electron gas with ferromagnetic gates, 2002, Appl. Phys. Lett. **81**, 4781.

Clifford, E., and J. M. D. Coey, Point contact Andreev reflection by nanoindentation of polymethyl methacrylate, 2006, **89**, 092506.

Climente, J. I., A. Bertoni, G. Goldoni, and E. Molinari, Phonon-induced electron relaxation in weakly confined single and coupled quantum dots, 2006, Phys. Rev. B **74**, 35313.

Climente, J. I., A. Bertoni, G. Goldoni, M. Rontani, and E. Molinari, Triplet-singlet spin relaxation in quantum dots with spin-orbit coupling, 2007, Phys. Rev. B **75**, R081303.

Cohen, M. L., and V. Heine, The fitting of pseudopotentials to experimental data and their subsequent application, 1970, in *Solid State Physics, Vol. 24*, edited by H. Ehrenreich, F. Seitz, and D. Turnbull (Academic Press, New York), p. 37.

Coish, W. A., and D. Loss, Singlet-triplet decoherence due to nuclear spins in a double quantum dot, 2005, Phys. Rev. B **72**, 125337.

Crooker, S. A., M. Furis, X. Lou, C. Adelmann, D. L. Smith, C. J. Palmstrøm, and P. A. Crowell, Imaging spin transport in lateral ferromagnet/semiconductor structures, 2005, Science **309**, 2191.

Crooker, S. A., M. Furis, X. Lou, P. A. Crowell, D. L. Smith, C. Adelmann, and C. J. Palmstrøm, Optical and electrical spin injection and spin transport in hybrid fe/gaas devices, 2007, J. Appl. Phys. **101**, 081716.

Crooker, S. A., and D. L. Smith, Imaging spin flows in semiconductors subject to electric, magnetic, and strain fields, 2005, Phys. Rev. Lett. **94**, 236601.

Dai, N., H. Luo, F. C. Zhang, N. Samarth, M. Dobrowolska, and J. K. Furdyna, Spin superlattice formation in ZnSe/$Zn_{1-x}Mn_xSe$ multilayers, 1991, Phys. Rev. Lett. **67**, 3824.

D'Amico, I., and G. Vignale, Theory of spin coulomb drag in spin-polarized transport, 2000, Phys. Rev. B **62**, 4853.

D'Amico, I., and G. Vignale, Spin coulomb drag in the two-dimensional electron liquid, 2003, Phys. Rev. B **68**, 045307.

Das, B., D. C. Miller, S. Datta, R. Reifenberger, W. P. Hong, P. K. Bhattacharya, J. Singh, and M. Jaffe, Evidence for spin splitting in $In_xGa_{1-x}As/In_{0.52}Al_{0.48}As$ heterostructures as $B \rightarrow 0$, 1989, Phys. Rev. B **39**, 1411.

Das Sarma, S., Spintronics, 2001, Am. Sci. **89**, 516.

Das Sarma, S., J. Fabian, X. Hu, and I. Žutić, Issues, concepts, and challenges in spintronics, 2000a, 58th DRC (Device Research Conference) Conference Digest (IEEE, Piscataway) , 95 cond-mat/0006369.

Das Sarma, S., J. Fabian, X. Hu, and I. Žutić, Spintronics: electron spin coherence, entanglement, and transport, 2000b, Superlattices Microstruct. **27**, 289.

Das Sarma, S., J. Fabian, X. Hu, and I. Žutić, Theoretical perspectives on spintronics and spin-polarized transport, 2000c, IEEE Trans. Magn. **36**, 2821.

Das Sarma, S., J. Fabian, X. Hu, and I. Žutić, Spin electronics and spin computation, 2001, Solid State Commun. **119**, 207.

Das Sarma, S., J. Fabian, and I. Žutić, Spin-polarized bipolar transport and its applications, 2003a, J. Supercond. **16**, 697.

Das Sarma, S., E. H. Hwang, and A. Kaminski, Temperature-dependent magnetization in diluted magnetic semiconductors, 2003b, Phys. Rev. B **67**, 155201.

Datta, S., 1995, *Electronic Transport in Mesoscopic Systems* (Cambridge University Press, Cambridge, England).

Datta, S., Proposal for a "spin capacitor", 2005, Appl. Phys. Lett. **87**, 013115.

Datta, S., and B. Das, Electronic analog of the electro-optic modulator, 1990, Appl. Phys. Lett. **56**, 665.

Daughton, J. M., GMR applications, 1999, J. Magn. Magn. Mater. **192**, 334.

Davydov, A. S., 1976, *Quantum Mechanics* (Pergamon Press, Oxford).

de Andrada e Silva, E. A., Conduction-subband anisotropic spin splitting in III-V semiconductor heterojunctions, 1992, Phys. Rev. B **46**, 1921.

de Andrada e Silva, E. A., G. C. La Rocca, and F. Bassani, Spin-split subbands and magneto-oscillations in III-V asymmetric heterostructures, 1994, Phys. Rev. B **50**, 8523.

de Andrada e Silva, E. A., G. C. La Rocca, and F. Bassani, Spin-orbit splitting of electronic states in semiconductor asymmetric quantum wells, 1997, Phys. Rev. B **55**, 16293.

de Carvalho, H. B., M. J. S. P. Brasil, V. Lopez-Richard, Y. Galvão Gobato, G. E. Marques, I. Camps, L. C. O. Dacal, M. Henini, L. Eaves, and G. Hill, Electric-field inversion asymmetry: Rashba and stark effects for holes in resonant tunneling devices, 2006, Phys. Rev. B **74**, 41305(R).

de Gennes, P. G., 1989, *Superconductivty of Metals and Alloys* (Addison-Wesley, Reading MA).

de Groot, R. A., F. M. Mueller, P. G. van Engen, and K. H. J. Buschow, New class of materials: Half-metallic ferromagnets, 1983, Phys. Rev. Lett. **50**, 2024.

de Jong, M. J. M., and C. W. J. Beenakker, Andreev reflection in ferromagnet-superconductor junctions, 1995, Phys. Rev. Lett. **74**, 1657.

De Teresa, J. M., A. Barthelemy, A. Fert, J. P. Contour, R. Lyonnet, F. Montaigne, P. Seneor, and A. Vaurès, Inverse tunnel magnetoresistance in Co/$SrTiO_3$/$La_{0.7}Sr_{0.3}MnO_3$: New ideas on spin-polarized tunneling, 1999a, Phys. Rev. Lett. **82**, 4288.

De Teresa, J. M., A. Barthélémy, A. Fert, J. P. Contour, F. Montaigne, and P. Seneor, Role of metal-oxide interface in determining the spin polarization of magnetic tunnel junctions, 1999b, Science **286**, 507.

Demler, E. A., G. B. Arnold, and M. R. Beasley, Superconducting proximity effects in magnetic metals, 1997, Phys. Rev. B **55**, 15174.

Deng, C., and X. Hu, Nuclear spin diffusion in quantum dots: Effects of inhomogeneous hyperfine interaction, 2005, Phys. Rev. B **72**, 165333.

Deng, C., and X. Hu, Analytical solution of electron spin decoherence through hyperfine interaction in a quantum dot, 2006, Phys. Rev. B **73**, 241303(R).

Dery, H., L. Cywinski, and L. J. Sham, Spin transference and magnetoresistance amplification in a transistor, 2006, Phys. Rev. B **73**, 161307(R).

Dery, H., P. Dalal, L. Cywinski, and L. J. Sham, Spin-based logic in semiconductors for reconfigurable large-scale circuits, 2007, Nature **447**, 573.

Dery, H., and L. J. Sham, Spin extraction theory and its relevance to spintronics, 2007, Phys. Rev. Lett. **98**, 046602.

Deutscher, G., Andreev-Saint-James reflections: A probe of cuprate superconductors, 2005, Rev. Mod. Phys. **77**, 109.

Dietl, T., Diluted magnetic semiconductors, 1994, in *Handbook of Semiconductors, Vol. 3*, edited by T. S. Moss and S. Mahajan (North-Holland, New York), p. 1251.

Dietl, T., Ferromagnetic semiconductors, 2002, Semicond. Sci. Technol. **17**, 377.

Dietl, T., Semiconductor spintronics, 2007, in *Modern Aspects of Spin Physics*, edited by W. Pötz, J. Fabian, and U. Hohenester (Springer, Berlin), pp. 1–46.

Dietl, T., A. Haury, and Y. M. d'Aubigné, Free carrier-induced ferromagnetism in structures of diluted magnetic semiconductors, 1997, Phys. Rev. B **55**, R3347.

Dietl, T., H. Ohno, F. Matsukura, J. Cibert, and D. Ferrand, Zener model description of ferromagnetism in zinc-blende magnetic semiconductors, 2000, Science **287**, 1019.

Dietzel, A., Hard Disk Drivers, 2003, in *Nanoelectronics and Information Technology*, edited by R. Waser (Wiley-VCH and Co. KGaA, Weinheim), pp. 617–631.

Djayaprawira, D. D., K. Tsunekawa, M. Nagai, H. Maehara, S. Yamagata, N. Watanabe, S. Yuasa, Y. Suzuki, and K. Ando, 230% room-temperature magnetoresistance in CoFeB/MgO/CoFeB magnetic tunnel junctions, 2005, Appl. Phys. Lett. **86**, 092502.

Dobrovitski, V. V., J. M. Taylor, and M. D. Lukin, Long-lived memory for electronic spin in a quantum dot: Numerical analysis, 2006, Phys. Rev. B **73**, 245318.

Dong, Z. C., D. Y. Xing, Z. D. Wang, Z. M. Zheng, and J. M. Dong, Anomaly of zero-bias conductance peaks in ferromagnet/d-wave superconductor junctions, 2001, Phys. Rev. B **63**, 144520.

Dorpe, P. V., Z. Liu, W. V. Roy, and V. F. Motsnyi, Very high spin polarization in GaAs by injection from a (Ga,Mn)As Zener diode, 2004, Appl. Phys. Lett. **84**, 3495.

Dresselhaus, G., Spin-orbit coupling effects in zinc blende structures, 1955, Phys. Rev. **100**, 580.

Dresselhaus, P. D., C. M. A. Papavassiliou, R. G. Wheeler, and R. N. Sacks, Observation of spin precession in GaAs inversion layers using antilocalization, 1992, Phys. Rev. Lett. **68**, 106.

Duke, C. B., Tunneling in solids, 1969, in *Solid State Physics, Supplement 10*, edited by F. Seitz, D. Turnbull, and H. Ehrenreich (Academic, New York).

D'yachenko, A. I., V. N. Krivoruchko, and V. Y. Tarenkov, Andreev spectroscopy of point contacts between a low-temperature superconductor and manganite, 2006, Low Temp. Phys. **32**, 824.

D'yakonov, M. I., and V. I. Perel', Feasibility of optical orientation of equilibrium electrons in semiconductors, 1971a, **13**, 206, [JETP Lett. **13**, 144-146 (1971)].

D'yakonov, M. I., and V. I. Perel', Spin relaxation of conduction electrons in noncentrosymmetric semiconductors, 1971b, Fiz. Tverd. Tela **13**, 3581, [Sov. Phys. Solid State **13**, 3023-3026 (1971)].

Dyson, F. J., Electron spin resonance absorption in metals. II. Theory of electron diffusion and the skin effect, 1955, Phys. Rev. **98**, 349.

Dzhioev, R. I., V. L. Korenev, B. P. Zakharchenya, D. Gammon, A. S. Bracker, J. G. Tischler, and D. S. Katzer, Optical orientation and the Hanle effect of neutral and negatively charged excitons in GaAs/Al$_x$Ga$_{1-x}$As quantum wells, 2002, Phys. Rev. B **66**, 153409.

Eble, B., O. Krebs, A. Lemaitre, K. Kowalik, A. Kudelski, P. Voisin, B. Urbaszek, X. Marie, and T. Amand, Dynamic nuclear polarization of a single charge-tunable InAs-GaAs quantum dot, 2006, Phys. Rev. B **74**, 081306(R).

Efros, A. L., and B. I. Shklovskii, Coulomb gap and low temperature conductivity of disordered systems, 1975, J. Phys. C **8**, L49.

Egues, J. C., Spin-dependent perpendicular magnetotransport through a tunable ZnSe/Zn$_{1-x}$Mn$_x$Se heterostructure: a possible spin filter?, 1998, Phys. Rev. Lett. **80**, 4578.

Egues, J. C., C. Gould, G. Richter, and L. W. Molenkamp, Spin filtering and magnetoresistance in ballistic tunnel junctions, 2001, Phys. Rev. B **64**, 195319.

Elliott, R. J., Theory of the effect of spin-orbit coupling on magnetic resonance in some semiconductors, 1954, Phys. Rev. **96**, 266.

Elsen, M., O. Boulle, J.-M. George, H. Jaffrés, R. Mattana, V. Cros, A. Fert, A. Lemaître, R. Giraud, and G. Faini, Spin transfer experiments on (Ga,Mn)As/(In,Ga)As/(Ga,Mn)As tunnel junctions, 2006, Phys. Rev. B **73**, 035303.

Elzerman, J. M., R. Hanson, L. H. Willem van Beveren, B. Witkamp, L. M. K. Vandersypen, and L. P. Kouwenhoven, Single-shot read-out of an individual electron spin in a quantum dot, 2004, Nature **430**, 431.

Engel, H.-A., V. N. Golovach, D. Loss, L. M. K. Vandersypen, J. M. Elzerman, R. Hanson, and L. P. Kouwenhoven, Measurement efficiency and n-shot readout of spin qubits, 2004, Phys. Rev. Lett. **93**(10), 106804.

Engel, H. A., and D. Loss, Detection of single spin decoherence in a quantum dot via charge currents, 2001, Phys. Rev. Lett. **86**, 4648.

Engel, H. A., E. I. Rashba, and B. I. Halperin, Theory of spin hall effects in semiconductors, 2007, in *Handbook of Magnetism and Advanced Magnetic Materials, Vol. 5*, edited by H. Kronmüller and S. Parkin (Wiley, in press), cond-mat/0603306.

Engels, G., J. Lange, T. Schäpers, and H. Lüth, Experimental and theoretical approach to spin splitting in modulation-doped In$_x$Ga$_{1-x}$As/InP quantum wells for $B \to 0$, 1997, Phys. Rev. B **55**, R1958.

Epstein, R. J., F. M. Mendoza, Y. K. Kato, and D. D. Awschalom, Anisotropic interactions of a single spin and dark-spin spectroscopy in diamond, 2005, Nature Physics **1**, 94.

Erlingsson, S. I., and Y. V. Nazarov, Hyperfine-mediated transitions between a Zeeman split doublet in GaAs quantum dots: The role of internal field, 2002, Phys. Rev. B **66**, 155327.

Erlingsson, S. I., Y. V. Nazarov, and V. I. Falko, Nucleus-mediated spin-flip transitions in GaAs quantum dots, 2001, Phys. Rev. B **64**, 195306.

Ertler, C., and J. Fabian, Proposal for a digital converter of analog magnetic signals, 2006a, Appl. Phys. Lett. **89**, 193507.

Ertler, C., and J. Fabian, Resonant tunneling magneto resistance in coupled quantum wells, 2006b, Appl. Phys. Lett. **89**, 242101.

Ertler, C., and J. Fabian, Theory of digital magneto resistance in ferromagnetic resonant tunneling diodes, 2007, Phys. Rev. B **75**, 195323.

Erwin, S. C., S.-H. Lee, and M. Scheffler, First-principles study of nucleation, growth, and interface structure of Fe/GaAs, 2002, Phys. Rev. B **65**, 205422.

Erwin, S. C., and I. Žutić, Tailoring ferromagnetic chalcopyrites, 2004, Nature Mater. **3**, 410.

Esaki, L., P. Stiles, and S. von Molnár, Magnetointernal field emission in junctions of magnetic insulators, 1967, Phys. Rev. Lett. **19**, 852.

Fabian, J., and S. Das Sarma, Phonon-induced spin relaxation of conduction electrons in aluminum, 1999a, Phys. Rev. Lett. **83**, 1211.

Fabian, J., and S. Das Sarma, Spin relaxation of conduction electrons, 1999b, J. Vac. Sci. Technol. B **17**, 1708.

Fabian, J., and S. Das Sarma, Spin transport in inhomogeneous magnetic fields: a proposal for Stern-Gerlach-like experiments with conduction electrons, 2002, Phys. Rev. B **66**, 024436.

Fabian, J., and U. Hohenester, Entanglement distilation by adiabatic passage in coupled quantum dots, 2005, Phys. Rev. B **72**, 201304(R).

Fabian, J., and I. Žutić, Spin-polarized current amplification and spin injection in magnetic bipolar transistors, 2004a, Phys. Rev. B **69**, 115314.

Fabian, J., and I. Žutić, Spin switch and spin amplifier: magnetic bipolar transistor in the saturation regime, 2004b, Acta Phys. Polonica **A 106**, 109.

Fabian, J., and I. Žutić, The Ebers-Moll model for magnetic bipolar transistors, 2005, Appl. Phys. Lett. **86**, 133506.

Fabian, J., I. Žutić, and S. Das Sarma, Theory of magnetic bipolar transistor, 2002a, cond-mat/0211639.

Fabian, J., I. Žutić, and S. Das Sarma, Theory of spin-polarized bipolar transport in magnetic p-n junctions, 2002b, Phys. Rev. B **66**, 165301.

Fabian, J., I. Žutić, and S. Das Sarma, Magnetic bipolar transistor, 2004, Appl. Phys. Lett. **84**, 85.

Fal'ko, V. I., C. J. Lambert, and A. F. Volkov, Andreev reflections and magnetoresistance in ferromagnet-superconductor mesoscopic structures, 1999, Zh. Eksp. Teor. Fiz. Pisma Red. **69**, 497, [JETP Lett. **69**, 532-538 (1999)].

Fang, Z. L., P. Wu, N. Kundtz, A. M. Chang, X. Y. Liu, and J. K. Furdyna, Spin dependent resonant tunneling through 6 micron diameter double barrier resonant tunneling diode, 2007, cond-mat/07051123 .

Faure-Vincent, J., C. Tiusan, E. Jouguelet, F. Canet, M. Sajieddine, C. Bellouard, E. Popova, M. Hehn, F. Montaigne, and A. Schuhl, High tunnel magnetoresistance of in epitaxial Fe/MgO/Fe tunnel junctions, 2003, Appl. Phys. Lett. **82**, 4507.

Fedichkin, L., and A. Fedorov, Decoherence rate of semiconductor charge qubit coupled to acoustic phonon reservoir, 2004, Phys. Rev. A **69**, 032311.

Fedorov, A. V., A. V. Baranov, I. D. Rukhlenko, and Y. Masumoto, New many-body mechanism of intraband carrier relaxation in quantum dots embedded in doped heterostructures, 2003, Sol. St. Comm. **128**, 219.

Feher, G., Electron spin resonance experiments on donors in silicon. I. electronic structure of donors by the electron nuclear double resonance technique, 1959, Phys. Rev. **114**, 1219.

Feher, G., and A. F. Kip, Electron spin resonance absorption in metals. I. Experimental, 1955, Phys. Rev. **98**, 337.

Felser, C., G. H. Fecher, and B. Balke, Spintronics: A challenge for material science and solid-state chemistry, 2007, Angew. Chem. Int. Ed. **46**, 668.

Feng, J.-F., and S.-J. Xiong, Tunneling resonances and Andreev reflection in transport of electrons through a ferromagnetic metal/quantum dot/superconductor system, 2003, Phys. Rev. B **67**, 045316.

Fernandez-Rossier, J., and L. Brey, Ferromagnetism mediated by few electrons in a semimagnetic quantum dot, 2004, Phys. Rev. Let. **93**, 117201.

Fernández-Rossier, J., and L. J. Sham, Theroy of ferromagnetism in planar heterostructures of (Mn,III)-V semiconductors, 2001, Phys. Rev. B **64**, 235323.

Fernández-Rossier, J., and L. J. Sham, Spin seperation in digital ferromagnetic heterostructures, 2002, Phys. Rev. B **66**, 73312.

Ferry, D. K., and S. M. Goodnick, 1997, *Transport in Nanostructures* (Cambridge University Press, Cambridge).

Fert, A., J. M. George, H. Jaffres, and R. Mattana, Semiconductors between spin-polarized sources and drains, 2007, IEEE Trans. Electronic Devices **54**, 921.

Fert, A., and H. Jaffres, Conditions for efficient spin injection from a ferromagnetic metal into a semiconductor, 2001, Phys. Rev. B **64**, 184420.

Fiederling, R., P. Grabs, W. Ossau, G. Schmidt, and L. W. Molenkamp, Detection of electrical spin injection by light-emitting diodes in top- and side-emission configurations, 2003, Appl. Phys. Lett. **82**, 2160.

Fiederling, R., M. Kleim, G. Reuscher, W. Ossau, G. Schmidt, A. Waag, and L. W. Molenkamp, Injection and detection of a spin-polarized current in a light-emitting diode, 1999, *Nature* **402**, 787.

Filip, A. T., P. LeClair, C. J. P. Smits, J. T. Kohlhepp, H. J. M. Swagten, B. Koopmans, and W. J. M. de Jonge, Spin-injection device based on EuS magnetic tunnel barriers, 2002, Appl. Phys. Lett. **81**, 1815.

Flatté, M. E., Spintronics, 2007, IEEE Trans. Electronic Devices **54**, 907.

Flatté, M. E., and G. Vignale, Unipolar spin diodes and transistors, 2001, Appl. Phys. Lett. **78**, 1273.

Flatté, M. E., Z. G. Yu, E. Johnston-Halperin, and D. D. Awschalom, Theory of semiconductor magnetic bipolar transistors, 2003, Appl. Phys. Lett. **82**, 4740.

Florescu, M., S. Dickman, M. Ciorga, A. Sachrajda, and P. Hawrylak, Spin-orbit interaction and spin relaxation in a lateral quantum dot, 2004, Physica E **22**, 414.

Florescu, M., and P. Hawrylak, Spin relaxation in lateral quantum dots: Effects of spin-orbit interaction, 2006, Phys. Rev. B **73**, 045304.

Fominov, Y. V., Proximity and Josephson effects in superconducting hybrid structures, 2003, Ph.D. Thesis (Universiteit Twente) .

Foreman, B. A., Effective-mass Hamiltonian and boundary conditions for the valence bands of semiconductor microstructures, 1993, Phys. Rev. B **48**, 4964.

Frait, Z., The g-factor in pure polycrystalline iron, 1977, Czech. J. Phys. **B 27**, 185.

Frenkel, D. M., Spin relaxation in GaAs-$Al_xGa_{1-x}As$ heterostructures in high magnetic fields, 1991, Phys. Rev. B **43**, 14228.

Frensley, W. R., Boundary conditions for open quantum systems driven far from equilibrium, 1990, Rev. Mod. Phys. **62**, 745.

Frustaglia, D., M. Hentschel, and K. Richter, Quantum transport in nonuniform magnetic fields: Aharonov-Bohm ring as a spin switch, 2001, Phys. Rev. Lett. **87**, 256602.

Frustaglia, D., M. Hentschel, and K. Richter, Aharonov-Bohm physics with spin. ii. spin-flip effects in two-dimensional ballistic systems, 2004, Phys. Rev. B **69**, 155327.

Frustaglia, D., and K. Richter, Spin interference effects in ring conductors subject to Rashba coupling, 2004, Phys. Rev. B **69**, 235310.

Fujisawa, T., D. G. Austing, Y. Tokura, Y. Hirayama, and S. Tarucha, Allowed and forbidden transitions in artificial hydrogen and helium atoms, 2002a, *Nature* **419**, 278.

Fujisawa, T., Y. Tokura, D. G. Austing, Y. Hirayama, and S. Tarucha, Spin-dependent energy relaxation inside a quantum dot, 2002b, Physica B **314**, 224.

Fujisawa, T., Y. Tokura, and Y. Hirayama, Transient current spectroscopy of a quantum dot in the coulomb blockade regime, 2001, Phys. Rev. B **63**, 081304(R).

Fukumura, T., H. Toyosaki, and Y. Yamada, Magnetic oxide semiconductors, 2005, Semicond. Sci. Technol. **20**, S103.

Furdyna, J. K., Diluted magnetic semiconductors, 1988, J. Appl. Phys. **64**, R29.

Galinon, C., K. Tewolde, R. Loloee, W. C. Chiang, S. Olson, H. Kurt, W. P. Pratt, Jr., J. B. andi P. X. Xu, K. Xia, and M. Talanana, Pd/Ag and Pd/Au interface specific resistances and interfacial spin flipping, 2005, Appl. Phys. Lett. **86**, 182502.

Galperin, Y. M., B. L. Altshuler, J. Bergli, and D. V. Shantsev, Non-gaussian low-frequency noise as a source of qubit decoherence, 2006, Phys. Rev. Lett. **96**, 97009.

Ganguly, S., A. H. MacDonald, L. F. Register, and S. Banerjee, Intrinsic Curie temperature bistability in ferromagnetic semiconductor resonant tunneling diodes, 2006a, Phys. Rev. B **73**, 33310.

Ganguly, S., A. H. MacDonald, L. F. Register, and S. K. Banerjee, Scattering dependence of bias-controlled magnetization switching in ferromagnetic resonant tunneling diodes, 2006b, Phys. Rev. B **74**, 153314.

Ganguly, S., L. F. Register, S. Banerjee, and A. H. MacDonald, Bias-voltage-controlled magnetization switch in ferromagnetic semiconductor resonant tunneling diodes, 2005, Phys. Rev. B **71**, 245306.

Ganichev, S. D., V. V. Bel'kov, L. E. Golub, E. L. Ivchenko, P. Schneider, S. Giglberger, J. Eroms, J. De Boeck, G. Borghs, W. Wegscheider, D. Weiss, and W. Prettl, Experimental separation of Rashba and Dresselhaus spin splittings in semiconductor quantum wells, 2004, Phys. Rev. Lett. **92**, 256601.

Ganichev, S. D., V. V. Belkov, S. A. Tarasenko, S. N. Danilov, S. Giglberger, C. Hoffmann, E. L. Ivchenko, D. Weiss, W. Wegscheider, C. Gerl, S. Schuh, J. Stahl, *et al.*, Zero-bias spin separation, 2006, Nature Physics **2**, 609.

Ganichev, S. D., E. L. Ivchenko, V. V. Belkov, S. A. Tarasenko, M. Sollinger, D. Weiss, W. Wegschelder, and W. Prettl, Spin-galvanic effect, 2002, *Nature* **417**, 153.

Ganichev, S. D., E. L. Ivchenko, S. N. Danilov, J. Eroms, W. Wegscheider, D. Weiss, and W. Prettl, Conversion of spin into directed electric current in quantum wells, 2001, Phys. Rev. Lett. **86**, 4358.

Ganichev, S. D., and W. Prettl, Spin photocurrents in quantum wells, 2003, J. Phys.: Condens. Matter. **15**, R935.

Ganichev, S. D., and W. Prettl, 2006, *Intense Terahertz Excitation of Semiconductors* (Oxford Univerity Press, Oxford).

Garcia-Calderon, G., Tunneling in semiconductor resonant structures, 1993, in *Physics of Low-Dimensional Semiconductor Structures*, edited by P. Butcher, N. H. March, and M. P. Tosi (Plenum Press, New York), p. 267.

Garraway, B. M., and S. Stenholm, Observing the spin of a free electron, 1999, Phys. Rev. A **60**, 63.

Garzon, S., I. Žutić, and R. A. Webb, Temperature-depenedent asymmetry of the nonlocal spin-injection resistance: evidence for spin nonconserving interface scattering, 2005, Phys. Rev. Lett **94**, 176601.

Ghazali, A., I. C. da Cunha Lima, and M. A. Boselli, Hole spin polarization in $Ga_{1-x}Al_xAs$:Mn structures, 2001, Phys. Rev. B **63**, 153305.

Giazotto, F., F. Taddei, R. Fazio, and F. Beltram, Ferromagnetic resonant tunneling diodes as spin polarimeters, 2003, Appl. Phys. Lett. **82**, 2449.

Giddings, A. D., M. N. Khalid, T. Jungwirth, J. Wunderlich, S. Yasin, R. P. Campion, K. W. Edmonds, J. Sinova, K. Ito, K.-Y. Wang, D. Williams, B. L. Gallagher, *et al.*, Very large tunneling anisotropic magnetoresistance of a (Ga,Mn)As/GaAs/(Ga,Mn)As stack, 2005, Phys. Rev. Lett. **94**, 127202.

Giglberger, S., L. E. Golub, V. V. Bel'kov, S. N. Danilov, D. Schuh, C. Gerl, F. Rohlfing, J. Stahl, W. Wegscheider, D. Weiss, W. Prettl, and S. D. Ganichev, Rashba and dresselhaus spin splittings in semiconductor quantum wells measured by spin photocurrents, 2007, Phys. Rev. B **75**, 035327.

Glazov, M. M., P. S. Alekseev, M. A. Odnoblyudov, V. M. Chistyakov, S. A. Tarasenko, and I. N. Yassievich, Spin-dependent resonant tunneling in symmetrical double-barrier structures, 2005, Phys. Rev. B **71**, 155313.

Glazov, M. M., and E. L. Ivchenko, Precession spin relaxation mechanism caused by frequent electron-electron collisions, 2002, Zh. Eksp. Teor. Fiz. Pisma Red. **75**, 476, [JETP Lett. **75**, 403-405 (2002)].

Glazov, M. M., and E. L. Ivchenko, Effect of electron-electron interaction on spin relaxation of charge carriers in semiconductors, 2004, Zh. Eksp. Teor. Fiz. **126**, 1465, [JETP **99**, 1279 (2004)].

Gnanasekar, K., and K. Navaneethakrishnan, Spin-polarized hole transport through a diluted magnetic semiconductor heterostructure with magentic-field modulations, 2006, Europhys. Lett. **73**, 768.

Godfrey, R., and M. Johnson, Spin injection in mesoscopic silver wires: Experimental test of resistance mismatch, 2006, Phys. Rev. Lett. **96**, 136601.

Goldman, V. J., D. C. Tsui, and J. E. Cunningham, Observation of intrinsic bistability in resonant tunneling structures, 1987, Phys. Rev. Lett. **58**, 1256.

Golovach, V. N., A. Khaetskii, and D. Loss, Phonon-induced decay of the electron spin in quantum dots, 2004, Phys. Rev. Lett. **93**(1), 16601.

Golovach, V. N., A. Khaetskii, and D. Loss, Spin relaxation at the singlet-triplet transition in a quantum dot, unpublished, cond-mat/0703427 .

Gorelenko, A. T., V. G. Gruzdov, V. A. Marushchank, and A. N. Titkov, Spin splitting of the conduction band of InP, 1986, Sov. Phys. Semicond. **20**, 216.

Gould, C., C. Rüster, T. Jungwirth, E. Girgis, G. M. Schott, R. Giraud, K. Brunner, G. Schmidt, and L. W. Molenkamp, Tunneling anisotropic magnetoresistance: a spin-valve-like tunnel magnetoresistance using a single magnetic layer, 2004a, Phys. Rev. Lett. **93**, 117203.

Gould, C., A. Slobodskyy, T. Slobodskyy, P. Grabs, C. R. Becker, G. Schmidt, and L. W. Molenkamp, Magnetic resonant tunneling diodes as voltage-controlled spin selectors, 2004b, phys. stat. sol. (b) **241**, 700.

Govorov, A. O., Thomas-Fermi model and ferromagnetic phases of magnetic semiconductor quantum dots, 2005, Phys. Rev. B **72**, 075358.

Gregg, J. F., I. Petej, E. Jouguelet, and C. Dennis, Spin electronics – a review, 2002, J. Phys. D: Appl. Phys. **35**, R121.

Griffin, A., and J. Demers, Tunneling in the normal-metal-insulator-superconductor geometry using the Bogoliubov equations of motion, 1971, Phys. Rev. B **4**, 2202.

Grodecka, A., L. Jacak, P. Machnikowski, and K. Roszak, Phonon impact on the coherent control of quantum states in semiconductor quantum dots, unpublished, cond-mat/0404364 .

Gruber, A., A. Dräbenstedt, C. Tietz, L. Fleury, J. Wrachtrup, and C. von Borczyskowski, Scanning confocal optical microscopy and magnetic resonance on single defect centers, 1997, Science **276**, 2012.

Gruber, T., M. Kein, R. Fiederling, G. Reuscher, W. Ossau, G. Schmidt, and L. W. Molenkamp, Electron spin manipulation using semimagnetic resonant tunneling diodes, 2001, Appl. Phys. Lett. **78**, 1101.

Grünberg, P., Layered magnetic structures: history, facts, and figures, 2001, J. Magn. Magn. Mater. **226-230**, 1688.

Grundler, D., Large Rashba splitting in InAs quantum wells due to electron wave function penetration into the barrier layers, 2000, Phys. Rev. Lett. **84**, 6074.

Guo, Y., B.-L. Gu, H. Wang, and Y. Kawazoe, Spin-resonant suppression and enhancement in ZnSe/$Zn_{1-x}Mn_x$Se multilayer heterostructures, 2001a, Phys. Rev. B **63**, 214415.

Guo, Y., J.-Q. Lu, B.-L. Gu, and Y. Kawazoe, Spin-resonant splitting in magnetically modulated semimagnetic semiconductor superlattices, 2001b, Phys. Rev. B **64**, 155312.

Guo, Y., C.-E. Shang, and X.-Y. Chen, Spin-dependent delay time and the Hartman effect in tunneling through diluted-magnetic-semiconductor/semiconductor heterostructures, 2005, Phys. Rev. B **72**, 45356.

Guo, Y., H. Wang, B.-L. Gu, and Y. Kawazoe, Spin-polarized transport through a ZnSe/$Zn_{1-x}Mn_x$Se heterostructure under an applied electric field, 2000, J. Appl. Phys. **88**, 6614.

Gurvitz, S. A., and G. Kalbermann, Decay width and the shift of a quasistationary state, 1987, Phys. Rev. Lett. **59**, 262.

Gurvitz, S. A., and Y. S. Prager, Microscopic derivation of rate equations for quantum transport, 1996, Phys. Rev. B **53**, 15932.

Gustavsson, F., J.-M. George, V. H. Etgens, and M. Eddrief, Structural and transport properties of epitaxial Fe/ZnSe/FeCo magnetic tunnel junctions, 2001, Phys. Rev. B **64**, 184422.

Guth, M., A. Dinia, G. Schmerber, and H. A. M. van den Berg, Tunnel magnetoresistance in magnetic tunnel junctions with a ZnS barrier, 2001, Appl. Phys. Lett. **78**, 3487.

Hall, K. C., W. H. Lau, K. Gündogdu, and M. E. Flatté, Nonmagnetic semiconductor spin transistor, 2003, Appl. Phys. Lett. **83**, 2937.

Halterman, K., and O. T. Valls, Proximity effects at ferromagnet-superconductor interfaces, 2001, Phys. Rev. B **65**, 014509.

Hanbicki, . A. T., O. M. J. van t Erve, R. Magno, G. Kioseoglou, C. H. Li, B. T. Jonker, G. Itskos, R. Mallory, M. Yasar, and A. Petrou, Analysis of the transport process providing spin injection through an Fe/AlGaAs Schottky barrier, 2003, Appl. Phys. Lett. **82**, 4092.

Hanbicki, A. T., B. T. Jonker, G. Itskos, G. Kioseoglou, and A. Petrou, Efficient electrical spin injection from a magnetic metal/tunnel barrier contact into a semiconductor, 2002, Appl. Phys. Lett. **80**, 1240.

Hanbicki, A. T., R. Mango, S.-F. Cheng, Y. D. Park, A. S. Bracker, and B. T. Jonker, Nonvolatile reprogrammable logic elements using hybrid resonant tunneling diode- giant magnetoresistance circuits, 2001, Appl. Phys. Lett. **79**, 1190.

Hanson, R., O. Gywat, and D. D. Awschalom, Room-temperature manipulation and decoherence of a single spin in diamond, 2006, Phys. Rev. B **74**, 161203(R).

Hanson, R., L. P. Kouwenhoven, J. R. Petta, S. Tarucha, and L. M. K. Vandersypen, Spins in few-electron quantum dots, 2007, Rev. Mod. Phys. (in press); cond-mat/0610433 .

Hanson, R., L. H. W. van Beveren, I. T. Vink, J. M. Elzerman, W. J. M. Naber, F. H. L. Koppens, L. P. Kouwenhoven, and L. M. K. Vandersypen, Single-shot readout of electron spin states in a quantum dot using spin-dependent tunnel rates, 2005, Phys. Rev. Lett. **94**, 196802.

Hanson, R., B. Witkamp, L. M. K. Vandersypen, L. H. W. van Beveren, J. M. Elzerman, and L. P. Kouwenhoven, Zeeman energy and spin relaxation in a one-electron quantum dot, 2003, Phys. Rev. Lett. **91**, 196802.

Hao, X., J. S. Moodera, and R. Meservey, Spin-filter effect of ferromagnetic europium sulfide tunnel barriers, 1990, Phys. Rev. B **42**, 8235.

Harrison, W. A., Tunneling from an independent-particle point of view, 1961, Phys. Rev. **123**, 85.

Haury, A., A. Wasiela, A. Arnoult, J. Cibert, S. Tatarenko, T. Dietl, and Y. M. d' Aubigné, Observation of a ferromagnetic transition induced by two-dimensional hole gas in modulation-doped CdMnTe quantum wells, 1997, Phys. Rev. Lett. **79**, 511.

Havu, P., N. Tuomisto, R. Väänänen, M. J. Puska, and R. M. Nieminen, Spin-dependent electron transport through a magnetic resonant tunneling diode, 2005, Phys. Rev. B **71**, 235301.

Hayashi, T., M. Tanaka, and A. Asamitsu, Tunneling magnetoresistance of a GaMnAs-based double barrier ferromagnetic tunnel junction, 2000, J. Appl. Phys. **87**, 4673.

Hazama, H., T. Sugimasa, T. Imachi, and C. Hamaguchi, Temperature dependence of the effective masses in IIIV semiconductors, 1986, J. Phys. Soc. Jpn. **55**, 1282.

Heersche, H. B., T. Schäpers, J. Nitta, and H. Takayanagi, Enhancement of spin injection from ferromagnetic metal into a two-dimensional electron gas using a tunnel barrier, 2001, Phys. Rev. B **64**, 161307(R).

Heida, J. P., B. J. van Wees, J. J. Kuipers, T. M. Klapwijk, and G. Borghs, Spin-orbit interaction in a two-dimensional electron gas in a InAs/AlSb quantum well with gate-controlled electron density, 1998, Phys. Rev. B **57**, 11911.

Heiliger, C., P. Zahn, B. Y. Yavorsky, and I. Mertig, Influence of the interface structure on the bias dependence of tunneling magnetoresistance, 2005, Phys. Rev. B **72**, 180406(R).

Heiliger, C., P. Zahn, B. Y. Yavorsky, and I. Mertig, Interface structure and bias dependence of Fe/MgO/Fe tunnel junctions: *Ab initio* calculations, 2006, Phys. Rev. B **73**, 214441.

Hentschel, M., H. Schomerus, D. Frustaglia, and K. Richter, Aharonov-Bohm physics with spin I: Geometric phases in one-dimensional ballistic rings, 2004, Phys. Rev. B **69**, 155326.

Hermann, C., and C. Weisbuch, k.p perturbation theory in III-V compounds and alloys: a reexamination, 1977, Phys. Rev. B **15**, 823.

Hershfield, S., and H. L. Zhao, Charge and spin transport through a metallic ferromagnetic-paramagnetic-ferromagnetic junction, 1997, Phys. Rev. B **56**, 3296.

Hirota, E., H. Sakakima, and K. Inomata, 2002, *Giant Magneto-Resistance Devices* (Springer-Verlag, Berlin).

Holub, M., and P. Bhattacharya, Spin-polarized light-emitting diodes and lasers, 2007, J. Phys. D **40**, R179.

Holub, M., S. Chakrabarti, S. Fathpour, P. Bhattacharya, Y. Lei, and S. Ghosh, Mn-doped InAs self-organized diluted magnetic quantum-dot layers with Curie temperatures above 300 K, 2004, Appl. Phys. Lett. **85**, 973.

Hong, J., R. Q. Wu, and D. L. Mills, Many-body effects on the tunneling magnetoresistance of spin valves, 2002, Phys. Rev. B **66**, 100406(R).

Hopkins, M., R. Nicholas, P. Pfeffer, W. Zawadzki, D. Gauthier, J. Portal, and M. DiForte-Poisson, A study of the conduction band non-parabolicity, anisotropy and spin splitting in GaAs and InP, 1987, Semicond. Sci. Technol. **2**, 568.

Hu, C.-M., and T. Matsuyama, Spin injection across a heterojunction: A ballistic picture, 2001, Phys. Rev. Lett. **87**, 066803.

Hu, C.-M., J. Nitta, T. Akazaki, H. Takayanagi, J. Osaka, P. Pfeffer, and W. Zawadzki, Zero-field spin splitting in an inverted $In_{0.53}Ga_{0.47}As/In_{0.52}Al_{0.48}As$ heterostructure: Band nonparabolicity influence and the subband dependence, 1999, Phys. Rev. B **60**, 7736.

Hu, C.-M., J. Nitta, A. Jensen, J. B. Hansen, and H. Takayanagi, Spin-polarized transport in a two-dimensional electron gas with interdigital-ferromagnetic contacts, 2001, Phys. Rev. B **63**, 125333.

Hu, C. R., Midgap surface states as a novel signature for $d_{x_a^2-x_b^2}$-wave superconductivity, 1994, Phys. Rev. Lett. **72**, 1526.

Hu, C.-R., Origin of the zero-bias conductance peaks observed ubiquitously in high-$T_c$ superconductors, 1998, Phys. Rev. B **57**, 1266.

Hu, C.-R., and X.-Z. Yan, Predicted giant magnetic moment on non-$\{$n0m$\}$ surfaces of d-wave superconductors, 1999, Phys. Rev. B **60**, R12573.

Hu, X., and S. Das Sarma, Charge-fluctuation-induced dephasing of exchange-coupled spin qubits, 2006, Phys. Rev. Lett. **96**, 100501.

Huang, B., D. Monsma, and I. Appelbaum, Coherent spin transport through an entire silicon wafer, 2007a, cond-mat/07060866.

Huang, B., D. Monsma, and I. Appelbaum, Experimental realization of a silicon spin-effect transistor, 2007b, cond-mat/07054260.

Huang, B., D. Monsma, and I. Appelbaum, Spin lifetime in silicon in the presence of parasitic electronic effects, 2007c, cond-mat/07043928.

Huang, B., L. Zhao, D. Monsma, and I. Appelbaum, 35 percent magnetocurrent with spin transport through si, 2007d, cond-mat/07043949.

Hübner, J., S. Döhrmann, D. Hägele, and M. Oestreich, Temperature dependent electron Landè g-factor and interband matrix element in GaAs, 2006, cond-mat/0608534.

Hüttel, A. K., J. Weber, A. W. Holleitner, D. Weinmann, K. Eberl, and R. H. Blick, Nuclear spin relaxation probed by a single quantum dot, 2004, Phys. Rev. B **69**, 73302.

Ikeda, S., J. Hayakawa, Y. Lee, R. Sakai, T. Meguro, F. Matsukura, and H. Ohno, Dependence of tunnel magnetoresistance in MgO based magnetic tunnel junctions on Ar pressure during MgO sputtering, 2005, Jpn. J. Appl. Phys. **44**, L1442.

Inoue, J., and S. Makeawa, Effects of spin-flip and magnon-inelastic scattering on tunnel magnetoresistance, 1999, J. Magn. Magn. Mater. **198**, 167.

Ishida, Y., D. D. Sarma, K. Okazaki, J. Okabayashi, J. I. Hwang, H. Ott, A. Fujimori, G. A. Medvedkin, T. Ishibashi, and K. Sato, In situ photoemission study of the room temperature ferromagnet $ZnGeP_2$:Mn, 2003, Phys. Rev. Lett. **91**, 107202.

Ivanov, V. A., T. G. Aminov, V. M. Novotortsev, and V. T. Kalinnikov, Spintronics and spintronic materials, 2004, Russ. Chem. Bull., Int. Ed. **53**, 2357.

Ivchenko, E. L., A. Y. Kaminski, and U. Rössler, Heavy-light hole mixing at zinc-blende (001) interfaces under normal incidence, 1996, Phys. Rev. B **54**, 5852.

Ivchenko, E. L., and G. E. Pikus, 1997, *Superlattices and Other Heterostructures, Symmetry and Optical Phenomena,* 2nd Ed. (Springer, New York).

Izyumov, Y. A., Y. N. Proshin, and M. G. Khusainov, Competition between superconductivty and magnetism in ferromagnet/superconductor heterostructures, 2002, Usph. Fiz. Nauk **172**, 113, [Phys. Usp. **45**, 109-148 (2002)].

Jackson, J. D., 1998, *Classical Electrodynamics, 3rd Ed.* (Wiley, New York).

Jancu, J. M., R. Scholz, E. A. de Andrada e Silva, and G. C. La Rocca, Atomistic spin-orbit coupling and $\mathbf{k} \cdot \mathbf{p}$ parameters in III-V semiconductors, 2005, Phys. Rev. B **72**, 193201.

Jansen, R., and J. S. Moodera, Magnetoresistance in doped magnetic tunnel junctions: Effect of spin scattering and impurity-assisted transport, 2000, Phys. Rev. B **61**, 9047.

Jantsch, W., Z. Wilamowski, N. Sandersfeld, and F. Schäffler, ESR investigations of modulation-doped Si/SiGe quantum wells, 1998, Phys. Stat. Sol. (b) **210**, 643.

Jaroszyński, J., G. Karczewski, J. Wróbel, T. Andrearczyk, A. Strycharczuk, T. Wojtowicz, G. Grabecki, E. Papis, E. Kamińska, A. Piotrowska, and T. Dietl, Temperature and size scaling of the QHE resistance: the case of large spin splitting, 2000, Physica E **6**, 790.

Jedema, F. J., H. B. Heersche, A. T. Filip, J. J. A. Baselmans, and B. J. van Wees, Electrical detection of spin precession in a metallic mesoscopic spin valve, 2002, *Nature* **416**, 713.

Jedema, F. J., M. S. Nijboer, A. T. Filip, and B. J. van Wees, Spin injection and spin accumulation in all-metal mesoscopic spin valves, 2003, Phys. Rev. B **67**, 085319.

Jedema, F. J., B. J. van Wees, B. H. Hoving, A. T. Filip, and T. M. Klapwijk, Spin-accumulation-induced resistance in mesoscopic ferromagnet-superconductor junctions, 1999, Phys. Rev. B **60**, 16549.

Jelezko, F., T. Gaebel, I. Popa, A. Gruber, and J. Wrachtrup, Observation of coherent oscillations in a single electron spin, 2004, Phys. Rev. Lett. **92**, 76401.

Jelezko, F., I. Popa, A. Gruber, C. Tietz, J. Wrachtrup, A. Nizovtsev, and S. Kilin, Single spin states in a defect center resolved by optical spectroscopy, 2002, Appl. Phys. Lett. **81**, 2160.

Jiang, J. H., M. Q. Weng, and M. W. Wu, Intense terahertz laser fields on a quantum dot with Rashba spin-orbit coupling, 2006, J. Appl. Phys. **100**, 63709.

Jin, D., Y. Ren, Z. Li, M. Xiao, G. Jin, and A. Hu, Spin-filter tunneling magnetoresistance in a magnetic tunnel junction, 2006, Phys. Rev. B **73**, 012414.

Johnson, A. C., J. R. Petta, J. M. Taylor, A. Yacobi, M. D. Lukin, C. M. Marcus, M. P. Hanson, and A. C. Gossard, Triplet-singlet spin relaxation via nuclei in a double quantum dot, 2005, Nature **435**, 925.

Johnson, M., Analysis of anomalous multilayer magnetoresistance within the thermomagnetic system, 1991, Phys. Rev. Lett. **67**, 3594.

Johnson, M., Bipolar spin switch, 1993, *Science* **260**, 320.

Johnson, M., Spintronics, 2005, J. Phys. Chem. B **109**, 14278.

Johnson, M., and R. H. Silsbee, Interfacial charge-spin coupling: Injection and detection of spin magnetization in metals, 1985, Phys. Rev. Lett. **55**, 1790.

Johnson, M., and R. H. Silsbee, Thermodynamic analysis of interfacial transport and of the thermomagnetoelectric system, 1987, Phys. Rev. B **35**, 4959.

Johnson, M., and R. H. Silsbee, Coupling of electronic charge and spin at a ferromagnetic-paramagnetic metal interface, 1988, Phys. Rev. B **37**, 5312.

Johnston-Halperin, E., D. Lofgreen, R. K. Kawakami, D. K. Young, L. Coldren, A. C. Gossard, and D. D. Awschalom, Spin-polarized Zener tunneling in (Ga,Mn)As, 2002, Phys. Rev. B **65**, 041306(R).

Jonker, B. T., S. C. Erwin, A. Petrou, and A. G. Petukhov, Electrical spin injection and transport in semiconductor spintronic devices, 2003, MRS Bull. **28**, 740.

Jonker, B. T., G. Kioseoglou, A. T. Hanbicki, C. H. Li, and P. E. Thompson, Electrical spin injection into silicon from ferromagnetic metal/tunnel barrier contact, 2007, Nature Physics **3**, 542.

Jonker, B. T., Y. D. Park, B. R. Bennett, H. D. Cheong, G. Kioseoglou, and A. Petrou, Robust electrical spin injection into a semiconductor heterostructure, 2000, Phys. Rev. B **62**, 8180.

Jonson, M., and A. Grincwaijg, Effect of inelastic scattering on resonant and sequential tunneling in double barrier heterostructures, 1987, Appl. Phys. Lett. **51**, 1729.

Jullierè, M., Tunneling between ferromagnetic films, 1975, Phys. Lett. **54 A**, 225.

Jungwirth, T., W. A. Atkinson, B. H. Lee, and A. H. MacDonald, Interlayer coupling in ferromagnetic semiconductor superlattices, 1999, Phys. Rev. B **59**, 9818.

Jungwirth, T., J. Sinova, J. Mašek, J. Kučera, and A. H. MacDonald, Theory of ferromagnetic (III,Mn)V semiconductors, 2006, Rev. Mod. Phys. **78**, 809.

Jungwirth, T., K. Y. Wang, J. Mašek, K. W. Edmonds, J. König, J. Sinova, M. Polini, N. A. Goncharuk, A. H. MacDonald, M. Sawicki, A. W. Rushforth, R. P. Campion, *et al.*, Prospects for high temperature ferromagnetism in (Ga,Mn)As semiconductors, 2005, Phys. Rev. B **72**, 165204.

Jusserand, B., D. Richards, G. Allan, C. Priester, and B. Etienne, Spin orientation at semiconductor heterointerfaces, 1995, Phys. Rev. B **51**, 4707.

Kalitsov, A., A. Coho, N. Kioussis, A. Vedyayev, M. Chshiev, and A. Granovsky, Impurity-induced tuning of quantum-well states in spin-dependent resonant tunneling, 2004, Phys. Rev. Lett. **93**, 46603.

Kane, E. O., Band structure of indium antimonide, 1957, J. Phys. Chem. Solids **1**, 249.

Kane, E. O., Band structure of narrow gap semiconductors, 1980, Lecture Notes in Physics **133**, 13.

Kashiwaya, S., and Y. Tanaka, Tunnelling effects on surface bound states in unconventional superconductors, 2000, Rep. Prog. Phys. **63**, 1641.

Kashiwaya, S., Y. Tanaka, N. Yoshida, and M. R. Beasley, Spin current in ferromagnet-insulator-superconductor junctions, 1999, Phys. Rev. B **60**, 3572.

Kasuya, T., and A. Yanase, Anomalous transport phenomena in Eu-chalcogenide alloys, 1968, Rev. Mod. Phys. **40**, 684.

Kato, Y. K., R. C. Myers, A. C. Gossard, and D. D. Awschalom, Observation of the spin Hall effect in semiconductors, 2004, Science **306**, 1910.

Kawakami, R. K., E. Johnston-Halperin, L. F. Chen, M. Hanson, N. Gubels, S. Speck, A. C. Gossard, and D. D. Awschalom, (Ga,Mn)As as a digital ferromagnetic heterostructure, 2000, Appl. Phys. Lett. **77**, 2379.

Keim, M., U. Lunz, C. Y. Hu, U. Zehnder, W. Ossau, A. Waag, and G. Landwehr, Semimagnetic II-VI heterostructures for resonant tunneling, 1999, J. Cryst. Growth **201/202**, 711.

Kennedy, T. A., and J. H. Pifer, Electron-paramagnetic-resonance study of metallic Si:P with iron, 1975, Phys. Rev. B **11**, 2017.

Kessler, J., 1985, *Polarized Electrons,* 2nd Ed. (Springer, New York).

Khaetskii, A. V., Spin relaxation in semiconductor mesoscopic systems, 2001, Physica E **10**, 27.

Khaetskii, A. V., D. Loss, and L. Glazman, Electron spin decoherence in quantum dots due to interaction with nuclei, 2002, Phys. Rev. Lett. **88**, 186802.

Khaetskii, A. V., and Y. V. Nazarov, Spin relaxation in semiconductor quantum dots, 2000, Phys. Rev. B **61**, 12639.

Khaetskii, A. V., and Y. V. Nazarov, Spin-flip transitions between Zeeman sublevels in semiconductor quantum dots, 2001, Phys. Rev. B **64**, 125316.

Kidner, C., I. Mehdi, J. R. East, and G. I. Haddad, Bias circuit instabilities and their effect on the d.c. current-voltage characteristics of double-barrier resonant tunneling diodes, 1991, Solid-State Electron. **34**, 149.

Kikkawa, J. M., and D. D. Awschalom, Resonant spin amplification in n-type GaAs, 1998, Phys. Rev. Lett. **80**, 4313.

Kikkawa, J. M., and D. D. Awschalom, Lateral drag of spin coherence in gallium arsenide, 1999, *Nature* **397**, 139.

Kikuchi, K., H. Imamura, S. Takahashi, and S. Maekawa, Conductance quantization and Andreev reflection in narrow ferromagnet/superconductor point contacts, 2001, Phys. Rev. B **65**, 020508(R).

Kim, H. J., and K. S. Yi, Magnetic phase structure of Mn-doped III-V semiconductor quantum wells, 2002, Phys. Rev. B **65**, 193310.

Kim, N., H. Kim, J. W. Kim, S. J. Lee, and T. W. Kang, Numerical self-consitent field calculation of a ferromagnetic ZnMnO quantum well, 2006, Phys. Rev. B **74**, 155327.

Kim, N., J. W. Kim, S. J. Lee, Y. Shon, T. W. Kang, G. Ihm, and T. George, Ferromagnetic properties of Mn-doped III-V semiconductor quantum wells, 2005, J. Supercond. **18**, 189.

Kim, N., S. J. Lee, T. W. Kang, and H. Kim, Curie-temperature modulation by electric fields in Mn $\delta$-doped asymmetric double quantum wells, 2004, Phys. Rev. B **69**, 115308.

Kioseoglou, G., A. T. Hanbicki, J. M. Sullivan, O. M. J. V. Erve, C. H. Li, S. C. Erwin, R. Mallory, M. Yasar, A. Petrou, and B. T. Jonker, Electrical spin injection from an n-type ferromagnetic semiconductor into a III-V device heterostructure, 2004, Nature Materials **3**, 799.

Kittel, C., 1996, *Introduction to Solid State Physics,* 7th Ed. (Wiley, New York).

Klauser, D., W. A. Coish, and D. Loss, Nuclear spin state narrowing via gate-controlled Rabi oscillations in a double quantum dot, 2006, Phys. Rev. B **73**, 205302.

Klimeck, G., R. Lake, R. C. Bowen, W. R. Frensely, and T. S. Moise, Quantum device simulation with a generalized tunneling formula, 1995, Appl. Phys. Lett. **67**, 2539.

Knap, W., C. Skierbiszewski, A. Zduniak, E. Litwin-Staszewska, D. Bertho, F. Kobbi, J. L. Robert, G. E. Pikus, F. G. Pikus, S. V. Iordanskii, V. Mosser, K. Zekentes, *et al.*, Weak antilocalization and spin precession in quantum wells, 1996, Phys. Rev. B **53**, 3912.

Knipp, P. A., and T. L. Reinecke, Coupling between electrons and acoustic phonons in semiconductor nanostructures, 1995, Phys. Rev. B **52**, 5923.

Knobel, R., I. P. Smorchkova, and N. Samarth, Fabrication and characterization of a two-dimensional electron gas in modulation doped ZnTe/$Cd_{1-x}Mn_x$Se quantum wells, 1999, J. Vac. Sci. Technol. B **17**, 1147.

Koga, T., J. Nitta, H. Takayanagi, and S. Datta, Spin-filter device based on the Rashba effect using a nonmagnetic resonant tunneling diode, 2002, Phys. Rev. Lett. **88**, 126601.

Kohda, M., T. Kita, Y. Ohno, F. Matsukura, and H. Ohno, Bias voltage dependence of the electron spin injection studied in a three-terminal device based on a (Ga,Mn)As/$n^+$–GaAs Esaki diode, 2006a, Appl. Phys. Lett. **89**, 12103.

Kohda, M., Y. Ohno, F. Matsukura, and H. Ohno, Effect of $n^+$–GaAs thickness and doping density on spin injection of GaMnAs/$n^+$–GaAs Esaki tunnel junction, 2006b, M. Kohda and Y. Ohno and F. Matsukura and H. Ohno **32**, 438.

Kohda, M., Y. Ohno, K. Takamura, F. Matsukura, and H. Ohno, A spin Esaki diode, 2001, Jpn. J. Appl. Phys. **40**, L1274.

Kolovos-Vellianitis, D., C. Herrmann, A. Trampert, L. Däweritz, and K. H. Ploog, Structural and magnetic properties of (Ga,Mn)As/AlAs multiple quantum wells grown by low-temperature molecular beam epitaxy, 2006, J. Vac. Sci. Technol. B **24**, 1734.

Kondo, T., J. Hayafuji, and H. Munekata, Investigation of spin voltaic effect in a p-n heterojunction, 2006, Jpn. J. Appl. Phys. **45**, L663.

König, J., J. Schliemann, T. Jungwirth, and A. H. MacDonald, Ferromagnetism in (III,Mn)V semiconductors, 2003, in *Electronic Structure and Magnetism of Complex Materials*, edited by D. J. Singh and D. A. Papaconstantopoulos (Springer Verlag, Berlin), p. 163.

König, M., A. Tschetschetkin, E. M. Hankiewicz, J. Sinova, V. Hock, V. Daumer, M. Schäfer, C. R. Becker, H. Buhmann, and L. W. Molenkamp, Direct observation of the Aharonov-Casher phase, 2006, Phys. Rev. Lett. **96**, 076804.

Koppens, F. H. L., C. Buizert, K. J. Tielrooij, I. T. Vink, K. C. Nowack, T. Meunier, L. P. Kouwenhoven, and L. M. K. Vandersypen, Driven coherent oscillations of a single electron spin in a quantum dot, 2006, Nature **442**, 766.

Kotissek, P., M. Bailleul, M. Sperl, A. Spitzer, D. Schuh, W. Wegscheider, C. H. Back, and G. Bayreuther, Cross-sectional imaging of spin injection into a semiconductor, 2007, Nature Physics (in press) .

Krebs, O., D. Rondi, J. L. Gentner, L. Goldstein, and P. Voisin, Inversion asymmetry in heterostructures of zinc-blende semiconductors: Interface and external potential versus bulk effects, 1998, Phys. Rev. Lett. **80**, 5770.

Krebs, O., W. Seidel, J. P. André, D. Bertho, C. Jouanin, and P. Voisin, Investigations of giant ‘forbidden’ optical anisotropy in GaInAs-InP quantum well structures, 1997, Semicond. Sci. Technol. **12**, 938.

Krebs, O., and P. Voisin, Light-heavy hole mixing and in-plane optical anisotropy of InP-$Al_xIn_{1-x}As$ type-II multiquantum wells, 2000, Phys. Rev. B **61**, 7265.

Kreuzer, S., J. Moser, W. Wegscheider, D. Weiss, M. Bichler, and D. Schuh, Spin polarized tunneling through single-crystal GaAs(001) barriers, 2002, Appl. Phys. Lett. **80**, 4582.

Krich, J. J., and B. I. Halperin, Cubic Dresselhaus spin-orbit coupling in 2D electron quantum dots, 2007, cond-mat/0702667.

Kroutvar, M., Y. Ducommun, D. Heiss, M. Bichler, D. Schuh, G. Absteiter, and J. J. Finley, Optically programmable electron spin memory using semiconductor quantum bits, 2004, Nature **432**, 81.

Krstajić, P., and F. M. Peeters, Spin-dependent tunneling in diluted magnetic semiconductors trilayer structures, 2005, Phys. Rev. B **72**, 125350.

Kuroda, S., N. Nishizawa, K. Takita, M. Mitome, Y. Bando, K. Osuch, and T. Dietl, Origin and control of high-temperature ferromagnetism in semiconductors, 2007, Nature Materials **6**, 440.

Kwok, S. H., H. T. Grahn, K. Ploog, and R. Merlin, Giant electropleochroism in GaAs-(Al,Ga)As heterostructures: The quantum-well pockels effect, 1992, Phys. Rev. Lett. **69**, 973.

Lai, C. W., P. Maletinsky, A. Badolato, and A. Imamoglu, Knight-field-enabled nuclear spin polarization in single quantum dots, 2006, Phys. Rev. Lett. **96**, 167403.

Lake, R., G. Klimeck, R. C. Bowen, and D. Jovanovic, Single and multiband modeling of quantum electron transport through layered semiconductor devices, 1997, J. Appl. Phys. **81**, 7845.

Lambert, C. J., and R. Raimondi, Phase-coherent transport in hybrid superconducting nanostructures, 1998, J. Phys.: Condens. Matter **10**, 901.

Lancaster, G., J. A. van Wyk, and E. E. Schneider, Spin-lattice relaxation of conduction electrons in silicon, 1964, Proc. Phys. Soc. **84**, 19.

Landauer, R., Spatial variation of currents and fields due to localized scatterers in metallic conduction, 1957, IBM J. Res. Develop. **1**, 223.

Landauer, R., Electrical resistance of disordered one-dimensional lattices, 1970, Phil. Mag. **21**, 863.

Lassnig, R., $\mathbf{k}.\mathbf{p}$ theory, effective mass approach, and spin splitting for two-dimensional electrons in GaAs-GaAlAs heterostructures, 1985, Phys. Rev. B **31**, 8076.

Lautenschlager, P., M. Garriga, S. Logothetidis, and M. Cardona, Interband critical points of GaAs and their temperature dependence, 1987, Phys. Rev. B **35**, 9174.

Lebedeva, N., and P. Kuivalainen, Modeling of ferromagnetic semiconductor devices for spintronics, 2003, J. Appl. Phys. **93**, 9845.

Lebedeva, N., and P. Kuivalainen, Spin-dependent current through a ferromagnetic resonant tunneling quantum well, 2005, phys. stat. sol. (b) **242**, 1660.

Lee, B., T. Jungwirth, and A. H. MacDonald, Theory of ferromagnetism in diluted semiconductor quantum wells, 2000, Phys. Rev. B **61**, 15606.

Lee, B., T. Jungwirth, and A. H. MacDonald, Ferromagnetism in diluted magnetic semiconductor heterojunction systems, 2002, Semi. Sci. Techn. **17**, 393.

Leger, Y., L. Besombes, J. Fernandez-Rossier, L. Maingault, and H. Mariette, Electrical control of a single Mn atom in a quantum dot, 2006, Phys. Rev. Lett. **97**, 107401.

Lepine, D. J., Spin resonance of localized and delocalized electrons in phosphorus-doped silicon between 20 and 30 K, 1970, Phys. Rev. B **2**, 2429.

Lepine, D. J., 2007, private communication .

Lev, S. B., V. I. Sugakov, and G. V. Vertsimakha, Polarization of electron spin in two barrier system based on semimagnetic semiconductors, 2006, phys. stat. sol. c **3**, 1091.

Leyland, W. J. H., G. H. John, R. T. Harley, M. M. Glazov, E. L. Ivchenko, D. A. Ritchie, I. Farrer, A. J. Shields, and M. Henini, Enhanced spin-relaxation time due to electron-electron scattering in semiconductor quantum wells, 2007, Phys. Rev. B **75**, 165309.

Li, F.-F., Z.-Z. Li, M.-W. Xiao, J. Du, W. Xu, and A. Hu, Bias dependence and inversion of the tunneling magnetoresistance in ferromagnetic junctions, 2004, Phys. Rev. B **69**, 054410.

Li, M. K., N. M. Kim, S. J. Lee, H. C. Jeon, and T. W. Kang, Characteristics of GaMnN based ferromagnetic resonant tunneling diode without external magnetic field, 2006a, Appl. Phys. Lett. **88**, 162102.

Li, T., M. Zhang, X. Song, B. Wang, and H. Yan, Rectifying characteristics of $La_{1-x}Sr_xMnO_3/TiO_2$ ($x$ = 0.2, 0.15, 0.04) heterostructures, 2006b, J. Appl. Phys. **100**, 063711.

Li, T., M. Zhang, B. Wang, and H. Yan, Effect of magnetic film thickness on rectifying properties of $La_{0.8}Sr_{0.2}MnO_3/TiO_2$ heterostructures, 2006c, Solid State Comm. **140**, 289.

Li, W., and Y. Guo, Dresselhaus spin-orbit effect on dwell time of electrons tunneling through double-barrier structures, 2006, Phys. Rev. B **73**, 205311.

Li, Y., B.-Z. Li, W.-S. Zhang, and D.-S. Dai, Tunneling conductance and magnetoresistance of ferromagnet/ferromagnetic insulator (semiconductor) /ferromagnet junctions, 1998, Phys. Rev. B **57**, 1079.

Liang, X.-T., Non-markovian dynamics and phonon decoherence of a double quantum dot charge qubit, 2005, Phys. Rev. B **72**, 245328.

Linder, J., M. S. Gronsleth, and A. Sudbo, Conductance spectra of ferromagnetic superconductors: Quantum transport in a ferromagnetic metal/non-unitary ferromagnetic superconductor junction, 2007, **75**, 054518.

Liu, C., F. Yun, and H. Morkoç, Ferromagnetism of ZnO and GaN, 2005, J. Mater. Sci. - Mater. Electron. **16**, 555.

Liu, S. S., and G. Y. Guo, Voltage-dependence of magnetoresistance in ferromagnetic tunneling junctions: a rigurous free electron model study, 2000, J. Magn. Magn. Mater. **209**, 135.

Lobenhofer, M., J. Moser, D. Schuh, W. Wegscheider, and D. Weiss, 2007, private communication .

Lommer, G., F. Malcher, and U. Rössler, Spin splitting in semiconductor heterostructures for $B \to 0$, 1988, Phys. Rev. Lett. **60**, 728.

Long, D., 1968, *Energy Bands in Semiconductors* (Wiley, New York).

Loss, D., and D. P. DiVincenzo, Quantum computation with quantum dots, 1998, Phys. Rev. A **57**, 120.

Lou, X., C. Adelmann, M. Furis, S. A. Crooker, C. J. Palmstrøm, and P. A. Crowell, Electrical detection of spin accumulation at a ferromagnet-semiconductor interface, 2006, Phys. Rev. Lett. **96**, 176603.

Lu, Y., X. W. Li, G. Xiao, R. A. Altman, W. J. Gallagher, A. Marley, K. Roche, and S. Parkin, Bias voltage and temperature dependence of magnetotunneling effect, 1998, J.Appl. Phys. **83**, 6515.

Luo, J., H. Munekata, F. F. Fang, and P. J. Stiles, Observation of the zero-field spin splitting of the ground electron subband in GaSb-InAs-GaSb quantum wells, 1988, Phys. Rev. B **38**, 10142.

Luo, P. S., H. Wu, F. C. Zhang, C. Cai, X. Y. Qi, X. L. Dong, W. Liu, X. F. Duan, B. Xu, L. X. Cao, X. G. Qiu, and B. R. Zhao, Andreev reflection in $YBa_2Cu_3O_7/La_{0.67}Ca_{0.33}MnO_3$: Existence of midgap states in the 110 surface of $YBa_2Cu_3O_7$, 2005, Phys. Rev. B **71**, 094502.

Luryi, S., Frequence limit of double-barrier resonant-tunneling oscillators, 1985, Appl. Phys. Lett. **47**, 490.

Lv, B., J. Wang, J. Yu, H. Mao, Y. Shen, Z. Zhu, and H. Xing, Enhancement of curie temperature under low electric fields in Mn selectively $\delta$- doped GaAs/AlGaAs wide quantum wells, 2007, Appl. Phys. Lett. **90**, 142513.

Lyanda-Geller, Y. B., I. L. Aleiner, and B. L. Altshuler, Coulomb blockade of nuclear spin relaxation in quantum dots, 2002, Phys. Rev. Lett. **89**, 107602.

MacDonald, A. H., T. Jungwirth, and M. Kasner, Temperature dependence of itinerant electron junction magnetoresistance, 1998, Phys. Rev. Lett. **81**, 705.

MacDonald, A. H., P. Schiffer, and N. Samarth, Ferromagnetic semiconductors: Moving beyond (Ga,Mn)As, 2005, Nature Materials **4**, 195.

Mackowski, S., T. Gurung, T. A. Nguyen, H. E. Jackson, L. M. Smith, G. Karczewski, and J. Kossut, Optically-induced magnetization of CdMnTe self-assembled quantum dots, 2004, Appl. Phys. Lett. **84**, 3337.

MacLaren, J. M., X.-G. Zhang, W. H. Butler, and X. Wang, Layer KKR approach to Bloch-wave transmission and reflection: Application to spin-dependent tunneling, 1999, Phys. Rev. B **59**, 5470.

Madelung, O., 1996, *Semiconductors Basic Data* (Springer, Berlin).

Maekawa, S., and U. Gäfvert, Electron tunneling between ferromagnetic leads, 1982, IEEE Trans. Magn. **18**, 707.

Maekawa, S., S. Takahashi, and H. Imamura, Theory of tunnel magnetoresistance, 2002, in *Spin Dependent Transport in Magnetic Nanostructures*, edited by S. Maekawa and T. Shinjo (Taylor and Francis, New York), pp. 143–236.

Maekawa (Ed.), S., 2006, *Concepts in Spin Electronics* (Oxford University Press).

Maezawa, K., and A. Förster, Quantum transport devices based on resonant tunneling, 2003, in *Nanoelectronics and Information Technology*, edited by R. Waser (Wiley-VCH, Weinheim), pp. 407–424.

Maezawa, K., and T. Mizutani, A new resonant tunneling logic gate employing monostable-bistable transition, 1993, Jpn. J. Appl. Phys. **32**, L42.

Maezawa, K., H. Sugiyama, S. Kishimoto, and T. Mizutani, 100 GHz operation of a resonant tunneling logic gate mobile, 2006, International Conference on InP and Related Materials Conference Proceedings , 46.

Magri, R., and A. Zunger, Anticrossing and coupling of light-hole and heavy-hole states in (001) GaAs/Al$_x$Ga$_{1-x}$As heterostructures, 2000, Phys. Rev. B **62**, 10364.

Mahan, G. D., 2000, *Many-particle Physics* (Kluwer, New York).

Mains, R. K., J. P. Sun, and G. I. Haddad, Observation of intrinsic bistability in resonant tunneling diode modeling, 1989, Appl. Phys. Lett. **55**, 371.

Majewski, J. A., and P. Vogl, Resonant spin-orbit interactions and phonon spin relaxation rates in superlattices, 2003, in *Physics of Semiconductors 2002*, edited by A. R. Long and J. H. Davis (IOP, Bristol), p. 305.

Makler, S. S., M. A. Boselli, J. Weberszpil, X. F. Wang, and I. da Cunha Lima, A resonant tunneling diode based on a Ga$_{1-x}$Mn$_x$As/GaAs double barrier structure, 2002, Physica B **320**, 396.

Malcher, F., G. Lommer, and U. Rössler, Electron states in GaAs/Ga$_{1-x}$Al$_x$As heterostructures: Nonparabolicity and spin-splitting, 1986, Superlatt. Microstruct. **2**, 267.

Malissa, H., W. Jantsch, M. Mühlberger, F. Schäffler, Z. Wilamowski, M. Draxler, and P. Bauer, Anisotropy of g-factor and electron spin resonance linewidth in modulation doped SiGe quantum wells, 2004, Appl. Phys. Lett. **85**, 1739.

Malkova, N., and U. Ekenberg, Spin properties of quantum wells with magnetic barriers. I. A k·p analysis for structures with normal band ordering, 2002, Phys. Rev. B **66**, 155324.

Mal'shukov, A. G., V. V. Shlyapin, and K. A. Chao, Quantum oscillations of spin current through a III-V semiconductor loop, 2002, Phys. Rev. B **66**, 081311(R).

Manassen, Y., R. J. Hamers, J. E. Demuth, and A. J. Castellano, Jr., Direct observation of the precession of individual paramagnetic spins on oxidized silicon surfaces, 1989, Phys. Rev. Lett. **62**, 2531.

Marquardt, F., and V. A. Abalmassov, Spin relaxation in a quantum dot due to Nyquist noise, 2005, Phys. Rev. B **71**, 165325.

Marushchak, V. A., M. N. Stepanova, and A. N. Titkov, Spin relaxation of conduction electrons in moderately doped gallium arsenide crystals, 1983, Sov. Phys. Solid State **25**, 2035.

Mathieu, R., P. Svedlindh, J. Sadowski, K. Światek, M. Karlsteen, J. Kanski, and L. Ilver, Ferromagnetism and interlayer exchange coupling in short-period (Ga,Mn)As/GaAs superlattices, 2002, Appl. Phys. Lett. **81**, 3013.

Mathon, J., Tight-binding theory of tunneling giant magnetoresistance, 1997, Phys. Rev. B **56**, 11810.

Mathon, J., and A. Umerski, Theory of tunneling magnetoresistance of an epitaxial Fe/MgO/Fe(001) junction, 2001, Phys. Rev. B **63**, 220403(R).

Matos-Abiague, A., and J. Fabian, Spin-orbit induced anisotropy in the tunneling magnetoresistance of magnetic tunnel junctions, 2007, cond-mat/0702387.

Matsukura, F., D. Chiba, T. Omiya, E. Abe, T. Dietl, Y. Ohno, K. Ohtani, and H. Ohno, Control of ferromagnetism in field-effect transistor of a magnetic semiconductor, 2002a, Physica E **12**, 351.

Matsukura, F., H. Ohno, and T. Dietl, 2002b, in *Handbook of Magnetic Materials*, edited by K. H. J. Buschow (Elsevier, Amsterdam), volume 14, p. 1.

Matsumoto, Y., M. Murakami, T. Shono, T. Hasegawa, T. Fukumura, M. Kawasaki, P. Ahmet, T. Chikyow, S. Koshihara, and H. Koinuma, Room-temperature ferromagnetism in transparent transition metal-doped titanium dioxide, 2001, Science **291**, 854.

Matsuyama, T., C.-M. Hu, D. Grundler, G. Meier, and U. Merkt, Ballistic spin transport and spin interference in ferromagnetic/InAs(2DES)/ferromagnetic devices, 2002, Phys. Rev. B **65**, 155322.

Matsuyama, T., R. Kürsten, C. Meißner, and U. Merkt, Rashba spin splitting in inversion layers on p-type bulk InAs, 2000, Phys. Rev. B **61**, 15588.

Mattana, R., M. Elsen, J.-M. George, H. Jaffrès, F. N. Van Dau, A. Fert, M. F. Wyczisk, J. Olivier, P. Galtier, B. Lépine, A. Guivarc'h, and G. Jézéquel, Chemical profile and magnetoresistance of $Ga_{1-x}Mn_xAs/GaAs/AlAs/GaAs/Ga_{1-x}Mn_xAs$ tunnel junctions, 2005, Phys. Rev. B **71**, 75206.

Mattana, R., J.-M. George, H. Jaffres, F. N. Van Dau, A. Fert, B. Lepine, A. Guivarc'h, and G. Jezequel, Electrical detection of spin accumulation in a p-type GaAs quantum well, 2003, Phys. Rev. Lett. **90**, 166601.

Mavropoulos, P., N. Papanikolaou, and P. H. Dederichs, Complex band structure and tunneling through ferromagnet/insulator/ferromagnet junctions, 2000, Phys. Rev. Lett. **85**, 1088.

May, S. J., and B. W. Wessels, Electronic and magnetotransport properties of ferromagnetic p(In,Mn)As/n-InAs heterojunctions, 2005, J. Vac. Sci. Technol. B **23**, 1769.

Mayer, H., and U. Rössler, Spin splitting and anisotropy of cyclotron resonance in the conduction band of GaAs, 1991, Phys. Rev. B **44**, 9048.

Mazin, I. I., How to define and calculate the degree of spin polarization in ferromagnets, 1999, Phys. Rev. Lett. **83**, 1427.

Mazin, I. I., A. A. Golubov, and B. Nadgorny, Probing spin polarization with Andreev reflection: A theoretical basis, 2001, J. Appl. Phys. **89**, 7576.

Medvedkin, G. A., T. Ishibashi, T. Nishi, K. Hayata, Y. Hasegawa, and K. Sato, Room temperature ferromagnetism in novel diluted magnetic semiconductor $Cd_{1-x}Mn_xGeP_2$, 2000, Jpn. J. Appl. Phys. **39**, L949.

Meier, F., and B. P. Zakharchenya (Eds.), 1984, *Optical Orientation* (North-Holand, New York).

Melko, R. G., R. S. Fishman, and F. A. Reboredo, Single layer of Mn in a GaAs quantum well: A ferromagnet with quantum fluctuations, 2007, Phys. Rev. B **75**, 115316.

Merkulov, I. A., A. L. Efros, and M. Rosen, Electron spin relaxation by nuclei in semiconductor quantum dots, 2002, Phys. Rev. B **65**, 205309.

Meunier, T., I. T. Vink, L. H. Willems van Beveren, K.-J. Tielrooij, R. Hanson, F. H. L. Koppens, H. P. Tranitz, W. Wegscheider, L. P. Kouwenhoven, and L. M. K. Vandersypen, Experimental signature of phonon-mediated spin relaxation in a two-electron quantum dot, 2007, Phys. Rev. Lett. **98**, 126601.

Miller, J. B., D. M. Zumbühl, C. M. Marcus, Y. B. Lyanda-Geller, D. Goldhaber-Gordon, K. Campman, and A. C. Gossard, Gate-controlled spin-orbit quantum interference effects in lateral transport, 2003, Phys. Rev. Lett. **90**, 076807.

Min, B.-C., K. Motohashi, C. Lodder, and R. Jansen, Tunable spin-tunnel contacts to silicon using low-work-function ferromagnets, 2006, Nature Mater. **5**, 817.

Miyazaki, T., Experiments of tunnel magnetoresistance, 2002, in *Spin Dependent Transport in Magnetic Nanostructures*, edited by S. Maekawa and T. Shinjo (Taylor and Francis, New York), pp. 143–236.

Miyazaki, T., and N. Tezuka, Giant magnetic tunneling effect in $Fe/Al_2O_3/Fe$ junction, 1995, J. Mag. Magn. Mater. **139**, L231.

Miyoshi, Y., Y. Bugoslavsky, and L. F. Cohen, Andreev reflection spectroscopy of niobium point contacts in a magnetic field, 2005, Phys. Rev. B **72**, 012502.

Miyoshi, Y., Y. Bugoslavsky, M. H. Syed, T. Robinson, L. F. Cohen, L. J. Singh, Z. H. Barber, C. E. A. Grigorescu, S. Gardelis, J. Giapintzakis, and W. L. Van Roy, Comparison of free surface polarization of NiMnSb and $Co_2MnSi$, 2006, Appl. Phys. Lett. **88**, 142512.

Mizuno, Y., S. Ohya, P. N. Hai, and M. Tanaka, Spin dependent transport properties in GaMnAs-based spin hot-carrier heterostructures, 2007, cond-mat/0702239.

Mizuta, H., and T. Tanoue, 1995, *The Physics and Applications of Resonant Tunneling Diodes* (Cambridge University Press, Cambridge).

Moodera, J. S., X. Hao, G. A. Gibson, and R. Meservey, Electron-spin polarization in tunnel junctions in zero applied field with ferromagnetic EuS barriers, 1988, Phys. Rev. B **42**, 8235.

Moodera, J. S., L. R. Kinder, T. M. Wong, and R. Meservey, Large magnetoresistance at room temperature in ferromagnetic thin film tunnel junctions, 1995, Phys. Rev. Lett. **74**, 3273.

Moodera, J. S., and G. Mathon, Spin polarized tunneling in ferromagnetic junctions, 1999, J. Mag. Magn. Mater. **200**, 248.

Moodera, J. S., J. Nassar, and G. Mathon, Spin-tunneling in ferromagnetic junctions, 1999, Annu. Rev. Mater. Sci. **29**, 381.

Moodera, J. S., J. Nowak, and R. J. M. van de Veerdonk, Interface magnetism and spin wave scattering in ferromagnet-insulator-ferromagnet tunnel junctions, 1998, Phys. Rev. Lett. **80**, 2941.

Moodera, J. S., T. S. Santos, and T. Nagahama, The phenomena of spin-filter tunnelling, 2007, J. Phys.: Condens. Matter **19**, 165202.

Moon, J. S., D. H. Chow, J. N. Schulman, P. Deelman, J. J. Zinck, and D. Z.-Y. Ting, Experimental demonstration of split side-gated resonant interband tunneling devices, 2004, App. Phys. Lett. **85**, 678.

Moser, J., A. Matos-Abiague, D. Schuh, W. Wegscheider, J. Fabian, and D. Weiss, Tunneling anisotropic magnetoresistance and spin-orbit coupling in Fe/GaAs/Au tunnel junctions, 2007, Phys. Rev. Lett. **99**, 056601.

Moser, J., M. Zenger, C. Gerl, D. Schuh, R. Meier, P. Chen, G. Bayreuther, C.-H. Lai, R.-T. Huang, M. Kosuth, and H. Ebert, Spin-filter tunneling magnetoresistance in a magnetic tunnel junction, 2006, Appl. Phys. Lett. **89**, 162106.

Motomitsu, E., M. Hirano, H. Yanagi, T. Kamiya, and H. Hosono, Bipolar room temperature ferromagnetic semiconductor LaMnOP, 2005, Jpn. J. Appl. Phys. **44**, L1344.

Mott, N. F., The scattering of fast electrons by atomic nuclei, 1929, Proc. R. Soc. London, Ser. A **124**, 425.

Mott, N. F., and H. S. W. Massey, 1965, *The Theory of Atomic Collisions,* 3rd Ed. (Clarendon, Oxford).

Munekata, H., H. Ohno, S. von Molnár, A. Segmüller, L. L. Chang, and L. Esaki, Diluted magnetic III-V semiconductors, 1989, Phys. Rev. Lett. **63**, 1849.

Murakami, S., N. Nagosa, and S.-C. Zhang, Dissipationless quantum spin current at room temperature, 2003, *Science* **301**, 1348.

Myers, R. C., M. Poggio, N. P. Stern, A. C. Gossard, and D. D. Awschalom, Antiferromagnetic $s - d$ exchange coupling in GaMnAs, 2005, Phys. Rev. Lett. **95**, 17204.

Nadgorny, B., I. I. Mazin, M. Osofsky, R. J. Soulen, Jr., P. Broussard, R. M. Stroud, D. J. Singh, V. G. Harris, A. Arsenov, and Y. Mukovskii, Origin of high transport spin polarization in $La_{0.7}Sr_{0.3}MnO_3$: Direct evidence for minority spin states, 2001, Phys. Rev. B **63**, 184433.

Nakagawa, N., M. Asai, Y. Mukunoki, T. Susaki, and H. Y. Hwang, Magnetocapacitance and exponential magnetoresistance in manganite-titanate heterojunctions, 2005, Appl. Phys. Lett. **86**, 082504.

Nazmul, A. M., T. Amemiya, Y. Shuto, S. Sugahara, and M. Tanaka, High temperature ferromagnetism in GaAs-based heterostructure with Mn $\delta$ doping, 2005, Phys. Rev. Lett. **95**, 17201.

Nazmul, A. M., S. Kobayashi, S. Sugahara, and M. Tanaka, Control of ferromagnetism in Mn delta-doped GaAs-based semiconductor heterostructures, 2004, Physica E **21**, 937.

Nazmul, A. M., S. Sugahara, and M. Tanaka, Ferromagnetism and high curie temperature in semiconductor heterostructures with Mn $\delta$-doped GaAs and $p$-type selective doping, 2003, Phys. Rev. B **67**, 241308(R).

Ngai, J., Y. C. Tseng, P. Morales, J. Y. T. Wei, F. Chen, and D. D. Perovic, Scanning tunneling spectroscopy under pulsed spin injection, 2004, Appl. Phys. Lett. **84**, 1907.

Nielsen, M. A., and I. L. Chuang, 2000, *Quantum Computation and Quantum Information* (Cambridge University, Cambridege/New York).

Nikolic, B. K., S. Souma, L. P. Zarbo, and J. Sinova, 2005, Phys. Rev. Lett. **95**, 046601.

Nikolic, B. K., L. Zarbo, and S. Souma, Imaging mesoscopic spin hall flow: Spatial distribution of local spin currents and spin densities in and out of multiterminal spin-orbit coupled semiconductor nanostructures, 2006, Phys. Rev. B **73**, 075303.

Nitta, J., T. Akazaki, H. Takayanagi, and T. Enoki, Gate control of spin-orbit interaction in an inverted $In_{0.53}Ga_{0.47}As/In_{0.52}Al_{0.48}As$ heterostructure, 1997, Phys. Rev. Lett. **78**, 1335.

Nitta, J., T. Akazaki, H. Takayanagi, and T. Enoki, Gate control of spin-orbit interaction in an InAs-inserted $In_{0.53}Ga_{0.47}/In_{0.52}Al_{0.48}As$ heterostructure, 1998, Physica E **2**, 527.

Nonoyama, S., and J. Inoue, Spin-dependent transport in a double barrier structure with a ferromagnetic material emitter, 2001, Physica E **10**, 283.

Ochiai, Y., and E. Matsuura, ESR in heavily doped n-type silicon near a metal-nonmetal transition, 1976, Phys. Stat. Sol. (a) **38**, 243.

Ochiai, Y., and E. Matsuura, Spin-lattice relaxation at high temperatures in heavily doped n-type silicon, 1978, Phys. Stat. Sol. (a) **45**, K101.

Oestreich, M., M. Römer, R. J. Haug, and D. Hägele, Spin noise spectroscopy in gaas, 2005, Phys. Rev. Lett. **95**, 216603.

Oestreich, M., and W. W. Rühle, Temperature dependence of the electron Landé g-factor in GaAs, 1995, Phys. Rev. Lett. **74**, 2315.

Ohkawa, F. J., and Y. Uemura, Quantized surface states of a narrow-gap semiconductor, 1974, J. Phys. Soc. Jpn. **37**, 1325.

Ohnishi, H., T. Inata, S. Muto, N. Yokoyama, and A. Shibatomi, Self-consistent analysis of resonant tunneling current, 1986, Appl. Phys. Lett. **49**, 1248.

Ohno, H., Properties of ferromagnetic III-V semiconductors, 1999, J. Magn. Magn. Mater. **200**, 110.

Ohno, H., Semiconductor spin electronics, 2002, JSAP International **5**, 4.

Ohno, H., N. Akiba, F. Matsukura, A. Shen, K. Ohtani, and Y. Ohno, Spontaneous splitting of ferromagnetic (Ga,Mn)As valence band observed by resonant tunneling spectroscopy, 1998, Appl. Phys. Lett. **73**, 363.

Ohno, H., D. Chiba, F. Matsukura, T. O. E. Abe, T. Dietl, Y. Ohno, and K. Ohtani, Electric-field control of ferromagnetism, 2000, *Nature* **408**, 944.

Ohno, H., H. Munekata, T. Penney, S. von Molnár, and L. L. Chang, Magnetotransport properties of p-type (In,Mn)As diluted magnetic III-V semiconductors, 1992, Phys. Rev. Lett. **68**, 2664.

Ohno, H., A. Shen, F. Matsukura, A. Oiwa, A. End, S. Katsumoto, and Y. Iye, (Ga,Mn)As: A new diluted magnetic semiconductor based on GaAs, 1996, Appl. Phys. Lett. **69**, 363.

Ohno, Y., I. Arata, F. Matsukura, and H. Ohno, Valance band barrier at (Ga,Mn)As/GaAs interfaces, 2002, Physica E **13**, 521.

Ohno, Y., D. K. Young, B. Beschoten, F. Matsukura, H. Ohno, and D. D. Awschalom, Electrical spin injection in a ferromagnetic semiconductor heterostructure, 1999, *Nature* **402**, 790.

Ohya, S., P. N. Hai, Y. Mizuno, and M. Tanaka, Resonant tunneling effect and tunneling magnetoresistance in GaMnAs quantum-well double barrier heterostructures, 2006, phys. stat. sol. (c) **3**, 4184.

Ohya, S., P. N. Hai, Y. Mizuno, and M. Tanaka, Quantum size effect and tunneling magnetoresistance in ferromagnetic-semiconductor quantum heterostructures, 2007, Phys. Rev. B **75**, 155328.

Ohya, S., P. N. Hai, and M. Tanaka, Tunneling magnetoresistance in GaMnAs/AlAs/InGaAs/ /AlAs/GaMnAs double-barrier magnetic tunnel junctions, 2005, Appl. Phys. Lett. **87**, 12105.

Oiwa, A., R. Moriya, Y. Kashimura, and H. Munekata, Formation of quantized states and spin dynamics in III-V based ferromagnetic quantum wells, 2004, J. Magn. Magn. Mater. **272-276**, 2016.

Oliveira, E. J. R., A. T. d. Lima, M. A. B. G. M. Sipahi, S. C. P. Rodrigues, and I. C. d. Lima, Spin-polarized transport in ferromagnetic multilayered semiconductor nanostructures, 2007, Appl. Phys. Lett. **90**, 112102.

Ono, K., and S. Tarucha, Nuclear-spin-induced oscillatory current in spin-blockaded quantum dots, 2004, Phys. Rev. Lett. **92**, 256803.

Onoda, M., and N. Nagaosa, Dynamics of localized spins coupled to the conduction electrons with charge and spin currents, 2006, Phys. Rev. Lett. **96**, 66603.

Osipov, V. V., and A. M. Bratkovsky, Spin accumulation in degenerate semiconductors near modified schottky contact with ferromagnetics: spin injection and extraction, 2005, Phys. Rev. B **72**, 115322.

Osipov, V. V., A. G. Petukhov, and V. N. Smelyankiy, Complete polarization of electrons in semiconductor layers and quantum dots, 2005, Appl. Phys. Lett. **87**, 202112.

Overberg, M. E., C. R. Abernathy, S. J. Pearton, N. A. Theodoropoulou, K. T. McCarthy, and A. F. Hebard, Indication of ferromagnetism in molecular-beam-epitaxy-derived n-type GaMnN, 2001, Appl. Phys. Lett. **79**, 1312.

Panguluri, R., B. Nadgorny, T. Wojtowicz, W. L. Lim, X. Liu, and J. K. Furdyna, Measurement of spin polarization by Andreev reflection in $In_{1-x}Mn_xSb$ epilayers, 2004, Appl Phys. Lett. **84**, 4947.

Panguluri, R. P., K. C. Ku, T. Wojtowicz, X. Liu, J. K. Furdyna, Y. B. Lyanda-Geller, N. Samarth, and B. Nadgorny, Andreev reflection and pair-breaking effects at the superconductor/magnetic semiconductor interface, 2005, Phys. Rev. B **72**, 054510.

Panguluri, R. P., G. Tsoi, B. Nadgorny, S. H. Chun, N. Samarth, and I. I. Mazin, Point contact spin spectroscopy of ferromagnetic MnAs epitaxial films, 2003, Phys. Rev. B **68**, 201307(R).

Pannetier, B., and H. Courtois, Andreev reflection and proximity effect, 2000, J. Low Temp. Phys. **118**, 599.

Papp, G., S. Borza, and F. M. Peeters, Spin transport in a Mn-doped ZnSe asymmetric tunnel structure, 2005, J. Appl. Phys. **97**, 113901.

Papp, G., S. Borza, and F. M. Peeters, Spin transport through a ZnSe-based diluted magnetic semiconductor resonant tunneling structure in the presence of electric and magnetic fields, 2006, phys. stat. sol (b) **243**, 1956.

Pappert, K., M. J. Schmidt, S. Hümpfner, C. Rüster, G. M. Schott, K. Brunner, C. Gould, G. Schmidt, and L. W. Molenkamp, Magnetization-switched metal-insulator transition in a (Ga,Mn)As tunnel device, 2006, Phys. Rev. Lett. **97**, 186402.

Park, Y. D., A. T. Hanbicki, S. C. Erwin, C. S. Hellberg, J. M. Sullivan, J. E. Mattson, T. F. Ambrose, A. Wilson, G. Spanos, and B. T. Jonker, A group IV ferromagnetic semiconductor: $Mn_xGe_{1-x}$, 2002a, Science **295**, 651.

Park, Y. D., A. T. Hanbicki, J. E. Mattson, and B. T. Jonker, Epitaxial growth of an $n$-type ferromagnetic semiconductor $CdCr_2Se_4$ on GaAs(001) and GaP(001), 2002b, Appl. Phys. Lett. **81**, 1471.

Parker, J. S., S. M. Watts, P. G. Ivanov, and P. Xiong, Spin polarization of $CrO_2$ at and across an artificial barrier, 2002, Phys. Rev. Lett. **88**, 196601.

Parkin, S. S. P., Applications of magnetic nanostructures, 2002, in *Spin Dependent Transport in Magnetic Nanostructures*, edited by S. Maekawa and T. Shinjo (Taylor and Francis, New York), pp. 237–271.

Parkin, S. S. P., C. Kaiser, A. Panchula, P. M. Rice, B. Hughes, M. Samant, and S. Yang, Giant tunneling magnetoresistance at room temperature with MgO (100) tunnel barriers, 2004, Nat. Mater. **3**, 862.

Pearton, S., C. Abernathy, G. T. Thaler, R. Frazier, F. Ren, A. F. Hebard, Y. D. Park, D. Norton, W. Tang, M. Stavola, J. M. Zavada, and R. Wilson, Effects of defects and doping on wide band gap ferromagnetic semiconductors, 2003a, Physica B **340-342**, 39.

Pearton, S. J., C. R. Abernathy, D. Norton, A. F. Hebard, Y. D. Park, L. A. Boatner, and J. D. Budai, Advances in wide bandgap materials for semicoductor spintronics, 2003b, Mat. Sci. Eng. R **40**, 137.

Pearton, S. J., C. R. Abernathy, M. E. Overberg, G. T. Thaler, D. P. Norton, N. Theodoropoulou, A. F. Hebard, Y. D. Park, F. Ren, J. Kim, and L. A. Boatner, Wide band gap ferromagnetic semiconductors and oxides, 2003c, J. Appl. Phys. **93**, 1.

Pearton, S. J., M. E. Overberg, G. T. Thaler, C. R. Abernathy, J. Kim, F. Ren, N. Theodoropoulou, A. F. Hebard, and Y. D. Park, Room temperature ferromagnetism in GaMnN and GaMnP, 2003d, phys. stat. sol.(a) **195**, 222.

Perel', V. I., S. A. Tarasenko, I. N. Yassievich, S. D. Ganichev, V. V. Bel'kov, and W. Prettl, Spin-dependent tunneling through a symmetric semiconductor barrier, 2003, Phys. Rev. B **67**, 201304(R).

Pershin, Y. V., Drift-diffusion approach to spin-polarized transport, 2004, Physica E **23**, 226.

Petković, I., N. M. Chtchelkatchev, and Z. Radović, Resonant amplification of the Andreev process in ballistic Josephson junctions, 2006, Phys. Rev. B **73**, 184510.

Petta, J. R., A. C. Johnson, J. M. Taylor, E. A. Laird, A. Yacoby, M. D. Lukin, C. M. Marcus, M. P. Hanson, and A. C. Gossard, Coherent manipulation of coupled electron spins in semiconductor quatum dots, 2005, Science **309**, 2180.

Petta, J. R., A. C. Johnson, A. Yacoby, C. M. Marcus, M. P. Hanson, and A. C. Gossard, Pulsed-gate measurement of the singlet-triplet relaxation time in a two-electron double quantum dot, unpublished, cond-mat/0412048 .

Petukhov, A. G., A. N. Chantis, and D. O. Demchenko, Resonant enhancement of tunneling magnetoresistance in double-barrier magnetic heterostructures, 2002, Phys. Rev. Lett. **89**, 107205.

Petukhov, A. G., D. O. Demchenko, and A. N. Chantis, Spin-dependent resonant tunneling in double-barrier magnetic heterostructures, 2000, J. Vac. Sci. Technol. B **18**, 2109.

Petukhov, A. G., D. O. Demchenko, and A. N. Chantis, Electron spin polarization in resonant interband tunneling devices, 2003, Phys. Rev. B **68**, 125332.

Petukhov, A. G., W. R. L. Lambrecht, and B. Segall, Spin-dependent resonant tunneling through semimetallic ErAs quantum wells in a magnetic field, 1996, Phys. Rev. B **53**, 3646.

Petukhov, A. G., I. Žutić, and S. C. Erwin, Thermodynamics of carrier-mediated magnetism in semiconductors, 2007, cond-mat/07054464.

Pfeffer, P., Spin splitting of conduction energies in GaAs-$Ga_{0.7}Al_{0.3}As$ heterostructures at $b = 0$ and $b \neq 0$ due to inversion asymmetry, 1997, Phys. Rev. B **55**, R7359.

Pfeffer, P., and W. Zawadzki, Conduction electrons in gaas: Five-level $k \cdot p$ theory and polaron effects, 1990, Phys. Rev. B **41**, 1561.

Pfeffer, P., and W. Zawadzki, Spin splitting of conduction subbands in GaAs-$Ga_{0.7}Al_{0.3}As$ heterostructures, 1995, Phys. Rev. B **52**, R14332.

Pfeffer, P., and W. Zawadzki, Five-level $k \cdot p$ model for the conduction and valence bands of GaAs and InP, 1996, Phys. Rev. B **53**, 12813.

Pfeffer, P., and W. Zawadzki, Spin splitting of conduction subbands in III-V heterostructures due to inversion asymmetry, 1999, Phys. Rev. B **59**, R5312.

Pfleiderer, C., G. J. McMullan, S. R. Julian, and G. G. Lonzarich, Magnetic quantum phase transition in MnSi under hydrostatic pressure, 1997, Phys. Rev. B **55**, 8330.

Pfund, A., D. Bercioux, and K. Richter, Coherent spin ratchets, 2006, cond-mat/0601118.

Picozzi, S., Engineering ferromagnetism, 2004, Nature Materials **3**, 349.

Pierce, D. T., and R. J. Celotta, Applications of polarized electron sources utilizing optical orientation in solids, 1984, in *Optical Orientation, Modern Problems in Condensed Matter Science, Vol. 8*, edited by F. Meier and B. P. Zakharchenya (North-Holland, Amsterdam), pp. 259–294.

Pierce, D. T., R. J. Celotta, J. Unguris, and H. C. Siegmann, Spin-dependent elastic scattering of electrons from a ferromagnetic glass, $Ni_{40}Fe_{40}B_{20}$, 1982, Phys. Rev. B **26**, 2566.

Pifer, J. H., Microwave conductivity and conduction-electron spin-resonance linewidth of heavily doped Si:P and Si:As, 1975, Phys. Rev. B **12**, 4391.

Pikus, G. E., V. A. Marushchak, and A. N. Titkov, Spin splitting of energy bands and spin relaxation of carriers in cubic III-V crystals, 1988, Sov. Phys. Semicond. **22**, 115, [Fiz. Tekh. Poluprovodn. 22, 185-200, (1988)].

Popescu, V., H. Ebert, N. Papanikolaou, R. Zeller, and P. H. Dederichs, Spin-dependent transport in ferromagnet/semiconductor/ferromagnet junctions: a fully relativistic approach, 2004, J. Phys.: Condens. Matter **16**, S5579.

Popescu, V., H. Ebert, N. Papanikolaou, R. Zeller, and P. H. Dederichs, Influence of spin-orbit interaction on the transport properties of magnetic tunnel junctions, 2005, Phys. Rev. B **72**, 184427.

Pötz, W., Self-consistent model of transport in quantum well tunneling structures, 1989, J. Appl. Phys. **66**, 2458.

Prinz, G. A., Magnetoelectronics applications, 1999, J. Magn. Magn. Mater. **200**, 57.

Priour, Jr., D. J., E. H. Hwang, and S. Das Sarma, Quasi-two-dimensional diluted magnetic semiconductor systems, 2005, Phys. Rev. Lett. **95**, 37201.

Qi, Y., D. Y. Xing, and J. Dong, Relation between Julliere and Slonczewski models of tunneling magnetoresistance, 1998, Phys. Rev. B **58**, 2783.

Quirt, J. D., and J. R. Marko, Absolute spin suceptibilities and other ESR parameters of heavily doped n-type silicon. i. metallic samples., 1972, Phys. Rev. B **5**, 1716.

Rashba, E. I., Theory of electrical spin injection: Tunnel contacts as a solution of the conductivity mismatch problem, 2000, Phys. Rev. B **62**, R16267.

Rashba, E. I., Diffusion theory of spin injection through resisitive contacts, 2002, Eur. Phys. J. B **29**, 513.

Rashba, E. I., Semiconductor spintronics: Progress and challenges, 2006, cond-mat/0611194.

Raychaudhuri, P., A. P. Mackenzie, J. W. Reiner, and M. R. Beasley, Transport spin polarization in $SrRuO_3$ measured through point-contact Andreev reflection, 2003, Phys. Rev. B **67**, 020411(R).

Reimann, S. M., and M. Manninen, Electronic structure of quantum dots, 2002, Rev. Mod. Phys. **74**, 1283.

Ren, C., J. Trbovic, R. L. Kallaher, J. G. Braden, J. S. Parker, S. von Molnár, and P. Xiong, Measurement of the spin polarization of the magnetic semiconductor EuS with zero-field and Zeeman-split Andreev reflection spectroscopy, 2007, Phys. Rev. B **75**, 205208.

Ricco, B., and M. Y. Azbel, Physics of resonant tunneling. the one-dimensional double-barrier case, 1984, Phys. Rev. B **29**, 1970.

Richards, D., B. Jusserand, G. Allan, C. Priester, and B. Etienne, Electron spin-flip Raman scattering in asymmetric quantum wells: spin orientation, 1996, Solid-State Electron. **40**, 127.

Rippard, W. H., A. C. Perrella, F. J. Albert, and R. A. Buhrman, Ultrathin aluminum oxide tunnel barriers, 2002, Phys. Rev. Lett. **88**, 046805.

Romo, R., and S. E. Ulloa, Dynamic polarization tunneling: A spin filtering mechanism, 2005, Phys. Rev. B **72**, 121305(R).

Rössler, U., Nonparabolicity and warping in the conduction band of GaAs, 1984, Solid State Commun. **49**, 943.

Rössler, U., 2004, *Solid State Theory* (Springer, Berlin).

Rössler, U., and J. Kainz, Microscopic interface asymmetry and spin-splitting of electron subbands in semiconductor quantum structures, 2002, Solid State Commun. **121**, 313.

Rudolph, J., D. Hägele, H. M. Gibbs, G. Khitrova, and M. Oestreich, Laser threshold reduction in a spintronic device, 2003, Appl. Phys. Lett. **82**, 4516.

Rugar, D., R. Budakian, H. J. Mamin, and B. W. Chui, Single spin detection by magnetic resonance force microscopy, 2004, Nature **430**, 329.

Rüster, C., C. Gould, T. Jungwirth, J. Sinova, G. M. Schott, R. Giraud, K. Brunner, G. Schmidt, and L. W. Molenkamp, Very large tunneling anisotropic magnetoresistance of a (Ga,Mn)As/GaAs/(Ga,Mn)As stack, 2005, Phys. Rev. Lett. **94**, 027203.

Saffarzadeh, A., Tunnel magnetoresistance in double spin filter junctions, 2003, J. Phys.: Condens. Matter **15**, 3041.

Saffarzadeh, A., M. Bahar, and M. Banihasan, Spin-dependent tunneling in ZnSe/ZnMnSe heterostructures, 2005, Physica E **27**, 462.

Saffarzadeh, A., and A. A. Shokri, Quantum theory of tunneling magnetoresistance in GaMnAs/GaAs/GaMnAs heterostructures, 2006, J. Magn. Magn. Mater. **305**, 141.

Saikin, S., M. Shen, M. C. Cheng, and V. Privman, Semiclassical Monte-Carlo model for in-plane transport of spin-polarized electrons in III-V heterostructures, 2003, J. Appl. Phys. **94**, 1769.

Saito, H., S. Yuasa, and K. Ando, Origin of the tunnel anisotropic magnetoresistance in $Ga_{1-x}Mn_xAs/ZnSe/Ga_{1-x}Mn_xAs$ magnetic tunnel junctions of II-VI/III-V heterostructures, 2005, Phys. Rev. Lett. **95**, 086604.

Saito, H., V. Zayets, S. Yamagata, and K. Ando, Room-temperature ferromagnetism in a II-VI diluted magnetic semiconductor $Zn_{1-x}Cr_xTe$, 2003, Phys. Rev. Lett. **90**, 207202.

Sakurai, J. J., 1963, *Advanced Quantum Mechanics* (Addison-Wesley, Reading, Massachusetts).

Sakurai, J. J., 1994, *Modern Quantum Mechanics* (Addison-Wesley, Reading, Massachusetts).

San-Jose, P., G. Zarand, A. Shnirman, and G. Schön, Geometrical spin dephasing in quantum dots, 2006, Phys. Rev. Lett. **97**, 76803.

Sánchez, D., C. Gould, G. Schmidt, and L. W. Molenkamp, Spin-polarized transport in II-VI magnetic resonant tunneling devices, 2007, IEEE Trans. Electron Devices **54**, 1.

Sánchez, D., A. H. MacDonald, and G. Platero, Field-domain spintronics in magnetic semiconductor multiple quantum wells, 2001, Phys. Rev. B **65**, 35301.

Sánchez, D., A. H. MacDonald, and G. Platero, Non-linear spin transport in magentic semiconductor multiple quantum wells, 2002, Physica E **13**, 525.

Sanderfeld, N., W. Jantsch, Z. Wilamowski, and F. Schäffler, ESR investigations of modulation-doped Si/SiGe quantum wells, 2000, Thin Solid Films **369**, 312.

Sankowski, P., P. Kacman, J. Majewski, and T. Dietl, Tight-binding model of spin-polarized tunneling in (Ga,Mn)As-based structures, 2006, Physica E **32**, 375.

Sankowski, P., P. Kacman, J. A. Majewski, and T. Dietl, Spin-dependent tunneling in modulated structures of (Ga,Mn)As, 2007, PRB **75**, 45306.

Santiago, R. B., and L. G. Guimarães, Spin dependent resonant tunneling between coupled levels in parabolic wells under crossed fields, 2003, Phys. Rev. B **67**, 193301.

Sasa, S., K. Anjiki, T. Yamaguchi, and M. Inoue, Electron transport in a large spin-spliiting 2DEG in InAs/AlGaSb heterostructures, 1999, Physica B **272**, 149.

Sasaki, S., T. Fujisawa, T. Hayashi, and Y. Hirayama, Electrical pump-and probe study of spin singlet-triplet relaxation in a quantum dot, 2005, Phys. Rev. Lett. **95**, 056803.

Sasaki, T., S. Sonoda, Y. Yamamoto, K. Suga, S. Shimizu, K. Kindo, and H. Hori, Magnetic and transport characteristics on high curie temperature ferromagnet of Mn-doped GaN, 2002, J. Appl. Phys. **91**, 7911.

Schallenberg, T., and H. Munekata, Preparation of ferromagnetic (In,Mn)As with a high curie temperature of 90 k, 2006, Appl. Phys. Lett. **89**, 042507.

Scheid, M., M. Wimmer, D. Bercioux, and K. Richter, Zeeman ratchets for ballistic spin currents, 2006, phys. stat. sol. (c) **3**, 4235.

Schiff, L. I., 1968, *Quantum Mechanics, 3rd (Ed).* (McGraw-Hill, New York).

Schliemann, J., Spin Hall effect, 2006, Int. J. Mod. Phys. B **20**, 1015.

Schliemann, J., J. C. Egues, and D. Loss, Nonballistic spin-field-effect transistor, 2003, Phys. Rev. Lett. **90**, 146801.

Schliemann, J., D. Loss, and R. M. Westervelt, Zitterbewegung of electronic wave packets in III-V zinc-blende semiconductor quantum wells, 2005, Phys. Rev. Lett. **94**, 206801.

Schliemann, J., D. Loss, and R. M. Westervelt, Zitterbewegung of electrons and holes in III-V semiconductor quantum wells, 2006, Phys. Rev. B **73**, 085323.

Schmidt, G., Concepts for spin injection into semiconductors - a review, 2005, J. Phys. D: Appl. Phys. **38**, R107.

Schmidt, G., D. Ferrand, L. W. Molenkamp, A. T. Filip, and B. J. van Wees, Fundamental obstacle for electrical spin injection from a ferromagnetic metal into a diffusive semiconductor, 2000, Phys. Rev. B **62**, R4790.

Schreiber, L., M. Heidkamp, T. Rohleder, B. Beschoten, and G. Güntherodt, Mapping of spin lifetimes to electronic states in n-type GaAs near the metal-to-insulator transition, 2007, cond-mat/07061884.

Semenov, Y. G., H. Enaya, and K. W. Kim, Bistability in a magnetic and nonmagnetic double-quantum-well structure mediated by the magnetic phase transition, 2005, App. Phys. Lett. **86**, 73107.

Semenov, Y. G., and K. W. Kim, Phonon-mediated electron spin phase diffusion in a quantum dot, 2004, Phys. Rev. Lett. **92**, 026601.

Semenov, Y. G., and K. W. Kim, Elastic spin-relaxation processes in semiconductor quantum dots, 2007, Phys. Rev. B **75**, 195342.

Semiconductor Industry Association, The international technology roadmap for semiconductors, 2004 edition. sematech: Austin, TX, 2004.

Semiconductor Industry Association, The international technology roadmap for semiconductors, 2005 edition. sematech: Austin, TX, 2005.

Semiconductor Industry Association, The international technology roadmap for semiconductors, 2006 update. sematech: Austin, TX, 2006.

Sengupta, K., I. Žutić, H.-J. Kwon, V. M. Yakovenko, and S. Das Sarma, Midgap edge states and pairing symmetry of quasi-one-dimensional organic superconductors, 2001, Phys. Rev. B **63**, 144531.

Shang, C. H., J. Nowak, R. Jansen, and J. S. Moodera, Temperature dependence of magnetoresistance and surface magnetization in ferromagnetic tunnel junctions, 1998, Phys. Rev. B **58**, R2917.

Sharma, M., S. X. Wang, and J. H. Nickel, Inversion of spin polarization and tunneling magnetoresistance in spin-dependent tunneling junctions, 1999, Phys. Rev. Lett. **82**, 616.

Sharma, P., A. Gupta, K. V. Rao, F. J. Owens, R. Sharma, R. Ahuja, J. M. O. Guillen, B. Johansson, and G. A. Gehring, Ferromagnetism above room temperature in bulk and transparent thin films of Mn-doped ZnO, 2003, Nature Materials **2**, 673.

Sharvin, Y. V., A possible method for studying Fermi surfaces, 1965, Zh. Eksp. Teor. Fiz. **48**, 984, [Sov. Phys. JETP **21**, 655-656 (1965)].

Shick, A. B., F. Máca, J. Mašek, and T. Jungwirth, Prospect for room temperature tunneling anisotropic magnetoresistance effect: density of states anisotropies in CoPt systems, 2006, Phys. Rev. B **73**, 024418.

Shimizu, H., and M. Tanaka, Blueshift of magneto-optical spectra and ferromagnetic ordering, 2002, J. Appl. Phys **91**, 7487.

Shirley, J. H., Solution of the Schrödinger equation with a Hamiltonian periodic in time, 1965, Phys. Rev. **138**, 979.

Shklovskii, B. I., Dyakonov-Perel spin relaxation near metal-insulator transition and in hopping transport, 2006, Phys. Rev. B **73**, 193201.

Sih, V., and D. D. Awschalom, Electrical manipulation of spin-orbit coupling in semiconductor heterostructures, 2007, J. Appl. Phys. **101**, 081710.

Sih, V., Y. Kato, and D. D. Awschalom, A hall of spin, 2005, Physics World **18**, 33.

Sih, V., W. H. Lau, R. C. Myers, V. R. Horowitz, A. C. Gossard, and D. D. Awschalom, Generating spin currents in semiconductors with the spin Hall effect, 2006, Phys. Rev. Lett. **97**, 096605.

Silsbee, R. H., Novel method for the study of spin transport in conductors, 1980, Bull. Magn. Reson. **2**, 284.

Singh, J., 1993, *Physics of Semiconductors and their Heterostructures* (McGraw-Hill, New York).

Sinova, J., D. Culcer, Q. Niu, N. A. Sinitsyn, T. Jungwirth, and A. H. MacDonald, Universal intrinsic spin Hall effect, 2004, Phys. Rev. Lett. **92**, 126603.

Slaughter, J. M., M. DeHerrera, and H. Dürr, Magnetoresistive RAM, 2003, in *Nanoelectronics and Information Technology*, edited by R. Waser (Wiley-VCH and Co. KGaA, Weinheim), pp. 591–606.

Slichter, C. P., 1996, *Principles of Magnetic Resonance,* 3rd Ed. (Springer, Berlin).

Slobodskyy, A., C. Gould, T. Slobodskyy, C. R. Becker, G. Schmidt, and L. W. Molenkamp, Voltage-controlled spin selection in a magnetic resonant tunneling diode, 2003, Phys. Rev. Lett. **90**, 246601.

Slobodskyy, A., C. Gould, T. Slobodskyy, G. Schmidt, L. W. Molenkamp, and D. Sánchez, Resonant tunneling diode with spin polarized injector, 2007, Appl. Phys. Lett. **90**, 122109.

Slonczewski, J. C., Conductance and exchange coupling of two ferromagnets separated by a tunneling barrier, 1989, Phys. Rev. B **39**, 6995.

Sluiter, M. H. F., Y. Kawazoe, P. Sharma, A. Inoue, A. R. Raju, C. Rout, and U. V. Waghmare, First principles based design and experimental evidence for a ZnO-based ferromagnet at room temperature, 2005, Phys. Rev. Lett. **94**, 187204.

Smirnov, S., D. Bercioux, and M. Grifoni, Bloch's theory in periodic structures with Rashba's spin-orbit interaction, 2007, cond-mat/07053830.

Smorchkova, I. P., N. Samarth, J. M. Kikkawa, and D. D. Awschalom, Spin transport and localizatin in a magnetic two-dimensional electron gas, 1997, Phys. Rev. Lett. **78**, 3571.

Smorchkova, I. P., N. Samarth, J. M. Kikkawa, and D. D. Awschalom, Giant magnetoresistance and quantum phase transitions in strongly localized magnetic two-dimensional electron gases, 1998, Phys. Rev. B **58**, R4238.

Sobkowicz, P., Theory of n-inversion layers in narrow gap semiconductors: the role of the boundary conditions, 1990, Semicond. Sci. Technol. **5**, 183.

Soulen Jr., R. J., J. M. Byers, M. S. Osofsky, B. Nadgorny, T. Ambrose, S. F. Cheng, P. R. Broussard, C. T. Tanaka, J. Nowak, J. S. Moodera, A. Barry, and J. M. D. Coey, Measuring the spin polarization of a metal with a superconducting point contact, 1998, *Science* **282**, 85.

Souma, S., and B. K. Nikolic, Modulating unpolarized current in quantum spintronics: Visibility of spin-interference effects in multichannel aharonov-casher mesoscopic rings, 2004, Phys. Rev. B **70**, 195346.

Souma, S., and B. K. Nikolic, Spin Hall current driven by quantum interferences in mesoscopic rashba rings, 2005, Phys. Rev. Lett. **94**, 106602.

Stano, P., and J. Fabian, Spin-orbit effects in single-electron states in coupled quantum dots, 2005, Phys. Rev. B **72**, 155410.

Stano, P., and J. Fabian, Orbital and spin relaxation in single and coupled quantum dots, 2006a, Phys. Rev. B **74**, 45320.

Stano, P., and J. Fabian, Theory of phonon-induced spin relaxation in laterally coupled quantum dots, 2006b, Phys. Rev. Lett. **96**, 186602.

Stano, P., and J. Fabian, Control of electron spin and orbital resonance in quantum dots through spin-orbit interactions, unpublished, cond-mat/0611228 .

Stavrou, V. N., and X. Hu, Charge decoherence in laterally coupled quantum dots due to electron-phonon interaction, 2005, Phys. Rev. B **72**, 075362.

Stepanenko, D., G. Burkard, G. Giedke, and A. Imamoglu, Enhancement of electron spin coherence by optical preparation of nuclear spins, 2006, Phys. Rev. Lett. **96**, 136401.

Stephens, J., J. Berezovsky, J. P. McGuire, L. J. Sham, A. C. Gossard, and D. D. Awschalom, Spin accumulation in forward-biased MnAs/GaAs Schottky diodes, 2004, Phys. Rev. Lett. **93**, 097602.

Stewart, D. A., and M. van Schilfgaarde, Digitally doped magnetic resonant tunneling devices, 2003, J. Appl. Phys. **93**, 7355.

Stich, D., J. Zhou, T. Korn, R. Schulz, D. Schuh, W. Wegscheider, M. W. Wu, and C. Schüller, Effect of initial spin polarization on spin dephasing and the electron g factor in a high-mobility two-dimensional electron systems, 2007, Phys. Rev. Lett. **98**, 176401.

Stone, A. D., and P. A. Lee, Effect of inelastic processes on resonant tunneling in one dimension, 1985, Phys. Rev. Lett. **54**, 1196.

Sugahara, S., and M. Tanaka, A spin metal-oxide-semiconductor field-effect transistor using half-metallic-ferromagnet contacts for the source and drain, 2004, Appl. Phys. Lett. **84**, 2307.

Sugahara, S., and M. Tanaka, A spin metal-oxide-semiconductor field-effect transistor (spin MOSFET) with a ferromagnetic semiconductor for the channel, 2005, J. Appl. Phys. **97**, 10D503.

Sugakov, V. I., and S. A. Yatskevich, Electron tunneling in parallel electric and magnetic fields through a double-barrier heterojunction doped with magnetic impurities, 1992, Sov. Tech. Phys. Lett. **18**, 134.

Sun, J. J., R. C. Sousa, T. T. P. Galvão, V. Soares, T. S. Plaskett, and P. P. Freitas, Tunneling magnetoresistance and current distribution effect in spin-dependent tunnel junctions, 1998, J. Appl. Phys. **83**, 6694.

Tamborenea, P. I., D. Weinmann, and R. A. Jalabert, Relaxation mechanism for electron spin in the impurity band of n-doped semiconductors, 2007, cond-mat/0701329.

Tanaka, M., and Y. Higo, Large tunneling magnetoresistance in GaMnAs/AlAs/GaMnAs ferromagnetic semiconductor tunnel junctions, 2001, Phys. Rev. Lett. **87**, 026602.

Tanaka, Y., and S. Kashiwaya, Theory of tunneling spectroscopy of $d$-wave superconductors, 1995, Phys. Rev. Lett. **74**, 3451.

Tanner, C. E., T. Williams, S. Schwall, S. T. Ruggiero, P. Shaklee, S. Potashnik, J. M. Shaw, and C. M. Falco, Magneto-optic effects in ferromagnetic films: Implications for spin devices, 2006, Optics Commun. , 704.

Tarasenko, S. A., and N. S. Averkiev, Interference of spin splittings in magneto-oscillation phenomena in two-dimensional systems, 2002, JETP Lett. **75**, 669.

Tedrow, P. M., and R. Meservey, Spin polarization of electrons tunneling from films of Fe, Co, Ni, Gd, 1973, Phys. Rev. B **7**, 318.

Tezuka, N., and T. Miyazaki, Barrier height dependence of MR ratio in Fe/Al-oxide/Fe junctions, 1998a, J. Magn. Magn. Mater. **177-181**, 1283.

Tezuka, N., and T. Miyazaki, Temperature and applied voltage dependence of magnetoresistance ratio in Fe/Al oxide/Fe junctions, 1998b, Jpn. J. Appl. Phys. **37**, L218.

Thaler, G. T., M. E. Overberg, B. Gila, R. Frazier, C. R. Abernathy, S. J. Pearton, J. S. Lee, S. Y. Lee, Y. D. Park, Z. G. Khim, J. Kim, and F. Ren, Magnetic properties of n-GaMnN thin films, 2002, Appl. Phys. Lett. **80**, 3964.

Theodoropoulou, N., A. F. Hebard, M. E. Overberg, C. R. Abernathy, S. J. Pearton, S. N. G. Chu, and R. G. Wilson, Unconventional carrier-mediated ferromagnetism above room temperature in ion-implanted (Ga,Mn)P:C, 2002, Phys. Rev. Lett. **89**, 107203.

Ting, D. Z.-Y., and X. Cartoixà, Resonant interband tunneling spin filter, 2002, Appl. Phys. Lett. **81**, 4198.

Ting, D. Z.-Y., and X. Cartoixà, Bulk inversion asymmetry enhancement of polarization efficiency in nonmagnetic resonant-tunneling spin filters, 2003, Phys. Rev. B **68**, 235320.

Ting, D. Z.-Y., and X. Cartoixà, Device concepts based on spin-dependent transmission in semiconductor heterostructures, 2005, J. Supercond. **18**, 411.

Tiusan, C., J. Faure-Vincent, C. Bellouard, M. Hehn, E. Jouguelet, and A. Schuhl, Interfacial resonance state probed by spin-polarized tunneling in epitaxial Fe/MgO/Fe tunnel junctions, 2004, Phys. Rev. Lett. **93**, 106602.

Tiwari, S., 1992, *Compound Semiconductor Device Physics* (Academic Press, San Diego).

Tokuyasu, T. A., J. A. Sauls, and D. Rainer, Proximity effect of a ferromagnetic insulator in contact with a superconductor, 1988, Phys. Rev. B **38**, 8823.

Trebin, H.-R., U. Rössler, and R. Ranvaud, Quantum resonances in the valence bands of zincblende semiconductors. I. Theoretical aspects., 1979, Phys. Rev. B **20**, 686.

Tse, W. K., and S. Das Sarma, Coulomb drag and spin drag in the presence of spin-orbit coupling, 2007, Phys. Rev. B **75**, 045333.

Tse, W.-K., J. Fabian, I. Žutić, and S. Das Sarma, Spin accumulation in the extrinsic spin hall effect, 2005, Phys. Rev.B **72**, 241303(R).

Tsu, R., and L. Esaki, Tunneling in a finite superlattice, 1973, Appl. Phys. Lett. **22**, 562.

Tsui, D. C., R. E. Dietz, and L. R. Walker, Multiple magnon excitation in NiO by electron tunneling, 1971, Phys. Rev. Lett. **27**, 1729.

Tsymbal, E. Y., A. Sokolov, I. F. Sabirianov, and B. Doudin, Resonant inversion of tunneling magnetoresistance, 2003, Phys. Rev. Lett. **90**, 186602.

Tyryshkin, A. M., S. A. Lyon, W. Jantsch, and F. Schäffler, Spin manipulation of free two-dimensional electrons in Si/SiGe quantum wells, 2005, Phys. Rev. Lett. **94**, 126802.

Ue, H., and S. Maekawa, Electron-spin-resonance studies of heavily phosphorus-doped silicon, 1971, Phys. Rev. **3**, 4232.

Ueda, K., H. Tabata, and T. Kawai, Magnetic and electric properties of transition-metal doped ZnO films, 2001, Appl. Phys. Lett. **79**, 988.

Upadhyay, S. K., A. Palanisami, R. N. Louie, and R. A. Buhrman, Probing ferromagnets with Andreev reflection, 1998, Phys. Rev. Lett. **81**, 3247.

Valenzuela, S. O., and M. Tinkham, Direct electronic measurement of the spin Hall effect, 2006, Nature **442**, 176.

Valet, T., and A. Fert, Theory of the perpendicular magnetoresistance in magnetic multilayers, 1993, Phys. Rev. B **48**, 7099.

van der Wiel, W. G., S. De Franceschi, J. M. Elzerman, T. Fujisawa, S. Tarucha, and L. P. Kouwenhoven, Electron transport through double quantum dots, 2003, Rev. Mod. Phys. **75**, 1.

van Dijken, S., X. Jiang, and S. S. P. Parkin, Comparison of magnetocurrent and transfer ratio in magnetic tunnel transistors with spin-valve bases containing Cu and Au spacer layers, 2003a, Appl. Phys. Lett. **82**, 775.

van Dijken, S., X. Jiang, and S. S. P. Parkin, Giant magnetocurrent exceeding 3400% in magnetic tunnel transistors with spin-valve base, 2003b, Appl. Phys. Lett. **83**, 951.

Van Dyck, R. S., P. B. Schwinberg, and H. G. Dehmelt, Electron magnetic moment from geonium spectra: early experiments and background concepts, 1986, Phys. Rev. D **34**, 722.

van Kesteren, H. W., E. C. Cosman, W. A. J. A. van der Poel, and C. T. Foxon, Fine structure of excitons in type-II GaAs/AlAs quantum wells, 1990, Phys. Rev. B **41**, 5283.

van Son, P. C., H. van Kempen, and P. Wyder, Boundary resistance of the ferromagnetic-nonferromagnetic metal interface, 1987, Phys. Rev. Lett. **58**, 2271.

Vandersypen, L. M. K., and I. L. Chuang, NMR techniques for quantum control and computation, 2005, Rev. Mod. Phys. **76**, 1037.

Vas'ko, V. A., K. Nikolaev, V. A. Larkin, P. A. Kraus, and A. M. Goldman, Differential conductance of the ferromagnet/superconductor interface of $DyBa_2Cu_3O_7/La_{2/3}Ba_{1/3}MnO_3$ heterostructures, 1998, Appl. Phys. Lett. **73**, 844.

Vassell, M. O., J. Lee, and H. F. Lockwood, Multibarrier tunneling in $Ga_{1-x}Al_xAs/GaAs$ heterostructures, 1983, J. Appl. Phys. **54**, 5206.

Viña, L., S. Logothetidis, and M. Cardona, Temperature dependence of the dielectric function of germanium, 1984, Phys. Rev. B **30**, 1979.

Vignale, G., and I. D'Amico, Coulomb drag, magnetoresistance, and spin-current injection in magnetic multilayers, 2003, Solid State Commun. **127**, 829.

Villegas-Lelovsky, L., Current spin-polarization in an inhomogeneous semiconductor, 2006a, Appl. Phys. Lett. **89**, 012108.

Villegas-Lelovsky, L., Transient spin dynamics in semiconductors, 2006b, Braz. J. Phys. **36**, 851.

Visani, C., V. Peña, J. Garcia-Barriocanal, D. Arias, Z. Sefrioui, C. Leon, J. Santamaria, N. M. Nemes, M. Garcia-Hernandez, J. L. Martinez, S. G. E. te Velthuis, and A. Hoffmann, Spin-dependent magnetoresistance of ferromagnet/superconductor/ferromagnet $La_{0.7}Ca_{0.3}MnO_3/YBa_2Cu_3O_7/La_{0.7}Ca_{0.3}MnO_3$ trilayers, 2007, Phys. Rev. B **75**, 054501.

Vodopyanov, B. P., Ballistic conductance of a point contact between a d-type superconductor and a ferromagnet, 2005, Zh. Eksp. Teor. Fiz. Pisma Red. **81**, 192, [JETP Lett. **81**, 151 (2005)].

von Ortenberg, M., Spin superlattice with tunable minigap, 1982, Phys. Rev. Lett. **49**, 1041.

Vorojtsov, S., E. R. Mucciolo, and H. U. Baranger, Phonon decoherence of a double quantum dot charge qubit, 2005, Phys. Rev. B **71**, 205322.

Voskoboynikov, A., S. S. Liu, and C. P. Lee, Spin-dependent electronic tunneling at zero magnetic field, 1998, Phys. Rev. B **58**, 15397.

Voskoboynikov, A., S. S. Liu, and C. P. Lee, Spin-dependent tunneling in double-barrier semiconductor heterostructures, 1999, Phys. Rev. B **59**, 12514.

Voskoboynikov, A., S. S. Liu, C. P. Lee, and O. Tretyak, Spin-polarized electronic current in resonant tunneling heterostructures, 2000, J. Appl. Phys. **87**, 387.

Vrijen, R., E. Yablonovitch, K. Wang, H. W. Jiang, A. Balandin, V. Roychowdhury, T. Mor, and D. DiVincenzo, Electron-spin-resonance transistors for quantum computing in silicon-germanium heterostructures, 2000, Phys. Rev. A **62**, 012306.

Vurgaftman, I., and J. R. Meyer, Ferromagnetic resonant interband tunneling diode, 2003a, Appl. Phys. Lett. **82**, 2296.

Vurgaftman, I., and J. R. Meyer, Spin-polarizing properties of the InAs/(AlSb)/GaMnSb//(AlSb)/InAs ferromagnetic resonant interband tunneling diode, 2003b, Phys. Rev. B **67**, 125209.

Vurgaftman, I., J. R. Meyer, and L. R. Ram-Mohan, Band parameters for III-V compound semiconductors and their alloys, 2001, J. Appl. Phys. **89**, 5815.

Wagner, K., D. Neumaier, M. Reinwald, W. Wegscheider, and D. Weiss, Dephasing in (Ga,Mn)As nanowires and rings, 2006, Phys. Rev. Lett. **97**, 56803.

Wang, B., Y. Guo, and B.-L. Gu, Tunneling time of spin-polarized electrons in ferromagnetic/insulator (semiconductor) double junctions under an applied electric field, 2002, J. Appl. Phys. **91**, 1318.

Wang, D., C. Nordman, J. M. Daughton, Z. Qian, and J. Fink, 70% TMR at room temperature for sdt sandwich junctions with CoFeB as free and reference layers, 2004, IEEE Trans. Magn. **40**, 2269.

Wang, J., D. Y. Xing, and H. B. Sun, Enhanced spin injection efficiency in ferromagnet/semiconductor tunnel junctions, 2003, J. Phys.: Condens. Matter **15**, 4841.

Wang, L. G., W. Yang, K. Chang, and K. S. Chan, Spin-dependent tunneling through a symmetric semiconductor barrier: The Dresselhaus effect, 2005, Phys. Rev. B **72**, 153314.

Weber, C. P., N. Gedik, J. E. Moore, J. Orenstein, J. Stephens, and D. D. Awschalom, Observation of spin Coulomb drag in a two-dimensional electron gas, 2005, *Nature* **437**, 1330.

Wei, J. Y. T., N.-C. Yeh, D. F. Garrigus, and M. Strasik, Directional tunneling and Andreev reflection on $YBa_2Cu_3O_{7-\delta}$ single crystals: Predominance of $d$-wave pairing symmetry verified with the generalized Blonder, Tinkham, and Klapwijk theory, 1998, Phys. Rev. Lett. **81**, 2542.

Weil, T., and B. Vinter, Equivalence between resonant tunneling and sequential tunneling in double-barrier diodes, 1987, Appl. Phys. Lett. **50**, 1281.

Weng, M. Q., and M. W. Wu, Spin dephasing in n-type gaas quantum wells, 2003, Phys. Rev. B **68**, 075312.

Weng, M. Q., M. W. Wu, and L. Jiang, Hot-electron effect in spin dephasing in n-type gaas quantum wells, 2004, Phys. Rev. B **69**, 245320.

Wetzels, W., G. E. W. Bauer, and M. Grifoni, Noncollinear single-electron spin-valve transistors, 2005, Phys. Rev. B **72**, 020407(R).

Wilamowski, Z., and W. Jantsch, ESR studies of the Bychkov-Rashba field in modulation doped Si/SiGe quantum wells, 2002, Physica E **12**, 439.

Wilamowski, Z., W. Jantsch, H. Malissa, and U. Rössler, Evidence and evaluation of the Bychkov-Rashba effect in SiGe/Si/SiGe quantum wells, 2002, Phys. Rev. B **66**, 195315.

Wilczyński, M., J. Barnaś, and R. Świrkowicz, Electron tunneling in planar double junctions with ferromagnetic barriers, 2003, J. Magn. Magn. Mater. **267**, 391.

Winkler, R., Rashba spin splitting in two-dimensional electron and hole systems, 2000, Phys. Rev. B **62**, 4245.

Winkler, R., 2003, *Spin-orbit coupling effects in two-dimensional electron and hole systems* (Springer, Berlin).

Winkler, R., Rashba spin splitting and Ehrenfest's theorem, 2004a, Physica E **22**, 450.

Winkler, R., Spin orientation and spin precession in inversion-asymmetric quasi-two-dimensional electron systems, 2004b, Phys. Rev. B **69**, 045317.

Witzel, W. M., R. de Sousa, and S. Das Sarma, Quantum theory of spectral-diffusion-induced electron spin decoherence, 2005, Phys. Rev. B **72**, 161306(R).

Wolf, S. A., D. D. Awschalom, R. A. Buhrman, J. M. Daughton, S. von Molnár, M. L. Roukes, A. Y. Chtchelkanova, and D. M. Treger, Spintronics: A spin-based electronics vision for the future, 2001, *Science* **294**, 1488.

Woods, L. M., T. L. Reinecke, and Y. Lyanda-Geller, Spin relaxation in quantum dots, 2002, Phys. Rev. B **66**, 161318(R).

Wrobel, J., T. Dietl, K. Fronc, A. Lusakowski, M. Czeczott, G. Grabecki, R. Hey, and K. H. Ploog, 2d and 1d electron transport in hybrid ferromagnet-semiconductor microstructures, 2001, Physica E **10**, 91.

Wrobel, J., T. Dietl, A. Lusakowski, G. Grabecki, K. Fronc, R. Hey, K. H. Ploog, and H. Shtrikman, Spin filtering in a hybrid ferromagnetic-semiconductor microstructure, 2004, Phys. Rev. Lett. **93**, 246601.

Wu, B. H., and K. H. Ahn, Proposal for an electrical spin cell with single barrier, 2006, Appl. Phys. Lett. **89**, 012108.

Wu, H., K. Chang, J. Xia, and F. M. Peeters, Resonant tunneling of holes in GaMnAs-related double-barrier structures, 2003a, J. Supercond. **16**, 279.

Wu, H.-C., Y. Guo, X.-Y. Chen, and B.-L. Gu, Rashba spin-orbit effect on traversal time in ferromagnetic/semiconductor/ ferromagnetic heterojunction, 2003b, J. Appl. Phys. **93**, 5316.

Wu, M. W., and C. Z. Ning, A novel mechanism for spin dephasing due to spin-conserving scatterings, 2002, Eur. Phys. J. B **18**, 373.

Wunderlich, J., B. Kaestner, J. Sinova, and T. Jungwirth, Experimental observation of the spin-hall effect in a two-dimensional spin-orbit coupled semiconductor system, 2005, Phys. Rev. Lett. **94**, 047204.

Wunnicke, O., P. Mavropoulos, R. Zeller, P. H. Dederichs, and D. Grundler, Ballistic spin injection from Fe(001) into ZnSe and GaAs, 2002, Phys. Rev. B **65**, 241306(R).

Xia, K., P. J. Kelly, G. E. W. Bauer, and I. Turek, Spin-dependent transparency of ferromagnet/superconductor interfaces, 2002, Phys. Rev. Lett. **89**, 166603.

Xiang, X. H., T. Zhu, J. Du, G. Landry, and J. Q. Xiao, Effects of density of states on bias dependence in magnetic tunnel junctions, 2002, Phys. Rev. B **66**, 174407.

Xu, P. X., V. M. Karpan, K. Xia, M. Zwierzycki, I. Marushchenko, and P. J. Kelly, Spin-filter tunneling magnetoresistance in a magnetic tunnel junction, 2006a, Phys. Rev. B **73**, 180402(R).

Xu, X. H., H. J. Blythe, M. Ziese, A. J. Behan, J. R. Neal, A. Mokhtari, R. M. Ibrahim, A. M. Fox, and G. A. Gehring, Carrier-induced ferromagnetism in n-type ZnMnAlO and ZnCoAlO thin films at room temperature, 2006b, New J. Phys. **8**, 135.

Yafet, Y., g factors and spin-lattice relaxation of conduction electrons, 1963, in *Solid State Physics, Vol. 14*, edited by F. Seitz and D. Turnbull (Academic, New York), p. 2.

Yamada, S., T. Kikutani, S. Gozu, Y. Sato, and T. Kita, Spontaneous spin-splitting observed in resonant tunneling diode with narrow band-gap asymmetric quantum well, 2002, Physica E **13**, 815.

Yamashita, T., K. Tanikawa, S. Takahashi, and S. Maekawa, Superconducting qubit with a ferromagnetic Josephson junction, 2005, Phys. Rev. Lett. **95**, 097001.

Yang, W., and K. Chang, Spin relaxation in diluted magnetic semiconductor quantum dots, 2005, Phys. Rev. B **72**, 75303.

Young, C. F., E. H. Poindexter, G. J. Gerardi, W. L. Warren, and D. J. Keeble, Electron paramagnetic resonance of conduction-band electrons in silicon, 1997, Phys. Rev. B **55**, 16245.

Yu, L., and O. Voskoboynikov, Time-resolved spin filtering in semiconductor symmetric resonant barrier structures, 2005, J. Appl. Phys. **98**, 23716.

Yu, P. Y., and M. Cardona, 2001, *Fundamentals of Semiconductors. 3rd Ed.* (Springer, Berlin).

Yuasa, S., A. Fukushima, H. Kubota, Y. Suzuki, and K. Ando, Giant tunneling magnetoresistance up to 410% at room temperature in fully epitaxial Co/MgO/Co magnetic tunnel junctions with bccc C(001) electrodes, 2006, Appl. Phys. Lett. **89**, 042505.

Yuasa, S., A. Fukushima, T. Nagahama, K. Ando, and Y. Suzuki, High tunnel magnetoresistance at room temperature in fully epitaxial Fe/MgO/Fe tunnel junctions due to coherent spin-polarized tunneling, 2004a, Jpn. J. Appl. Phys. **43**, L588.

Yuasa, S., T. Nagahama, A. Fukushima, Y. Suzuki, and K. Ando, Giant room-temperature magnetoresistance in single-crystal Fe/MgO/Fe magnetic tunnel junctions, 2004b, Nat. Mater. **3**, 868.

Yuasa, S., T. Nagahama, and Y. Suzuki, Spin-polarized resonant tunneling in magnetic tunnel junctions, 2002, Science **297**, 234.

Zaitsev, O., D. Frustaglia, and K. Richter, The role of orbital dynamics in spin relaxation and weak antilocalization in quantum dots, 2005, Phys. Rev. Lett. **94**, 026809.

Zareyan, M., W. Belzig, and Y. V. Nazarov, Superconducting proximity effect in clean ferromagnetic layers, 2002, Phys. Rev. B **65**, 184505.

Zarifis, V., and T. G. Castner, ESR linewidth behavior for barely metallic n-type silicon, 1987, Phys. Rev. B **36**, 6198.

Zarifis, V., and T. G. Castner, Observation of the conduction-electron spin resonance from metallic antimony-doped silicon, 1998, Phys. Rev. B **57**, 14600.

Zaslavsky, A., V. J. Goldman, D. C. Tsui, and J. E. Cunningham, Resonant tunneling and intrinsic bistability in asymmetric double-barrier heterostructures, 1988, Appl. Phys. Lett. **53**, 1408.

Zawadzki, W., and P. Pfeffer, Average forces in bound and resonant quantum states, 2001, Phys. Rev. B **64**, 235313.

Zawadzki, W., and P. Pfeffer, Spin splitting of subband energies due to inversion asymmetry in semiconductor heterostructures, 2004, Semicond. Sci. Technol. **19**, R1.

Zega, T. J., A. T. Hanbicki, S. C. Erwin, I. Žutić, G. Kioseoglou, C. H. Li, B. T. Jonker, and R. M. Stroud, Determination of interface atomic structure and its impact on spin transport using Z-contrast microscopy and density-functional theory, 2006, Phys. Rev. Lett. **96**, 196101.

Zeng, Z. Y., B. Li, and F. Claro, Non-equilibrium Green's-function approach to electronic transport in hybrid mesoscopic structures, 2003, Phys. Rev. B **68**, 115319.

Zenger, M., J. Moser, W. Wegscheider, D. Weiss, and T. Dietl, High-field magnetoresistance of Fe/GaAs/Fe tunnel junctions, 2004, J. Appl. Phys. **96**, 2400.

Zhai, F., Y. Guo, and B.-L. Gu, Effects of conduction band offset on spin-polarized transport through a semimagnetic semiconductor heterostructure, 2001, J. Appl. Phys. **90**, 1328.

Zhai, F., Y. Guo, and B.-L. Gu, Current and spin-filtering dual diodes based on diluted magnetic semiconductor heterostructures with a nonmagnetic barrier, 2003, J. Appl. Phys. **94**, 5432.

Zhang, J., and R. M. White, Voltage dependence of magnetoresistance in spin dependent tunneling junctions, 1998, J. Appl. Phys. **83**, 6512.

Zhang, S., and P. M. Levy, Magnetoresistance of magnetic tunnel juctions in the presence of a nonmagnetic layer, 1998, Phys. Rev. Lett. **81**, 5660.

Zhang, S., P. M. Levy, A. C. Marley, and S. S. P. Parkin, Quenching of magnetoresistance by hot electrons in magnetic tunnel junction, 1997, Phys. Rev. Lett. **79**, 3744.

Zhao, E. H., and J. A. Sauls, Dynamics of spin transport in voltage-biased Josephson junctions, 2007, Phys. Rev. Lett. **98**, 206601.

Zhao, J. H., F. Matsukara, E. Abe, D. Chiba, Y. Ohno, K. Takamura, and H. Ohno, Growth and properties of (Ga,Mn)As on Si (100) substrate, 2002, J. Cryst. Growth **237-239**, 1349.

Zhu, J.-X., B. Friedman, and C. S. Ting, Spin-polarized quasiparticle transport in ferromagnet-d-wave-superconductor junctions with a 110 interface, 1999, Phys. Rev. B **59**, 9558.

Zhu, Y., Q.-F. Sun, and T.-H. Lin, Andreev reflection through a quantum dot coupled with two ferromagnets and a superconductor, 2001, Phys. Rev. B **65**, 024516.

Zhu, Z.-G., and G. Su, Magnitude of magnetic field dependence of a possible spin selective spin filter in ZnSe/$Zn_{1-x}Mn_xSe$ multilayer heterostructure, 2004, Phys. Rev. B **70**, 193310.

Žikić, R., and L. Dobrosavljević-Grujić, Superharmonic Josephson relations in unconventional superconductor junctions wiht a ferromagnetic barrier, 2007, Phys. Rev. B **75**, 100502(R).

Žutić, I., Novel aspects of spin-polarized transport and spin dynamics, 2002, J. Supercond. **15**, 5.

Žutić, I., Gadolinium makes good spin contacts, 2006, Nature Mater. **5**, 771.

Žutić, I., and S. Das Sarma, Spin-polarized transport and Andreev reflection in semiconductor/superconductor hybrid structures, 1999, Phys. Rev. B **60**, R16322.

Žutić, I., and J. Fabian, Spin-voltaic effect and its implications, 2003, Mater. Trans., JIM **44**, 2062.

Žutić, I., and J. Fabian, Silicon twists, 2007, Nature **447**, 269.

Žutić, I., J. Fabian, and S. Das Sarma, A proposal for a spin-polarized solar battery, 2001a, Appl. Phys. Lett. **79**, 1558.

Žutić, I., J. Fabian, and S. Das Sarma, Spin injection through the depletion layer: a theory of spin-polarized p-n junctions and solar cells, 2001b, Phys. Rev. B **64**, 121201(R).

Žutić, I., J. Fabian, and S. Das Sarma, Spin-polarized transport in inhomogeneous magnetic semiconductors: theory of magnetic/nonmagnetic p-n junctions, 2002, Phys. Rev. Lett. **88**, 066603.

Žutić, I., J. Fabian, and S. Das Sarma, Proposal for all-electrical measurement of $T_1$ in semiconductors, 2003, Appl. Phys. Lett. **82**, 221.

Žutić, I., J. Fabian, and S. Das Sarma, Spintronics: Fundamentals and applications, 2004, Rev. Mod. Phys. **76**, 323.

Žutić, I., J. Fabian, and S. C. Erwin, Bipolar spintronics: Fundamentals and applications, 2006a, IBM. J. Res. & Dev. **50**, 121.

Žutić, I., J. Fabian, and S. C. Erwin, Spin injection and detection in silicon, 2006b, Phys. Rev. Lett. **97**, 026602.

Žutić, I., J. Fabian, and S. C. Erwin, Bipolar spintronics: from spin injection to spin-controlled logic, 2007, J. Phys.: Condens. Matter **19**, 165219.

Žutić, I., and I. Mazin, Phase-sensitive tests of the pairing state symmetry in $Sr_2RuO_4$, 2005, Phys. Rev. Lett. **95**, 217004.

Žutić, I., and O. T. Valls, Spin polarized tunneling in ferromagnet/unconventional superconductor junctions, 1999, Phys. Rev. B **60**, 6320.

Žutić, I., and O. T. Valls, Tunneling spectroscopy for ferromagnet/superconductor junctions, 2000, Phys. Rev. B **61**, 1555.